\RequirePackage[l2tabu]{nag}		
\documentclass[a4paper,11pt,twoside,openright]{memoir} 

\usepackage{tikz}
\usepackage{algorithm}
\usepackage{algorithmic}
\usepackage[utf8]{inputenc} 
\usepackage[T1]{fontenc}

\usepackage{datetime}
\usepackage{ifpdf}

\usepackage{bm}
\usepackage{bbm}
\usepackage{sidecap}
\usepackage{enumitem}
\usepackage{scrextend}

\def\1{\bm{1}}

\def\vtheta{{\bm{\theta}}}

\def\vg{{\bm{g}}}

\def\vv{{\bm{v}}}
\def\vw{{\bm{w}}}
\def\vx{{\bm{x}}}
\def\vy{{\bm{y}}}

\def\vphi{{\bm{\phi}}}

\def\mB{{\bm{B}}}
\def\mC{{\bm{C}}}

\def\mE{{\bm{E}}}

\def\mH{{\bm{H}}}
\def\mI{{\bm{I}}}

\def\mX{{\bm{X}}}
\def\mY{{\bm{Y}}}

\def\mLambda{{\bm{\Lambda}}}

\DeclareMathAlphabet{\mathsfit}{\encodingdefault}{\sfdefault}{m}{sl}
\SetMathAlphabet{\mathsfit}{bold}{\encodingdefault}{\sfdefault}{bx}{n}

\newcommand{\E}{\mathbb{E}}

\newcommand{\R}{\mathbb{R}}

\newcommand{\Var}{\mathrm{Var}}

\newcommand{\Cov}{\mathrm{Cov}}

\usetikzlibrary{decorations.pathmorphing}
\usetikzlibrary{shapes.geometric, arrows}
\usetikzlibrary{calc}
\tikzset{
  branch cut/.style={
    decorate,decoration=snake,
    to path={
      (\tikztostart) -- (\tikztotarget) \tikztonodes
    },
  }
}
\tikzstyle{io} = [rectangle,  minimum width=3cm, minimum height=1cm, rounded corners, text centered, draw=black, fill=blue!30]

\tikzstyle{process} = [rectangle, minimum width=3cm, minimum height=1cm, text centered, draw=black, fill=orange!30]
\tikzstyle{arrow} = [thick,->,>=stealth]
\usepackage{csquotes}
\usepackage[section]{placeins}

\ifpdf
\fi
\ifdraftdoc 
	\usepackage{draftwatermark}				
	\SetWatermarkScale{0.3}
	\SetWatermarkText{\textbf Draft: \today}
\fi
\settrimmedsize{297mm}{210mm}{*}
\settypeblocksize{634pt}{448.13pt}{*} 
\setulmargins{4cm}{*}{*} 
\setlrmargins{*}{*}{1.5} 
\setmarginnotes{17pt}{51pt}{\onelineskip} 
\setheadfoot{\onelineskip}{2\onelineskip} 
\setheaderspaces{*}{2\onelineskip}{*} 
\checkandfixthelayout 
\usepackage[p]{baskervillef}
\usepackage[T1]{fontenc}

\OnehalfSpacing 
\setsecnumdepth{subsection} 
\maxsecnumdepth{subsubsection}
\usepackage{calc,soul,fourier}
\makeatletter 
\newlength\dlf@normtxtw 
\newsavebox{\feline@chapter} 
\newcommand\feline@chapter@marker[1][4cm]{%
	\sbox\feline@chapter{%
		\resizebox{!}{#1}{\fboxsep=1pt%
			\colorbox{gray}{\color{white}\thechapter}%
		}}%
		\rotatebox{90}{%
			\resizebox{%
				\heightof{\usebox{\feline@chapter}}+\depthof{\usebox{\feline@chapter}}}%
			{!}{\scshape\so\@chapapp}}\quad%
		\raisebox{\depthof{\usebox{\feline@chapter}}}{\usebox{\feline@chapter}}%
} 
\newcommand\feline@chm[1][4cm]{%
	\sbox\feline@chapter{\feline@chapter@marker[#1]}%
	\makebox[0pt][c]{
		\makebox[1cm][r]{\usebox\feline@chapter}%
	}}
\makechapterstyle{daleifmodif}{

	\renewcommand\printchapternum{\null\hfill\feline@chm[2.5cm]\par}

} 
\makeatother 
\chapterstyle{daleifmodif}
\makepagestyle{myvf} 
\makeoddfoot{myvf}{}{\thepage}{} 
\makeevenfoot{myvf}{}{\thepage}{} 
\makeheadrule{myvf}{\textwidth}{\normalrulethickness} 
\makeevenhead{myvf}{\small\textsc{\leftmark}}{}{} 
\makeoddhead{myvf}{}{}{\small\textsc{\rightmark}}
\newcommand{\clearemptydoublepage}{\newpage{\thispagestyle{empty}\cleardoublepage}}
\makeindex
\usepackage{import}
\usepackage{subcaption}
\usepackage{lipsum}					
\usepackage{amsfonts} 					
\usepackage{amsmath}			
\usepackage{amssymb}					
\usepackage{amsthm}		
\usepackage{thmtools, thm-restate}
\usepackage{layouts}					
\usepackage{graphicx}					
\usepackage{longtable,rotating}			

\usepackage{pdfpages}				
\usepackage{colortbl}					
\usepackage{mathrsfs}					
\usepackage{float}						
\usepackage{verbatim}					

\usepackage{url}						
\usepackage[backend=biber,style=alphabetic,texencoding=utf8]{biblatex}  

\usepackage{color}                    				

\usepackage[colorlinks=true,
		     allcolors=black]{hyperref}              
\usepackage{memhfixc}					
\usepackage{enumerate}					
\usepackage{footnote}					
\usepackage{microtype}					
\usepackage{rotfloat}					
\usepackage{mathtools}								
\usepackage{alltt}						

\newdateformat{monthyeardate}{%
  \monthname[\THEMONTH], \THEYEAR}
         
\usepackage{pgf}			
\usepackage{tikz}						
\usetikzlibrary{arrows,shapes,snakes,
		       automata,backgrounds,
		       petri,topaths}				
\newcommand{\pgftextcircled}[1]{                                                                    
    \setbox0=\hbox{#1}%
    \dimen0\wd0%
    \divide\dimen0 by 2%
    \begin{tikzpicture}[baseline=(a.base)]%
        \useasboundingbox (-\the\dimen0,0pt) rectangle (\the\dimen0,1pt);
        \node[circle,draw,outer sep=0pt,inner sep=0.1ex] (a) {#1};
    \end{tikzpicture}
}

\newcounter{proofcount}
\renewenvironment{proof}[1][\proofname.]{\par
 \ifnum \theproofcount>0 \pushQED{\whiteged} \else \pushQED{\blackged} \fi%
 \refstepcounter{proofcount}
 \normalfont 
 \trivlist
 \item[\hskip\labelsep
       \itshape
   {\textbf\em #1}]\ignorespaces
}{%
 \addtocounter{proofcount}{-1}
 \popQED\endtrivlist
}
\let\oldsqrt\sqrt
\def\sqrt{\mathpalette\DHLhksqrt}
\def\DHLhksqrt#1#2{%
\setbox0=\hbox{$#1\oldsqrt{#2\,}$}\dimen0=\ht0
\advance\dimen0-0.2\ht0
\setbox2=\hbox{\vrule height\ht0 depth -\dimen0}%
{\box0\lower0.4pt\box2}}
\usepackage{lettrine}
\newcommand{\initial}[1]{%
	\lettrine[lines=3,lhang=0.33,nindent=0em]{
		\color{gray}
     		{\textsc{#1}}}{}}
\newcommand{\jstat}[1]{#1}
\theoremstyle{plain}
\newtheorem{theorem}{Theorem}[chapter]
\theoremstyle{plain}
\newtheorem{prop}{Proposition}[chapter]
\theoremstyle{plain}
\theoremstyle{definition}
\newtheorem{defn}{Definition}[chapter]
\theoremstyle{plain}
\newtheorem{lemma}{Lemma}[chapter]
\theoremstyle{plain}
\newtheorem{corollary}{Corollary}[chapter]
\theoremstyle{plain}

\theoremstyle{remark}
\newtheorem{exmp}{Example}[chapter]
\theoremstyle{remark}
\newtheorem{remark}{Remark}[chapter]
\newtheorem{assump}[theorem]{Assumption}

\newcommand{\defeq}{\vcentcolon=}
\newcommand{\expect}{\mathbb{E}}
\newcommand{\prob}{\mathbb{P}}
\renewcommand{\vec}[1]{\bm{#1}}
\newcommand{\indic}{\mathbbm{1}}

\newcommand{\expectGOE}{\expect_{GOE}^{N}}
\newcommand{\Tr}{\text{Tr}}
\newcommand{\wD}{w^{(D)}}
\newcommand{\vwD}{\vec{w}^{(D)}}
\newcommand{\wG}{w^{(G)}}
\newcommand{\vwG}{\vec{w}^{(G)}}
\newcommand{\lD}{\ell^{(D)}}
\newcommand{\lG}{\ell^{(G)}}
\newcommand{\LD}{L^{(D)}}
\newcommand{\LG}{L^{(G)}}

\renewcommand{\P}{\mathbb{P}}
\newcommand{\C}{\mathbb{C}}
\newcommand{\N}{\mathbb{N}}

\newcommand{\pD}{\partial^{(D)}}
\newcommand{\pG}{\partial^{(G)}}

\newcommand{\datax}{\mathcal{X}}
\newcommand{\CNk}{C_{N,k}}
\newcommand{\CN}{C_{N}}

\newcommand{\ind}[1]{i_{\leq {#1}}}
\newcommand{\grad}{\nabla}

\newcommand{\trg}{\text{trg}}

\newcommand{\reff}{\mathcal{R}_\text{est-curv}}

\newcommand{\Pdata}{\mathbb{P}_{\text{data}}}

\DeclareMathOperator*{\argmin}{arg\,min}

\renewcommand{\eqref}[1]{(\ref{#1})}
\newcommand{\blackged}{\hfill$\blacksquare$}
\newcommand{\whiteged}{\hfill$\square$}
\newcommand{\vepsilon}{\vec{\epsilon}}

\renewcommand{\j}{\mathfrak{j}}

\newcommand{\B}{\mathscr{B}}

\newcommand{\bX}{\bar{X}}
\newcommand{\TN}{\mathbb{T}_N}
\newcommand{\Vs}{\mathbb{V}_s}
\newcommand{\hQUE}{\widehat{\text{QUE}}}
\newcommand{\sbplain}{\textsf{s}}
\newcommand{\sbf}{\sbplain(b)}
\newcommand{\rmes}{\texttt{r}}
\newcommand{\lmes}{\texttt{l}}
\newcommand{\jmlrreview}[1]{#1}
\newcommand{\barlam}{\bar{\lambda}^{(i, e)}_b}
\newcommand{\barlamnoe}{\bar{\lambda}^{(i)}_b}
\newcommand{\lam}{\lambda^{(i, j, e)}_b}

\newcommand{\vivacom}[1]{\textcolor{black}{#1} }
\DeclareMathOperator{\supp}{supp}

\renewcommand{\epsilon}{\varepsilon}

\renewcommand{\leq}{\leqslant}
\renewcommand{\geq}{\geqslant}

\renewcommand{\supp}{\text{supp}}

\newcommand{\review}[1]{#1}

\def\Xint#1{\mathchoice
{\XXint\displaystyle\textstyle{#1}}%
{\XXint\textstyle\scriptstyle{#1}}%
{\XXint\scriptstyle\scriptscriptstyle{#1}}%
{\XXint\scriptscriptstyle\scriptscriptstyle{#1}}%
\!\int}
\def\XXint#1#2#3{{\setbox0=\hbox{$#1{#2#3}{\int}$}
\vcenter{\hbox{$#2#3$}}\kern-.5\wd0}}

\def\dashint{\Xint-}

\def\mB{{\bm{B}}}
\def\mC{{\bm{C}}}

\def\mE{{\bm{E}}}

\def\mH{{\bm{H}}}
\def\mI{{\bm{I}}}

\def\mX{{\bm{X}}}
\def\mY{{\bm{Y}}}

\def\vg{{\bm{g}}}

\def\vv{{\bm{v}}}
\def\vw{{\bm{w}}}
\def\vx{{\bm{x}}}
\def\vy{{\bm{y}}}

\DeclareMathAlphabet{\mathcal}{OMS}{cmsy}{m}{n}
\begin{document}

%
%
%
%
%
\frontmatter
\pagenumbering{roman}
%
%
%
%
%
%
\begin{titlingpage}
\begin{SingleSpace}
\calccentering{\unitlength} 
\begin{adjustwidth*}{\unitlength}{-\unitlength}
\vspace*{-18mm}
\begin{center}
{\LARGE \textsc{University of Bristol}}
\vspace*{35mm}

\rule[0.5ex]{\linewidth}{1pt}\\[\baselineskip]
{\HUGE Random matrix theory and the loss surfaces of neural networks}\\[4mm]
\rule[0.5ex]{\linewidth}{1pt}

\vspace*{26.5mm}
{\LARGE\textsc{Nicholas P. Baskerville}}\\
{\normalsize{MMath MA (Cantab), CMath MIMA}}\\
\vspace{17mm}
\includegraphics[scale=1.3]{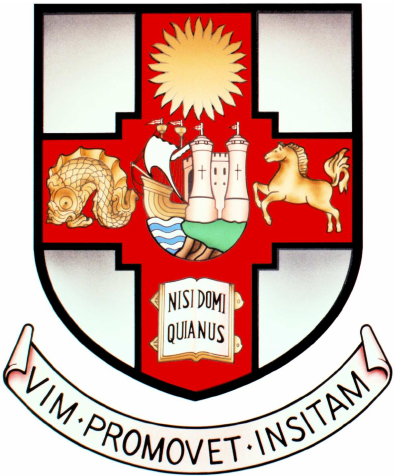}\\
\vspace{17mm}

{\monthyeardate\today}

\vspace{11mm}

\begin{minipage}{15cm}
A dissertation submitted to the University of Bristol in accordance with the requirements for award of the degree of Doctor of Philosophy in the Faculty of Science, School of Mathematics.
\end{minipage}\\
\end{center}
\begin{flushright}
\end{flushright}
\end{adjustwidth*}
\end{SingleSpace}
\end{titlingpage}
\clearemptydoublepage

\chapter*{Abstract}
\begin{SingleSpace}
\initial{N}eural network models are one of the most successful approaches to machine learning, enjoying an enormous amount of development and research over recent years and finding concrete real-world applications in almost any conceivable area of science, engineering and modern life in general.
The theoretical understanding of neural networks trails significantly behind their practical success and the engineering heuristics that have grown up around them.
Random matrix theory provides a rich framework of tools with which aspects of neural network phenomenology can be explored theoretically.
In this thesis, we establish significant extensions of prior work using random matrix theory to understand and describe the loss surfaces of large neural networks, particularly generalising to different architectures.
Informed by the historical applications of random matrix theory in physics and elsewhere, we establish the presence of local random matrix universality in real neural networks and then utilise this as a modeling assumption to derive powerful and novel results about the Hessians of neural network loss surfaces and their spectra.
In addition to these major contributions, we make use of random matrix models for neural network loss surfaces to shed light on modern neural network training approaches and even to derive a novel and effective variant of a popular optimisation algorithm.

\medskip

Overall, this thesis provides important contributions to cement the place of random matrix theory in the theoretical study of modern neural networks, reveals some of the limits of existing approaches and begins the study of an entirely new role for random matrix theory in the theory of deep learning with important experimental discoveries and novel theoretical results based on local random matrix universality.
\end{SingleSpace}
\clearpage
\clearemptydoublepage
\chapter*{Dedication and acknowledgements}
\begin{SingleSpace}
    \initial{I} dedicate this thesis to Charlotte, without whose love, encouragement and support it would not have been possible.
    Doing a PhD essentially as a hobby was a bit deranged and I had no right to expect you to put up with the silly hours nor my doing maths on the beach or other such nonsense, but you always did.
    As romantic gestures go, 200-odd pages of maths is a poor substitute for a symphony, but for what it's worth, it is yours.
    
    \medskip
    My parents have never failed to support my endeavours, academic or otherwise, and have ever been a source of steadfast encouragement, for which I am truly grateful. This thesis is really the conclusion of something we began together in the Cambridge branch of \emph{Eat} over ten years ago.
    
    \medskip
    I thank Jon Keating, Francesco Mezzadri and Joseph Najnudel for the years of enjoyable collaboration and for taking the punt on me and my project in the first place. Their advice and ideas were integral to the success of this research. They provided just the right amount of guidance to keep projects moving, while allowing me to develop my own ideas and approaches.
    
    \medskip
    I thank Diego Granziol for several years of enjoyable collaboration. Much of the second half of this thesis was influenced by our discussions and is, I am sure, much the better for it.
    
     \medskip
    Thanks are due to the RMT group in Bristol for making me feel part of things, despite only being in the office once a week.
    I gratefully acknowledge financial support for tuition fees from the School of Mathematics in the University of Bristol.
    
    \medskip
    Finally, I thank my former team at GCHQ and particularly Tom L. His initial support and practical help were instrumental in getting my PhD off the ground and I shall always be grateful for the three years during which he gladly encouraged me as I juggled PhD work and our day-jobs. I also gratefully acknowledge the three years of funding for tuition fees and travel expenses that came from AR/RIF/STR and the DRE scholarship scheme.

\end{SingleSpace}
\clearpage
\clearemptydoublepage
\clearemptydoublepage

\chapter*{Author's declaration}
\begin{SingleSpace}
\begin{quote}
\initial{I} declare that the work in this dissertation was carried out in accordance with the requirements of  the University's Regulations and Code of Practice for Research Degree Programmes and that it  has not been submitted for any other academic award. Except where indicated by specific  reference in the text, the work is the candidate's own work. Work done in collaboration with, or with the assistance of, others, is indicated as such. Any views expressed in the dissertation are those of the author.
\end{quote}
\vspace{0.5cm}
\begin{flushright}
Nicholas P. Baskerville\\
\today, Bristol
\end{flushright}

\end{SingleSpace}
\clearpage

\clearemptydoublepage
\renewcommand{\contentsname}{Table of Contents}
\maxtocdepth{subsection}
\tableofcontents*
\addtocontents{toc}{\par\nobreak \mbox{}\hfill{\textbf Page}\par\nobreak}
\newpage
\listoftables*
\newpage
\listoffigures*

\chapter*{Abbreviations}
The following abbreviations are used throughout this thesis.

\begingroup
\setlength{\tabcolsep}{10pt} 
\renewcommand{\arraystretch}{1.5} 
\begin{center}
\begin{tabular}{ll}
     NN & neural network\\
     ANN & artificial neural network\\
     DNN &  deep neural network\\
     SGD & stochastic gradient descent\\
     MLP & multi-layer perceptron\\
     CNN & convolutional neural network \\
     RNN & recurrent neural network \\
     GAN & generative adversarial network\\
     RMT & random matrix theory\\
     GOE & Gaussian orthogonal ensemble\\
     LSD & limiting spectral density\\
     ESD & empirical spectral density\\
     NNSD & nearest neighbour spacing distribution\\
     WLOG & without loss of generality\\
     a.s. & almost surely
\end{tabular}
\end{center}
\endgroup
\chapter*{Notation}
The following notation will be used consistently throughout this thesis unless stated otherwise.

\begingroup
\setlength{\tabcolsep}{10pt} 
\renewcommand{\arraystretch}{1.5} 
\begin{center}
\begin{tabular}{p{0.2\linewidth}p{0.8\linewidth}}
    $\delta_x$ & A Dirac $\delta$-function centred at the point $x$\\
    $\hat{\mu}_N$ & The empirical spectral measure of an $N\times N$ matrix\\
    $\rho_{SC}$ & The semi-circle density\\
    $g_{\mu}$ & The Stieljtes transform of the measure $\mu$ \\
    $R_{\mu}$ & The $R$-transform of the measure $\mu$ \\
    $\mu\boxplus\nu$ & Additive free convolution of measure $\mu$ and $\nu$\\
    $I_N$ & The $N\times N$ identity matrix \\
    $O(N)$ & The orthogonal group on $N\times N$ matrices\\
    $\Delta(\vec{x})$ & The Vandermonde determinant over $N$ symbols $\{x_1,\ldots, x_N\}$\\
    $\mathcal{N}(\mu, \Sigma)$ & A Gaussian random variable with mean $\mu$ and covariance $\Sigma$\\
    $\Re z$ & The real part of $z\in\C$\\
    $\Im z$ & The imaginary part of $z\in\C$\\
    $i(X)$ & The index of an Hermitian matrix $X$\\
    $\mu_{Haar}$ & The Haar measure on $O(N)$ \\
    $\mathcal{O}(\cdot)$ & Asymptotic ``big-o'' notation. $f(x)=\mathcal{O}(g(x))$ if $\exists$ some constant $c>0$ such that $|f(x)| \leq c|g(x)|$ for all large enough $x$.\\
    $o(\cdot)$ & Asymptotic ``little-o'' notation. $f(x)=o(g(x))$ if $f(x)/g(x)\rightarrow 0$ as $x\rightarrow \infty$.\\
    $f \sim g$ & Asymptotic equivalence. $f\sim g$ if $|f(x)/g(x)| \rightarrow 1$ as $x\rightarrow\infty$.\\
    $[N]$ & The set of integers from $1$ to $N$: $\{1,2,\ldots, N\}$.
\end{tabular}
\end{center}

\endgroup
\clearemptydoublepage

\mainmatter
\chapter{Introduction}\label{chap:intro}
In this chapter we introduce the central objects of study for this thesis, namely deep neural networks and their loss surfaces.
Deep neural networks are an important sub-field of machine learning, so we begin with some introductory material and context for machine learning.
We make no attempts to be exhaustive, but aim to provide a self-contained introduction, accessible for any mathematically literate reader, to the key ideas from machine learning relevant to our investigations.
We will provide a rather more detailed introduction to deep neural networks specifically, again aiming to be accessible to any mathematical reader.
The reader familiar with machine learning and deep neural networks may well safely skip these introductory sections, however they do establish some conventions and points of view, which may be more or less familiar depending on the reader's background.
Following these broad introductory sections, we will sharpen the focus to provide a summary of the prior literature on deep neural network loss surfaces, particular focusing on the mathematical works upon which this thesis is built.
We will also take this opportunity to draw out and summarise the existing connections between deep neural network loss surfaces and random matrix theory, but an introduction to random matrix theory itself is postponed until the next chapter.
We conclude this introductory chapter with a summary of the new results which make up the principal intellectual contribution of this thesis and a literature review of related work.

\section{Machine learning}
Machine learning encompasses to a great variety of areas of study and practical application in computer science, statistics, data science, engineering, economics, genomics etc. See, for example, Chapter 5 of \cite{lecun2015deep} for a high-level summary of many applications.
One could summarise the essential aspects of  machine learning as: \emph{data} and a \emph{model}.
Data could refer to traditional tabular numeric values (e.g. stock market indices or weather readings), natural language, digital imagery, digital voice recordings, internet search engine logs etc.
All of these fields (and many more besides) make use of data of one form or another.
Researchers and practitioners typically wish to use data they have acquired to address questions such as:
\begin{enumerate}
    \item Do these data support a particular hypothesis?
    \item What underlying structure or dynamics are suggested by the observations in these data?
    \item Can one use past data to predict future events?
    \item Can one algorithmically find certain interesting subsets of a dataset?
\end{enumerate}
None of these questions are unique to the field of machine learning.
Indeed, many such questions have been asked by statisticians and physical and biological scientists for centuries.
The lines between machine learning and other, as it were, traditional statistical or mathematical modeling techniques are not entirely clear.
Generally speaking, a machine learning approach to a problem is driven more by the data than any particular model.
Motivated by intuition, prior observations or theoretical work, a physicist would traditionally start by proposing a model for the physical system under consideration and then obtain predictions to be tested theoretically.
The physical model may well contain a number of parameters, such as physical constants, which should be estimated from data, however these parameters are typically few in number and possess meaningful physical interpretations.
The physicist's model is as much a tool for making useful predictions about the world as it is a tool with which the underlying physical reality may be studied.
A physicist may be able to improve the \emph{predictive} power of their model, say, by introducing more parameters that can be tuned to the available data, but doing so would compromise its physical foundations and degrade its \emph{explanatory} power.
To the machine learning practitioner, there is no tension here: data is king and, crudely speaking, a model that better fits and predicts the data is a superior model.

\medskip
The preceding description certainly does not precisely define machine learning and there are doubtless examples of machine learning applications that lie outside of what we have presented, however our focus is exclusively on neural networks which, as we shall see, fall well within the boundary of machine learning as we have presented it.
In the following subsections, we will outline sub-fields within machine learning.
Such is the success of deep neural networks in modern machine learning, they are to be found in use in all of these sub-fields and, in many cases, they are the best available approach.

\subsection{Supervised learning}
A very common problem in machine learning is that of constructing a model from a \emph{labeled dataset}.
Consider a dataset of the form $\{\vec{x}_i, y_i\}_{i=1}^N$, where $\vec{x}_i$ are the data points and the $y_i$ are the \emph{labels}.
The $\vec{x}_i$ may have come from any source and may or may not have a natural numerical representation as column vectors in some $\mathbb{R}^d$, however we assume that a representation of that form has been found.
In some cases, the $\vec{x}_i$ may have genuine geometrical meaning, while in other cases they may simply be numerical values stacked into vectors.
The labels can be categorical, in which case the problem is called \emph{classification}, or continuous, in which case the problem is \emph{regression}.
Here are two specific examples:
\begin{enumerate}
    \item $\vec{x}_i = (\# \text{bedrooms}_i, \# \text{bathrooms}_i, 
    \text{floor area}_i, \text{latitude}_i, \text{longitude}_i)$ and $y_i = \text{market value (\textsterling)}$ for a set of houses in the UK.
    \item $\vec{x}_i = (\text{pixel}_{i1}, \ldots, \text{pixel}_{id})$ and $y_i \in \{ (0,0,1), (0,1,0), (1, 0, 0)\}$ for a set of images categorised into three disjoint classes: cat, dog and rabbit.
\end{enumerate}

\includegraphics[width=\linewidth]{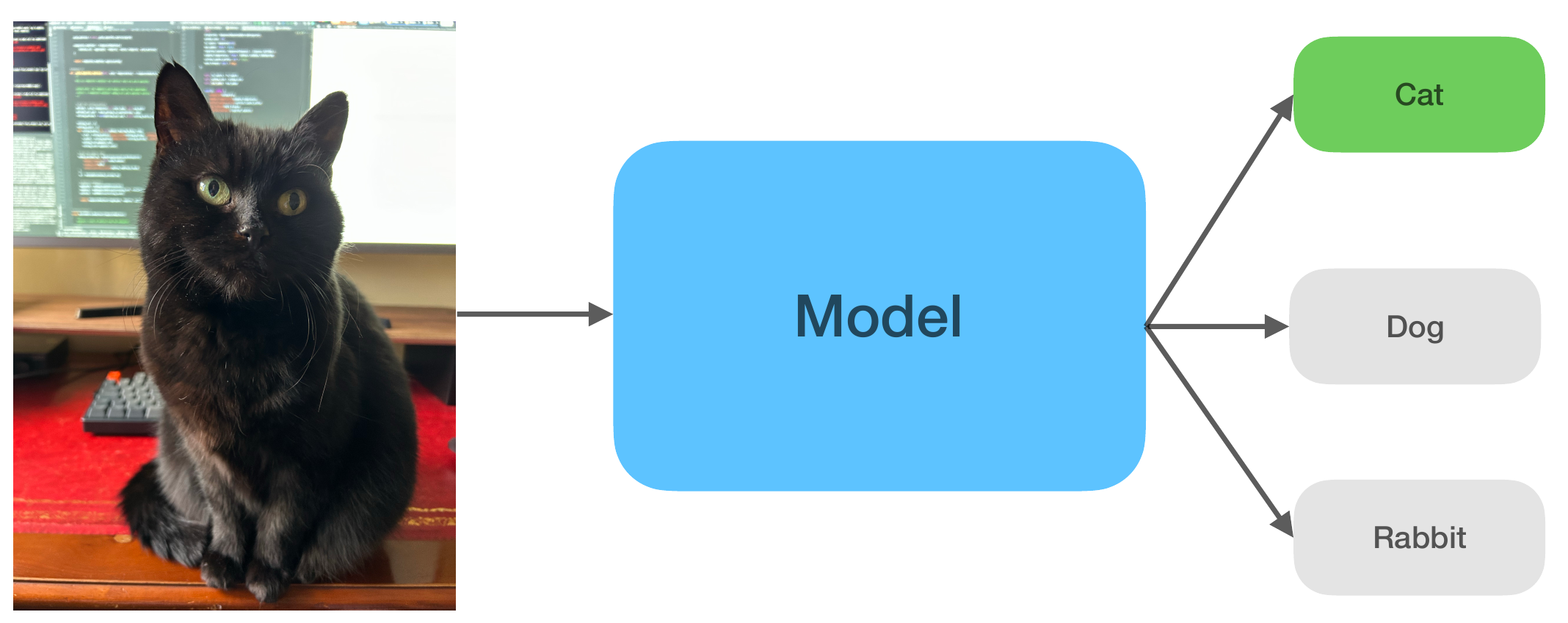}

A model in this context is a function $f$ that is a good approximation to $\vec{x}_i \mapsto y_i$.
$f$ should not simply be a memorisation of the pairs $\{(\vec{x}_i, y_i)\}_{i=1}^N$, since applications typically require $f$ to be useful on other datasets $\{(\hat{\vec{x}}_i, \hat{y}_i)\}_{i=1}^M$ generated from the same underlying distribution as  $\{(\vec{x}_i, y_i)\}_{i=1}^N$, or else to reveal something of the underlying distribution.
$f$ can be deterministic or stochastic and must be computable by some algorithm, preferably quite efficiently, though this is not a universal necessity.
To be more precise, let us introduce a data generating distribution $\Pdata$ supported on $\mathcal{X}\times \mathcal{Y}$, where $\mathcal{X}$ is in the majority of cases some $\mathbb{R}^d$ or a subset thereof.
$\mathcal{Y}$ may be a subset of some $\mathbb{R}^c$ in the regression case, or a countable or even finite set in the classification case.
A single sample $(\vec{x}, y)$ from $\Pdata$ is a single data point and its corresponding label, while a dataset $\mathcal{D}$ is some finite sample from $\Pdata$ (usually taken to be sampled i.i.d.).
Let $\mathcal{D}_{\text{train}}$ and $\mathcal{D}_{\text{test}}$ be two separate finite datasets sampled from $\Pdata$.
Supervised learning consists of using the \emph{training set} $\mathcal{D}_{\text{train}}$ to construct a model $f:\mathcal{X}\rightarrow\mathcal{Y}$ such that $(\vec{x}, f(\vec{x}))$ is close in distribution, in some sense, to $\Pdata$, and practically this is measured using the \emph{test set} $\mathcal{D}_{\text{test}}$.

\medskip
No modern summary of supervised learning would be complete without mentioning \emph{semi-supervised learning}.
Within the context of this thesis, the distinction between supervised and semi-supervised learning is not of much importance; the difference lies in how the labels are obtained.
Standard supervised learning datasets are often constructed by expending human effort to assign labels to data points.
For example, people may be paid to label images with which they are presented as containing some objects of interest.
In some cases, labels can be obtained systematically without any human labeling, for instance in the example of house prices above, the data already exist in some database (though of course human effort was almost certainly required at some point to generate the data and input them to the database).
In semi-supervised learning, labels are derived directly from the data points in some algorithmic manner.
A quite natural example is that of time series, where a model may be constructed to predict, say, the temperature in Bristol tomorrow given the observed temperate today and for every day in the previous week.
Thus the $\mathcal{X}$ is $7$ dimensional (one dimension for each day), and $\mathcal{Y}$ is one dimensional (the temperature tomorrow).
Given a dataset of historical temperatures in Bristol, simply a univariate time series $T_i$ where $i$ indexes the day, one can automatically construct a labelled dataset: $\vec{x}_i = ((T_{i-7},\ldots, T_{i-1}), T_i)$, for all $i$ for which the indices are valid.
Any supervised learning method can then be applied to the resulting labelled data set to produce a model capable of predicting tomorrow's temperature.
Again, from the perspective of this thesis, semi-supervised learning is indistinguishable from supervised learning, so we will not discuss it further.

\subsection{Unsupervised learning}
\emph{Unsupervised learning} considers the case where one only has data points $\vec{x}$ and no labels $y$.
Returning again to the house prices example, given only a dataset of data points $\vec{x}$ containing key parameters about houses, but no labels giving their market value, what can one learn about houses in UK?
For example, one might imagine that using only the key parameters contained in $\vec{x}$ from a large dataset of houses, one could discover useful structure about broad categories of houses.
One common strategy that is particularly relevant in the context of deep learning is \emph{embedding}.
Given only a data set of data points $\{\vec{x}_i\}_{i=1}^N$, an embedding model is some map $f:\mathbb{R}^d \rightarrow \mathbb{R}^e$ where typically $e < d$.
Whatever the meaning or structure of the native data points $\vec{x}_i \in \mathbb{R}^d$, the embedding model $f$ will usually be constructed so that the embeddings $\{f(\vec{x}_i)\}_{i=1}^N$ have some useful geometrical meaning.
The canonical example of embedding models are word embedding models, for example see \cite{mikolov2013efficient, pennington2014glove, bojanowski2017enriching}, where the data sets are just large collections of natural language, and the embedding models aim to represent words in some Euclidean space such that the geometry of Euclidean space has semantic meaning.

\subsection{Generative modelling}
Consider a dataset $\{\vec{x}_i\}_{i=1}^N$ sampled from some underlying distribution $\mathbb{P}$. We wish to construct an approximating distribution $\tilde{\mathbb{P}}$ from which samples can be easily drawn. In this case, $\tilde{\mathbb{P}}$ would be the model. 
A very elementary example of a generative modeling problem would be heights of people in some population, say $x_i = \text{height of person } i$.
In this case, we expect $\mathbb{P}$ to be Gaussian and so $\tilde{\mathbb{P}}$ can be obtained simply by estimating the mean and variance.
We can extend to produce a less trivial example, where the population is a co-educational school.
Rather than fitting a single Gaussian to the whole population, it would clearly be sensible to split into boys and girls and by year groups, and fit a Gaussian to each.
Sampling a height from the population then consists of sampling boy/girl from a Bernoulli random variable, sampling year group from a Categorical random variable, and then sampling the height from a Gaussian.
Clearly, even in the still rather modest example, the problem of appropriately estimating all of the Gaussian means and variances and the Bernoulli and categorical probabilities is much harder than estimating a single Gaussian, but the model is more expressive and will likely better represent the data.
A much more complicated and modern example, is $\vec{x}_i = (\text{pixel}_{i1}, \ldots, \text{pixel}_{in})$ for some set of images of faces.
Constructing an adequate parametric model is likely to be infeasible in this case, with the overwhelmingly most successful modern approach being \emph{generative adversarial models} (GANs) \cite{goodfellow2014generative} (see below).

\subsection{Loss surfaces and the training of machine learning models}
As some of the above examples have already hinted, constructing a machine learning model has two distinct stages: model design and model training.
In the height example above, model design is simply the choice to use a Gaussian distribution and model training is just estimating the mean and variance, e.g. by taking the sample mean and the unbiased estimate of the population variance.
Increasing in complexity, let us consider a linear regression model $f(\vec{x}) = W\vec{x} + \vec{b}$, where the matrix $W$  and the vector $\vec{b}$ contain the parameters of the model.
Here model design is the choice of the form of $f$, namely as a linear map, while model training consists of choosing $W$ and $\vec{b}$ to obtain $f$ that best fits the data out of all possible models of the same linear form.
It happens that the linear regression models, like Gaussian models, are one of the few model types for which optimal parameters can be computed exactly and in closed-form. 

Let us discuss how more general machine learning models are constructed and trained. We will describe the supervised case, for the sake definiteness, but much of what we say applies, \emph{mutatis mutandis}, to unsupervised and generative modeling. Consider again a dataset $\{\vec{x}_i, y_i\}_{i=1}^N$ where $\vec{x}_i\in\mathbb{R}^d, y_i\in\mathbb{R}^c$, for some positive integers $d, c$. Denote again by $\mathbb{P}$ the underlying distribution from which the pairs $(\vec{x}_i, y_i)$ are sampled; all expectations below are taken with respect to $\mathbb{P}$. We fix some \emph{loss function}\footnote{Also known as an \emph{objective function}, or simply `loss' or `objective'.} \begin{align}
    \mathcal{L}\colon \mathbb{R}^c\times \mathbb{R}^c &\rightarrow \mathbb{R}\notag\\
y, \hat{y} &\mapsto \mathcal{L}(y, \hat{y})\notag
\end{align}
which is some typically simple function chosen to measure the performance of a model. Typically there is some quantity of practical interest that one wishes to optimise a model with respect to, for example classification accuracy or mean-squared-error. $\mathcal{L}$ will either be directly the quantity of interest (e.g. mean-squared-error) or will be chosen to correlate with the quantity of interest (e.g. mutual entropy in the case of accuracy). We can now state the central aim of machine learning as an optimisation problem: \begin{align}
    \text{argmin}_{f\in\mathscr{F}} \mathbb{E}\mathcal{L}(y, f(\vec{x}))
\end{align}
where $\mathscr{F}$ is some class of functions. Of course, in any non-trivial case, one does not have access to $\mathbb{P}$ but only the finite sample $\{\vec{x}_i, y_i\}_{i=1}^N$. The training set is used to optimise the function $f$, while the test set is reserved for estimating $\mathbb{E}\mathcal{L}(y, f(\vec{x}))$ so that the quality of the training procedure can be measured. Here are some examples of loss functions:\begin{enumerate}
    \item $L_2$: $\mathcal{L}(y, \hat{y}) = (y - \hat{y})^2$.
    \item $L_1$: $\mathcal{L}(y, \hat{y}) = |y - \hat{y}|$.
    \item Cross-entropy:  $\mathcal{L}(y, \hat{y}) = -\sum_j y_j \log\hat{y}_j$.
\end{enumerate}
The set of functions $\mathscr{F}$ can be defined in a variety of ways, but will always have some set of parameters which are tuned to minimise the training loss \begin{align}
  \sum_i  \mathcal{L}(y_i, f(\vec{x}_i)).
\end{align}
Here are some examples of $\mathscr{F}$:

\begin{enumerate}
    \item Linear regression: $\mathscr{F} = \{f(\vec{x}) = W\vec{x} + \vec{b} ~:~ W\in\mathbb{R}^{c\times d}, \vec{b}\in\mathbb{R}^c\}$. Parameters are $\vec{w}$ (regression coefficients) and $\vec{b}$ (bias). $\mathscr{F}$ is isomorphic to $\mathbb{R}^{dc+c}$ as a vector space.
    \item Gaussian process regression: $\mathscr{F}$ consists of the posteriors given the data $\{(\vec{x}_i y_i)\}_i$ and a prior with mean function $m\colon \mathbb{R}^d\rightarrow\mathbb{R}$ and covariance function $k\colon \mathbb{R}^d\times \mathbb{R}^d\rightarrow \mathbb{R}$. $m$ and $k$ may be simple functions possessing of a small number of \emph{hyper-parameters}. $\mathscr{F}$ is infinite-dimensional, though it is possible to consider the posterior for a fixed data set and then there are simply the prior hyperparameters to tune, giving again a space isomorphic to some $\mathbb{R}^K$.
    \item Neural networks, a full discussion of which is given below in Section \ref{sec:nn_intro}.
\end{enumerate}
Henceforth, we shall consider only finite-dimensional $\mathscr{F}$ isomorphic to some $\mathbb{R}^N$ and we assume a given parametrisation of $\mathscr{F}$ with some vector parameter denoted by $\vec{w}$. For $\vec{w}\in\mathbb{R}^N$, $f_{\vec{w}}\in\mathscr{F}$ denote the member of $\mathscr{F}$ corresponding the vector of parameters $\vec{w}$.

Having defined $\mathscr{F}$ and $\mathcal{L}$, we obtain the notion of the \emph{loss surface} \begin{align}
    \{\mathbb{E} \mathcal{L}(y, f(\vec{x})) ~:~ f\in\mathscr{F}\}.
\end{align}
Finding the global minimum, or some sufficiently good local minimum or saddle point, on the loss surface is a matter of tuning a finite number of parameters.
As mentioned above, there are some special cases for which the globally optimal parameters can be computed in closed-form.
For linear regression with $L_2$ loss, one can straightforwardly compute derivatives and solve $\partial \mathcal{L} / \partial \vec{w} = 0$ to find a unique global minimum.
In almost all cases, however, no such exact solution will be possible and one must resort of approximate algorithmic approaches.
A simple approach which nevertheless turns out to be extremely powerful and the basis of much of modern machine learning is \emph{gradient descent}.
Suppose that one can compute the gradient \begin{align}
    \frac{\partial }{\partial \vec{w}} \mathcal{L}(y ,f_{\vec{w}}(\vec{x})),
\end{align}
where this may be exact or in some cases approximate.
Defining a small \emph{learning rate} $\eta>0$, a natural way to slightly improve the parameters is \begin{align}\label{eq:grad_desc_1}
    \vec{w}_{t+1} = \vec{w}_t - \eta \sum_i \frac{\partial }{\partial \vec{w}} \mathcal{L}(y_i , f_{\vec{w}}(\vec{x}_i))
\end{align}
where $\vec{w}_t$ our the current parameter estimates and $\vec{w}_{t+1}$ are the updated parameters.
One could imagine repeatedly iterating to update the parameters and obtaining optimal, or at least sufficiently performant, parameters.
$\sum_i$ can refer to a sum over the whole training set, some subset, or a a single item.
In the the first case, the described algorithm is precisely gradient descent, whereas in the latter two cases, if the subset is randomly sampled, the algorithm is stochastic gradient descent, since at each iteration a noisy estimate of the gradient is computed.

\section{Neural networks}\label{sec:nn_intro}
In this thesis, a \emph{neural network} shall refer exclusively to a particular type of machine learning model that was originally coined as \emph{artificial neural network} (ANN) \cite{jain1996artificial} to draw distinction between the machine learning models and the biological systems by which they are inspired. For our purposes, and typically for the purposes of modern machine learning, any historical connection with biological neural networks is of limited value (despite being historically important) and so we adopt the common terminology of merely \emph{neural network}, with the `artificial' being implicit. The distinction between neural networks and \emph{deep neural networks} is important, practically and theoretically, and will be made clear in the following discussion.

\medskip 
Conceptually, neural networks are non-linear functions from $\mathbb{R}^d$ to $\mathbb{R}^c$ parameterised by some $\vec{w}\in\mathbb{R}^N$ and formed as the composition of simple affine-linear maps and simple pointwise non-linearities in a layered structure.
Being composed of simple, modular components, neural networks provide an elegant and efficient way of effectively arbitrarily scaling the capacity of models.
Heuristically, the number of parameters $N$ of a parameterised model determines its capacity to learn patterns in data: the larger $N$ is, the more complicated and diverse the patterns that can be learned. 
Naturally one then wishes to define models with many parameters and easily scale up the number of parameters to obtain better results on complicated datasets.
With traditional statistical models, the parameters typically have some interpretation, being attached to some distribution for example, and so substantially increasing the number of parameters will typically require complete redesign of the model.
Even with non-neural machine learning models, it is typically not possible to arbitrarily scale the number of model parameters, as they are typically constrained by the design of the model and/or the data.
For example, a linear regression model has no freedom: the number of parameters is determined entirely by the data dimensionality.
Neural networks immediately solve this issue, essentially providing a simple recipe for constructing arbitrarily large models of some fixed type.
Neural networks are defined by their \emph{architecture} and their parameters.
The architecture is the specification of how the parameters $\vec{w}$ are used to define a function $f_{\vec{w}}$.
There are many different architectures in the machine learning literature and in practical use \cite{lecun2015deep}, however there are a small number of standard types of architecture that cover the vast majority of architectures - we shall describe a few of the most significant types below.
Finally, we note that it is near-ubiquitous in the machine learning literature to use the term neural network to refer to specific architectures with arbitrary parameters (which are families of functions) \emph{and} specific architectures with specific parameters (which are bona fide functions).

\paragraph{Multi-layer perceptrons (MLPs)} The simplest and oldest type of neural network is the MLP\footnote{MLPs are also commonly called fully-connected networks.} \cite{gardner1998artificial,murtagh1991multilayer}.
Let $L>0$ be an integer, and let $n_0,n_1,\ldots, n_L > 0$ be integers, with $n_0=d$ and $n_L = c$.
Define matrices $W^{(i)}\in\mathbb{R}^{n_{i-1}\times n_i}$ and vectors $\vec{b}^{(i)}\in\mathbb{R}^{n_i}$; these are the \emph{weights} and \emph{biases} respectively.
Let $\sigma:\mathbb{R}\rightarrow\mathbb{R}$ be a non-linear function\footnote{Note that the definition of any neural network works if $\sigma$ is linear, but this case is not generally interesting (as it results in linear neural networks), so we exclude it by definition.} - the \emph{activation function}.
Theoretically, $\sigma$ is often assumed to be differentiable, though this assumption is not required by some of our results.
In all practical cases, $\sigma$ will be twice-differentiable except possibly at a finite set of points at which it is merely continuous.
We shall use this latter, weaker, condition, with the convention that, whenever expressions involving derivatives of $\sigma$ are encountered, they implicitly exclude the finite set of points at which the derivative does not exists. 
This convention mirrors what is seen in practice, where $\sigma'(x_*) = \lim_{x\rightarrow x_*^-}\sigma'(x)$ for any non-differentiable point $x_*$.
An MLP with $L$ layers is now defined as
\begin{align}\label{eq:mlp_def}
    f_{\vec{w}}(\vec{x}) = \vec{z}^{(L)}, ~~ \vec{z}^{(l)} = W^{(l)}\sigma(\vec{z}^{(l-1)}) + \vec{b}^{(l)}, ~ l=1, \ldots, L, ~~ \vec{z}^{(0)} = \vec{x},
\end{align}
where $\sigma(\vec{x})$ for vector $\vec{x}$ is defined as the vector with components $\sigma(x_i)$, i.e. $\sigma$ is applied element-wise.
There may optionally be another non-linearity applied to $\vec{z}^{(L)}$, which may be different from $\sigma$, but we will not need to consider that case here.
Note that if $L>1$, all layers apart from the final layer are called \emph{hidden layers}.
\emph{Deep neural networks} are usually defined to be networks with at least one hidden layer, though most of the major practical successes of neural networks comes from models with tens, or even hundreds, or hidden layers.
Machine learning using deep neural network is commonly referred to as \emph{deep learning}.

\paragraph{Convolutional neural networks (CNNs)} MLPs are a very general form of neural network that can be applied to data of any structure, given some strategy for converting each data point to a single vector representation.
If the data are not naturally represented as vectors, forcing them into such a representation so that an MLP can be used is likely to be sub-optimal.
The classical motivating example is that of image data, where each data point is an image and so naturally represented as a rank 3 array of pixels: (width, height, channels).
By flattening the pixel arrays in vectors and applying an MLP, we would almost certainly be making the learning problem more difficult than it really is.
For example, if the network's only objective is to detect cats in images, a picture of a cat located in the top left of the image should appear the same to the network as a picture of a cat in the bottom right of the image, but an MLP presented with flattened vectors must learn separately to identify cats in all possible locations.
CNNs are the standard solution to this kind of problem, particularly for image data \cite{lecun1989backpropagation,lecun1995convolutional}, but also for other data types such as time series and even natural language text \cite{dos2014deep}.
We can write a basic CNN as:
\begin{align}
    f_{\vec{w}}(\vec{x}) = \vec{z}^{(L)}, ~~ \vec{z}^{(l)} = g(\sigma(\vec{z}^{(l-1)}); W^{(l)}, \vec{b}^{(l)}), ~ l=1, \ldots, L, ~~ \vec{z}^{(0)} = \vec{x},
\end{align}
where $g(\cdot; W, \vec{b})$ is an affine-linear function with respect to its input and also its parameters $W, \vec{b}$, and the shape of the weights and biases are entirely general.
This definition is clearly a strict generalisation of the MLP, which is given by $g(\vec{x}; W, \vec{b}) = W\vec{x} + \vec{b}$.
CNNs take $g$ to be a \emph{convolution} operation.
Let $W\in\mathbb{R}^{2k+1\times 2k + 1 \times c_1 \times c_2}$ be a \emph{kernel} and let $\vec{x}\in\mathbb{R}^{h\times l \times c_1}$, then 
\begin{align}
    g(\vec{x}; W_{ijk}) = \sum_{p = i-k}^{i+k}\sum_{q=j-k}^{j+k} \sum_{r=1}^{c_1} W_{pqrk}x_{pqr}.\label{eq:cnn_def}
\end{align}
$g$ can be similarly defined to include biases, care must be taken with the definition at the edges (e.g. when $i-k < 0$) and the first two indices of $W$ needn't have odd dimension, but for our purposes there is no need to consider these details.
Here $2k+1$ is the \emph{filter size}, $c_1$ is the number of input channels and $c_2$ the number of output channels.
$c_1, c_2$ are the analogue of the input and output size of each layer of an MLP.
Typically, in the first layer of a CNN, $k$ is much less than $h$ and $l$, so that the number of parameters in $W$ is much less than the number of parameters in the a corresponding weight matrix of an MLP: $(2k+1)^2 c_1c_2$ compared to $hlc_1c_2$.

Note also that the convolutional structure of (\ref{eq:cnn_def}) reuses entries of $W$ in multiple location on the input $\vec{x}$.
As well as reducing the number of parameters compared to equivalent MLPs, CNNs also restrict to functions which are translation invariant in the desired sense motivated by the above example of cat detection in images.
Finally, note that CNNs are special case of MLPs; the operation defined in (\ref{eq:cnn_def}) is affine-linear and so for any index flattening transformation $\phi(\vec{x})$ there exists a matrix $\hat{W}$ such that $g(\vec{x}; W) = \hat{W}\phi(\vec{x})$.
Nevertheless, CNNs are preferred to MLPs on any data for which the convolution operation is appropriate, as they provide a beneficial \emph{inductive bias}, essentially encouraging the optimisation procedure (recall (\ref{eq:grad_desc_1})) to find superior local optima than would be found for an MLP.

\paragraph{Sequential modelling architectures} CNNs are well-adapted to image data and, loosely speaking, data which can reasonably be represented as images (e.g. spectrograms \cite{badshah2017speech}). CNNs have also been successfully applied to natural language data \cite{dos2014deep}, however there are a few other architecture types designed for natural language data and other sequential data. In particular, \emph{recurrent neural networks} (RNNs) \cite{medsker2001recurrent} and later variants such as long short-term memory (LSTM) \cite{hochreiter1997long} networks and gated recurrent units (GRUs) \cite{chung2014empirical} have architectures designed to respect the time-ordering of the data (e.g. the order of words in a sentence) while possessing the appropriate time re-parametrisation invariance.
More recently, transformer models \cite{devlin-etal-2019-bert,brown2020language} have been proposed and enjoyed considerably practical success over RNN and CNN architectures.

\paragraph{Architecture combinations} The different architecture types outlined above need not be used in isolation, but can be combined.
For instance, it is standard practice to construct architectures as a concatenation of a CNN and an MLP, with the MLP acting on the flattened output of the CNN\footnote{Historically, such concatenations of CNNs and MLPs were the standard approach, so are universally referred to simply as CNNs and networks with only convolutional layers are often called \emph{fully-convolutional networks}.}.
RNNs and transformer architectures are usually built as extensions of MLPs, though there are also convolutional examples (see e.g. convolutional RNNs).

\paragraph{Generative adversarial networks (GANs)} MLPs, CNNs and the various sequential modelling architectures are the most common neural network architecture types in practical use and, between them, provide the basis of the vast majority of applications of deep learning to supervised, unsupervised and semi-supervised learning problems.
GANs \cite{goodfellow2014generative} are the canonical basic approach to generative modelling using neural networks.
 GANs are composed of two neural networks: \emph{generator} ($G$) and \emph{discriminator} ($D$). $G$ is a map $\mathbb{R}^m\rightarrow\mathbb{R}^d$ and $D$ is a map $\mathbb{R}^d\rightarrow\mathbb{R}$. $G$'s purpose is to generate synthetic data samples by transforming random input noise, while $D$'s is to distinguish between real data samples and those generated by $G$. Given some probability distribution $\mathbb{P}_{data}$ on some $\mathbb{R}^d$, GANs have the following minimax training objective \begin{align}
    \min_{\vec{w}_G}\max_{\vec{w}_D}\left\{\mathbb{E}_{\vec{x}\sim \mathbb{P}_{data}} \log D(\vec{x}) + \mathbb{E}_{\vec{z}\sim \mathcal{N}(0, \sigma_z^2)}\log(1 - D(G(\vec{z})))\right\},
\end{align}
where $\vec{w}_D, \vec{w}_G$ are the parameters of the discriminator and generator respectively.
Given a well optimised generator model, one can sample approximately from the data distribution by sampling latent vectors in the space $\mathbb{R}^l$ and passing them through the generator.

\paragraph{Training neural networks} By defining neural network architecture suitable for some data and by varying the number of layers, or the size of the layers (i.e. the size dimensions of the weights), one can specify very large families of parameterised non-linear functions with essentially arbitrary expressivity and complexity.
Indeed, there are many results beginning with shallow networks \cite{barron1993universal,cybenko1989approximation,hornik1989multilayer} that establish neural networks as universal function approximators within certain classes of functions and considerable amounts of more recent work that establish the representational power of deep networks \cite{daubechies2022nonlinear,telgarsky2015representation,petersen2018optimal,lu2017expressive,lu2020universal}.
Therefore, given any data, any learning task defined on that data and any theoretically possible level of performance, one can be quite sure of constructing an neural network architecture, and hence a family of parametrised functions, such that there exist some parameter values giving the specified level of performance at the task on the data.
If neural networks are to be practically useful, however, there must exist some feasible algorithm to find such parameter values.
Feasible here has at least two meanings:
\begin{itemize}
    \item computationally feasible, i.e. the algorithm must terminate in a reasonable time using a reasonable amount of computational resource;
    \item the algorithm must be general-purpose, i.e. one requires algorithms that apply to a wide variety of datasets and architectures - it would be infeasible if a bespoke algorithm were required for every (dataset, architecture) combination.
\end{itemize}
We have already seen how the layered structure of neural network, building complicated functions from the composition of simple primitives, makes feasible the specification of models with arbitrary capacity and complexity, the layered structure is also essential for feasible training.
In particular, despite their potentially enormous size and considerable complexity, most neural networks are efficient to evaluate, as the vast majority of the computational work in their evaluation comprises linear algebraic operations which have been well-optimised for many computational architectures \cite{lecun2015deep,paszke2017automatic,abadi2016tensorflow,berjon2015optimal}.
Moreover, the layered structure makes possible the efficient and automatic computation of derivatives of neural networks with respect to their parameters.
Indeed, consider the form an MLP in (\ref{eq:mlp_def}).
Differentiating $f_{\vec{w}}(\vec{x})$ with respect to any of the $W^{(l)}$ is a mathematically simple matter: one simply applies the chain rule.
Let us define $\vec{y}^{(l)} = \sigma(\vec{z}^{(l)})$, so $\vec{z}^{(l)} = W^{(l)}\vec{y}^{(l-1)} + \vec{b}^{(l)}$.
Then \begin{align}
    \frac{\partial \vec{z}^{(l)}}{\partial \vec{y}^{(l-1)}} = W^{(l)}, ~~ \frac{\partial \vec{y}^{(l)}}{\partial \vec{z}^{(l)}} = \sigma'(\vec{z}^{(l)}), ~~ \frac{\partial \vec{z}^{(l)}}{\partial  W^{(l)}} = \vec{y}^{(l)},
\end{align}
so observe that, if $\sigma'$ is known in closed-form and an implementation provided, a computer can implement the chain rule to automatically compute exact derivatives of $f_{\vec{w}}$.
If derivatives of $\mathcal{L}$ are also implemented, then the full derivatives $\partial_{\vec{w}} \mathcal{L}(y, f_{\vec{w}}(\vec{x}))$ can be computed for any $\vec{x}, y$ and at any $\vec{w}$.
Moreover, all these gradient computations also benefit from the optimised implementations of linear algebraic primitives.
In the machine learning literature, computing $f_{\vec{x}}(\vec{x})$ is called a \emph{forward pass} and computing $\partial_{\vec{w}} f_{\vec{w}}(\vec{x})$ is called a \emph{backward pass}.
Since neural networks allow for efficient automatic computation of loss gradients  $\partial_{\vec{w}} \mathcal{L}(y, f_{\vec{w}}(\vec{x}))$, the simplest algorithm one could imagine to optimise the parameters $\vec{w}$ for a dataset is stochastic gradient descent (\ref{eq:grad_desc_1}).
So far it is clear that using SGD in combination with neural network backward pass represents a feasible optimisation algorithm for general neural networks and it quite feasible to perform hundreds of thousands of steps of SGD in an acceptable time-frame, though obviously this varies with model and dataset size, as do the requirements on the computational hardware.
However this discussion does not address the quality of the optimisation.
That is to say, we have described a procedure for neural network optimisation that is general-purpose,  feasible to implement and apply to any architecture and dataset, and simple computational experiments would be sufficient to determine how many SGD steps can be performed per second for a given model and given hardware.
For this procedure to be of value, however, it must, with sufficient probability, find parameter values $\vec{w}$ that give sufficiently good performance of the neural network on the defined task.
While SGD is an intuitive and appealing algorithm, the cases for which it can be proven to find, say, global minima are narrow \cite{polyak1992acceleration,varre2021last} and certainly cannot be expected to generically apply to deep neural networks.
Indeed, a priori, for large neural networks with many parameters, one should expect there to be a great many saddle points and local optima of the loss surface around which SGD could get stuck.
Algorithmic innovations can somewhat mitigate the problem of saddle points, such as endowing the gradient descent trajectory with momentum \cite{nesterov2013introductory} or adjusting the learning rates in different directions on the loss surface \cite{duchi2011adaptive,kingma2014adam}, and these techniques can greatly improve practical performance of neural networks \cite{Bottou2012}.
In very high dimensions the intuition of such techniques does not necessarily apply and if there are a great many local minima, then we should expect SGD to converge to, at best, some local minimum determined by the random initialisation of $\vec{w}$.
In general, there is no reason to expect that these local minima will provide network performance anywhere near the global optimum, or even useful performance at all.
In bold defiance of these arguments, neural networks continue to have substantial success when applied to an increasingly long list of machine learning problems: computer vision, speech processing, natural language processing, reinforcement learning, media generation etc. We refer the interested reader to the excellent website \cite{paperswithcode} where they will find links to published literature detailing the success of neural networks in all fields of machine learning. Networks are trained using stochastic gradient-based optimisation methods on very high-dimensional, strongly non-convex surfaces for which no formal convergence or performance guarantees exist and yet excellent practical performance is routinely obtained with little concern for whether the optimisation problem has been solved. Extremely over-parametrised models can be trained with large numbers of passes through the data without overfitting. Models with equivalent training performance can have radically different generalisation performance depending on complicated interactions between design choices such as learning rate size (and scheduling) and weight-decay \cite{loshchilov2018decoupled}.

\section{Structure of neural network loss surfaces}

\jstat{One strand of theoretical work focuses on studying properties of the loss surfaces of large neural networks and the behaviour of gradient descent algorithms on those surfaces. Much of the content of this thesis sits within this line of research. \cite{sagun2014explorations} presented experimental results pointing to a similarity between the loss surfaces of multi-layer networks and spherical multi-spin glasses \cite{mezard1987spin}. \cite{choromanska2015loss} built on this work by presenting modeling assumptions under which the training loss of multi-layer perceptron neural networks with \texttt{ReLU} activations can be shown to be equivalent to a spherical multi-spin glass (with network weights corresponding to spin states). The authors then applied spin glass results of \cite{auffinger2013random} to obtain precise asymptotic results about the complexity\footnote{\emph{Complexity} will be given a formal definition in Chapter \ref{chap:maths}.} of the training loss surfaces. Crudely, the implication of this work is that the unreasonable efficacy of gradient descent on the high-dimensional and strongly non-convex loss surfaces of neural network models can in part be explained by favourable properties of their geometry that emerge in high dimensions. Relationships between simpler neural networks and spin glasses have been known since \cite{kanter1987associative,gardner1988optimal,engel2001statistical} and, more generally, connections between spin glass theory and computer science were studied in \cite{nishimori2001statistical} in the context of signal processing (image reconstruction, error correcting codes).}

\jstat{More recent work has dispensed with deriving explicit links between neural networks and spin glasses, instead taking spin glass like objects as a tractable playground for gradient descent in complex high-dimensional environments. In particular, \cite{baity2019comparing} compare empirically the dynamics of state-of-the-art deep neural networks and glassy systems, while \cite{mannelli2019afraid, ros2019complex, arous2019landscape, mannelli2019passed} study random tensor models containing some `spike' to represent other features of machine learning problems (some `true signal' to be recovered) and perform explicit complexity calculations as well as gradient descent dynamical calculations revealing phase transitions and landscape trivialisation. \cite{maillard2019landscape} simplify the model in favour of explicitly retaining the activation function non-linearity and performing complexity calculations \`{a} la \cite{auffinger2013random, fyodorov2007replica,  fyodorov2004complexity} for a single neuron. \cite{pennington2017geometry} study the loss surface of random single hidden layer neural networks by applying the generalised Gauss-Newton matrix decomposition to their Hessians and modelling the two components as freely-additive random matrices from certain ensembles. \cite{pennington2017nonlinear, benigni2019eigenvalue} consider the loss surfaces of single layer networks by computing the spectrum of the Gram matrix of network outputs. These works demonstrate the value of studying simplified, randomised neural networks for understanding networks used in practice. }
\jstat{The situation at present is far from clear. The spin glass correspondence and consequent implications for gradient descent based learning from \cite{choromanska2015loss, sagun2014explorations} are tantalising, however there are significant challenges. Even if the mean asymptotic properties of deep neural network loss surfaces were very well described by corresponding multi-spin glass models, the question would still remain whether these properties are in fact relevant to gradient-based algorithms running for sub-exponential time, with some evidence that the answer is negative \cite{baity2019comparing, mannelli2019passed, folena2019rethinking}. Another challenge comes from recent experimental studies  of deep neural network Hessians \cite{papyan2018full, ghorbani2019investigation, granziol2020beyond, granziol2019towards} which reveal spectra with several large outliers and considerable rank degeneracy, deviating significantly from the Gaussian Orthogonal Ensemble semi-circle law implied by a spin glass model. Bearing all this in mind, there is a long and illustrious history in the physics community of fruitfully studying quite unrealistic simplified models of complicated physical systems and still obtaining valuable insights into aspects of the true systems.}

\jstat{Several of the assumptions used in \cite{choromanska2015loss} to obtain a precise spherical multi-spin glass expression are undesirable, as outlined clearly in \cite{choromanska2015open}. Assuming i.i.d. Gaussian data and random labels is clearly a going to greatly simplify the problem, however it is also the case that many of the properties of deep neural networks during training are not specific to any particular dataset, and there may well be phases of training to which such assumptions are more applicable than one might first expect. Gaussian and independence assumptions are commonplace when one is seeking to analyse theoretically very complicated systems, so while they are strong, they are not unusual and it is not unreasonable to expect some important characteristics of real networks to persist. By contrast, the restriction of the arguments in \cite{choromanska2015loss} to exclusively \texttt{ReLU} activations seems innocuous, but we argue quite the opposite is true. There are deep mathematical reasons why Gaussian and independence assumptions are required to make progress in the derivation in \cite{choromanska2015loss}, while the restriction to \texttt{ReLU} activations appears to be an obscure peculiarity of the calculations. The \texttt{ReLU} is certainly a very common choice in practice, but it is by no means the only valid choice, nor always the best; see e.g. leaky \texttt{ReLU} in state-of-the-art image generation \cite{karras2019style} and GELU in state-of-the-art language models \cite{devlin2018bert}. It would not be at all surprising if a spin glass correspondence along the lines of \cite{choromanska2015loss} were impossible without Gaussian and/or independence assumption on the data, however it would be extremely concerning if such a correspondence specifically required \texttt{ReLU} activations. If the conclusions drawn in \cite{choromanska2015loss} about deep neural networks from this correspondence are at all relevant in practice, then they must apply equally to all activation functions used in practice. On the other hand, if the conclusions were \emph{precisely} the same for all reasonable activation functions, it would reveal a limitation of the multi-spin glass correspondence, since activation function choice can have significant implications for training neural networks in practice.}

\section{Contributions of this thesis}
\begin{figure}[h]
    \centering

\begin{tikzpicture}
    \node (choro) [io, xshift=3cm] {\cite{choromanska2015loss,auffinger2013random}};
    \node (gen) [process, below of=choro, yshift=-1cm] {\cite{baskerville2021loss} - Chapter \ref{chap:general_activation_functions}};
        \node (gan) [process, below of=gen, yshift=-1cm] {\cite{baskerville2022spin} - Chapter \ref{chap:spin_glass_gans}};
        
        \node (spec_mot) [io, xshift=7cm] {\cite{sagun2017empirical,papyan2018full,granziol2020beyond}};
        
        \node (spac) [process, below of=spec_mot, yshift=-1cm] {\cite{baskerville2022appearance} - Chapter \ref{chap:spacings}};
        
    \node (gadam) [process, yshift=-4cm, xshift=-1cm] {\cite{ia} - Chapter \ref{chap:gadam}};
    \node (lr) [io,xshift=11cm]{\cite{JMLR:v23:20-1258}};
    \node (univ) [process, yshift=-3cm, xshift=0cm, below of=spac] {\cite{Baskerville_2022} - Chapter \ref{chap:univ}};
        
    \node (damp) [process,yshift=-1cm, below of=lr]{\cite{granziol22damping} - Chapter \ref{chap:damp}};
      \begin{pgfonlayer}{background}

    \draw [arrow] (choro) -- (gen);
    \draw [arrow] (gen) -- (gan);
    \draw [arrow] (spec_mot) -- (spac);
    \draw [arrow] (choro) -- (spac);
    \draw [arrow] (choro) -- (gadam);
    \draw [arrow] (gen) -- (gadam);
    \draw [arrow] (lr) -- (univ);
    \draw [arrow] (gen) -- (univ);
    \draw [arrow] (gan) -- (univ);
    \draw [arrow] (spac) -- (univ);
    \draw [arrow] (lr) -- (damp);
      \end{pgfonlayer}
\end{tikzpicture}
    
    \caption{Schematic of the contributions of this thesis }
    \label{fig:contributions_diag}
\end{figure}
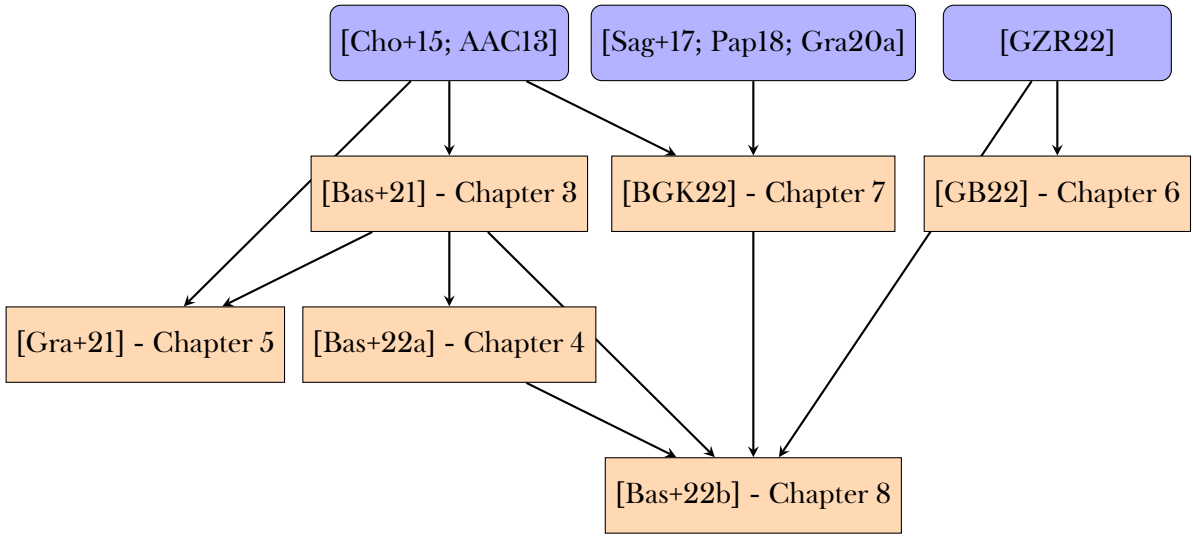

In Figure \ref{fig:contributions_diag} below we give a diagram that outlines the contributions of this thesis and their position within the literature.
Rounded purple boxes denote antecedents and influences of our contributions from the literature. The references given in these boxes are not intended to be exhaustive but simply indicators.
Rectangular orange boxes denote our contributions, where we display both the published papers and the corresponding Chapter in this thesis. We expand further on the context of this thesis and its contributions in the following subsections. 
Chapters \ref{chap:general_activation_functions} and \ref{chap:spin_glass_gans} form the first major contribution and are discussed in section \ref{sec:major1_intro}. Chapters \ref{chap:spacings} and \ref{chap:univ} form a distinct major contribution but are nevertheless related to the the earlier chapters, as indicated in the diagram. Chapters \ref{chap:gadam} and \ref{chap:damp} are distinct contributions that are certainly connected to the major parts of the thesis, but are more peripheral in their contribution; they are discussed in section \ref{sec:minor1_intro} and \ref{sec:minor2_intro} respectively.

\medskip

\subsection{Generalisation of spin glass models for neural networks loss surfaces}\label{sec:major1_intro}
The first major contribution of this thesis is a significant generalisation of the understanding of spin glass models for neural network loss surfaces. Beginning with Chapter \ref{chap:general_activation_functions}, we return to the modeling assumptions and methodology of \cite{choromanska2015loss} and extend their results to multi-layer perceptron models with any activation function. We demonstrate that the general activation function has the effect of modifying the exact multi-spin glass by the addition of a new deterministic term in the Hamiltonian. We then extend the results of \cite{auffinger2013random} to this new high-dimensional random function. At the level of the logarithmic asymptotic complexity of the loss surface, we obtain precisely the same results as \cite{choromanska2015loss}, however the presence of a general activation function is felt in the sharp asymptotic complexity. On the one hand, our results strengthen the case for \cite{choromanska2015loss} by showing that their derivation is not just an accident in the case of \texttt{ReLU} networks. On the other hand, we have shown that this line of reasoning about neural networks is insensitive to an important design feature of real networks that can have significant impacts on training in practice, namely the choice of activation function. The main calculation for our result  uses a Kac-Rice formula to compute landscape complexity of the modified multi-spin glass model we encounter. Kac-Rice formulae have a long history in the Physics literature \cite{bray1980metastable, bray1981metastable} and more specifically to perform complexity calculations \cite{fyodorov2004complexity, fyodorov2005counting, auffinger2013random}. Complexity calculations in spiked matrix and tensor models in \cite{ros2019complex, arous2019landscape} have addressed spin glass objects with specific rank-1  deterministic additive terms, however those calculations do not extend to the case encountered here since those deterministic terms create a single distinguished direction --- parallel to the gradient of that term everywhere on the sphere --- which is critical to their analysis; our extra deterministic term creates no such single distinguished direction. We chart a different course using supersymmetric methods in Random Matrix Theory. Supersymmetric methods have been used before in spin glass models and complexity calculations \cite{cavagna1999quenched,annibale2003supersymmetric,crisanti2003complexity,fyodorov2004complexity}, often using the replica trick. We show how the full logarithmic complexity results of \cite{auffinger2013random} can be obtained using a supersymmetric approach quite different to the approach used in that and similar works. By moving to this approach, we can make progress despite the presence of the extra deterministic term in the multi-spin glass. Our approach to the supersymmetric calculations most closely follows \cite{fyodorov2015random,nock}, but several steps require approximations due to the extra term. Some of our intermediate results in the supersymmetric and RMT calculations are stronger than required here, but may well be useful in future calculations, e.g. spiked spherical multi-spin glass models with any fixed number of spikes. Finally, our approach computes the total complexity summed over critical points of any index and then uses large deviations principles to obtain the complexity with specified index. This is the reverse order of the approach taken in \cite{auffinger2013random} and may be more widely useful when working with perturbations of matrices with known large deviations principles.

\medskip
Motivated by our results in Chapter \ref{chap:general_activation_functions}, we ask if it is possible to further extend the spin glass modeling approach to capture yet further peculiarities and details of modern neural networks. We seek, in particular, a model that is capable of revealing the influence of architectural details \emph{at leading order} in the annealed complexity, unlike the relatively weak effect of the activation function seen in Chapter \ref{chap:general_activation_functions}. Modern deep learning contains a very large variety of different design choices in network architecture, such as convolutional networks for image and text data (among others) \cite{goodfellow2016deep,conneau-etal-2017-deep}, recurrent networks for sequence data \cite{hochreiter1997long} and self-attention transformer networks for natural language \cite{devlin-etal-2019-bert,radford2018improving}. Given the ubiquity of convolutional networks, one might seek to study those, presumably requiring consideration of local correlations in data. One could imagine some study of architectural quirks such as residual connections \cite{he2016deep}, and batch-norm has been considered to some extent by \cite{pennington2017nonlinear}. In Chapter \ref{chap:spin_glass_gans}, we propose a novel model for \emph{generative adversarial networks} (GANs) \cite{NIPS20145423} as two interacting spherical spin glasses. GANs have been the focus of intense research and development in recent years, with a large number of variants being proposed \cite{radford2015unsupervised,zhang2018self,liu2016coupled,karras2020a,mirza2014conditional,arjovsky2017wasserstein,zhu2017unpaired} and rapid progress particularly in the field of image generation. From the perspective of optimisation, GANs have much in common with other deep neural networks, being complicated high-dimensional functions optimised using local gradient-based methods such as stochastic gradient descent and variants. On the other hand, the adversarial training objective of GANs, with two deep networks competing, is clearly an important distinguishing feature, and GANs are known to be more challenging to train than single deep networks. Our objective is to capture the essential adversarial aspect of GANs in a tractable model of high-dimensional random complexity which, though being a significant simplification, has established connections to neural networks and high dimensional statistics. 

\medskip
Our model is inspired by \cite{choromanska2015loss,ros2019complex,mannelli2019afraid,arous2019landscape} with spherical multi-spin glasses being used in place of deep neural networks. We thus provide a complicated, random, high-dimensional model with the essential feature of GANs clearly reflected in its construction. By employing standard Kac-Rice complexity calculations \cite{fyodorov2004complexity,fyodorov2007replica,auffinger2013random} we are able to reduce the loss landscape complexity calculation to a random matrix theoretic calculation. We then employ various Random Matrix Theory techniques as in \cite{baskerville2021loss} to obtain rigorous, explicit leading order asymptotic results. Our calculations rely on the supersymmetric method in Random Matrix Theory, in particular the approach to calculating limiting spectral densities follows \cite{Verbaarschot_2004} and the calculation also follows \cite{guhr1990isospin,guhr1991dyson} in important ways. The greater complexity of the random matrix spectra encountered present some challenges over previous such calculations, which we overcome with a combination of analytical and numerical approaches. Using our complexity results, we are able to draw qualitative implications about GAN loss surfaces analogous to those of \cite{choromanska2015loss} and also investigate the effect of a few key design parameters included in the GAN. We compare the effect of these parameters on our spin glass model and also on the results of experiments training real GANs. Our calculations include some novel details, in particular, we use precise sub-leading terms for a limiting spectral density obtained from supersymmetric methods to prove a required concentration result to justify the use of the Coulomb gas approximation. We note that our complexity results could be also be obtained in principle using the methods developed in \cite{arous2021exponential}, however our work was completed several months before this pre-print appeared. Our approach for computing the limiting spectral density may nevertheless be the simplest and would be used as input to the results of \cite{arous2021exponential}.

\medskip
The role that statistical physics models such as spherical multi-spin glasses are to ultimately play in the theory of deep learning is not yet clear, with arguments both for and against their usefulness and applicability.
Before our contributions, the major result was \cite{choromanska2015loss} which, though influential, has received considerable criticism and could have reasonably been considered a parochial curiosity, rather than profound insight into neural network loss surfaces.
Our work in Chapter \ref{chap:general_activation_functions} considerably weakens the case against \cite{choromanska2015loss}, and our work in Chapter \ref{chap:spin_glass_gans} clearly demonstrates the potential of spin glass models (and statistical physics based models in general) to capture and explain phenomena in deep neural networks.
Indeed, to the best of our knowledge, Chapter \ref{chap:spin_glass_gans} provides the first attempt to model an important architectural feature of modern deep neural networks within the framework of spin glass models. Our analysis reveals potential explanations for observed properties of GANs and demonstrates that it may be possible to inform practical hyperparameter choices using models such as ours. Much of the  advancement in practical deep learning has come from innovation in network architecture, so if deep learning theory based on simplified physics models like spin-glasses is to keep pace with practical advances in the field, then it will be necessary to account for architectural details within such models; our work is a first step in that direction.

\subsection{Discovery of RMT universality in loss surfaces and consequences for loss surface models}\label{sec:major2_intro}
The other major contribution of this thesis is the instigation of the study of the role of random matrix theory statistics in deep learning at the local (i.e. microscopic) scale and the building of a strong case that the results which characterise the first half of the thesis, and other RMT-based results from the literature besides, can be expected to be much more general in applicability than their very restrictive modeling assumptions would suggest.

\begin{figure*}[t!]
    \centering
    \begin{subfigure}{0.32\linewidth}
        \includegraphics[width=1\linewidth,trim={0 0 0 0},clip]{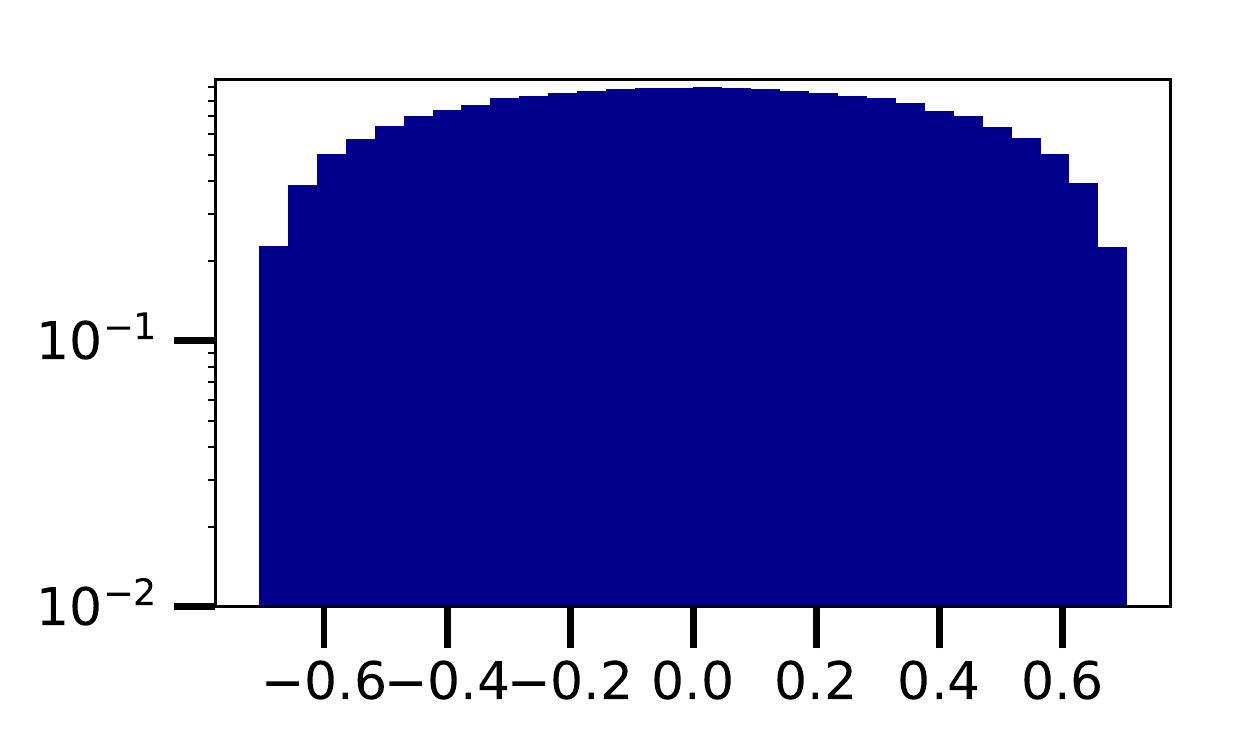}
        \caption{Wigner semicircle}
        \label{subfig:wignerstemintro}
    \end{subfigure}
    \begin{subfigure}{0.32\linewidth}
        \includegraphics[width=1\linewidth,trim={0 0 0 0},clip]{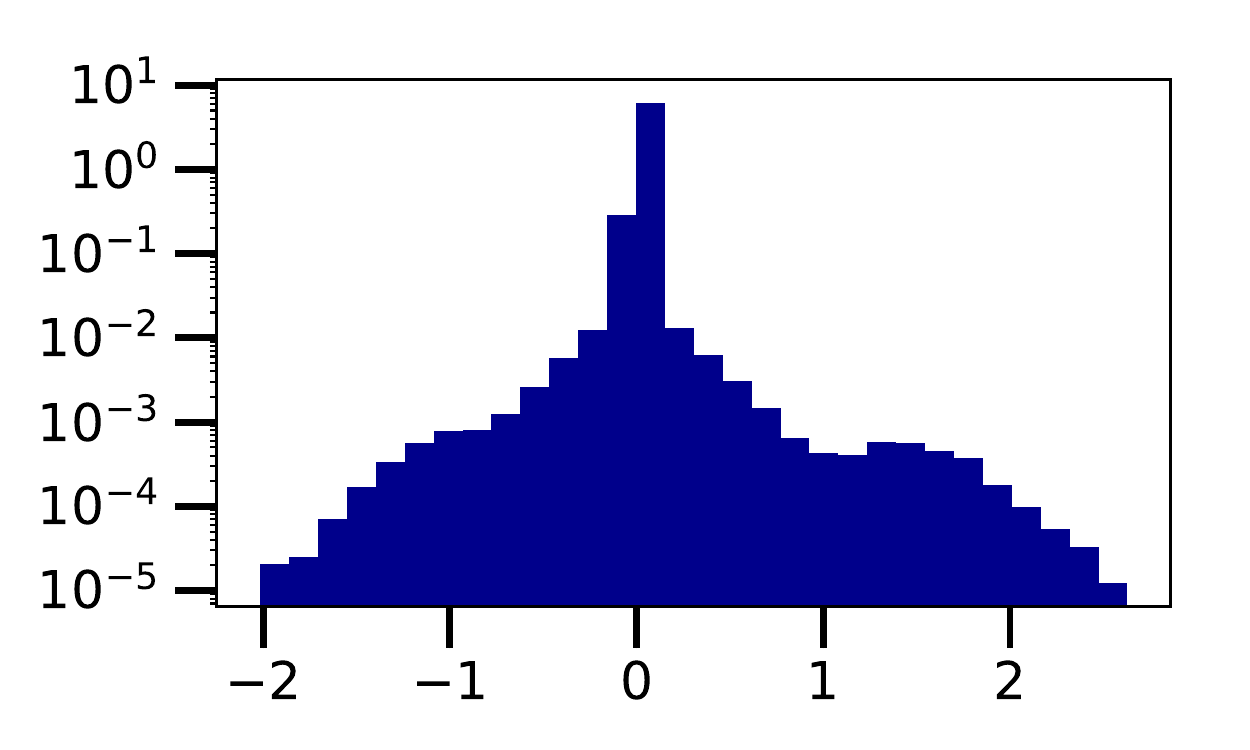}
        \caption{MLP}
        \label{subfig:mlp}
    \end{subfigure}
    \begin{subfigure}{0.32\linewidth}
        \includegraphics[width=1\linewidth,trim={0 0 0 0},clip]{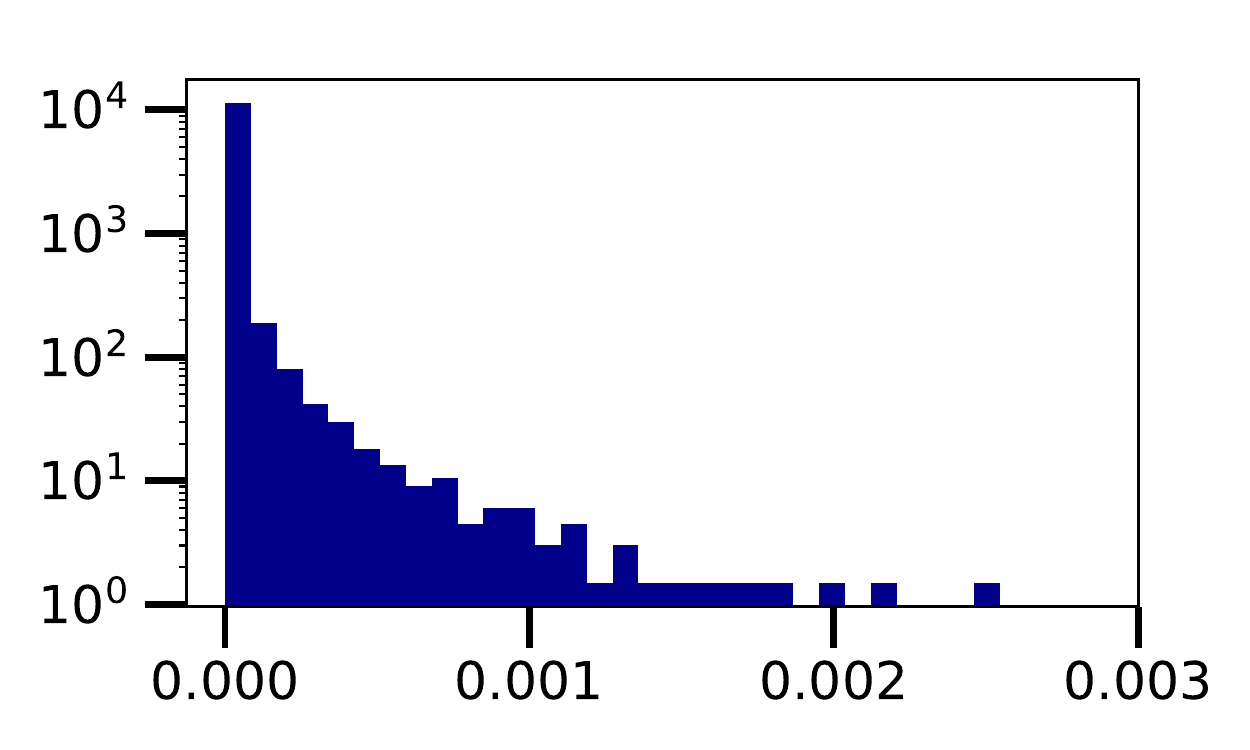}
        \caption{Logistic Regression}
        \label{subfig:logisticregression}
    \end{subfigure}
    \caption{Comparison of different global spectral statistics (spectral densities). (a) We show actual GOE data to demonstrate the form of the Wigner semicircle. (b) Hessian of cross entropy loss for MLP on MNIST. (c) Hessian of cross entropy loss for logistic regression on MNIST. Note the log-scale on the y-axis. A few outliers have been clipped from logistic regression to aid visualisation.}\label{fig:goeisabadfit}
\end{figure*}

\medskip
An important and fundamental problem with Chapters \ref{chap:general_activation_functions} and \ref{chap:spin_glass_gans} and related work in the literature is that typically the average spectral density of the Hessian of neural networks does not match that of the associated canonical random matrix ensembles that results from the modeling assumptions and are crucial in the technicalities of the calculations.  This is illustrated in Figure \ref{fig:goeisabadfit}. 
Put simply, \emph{one does not observe the Wigner semicircle or Marchenko-Pastur eigenvalue distributions, implied by the Gaussian Orthogonal or Wishart Ensembles}. 
As shown in \cite{granziol2020beyond,granziol2019towards,papyan2018full,papyan2019measurements,ghorbani2019investigation,sagun2016eigenvalues,sagun2017empirical} the spectral density of neural network Hessians contain outliers and a large number of near zero eigenvalues, features not seen in canonical random matrix ensembles. 
Furthermore, even allowing for this, as shown in \cite{granziol2020towards} by specifically embedding outliers as a low rank perturbation to a random matrix, the remaining bulk spectral density still does not match the Wigner semicircle or Marchenko-Pastur distributions \cite{granziol2020beyond}, bringing into question the validity of the underlying modelling.  
The fact that the experimental results differ markedly from the theoretical predictions has called into question the validity of neural network analyses based on canonical random matrix ensembles. Moreover, the compelling results of works such as \cite{choromanska2015loss, pennington2017geometry} are obtained using very particular properties of the canonical ensembles, such as large deviation principles, as pointed out in \cite{granziol2020beyond}. The extent to which such results can be generalised is an open question.
Hence, further work is required to better understand to what extent random matrix theory can be used to analyse the loss surfaces of neural networks.
In Chapter \ref{chap:spacings}, we show that the  {\em local spectral statistics} (i.e.~those measuring correlations on the scale of the mean eigenvalue spacing) of neural network Hessians are well modelled by those of GOE random matrices, even when the mean spectral density is different from the semicircle law. We display these results experimentally on MNIST trained multi-layer perceptrons and on the final layer of a ResNet-$34$ on CIFAR-$10$. The objective of Chapter \ref{chap:spacings} is to motivate a new use for Random Matrix Theory in the study of the theory of deep neural networks.  In the context of more established applications of random matrix theory, this conclusion may not be so surprising -- it has often been observed that the local spectral statistics are universal while the mean density is not -- however, in the context of machine learning this important point has not previously been made, nor its consequences explored. In Chapter \ref{chap:spacings} we illustrate it in that setting, through numerical experiments, and start to examine some of its implications.   

\medskip
Having established experimentally the presence of universal local random matrix statistics in real-world neural networks (though admittedly very small ones by modern standards), we proceed in Chapter \ref{chap:univ} to demonstrate how local random matrix statistics can be used as modeling assumptions for models of deep neural network Hessians to obtain surprisingly strong generalisations of prior spectral results. 
    Works such as \cite{auffinger2013random,choromanska2015loss, fyodorov2005counting} and our own contributions in Chapters \ref{chap:general_activation_functions} and \ref{chap:spin_glass_gans} show how detailed calculations can be completed beyond in and beyond the standard spin glass case, however these results all depend on important properties of the GOE, to which the Hessians in those cases are closely related.
    In a recent work, \cite{granziol2020learning} showed how valuable practical insights about DNN optimisation can be obtained by considering the outliers in the spectrum of the loss surface Hessian.
    Once again, this work relies on special properties from random matrix theory, indeed an expression for the outliers follows from a known phase transition result whereby the largest eigenvalue ``pops out'' of the bulk. This result has been proven only for rotationally invariant matrix ensembles in \cite{benaych2011eigenvalues}, itself a generalisation of the celebrated BBP phase transition \cite{baik2005phase}, though it was conjectured in \cite{benaych2011eigenvalues} to be more general (a point which we clarify in Chapter \ref{chap:univ}, section \ref{sec:interlude}).
    In addition, the explicit form of a Wigner semi-circle density was used to obtain the concrete outlier expression used in practice.
    
        \review{Microscopic random matrix universality is known to be far more robust than universality on the macroscopic scale. Indeed, such results are well established for invariant ensembles and can be proved using Riemann-Hilbert methods \cite{deift1999orthogonal}.}
    \review{For more general random matrices, microscopic universality} has been proved \review{by quite different methods} in a series of works over the last decade or so, of which a good review is \cite{erdos2017dynamical}.
    Crucial in these results is the notion of a \emph{local law} for random matrices.
    The technical statement of local laws is given later in section \ref{sec:universal_intro}, but roughly they assert that the spectrum of a random matrix is, with very high probability, close to the deterministic spectrum defined by its limiting spectral density (e.g. the semicircle law for Wigner matrices).
    Techniques vary by ensemble, but generally a local law for a random matrix ensemble provides the control required to demonstrate that certain matrix statistics are essentially invariant under the evolution of the Dyson Brownian motion.
    In the case of real symmetric matrices, the Dyson Brownian motion converges in finite time to the GOE, hence the statistics preserved under the Dyson Brownian motion must match the GOE.
    The $n$-point correlation functions of eigenvalues are one such preserved quantity, from which follows, amongst other properties, \review{that the Wigner surmise is a good approximation to the adjacent spacings distribution.}
    
    \medskip
    At the macroscopic scale, there are results relevant to neural networks, for example \cite{pennington2018emergence,pastur2020random} consider random neural networks with Gaussian weights and establish results that are generalised to arbitrary distributions with optimal conditions, so demonstrating universality.
    On the microscopic scale, our work in Chapter \ref{chap:spacings} provided the first evidence of universal random matrix theory statistics in neural networks and was subsequently to the weight matrices of neural networks in \cite{thamm2022random}, but no prior work has considered the implications of these statistics, that being the central contribution of Chapter \ref{chap:univ}.
    Our main mathematical result is a significant generalisation of the Hessian spectral outlier result recently presented by \cite{granziol2020learning}.
    This generalisation removes any need for GOE or Wigner forms of the Hessian and instead leverages much more universal properties of the eigenvectors and eigenvalues of random matrices which we argue are quite likely to hold for real networks.
    Our results make concrete predictions about the outliers of DNN Hessians which we compare with experiments on several real-world DNNs.
    These experiments provide indirect evidence of the presence of universal random matrix statistics in the Hessians of large DNNs, which is noteworthy as certainly these DNNs are far too large to permit exact eigendecomposition of their Hessians as done in Chapter \ref{chap:spacings}.
    Along a similar line, we show how local random matrix laws in DNNs can dramatically simplify the dynamics of certain gradient descent optimisation algorithms and may be in part responsible for their success.
    Finally, we highlight another aspect of random matrix universality relevant to DNN loss surfaces.
    Recent work \cite{arous2021exponential} has shown that the so-called `self averaging' property of random matrix determinants is very much more universal than previously thought.
    The self-averaging of random matrix determinants has been used in the spin glass literature both rigorously and non-rigorously (e.g. \cite{fyodorov2004complexity,fyodorov2005counting,auffinger2013random,baskerville2021loss,baskerville2022spin} inter alia) and is the key property that produces the exponentially large/small number of local optima repeatedly observed.
    We argue that insights into the geometry of DNN loss surfaces can be conjectured from quite general assumptions about the Hessian and gradient noise and from the general self-averaging effect of random matrix determinants.

\subsection{Correlated noise models for neural network loss surfaces}\label{sec:minor1_intro}
Spin glass and statistical physics based models provide an important perspective on the geometric and statistical properties of neural network loss surfaces, as is extensively explored in Chapters \ref{chap:general_activation_functions} and \ref{chap:spin_glass_gans}, alongside prior work in the literature.
Part of the appeal of these approaches is their ontological separation from classical approaches to analysis and methods of proof in statistical learning theory.
Having defined a model and settled on stochastic gradient descent (or a variant) as the optimisation approach, a natural question is: does stochastic gradient descent converge under some assumptions and what, if any, guarantees are there on the parameters to which it converges?
Questions like this are well-studied in the statistical learning and optimisation literature \cite{polyak1992acceleration,varre2021last}, but in the context of neural network this work is of limited applicability as the known results all rely on properties that are not possessed by neural networks, such as convexity of the loss surface (as a function of the network weights).
Some recent work has established convergence results using weaker assumptions like the PL inequality \cite{belkin2021fit} and \cite{PDLT-2022} has developed a theory of neural network training dynamics based on perturbation analysis.
In aggregate, there are many separate results giving guarantees on SGD under a variety of assumptions, some of which are plausible for neural networks but what exists is far short of a complete theory.
The results based on spin glass models, as seen in Chapter \ref{chap:general_activation_functions} and \ref{chap:spin_glass_gans}, are quite different in nature from these SGD convergence results, providing insight into the overall structure and complexity of the loss surfaces on which SGD operates.
These approaches are able to capture much more of the genuine complexity of the loss surfaces of real neural networks than the classical SGD convergence analyses, however the results they provide are somewhat like descriptive sketches of the the loss surfaces, unlike the precise convergence guarantees of the classical analyses.
In Chapter \ref{chap:gadam} we present work that bridges that gap between these two parallel streams of thought.
Concretely, we obtain several variations on SGD convergence results, particularly in the case of \emph{iterate averaging}.
Iterate averaging is a well-known technique in stochastic optimisation, where the parameter iterates $\vec{w}_k$ are simply averaged to produce the new sequence \begin{align}
    \hat{\vec{w}}_t = \frac{1}{t}\sum_{k=1}^t \vec{w}_k.
\end{align}
Intuitively, this simple averaging should have the effect of reducing the variance in the parameter estimates, and indeed this very fact is critical in some convergence proofs, such as that for Adam \cite{kingma2014adam} given in \cite{reddi2019convergence}.
That being said, to the best of our knowledge there has been no explicit theoretical work analysing the generalisation benefit of iterate averaging. Whilst \cite{izmailov2018averaging} propose that iterate averaging leads to ``flatter minima which generalise better'', flatness metrics are known to have limitations as a proxy for generalisation \cite{dinh2017sharp}.
\cite{martens2014new} show that the iterate average convergence rate for both SGD and second-order methods are identical, but argue that second-order methods have an optimal pre-asymptotic convergence rate on a quadratic loss surface. Here, pre-asymptotic means before taking the number of iterations $t \rightarrow \infty$ and quadratic means that the Hessian is constant at all points in weight-space.
The analysis does not extend to \textit{generalisation} and no connection is made to adaptive gradient methods, nor to the importance of the high parameter-space dimensionality of the problem, both of which are addressed by our approach in Chapter \ref{chap:gadam}.
Amendments to improve the generalisation of adaptive methods include switching between Adam and SGD \cite{keskar2017improving} and decoupled weight decay \cite{loshchilov2018decoupled}, limiting the extent of adaptivity \cite{chen2018closing,zhuang2020adabelief}. We incorporate these insights into our algorithms but significantly outperform them experimentally. The closest algorithmic contribution to our work is \textit{Lookahead} \cite{zhang2019lookahead}, which combines adaptive methods with an exponentially moving average scheme.

\medskip
The key contribution of Chapter \ref{chap:gadam} is to introduce spin glass like statistical models for neural network loss into the realm of SGD convergence results.
In particular, we make use of a general stationary Gaussian process model for the noise of the loss surface which is a generalisation of the spin glass models used prior work and our own and bring two important benefits.
Firstly, these models are intrinsically amenable to asymptotic analysis in the regime of very large parameter dimensionality, indeed this kind of asymptotic analysis is our focus in Chapters \ref{chap:general_activation_functions} and \ref{chap:spin_glass_gans}.
As there, this is an important feature of any analysis of neural networks, as virtually all successful modern applications use large networks with very many parameters.
Secondly, these loss surface models are inherently models of statistical dependence between the noise on loss surface gradient iterates, a feature which, again, is central to the calculations in Chapters \ref{chap:general_activation_functions} and \ref{chap:spin_glass_gans}.
In the context of SGD convergence results and iterate averaging, statistical dependence between gradient iterates is essential for a realistic analysis, as the weights, and hence gradients, at each iteration of stochastic gradient descent are clearly not independent.
Beginning with a simple model of independent, isotropic Gaussian gradient noise, we first establish a basic result for SGD with iterate averaging in the high-dimensional regime, exhibiting explicitly the variance reduction effect of iterate averaging compared to standard SGD.
We then replace the inadequate and na\"{i}ve assumption of independent gradient noise with a Gaussian process model for the loss noise, from which we derive a dependent model for the gradient noise.
In this setting, we prove a generalised convergence result for SGD and SGD with iterate averaging, again demonstrating the variance reducing effect of iterate averaging but also providing insights into the effect of learning rate which derives directly from the dependence between gradient iterates.
We additionally establish a sequence of results for variations on the basic Gaussian process noise model and also for certain adaptive gradient descent algorithms.
Overall, our work provides an entirely novel approach to the modeling and analysis SGD algorithms which incorporates important properties of modern neural networks and creates connections between two previously separate approaches in the study of their training. Our novel perspective on the issue of SGD convergence and iterate averaging provides insight into the interaction between iterate averaging, adaptive gradient descent methods and learning rates, which helps to explain why most experimental results with iterate averaging may have historically been poor.

\subsection{Practical application of random matrix loss surface models for hyperparameter tuning}\label{sec:minor2_intro}
A unifying feature of all work in this thesis is the study of neural networks via models of their loss surfaces.
Our work shows how such models can be developed and analysed to shed light on the important features such as the configuration of local optima and the spectral outliers of loss surface Hessians, both of which are relevant to gradient-based optimisation of f neural networks' parameters.
As important as these studies are for advancing the relatively primitive theoretical understanding of what has become a ubiquitous and indispensable approach to machine learning, the immediate practical applications are quite limited.
The spin-glass models of Chapters \ref{chap:general_activation_functions} and \ref{chap:spin_glass_gans} are largely without any direct practical application, being too crude a statistical modern for practical neural networks.
We demonstrate in Chapters \ref{chap:spacings} and \ref{chap:univ} that universal local random matrix theory statistics can be used to build much more realistic models of neural network loss surfaces and yield detailed predictions about spectral outliers of their Hessians.
It is beyond doubt that such results about spectral outliers are of practical use, as clearly demonstrated in \cite{JMLR:v23:20-1258}, where the results are used to derive practical and effective scaling rules for learning rates.
Our results considerably expand and substantiate those of \cite{JMLR:v23:20-1258}, but it has not been demonstrated that these much more precise results add anything practically over the cruder and less rigorous approach of their antecedents.
Chapter \ref{chap:damp} introduces an entirely new application of random matrix theory techniques to neural network loss surfaces, producing immediate practical benefit to the training of real-world networks.
\medskip

The founding idea of Chapter \ref{chap:damp} is a simple observation about a very common numerical `hack' used in several standard variants of stochastic gradient descent.
Let $L(\vec{w})$ be the loss surface of some neural network with parameters $\vec{w}\in\R^N$ and let $H = \nabla^2 L$ be its Hessian.

Stochastic gradient descent updates weights according to the rule
\begin{align}
    \vw_{k+1} = \vw_{k} - \alpha_k\nabla L
\end{align}
where $\vec{w}_k$ are the network parameters after $k$ iterations of SGD and at each iteration a different batch is used. $\alpha_k>0$ is the \emph{learning rate} which, in the simplest setting for SGD, does not depend on $k$, but in general can be varied throughout training to achieve superior optimisation and generalisation.
The general form of adaptive optimiser updates is 
\begin{align}
    \vw_{k+1} = \vw_{k} - \alpha_kB^{-1}\nabla L
\end{align}
where $B$ is a \emph{pre-conditioning matrix}.
The essential idea of adaptive methods is to use the pre-conditioning matrix to make the geometry of $L$ more favourable to SGD.

One approach is to take $B$ to be diagonal, which can be thought of as having per-parameter learning rates adapted to the local loss surface geometry.
More generally, one might seek an approximation $B$ to the local loss surface Hessian, effectively changing the basis of the update rule to a natural one, with per-direction learning rates.
Alternatively, if $B \approx H$ then the local quadratic approximation to the loss surface, i.e. the second-order term in a Taylor expansion, is isotropic in weight space.
What both of these approaches have in common is that they in principle allow for bigger steps (i.e. larger $\alpha_k$, as the different scales of the $\nabla L$ in the different parameters are normalised. Indeed, a standard approach for diagonal $B$ is to construct a diagonal approximation to $H$. Without this, $\alpha_k$ must essentially be tuned to be so small that the change of $\vw$ in the direction of the largest component of $\nabla L$ is not too large.
For Adam \cite{kingma2014adam}, the most commonplace adaptive optimiser in the deep learning community, $B$ is given by the  diagonal matrix with entries $\frac{\sqrt{\langle{g}^{2}_{k}\rangle}+\epsilon}{\langle{g}_{k}\rangle}$. Here $\vg$ is the loss gradient and $\langle\cdot\rangle$ denotes an empirical exponential moving average or iterations.

For many practical problems of interest, the test set performance of adaptive gradient methods is significantly worse than SGD \cite{wilson2017marginal}---a phenomenon that we refer to as the \textit{adaptive generalisation gap}. As a consequence of this effect, many state-of-the-art models, especially for image classification datasets such as CIFAR \cite{yun2019cutmix} and ImageNet \cite{xie2019selftraining,cubuk2019randaugment}, are still trained using SGD with momentum. Although less widely used, another class of adaptive methods which suffer from the same phenomenon \cite{tornstad2020evaluating} are \emph{stochastic second order methods}, which seek to alter the learning rate along the eigenvectors of the Hessian of the loss function. KFAC~\cite{martens2015optimizing} uses a Kroenecker factored approximation of the Fisher information matrix (which can be seen as a positive definite approximation to the Hessian \cite{martens2014new}). Other methods use Hessian--vector products \cite{dauphin2014identifying,martens2010deep} in conjunction with Lanczos methods and conjugate gradients \cite{meurant2006lanczos}. 
All second order and adaptive gradient methods, are endowed with an extra hyper-parameter called the damping or numerical stability co-efficient respectively. This parameter limits the maximal learning rate along the eigenvectors or unit vectors in the parameter space respectively and is typically set to a very small value by practitioners.

In principle there is no reason why a certain parameter gradient should not be zero (or very small) and hence the inversion of $B$ could cause numerical issues. This is the original reason given by \cite{kingma2014adam} for the numerical stability coefficient $\epsilon$. Similarly so for KFAC for which $B = \sum_{i}^{P}\lambda_{i}\vphi_{i}\vphi_{i}^{T}$ where $\{\lambda_{i},\vphi_{i}\}_{i=1}^P$ are the eigenvalue, eigenvector pairs of the kronecker factored approximation to the Hessian.
Hence to each eigenvalue a small damping coefficient $\delta$ is added. Whilst for both adaptive and second order gradient methods, the numerical stability and damping coefficients are typically treated in the literature as extra nuisance parameters which are required to be non-zero but not of great theoretical or practical importance, we strongly challenge this view. In Chapter \ref{chap:damp}, we relate these coefficients to the well known linear shrinkage method from random matrix theory. It is clear from a random matrix theory perspective, that the sub-sampling of the Hessian will lead to the creation of a noise bulk in its spectrum around the origin, precisely the region where the damping coefficient is most relevant.
We show, both experimentally and theoretically, that these coefficients should be considered as extremely important hyper-parameters whose tuning has a strong impact on generalisation. Furthermore, we derive from a random matrix theory additive noise model of the loss surface Hessian a novel algorithm for their online estimation, which we find effective in experiments on real networks and datasets.

\subsection{Mathematical contributions}
We end this section with a brief summary of the purely mathematical contributions of this thesis, much of which has been covered above but in the context of their applications.

Due to the presence of an additive term deforming the GOE matrix, in Chapter \ref{chap:general_activation_functions} we are forced to use different methods to obtain the complexity results analogous to \cite{auffinger2013random} and in so doing provide a novel approach to these calculations.
\cite{auffinger2013random} starts by computing the index-specific complexity and then sums over index to obtain the non-specific complexity.
By contrast, we use supersymmetric methods to first obtain the non-specific complexity and then use the large deviations principle to reintroduce the index dependence.
To the best of our knowledge, this approach has not been used before, though there are of course many works that perform the first part of this calculation for various models.

In Chapter \ref{chap:spin_glass_gans} we make use of the Coulomb gas method to calculate a random matrix determinant as part of the complexity calculation, which is entirely routine, however we also provide a proof of the validity of the Coulomb gas method for the relevant matrix ensemble.
The proof structure is a standard matter of establishing complementary upper and lower bounds.
The proof of the upper bound makes use of standard probabilistic inequalities and properties of Gaussians, however we use the supersymmetric method integral representations to derive error bounds on the mean spectral density which are the key ingredient in the proof of the lower bound.

Finally, in Chapter \ref{chap:univ} we prove a novel result for the limiting spectral measures of additions of random matrices.
It is well known \cite{anderson2010introduction,voiculescu1992free} that the sum of two free independent random matrices with well defined limiting spectral measures has a limiting spectral measure given by the free convolution of the two.
We are able to establish the same free convolutional limiting spectral measure but requiring only that one of the matrices obeys quantum unique ergodicity.
The proof of this result is also a novel application of quantum unique ergodicity, as we leverage a supersymmetric representation to compute the limiting spectral and use the defining quantum unique ergodicity property to compute the integral over the matrix eigenvectors.

\section{Literature review of deep learning theory}
We close this chapter with a broad review of the literature on deep learning theory.
This is a field experiencing a tremendous amount of activity so our review shall be far from exhaustive.
We will give particular attention to the literature related to random matrix theory, but shall also seek to highlight the other broad approaches that have attained some prevalence.

\subsection{Random matrix theory}
\paragraph{Random and complex landscapes}The work most closely related to our own began with \cite{choromanska2015loss,choromanska2015open,sagun2014explorations} where the connections between neural network loss surfaces and spin glasses were first introduced and studied, with the underpinning mathematical results being drawn from the random matrix theory literature such as \cite{fyodorov2004complexity,fyodorov2005counting,auffinger2013random}; we discuss these works in detail elsewhere in this chapter and the next.
In the same lineage of work are more recent notable examples such as \cite{ros2019complex,mannelli2019afraid,arous2019landscape} which can be summarised as the study of high-dimensional signal-plus-noise models.
These works avoid any direct connection to neural networks, instead focusing on much simpler random matrix and tensor models that act as playgrounds for stochastic gradient descent on high-dimensional loss surfaces.
This approach is of course inspired by \cite{choromanska2015loss} and these works similarly consider issues of loss surface complexity, but with the explicit inclusion of extra structure, or `signal'.
This signal was notably lacking from \cite{choromanska2015loss}, as the spin-glass is really just a model of pure noise.
Intuitively, one expects that the loss surfaces of real neural networks contain some underlying structure induced by the structure of the data and the network itself, but that a considerable component of noise is also induced on the surface by the noise on the data and also possibly the weights and biases themselves.
By creating simple, paired-down loss surface models containing the same kind of high-dimensional noise present in the spin glass, but with some signal (or structure) injected, these works are able to study questions about the presence and prevalence of \emph{spurious minima} i.e. local minima of the noisy loss surface that are uncorrelated with the true minima of the noise-less surface. They uncover phase transitions between chaotic surfaces on which the structure-induced minima are swamped by spurious minima and surfaces which, though they contain many noise-induced minima, the structure of the minima is such that the signal is still recoverable. 

\paragraph{Random neural networks}
In the line of work discussed above, random matrices arise somewhat indirectly in the study of neural networks via the Kac-Rice approach to landscape complexity analysis.
Since neural networks are constructed using, and parametrised by, weight matrices in each of their layers, one can naturally seek a theory of \emph{random} neural networks by considering these weight matrices to be random.
\cite{pennington2017geometry} bridged the gap between studies of landscape complexity and random neural networks by considering networks with i.i.d. normal weights applied to i.i.d. normal data and computing the limiting spectral density of their Hessians in the large parameter number limit.
They decompose the Hessian as the sum of a positive semi-definite matrix (often called to Gauss-Newton matrix elsewhere \cite{martens2016second}) and a matrix that contains all the dependence on the residuals (i.e. the error terms between the network predictions and the truth values).
With this decomposition, they make assumptions of free independence to enable the use of tools from free probability to compute the limiting spectral densities.
By assuming also an i.i.d. Gaussian form of the residuals parameterised by some variance $\epsilon$, they are able to describe the spectra of neural networks Hessian at different loss values a compare with experiment.
Random networks were also considered in \cite{louart2018random} in the context of random feature ridge regression i.e. a 1-layer neural network with MSE loss and an L2 ridge regularisation penalty for which only the final layer is trained.
The first layer, being untrained, acts as a random transformation of the the input data and then the weights of the final layer have a unique solution known in closed form, since the final layer is simply a linear ridge regression on the random features.
Since the final layer weights can be solved in closed form, a closed form is available for the training error which is found to be given in terms of the resolvent $Q = (N^{-1}\Sigma^T\Sigma + \gamma I)^{-1}$ where $\Sigma = \sigma(WX)$ are the random features produced by the random weights $W$ and input data points $X$ and $N$ is the number of random features (i.e. the width of the hidden layer).
The proofs rely largely on concentration properties of sub-Gaussian random variables to establish that various random matrix quantities concentrate on their expectations.
In a related work \cite{pennington2017nonlinear} 1-layer neural networks with random weights were considered.
The authors compute the limiting spectral density of the Gram matrix $Y^TY$ of the network output $Y$.
This work was the first in which the non-linearities introduced by neural network activation functions were handled directly and analytically in the setting of random matrix theory, since \cite{louart2018random} was restricted to polynomial activation functions.
The weight entries and the data entries are assumed to i.i.d. Gaussians and the proof of the limiting spectral density uses the moment method of random matrix theory.
An interesting consequence of the results is that there exist certain non-linear activation functions for which the Gram matrix spectrum is the Marcenko-Pastur distribution, so that the spectrum is preserved through the non-linear activation function.
The authors conjecture that these ``isospectral'' activation functions may have beneficial practical properties for training, as the spectral statistics remain constant through the layers, an idea somewhat reminiscent of batch norm \cite{ioffe2015batch}.
\cite{benigni2019eigenvalue} extends the results of \cite{pennington2017nonlinear} to more general (i.e. sub-Gaussian) entry distributions on the network weights and the data, using again a moment method proof.
They also extend to the case of multiple layers, though the results in that case are very intricate and opaque.
Continuing again in this line of work, \cite{adlam2022random} extends the analysis to 1-layers random networks with random biases ans shows that the distribution of the biases induces something like a mixture over activation functions.
\cite{pennington2018emergence} considers the input-output Jacobian $J$ of random multi-layer networks using the techniques of free probability theory to derive the spectrum of the Gram matrix $JJ^T$.
Using these results, they are able to derive necessary and sufficient conditions on the spectra of the weight matrices to give a stable spectrum (i.e. not no explosion nor collapse) in the large network depth limit.
These results were subsequently generalised and given a fully-rigorous proof in a series of papers by Pastur and collaborators \cite{pastur2020random,pastur22_gen_iid,pastur22_ortho}.
The first paper in the sequence considers the Gaussian case, as in \cite{pennington2018emergence}, with the chief difficulty being that the free independence that is required to apply the streamlined free probability argument given in \cite{pennington2018emergence} is not apparent.
The second paper extends to general i.i.d. distributions with at least four finite moments and the third extends to weights matrices with orthogonal distributions (so not i.i.d. entries).
Another perspective on random neural networks is given in the works \cite{schoenholz2017correspondence,yang2019mean}, where the techniques of mean field theory are applied to the standard multi-layer perceptron architectures, firstly with linear or ReLU activations and then with more general activations and batch normalisation.
The training loss of the network plays the role of the Lagrangian and the partition function is computed by explicitly integrating out the random (i.i.d. Gaussian) weights and biases. 
In the case of batch normalisation, the authors are able to use the mean field techniques to make predictions about instabilities (e.e. due to gradient explosion) of very deep networks in the presence of batch normalisation.
Beyond the question of why does SGD work at all for deep neural networks, there are various phenomena observed in their training and use that lack adequate theoretical explanations.
One such is the \emph{double/triple descent phenomenon}, which is commonly observed in large modern deep neural networks but is at odds with classical statistical learning theory.
Standard results from statistical learning theory dictate that the best attainable test loss of a particular model decreases as the number of parameters $N$ of the model increases, but only up to a point beyond which the loss increases again.
This is a reflection of the classical \emph{bias-variance} trade-off  \cite{hastie2009elements} which states that the expected test error of a machine learning model can be decomposed into two additive terms, bias and variance, which account for different sources of error in the fitting process.
High variance means that there is high variation in the estimated parameters between different sampled instances of the training set which indicates that the model tends to systematically fit to the noise in the training data, rather than the underlying structure (called \emph{overfitting}).
High bias means that the test error over different sampled instances of the training set is biased away from zero, indicating that the model tends to systematically fail to identify meaningful generalisable structure in the data (called \emph{underfitting}).
It is intuitive that a model with too few parameters will tend to underfit, as the model lacks the expressive capacity to capture the structure in the data. 
On the other hand, a model with too many parameters (i.e. more than are really needed to capture the structure in the data) will tend to overfit as it is has spare capacity that can be used to interpolate noise in the training data, which of course drives down the training loss, but at the expense of increasing the test loss.
All of this holds for classical approaches to machine learning, i.e. broadly those before the deep learning revival of the 2010s, however repeated empirical observations with increasingly larger deep networks have revealed that this classical picture has its limits.
Modern deep networks used in computer vision applications are routinely chosen to have 10s of millions of parameters, which by any reasonable measure is considerably more than would be required to express the true structure in the data and is indeed sufficient to allow for perfect interpolation of the training data \cite{he2016deep}.
Modern transformer networks used extensively in natural language processing are larger still \cite{brown2020language} with 100s of \emph{billions} of parameters.
Repeatedly and in multiple domains, it has been observed that dramatically increasing the number of networks parameters and also the training time can lead to ever better test set performance \emph{even when training data are near perfectly interpolated}.
This phenomenon was dubbed the \emph{double descent}, referring to the shape of the graph of test error against number of parameters.
Classically, this graph has a single local minimum at the point of bias-variance balance, but very large deep neural networks have revealed a second, lower minimum in the greatly (``abundantly'') over-parameterised region \cite{zhang2016understanding,zhang2021understanding,belkin2019does,belkin2019reconciling}.
Prior works attempted to analyse this phenomenon in the simplest cases of linear regression models \cite{belkin2018overfitting}, but the key contribution of \cite{adlam2020understanding} was to analyse the effect of parameter number on single hidden layer random networks.
Neural networks with a single hidden layer are the simplest example of a model in which the number of trainable parameters $N$ can be specified separately from the input data dimension $d$ and target data dimension $C$, since in a  model with no hidden layers (e.g. linear or logistic regression) $N$ is necessarily equal to $dC$, whereas the width of even a single hidden layer can be specified arbitrarily.
The authors were able to show that single hidden layer networks with random i.i.d. Gaussian weights trained on entirely random data with random labels display a double descent, even a \emph{triple descent}, with a third test error minima in an extreme ``hyperabdundant'' parametrisation region.
Much like the earlier work \cite{couillet2019random}, the test error is expressed as a certain random matrix resolvent which is in turn computed by determining the limiting spectral density of a certain random matrix via tools from free probability theory and invoking notions of random matrix universality to replace the complicated, intractable matrix ensembles arising from the network with certain independent Gaussian matrices.
This work produces an immediate insight: the double (triple) descent phenomenon is not unique to deep neural networks, nor even to the type of data on which they are typically trained or the training procedure, but rather it is a ``background'' property of over-parametrised non-linear models and generic data.

\paragraph{Spectra of neural networks}
The works discussed so far consider random neural networks and random matrices in neural networks \emph{ex-ante}, i.e. modeling assumptions are made, or models constructed, that explicitly introduce randomness to neural networks or their loss surfaces.
Their is another line of work which is better characterised as \emph{ex-post} randomisation, wherein neural networks are directly studied and, for example, spectral properties of their loss surface Hessians or weights are analysed.
For the fist time in \cite{papyan2018full,papyan2019measurements}, the spectra of loss surface Hessians of real-world neural networks were approximated and analysed.
For practical modern neural networks, the loss surface Hessian is of course far too large to even store in memory, let alone compute via automatic differentiation or eigen-decompose, having $N^2$ entries, where the number of network parameters $N$ is typically $10^7$ or more.
The key numerical advance in these works is the application of Lanczos iteration methods \cite{lanczos1950iteration,meurant2006lanczos} to compute high-quality approximations to the spectral density of very large matrices given only the matrix-vector multiplication function $\mathcal{M}_H : \mathbb{R}^N\rightarrow \R^N$ with $\mathcal{M}_H(\vec{v}) = H\vec{v}$ and not the whole matrix $H$.
This can be combined with the Pearlmutter trick \cite{pearlmutter1994fast} which computes \begin{align*}
    \frac{\partial^2 l}{\partial w_i \partial w_j}\vec{v} = \frac{\partial}{\partial w_i} \left(\vec{v}^T\frac{\partial l}{\partial w_j}\right)
\end{align*}
which is very much amenable to automatic differentiation in modern deep learning frameworks.
Actually, this approach was pioneered contemporaneously by Granziol and collaborators in a sequence of pre-prints for which the best reference is \cite{JMLR:v23:20-1258}.
One of the key insights in those works was to highlight the very considerable discrepancy between the spectra of real neural network Hessians and those of standard canonical random matrix models such as the GOE that is assumed by spin glass models such as \cite{choromanska2015loss} and in \cite{granziol2020beyond}, it was proposed that the spectra of products of canonical random matrix ensembles can be used to obtain agreement with certain aspects of the spectra of real neural networks, in particular their considerable rank degeneracy.

These empirical analyses uncover rich and interesting structure in the spectra of real deep neural networks, in particular the spectra clearly display a bulk and some large outliers.
The outliers appear to be directly attributable to the classes in a typical classification problem (i.e. one outlier per class) and naturally one expects from random matrix theory that the bulk corresponds to noise \cite{potters2020first}.
There is further structure still, with the discovery in an later work \cite{papyan2020traces} of a group of eigenvalues outside of the bulk\footnote{Though not stated by the author, this extra group of outlier eigenvalues must clearly be outside the Tracy-Widom region as well.} but much smaller than the main outliers.
There are typically $C(C-1)$ of these outliers, for a $C$ class classification problem, so they appear to correspond somehow to inter-class correlations.

Rather than considering loss surface Hessians, another line of inquiry has directly analysed the spectra of neural network weight matrices before, during and after training.
\cite{martin2018implicit} consider several types of network trained on real datasets and look at the spectra of their weights matrices at initialisation and as training progresses.
They identify several distinct phases of training from these spectra, beginning with full classical random matrix behaviour at initialisation and developing towards some heavy-tailed distribution leading to the conjecture that neural networks are implicitly regularised by some process inducing these heavy tailed spectra as training proceeds. Note that the idea of implicit regularisation of neural networks via stochastic gradient descent pre-dates this work by several years \cite{neyshabur2014search,neyshabur2017exploring,neyshabur2017geometry,neyshabur2017implicit}.
Finally, we mention \cite{thamm2022random} in which the spectra of random and trained neural network weight matrices was analysed but on the \emph{local} scale, rather than the global scale pursued by \cite{martin2018implicit}.
This work followed on from our own in Chapter \ref{chap:spacings} \cite{baskerville2022appearance} and similarly discovered the robust presence of universal GOE random matrix spacing statistics in the spectra. 

\subsection{Other approaches}
We mentioned above some mean-field approaches to the analysis of neural networks, but this would not be complete without also mentioning the recent work of Roberts and Yaida \cite{roberts2021principles} in which this subject is developed in considerable depth.
The authors proceed incrementally from linear networks at initialisation (the simplest case), to non-linear networks and ultimately training dynamics via a perturbation theory approach.
This analysis relies heavily on the \emph{neural tangent kernel} which can be introduced quite simply by considering the loss derivatives via the chain rule:
\begin{align*}
    \frac{\partial L}{\partial \theta_a} = \sum_i \frac{\partial L}{\partial z_i} \frac{\partial z_i}{\partial \theta_a}
\end{align*}
where $\vec{z}$ is the network output which is fed into the loss $L$.
A single step of stochastic gradient descent will update the weights $\vec{\theta}$ by taking a small step of scale $\eta$ along the negative gradient direction, so that the leading order (in $\eta$) change in the loss is \begin{align*}
    \Delta L = -\eta \sum_{i,j}\sum_{a,b} \frac{\partial L}{\partial z_i}\frac{\partial L}{\partial z_i} \frac{\partial z_i}{\partial \theta_a}\ \frac{\partial z_i}{\partial \theta_b}
\end{align*}
which leads to the identification of the neural tangent kernel \begin{align*}
    K_{i,j} = \sum_{a,b}\frac{\partial z_i}{\partial \theta_a}\ \frac{\partial z_i}{\partial \theta_b}.
\end{align*}
The neural tangent kernel can be seen to largely govern the dynamics of stochastic gradient descent for very wide networks (i.e. those with some fixed number of layer but very many parameters in each layer), see e.g. \cite{jacot2018neural,adlam2020neural}.

\medskip
Building on the above-mentioned decomposition of neural network Hessian spectra into components attributable to class centres and inter-class correlations \cite{papyan2020traces}, the concept of \emph{neural collapse} has been advanced.
Empirical studies of network pre-activations in \cite{papyan2020prevalence} discovered that, in networks trained to good accuracy, the pre-activations coalesce around $C$ clusters, one for each class in the classification problem.
Indeed, as training progresses the pre-activations converge to very low variance around the class cluster centres and the cluster centres themselves converge to an equiangular tight frame.

Another recent line of work studies neural networks in their capacity as function approximators \cite{https://doi.org/10.48550/arxiv.2009.10713} and attempts to characterise using the tools of mathematical analysis the sets of functions that can be well approximate by neural networks.
A 2-layer (i.e. 1 hidden layer) network can be expressed as a random feature model \begin{align*}
    f(\vec{x}, \vec{a}) = \frac{1}{m}\sum_{j=1}^m a_j \phi (\vec{x}; \vec{w}_j), ~~ \phi(\vec{x}, \vec{w}) = \sigma(\vec{x}^T\vec{w}).
\end{align*}
This expression can be rewritten as an integral by defining an atomic probability measure $\pi = m^{-1}\sum_{j=1}^m \delta_{\vec{w}_j}$ over the first layer weights $\{\vec{w}_j\}_j$ \begin{align*}
    f(\vec{x}, \vec{a}) = \int a(\vec{w}) \phi (\vec{x}; \vec{w}) d\pi(\vec{w}),
\end{align*}
which suggests the generalisation of this expression to any probability measure $\pi$, so producing a type of random neural network with marginalised first layer weights.  
In this construction, the 2-layer MLP network can be viewed as a Monte Carlo integration approximation to this more general object.
An important insight about the role of the curse of dimensionality in deep learning is revealed by this formalism.
Classical function approximation theory typically constructs approximations of a function $f$ by defining some Sobolev space with a convenient basis, say of polynomials. If $m$ is the number of free parameters in the approximation (e.g. the maximum degree of the polynomial basis) and $d$ is the input dimension of $f$, then one obtains an approximation error that scales something like $m^{-\alpha/d}$ for some $\alpha>0$  defined by the details of the chosen approximation space.
As the input dimension $d$ grows, this error term becomes less and less favourable, requiring exponentially more free parameters $m$ to achieve the same approximation error.
This contrasts sharply with the above Monte Carlo integration interpretation of a 2-layer MLP, which has an error term with the standard MC scaling of $m^{-1/2}$, which crucially is independent of the input dimension $d$.
This analysis approach provides some insight into how neural networks appear to overcome the curse of dimensionality in their input space faced by other approaches to machine learning.
The results in \cite{https://doi.org/10.48550/arxiv.2009.10713} go further and in fact identify precisely the function spaces for which 2 layer MLPs can provide good approximations.

\cite{belkin2021fit} considers the success of stochastic gradient descent at finding high quality minima for deep neural networks.
As we have already discussed, classical optimisation theory holds that finding global minima of non-convex functions is generally intractable and \cite{belkin2021fit} argues that the considerable over-parametrisation of modern neural networks implies that their loss surfaces are filled with many local minima and they are generically not even locally convex around those minima.
The \emph{PL inequality} \cite{polyak1964gradient,lojasiewicz1963topological} for a loss function $L$ with constant $\mu$ is $\frac{1}{2}\|\nabla L(\vec{w})\|^2 \geq \mu L(\vec{w})$ and, combined with a smoothness condition, is sufficient to guarantee exponential convergence of stochastic gradient descent \cite{belkin2021fit}, but the PL condition is much weaker than even local convexity.
The conclusion of this line of work is broadly that the classical picture that lack of convexity and numerous local minima mean that stochastic gradient descent on neural networks is doomed to fail is overly pessimistic and weaker, more plausible conditions may suffice to provide expectation of convergence.

\chapter{Mathematical tools}\label{chap:maths}
This chapter aims to provide a self-contained introduction to the main mathematical tools required in the subsequent chapters, intended to be accessible to a mathematical audience with no previous familiarity with random matrix theory.

\section{Introduction to random matrix theory}
Random matrix theory provides much of the mathematical context and insight for the results in this thesis, as well as providing most of the techniques used in the calculations.
It is a large a diverse field touching many areas of pure and applied mathematics and physics and we shall not attempt to provide comprehensive introduction.
The classic introduction is Mehta's book \cite{mehta2004random}.
Thorough and mathematically orientated modern treatments can be found in the books by Anderson, Guionnet and Zeitouni \cite{anderson2010introduction}, Tau \cite{tao2012topics} and Meckes \cite{meckes2019random}.
Accessible and application orientated introductions are given by \cite{livan2018introduction} and \cite{potters2020first}.
A detailed introduction to modern topics in a mathematically rigorous style can be found in \cite{erdos2017dynamical}.
Given the breadth of random matrix theory, only a fraction of its concepts and tools are required in this thesis and so we restrict this introduction to those.

\subsection{Random matrices}
A random matrix is no more nor less than one would expect, namely a matrix-valued random variable.
Such objects are entirely natural in almost any branch of applied mathematics or statistics.
Consider for example a sample of $N$ data points each being represented as a tuple of $M$ real values, such as 2-tuples of latitude and longitude for locations of house or 500-long tuples of returns data for the S\&P 500 index.
It is natural, at least from the perspective of computational convenience, to stack these data points into an array $X$ of shape $N\times M$ with each row corresponding to a single sample.
From the perceptive of a pure mathematician thinking of matrices as representations of linear maps on vector spaces, $X$ does not appear to be a matrix, but just a collection of number conveniently packed into a array.
Suppose that the $N$ samples are $\vec{x}_1, \ldots, \vec{x}_N$ drawn from a multivariate Gaussian distribution $\mathcal{N}(0, \Sigma)$.
The information contained in the sample is entirely represented by this sequence in $\mathbb{R}^N$, so what is the purpose of stack them into a `matrix' $X$?
One answer is, of course, numerical convenience and efficiency. 
For example, suppose that $\Sigma$ is known and we wish to construct the standardised variables $\vec{z}_i = \Sigma^{-1/2} \vec{x}_i$.
One can view this as a sequence of $N$ matrix-vector operations, but it is more mathematically compact and numerically efficient to instead view it as a single matrix-matrix operation $Z = \Sigma^{-1/2}X$.
There are, however, deeper and richer reasons to consider $X$. 
Consider the matrix $S = \frac{1}{N}X^T X$ - an $M\times M$ positive semi-definite symmetric matrix.
One can clearly write \begin{align*}
    S_{ij} = \frac{1}{N} \sum_{k=1}^N (\vec{x}_k)_i (\vec{x}_k)_j 
\end{align*}
and so $S_{ij}$ is an empirical estimate from $N$ samples of the covariance between the $i$-th and $j$-th coordinates in the data distribution.
The eigenvalues and eigenvectors of $S$ clearly have meaning, for example the eigenvector corresponding to the largest eigenvalue is the direction in $\mathbb{R}^M$ responsible for the most variance in the data.
In the S\&P 500 example above, this direction would correspond to `the market', and in the coordinates example, it may correspond to a major river along whose banks most settlements are found.
We need not restrict ourselves to matrices of the form of $N$ samples of $M$ dimensional variables.
Consider data collected from a telecommunications network on $N$ end-points (or nodes), examples of which include telephone numbers or registered users of instant messaging services.
Let $X_{ij}$ be the number of communication events between end-point $i$ and end-point $j$ observed over some time period.
Properly normalised by the total number of events in the same period, $X_{ij}$ could instead be an empirical estimate of the probability of communication between end-points $i$ and $j$.
Viewing $X$ as a symmetric matrix, not merely and array, and computing its spectral decomposition, one will find that the eigenvectors corresponding to meaningful communities in the network, with the eigenvalues giving an estimate of the relative importance of each community in the network.

\medskip
These examples illustrate a critical point: viewing arrays of random variables (or data) as matrices is not a mere numerical convenience, for one finds that bona fide linear algebraic objects such as eigenvalues and eigenvectors have meaning and structure.
Let us return to the example of a matrix $X$ containing financial data, e.g. share prices or returns, for $M$ assets sampled over $N$ days. 
If $M$ is small compared to a large sample size $N$, then we can expect much of the noise in the samples to average out to produce a matrix $S$ with $M$ meaningful eigenvector representing genuine correlations between the $M$ assets.
In the opposite extreme where $M$ is much larger that $N$, we expect that many of the genuine correlations in the data will be lost in the noise.
But what of the intermediate case, where $M$ and $N$ are of comparable size?
Intuitively, one expects that the strongest signals in the data (such as the the market) will be preserved and clearly visible through the noise in the data, while more subtle signals will be lost.
Translating this into the language of random matrices, the largest eigenvalues (and their eigenvectors) correspond to genuine signal in the data, while the smallest correspond to sample noise.
The obvious question is whether one can separate the signal from the noise, i.e. how many of the largest eigenvalues are signal?
This question can be seen, conceptually, as motivating much of the work in random matrix theory.
Consider any linear algebraic property of a matrix: eigenvalues, eigenvectors, determinant, trace, characteristic polynomial, condition number, etc.
Given a distribution on a matrix, what is the distribution on any of these objects?
If one can answer this question for pure noise random matrices, then one can easily identify matrices that contain signal.
If one can answer the question in the case of signal-plus-noise random matrices, then one can separate the signal from the noise.

\medskip
The above discussion has been rather statistically-focused, but historically random matrix theory was used by Wigner and Dyson \cite{wigner1967random,dyson1962brownian,dyson1962statistical,dyson1970correlations} to provide elegant and powerful models for atomic nuclei.
The governing quantum mechanical equation for an atomic nucleus is the Schr\"{o}dinger equation
\begin{align}
    H \psi_i = E_i \psi_i\label{eq:schrodinger}
\end{align}
where $H$ is an Hermitian operator (the \emph{Hamiltonian}) on an Hilbert space, $\{\psi_i\}$ is a wave functions and $E_i$ are corresponding energy levels.
The physical observables here are the energy levels, but in all but the very simplest of cases (such as a Hydrogen nucleus) they cannot be computed analytically, or even numerically, owing to the complexity of the interaction between the nucleons.
Dyson and Wigner's insight was that the general appearance of energy levels \emph{on average} can be described by simple statistical models of (\ref{eq:schrodinger}) not requiring detailed knowledge of the equation or its solution. To quote Dyson \cite{dyson1962statistical}:

\begin{addmargin}[2em]{2em}%
\emph{The statistical theory will not predict the detailed sequence of levels in any one nucleus, but it will describe the general appearance and the degree of irregularity of the level structure, that is expected to occur in any nucleus which is too complicated to be understood in detail.}
\end{addmargin}

This aspect of random matrix theory will be of particular value in this thesis.
We endeavour to understand properties of very large deep neural networks applied to complicated high-dimensional tasks on real-world data. 
Such models may contain millions of free parameters operating on datasets of millions of samples with many thousands of dimensions per sample and complicated statistical dependence between dimensions.
The dynamics of the model parameters as they are trained are far too complicated to be studied directly.
As with atomic nuclei many decades earlier, the central hypothesis of this thesis and other related contemporary work is that statistical theories of deep neural networks can describe their general properties and be used to understand their behaviour without reference to the intractable details of their training dynamics.

\subsection{Random matrix ensembles}
Probability distributions on matrices are commonly referred to as \emph{ensembles} in random matrix theory.
There are modest number of canonical random matrix ensembles that form the foundation of much of the work in random matrix theory and about which a great deal is known in considerable mathematical detail. 
The importance of each of the canonical ensembles tends to vary between application areas, so we shall restrict ourselves in this section to only those ensembles that feature in the coming chapters. We shall be exclusively interested in real matrices, as the matrices that arise when studying neural networks and machine learning are almost always real.
Moreover, many of the matrices we shall be interested in will be symmetric.
The most important random matrix ensemble for this thesis is the \emph{Gaussian orthogonal ensemble} (GOE).
There is some variation between authors on unimportant normalisation, but we shall say that a $N\times N$ matrix $X \in \R^{N\times N}$ is a GOE matrix, $X\sim \text{GOE}^N$ if 
\begin{align}
    X_{ij} \overset{\text{i.i.d.}}{\sim} \mathcal{N}\left(0, \frac{1 + \delta_{ij}}{2N}\right) \text{ for } i \leq j \text{ and } X_{ij} = X_{ji} \text{ for } i> j,
\end{align}
i.e. $X$ has Gaussian entries, independent up-to symmetry and with twice the variance on the diagonal as off-diagonal.
This specific variance structure allows for a powerful closed-form expression of the law of $X$:
\begin{align}\label{eq:goe_measure_def}
   d\mu(X) = \frac{1}{Z_N}\exp\left(-N\frac{\Tr X^TX}{2}\right) dX
\end{align}
where $Z_N$ is a normalisation constant and $dX$ is simply the standard Lebesgue product measure on the upper-diagonal and diagonal entries of $X$.
Note that the GOE is called an orthogonal ensemble because it possesses symmetry with respect to the real orthogonal group $O(N)$ on orthogonal matrices.
Sampling a matrix $X$ from $\text{GOE}^N$ can be done with a very simple algorithm:
\begin{align*}
    Y_{ij} \overset{\text{i.i.d.}}{\sim} \mathcal{N}(0, 1), ~~ X = \frac{Y + Y^T}{2N}.
\end{align*}

The GOE is a specific case of the more general class of \emph{Wigner} matrices, which have independent (up-to symmetry) Gaussian entries with variance $\sigma_d^2/N$ on the diagonal and $\sigma_u^2/N$ off the diagonal.
Generalising even further, generalised Wigner matrices take the form \begin{align*}
    X_{ij} \overset{\text{i.i.d.}}{\sim} \mu \text{ for } i < j, ~~ X_{ii}  \overset{\text{i.i.d.}}{\sim} \mu_d \text{ and } X_{ij} = X_{ji} \text{ for i > j}
\end{align*}
for any sufficiently well-behaved measures $\mu$ and $\mu_d$ on $\mathbb{R}$.
There are complex Hermitian and quaternionic version of GOE and Wigner matrices for details of which we refer the reader to any standard reference on random matrix theory.

\medskip
An alternative generalisation of the GOE is born of (\ref{eq:goe_measure_def}), which we rewrite as \begin{align*}
       d\mu(X) = \frac{1}{Z_N}\exp\left(-N\Tr V(X)\right) dX
\end{align*}
where $V:\mathbb{R}^{N\times N} \rightarrow \mathbb{R}^{N\times N}$ is defined to be $V(X) = \frac{1}{2}X^TX$.
In deference to its origins in statistical physics, $V$ is often referred to as a potential.
With this rewritten form of the GOE density, one can simply change the definition of $V$ and so obtain different distributions on real symmetric matrices.

\medskip
We shall also encounter matrices distributed with \emph{Haar measure} on the orthogonal group $O(N)$.
The Haar measure on any compact group $G$ \cite{meckes2019random} is the unique measure $\mu_{Haar}$ finite on all subsets of $G$ such that \begin{align*}
    \mu(gS) = \mu(S) ~ \forall g\in G \text{ and } S\subset G.
\end{align*}
The Haar measure can be viewed as the `flat random' measure on $G$ and, in the case $G = O(N)$, a matrix distributed with Haar measure is a uniform random matrix on the real orthogonal group.
Geometrically, a matrix with Haar measure on $O(N)$ is a uniform random basis rotation.
Haar random orthogonal matrices $O$ can be sampled quite simply by sampling $N$ i.i.d. vectors $\vec{x}_i$ with i.i.d. $\mathcal{N}(0,1)$ entries and then applying the Gram-Schmidt algorithm the obtain an orthonormal set of vectors $\vec{o}_1, \ldots, \vec{o}_N$ which are the rows of the Haar-distributed matrix \cite{mezzadri2006generate}.

\medskip
Finally, we mention the real \emph{Ginibre} \cite{anderson2010introduction} ensemble on $N\times M$ matrices which are simply matrices with i.i.d. entries and no symmetry constraint. 
The Ginibre analogue of the GOE is and ensemble of matrices with i.i.d. $\mathcal{N}(0, 1/N)$ entries.

\subsection{Eigenvalues and spectral measures}
Let $X_N$ be any real symmetric random matrix of shape $N\times N$.
The following discussion could equally be presented for any Hermitian random matrix and could be generalised much further at the expense of having to account for non-real eigenvalues.
For the purposes of this thesis, we may restrict our discussion to matrices with real eigenvalues and, to be concrete, let us stick to real symmetric matrices.
Let $\lambda_1 < \lambda_2 < \ldots < \lambda_N$ be the eigenvalues of $X_N$. The \emph{empirical spectral measure} of $X_N$ is defined as \begin{align}
    \hat{\mu}_N = \frac{1}{N}\sum_{i=1}^N \delta_{\lambda_i}
\end{align}
where $\delta_{\lambda}$ is a Dirac $\delta$-function mass at location $\lambda$, i.e. defined by \begin{align*}
    \int_A \delta_{\lambda} = \1\{\lambda \in A\}
\end{align*}
for any set $A \subset \mathbb{R}$.
Since $X_N$ is random, its eigenvalues $\{\lambda_1, \ldots, \lambda_N\}$ are random variable with some joint probably density $p(\lambda_1, \ldots, \lambda_N)$.
$\hat{\mu}_N$ is a probability measure on $\mathbb{R}$ and moreover, it is a \emph{random} probability measure, its distribution being induced by $p(\lambda_1, \ldots, \lambda_N)$.
Imagine constructing many independent samples of $X_N$, hence from $p(\lambda_1, \ldots, \lambda_N)$ and hence of $\hat{\mu}_N$.
Once could imagine averaging the samples of $\hat{\mu}_N$ \begin{align*}
    \frac{1}{m} \sum_{j=1}^m \hat{\mu}_N
\end{align*}
and so obtaining some indication of the average location of the eigenvalues of $X_N$.
Intuitively, one would imagine this average measure becoming a better and better approximation to some absolutely continuous measure (though there is, of course, no general guarantee of such convergence).
Extending this to a concrete mathematical question: does $\E \hat{\mu}_N$ exist, and what is it?
In the same way that one can imagine growing the number of sampled eigenvalues by increasing the number of independent samples of $X_N$, one can also consider a family of distributions on $X_N$, parametrised by dimension $N\in \N$, and let the dimension $N$ for a single sample grow.
In this context, there is another natural question: does $\lim_{N\rightarrow\infty}\hat{\mu}_N$ exist, how strong is the convergence, and what it the limit measure?
When it exists, we shall define \begin{align}
    \mu_{\infty} = \lim_{N\rightarrow\infty} \hat{\mu}_N
\end{align}
to be the \emph{limiting spectral measure} of $X_N$ (being intentionally vague about the strength of convergence, for now).
Likewise, when the expectation exists, we define \begin{align}
    \mu_N = \E \hat{\mu}_N
\end{align}
to be the \emph{mean spectral measure} of $X_N$.
When either of these measure are absolutely continuous with respect to Lebesgue measure, we define \begin{align*}
    \rho_{\infty}(\lambda) = \frac{d\mu_{\infty}}{d\lambda}
\end{align*}
to be the \emph{limiting spectral density} (LSD) and similarly \begin{align*}
      \rho_{N}(\lambda) = \frac{d\mu_{N}}{d\lambda}
\end{align*}
is the \emph{mean spectral density}.
Let us now be concrete and consider some specific examples, beginning with the most famous.

\begin{exmp}[GOE]
Recalling the form (\ref{eq:goe_measure_def}) of the GOE measure on $N \times N$ matrices, we can now explore why it was described as ``powerful''.
Let $X$ be an $N\times N$ GOE random matrix.
Since $X$ is real symmetric, it is an elementary result of linear algebra that $X$ can be written in the form $X = U^T\Lambda U$, where $U\in O(N)$ is a real orthogonal matrix and $\Lambda$ is a real diagonal matrix.
Of course, the diagonal entries of $\Lambda$ are the eigenvalues of $X$ and the rows of $U$ are the corresponding eigenvectors.
But now \begin{align*}
       \exp\left(-\frac{N\Tr X^TX}{2}\right) = \exp\left(-\frac{N\Tr U^T\Lambda U U^T \Lambda U}{2}\right) =  \exp\left(-\frac{N\Tr U^T\Lambda^2 U}{2}\right)=\exp\left(-\frac{N\Tr \Lambda^2}{2}\right)
\end{align*}
which depends only on the eigenvalues of $X$ and not on the eigenvectors.
We must deal with the Jacobian of the change of variables from $X$ to $(\Lambda, U)$.
A standard calculation found in any introductory text on random matrix theory shows that \begin{align*}
    \prod_{1\leq i \leq j\leq N} dX_{ij} =  \Delta(\{\lambda_i\}_{i=1}^N)d\mu_{Haar}(U)\prod_{i=1}^N d\lambda_i 
\end{align*}
where the \emph{Vandermonde} determinant is defined by \begin{align}
    \Delta(\{\lambda_1, \ldots, \lambda_N\}) =\left|\begin{array}{cccc}
        1 & 1 & \cdots & 1 \\
        \lambda_1 & \lambda_2 & \cdots &  \lambda_N\\
        \lambda_1^2 & \lambda_2^2 & \cdots &  \lambda_N^2\\
        \vdots &\vdots &\vdots &\vdots \\ 
\lambda_1^{N-1} & \lambda_2^{N-1} & \cdots &  \lambda_N^{N-1}\\
    \end{array}\right|= \prod_{1\leq i < j\leq N} |\lambda_i - \lambda_j|.
\end{align}
Overall, we see that \begin{align}\label{eq:goe_measure_spectral_decomp}
    d\mu(X) = d\mu_{Haar}(U) \Delta(\{\lambda_i\}_{i=1}^N)\prod_{i=1}^N d\lambda_i \frac{e^{-\frac{N\lambda_i^2}{2}}}{\sqrt{2\pi N}}
\end{align}
where there was of course no need to compute the normalisation constant, as it can simply be written down from the simple Gaussian product measure form in the $\lambda_1, \ldots, \lambda_N$.
The form (\ref{eq:goe_measure_spectral_decomp}) already reveals much about the statistics of the eigenvalues and eigenvectors of $X$.
It is immediately obvious that the eigenvalues are independent of the eigenvectors.
The eigenvectors are Haar-distributed, so they are simply a flat random orthonormal basis of $\R^N$.
The eigenvalues have richer structure, but we can immediately make some heuristic comments on their statistics.
Absent the Vandermonde term, the eigenvalues would be i.i.d. centred Gaussians with variance $1/N$, so the larger $N$ is, the less dispersed the eigenvalues will be around $0$.
The Vandermonde term introduces dependence between all of the eigenvalue, and specifically it introduces \emph{repulsion}, as $\Delta({\lambda}_i)$ is a decreasing function of the distance between the eigenvalues.
We can predict therefore that the distribution of the $\lambda_1, \ldots, \lambda_N$ is some equilibrium balancing the repulsion between all eigenvalues and the independent confining Gaussian potentials on each eigenvalue.

\medskip
We shall now turn our attention to the mean and limiting spectral measures of the GOE.
There are several quite different routes by which one can obtain these results.
For a general and entirely rigorous approach, which in fact applies to \emph{any} generalised Wigner matrix, we direct the reader to \cite{anderson2010introduction}.
We present an approach using supersymmetric methods later in this chapter.
For now, we shall present a derivation using the \emph{Coulomb gas method} \cite{cunden2016shortcut} which, in addition to the supersymmetric method, is of great relevance to the central calculations of this thesis.

Let us introduce the following reformulation of the eigenvalue joint density function of the GOE:
\begin{align*}
    p(\lambda_1, \ldots, \lambda_N) = \Delta(\{\lambda_i\}_{i=1}^N)\prod_{i=1}^N \frac{e^{-\frac{N\lambda_i^2}{2}}}{\sqrt{2\pi N}} = \frac{1}{(2\pi N)^{N/2}}\exp\left(-\frac{N}{2} \sum_{i=1}^N \left\{\lambda_i^2- \frac{1}{N}\sum_{j\neq i} \log|\lambda_i - \lambda_j| \right\}\right).
\end{align*}
Further, using the definition of the empirical spectral density, we can write \begin{align}
      p(\lambda_1, \ldots, \lambda_N) = \frac{1}{(2\pi N)^{N/2}}\exp\left(-\frac{N^2}{2} \int d\hat{\mu}_N(\lambda)\left\{\lambda^2- \int_{\lambda' \neq \lambda} d\hat{\mu}_N(\lambda') \log|\lambda' - \lambda| \right\}\right)
\end{align}
from which the repulsion vs confinement statistics of the eigenvalues is made most clear.
The logarithmic (Coulomb) potential has a singularity and $0$ which penalises eigenvalue configurations with insufficient space between eigenvalues, whereas the quadratic potential penalises configurations with any eigenvalues too far from the origin.
Continuing in the parlance of statistical physics, define the Lagrangian  \begin{align*}
    \mathcal{E}(\lambda; \mu) = \lambda^2 - \int_{\lambda'\neq \lambda} d\mu(\lambda') \log|\lambda - \lambda'|
\end{align*}
and thence the action \begin{align*}
    \mathcal{I}[\mu] = \int d\mu(\lambda) \mathcal{E}(\lambda; \mu)
\end{align*}
with which we have \begin{align}\label{eq:jpdf_as_functional_intro}
      p(\lambda_1, \ldots, \lambda_N) = \frac{1}{Z_N}\exp\left(-\frac{N^2}{2} \mathcal{I}[\hat{\mu}_N]\right).
\end{align}
Let us consider $N\rightarrow\infty$ and for now assume that $\hat{\mu}_N$ converges, in some sense, to $\mu_{\infty}$.
It is clear from the action principle (or Laplace's method for asymptotic evaluation of integrals) that $\mu_{\infty}$ must be a global minimiser of $\mathcal{I}$.
As such, $\mu_{\infty}$ must be a deterministic probability measure on $\R$, so we shall assume weak almost sure convergence of $\hat{\mu}_N$ to $\mu_{\infty}$.
It remains just to solve the variational problem \begin{align*}
    \text{argmin}_{\mu \in \mathcal{P}(\R)}\mathcal{I}[\mu]
\end{align*}
where $\mathcal{P}(\R)$ is the set of all probability measures on $\R$.
The solution  for $\rho_{\infty}$ can be found e.g. in \cite{arous1997large} and is \begin{align}
    \rho_{\infty}(\lambda) = \rho_{SC}(\lambda) = \frac{1}{\pi} \sqrt{2 - \lambda^2}
\end{align}
which is the celebrated Wigner \emph{semi-circle} density.
\begin{figure}
    \centering
    \includegraphics[width=\linewidth]{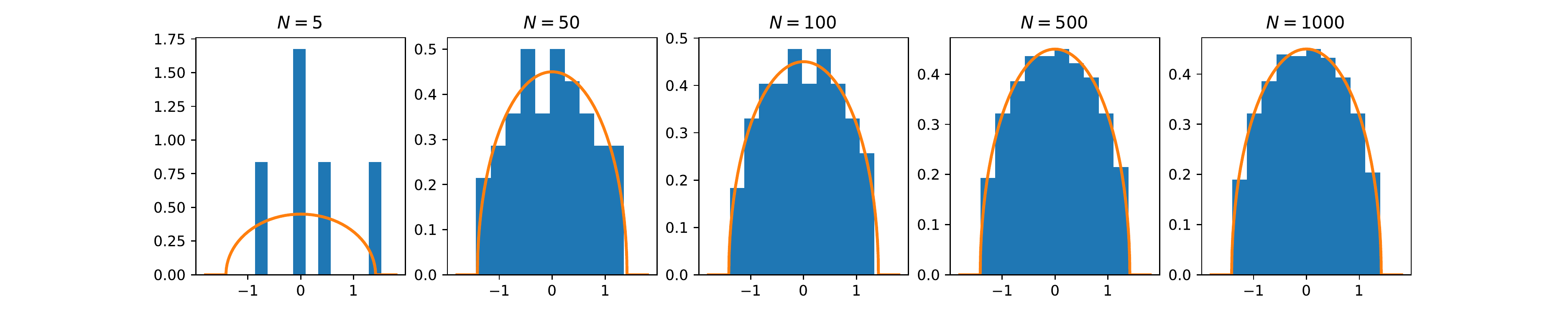}
    \caption{The Wigner semi-circle density (orange line plot) compared to histograms of eigenvalues computed from single samples of $N\times N$ GOE matrices (blue) for various values of $N$}
    \label{fig:sc_hist_1}
\end{figure}

The semi-circle density is striking by its simplicity and elegance which, in fact, hint at a much deeper role in random matrix theory than just the limiting spectral density of a particular random matrix ensemble.
Firstly, the semi-circle is not unique to the GOE but is shared by all generalised Wigner matrices (though, of course, the derivation above is possible only for the three canonical Gaussian Wigner ensembles: GOE, GUE, and GSE).
More importantly, the semi-circle takes the place of the Gaussian in an analogue of the the central limit theorem for random matrices, about which we provide more discussion in section \ref{sec:free_prob}.

So far, we have spoken only of the LSD, but what of the mean spectral density?
We shall defer an explicit calculation for the GOE to section \ref{sec:susy_intro}, but we shall see that the density $\rho_N$ of the mean spectral measure $\mu_N$ can be written as \begin{align*}
    \rho_N(\lambda) = \rho_{SC}(\lambda) + o(1)
\end{align*}
where the $o(1)$ term is uniformly small in $N$ for $\lambda\in \R$.
Once again, this property of the very special GOE ensemble points to a much deeper phenomenon in random matrix theory: \emph{self-averaging}.
To leading order in large $N$, the spectral density of a single random GOE matrix of size $N\times N$ is deterministic and identical to the mean spectral density, which is an average of the whole GOE ensemble of random matrices.
\begin{figure}
    \centering
    \includegraphics[width=\linewidth]{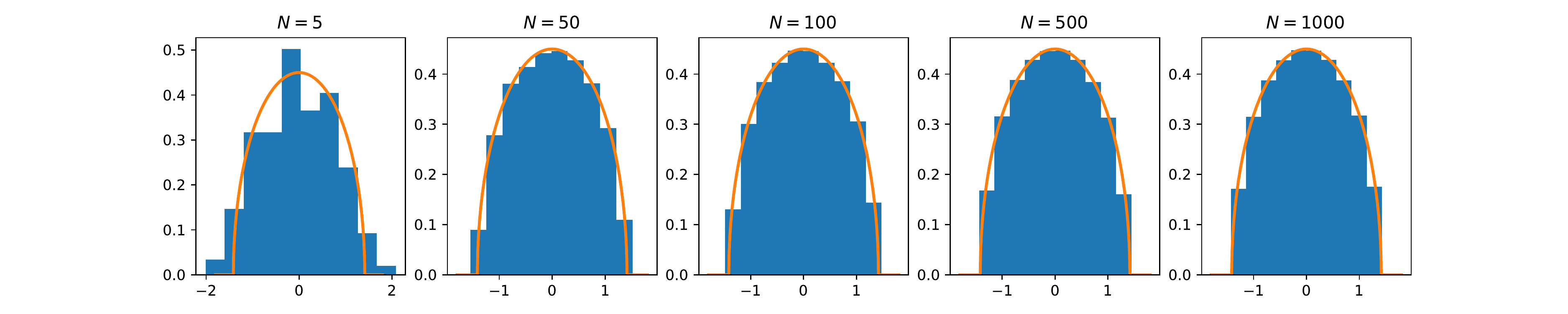}
    \caption{The Wigner semi-circle density (orange line plot) compared to histograms of eigenvalues computed from 100 i.i.d. samples of $N\times N$ GOE matrices (blue) for various $N$ values.}
    \label{fig:sc_hist_100}
\end{figure}

\end{exmp}
\begin{exmp}[An invariant ensemble]
Recall the Lagrangian defined above \begin{align*}
    \mathcal{E}(\lambda; \mu) = \lambda^2 - \int_{\lambda \neq \lambda'} d\mu(\lambda') \log|\lambda - \lambda'|
\end{align*}
with which we were able to express the GOE joint eigenvalue density as \begin{align}
    p(\lambda_1, \ldots, \lambda_N) = \frac{1}{Z_N} \exp\left(-\frac{N^2}{2} \int d\hat{\mu}_N(\lambda) \mathcal{E}(\lambda; \hat{\mu}_N)\right).
\end{align}
The origin of the two terms in $\mathcal{E}$ is quite plain: $\lambda^2$ simply comes from the Gaussian distribution of the GOE entries, while the logarithmic term comes from the Vandermonde determinant.
The Vandermonde term is therefore universal to any real symmetric matrix ensemble, as it follows simply from the matrix change of variables. Similarly, if we were discussion complex Hermitian matrices there would be a universal Vandermonde term simply twice that for real symmetric matrices.
So for any real symmetric matrix ensemble, we could in principle repeat the above procedure and arrive at a Lagrangian with exactly the same logarithmic Vandermonde term, along with some ensemble specific term.
Of course, in general this term would not factorise nicely over the eigenvalues, so the above reduction to simply $\int \hat{\mu}_N(\lambda) \mathcal{E}(\lambda, \hat{\mu}_N)$ would not be possible.
Let us then just consider ensembles for which this factorisation \emph{does} occur, so that one would obtain the same Lagrangian form of the eigenvalues density but with Lagrangian \begin{align*}
     \mathcal{E}_V(\lambda; \mu) =V(\lambda) - \int_{\lambda \neq \lambda'} d\mu(\lambda') \log|\lambda - \lambda'|
\end{align*}
where $V \colon \R \rightarrow\R$ is some function with sufficient smoothness and sufficiently fast growth at infinity to define a normalisable probability density.
Such a random matrix ensemble is known as an \emph{invariant ensemble} because it retains the same orthogonal invariance possessed by the GOE.
The matrix density for an invariant ensemble can be simply written as \begin{align*}
    p(X)dX \propto e^{-\frac{N}{2} \Tr V(X)}dX.
\end{align*}
For a real symmetric matrix argument $X = O^T\Lambda O$ one has the power series definition \begin{align*}
   \Tr V(X) = \sum_{r\geq 0} a_r \Tr X^r = \sum_{r\geq 0} a_r \Tr O^T\Lambda O\ldots O^T \Lambda O = \sum_{r\geq 0} a_r \Tr \Lambda^r = V(\Lambda) = \Tr V(\Lambda),
\end{align*}
so
\begin{align*}
     p(X) dX \propto e^{-\frac{N}{2}\sum_{j=1}^N V(\lambda_j)} \Delta(\{\lambda_i\}_{i=1}^N) d\lambda_1\ldots d\lambda_N d\mu_{Haar}(O)
 \end{align*}
 which confirms the Lagrangian expression given above.
\end{exmp}

\subsection{The Wigner surmise}
As we have seen in the preceding section, though the semi-circle plays a deep role in random matrix theory, it is by no means a universal spectral density for random matrix ensembles.
Simply change the potential to deviate from the simple quadratic case was sufficient to produce entirely different spectral densities with invariant ensembles.
So, at the level of the mean (or limiting) spectral measure, the semi-circle is more general that the GOE and the Gaussian Wigner ensembles, but is specific to Wigner matrices.
One of the most astonishing results in random matrix theory is that there are properties of GOE matrices that are, in fact, \emph{universal} in the sense that they are properties shared by a very wide class of matrices beyond the GOE and Wigner ensembles.
A full discussion of this kind of random matrix universality is deferred to the later Section \ref{sec:universal_intro}.

\medskip
Random matrix theory was first developed in physics to explain the statistical properties of nuclear energy levels, and later used to describe the spectral statistics in atomic spectra, condensed matter systems, quantum chaotic systems etc; see, for example \cite{weidenmuller2008random, beenakker1997random, berry1987quantum, bohigas1991random}. \emph{None of these physical systems exhibits a semicircular empirical spectral density}. However they all generically show agreement with random matrix theory at the level of the mean eigenvalue spacing when local spectral statistics are compared.  
The key insight here is that while almost any realistic physical system, model or even the machine learning systems which are the central objects of study for this thesis, will certainly not posses semi-circular densities at the macroscopic scale of the mean spectral density, but nevertheless random matrix theory can still describe spectral fluctuations on the microscopic scale of the mean eigenvalue spacing.

\medskip
It is worth noting in passing that possibilities other than random-matrix statistics exist and occur. For example, in systems that are classically integrable, one finds instead Poisson statistics \cite{berry1977level, berry1987quantum}; similarly, Poisson statistics also occur in disordered systems in the regime of strong Anderson localisation \cite{efetov1999supersymmetry}; and for systems close to integrable one finds a superposition of random-matrix and Poisson statistics \cite{berry1984semiclassical}. So showing that random matrix theory applies is far from being a trivial observation.  Indeed it remains one of the outstanding challenges of mathematical physics to prove that the spectral statistics of any individual Hamiltonian system are described by it in the semi-classical limit.

\medskip
Random matrix calculations in physics re-scale the eigenvalues to have a mean level spacing of $1$ and then typically look at the \emph{nearest neighbour spacings distribution} (NNSD), i.e. the distribution of the distances between adjacent pairs of eigenvalues.  One theoretical motivation for considering the NNSD is that it is independent of the Gaussianity assumption and reflects the symmetry of the underlying system. It is the NNSD that is universal (for systems of the same symmetry class) and not the average spectral density, which is best viewed as a parameter of the system. The aforementioned transformation to give mean spacing $1$ is done precisely to remove the effect of the average spectral density on the pair correlations leaving behind only the universal correlations.

\begin{figure}[h]
    \centering
    \includegraphics[width=0.8\textwidth]{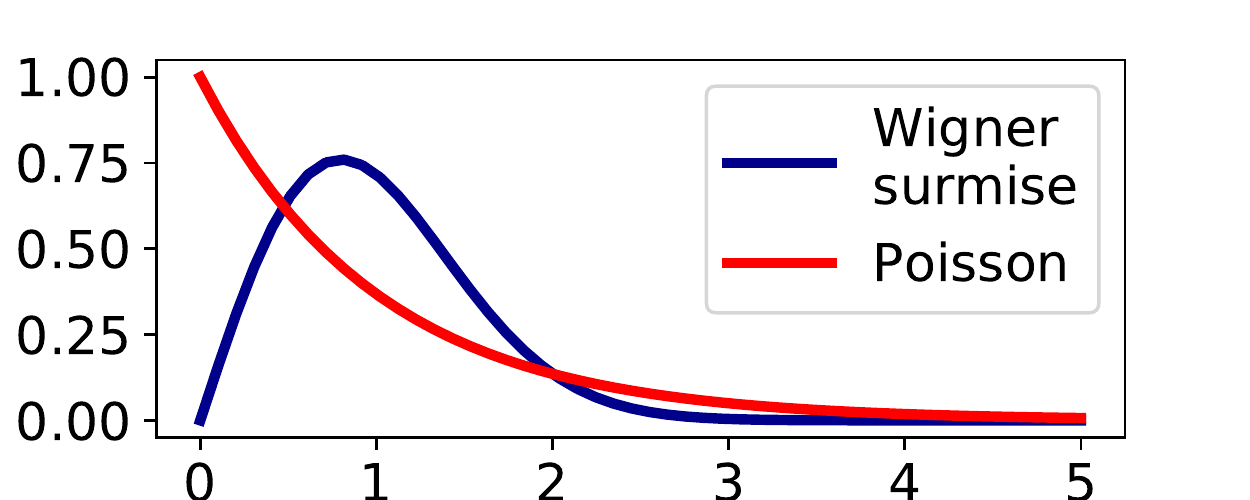}
    \caption{The density of the Wigner surmise.}
    \label{fig:wigner_vs_poisson}
\end{figure}

In contrast to the LSD, other $k$-point correlation functions are also normalised such that the mean spacing between adjacent eigenvalues is unity. At this \emph{microscopic} scale, the LSD is locally constant and equal to 1 meaning that its effect on the eigenvalues' distribution has been removed and only microscopic correlations remain. In the case of Wigner random matrices, for which the LSD varies slowly across the support of the eigenvalue distribution, this corresponds to scaling by $\sqrt{P}$. On this scale the limiting eigenvalue correlations when $P\to\infty$ are {\em universal}; that is, they are the same for wide classes of random matrices, depending only on symmetry \cite{guhr1998random}.  For example, this universality is exhibited by the NNSD. Consider a $2\times 2$ GOE matrix, in which case the j.p.d.f has a simple form: \begin{equation}
    p(\lambda_1, \lambda_2) \propto |\lambda_1 - \lambda_2| e^{-\frac{1}{2}(\lambda_1^2 + \lambda_2^2)}.
\end{equation}
Making the change of variables $\nu_1 = \lambda_1 - \lambda_2, \nu_2 = \lambda_1 + \lambda_2$, integrating out $\nu_2$ and setting $s = |\nu_1|$ results in a density $\rho_{Wigner}(s) = \frac{\pi s}{2}e^{-\frac{\pi}{4}s^2}$, known as the \emph{Wigner surmise} (see Figure \ref{fig:wigner_vs_poisson}). For larger matrices, the j.p.d.f must include an indicator function $\mathbbm{1}\{\lambda_1\leq \lambda_2\leq \ldots \lambda_P\}$ before marginalisation so that one is studying pairs of \emph{adjacent} eigenvalues. While the Wigner surmise can only be proved exactly, as above, for the 2 × 2 GOE, it holds to high accuracy for the NNSD of GOE matrices of any size provided that the eigenvalues have been scaled to give mean spacing 1.\footnote{An exact formula for the NNSD of GOE matrices of any size, and one that holds in the large $P$ limit, can be found in \cite{mehta2004random}.} The Wigner surmise density vanishes at $0$, capturing `repulsion' between eigenvalues that is characteristic of RMT statistics, in contrast to the distribution of entirely independent eigenvalues given by the \emph{Poisson law} $\rho_{Poisson}(s) = e^{-s}$. The Wigner surmise is universal in that the same density formula applies to all real-symmetric random matrices, not just the GOE or Wigner random matrices. 

\subsection{Eigenvectors}
What of the eigen\emph{vectors} of random matrices?
We have already seen that GOE matrices, and invariant ensembles in general, have Haar-distributed eigenvectors entirely independent of the eigenvalues.
Just as the semi-circle is unique to Wigner matrices but the GOE Wigner surmise is seen in all matrices with orthogonal group symmetry, so Haar-distributed eigenvectors independent of the eigenvalues are seen only in invariant ensembles (not even in non-Gaussian Wigner matrices) but certain properties of Haar matrices are universal across a similarly wide class of random matrices.
Once again, the discussion of these deep universality results will be given in Section \ref{sec:universal_intro}, but we shall set the scene by first describing the \emph{delocalisation} property of Haar-distributed eigenvectors.

\medskip
Let $U$ be an $N\times N$ Haar-distributed orthogonal matrix and let $\vec{u}_1, \ldots, \vec{u}_N$ be its rows.
Recall from the discussion above wherein the Haar distribution was introduced the following construction:

\begin{align}
    &\text{Let }\vec{g}_1, \ldots, \vec{g}_N \text{ be i.i.d. vectors from } \mathcal{N}(0, I_N);\\
    &\text{let } \vec{v}_1, \ldots, \vec{v}_N \text{ be the results of a Gram-Schmidt algorithm};\\
    &\text{then, in distribution, } \{\vec{u}_i\}_{i=1}^N = \{\vec{v}_i/\|\vec{v}_i\|_2\}_{i=1}^N.
\end{align}
Fix some $r < N$ and introduce the event \begin{equation}
    B_N(\upsilon) \defeq\left\{| N^{-1}\langle \vec{g}_i, \vec{g}_j\rangle - \delta_{ij}| \leq N^{-\upsilon}, ~~~ 1\leq i, j \leq r\right\}.
\end{equation}
Then it is an exercise in Gaussian calculations and asymptotics, as given in \cite{guionnet2005fourier}, to conclude that under the i.i.d Gaussian law of the $(\vec{g}_j)_{j=1}^{N}$ the complementary event has low probability for large $N$: \begin{equation}
    \mathbb{P}(B_N(\upsilon)^c) =\mathcal{O}( C(\upsilon) e^{-\alpha N^{1-2\upsilon}}),
\end{equation}
where $\alpha, C(\upsilon) > 0$ and we take $0<\upsilon < \frac{1}{2}$ to make this statement meaningful.
What's more, one can directly obtain that, given $B_N$, \begin{align}
    \|\vec{g}_i\|_2^2 = N^{1 - \upsilon}
\end{align}
for any $\upsilon > 0$.
So, restricting to only a fixed subset of the eigenvectors as $N\rightarrow\infty$, the simply i.i.d. Gaussian vectors $\vec{g}_i$ from which they are constructed are, with high probability, close to being orthogonal even before applying Gram-Schmidt algorithm and they all have the same $L_2$ norm to leading order in $N$.
This line of reasoning leads to the fact that, with high probability,
\begin{align}
   ||N^{-1/2}\tilde{\vec{g}}_j - \vec{u}_j|| \leq N^{-\frac{\upsilon}{2}},\label{eq:intro_haar_gauss}
\end{align}
so, indeed, in the above precise probabilistic sense, any subset of Haar-distributed eigenvectors are extremely close to an corresponding set of i.i.d. standard Gaussian vectors, re-scaled by $N^{-1/2}$.

\section{Kac-Rice formulae}
The majority of chapters \ref{chap:general_activation_functions} and \ref{chap:spin_glass_gans} is concerned with computing the expected complexity of certain loss surfaces in the limit as the number of parameters $N\rightarrow\infty$.
Let us recall a basic definition of complexity as introduced above.
Let $\mathcal{M}$ be a compact, oriented, N-dimensional $C^1$ manifold with a $C^1$ Riemannian metric $g$. Let $\psi:\mathcal{M}\rightarrow \mathbb{R}$ be a random field on $\mathcal{M}$. For an open set $A\subset\mathbb{R}$, let \begin{equation}
    C(A) \equiv \left|\{x\in\mathcal{M} ~|~ \grad\psi(x) = 0, ~ \psi(x)\in A\}\right|.
\end{equation}
$C(A)$ simply counts the number of local optima of $\psi$ for which $\psi$ lies in the set $A$.
Note that the condition of a compact manifold $\mathcal{M}$ is important here; without other constraints (for which see e.g. \cite{arous2021landscape}) there is no guarantee of a finite value for $C(A)$ given a non-compact $\mathcal{M}$.
Computing anything about $C(A)$ appears extremely challenging, but one can make some informal progress rather directly with an integral expression
\begin{align}
 C(A) = \int_{\nabla\psi(\mathcal{M})} d\vec{u} ~\delta(\vec{u}) \1\{\psi(\nabla\psi^{-1}(\vec{u}))\in A\}
\end{align}
which one can write down simply from the sampling property of the $\delta$-function.
The the composition property of the $\delta$-function gives \begin{align}
    C(A) = \int_{\mathcal{M}} d\vec{x} ~ |\det \nabla^2\psi(\vec{x})|\delta(\nabla\psi(\vec{x})) \1\{\psi(\vec{x})\in A\}.
\end{align}
From this simple argument, we see that the \emph{Hessian} of $\psi$, and in particular the absolute value of its determinant, will be central to calculation of $C(A)$.
Recall that $\psi$ is a random field, so its Hessian $\nabla^2\psi$ is a random matrix of size $N\times N$, so one can see already that the complexity of random functions is connected with random matrix theory.
What these simple arguments lack is any reference to the probability density of $\psi$.
Since $\psi$ is random, so also is $C(A)$, so we must be more precise about what `calculating $C(A)$' means.
One could attempt to compute the entire density of $C(A)$, but this is clearly the most difficult objective.
Let us restrict our consideration to simple statistics of $C(A)$ and, in particular, its expected value.
Proceeding informally, we have \begin{align}
    \E C(A) = \E \left\{\int_{\mathcal{M}}d\vec{x} ~ |\det \nabla^2\psi(\vec{x})|\delta(\nabla\psi(\vec{x})) \1\{\psi(\vec{x})\in A\} \Bigg| \nabla\psi(\vec{x}) = 0 \right\} p_{\vec{x}}(0)
\end{align} 
where $p_{\vec{x}}$ is the density of $\nabla \psi$ at the point $\vec{x}\in\mathcal{M}$.
Within the conditional expectation, the delta function can be dropped, giving simply
 \begin{align}
    \E C(A) = \E \left[\int_{\mathcal{M}}d\vec{x} ~ |\det \nabla^2\psi(\vec{x})| \1\{\psi(\vec{x})\in A\} \Bigg| \nabla\psi(\vec{x}) = 0 \right] p_{\vec{x}}(0)
\end{align} 
and finally swapping the order of integration informally gives 
\begin{align}\label{eq:informal_kr}
    \E C(A) =  \int_{\mathcal{M}}d\vec{x} ~  p_{\vec{x}}(0)\E\left[|\det \nabla^2\psi(\vec{x})| \1\{\psi(\vec{x})\in A \}\Bigg| \nabla\psi(\vec{x}) = 0 \right].
\end{align}
We see now that $\E C(A)$ will be tractable if we can compute the joint distribution of $\psi, \nabla^2\psi$ conditional on $\nabla\psi$, and subsequently evaluate the random determinant's expected value.
The expression (\ref{eq:informal_kr}) is an example of a \emph{Kac-Rice formula} \cite{kac1943average,rice1944mathematical}.
These kind of informal arguments have been extensively used in the mathematical physics literature to compute quantities such as expected landscape complexities and cardinalities of other level-sets of random functions \cite{berry2002statistics,berzin1998level,fyodorov2002characteristic,fyodorov2004complexity,fyodorov2005counting}.
These arguments can be made fully rigorous and cast in a more general setting as is shown in the important book by Adler and Taylor \cite{adler2009random}.
We repeat here the foundational Kac-Rice result  from that work which is central to our complexity calculations in the coming chapters.

\begin{theorem}[\cite{adler2009random} Theorem 12.1.1]\label{thm:adler_kac_rice}
Let $\mathcal{M}$ be a compact , oriented, N-dimensional $C^1$ manifold with a $C^1$ Riemannian metric $g$. Let $\phi:\mathcal{M}\rightarrow\mathbb{R}^N$ and $\psi:\mathcal{M}\rightarrow \mathbb{R}^K$ be random fields on $\mathcal{M}$. For an open set $A\subset\mathbb{R}^K$ for which $\partial A$ has dimension $K-1$ and a point $\vec{u}\in\mathbb{R}^{N}$ let \begin{equation}
    N_{\vec{u}} \equiv \left|\{x\in\mathcal{M} ~|~ \phi(x) = \vec{u}, ~ \psi(x)\in A\}\right|.
\end{equation}

Assume that the following conditions are satisfied for some orthonormal frame field E:
\begin{enumerate}[label=(\alph*)]
\item
All components of $\phi$, $\nabla_E \phi$, and $\psi$ are a.s. continuous and have finite variances (over $\mathcal{M}$).
\item
 For all $x\in\mathcal{M}$, the marginal densities $p_{x}$  of $\phi(x)$ (implicitly assumed to exist) are continuous at  $\vec{u}$.
 \item
 The conditional densities $p_{x}(\cdot|\nabla_E\phi(x),\psi(x))$ of $\phi(x)$ given $\psi(x)$ and $\nabla_E\phi(x)$ (implicitly assumed to exist) are bounded above and continuous at $\vec{u}$, uniformly in $\mathcal{M}$.
 \item
 The conditional densities $p_x (\cdot|\phi(x) = \vec{z})$ of $\det(\nabla_{E_j}\phi^i  (x))$ given are continuous in a neighbourhood of $0$ for $\vec{z}$ in a neighbourhood of $\vec{u}$  uniformly in $\mathcal{M}$.
 \item
 The conditional densities $p_x (\cdot|\phi (x) = \vec{z})$ are continuous for $\vec{z}$ in a neighbourhood of $\vec{u}$ uniformly in $\mathcal{M}$.
 \item
 The following moment condition holds \begin{equation}
     \sup_{x\in\mathcal{M}}\max_{1\leq i,j\leq N}\expect\left\{\left|\nabla_{E_j}\phi^i(x)\right|^N\right\}< \infty
 \end{equation}
 \item
 The moduli of continuity with respect to the (canonical) metric induced  by $g$ of each component of $\psi$, each component of $\phi$ and each $\nabla_{E_j}\phi^i$ all satisfy, for any $\epsilon > 0$ \begin{equation}\label{eq:moduli_condition}
    \mathbb{P}( \omega(\eta) >\epsilon) = o(\eta^N), ~~ \text{as } \eta\downarrow 0
 \end{equation}
 where the \emph{modulus of continuity} of a real-valued function $G$ on a metric space $(T, \tau)$ is defined as (c.f. \cite{adler2009random} around (1.3.6)) \begin{equation}
     \omega(\eta) \coloneqq \sup_{s,t : \tau(s,t)\leq\eta}\left|G(s) - G(t)\right|
 \end{equation} 
\end{enumerate}
Then \begin{equation}\label{eq:adler_taylor_kac_rice}
    \expect N_{\vec{u}} = \int_{\mathcal{M}}\expect \left\{|\det \nabla_E\phi(x)|\indic\{\psi(x)\in A\} ~| ~ \phi(x) = \vec{u}\right\}p_x(\vec{u}) \text{Vol}_g(x)
\end{equation}
where $p_x$ is the density of $\phi$ and $\text{Vol}_g$ is the volume element induced by $g$ on $\mathcal{M}$.
\end{theorem}
Note the greater generality of this theorem compared to the heuristic derivation above. The required result for complexity can be obtained as a special case by taking $\phi = \nabla\psi$ and $\vec{u} = 0$.

\section{Supermathematics}\label{sec:susy_intro}
\emph{Grassmann variables} are entirely algebraic objects defined by an \emph{anti-commutation rule}.
Let $\{\chi_i\}_i$ be a set of Grassmann variables, then by definition
 \begin{align}
    \chi_i\chi_j = - \chi_j\chi_i, ~~ \forall i,j\label{eq:anticom_def}.
\end{align}
The complex conjugates $\chi_i^*$ are separate objects, with the complex conjugation unary operator ${}^*$ defined so that $\left(\chi_i^*\right)^* = -\chi_i^*,$ and Hermitian conjugation is then defined as usual by $\chi^{\dagger} = (\chi^T)^*.$ The set of variables $\{\chi_i, \chi_i^*\}_{i=1}^N$ generate a \emph{graded algebra} over $\mathbb{C}$. Mixed vectors of commuting and anti-commuting variables are called \emph{supervectors}, and they belong to a vector space called \emph{superspace}. The integration symbol $\int d\chi_id\chi^*$ is defined as a formal algebraic linear operator by the properties \begin{align}\label{eq:berezin_def}
    \int d\chi_i = 0, ~~~~~ \int d\chi_i ~\chi_j = \delta_{ij},
\end{align}
and these are called \emph{Berezin} integrals.
Functions of the the Grassmann variables are defined by their formal power series, e.g. \begin{align}
    e^{\chi_i} = 1 + \chi_i + \frac{1}{2}\chi_i^2 + \ldots = 1 + \chi_i
\end{align}
where the termination of the series follows from $\chi_i^2 = 0 ~~ \forall i$, which is an immediate consequence of (\ref{eq:anticom_def}). From this it is apparent that (\ref{eq:berezin_def}), along with (\ref{eq:anticom_def}), is sufficient to define Berezin integration over arbitrary functions of arbitrary combinations of Grassmann variables. Finally we establish our notation for supersymmetric (or \emph{graded}) traces of supermatrices. We will encounter supermatrices of the form \begin{align*}
    M = \left(\begin{array}{cc}
         A & B \\
         C & D
    \end{array}\right)
\end{align*}
where $A, D$ are square block matrices of commuting variables and $B, C$ are rectangular block matrices of Grassmann variables. In this case, the graded trace is given by $\trg M = \Tr A - \Tr D $ and such matrices are referred to as $(B+F)\times (B + F)$, where $A$ is shape $B\times B$ and $D$ is shape $F\times F$. We refer the reader to \cite{efetov_1996} for a full introduction to supersymmetric variables and methods.

\medskip
Grassmann variables play an important role in quantum field theory and related fields \cite{peskin2018introduction,green_schwarz_witten_2012}, being the algebraic representation of fermions, with bosons being represented by commuting variables.
As such, even in applications unrelated to quantum physics, the particle nomenclature may be used; for example the diagonal blocks of the matrix $M$ above may be referred to as \emph{bosonic blocks} and the off-diagonals referred to as \emph{fermionic blocks}.
There are important connections between random matrix theory and quantum field theory in which the role of supersymmetry in both is made quite plain \cite{Verbaarschot_2004}, but for the purposes of this thesis, Grassmann variables and supersymmetric methods are simply mathematical tools that we use to compute certain random matrix quantities.
Supersymmetric methods provide a powerful way of computing random matrix determinants, which in turn can have many applications to compute various quantities of interest \cite{Verbaarschot_2004,nock}.
We will focus on two such applications that are used in chapters \ref{chap:general_activation_functions}, \ref{chap:spin_glass_gans} and \ref{chap:univ}.

\medskip
Consider a random $N\times N$ matrix $X$ and suppose if has a limiting spectral measure $\mu$ with density $\rho$ and Stieljtes transform $g$.
Given a density on $X$, an important question is to determine the spectral density $\rho$, by the Stieljtes inversion formula, it is sufficient to compute $g$:
 \begin{align}
     \rho(x) = \frac{1}{\pi}\lim_{\epsilon \rightarrow 0} \Im  g(x + i\epsilon).
 \end{align}
Let $G(z)$ be the Stieljtes transform of the empirical spectral measure of $X$, i.e. 
\begin{align}
    G(z) = \frac{1}{N} \sum_{i=1}^N \frac{1}{z - \lambda_i}
\end{align}
where $\lambda_i$ are the eigenvalues of $X$.
$G$ is a random function and for many matrix ensembles will have the convergence property $G\rightarrow g$ weakly almost surely as $N\rightarrow\infty$.
Similarly, $G$ will typically have the self-averaging property $\E G \rightarrow g$ in the sense of deterministic functions.
It follows that computing $\rho$ can be achieved by computing the leading order term in an asymptotic expansion for $\E G$ in the limit $N\rightarrow\infty$.
The key to this approach is that, if the average $\E G$ can be computed over the matrix ensemble $X$, then the asymptotic analysis for large $N$ can be performed on deterministic objects to obtain $\rho$, rather than having to deal with asymptotics of random functions.
To see the connection with random determinants and thence supersymmetry, we can rewrite \begin{align}
    G(z) = \frac{1}{N}\frac{1}{\det(z\vivacom{I} - X)}\sum_{i=1}^N\prod_{\vivacom{k}\neq i}^N (z - \lambda_{\vivacom{k}})=\frac{1}{N} \frac{\partial}{\partial \mathfrak{j}}\Bigg|_{\mathfrak{j} =0}\frac{\det(z\vivacom{I} - X + \mathfrak{j}\vivacom{I})}{\det(z\vivacom{I} - X)}
\end{align}
where the first equality is a simple algebraic identity and the second follows from the product rule of differentiation. It follows that computing $\E G$ is equivalent to computing the random matrix average \begin{align}
    \E \frac{\det(z\vivacom{I} - X + \mathfrak{j}\vivacom{I})}{\det(z\vivacom{I} - X)}
\end{align}
followed by some differentiation. Note that this `trick' involving the introduction of the dummy variable $\mathfrak{j}$ is widely used as well in the perturbation theory approach to quantum field theory \cite{peskin2018introduction}.

\medskip
We have seen in the previous section that random matrix determinants are to be expected for example in complexity calculations, i.e. one needs to compute $\E |\det X|$ for a random $N\times N$ matrix $X$, and doing so in the large $N$ limit may be sufficient.
The presence of the absolute value here is a particular nuisance, but just like the Stieljtes transform above, ratios of determinants can be used to provide an alternate formulation:
\begin{align}
    |\det X| = \frac{(\det X)^2}{( \sqrt{\det X})^2} = \frac{\det X \det X}{\sqrt{\det X}\sqrt{\det X}}
\end{align}
where the principal branch of the square root is taken.

\medskip
The general challenge here is to compute expectations of ratios of integer and half integer powers of random matrix determinants.
This topic has been much explored in the literature, see e.g. \cite{fyodorov2005counting,Verbaarschot_2004,nock,nock2017characteristic}.
The role of supersymmetric methods in this approach stems from a familiar change of variables result. For a non-singular Hermitian $N\times N$ matrix $X$ 
\begin{align}
    \frac{1}{(-i)^N \pi^N}\int_{\R^N} d\vec{z} e^{-i\vec{z}^T X \vec{z}} = \frac{1}{\sqrt{\det X}}
\end{align}
where the determinant is the simply the Jacobian of the transformation from variables $\vec{z}$ to $X^{1/2}\vec{z}$.
Similarly, using complex integration variables one can obtain \begin{align}
   \frac{1}{(2\pi)^N} \int_{\C^N} d\vec{z}d\vec{z}^* e^{-i\vec{z}^{\dagger} X \vec{z}} = \frac{1}{\det X}.
\end{align}
The final ingredient is to use Grassmann integration variables to obtain an analogous expression for $\det X$, as opposed to powers of its reciprocal. Indeed, by introducing Grassmann variables $\chi_i, \chi_i*$ and a Berezin integral, we obtain \begin{align}
     \frac{1}{(-i)^N}\int\prod_{i=1}^N d\chi_i d\chi_i^* e^{-i\chi^{\dagger}X\chi} = \det{X}.\end{align}
Rather than a change of variable result from multivariate calculus, this result is proved by expanding the exponential.
Recall that $\int d\chi_i ~ 1 = 0$, so the only therm in the exponential expansion that can be non-zero after Berezin integration are those that contain each $\chi_i$ and $\chi_i^*$ at least once. But also, since $\chi_i^2 = 0 = (\chi_i^*)^2$, the only non-zero terms are those that contain each $\chi_i, \chi_i^*$ \emph{exactly} once, hence \begin{align}
    \int \prod_{i=1}^N d\chi_i d\chi_i^* e^{-i\chi^{\dagger}X\chi} &=\frac{1}{N!} \int\prod_{i=1}^N d\chi_i d\chi_i^* (-i\chi^{\dagger}X\chi)^N \notag\\
    &= (-i)^N\frac{1}{N!}\int\prod_{i=1}^N d\chi_i d\chi_i^* \sum_{j_1, k_1,\ldots, j_N, k_N=1}^N \chi^*_{j_1}X_{j_1k_1}\chi_{k_1}\ldots \chi^*_{j_N}X_{j_Nk_N}\chi_{k_N}
\end{align}
The only non-zero terms from the sum must have $j_1, \ldots, j_N$ equal to a permutation of $1, \ldots, N$ and similarly $k_1, \ldots, k_N$, so we can write 
\begin{align}
    \int \prod_{i=1}^N d\chi_i d\chi_i^* e^{-i\chi^{\dagger}X\chi} &=(-i)^N\frac{1}{N!} \int\prod_{i=1}^N d\chi_i d\chi_i^* \sum_{\sigma, \tau\in S_N} \chi^*_{\sigma(1)}X_{\sigma(1)\tau(1)}\chi_{\tau(1)}\ldots \chi^*_{\sigma(N)}X_{\sigma(N)\tau(N)}\chi_{\tau(N)}.
\end{align}
Re-indexing the sum over the symmetric group by defining $\sigma = \sigma' \circ \tau$, we see that the sum over $\tau$ can be rendered trivial, giving just a constant factor of $N!$, so 
\begin{align}
    \int \prod_{i=1}^N d\chi_i d\chi_i^* e^{-i\chi^{\dagger}X\chi} &=(-i)^N \int\prod_{i=1}^N d\chi_i d\chi_i^* \sum_{\sigma'\in S_N} \chi^*_{\sigma'(1)}X_{\sigma'(1)1}\chi_{1}\ldots \chi^*_{\sigma'(N)}X_{\sigma'(N)N}\chi_{N}.
\end{align}
Finally, the Grassmann terms must be commuted to render them in the correct order to agree with the differentials, i.e. \begin{align}
        \int \prod_{i=1}^N d\chi_i d\chi_i^* e^{-i\chi^{\dagger}X\chi} &=(-i)^N \int\prod_{i=1}^N d\chi_i d\chi_i^*   \prod_{j=N}^1 \chi_j^* \chi_j \sum_{\sigma\in S_N} \text{sgn}(\sigma) \prod_{k=1}^N X_{\sigma(k)k} =(-i)^N \det X.
\end{align}
To conclude, ratios of certain powers of random matrix determinants can be written as Gaussian integrals over supersymmetric (i.e. mixed commuting and Grassmann) vectors.
While this may seem at first like an increase in complexity, the supersymmetric representations have certain advantages, such as linearity since $e^{-i\phi^{\dagger} (X + Y) \phi}=e^{-i\phi^{\dagger} X  \phi} e^{-i\phi^{\dagger} Y \phi}$. For example, if $X$ and $Y$ are independent, then a supersymmetric representation allows $\E|\det(X+Y)|$ to be computed as two separate and independent expectations of $X$ and $Y$.
It is this linearisation effect of supersymmetric representations that is at the heart of its application to many calculations, including those in chapters  \ref{chap:general_activation_functions} and \ref{chap:spin_glass_gans}.
In all applications of the supersymmetric method in random matrix theory, the random matrix calculation is reduced to `simply' $\E e^{-i \Tr XK}$ where the matrix $K = \phi\phi^{\dagger} + \chi\chi^{\dagger}$ for some commuting vector $\phi$ and Grassmann $\chi$ of dimension $N$. The distribution of $X$ is then encoded in this Fourier transform like object, and the remainder of the calculation is then an exercise in evaluating supersymmetric integrals.
In the case of the GOE, this average is particular easy to compute:
\begin{align}
    \E e^{-i\Tr XK} &= \frac{N^N}{(2\pi)^{N/2}}\int dX \exp\left\{-\frac{N}{2}\Tr X^2 - i\Tr XK\right\}\notag\\
   &= \frac{N^N}{(2\pi)^{N/2}}\int dX \exp\left\{-\frac{N}{2}\Tr \left(X+ i\frac{1}{N}K\right)^2 - \frac{1}{2N}\Tr K^2\right\}\notag\\
   &= = e^{-\frac{1}{2N}\Tr K^2}.
\end{align}

\medskip
The final technique we need to introduce for supersymmetric methods is the \emph{Hubbard-Stratonovich} transformation.
Consider complex commuting integration variables $\phi\in \C^N$ and Grassmann variables $\chi$. Then \Tr $(\phi\phi^{\dagger} + \chi\chi^{\dagger})^2 = \trg Q^2$ where
\begin{align}
    Q = \left(\begin{array}{cc} \phi^{\dagger}\phi & \phi^{\dagger}\chi \\ \chi^{\dagger}\phi & \chi^{\dagger}\chi\end{array}\right).
\end{align}
The Hubbard-Stratonovich transformation introduces a $2\times 2$ matrix integration variable $\sigma$ which is of the same supersymmetric $1+1$ type as $Q$ then
\begin{align}
    e^{-\frac{1}{2N} \trg Q^2} = \int d\sigma ~ e^{-\frac{N}{2}\trg \sigma^2 - i\psi^{\dagger}\sigma \psi}
\end{align}
where \begin{align}
      \psi = \phi \otimes \left(\begin{array}{c} 1 \\ 0 \end{array}\right) + \chi \otimes \left(\begin{array}{c} 0 \\ 1 \end{array}\right).
\end{align}
The power of the Hubbard-Stratonovich transformation is that it linearises the dependence of the supersymmetric integrand on the supersymmetric $N$-dimensional vectors at the cost only of introducing an integral over an extra $2\times 2$ (or in general $k\times k$) supersymetric matrix. In many calculations, this transformation makes the $N$-dimensional supersymmetric integral easy to compute, leaving only an integration of a fixed number of supersymmetric variables, which is precisely the conditions required for applying standard techniques from asymptotic analysis for $N\rightarrow\infty$.

\section{Large deviations principles}
\begin{figure}
    \centering
    \includegraphics[scale=0.8]{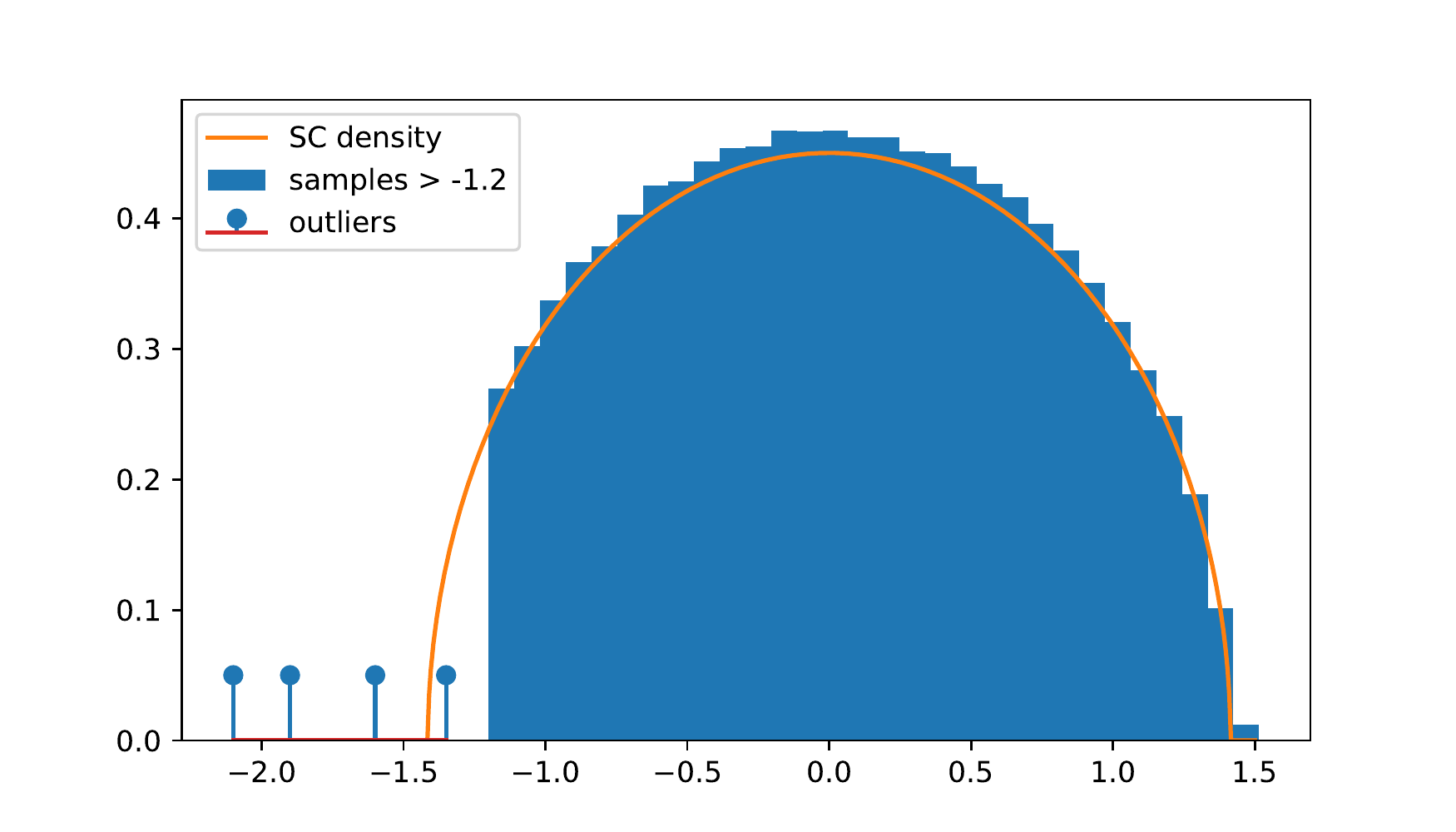}
    \caption{The samples show an example spectrum of a very large random matrix with only a small number of eigenvalues less than $-1.2$. One can see that the empirical spectral density $\hat{\mu}_N$ deviates in a non-negligible manner from $\mu_{SC}.$}
    \label{fig:ldp_demo_intro}
\end{figure}

Consider as an example a $N\times N$ GOE matrix $X$ normalised so that the semi-circular radius is $2$.
By the very existence of such a compact limiting spectral density, eigenvalues greater than $2$ or less than $-2$ are in some sense unlikely for large $N$.
Large deviations principles (LDPs) answer the question of precisely \emph{how unlikely} these eigenvalues are.
Let $\lambda_1 \leq \ldots \leq \lambda_N$ be the eigenvalues of $X$.
The large deviation event for $\lambda_1$ is $\{\lambda_1 < x\}$ for $x < -2$, and similarly for $\lambda_N$ is $\{\lambda_N > x\}$ for $x > 2$.
Fixing an integer $k\vivacom{\geq} 1$ as $N\rightarrow\infty$ there are also the large deviations events $\{\lambda_k < x\}$ for $x<-2$ (and similarly for $\lambda_{N-k}$).
Formally, a large deviations principle for $\lambda_k$ with speed $\alpha(N)$ and rate function $I_k(x)$ requires \begin{align}
    &\limsup_{N\rightarrow\infty}\frac{1}{\alpha(N)}\log \P(\lambda_k < x) = -I_k(x) ~~~ \text{for } x \leq -2,\label{eq:ldp_intro_1}\\
    &\limsup_{N\rightarrow\infty}\frac{1}{\alpha(N)}\log \P(\lambda_k \geq x) = -\infty ~~~ \text{for } x \in (-2, 2).\label{eq:ldp_intro_2}
\end{align}
For $x\in (-2, 2)$, if $\lambda_k\geq x$, then there is an non-empty interval $(-2, x)$ in which there are at most $k$ eigenvalues, so this represents a configuration of eigenvalues for which the difference between $\hat{\mu}_N$ and $\mu$ is not negligible, which must be extremely unlikely since $\hat{\mu}_N$ converges to $\mu$.
The large deviations principle encodes this as the infinity in (\ref{eq:ldp_intro_2}), which says that $\{\lambda_k \geq x\}$ is very much more unlikely than even $\{\lambda_k < -2\}$.
In the case of the GOE \cite{arous2001aging,auffinger2013random}, $\alpha(N) = N$ and \begin{align*}
I_k(x) = kI_1(x) = \begin{cases}
    \frac{k}{2} \int_{x}^{\vivacom{-}2} dz\sqrt{z^2 - 4},~~~ &\text{for } x \leq -2,\\
    \infty &\text{otherwise}.
\end{cases}
\end{align*}
Note that the infinity in (\ref{eq:ldp_intro_2}) should be expected from the expression (\ref{eq:jpdf_as_functional_intro}) of the eigenvalue j.p.d.f. as $e^{-N^2 \mathcal{I}[\hat{\mu}_N]}$, since $\lambda_k \geq x$ for $x\in(-2, 2)$ implies $\mathcal{I}[\hat{\mu}_N] > 0$; figure \ref{fig:ldp_demo_intro} shows this argument pictorially.

Indeed, this intuition is a good representation of the full rigorous argument to prove this LDP.
Note that that $\hat{\mu}_N$ is a random probability measure, so it appears as though $\hat{\mu}_N$ obeys a LDP with speed $N^2$ and rate function something like $\mathcal{I}$, where recall 
\begin{align*}
    \mathcal{I}[\mu] = \int d\mu(\lambda) \mathcal{E}(\lambda; \mu) = \int d\mu(\lambda)  \lambda^2 - \int d\mu(\lambda)d\mu(\lambda') \log|\lambda - \lambda'|.
\end{align*}
This is in fact the case and was established in \cite{arous1997large}, where the rate function was found for for $\beta=1,2,4$ to be \begin{align}
    J_{\beta}[\mu]=\frac{1}{2}\left( \int d\mu(\lambda)  \lambda^2 -\beta \int d\mu(\lambda)d\mu(\lambda') \log|\lambda - \lambda'| + \frac{\beta}{2}\log\frac{\beta}{2} - \frac{3}{4}\beta\right),
\end{align}
which has a unique minimiser, with with value 0, among all probability measures on $\R$ at the semi-circle measure with radius $\sqrt{2\beta}$.
This fact alone is sufficient to establish (\ref{eq:ldp_intro_2}) for the GOE.
Indeed, consider the bounded Lipschitz distance on probability measure on $\R$
\begin{align*}
    d_{Lip}(\mu, \nu) = \sup_{\|f\|_{Lip}\leq 1}\left| \int f(x) d(\mu - \nu)(x)\right|
\end{align*}
where the supremum is taken over all Lipschitz function with Lipschitz constant at most $1$.
Using this metric, one can define a ball  $B_{\epsilon}(\mu_{SC})$ of radius $\epsilon$ centred on the minimiser $\mu_{SC}$ of $J_1$.
For $x\in (-2, 2)$, if $\lambda_k \geq x$ then \begin{align*}
    d_{Lip}(\mu_{SC}, \hat{\mu}_N) \geq \left| \int_{-2}^{x} d\mu_{SC}(\lambda) - \frac{k}{N}\right| > \epsilon
\end{align*}
for all large enough $N$ and for some fixed $\epsilon>0$ (independent of $N$).
So if $\lambda_k  \geq x$, then $\hat{\mu}_N$ lies outside the ball of radius $\epsilon$ centred on $\mu_{SC}$, but since $J_1$ has a unique minimiser, it follows that $J_1[\hat{\mu}_N] > \delta$ for all large enough $N$ and for some $\delta >0$ (independent of $N$), so the LDP on $\hat{\mu}_N$ yields the infinite limit (\ref{eq:ldp_intro_2}).
The proof to establish the complementary limit (\ref{eq:ldp_intro_1}) also makes use of the LDP on $\hat{\mu}_N$.
The joint density $p(\lambda_1, \ldots, \lambda_N)$ can be split as \begin{align*}
    p(\lambda_1, \ldots, \lambda_N) &= p(\lambda_{k+1}, \ldots, \lambda_N) f(\lambda_1, \ldots, \lambda_k; \lambda_{k+1}, \ldots, \lambda_N)\\
    &\propto \Delta(\{\lambda_i\}_{i=1}^k) p(\lambda_{k+1}, \ldots, \lambda_N) \exp\left(-N\sum_{j=1}^k\left\{\frac{\lambda_j^2}{2} - \int d\hat{\mu}_{N-k}(\lambda) \log |\lambda - \lambda_j|\right\}\right).
\end{align*}
By placing all eigenvalues inside large ball of radius $M$, the left-over Vandermonde term $\Delta(\{\lambda_i\}_{i=1}^k)$ can be bounded by $2M^{k^2}$, say. 
Since $N$ is very large and $k$ fixed, the LDP for the empirical spectral density $\hat{\mu}_{N-K}$ of $\lambda_{k+1},\ldots, \lambda_N$ applies and $\hat{\mu}_{N-k}$ can be effectively replaced by $\mu_{SC}$, incurring only an error term suppressed by an LDP bound of size $e^{-cN^2}$ for some constant $c>0$.
It then remain to bound the contribution from eigenvalues outside the ball of radius $M$ and to evaluate the supremum over $\lambda < x$ of $\int d\mu_{SC}(\lambda') \log |\lambda' - \lambda| - \frac{1}{2}\lambda^2$, which gives precisely the result $I_k(x)$ stated above.

\section{Random determinants}
We have discussed how the supersymmetric method can be used in random determinant calculations such as $\E_X\log |\det X|$ and this in fact provides the basis for much of our work with random determinants arising in Kac-Rice formulae in chapters \ref{chap:general_activation_functions} and \ref{chap:spin_glass_gans}.
In this section, we provide some broader background on random determinant calculations using different techniques and for other statistics.

\medskip
A foundational work in random determinant calculations is \cite{auffinger2013random}.
The focus of that work is the calculation of the complexity of the basic spherical p-spin glass model.
Let $f\colon \R^N \rightarrow \R$ be the spin glass and \vivacom{consider the following sets of points on the $N$-sphere:
\begin{align*}
    &\{\vec{x}\in S^N ~\mid~ \nabla f(\vec{x}) = 0, ~ f(\vec{x})< \sqrt{N}u\},\\
    &\{\vec{x}\in S^N ~\mid~ \nabla f(\vec{x}) = 0, ~ f(\vec{x})< \sqrt{N}u, ~ i(\nabla^2 f(\vec{x})) = k\},
\end{align*}}
where $i(\cdot)$ is the \emph{index}, which simply counts the number of negative eigenvalues of a real symmetric matrix.
\vivacom{For fixed $N$, note that these sets are almost surely finite and so one can unambiguously define the following notions of complexity simply as the cardinality of these sets:}
\begin{align*}
    \E C_N(u) &= |\{\vec{x}\in\R^N ~\mid~ \nabla f(\vec{x}) = 0, ~ f(\vec{x})< \sqrt{N}u\}|,\\
    \E C_{N,k}(u) &= |\{\vec{x}\in\R^N ~\mid~ \nabla f(\vec{x}) = 0, ~ f(\vec{x})< \sqrt{N}u, ~ i(\nabla^2 f(\vec{x})) = k\}|,
\end{align*}

The appropriateness of the scale $\sqrt{N}$ of the upper bound on $f$ will become apparent below.
The argument proceeds in the following steps:
\begin{enumerate}
    \item Apply a Kac-Rice formula to express the complexity as integral involving the absolute value of the determinant of a random Hessian:
    \begin{align*}
        \E C_{N,k}(u) = \int_{S^N} d\vec{x} ~ \E \left[|\det\nabla^2f| \1\{f(\vec{x}) \leq \sqrt{N}u\}\1\{i(\nabla^2 f(\vec{x})) = k\} ~\mid~ \nabla f(\vec{x}) = 0\right] p_{\vec{x}}(0)
    \end{align*}
    where $p_{\vec{x}}$ is the density of $\nabla f$ at $\vec{x}$.
    \item Exploit spherical symmetry of the integrand to dispense with the integral over the $N$-sphere.
    \item Use the covariance function of $f$ to derive the joint distribution of $f$, its derivatives and its Hessian. The derivatives must be taken parallel to the $N$-sphere, so the Hessian is an $N-1\times N-1$ matrix. One discovers that $\nabla f$ is independent of $f$ and $\nabla f$, which greatly simplifies the above expectation. Moreover, $\nabla^2 f$ just has Gaussian entries and Gaussian conditioning laws can be used to derive the distribution of $\nabla^2 f ~\mid ~ f(\vec{x}) = y$. One finds that it is a shifted GOE $X - yI$, where $X$ is a standard GOE. In addition, $\nabla f$ is an isotropic Gaussian vector with variance $p$.
    
    \item The complexity is then given by \begin{align}
        \E C_{N,k} \propto \int_{-\infty}^u dy ~ e^{-\frac{Ny^2}{2}} \E\left[ |\det(X - yI)| ~\mid~ \1\{i(X-yI) = k\}\right].
    \end{align}
    \item The determinant simplifies greatly and can be written as a product over eigenvalues. Then the expectation can be rewritten as \begin{align*}
        \E\left[ |\det(X - yI)| ~\mid~ \1\{i(X-yI) = k\}\right] \propto \int &d\lambda_1\ldots\lambda_{N-1} e^{-\frac{Ny^2}{2}} \prod_{j=1}^{N-1} e^{-\frac{(N-1)\lambda_j^2}{2}}\Delta(\{\lambda_i\}_{i=1}^{N-1}) \prod_{j=1}^{N-1} |\lambda_j - y| \\
        &\1\{\lambda_1 \leq \ldots \leq y \leq \lambda_{N-1}\}.
    \end{align*}
    Note that from the above expression it is clear that $\sqrt{N}$ is the correct scaling to make the density of $f$ agree with that of the eigenvalues of $\nabla^2 f$.
    With some re-scaling of variables, the determinant and the Vandermonde terms combine to give an $N\times N$ Vandermonde, so overall \begin{align*}
        \E C_{N, k}(u) \propto \P(\lambda_k \leq A_Nu)
    \end{align*}
    with the probability taken over an $N\times N$ GOE and for some constant $A_N$.
    $\E C_{N,k}$ can then be computed using a large deviations principle for the $k$-th eigenvalue of an $N\times N$ GOE.
    \item $C_N$ can be derived from $C_{N,k}$ by summing over all $k$.
\end{enumerate}
In reality, the main results of \cite{auffinger2013random} and related work (such as our own) focus on computing the leading order term in a large $N$ asymptotic expansion of $\log \E |\det (X - yI)|$, though in some cases it is possible to compute the sharp leading order term in $\E |\det (X - yI)|$, as done in \cite{auffinger2013random} and also in chapter \ref{chap:general_activation_functions}.
To state the precise results from \cite{auffinger2013random}, we require the following definitions:

 \begin{align}
    \Theta_p(u) = \begin{cases} \frac{1}{2}\log(p-1) - \frac{p2-2}{4(p-1)}u^2 - I_1(u; E_{\infty}) ~~ &\text{if } u\leq -E_{\infty},\\
    \frac{1}{2}\log(p-1) - \frac{p-2}{4(p-1)}u^2 &\text{if } -E_{\infty} \leq u \leq 0,\\
    \frac{1}{2}\log(p-1) &\text{if } 0\geq u,\end{cases}
    \end{align}
    where $E_{\infty} = 2\sqrtsign{\frac{p-1}{p}}$, and $I_1(\cdot; E)$ is defined on $(-\infty, -E]$ by \begin{equation}
        I_1(u; E) = \frac{2}{E^2}\int_{u}^{-E} (z^2 - E^2)^{1/2} dz = -\frac{u}{E^2}\sqrtsign{u^2 - E^2} - \log\left(-u + \sqrtsign{u^2 - E^2}\right) + \log E,
    \end{equation} 
    and  \begin{align}
    \Theta_{p,k}(u) = \begin{cases} \frac{1}{2}\log(p-1) - \frac{p-2}{4(p-1)}u^2 - (k+1)I_1(u; E_{\infty}) ~~ &\text{if } u\leq -E_{\infty},\\
     \frac{1}{2}\log(p-1) - \frac{p-2}{p}
     &\text{if } u > -E_{\infty}. \end{cases}
    \end{align}
    
    Then we have the following limit results \begin{align}
        \lim_{N\rightarrow\infty} \frac{1}{N}\log\expect C_{N}(u) = \Theta_p(u),~~
        \lim_{N\rightarrow\infty} \frac{1}{N}\log\expect C_{N,k}(u) = \Theta_{p,k}(u).
    \end{align}

\begin{figure}
\centering
\begin{subfigure}{0.45\textwidth}
    \centering
    \includegraphics[width=\textwidth]{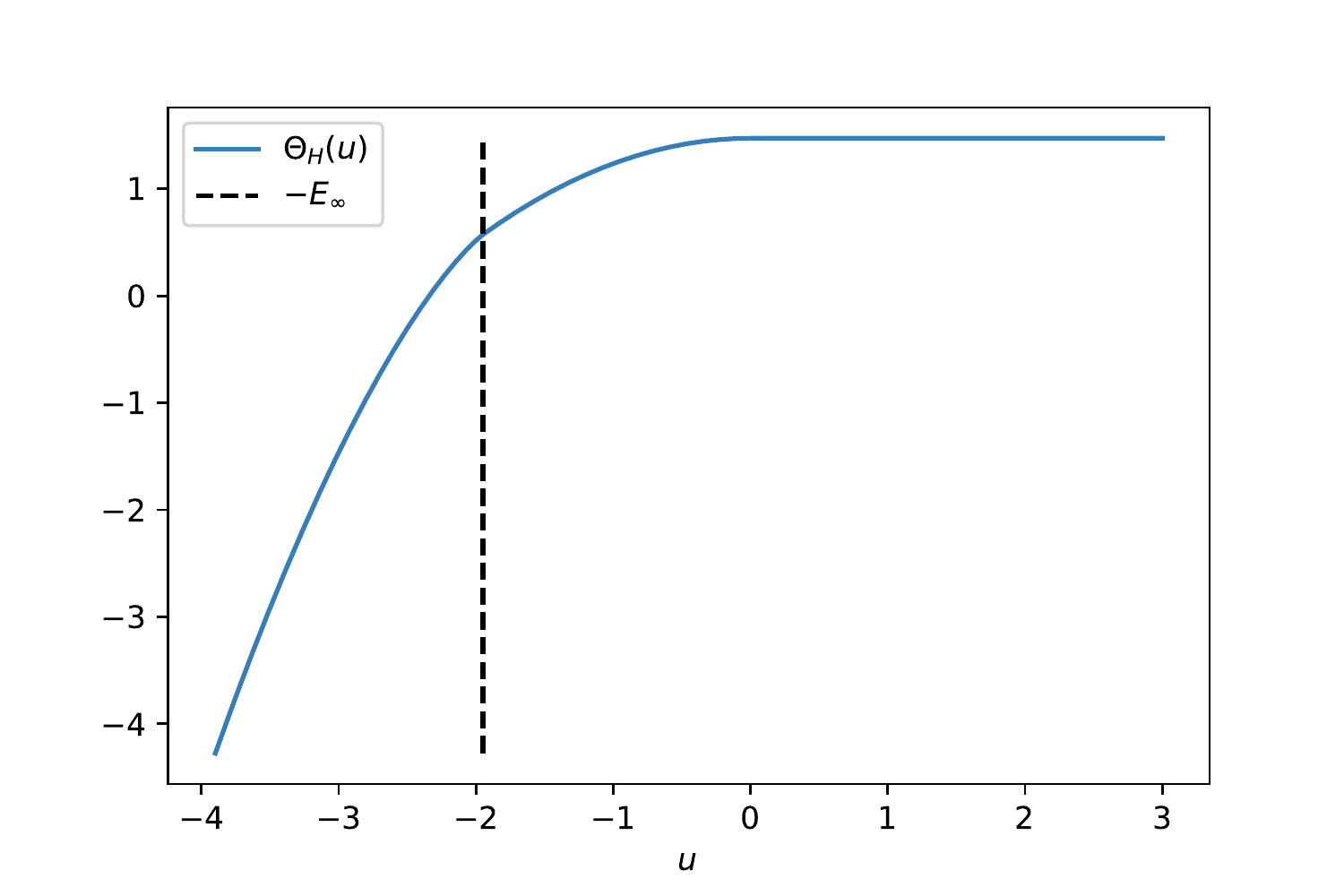}
    \caption{Plot of  $\Theta_H$.}
\end{subfigure}
\begin{subfigure}{0.45\textwidth}
    \centering
    \includegraphics[width=\textwidth]{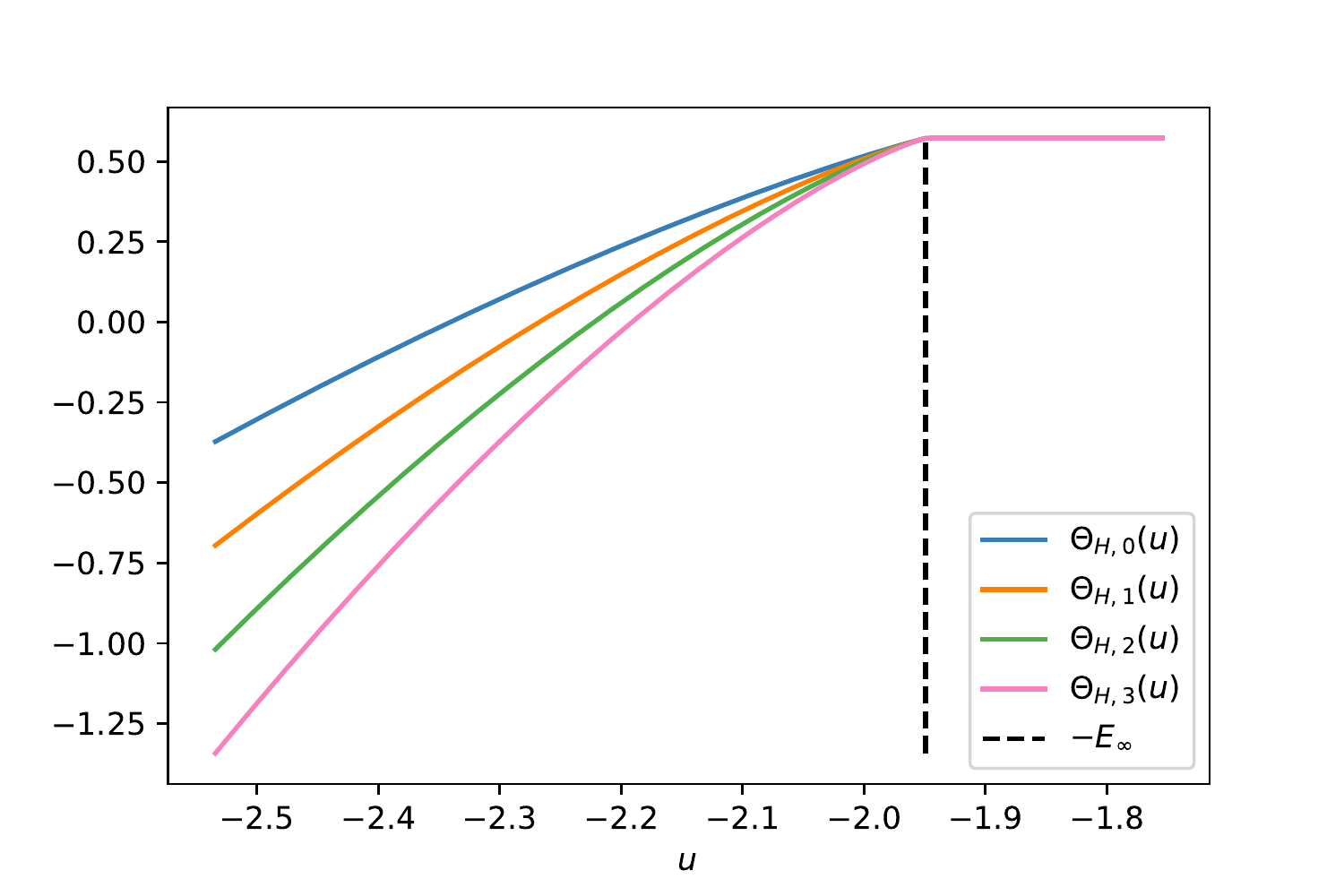}
    \caption{Plot of $\Theta_{H,k}$  $k=0,1,2,3$.}
\end{subfigure}
\caption{Plots of the functions $\Theta_H$ and $\Theta_{H,k}$ for $H=20$.}
\label{fig:auff_thetas_intro}
\end{figure}

There are some important features to highlight about these results.
Note that $E_{\infty}$ plays the role of the left edge of the support of a semi-circle density which, of course, has its origin in the GOE distribution of $f$'s Hessian.
In particular, note that $\Theta_{p, k}$ includes large deviations terms for $u$ below $-E_{\infty}$, the effective left edge of a semi-circle, but not above it.
We also note the structure of stationary points of $f$ that is encoded in $\Theta_p$ and $\Theta_{p,k}$ for which we show plots in Figure \ref{fig:auff_thetas_intro}.
Negative values of $\Theta_{p,k}(u)$ correspond to upper bounds on $f$ below which it has `exponentially few' stationary points of index $k$ i.e. effectively none.
Positive values, by contrast, correspond to exponentially many stationary points of index $k$.
This therefore is the mathematical description of the `layered structure' of spin glass stationary points on which \cite{choromanska2015loss} and our results in chapters \ref{chap:general_activation_functions} and \ref{chap:spin_glass_gans} depend.
There exist critical values ${E_i}_{i=1}^{\infty}$ such that $\Theta_{p, i}(-E_i) = 0$.
For $f$ below the critical value $-E_0$, there are effectively no stationary points of $f$.
Between $-E_0$ and $-E_1$, there are exponentially many local minima, but effectively no stationary points of any other index.
Between $-E_1$ and $-E_2$ there are exponentially many local minima and stationary points of index 1, but effectively none of any higher indices.
The final critical value is $-E_{\infty}$, above which stationary points of all indices are found 

The quantity $\log\E C_N$, where the logarithm is taken \emph{after} the expectation is known as the \emph{annealed average}, and so the corresponding complexity is known as the annealed complexity.
The alternative is known as the \emph{quenched complexity}, in which the expectation is taken after the logarithm. We shall discuss the differences between the two below.
The first few steps outlined above are quite general and we shall see them repeated, mutatis mutandis, in chapters \ref{chap:general_activation_functions} and \ref{chap:spin_glass_gans}.
The later steps, however, are clearly highly specific to the precise conditional Hessian distribution of the spin-glass.
In particular, if the Hessian were a GOE shifted by a some matrix other than a multiple of the identity, then one would be unable to so easily dispense with the eigenvector component of the matrix expectation.
Further, step 5 is an miraculous simplification wherein the conditional value of $f$ is effectively inserted as an extra eigenvalue of the GOE, so reducing the whole calculation to a tail probability of the $k$-th eigenvalue of a GOE.
We shall in chapters \ref{chap:general_activation_functions}, \ref{chap:spin_glass_gans} and \ref{chap:univ} how supersymmetric techniques, among others, can be be employed to generalise these steps in more complicated settings.
In a sequence of recent works \cite{arous2021exponential,arous2021landscape,mckenna2021complexity} the question of random determinants was considered for considered for very general random matrices.
Indeed, there is every reason to believe that the general framework developed particularly in \cite{arous2021exponential} provides close to optimal conditions under which the annealed average over absolute values of random matrix determinants can be computed. 
The method developed in that work is, in essence, a rigorous mathematically justified version of a general mathematical physics approach known as the \emph{Coulomb gas method} \cite{cunden2016shortcut,forrester2010log}.
Consider a random $N\times N$ matrix $X$ with (random) eigenvalues $\lambda_1,\ldots, \lambda_N$, empirical spectral density $\hat{\mu}_N$ and assume a limiting spectral density $\mu$.
Let us consider real symmetric $X$, but of course what we describe can be equally well presented for Hermitian $X$.
One can simply express the determinant of $X$ in terms of its eigenvalues alone and then use the definition of $\hat{\mu}_N$ to write \begin{align*}
    \E |\det X| = \E \prod_{j=1}^N |\lambda_j| = \E \exp\left\{N\int d\hat{\mu}_N(\lambda) \log|\lambda|\right\}.
\end{align*}
Recall from (\ref{eq:jpdf_as_functional_intro}) that the eigenvalue density can be written in the form \begin{align*}
      p(\lambda_1, \ldots, \lambda_N) = \frac{1}{Z_N}\exp\left(-\frac{N^2}{2} \mathcal{I}[\hat{\mu}_N]\right),
\end{align*}
so that \begin{align*}
     \E |\det X| = \frac{1}{Z_N}\int d\lambda_1\ldots d\lambda_N  \exp\left\{N\int d\hat{\mu}_N(\lambda) \log|\lambda|\right\}\exp\left(-\frac{N^2}{2} \mathcal{I}[\hat{\mu}_N]\right).
\end{align*}
Heuristically, the Laplace method can be applied to conclude that the dominant leading order contribution to this integral as $N\rightarrow\infty$ comes from $\hat{\mu}_N$ in a small ball around $\mu$, so \begin{align*}
    \E |\det X| \sim \exp\left\{N\int d\mu(\lambda) \log|\lambda|\right\}
\end{align*}
and then \begin{align*}
    \frac{1}{N} \log\E |\det X| \sim \int d\mu(\lambda) \log|\lambda|.
\end{align*}
This approach gives solid intuition for the asymptotic behaviour of $\E |\det X|$ in general, but is of course only heuristic.
In chapter \ref{chap:spin_glass_gans}, the Coulomb gas method plays an important part in the Kac-Rice calculation of complexity, however we have to expend some effort to provide the rigorous justification for its use in that particular case and these arguments are quite specific to the matrix ensemble in question.
The main theorems of \cite{arous2021exponential} provide a general justification for the Coulomb gas method, or really the result above that can be derived using it.
The theorems are quite general but rely on a number of technical conditions on the matrix ensemble and much of the effort in that paper and its companions \cite{arous2021landscape,mckenna2021complexity} is devoted to proving satisfaction of these conditions for some particular matrix ensembles of interest.
Interestingly, parts of the argument in \cite{arous2021exponential} are not dissimilar to the Laplace method heuristic above, as one of the key ingredients is a condition on $X$ giving good enough bounds on the convergence rate of $\hat{\mu}_N$ to $\E\hat{\mu}_N$ and $\E\hat{\mu}_N$ to $\mu$.
At the time of writing, these results are the most general and powerful tools for calculating $\E|\det X|$, however establishing satisfaction of their conditions is by no means straightforward so for some matrix ensembles, less general techniques may be easier to apply.

\medskip
We close this section by mentioning the differences between the annealed averages that we have discussed in some detail and the alternative quenched averages.
The Jensen-Shannon theorem gives the inequality \begin{align*}
   \E  \log |\det X| \leq \log \E |\det X|
\end{align*}
so the annealed average is an upper bound for the quenched average and likewise the annealed complexity of a random function is an upper bound for the quenched complexity.
The annealed complexity has received much more attention in the literature, in part because it is more analytically tractable.
At least heuristically, one can see why this should be by just trying to repeat the simple Coulomb gas argument above.
Recall that the key to the argument's success (and, in some real sense, the success of \cite{arous2021exponential}) is expressing $|\det X|$ as $\exp\left(N\int d\hat{mu}_N(\lambda) \log|\lambda|\right)$.
This expression, written as a functional of $\hat{\mu}_N$ and in the form $e^{N\ldots}$ is exactly what is required for Laplace style asymptotic analysis when combined with the eigenvalue density inside the expectation.
By contrast, $\log |\det X| = N\int d\hat{\mu}_N(\lambda)\log|\lambda|$, which cannot be expressed in the above Laplace-amenable form.
\cite{ros2019complex} is an important recent work that begins the extension of the Kac-Rice approach to quenched complexity via the non-rigorous replica method.
The authors highlight that the quenched and annealed complexities do not in general agree even to leading order and argue that the quenched complexity is, in some sense, the better representation of a surface's complexity.
In chapters \ref{chap:general_activation_functions} and \ref{chap:spin_glass_gans} in which annealed complexity calculations feature significantly, we use highly simplified statistical physics models of much more complicated objects (deep neural networks), attempting to retain just enough of the original structure to provide some insight while still having an analytically tractable complexity.
What's more, the complexity itself, annealed or quenched, is just a static snapshot of the already much simplified loss landscape, whereas real-world neural networks are trained over some complex stochastic trajectory in parameter space.
As with any model of a complex system, these complexity calculations can only ever be expected to provide some limited insight into aspects of the underlying system. Given that the models themselves are very simplified and a focus on just their complexity is a considerable simplification of real training dynamics, we argue that the distinction between annealed on quenched complexity in this context, while important, is not the most significant factor affecting ecological validity.

\medskip
Finally, we note that quantities other than the expectation of complexity (equivalently: absolute values of determinants) have been considered.
In the context of spherical $p$-spin glass considered in \cite{auffinger2013random}, the \emph{variance} of the complexity is obtained in \cite{Sub2017} which is necessary to determine whether the expected value is typical.
The proofs in this case are much more technical than those for the expectation and extensions to more complicated models such as those considered in Chapters \ref{chap:general_activation_functions} and \ref{chap:spin_glass_gans} appears out of reach.

\section{Free probability}\label{sec:free_prob}
Free probability theory is a rich and deep field describing probability distributions on non-commuting algebras. The notation of \emph{freeness} itself provides the generalisation of the concept of independence from standard probability theory to non-commuting algebras.
The theory extends beyond the boundaries of random matrix theory to probability distributions on more general algebras \cite{voiculescu1992free}, but its connection to random matrix theory is immediately clear: random matrices are non-commuting objects endowed with probability distributions.
For the purposes of this thesis, we will need only a basic introduction to free probability in the context of random matrices.

\medskip
Consider two $N\times N$ real matrices $A$ and $B$, where $A$ is random and $B$ may be random or deterministic.
Suppose that $A$ is rotationally invariant, i.e. its eigenvectors follow Haar measure on the orthogonal group.
$A$ is then said to be \emph{in general position} compared to $B$, which means roughly that there is entirely no correlation or dependence between their eigenspaces.
In this case, $A$ and $B$ can be shown to be free independent of each other.
Suppose that both $A$ and $B$ have limiting spectral measures $\mu$ and $\nu$ respectively and let $C = A + B$. 
Since $A$ and $B$ are free independent, it is known \cite{voiculescu1992free,anderson2010introduction} that $C$ has the limiting spectral measure $\mu \boxplus \nu$, which is known as the free additive convolution between the measures $\mu$ and $\nu$.
To define the free additive convolution, we must introduce some integral transforms.
Let $g_{\mu}, g_{\nu}$ be the Stieljtes transforms of $\mu$ and $\nu$ and let $B_{\mu} = g^{-1}_{\mu}$ and $B_{\nu} = g^{-1}_{\nu}$ be their inverses.
The $R$-transforms are then defined as $R_{\mu}(z) = B_{\mu}(z) - z^{-1}$ and  $R_{\nu}(z) = B_{\nu}(z) - z^{-1}$.
The $R$-transforms play the role of Fourier transforms for probability measures, as one has the result \begin{align}\label{eq:intro_r_trans_plus}
    R_{\mu\boxplus \nu} = R_{\mu} + R_{\nu}.
\end{align}
In fact, one must take care with the definitions of these transforms. The above expressions are just a consequence of their true definitions as formal power series in the complex plane.

The Stieljtes transform of a measure is given by the power series

\begin{align}
    g_{\mu}(z) = \sum_{n\geq 0} m_n^{(\mu)} z^{-(n+1)}
\end{align}
where $m_n^{(\mu)}= \int d\mu(x) ~ x^n$ is the $n$-th moment of $\mu$ (likewise for $\nu$).
The $R$-transform of a measure is defined as a formal power series \cite{anderson2010introduction}
\begin{align}
   R_{\mu}(z) = \sum_{n=0}^{\infty} k_{n+1}^{(\mu)} z^n
\end{align}
where $k_n^{(\mu)}$ is the $n$-th cumulant of the measure $\mu$.
It is known \cite{anderson2010introduction} that $k_n^{(\mu)}=C_n^{(\mu)}$ where the functional inverse of the Stieljtes transform of the measure is given by the formal power series
\begin{align}
    B_{\mu}(z) = \frac{1}{z} + \sum_{n=1} C_n^{(\mu)} z^{n-1}.
\end{align}

So the key result (\ref{eq:intro_r_trans_plus}) is really a statement about the cumulants of $\mu, \nu$ and $\mu\boxplus\nu$, namely $k_n^{(\mu\boxplus \nu)} = k_n^{(\mu)} + k_n^{(\nu)}$.

There is a useful relation between cumulants and moments which can be found, for example, in the proof of Lemma 5.3.24 in \cite{anderson2010introduction}:\begin{align}
    m_n &= \sum_{r=1}^n ~\sum_{\substack{0\leq i_1,\ldots, i_r\leq n-r \\ i_1+\ldots + i_r = n-r}} k_r m_{i_1}\ldots m_{i_r}.
\end{align}

The final concept we need from free probability theory is \emph{subordination functions}.
Given measures $\mu, \nu$ there exists a subordination function $\omega:\C \rightarrow\C$ such that $g_{\mu\boxplus \nu}(z) = g_{\nu}(\omega(z))$  \cite{biane1997free}.
Depending on the context, the subordination function formulation relating $\mu\boxplus\nu$ to $\mu$ and $\nu$ can prove more convenient than the formulation via sums of $R$-transforms, see e.g. \cite{biane1997free,capitaine2016spectrum} and chapter \ref{chap:univ} below.

\medskip
We conclude this briefest of introduction to free probability by providing a few concrete results for integral transforms of the a specific measure, namely the semi-circle $\mu_{SC}$ with density $\rho_{SC}(x) = \pi^{-1}\sqrt{2 - x^2}$. We shall include the calculations as we have been repeatedly frustrated to find them absent from the literature. Henceforth $\mu = \mu_{SC}$ and we will drop all $\mu$ and $SC$ labels.

\paragraph{Stieltjes transform} For odd $n$ clearly $m_n =0$ by the symmetry of the semi-circle measure. Now consider the even moments:
\begin{align}
    m_{2n} = \pi^{-1}\int dx ~ x^{2n} \sqrt{2 - x^2} &= 2^{1 + n}\pi^{-1}\int_{-\pi/2}^{\pi/2} d\theta \cos^2\theta \sin^{2n}\theta\notag\\
    &= 2^{1 + n}\pi^{-1}\int_{-\pi/2}^{\pi/2} d\theta (\sin^{2n}\theta - \sin^{2(n+1)}\theta)
\end{align}

The trigonometric integrals are standard exercises in basic calculus\footnote{The usual approach is to write $\sin^{2n}\theta = \sin^{2n-2}\theta - \cos^2\theta\sin^{2n-2}\theta$, apply integration by parts to the second term and then iterate.}: \begin{align}
    \int_{-\pi/2}^{\pi/2} d\theta \sin^{2n}\theta = \pi \frac{2n - 1}{2n}\frac{2n - 3}{2n-2}\ldots \frac{1}{2}
\end{align}
so \begin{align}
    m_{2n} = 2^{1 + n} \frac{2n - 1}{2n}\frac{2n - 3}{2n-2}\ldots \frac{1}{2}\left(1-\frac{2n+1}{2n+2}\right) = 2^{1 + n} \frac{2n - 1}{2n}\frac{2n - 3}{2n-2}\ldots \frac{1}{2}\frac{1}{2n + 2}.
\end{align}
Thus we have the Stieltjes transform 
\begin{align}
    g(z)& = \sum_{n=0}^{\infty} z^{-(2n + 1)} 2^{1+n}\frac{1}{2n + 2} \frac{2n - 1}{2n}\frac{2n - 3}{2n-2}\ldots \frac{1}{2}\notag\\
    & = z \sum_{n=0}^{\infty} \left(\frac{z^2}{2}\right)^{-(n+1)}\frac{1}{2n + 2} \frac{2n - 1}{2n}\frac{2n - 3}{2n-2}\ldots \frac{1}{2}\notag\\
        & = z \sum_{n=0}^{\infty} \left(\frac{z^2}{2}\right)^{-(n+1)}\frac{1}{(n+1)!} \frac{(2n - 1)(2n-3)\ldots 1}{2^{n+1}}\notag\\
        & = z \sum_{n=0}^{\infty} \left(\frac{-z^2}{2}\right)^{-(n+1)}\frac{1}{(n+1)!} \frac{(2n - 1)(2n-3)\ldots 1}{2^{n+1}}(-1)^{n+1}
\end{align}
and we can now identity the Taylor expansion of a familiar function, so \begin{align}
    g(z) = z \left(1- \sqrt{1 - \frac{2}{z^2}} \right) = z-\sqrt{z^2 - 2}.
\end{align}
For a general semi-circle with radius $r$, we can thence immediately write down its Stieltjes transform \begin{align}
    \frac{2}{r} \left(z-\sqrt{z^2 - r^2}\right)
\end{align}
where the pre-factor comes simply from the appropriate normalisation of the density $\sqrt{r^2 - x^2}$ relative to $\sqrt{2 - x^2}$.
Inverting this Siteltjes transform is simple. Let $y(z) = g_r^{-1}(z)$, then \begin{align*}
    &rz = 2y - 2\sqrt{y^2 - r^2}\\
    \iff & 4y^2 -4r^2 = 4y^2 - 4zry + z^2r^2\\
    \iff & g_r^{-1}(z) = y(z) = \frac{1}{z} + \frac{rz}{4}
\end{align*}
from which it follows that $R_r(z) = \frac{rz}{4}$.

\section{Local laws and universality}\label{sec:universal_intro}
Earlier in this chapter, we introduced the Wigner surmise and the rough notion of local universality in random matrices. This section provides further details about universality, with particular emphasis on the rather stunning sequence of papers beginning around \cite{erdHos2012bulk} that are well on the way to answering quite definitively the question of local universality.\\

  Broadly speaking, universality refers to the phenomenon that certain properties of special random matrix ensembles (such as the GOE) remain true for more general random matrices that share some key feature with the special ensembles.
    For example, the Wigner semicircle is the limiting spectral density of the Gaussian Wigner ensembles, i.e. matrices with Gaussian entries, independent up \review{to} symmetry (symmetric real matrices, Hermitian complex matrices) \cite{mehta2004random}.
    The Gaussian case is the simplest to prove, and there are various powerful tools not available in the non-Gaussian case, however the Wigner semicircle has been established as the limiting spectral density for Wigner matrices with quite general distributions on their entries \cite{anderson2010introduction,tao2012topics}.
    While surprisingly general is some sense, the Wigner semicircle relies on independence (up to symmetry) of matrix entries, a condition which is not typically satisfied in real systems.
    The limiting form of the spectral density of a random matrix ensemble is a \emph{macroscopic} property, i.e. the matrix is normalised such that the average distance between adjacent eigenvalues is on the order of $1/\sqrt{N}$, where $N$ is the matrix size.
    At the opposite end of the scale is the \emph{microscopic}, where the normalisation is such that eigenvalues are spaced on a scale of order $1$; at this scale, random matrices display a remarkable universality.
    For example, any real symmetric matrix has a set of orthonormal eigenvectors and so the set of all real symmetric matrices is closed under conjugation by orthogonal matrices.
    Wigner conjectured that certain properties of GOE matrices hold for very general random matrices that share the same (orthogonal) symmetry class, namely symmetric random matrices (the same is true of Hermitian random matrices and the unitary symmetry class).
    The spacings between adjacent eigenvalues should follow a certain explicit distribution, the Wigner surmise, and the eigenvectors should be \emph{delocalised}, i.e. the entries should all be of the same order as the matrix size grows.
    Both of these properties are true for the GOE and can be proved straightforwardly with quite elementary techniques.
    \review{Indeed, in the case of $2\times 2$ GOE, it is a standard first exercise in random matrix theory to prove that the eigenvalue spacing distribution is precisely the Wigner surmise (for $N\times N$ GOEs it is only a good approximation and improves as $N\rightarrow\infty$).}
    \review{Microscopic random matrix universality is known to be far more robust than universality on the macroscopic scale. Indeed, such results are well established for invariant ensembles and can be proved using Riemann-Hilbert methods \cite{deift1999orthogonal}.}
    \review{For more general random matrices, microscopic universality} has been proved \review{by quite different methods} in a series of works over the last decade or so, of which a good review is \cite{erdos2017dynamical}.
    Crucial in these results is the notion of a \emph{local law} for random matrices.
    The technical statements of some local laws are given below, but roughly they assert that the spectrum of a random matrix is, with very high probability, close to the deterministic spectrum defined by its limiting spectral density (e.g. the semicircle law for Wigner matrices).
    Techniques vary by ensemble, but generally a local law for a random matrix ensemble provides the control required to demonstrate that certain matrix statistics are essentially invariant under the evolution of the Dyson Brownian motion.
    In the case of real symmetric matrices, the Dyson Brownian motion converges in finite time to the GOE, hence the statistics preserved under the Dyson Brownian motion must match the GOE.
    The $n$-point correlation functions of eigenvalues are one such preserved quantity, from which follows, amongst other properties, \review{that the Wigner surmise is a good approximation to the adjacent spacings distribution.}
    The process we have just outlined is known as the `three step strategy', which we now state in its entirety for real Wigner matrices, though the essence of the strategy is much more general.
    \begin{enumerate}
        \item Establish a local semi-circle law for the general Wigner ensemble $X$.
        \item Universality for Gaussian divisible ensembles. Consider a random matrix $X_t= e^{-t/2}X + \sqrt{1 - e^{-t}} G$, where $G$ is a standard GOE matrix. One must show that $X_t$ has universality for $t=N^{-\tau}$ for any $0< tau < 1$. The clearest interpretation of this result is that, as $X$ evolves under a matrix Ornstein-Uhlenbeck process, its local eigenvalue statistics have `relaxed' to those of the GOE after any timescales greater than $N^{-1}$. Concretely this process is \begin{align*}
            dX_t = \frac{1}{\sqrt{N}}dB_t - \frac{1}{2} X_t dt
        \end{align*}
        where $B_t$ is a standard symmetric Brownian motion and the initial data is $X_0 = X$.
        The local law on $X$ is a key ingredient in establishing this result.
        \item Approximation by a Gaussian divisible ensemble. This final step, sometimes called the `comparison step', has to show that the local statistics of the matrix $X$ can be well approximated by those of the Gaussian divisible ensemble $X_t$ for short times scale $N^{-\tau}$ where $\tau < 1$. Combining with step 2, one then obtain universality for $X$.
    \end{enumerate}

\medskip
We now make the preceding statements about correlation functions precise, following the treatment in \cite{erdHos2017dynamical}. For an $N\times N$ matrix $X$, let $p_N^{(k)}$ be its $k$-point correlation function, i.e. \begin{align*}
    p_N^{(k)}(x_1, \ldots, x_k) = \int d\lambda_{k+1}\ldots d\lambda_N p_N(x_1, \ldots, x_k, \lambda_{k+1}, \ldots, \lambda_N)
\end{align*}
where $p_N$ is simply the symmetrised joint probability density of the eigenvalues of $X$ (i.e. the joint density of the unordered eigenvalues).
Assume that $X$ has a limiting spectral density $\rho$ with compact support and is normalised so that it the support is $[-\sqrt{2}, \sqrt{2}]$.
Assume also that the symmetry group of $X$ is $O(N)$, i.e. $X$ is real-symmetric.
One statement of spectral universality for $X$ is that for any $\kappa>0$ and for any $E\in [-\sqrt{2} + \kappa, \sqrt{2}-\kappa]$ we have
\begin{align*}
    \lim_{N\rightarrow\infty} \frac{1}{\rho(E)^k} \int_{\R^k} d\vec{\alpha} F(\vec{\alpha}) p_N^{(n)} \left( E + \frac{\vec{\alpha}}{N\rho(E)}\right) = \int_{\R^k} d\vec{\alpha} F(\vec{\alpha}) q_{GOE}^{(k)} (\vec{\alpha})
\end{align*}
for any smooth and compactly supported function $F\colon \R^k \rightarrow \R$.
Here $ q_{GOE}^{(k)}$ is simply the $k$-point correlation function for a GOE scaled so that its semi-circular radius is $\sqrt{2}$. 
This is so-called \emph{spectral universality in the bulk}.
From this statement, the local nature of spectral universality is quite plain.
One fixes some location inside the bulk of the limiting spectral density of $X$, referred to as an \emph{energy} $E$\footnote{The physics terminology is due to the historical origins of spectral universality in the Wigner surmise within the context of random matrix models for quantum mechanical Hamiltonians.}, then ones takes an fixed number $k$ of eigenvalues and looks at their marginal joint probability density in a region of the spectrum centred tightly on $E$.
As the matrix size $N$ diverges, so the small region around $E$ shrinks and the joint distribution of the $k$ eigenvalues in the small region converges to simply the joint distribution of $k$ eigenvalues of a standard GOE matrix.
Note that the `small region' around the location $E$ in the spectral bulk has a precisely prescribed scaling of $1/N$, which is the scaling so that, with overwhelming probability, the number of eigenvalues in the small region is of order 1.
Spectral universality as presented above is clearly good deal stronger than the Wigner surmise and is describing at least a similar phenomenon.
We can go further however, an consider a different formulation of spectral universality that is a direct generalisation of the Wigner surmise, namely \emph{spectral gap universality in the bulk}.
Of course, we note that all of the above has been stated for real symmetric matrices and the GOE, but could equally well have been stated for Hermitian matrices and the GUE.

For an $0<\alpha<1$ and any integers $r,s\in [\alpha N, (1-\alpha)N]$
\begin{align*}
    \lim_{N\rightarrow\infty} \Bigg|&\E_{X} F\left(N\rho(\lambda_r)(\lambda_r - \lambda_{r+1}), \ldots, N\rho(\lambda_r)(\lambda_r- \lambda_{r+k})\right) \\ &- 
    \E_{GOE} F\left(N\rho_{SC}(\lambda_s)(\lambda_s - \lambda_{s+1}), \ldots, N\rho_{SC}(\lambda_s)(\lambda_s- \lambda_{s+K})\right)  \Bigg| = 0 
\end{align*}
where $F$ is an arbitrary function as before.
These two formulations of spectral universality are known to be equivalent \cite{erdHos2017dynamical}.
To recover the Wigner surmise, take $n=1$ and then one obtains
\begin{align}\label{eq:intro_prec_wigner_gap}
    \lim_{N\rightarrow\infty} \Bigg|\E_{X} F\left(N\rho(\lambda_r)(\lambda_r - \lambda_{r+1})\right) - \E_{GOE} F\left(N\rho_{SC}(\lambda_s)(\lambda_s - \lambda_{s+1})\right)\Bigg| = 0.
\end{align}
Note that $\rho_{SC}(\lambda_s)N$ is precisely the scaling required around $\lambda_s$ to bring the GOE eigenvalues onto the scale on which the mean spacing is unity, thus for large $N$ \begin{align*}
    \E_{GOE} F\left(N\rho_{SC}(\lambda_s)(\lambda_s - \lambda_{s+1})\right) = \int dr \rho_{\text{Wigner}}(r) F(r) + o(N),
\end{align*}
and so (\ref{eq:intro_prec_wigner_gap}) is indeed the precise statement of the universality of the Wigner surmise for $X$.

\medskip
There are several forms of local law, but \review{all provide high probability control on} the error between the (random) matrix Green's function $G(z) = (z - X)^{-1}$ and certain deterministic equivalents.
In all cases we use the set 
\begin{align}
    \vec{S} = \left\{E + i\eta \in \C \mid |E| \leq \omega^{-1}, ~ N^{-1 + \omega} \leq \eta \leq \omega^{-1}\right\}
\end{align}
for $\omega\in(0, 1)$ and the local law statements holds for all (large) $D>0$ and (small) $\xi > 0$ and for all large enough $N$.
The \emph{averaged local law} states:
\begin{align}
   \sup_{z\in\vec{S}} \P\left(\left|\frac{1}{N}\Tr G(z) - g_{\mu}(z)\right| > N^{\xi}\left(\frac{1}{N\eta} + \sqrtsign{\frac{\Im g_{\mu}(z)}{N\eta}}\right)\right) \leq N^{-D}.
\end{align}
The \emph{isotropic local law} states:
\begin{align}\label{eq:iso_local_law_intro}
    \sup_{\|\vec{u}\|,\|\vec{v}\|  = 1, z\in\vec{S}}\P\left( |\vec{u}^TG(z)\vec{v} - g_{\mu}(z)| > N^{\xi}\left(\frac{1}{N\eta} + \sqrtsign{\frac{\Im g_{\mu}(z)}{N\eta}}\right)\right) \leq N^{-D}.
\end{align}
The \emph{anisotropic local law} states:
\begin{align}\label{eq:aniso_local_law_intro}
    \sup_{\|\vec{u}\|,\|\vec{v}\|  = 1, z\in\vec{S}}\P\left( |\vec{u}^TG(z)\vec{v} - \vec{u}^T\Pi(z)\vec{v}| > N^{\xi}\left(\frac{1}{N\eta} + \sqrtsign{\frac{\Im g_{\mu}(z)}{N\eta}}\right)\right) \leq N^{-D}
\end{align}
where $\Pi(\cdot)$ is an $N\times N$ deterministic matrix function on $\mathbb{C}$.
The \emph{entrywise local law} states:
\begin{align}\label{eq:entry_local_law_intro}
    \sup_{z\in\vec{S}, 1\leq i,j\leq N}\P\left( |G_{ij}(z) - \Pi_{ij}(z)| > N^{\xi}\left(\frac{1}{N\eta} + \sqrtsign{\frac{\Im g_{\mu}(z)}{N\eta}}\right)\right) \leq N^{-D}.
\end{align}

The anisotropic local law is a stronger version of the entrywise local law.
The anisotropic local law is a more general version of the isotropic local law, which can be recovered in the isotropic case by taking $\Pi = g_{\mu} I$.
The entrywise local law can also be applied in the isotropic case by taking $\Pi = g_{\mu} I$.
The averaged local law is weaker than all of the other laws.
General Wigner matrices are known to obey isotropic local semi-circle laws \cite{erdHos2017universality}.
Anisotropic local laws are known for general deformations of Wigner matrices and general covariance matrices \cite{knowles2017anisotropic} as well as quite general classes of correlated random matrices \cite{erdHos2019random}.

\medskip
Local universality is not limited to the eigenvalues of random matrices.
Recall that the eigenvectors of the canonical Gaussian orthogonal, unitary and symplectic ensembles are distributed with Haar measure on their respective symmetry groups.
We have seen the precise and deep sense in which the eigenvalues of very general random matrices are similar to those of the very special canonical Gaussian orthogonal ensemble of the same symmetry class, but what of the eigenvectors? Is there some precise sense in which the eigenvectors of quite general random matrices are similar to Haar-distributed sets of vectors on their corresponding symmetry group?
The first steps in this direction can be found in \cite{bourgade2017eigenvector} where \emph{quantum unique ergodicity} (QUE) is proved for generalised Wigner matrices.
It is well known that the eigenvectors of quite general random matrices display a universal property of \emph{delocalisation}, namely
\begin{align}
    |u_k|^2 \sim \frac{1}{N}
\end{align}
for any component $u_k$ of an eigenvector $\vec{u}$.
Universal delocalisation was conjectured by Wigner along with the Wigner surmise for adjacent eigenvalue spacing.
QUE states that the eigenvectors of a random matrix are approximately Gaussian in the following sense (\cite{bourgade2017eigenvector} Theorem 1.2):
\begin{align*}
    \sup_{||\vec{q}|| = 1}\sup_{\substack{I\subset [N],\\ |I| = n}} \left|\E P\left(\left(N|\vec{q}^T\vec{u}_k|^2\right)_{k\in I}\right) - \E P\left(\left(|\mathcal{N}_j|^2\right)_{j=1}^{\review{n}}
    \right)\right| \leq N^{-\epsilon},
\end{align*}
for large enough $N$, where $\mathcal{N}_j$ are i.i.d. standard normal random variables, $(\vec{u}_k)_{k=1}^N$ are the normalised eigenvectors, $P$ is any polynomial in $n$ variables and $\epsilon > 0$. \review{Note that the set $I$ in this statement is a subset of $[N]\equiv\{1,2,\ldots, N\}$ of \emph{fixed size} $n$; $n$ is not permitted to depend on $N$.}
Recall from earlier in this chapter, around (\ref{eq:intro_haar_gauss}), that fixed size subsets of Haar distributed eigenvectors of large random matrices can be well approximated by vectors of independent Gaussian entries.
Note that the statement of QUE given above is of precisely the same character


%
\let\textcircled=\pgftextcircled

\chapter{Neural networks with general activation functions}
\label{chap:general_activation_functions}
The content of this chapter was published first as a pre-print in April 2020 (\url{https://arxiv.org/abs/2004.03959}) and later as a journal article: ``The loss surfaces of neural networks with general activation functions''. \textbf{Nicholas P
Baskerville}, Jonathan P Keating, Francesco Mezzadri and Joseph Najnudel. \emph{Journal
of Statistical Mechanics: Theory and Experiment}, 2021(6):064001, 2021.
\medskip

\textbf{NPB} suggested general activation functions as a focus, performed all of the calculations
and experiments and wrote the paper. The other authors contributed ideas for possible
approaches, provided feedback on results throughout and made small revisions to the drafts.
Anonymous reviewers spotted some minor errors, advised on changes of presentation and
provided useful references.

\section{Introduction}
\subsection{Multi-layer perceptron neural networks}
Let $f:\mathbb{R}\rightarrow\mathbb{R}$ be a suitably well-behaved (e.g. differentiable almost everywhere and with bounded gradient) non-linear \emph{activation function} which is taken to applied entry-wise to vectors and matrices. 
We study multi-layer perceptron neural networks of the form \begin{equation}
    \vec{y}(\vec{x}) = f(W^{(H)}f(W^{(H-1)}f(\ldots f(W^{(1)}\vec{x})\ldots)))\label{eq:nn_def}
\end{equation}
where the input data vectors $\vec{x}$ lie in $\mathbb{R}^d$ and the \emph{weight matrices} $\{W^{(\ell)}\}_{\ell=1}^H$ have any shapes compatible with $\vec{x}\in\mathbb{R}^d$ and $\vec{y}(\vec{x})\in\mathbb{R}^c$. 
\vivacom{As discussed in Chapter \ref{chap:intro}, the matrices $W^{(\ell)}$ are parameters of the neural network $f$ and in practice they will be randomly initialised with some standard distribution and then ``learned'' using some gradient descent algorithm on a data set. Their shapes are essentially arbitrary up-to compatibility constraints and the choice of \emph{hidden layer} widths (i.e. the number of rows in each $W^{\ell)}$) is an engineering decision unique to each concrete application. }
Note that, as in \cite{choromanska2015loss}, we do not consider biases in the network.

\subsection{Outline of results and methods}\label{subsec:outline_results}
Following \cite{choromanska2015loss}, we view $\vec{y}$ as a random function over a high-dimensional weight-space and explore its critical points, i.e. vanishing points of its gradient. The randomness will come from taking the input data to be random. We define the following key quantities\footnote{Recall that the \emph{index} of a critical points is the number of negative eigenvalues of the Hessian at that point.}:

\begin{align}
        C_{k,H}(u) =  &\textrm{expected number of critical points of } \vec{y}  \textrm{ of index } k \textrm{ taking values at most } u, \label{eq:c_kh_rough}\\
        C_{H}(u) =  &\textrm{expected number of critical points of } \vec{y} \textrm{ taking values at most } u.\label{eq:c_h_rough}
\end{align}

In Section \ref{sec:nns_random_funcs} we make precise our heuristic definitions in (\ref{eq:c_kh_rough})-(\ref{eq:c_h_rough}). Following \cite{auffinger2013random} we obtain precise expressions for $C_{k,H}$ and $C_H$ as expectations under the Gaussian Orthogonal Ensemble (GOE) and use them to study the asymptotics in the large-network limit. Our results reveal almost the same `banded structure' of critical points as first found in \cite{choromanska2015loss}. In particular we establish the existence of the same critical values $E_0 > E_1 >  ... >E_{\infty}$ such that, with overwhelming probability, critical points taking (scaled) values in $(-E_k, -E_{k+1})$ have index at-most $k+2$, and that there are exponentially many such critical points. We further obtain the exact leading order terms in the expansion of $C_H(u)$, this being the only point at which the generalised form of the activation function $f$ affects the results. In passing, we also show that the network can be generalised to having any number of output neurons without much affecting the calculations of \cite{choromanska2015loss} who only consider single-output networks.

\medskip
In Section \ref{sec:nns_random_funcs} we extend the derivation of \cite{choromanska2015loss} to general activation functions by leveraging piece-wise linear approximations, and we extend to multiple outputs and new loss functions with a simple extension of the corresponding arguments in \cite{choromanska2015loss}. In Section \ref{sec:goe_expressions} we obtain expressions for the complexities $C_{k,H}, C_H$ using a Kac-Rice formula as in \cite{auffinger2013random, fyodorov2007replica, fyodorov2004complexity} but are forced to deal with a perturbed GOE matrix, preventing the replication of the remaining calculations in that work. Instead, in Section \ref{sec:asymptotic_evaluation} we use the supersymmetric method following closely the work of \cite{nock, fyodorov2015random} and thereby reach the asymptotic results of \cite{auffinger2013random} by entirely different means.

\section{Neural networks as random functions}\label{sec:nns_random_funcs}
In this section we show that, under certain assumptions, optimising the loss function of a neural network is approximately equivalent to minimising the value of a random function on a high dimensional hypersphere, closely related to the spin glass. Our approach is much the same as \cite{choromanska2015loss} but is extended to a general class of activation functions and also to networks with multiple output neurons.

\subsection{Modelling assumptions}\label{subsec:modelling_assumptions}
We make the following assumptions, all of which are required for the specific analytic framework of the results in this chapter and are taken either exactly from, or by close analogy with \cite{choromanska2015loss}. We defer a discussion of their plausibility  and necessity to Section \ref{subsec:discussion_assumptions}.

\begin{enumerate}
    \item Components of data vectors are i.i.d. standard Gaussians.\label{item: assumption_gaussian}
    \item \label{item: assumption_sparse}The neural network can be well approximated as a much sparser\footnote{As in \cite{choromanska2015loss}, a network with $N$ weights is sparse if it has $s$ unique weight values and $s\ll N$.} network that achieves very similar accuracy.
    \item \label{item: assumption_uniform_weights}The unique weights of the sparse network are approximately uniformly distributed over the graph of weight connections.
    \item \label{item: assumption_diff}The activation function is twice-differentiable almost everywhere in $\mathbb{R}$ and can be well approximated as a piece-wise linear function with finitely many linear pieces.
        \item \label{item: assumption_bernoulli}The action of the piece-wise linear approximation to the activation function on the network graph can be modelled as i.i.d. discrete random variables, independent of the data at each node, indicating which linear piece is active.
    \item \label{item: assumption_sphere}The unique weights of a the sparse neural network lie on a hyper-sphere of some radius.
\end{enumerate}
\begin{remark}
An alternative to assumption \ref{item: assumption_bernoulli} would be to take the activation function to be \emph{random} (and so too its piece-wise linear approximation). In this paradigm, we consider the ensuing analysis of this chapter to be a study of the \emph{mean properties} of the induced ensemble of neural networks. Resorting to studying mean properties of complicated stochastic systems is a standard means of simplifying the analysis. We do not develop this remark further, but claim that the following calculations are not much affected by switching to this interpretation.
\end{remark}

\subsection{Linearising loss functions}\label{subsec:linear_loss}
In \cite{choromanska2015loss} the authors consider networks with a single output neuron with either $L_1$ or hinge loss and show that both losses are, in effect, just linear in the network output and with positive coefficient, so that minimising the loss can be replaced with minimising the network output. Our ensuing analysis can just as well be applied to precisely these situations, but here we present arguments to extend the applicability to multiple output neurons for $L_1$ regression loss and the widely-used cross-entropy loss \cite{DBLP:journals/corr/JanochaC17} for classification.\\

\textbf{$\textrm{L}_1$ loss.}
The $L_1$ loss is given by \begin{equation}
    \mathcal{L}_{L_1}(\vec{y}(\vec{X}), \vec{Y}) \defeq \sum_{i=1}^c| y_i(\vec{X}) - Y_i|\label{eq:l1_def}
\end{equation}
where $\vec{X}$ is a single random data vector and $\vec{Y}$ a single target output.
Following \cite{choromanska2015loss}, we assume that the absolute values in (\ref{eq:l1_def}) can be modelled 
\vivacom{by using Bernoulli random variables, }$M_i$ say, taking values in $\{-1, 1\}$.
\vivacom{Precisely, we replace $|y_i(\vec{X}) - Y_i|$ with $M_i(y_i(\vec{X}) - Y_i)$, so that the Bernoulli variables $M_i$ model which section of the absolute value function $y_i(\vec{X}) - Y_i$ lies in.}
We do not expect $\vec{X}, \vec{Y}$ and the $M_i$ to be independent, however it may be reasonable to assume that $\vec{X}$ and the $M_i$ are conditionally independent conditioned on $\vec{Y}$. We then have  \begin{align}
    \expect_{M | \vec{Y}}\mathcal{L}_{L_1}(\vec{y}(\vec{X}), \vec{Y}) = \expect_{M | \vec{Y}}  \sum_{i=1}^c M_i (y_i(\vec{X}) - Y_i) &=  \sum_{i=1}^c (2\pi_i-1) y_i(\vec{X}) -   \sum_{i=1}^c \expect_{M | \vec{Y}} M_iY_i\notag\\
    &=  \sum_{i=1}^c (2\pi_i-1) y_i(\vec{X}) -   \sum_{i=1}^c (2\pi_i - 1)Y_i
    \label{eq:l1_linear_result}
\end{align}
where the $M_i$ are Bernoulli random variables with $\prob(M_i = 1) = \pi_i$. Observe that the second term in (\ref{eq:l1_linear_result}) is independent of the parameters of the network.

\textbf{Cross-entropy loss.}
The cross-entropy loss is given by \begin{equation}
    \mathcal{L}_{\text{entr}}(\vec{y}(\vec{X}), \vec{Y}) \defeq -\sum_{i=1}^c Y_i \log\left(\text{SM}[\vec{y}(\vec{X})]_i\right)\label{eq:ce_def}
\end{equation}
where $\text{SM}$ is the \emph{soft-max} function: \begin{align}
    \text{SM} : &\mathbb{R}^c \rightarrow \mathbb{R}^c,\notag\\
    &\vec{z} \mapsto \frac{\exp(\vec{z})}{\sum_{i=1}^m \exp(z_i)} \label{eq:softmax_def}
\end{align}
and $\exp(\cdot)$ is understood to be applied entry-wise. Note that we are applying the standard procedure of mapping network outputs onto the simplex $\Delta^{c-1}$ to allow us to calculate a mutual entropy. Restricting to $c$-class classification problems and using one-hot label vectors \cite{one-hot}, we obtain \begin{align}
   \mathcal{L}_{\text{entr}}(\vec{y}(\vec{X}), \vec{Y}) &= -\sum_{i=1}^c Y_i\left\{y_i(\vec{X}) - \log\left(\sum_{j=1}^c \exp(y_j(\vec{X}))\right)\right\}\label{eq:ce_linear_part}
\end{align}
We note that classification networks typically produce very `spiked' soft-max outputs \cite{DBLP:journals/corr/GuoPSW17}, therefore we make the approximation \begin{equation}
    \sum_{i=1}^c \exp(y_i(\vec{X})) \approx \max_{i=1,\ldots, c} \{\exp(y_i(\vec{X}))\}\label{eq:ce_approx}
\end{equation}
and so we obtain from (\ref{eq:ce_linear_part}) and (\ref{eq:ce_approx}) \begin{align}
  \mathcal{L}_{\text{entr}}(\vec{y}(\vec{X}), \vec{Y}) &\approx -\sum_{i=1}^c\left\{ Y_i y_i(\vec{X}) - Y_i\max_{j=1,\ldots,c}\{y_j(\vec{X})\}\right\}\label{eq:ce_linear_approx_pre_categorical}
\end{align}
We now model the max operation in (\ref{eq:ce_linear_approx_pre_categorical}) with a categorical variable, $M''$ say, over the indices $i=1,\ldots, c$ and take expectations (again assuming conditional independence of $\vec{X}$ and $M''$) to obtain \begin{equation}
    \expect_{M'' | \vec{Y} }\mathcal{L}_{\text{entr}}(\vec{y}(\vec{X}), \vec{Y}) = -\sum_{i=1}^c  Y_i \left(y_i(\vec{x}) - \sum_{j=1}^c \pi_j'' y_j(\vec{X})\right)\label{eq:ce_linear_final}
\end{equation}
Now $\vec{Y}$ is a one-hot vector and so (\ref{eq:ce_linear_final}) in fact reduces to \begin{equation}\label{eq:ce_linear_result}
      \expect_{M'' | \vec{Y} }\mathcal{L}_{\text{entr}}(\vec{y}(\vec{X}), \vec{Y}) = \sum_{j=1}^c \pi_j'' y_j(\vec{x}) - y_i(\vec{x})
\end{equation} 
for some $i$.\\

\begin{remark}
The arguments in this section are not intended to be anything more than heuristic, so as to justify our study of $\vec{a}^T\vec{y}$ for some constant $\vec{a}$ instead of the actual loss function of a neural network. The modelling assumptions required are no stronger than those used in \cite{choromanska2015loss}.
\end{remark}

\subsection{Network outputs as spin glass-like objects}
We assume that the activation function, $f$, can be well approximated by a piece-wise linear function with finitely many linear pieces. To be precise, given any $\epsilon > 0$ there exists some positive integer $L$ and real numbers $\{\alpha_i, \beta_i\}_{i=1}^L$ and real $a_1 < a_2 < \ldots < a_{L-1}$ such that \begin{align}
    |f(x) - (\alpha_{i+1} x + \beta_{i+1})| &< \epsilon ~~~ \forall x\in(a_i, a_{i+1}], ~ 1 \leq i \leq L-2,\notag\\
    |f(x) - (\alpha_1 x + \beta_1)| &< \epsilon ~~~ \forall x\in(-\infty, a_1],\label{eq:piecewise_lin_def}\\
    |f(x) - (\alpha_L x + \beta_L)| &< \epsilon ~~~ \forall x\in(a_{L-1}, \infty).\notag
\end{align}
Note that the $\{\alpha_i, \beta_i\}_{i=1}^L$ and $\{a_i\}_{i=1}^{L-1}$ are constrained by $L-1$ equations to enforce continuity, viz. \begin{align}\label{eq:pwise_cont_constraint}
    \alpha_{i+1}a_i + \beta_{i+1} = \alpha_{i}a_i + \beta_i, ~~~~ 1\leq i \leq L-1
\end{align}
\begin{defn}
A continuous piece-wise linear function with $L$ pieces $\hat{f}\left(x; \left\{\alpha_i, \beta_i\right\}_{i=1}^L , \left\{a_i\right\}_{i=1}^{L-1}\right)$ is an $(L,\epsilon)$-\emph{approximation} to to a function $f$ if $\left|f(x) - \hat{f}\left(x; \left\{\alpha_i, \beta_i\right\}_{i=1}^L , \left\{a_i\right\}_{i=1}^{L-1}\right)\right| < \epsilon$ for all $x\in\mathbb{R}$.
\end{defn}

Given the above definition, we can establish the following.

\begin{lemma}\label{lemma:linear_approx}
Let $\hat{f}\left(\cdot ; \left\{\alpha_i, \beta_i\right\}_{i=1}^L , \left\{a_i\right\}_{i=1}^{L-1}\right)$ be a $(L, \epsilon)$-approximation to $f$. Assume that all the $W^{(i)}$ are bounded in Frobenius norm\footnote{Recall assumption \ref{item: assumption_sphere}, which is translated here to imply bounded Frobenius norm.}. Then there exists some constant $K>0$, independent of all $W^{(i)}$, such that\begin{equation}
    \left\Vert f(W^{(H)}f(W^{(H-1)}f(\ldots f(W^{(1)}\vec{x})\ldots))) -  \hat{f}(W^{(H)}\hat{f}(W^{(H-1)}\hat{f}(\ldots \hat{f}(W^{(1)}\vec{x})\ldots)))\right\Vert_2 < K\epsilon\label{eq:pwise_lemma}
\end{equation}
for all $\vec{x}\in\mathbb{R}^d.$
\end{lemma}

\begin{proof}
Suppose that (\ref{eq:pwise_lemma}) holds with $H-1$ in place of $H$.
Because $\hat{f}$ is piece-wise linear and continuous then we clearly have \begin{equation}
   |\hat{f}(x) - \hat{f}(y)| \leq \max_{i=1,\ldots, L}\{|\alpha_i|\} |x-y|\equiv K'|x-y|\label{eq:fhat_lipschitz}
\end{equation} 
which can be seen by writing \begin{equation}
    \hat{f}(x) - \hat{f}(y)  = (\hat{f}(x) - \hat{f}(a_i)) + (\hat{f}(a_i) - \hat{f}(a_{i-1})) + \ldots + (\hat{f}(a_{j+1}) - \hat{f}(a_j)) + (\hat{f}(a_j) - \hat{f}(y))
\end{equation}
for all intermediate points $a_j, \ldots, a_i \in (y, x)$. Using (\ref{eq:fhat_lipschitz}) and our induction assumption we obtain \begin{align}
    &\left\Vert \hat{f}(W^{(H)}f(W^{(H-1)}f(W^{(H-2)}f(\ldots f(W^{(1)}\vec{x})\ldots))) - \hat{f}(W^{(H)}\hat{f}(W^{(H-1)}\hat{f}(W^{(H-2)}\hat{f}(\ldots \hat{f}(W^{(1)}\vec{x})\ldots)))\right\Vert_2\notag \\
    \leq &cK'\left\Vert W^{(H)}\left[f(W^{(H-1)}f(W^{(H-2)}f(\ldots f(W^{(1)}\vec{x})\ldots))) - \hat{f}(W^{(H-1)}\hat{f}(W^{(H-2)}\hat{f}(\ldots \hat{f}(W^{(1)}\vec{x})\ldots)))\right]\right\Vert_2\notag\\
    \leq &cKK'\left\Vert W^{(H)}\right\Vert_F \epsilon \notag \\
    \leq &K''\epsilon,\notag
\end{align}
for some $K''$, where on the last line we have used the assumption that the network weights are bounded to bound $\Vert W^{(H)}\Vert_F$. The result for $H=1$ follows immediately from (\ref{eq:fhat_lipschitz}).
\end{proof}

\begin{remark}
One could be more explicit in the construction of the piece-wise linear approximation $\hat{f}$ from $f$ given the error tolerance $\epsilon$ by following e.g. \cite{berjon2015optimal}. We do not develop this further here as we do not believe it to be important to the practical implications of our results.
\end{remark}

In much the same vein as \cite{choromanska2015loss} (c.f. Lemma 8.1 therein), we now use the following general result for classifiers to further justify our study of approximations to a neural network in the rest of the chapter.

\begin{theorem}\label{thm:correlation}
Let $Z_1$ and $Z_2$ be the outputs of two arbitrary $c$-class classifiers on a dataset $\mathcal{X}$. That is, $Z_1(x),Z_2(x)$ take values in $\{1,2,\ldots, c\}$ for $x\in \mathcal{X}$. If $Z_1$ and $Z_2$ differ on no more than $\epsilon|\mathcal{X}|$ points in $\mathcal{X}$, then \begin{equation}
    \text{corr}(Z_1, Z_2) = 1 - \mathcal{O}(\epsilon)
\end{equation}
where, recall, the correlation of two random variables is given by \begin{equation}
   \frac{ \mathbb{E}(Z_1Z_2) - \mathbb{E}Z_1\mathbb{E}Z_2}{std(Z_1)std(Z_2)}.
\end{equation}
\end{theorem}
\begin{proof}
Let $\datax_i\subset\datax$ be the set of data points for which $Z_1=i$ for $i=1,2, \ldots, c$. Let $\datax_{i,j}\subset\datax_i$ be those points for which $Z_1 = i$ but $Z_2 = j$ where $j\neq i$. Define the following: \begin{equation}
    p_i = \frac{|\datax_i|}{|\datax|}, ~~~ \epsilon_i^+ = \sum_{j\neq i}\frac{|\datax_{i,j}|}{|\datax|}, ~~~ \epsilon_i^- = \sum_{j\neq i}\frac{|\datax_{j,i}|}{|\datax|}.
\end{equation}
We then have \begin{align}
    \expect Z_1 &= \sum_{i=1}^c i p_i,\label{eq:corr_Ez1}\\
    \expect Z_2 &= \sum_{i=1}^c i (p_i - \epsilon^+_i + \epsilon^-_i)\label{eq:corr_Ez2}\\
    \expect Z_1Z_2 &= \sum_{i=1}^c i^2 (p_i - \epsilon_i^+) + \sum_{1\leq i < j \leq c} ij\frac{|\datax_{i,j}| + |\datax_{j,i}|}{|\datax|}\label{eq:corr_Ez1z2} \\
    std(Z_1) &= \left[\sum_{i=1}^c i^2p_i - \sum_{i,j}ijp_ip_j\right]^{1/2}\label{eq:corr_stdZ1}\\
     std(Z_2) &= \left[\sum_{i=1}^c i^2(p_i -\epsilon_i^+ + \epsilon_i^-) - \sum_{i,j}ij(p_i - \epsilon_i^+ + \epsilon_i^-)(p_j - \epsilon_j^+ + \epsilon_j^-)\right]^{1/2}.\label{eq:corr_stdZ2}
\end{align}

Now, by assumption $\sum_i \epsilon_i^{\pm} \leq \mathcal{O}(\epsilon)$ and so $\epsilon_i^{\pm} \leq \mathcal{O}(\epsilon)$ for all $i$. Similarly, $|\datax_{i,j}|/|\datax| \leq \mathcal{O}(\epsilon)$ and so we quickly obtain from (\ref{eq:corr_Ez1})-(\ref{eq:corr_Ez1z2}) \begin{equation}
    cov(Z_1, Z_2) = \sum_{i=1}^c i^2p_i - \sum_{i,j}ijp_ip_j + \mathcal{O}(\epsilon).\label{eq:corr_cov}
\end{equation}
Finally, combining (\ref{eq:corr_stdZ1}) - (\ref{eq:corr_cov}) we obtain \begin{equation}
    corr(Z_1, Z_2) = \frac{1 + \mathcal{O}(\epsilon)}{(1  + \mathcal{O}(\epsilon))^{1/2}} = 1 + \mathcal{O}(\epsilon).
\end{equation}
\end{proof}

The final intermediate result we require gives an explicit expression for the output of a neural network with a piece-wise linear activation function.

\begin{lemma}\label{lemma:pwise_network}
Consider the following neural network \begin{equation}
    \hat{\vec{y}}(\vec{x}) = \hat{f}(W^{(H)}\hat{f}(\ldots \hat{f}(W^{(1)}\vec{x})\ldots))
\end{equation}
where $\hat{f}\left(\cdot; \left\{\alpha_i, \beta_i\right\}_{i=1}^L , \left\{a_i\right\}_{i=1}^{L-1}\right)$ is a piece-wise linear function with $L$ pieces. Then there exist $A_{i,j}$ taking values in \begin{equation}
   \mathcal{A} \defeq \left\{\prod_{i=1}^H \alpha_{j_i}\ ~:~ j_1,\ldots, j_H \in \{1,\ldots, L\}\right\}
\end{equation} and $ A^{(\ell)}_{i,j}$ taking values in \begin{equation}
    \mathcal{A}^{(\ell)} \defeq\left\{\beta_k\prod_{r=1}^{H-\ell} \alpha_{j_r} ~:~ j_1,\ldots, j_{H-\ell}, k \in \{1,\ldots, L\}\right\}\end{equation}
    such that \begin{equation}
        \hat{y_i}(\vec{x}) = \sum_{j=1}^d \sum_{k\in\Gamma_i}x_{j,k} A_{j,k} \prod_{l=1}^H w_{j,k}^{(l)} + \sum_{\ell=1}^H\sum_{j=1}^{n_{\ell}} \sum_{k\in\Gamma_i^{(\ell)}} A_{j,k}^{(\ell)} \prod_{r=\ell +1}^H w_{j,k}^{(r)}
    \end{equation}
    where $\Gamma_i$ is an indexing of all paths through the network to the $i$-th output neuron, $\Gamma_i^{(\ell)}$ is an indexing of all the paths through the network from the $\ell$-th layer to the $i$-th output neuron, $w_{j,k}^{(l)}$ is the weight applied to the $j$-th input on the $k$-th path in the $l$-th layer, $x_{j,k} = x_j$, and $n_{\ell}$ is the number of neurons in layer $\ell$.
\end{lemma}
\begin{proof}
Firstly, for some $j=1,\ldots, L$ \begin{equation}
    \hat{f}(W^{(1)}\vec{x})_i = \alpha_j (W^{(1)}\vec{x})_i + \beta_j
\end{equation}
and so there exist $j_1, j_2,\ldots \in\{1,\ldots, L\}$ such that  \begin{equation}
    [W^{(2)} \hat{f}(W^{(1)}\vec{x})]_i = \sum_k W^{(2)}_{ik}( \alpha_{j_k}(W^{(1)}\vec{x})_k  + \beta_{j_k}) = \sum_k \alpha_{j_k}W^{(2)}_{ik} \sum_l W^{(1)}_{kl}x_l + \sum_k  W^{(2)}_{ik}\beta_{j_k}.\label{eq:pwise_network}
\end{equation}
Continuing in the vein of (\ref{eq:pwise_network}), there exist $k_1,k_2, \ldots\in\{1, \ldots, L\}$ such that \begin{equation}
\hat{f}(W^{(2)} \hat{f}(W^{(1)}\vec{x}))_i = \alpha_{k_i}\sum_r \alpha_{j_r}W_{ir}^{(2)}\sum_l W_{kl}^{(1)}x_l + \alpha_{k_i}\sum_{r} W^{(2)}_{ir} \beta_{j_r} + \beta_{k_i}
\end{equation}
from which we can see that the result follows by re-indexing and induction.
\end{proof}

We now return to the neural network $\vec{y}(\cdot)$. Fix some small $\epsilon>0$, let $\hat{f}\left(\cdot; \{\alpha_i, \beta_i\}_{i=1}^L, \{x_i\}_{i=}^{L-1}\right)$ be a $(L,\epsilon)$-approximation to $f$ and let $\hat{\vec{y}}$ be the same network as $\vec{y}$ but with $f$ replaced by $\hat{f}$. By Lemma \ref{lemma:linear_approx}, we have\footnote{Here we use the standard notation that, for a function $p$ on $\mathcal{B}$, $p\lesssim \epsilon$ if there exists a constant $K$ such that $p(x) \leq K\epsilon$ for all $x\in\mathcal{B}$.}  \begin{equation}\label{eq:y_yhat_epsilon}
\Vert\vec{y}(\vec{x}) - \hat{\vec{y}}(\vec{x})\Vert_2 \lesssim \epsilon\end{equation} for all $\vec{x}\in\mathbb{R}^d$, and so we can adjust the weights of $\hat{\vec{y}}$ to obtain a network with accuracy within $\mathcal{O}(\epsilon)$ of $\vec{y}$. We then apply Lemma \ref{lemma:pwise_network} to $\hat{\vec{y}}$ and assume\footnote{This assumption is the natural analogue of the assumption used in \cite{choromanska2015loss}.} that the $A_{i,j}$ and $A_{i,j}^{(\ell)}$ can be modelled as i.i.d. discrete random variables with \begin{align}\label{eq:rho_def}
    \expect A_{i,j} = \rho, ~~~ \expect A_{i,j}^{(\ell)} = \rho_{\ell}
\end{align}
and then \begin{equation}\label{eq:spin_glass_pre_sparse}
\expect \hat{y}_i(\vec{X}) = \rho \expect_{\vec{x}} \sum_{j=1}^d \sum_{k\in\Gamma_i}X_{j,k} \prod_{l=1}^H w_{j,k}^{(l)} + \sum_{\ell=1}^H \rho_{\ell}\sum_{j=1}^{n_{\ell}} \sum_{k\in\Gamma_i^{(\ell)}} \prod_{r=\ell +1}^H w_{j,k}^{(r)}.
\end{equation}
Our reasoning is now identical to that in Section 3.3 of \cite{choromanska2015loss}. We use the assumptions of sparsity and uniformity (Section \ref{subsec:modelling_assumptions}, assumptions \ref{item: assumption_sparse}, \ref{item: assumption_uniform_weights}) and some further re-indexing to replace (\ref{eq:spin_glass_pre_sparse}) by \begin{equation}\label{eq:y_tilde_results}
    \expect \tilde{y}_i(\vec{X}) = \rho \expect_{\vec{X}} \sum_{i_1, \ldots, i_H = 1}^{\Lambda} X_{i_1, \ldots, i_H} \prod_{k=1}^H w_{i_k} + \sum_{\ell=1}^H\rho_{\ell} \sum_{i_{\ell + 1}, \ldots, i_H=1}^{\Lambda} \prod_{k=\ell +1}^H w_{i_k}
\end{equation}
where $\Lambda$ is the number of unique weights of the network and, in particular, the sparsity and uniformity assumptions are chosen to give \begin{align}\label{eq:sparse_uniform_epsilon}
    \expect_{\vec{X}}\left\Vert \tilde{\vec{y}}(\vec{X}) -  \hat{\vec{y}}(\vec{X})\right\Vert_2 \lesssim \epsilon.
\end{align}
(\ref{eq:y_yhat_epsilon}) and (\ref{eq:sparse_uniform_epsilon}) now give \begin{equation}\label{eq:y_ytwid_eps}
       \expect_{\vec{X}} \left\Vert   \tilde{\vec{y}}(\vec{X}) -  \vec{y}(\vec{X})\right\Vert_2 \lesssim \epsilon
\end{equation}
and in the case of classifiers, (\ref{eq:y_ytwid_eps}) ensures that the conditions for  Theorem \ref{thm:correlation} are met, so establishing that \begin{equation}\label{eq:spin_glass_corr}
corr(\tilde{\vec{y}}(\vec{X}) ,\vec{y}(\vec{X})) = 1 - \mathcal{O}(\epsilon).
\end{equation}
As in \cite{choromanska2015loss}, we use these heuristics to justify studying $\tilde{\vec{y}}$ hereafter in place of $\vec{y}$.

Recalling the results of Section \ref{subsec:linear_loss}, in particular (\ref{eq:l1_linear_result}) and (\ref{eq:ce_linear_result}) we conclude that to study the loss surface of $\tilde{\vec{y}}$ under some loss function it is sufficient to study quantities of the form $\sum_{i=1}^c \eta_i \tilde{y}_i$ and, in particular, we study the critical points. The $X$ are centred Gaussian random variables and so any finite weighted sum of some $X$ is a centred Gaussian variable with some variance. We can re-scale variances and absorb constants into the $\rho_{\ell}$ and thereby replace $\sum_i \eta_i \tilde{y}_i(\vec{X})$ with $\tilde{y}_i(\vec{X})$.

Note that we assumed an $L_2$ constraint on the network weights (Section \ref{subsec:modelling_assumptions}, point 6)  and that now carries forward as \begin{equation}\label{eq:weight_norm}
    \frac{1}{\Lambda}\sum_{i=1}^{\Lambda} w_i^2 = \mathcal{C}
\end{equation} 
for some constant $\mathcal{C}$. For ease of notation in the rest of the chapter, we define \begin{equation}\label{eq:g_def}
    g(\vec{w}) = \sum_{i_1, \ldots, i_H = 1}^{\Lambda} X_{i_1, \ldots, i_H} \prod_{k=1}^H w_{i_k} + \sum_{\ell=1}^H\rho_{\ell}' \sum_{i_{\ell + 1}, \ldots, i_H=1}^{\Lambda} \prod_{k=\ell +1}^H w_{i_k}
\end{equation}
where $\rho_{\ell}' \defeq  \rho_{\ell}/\rho$. Finally, recall that we assumed the data entries  $X_i$ are i.i.d standard Gaussians. To allow further analytic progress to be made, we follow \cite{choromanska2015loss} and now extend this assumption to $X_{i_1, \ldots, i_H}\overset{\text{i.i.d}}{\sim} \mathcal{N}(0,1)$. The random function $g$ is now our central object of study and, without loss of generality, we take $\mathcal{C}=1$ in (\ref{eq:weight_norm}) so that $g$ is a random function on the ($\Lambda$-1)-sphere of radius $\sqrtsign{\Lambda}$.

Observe that the first term in (\ref{eq:g_def}) is precisely the form of an $H$-spin glass as found in \cite{choromanska2015loss} and the second term is deterministic and contains (rather obliquely) all the dependence on the activation function. Having demonstrated the link between our results and those in \cite{choromanska2015loss}, we now set $\Lambda = N$ for convenience and to make plain the similarities between what follows and \cite{auffinger2013random}. We also drop the primes on $\rho_{\ell}'$.

\subsection{Validity of the modelling assumptions.}\label{subsec:discussion_assumptions}
The authors of \cite{choromanska2015loss} discuss the modelling assumptions in \cite{choromanska2015open}. We add to their comments that the hyper-sphere assumption \ref{item: assumption_sphere} seems easily justifiable as merely $L_2$ weight regularisation.\\

Assumption \ref{item: assumption_bernoulli} from Section \ref{subsec:modelling_assumptions} is perhaps the least palatable, as the section of a piece-wise linear activation function in which a pre-activation value lies is a deterministic function of that pre-activation value and so certainly not i.i.d. across the network and the data items. It is not clear how to directly test the assumption experimentally, but we can certainly perform some experiments to probe its plausibility.

For the sake of clarity, consider initially a  \texttt{ReLU} activation function. Let $\mathscr{N}$ be the set of all nodes (neurons) in a neural network, and let $\mathscr{D}$ be a dataset of inputs for this network. Assumption \ref{item: assumption_bernoulli} says that we can model the action of the activation function at any neuron  $\mathfrak{n}\in\mathscr{N}$ and any data point $\vec{x}\in\mathscr{D}$ as i.i.d. Bernoulli random variables. In particular, this is why the the expectations over the activation function indicators and the data distribution can be taken independently in (\ref{eq:spin_glass_pre_sparse}). If one fixes some neuron $\mathfrak{n}\in\mathscr{N}$, and observes its pre-activations over all data points in $\mathscr{D}$, one will observe some proportion $\rho^{\mathfrak{n}}$ of positive values. Assumption \ref{item: assumption_bernoulli} implies that this proportion should be approximately the same for each $\mathfrak{n}\in\mathscr{N}$, namely $p$, where $p$ is the success probability of the Bernoulli. Taking all of the $\rho^{\mathfrak{n}}$ together, their empirical distribution should have low variance and be centred on $p$. More precisely, for large $|\mathscr{D}|$ each $\rho^{\mathfrak{n}}$ should be close in distribution to i.i.d. Gaussian with mean $p$ and variance of order $|\mathscr{D}|^{-1}$, a fact that can be derived simply from the central limit theorem applied to i.i.d. Bernoulli random variables. Similarly, assumption \ref{item: assumption_bernoulli} implies that one can exchange data points and neurons in the previous discussion and so observe proportions $\bar{\rho}^{\vec{x}}$ for each $\vec{x}\in\mathscr{D}$, which again should have an empirical distribution centred on $p$ and with low variance. The value of $p$ is not prescribed by any of our assumptions and nor is it important, all that matters is that the distributions of $\{\rho^{\mathfrak{n}}\}_{\mathfrak{n}\in\mathscr{N}}$  and $\{\bar{\rho}^{\vec{x}}\}_{\vec{x}\in\mathscr{D}}$ are strongly peaked around some common mean.

We will now generalise the previous discussion to the case of any number of linear pieces of the activation function. Suppose that the activation function is piece-wise linear in $L$ pieces and denote by $I_1, \ldots, I_L$ the disjoint intervals on which the activation function is linear; $\{I_i\}_{i=1}^L$ partition $\mathbb{R}$. Let $\iota(\vec{x}, \mathfrak{n})$ be defined so that the pre-activation to neuron $\mathfrak{n}\in\mathscr{N}$ when evaluating at $\vec{x}\in\mathscr{D}$ lies in $I_{\iota(\vec{x}, \mathfrak{n})}.$ We consider two scenarios, \emph{data averaging} and \emph{neuron averaging}. Under data averaging, we fix a neuron and observe the pre-activations observed over all $\mathscr{D}$, i.e. define for $j=1,\ldots, L$ the counts \begin{align}
    \chi_j^{\mathfrak{n}} = |\{\vec{x}\in\mathscr{D} ~:~ \iota(\vec{x}, \mathfrak{n}) = j\}|
\end{align}
and thence the $L-1$ independent ratios \begin{align}
    \rho_j^{\mathfrak{n}} = \frac{ \chi_j^{\mathfrak{n}} }{\sum_{i=1}^L \chi_1^{\mathfrak{n}}}
\end{align}
for $j=2,\ldots, L$. Similarly, in neuron averaging we define \begin{align}
     \bar{\chi}_j^{\vec{x}} &= |\{\mathfrak{n}\in\mathscr{N} ~:~ \iota(\vec{x}, \mathfrak{n}) = j\}|,\\
       \bar{\rho}_j^{\vec{x}} &= \frac{ \bar{\chi}_j^{\vec{x}} }{ \sum_{i=1}^L\bar{\chi}_1^{\vec{x}} }.
\end{align}

We thus have the sets of observed real quantities \begin{align}
   R_j &= \{ \rho_j^{\mathfrak{n}} ~:~ \mathfrak{n}\in\mathscr{N}\},\\
   \bar{R}_j &= \{ \bar{\rho}_j^{\vec{x}} ~:~ \vec{x}\in\mathscr{D}\}.\\
\end{align}
Under assumption \ref{item: assumption_bernoulli}, the empirical variance of the values in $R_j$ and $\bar{R}_j$ should be small. We run experiments to interrogate this hypothesis under a variety of conditions. In particular: \begin{enumerate}
    \item  Standard Gaussian i.i.d. data vs. `real' data (MNIST digits \cite{lecun-mnisthandwrittendigit-2010}).
    \item Multi-layer perceptron (MLP) vs. convolutional (CNN) architecture.
    \item Trained vs. randomly initialised weights.
    \item Various piece-wise linear activation functions.
\end{enumerate}

In particular:

\begin{enumerate}
    \item We generate 10000 i.i.d. Gaussian data vectors of length 784 (to match the size of MNIST digits).
    \item We fix a MLP architecture of 5 layers and a CNN architecture with 3 convolutional layers and 2 fully-connected. The exact architecture details are given in the Appendix.
    \item We train all networks to test accuracy of at least $97\%$ and use dropout with rate $0.1$ during training.
    \item We test \texttt{ReLU} (2 pieces), \texttt{HardTanh} (3 pieces) and a custom 5 piece function. Full details are given in Appendix \ref{ap:experiments}.
\end{enumerate}

To examine the $R_j$ and $\bar{R}_j$, we produce histograms of $R_2$ for $L=2$ (i.e. \texttt{ReLU}), joint density plots of $(R_2, R_3)$ for $L=3$ (i.e. \texttt{HardTanh}) and pair-plots of $(R_2, R_3, R_4, R_5)$ for $L=5$.  We are presently only interested in the size of the variance shown, but these full distribution plots are included in-case any further interesting observations can be made in the future. Figures \ref{fig:probe_agg_data_random_weights}-\ref{fig:probe_agg_neuron_trained_weights} show the results for \texttt{ReLU} activations and Figures \ref{fig:probe_agg_data_random_weights_tanh}-\ref{fig:probe_agg_neuron_trained_weights_tanh} show the results for \texttt{HardTanh}. The qualitative trends are much the same for all three activation functions, but the plots for the 5-piece function are very large and so are relegated to the supplementary material\footnote{\url{https://github.com/npbaskerville/loss-surfaces-general-activation-functions/blob/master/Loss_surfaces_of_neural_networks_with_general_activation_functions___supplimentary.pdf}}. We make the following observations:
  \begin{figure*}[p]
        \centering
        \begin{subfigure}[b]{0.236\textwidth}
            \centering
            \includegraphics[width=\textwidth]{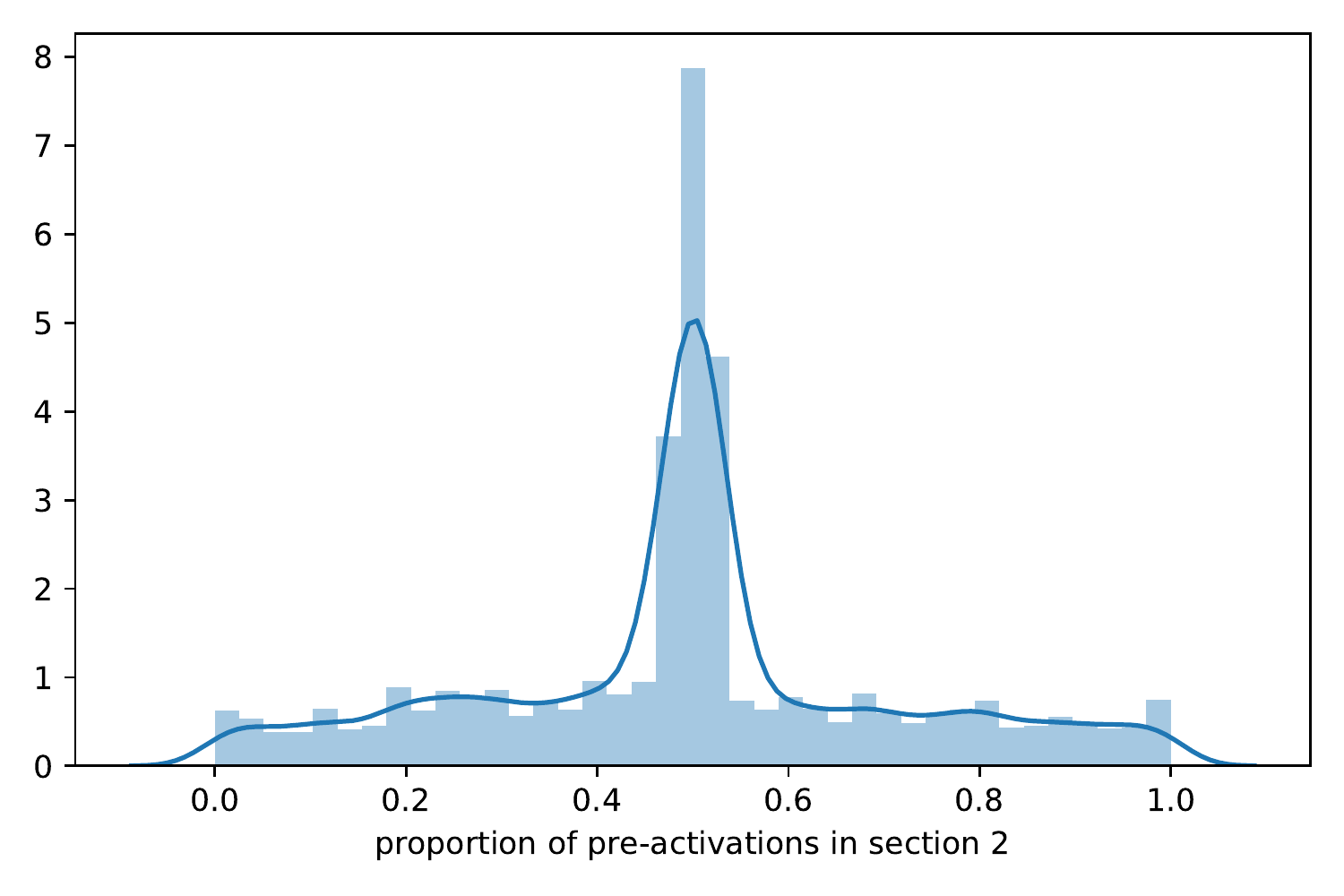}
            \caption{MLP, i.i.d. data.} 
            \label{fig:probe_agg_data_random_weights_mlp_iid}
        \end{subfigure}
        \begin{subfigure}[b]{0.236\textwidth}
            \centering
            \includegraphics[width=\textwidth]{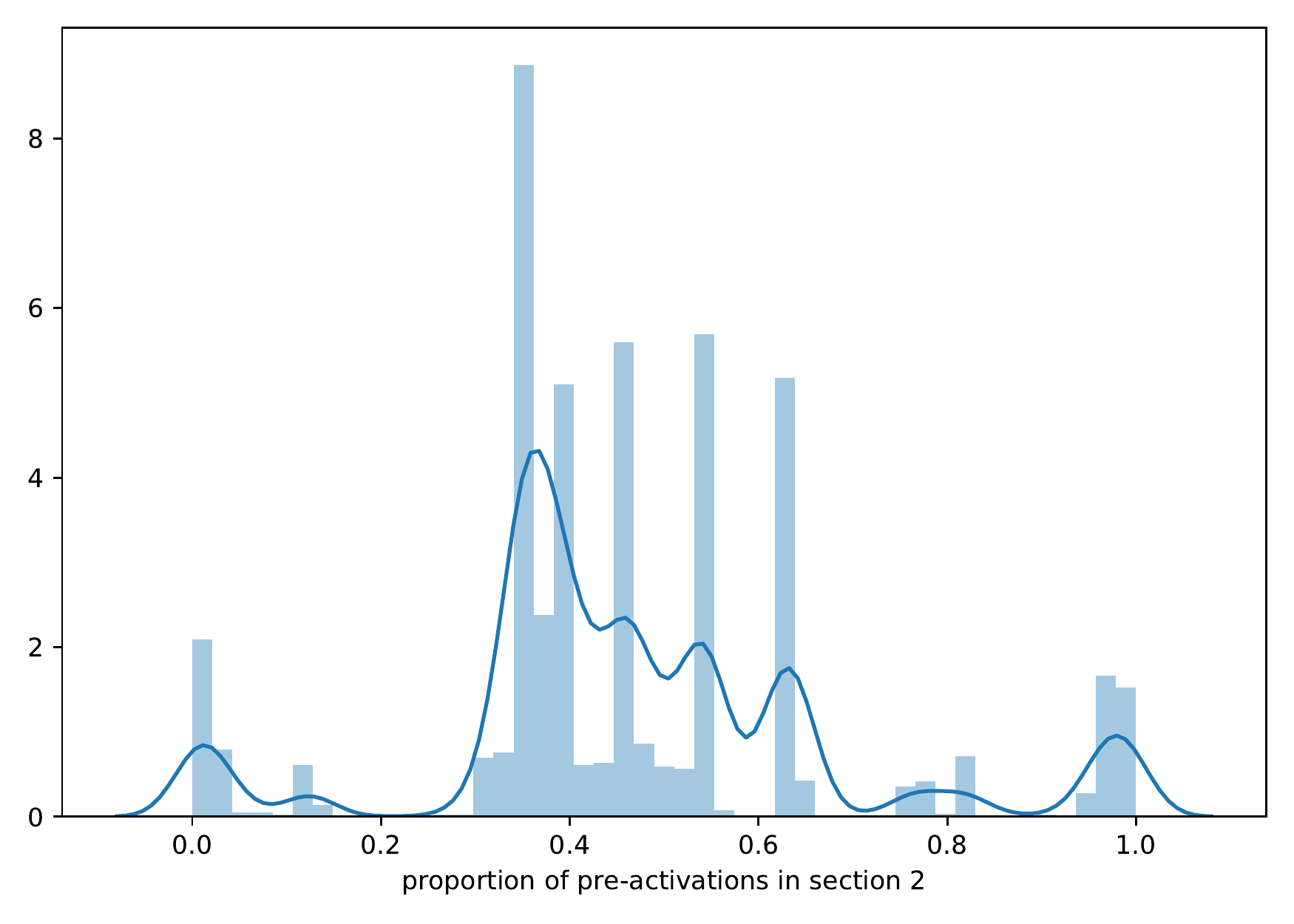}
            \caption{LeNet, i.i.d. data.} 
            \label{fig:probe_agg_data_random_weights_lenet_iid}
        \end{subfigure}
        \begin{subfigure}[b]{0.236\textwidth}
            \centering
            \includegraphics[width=\textwidth]{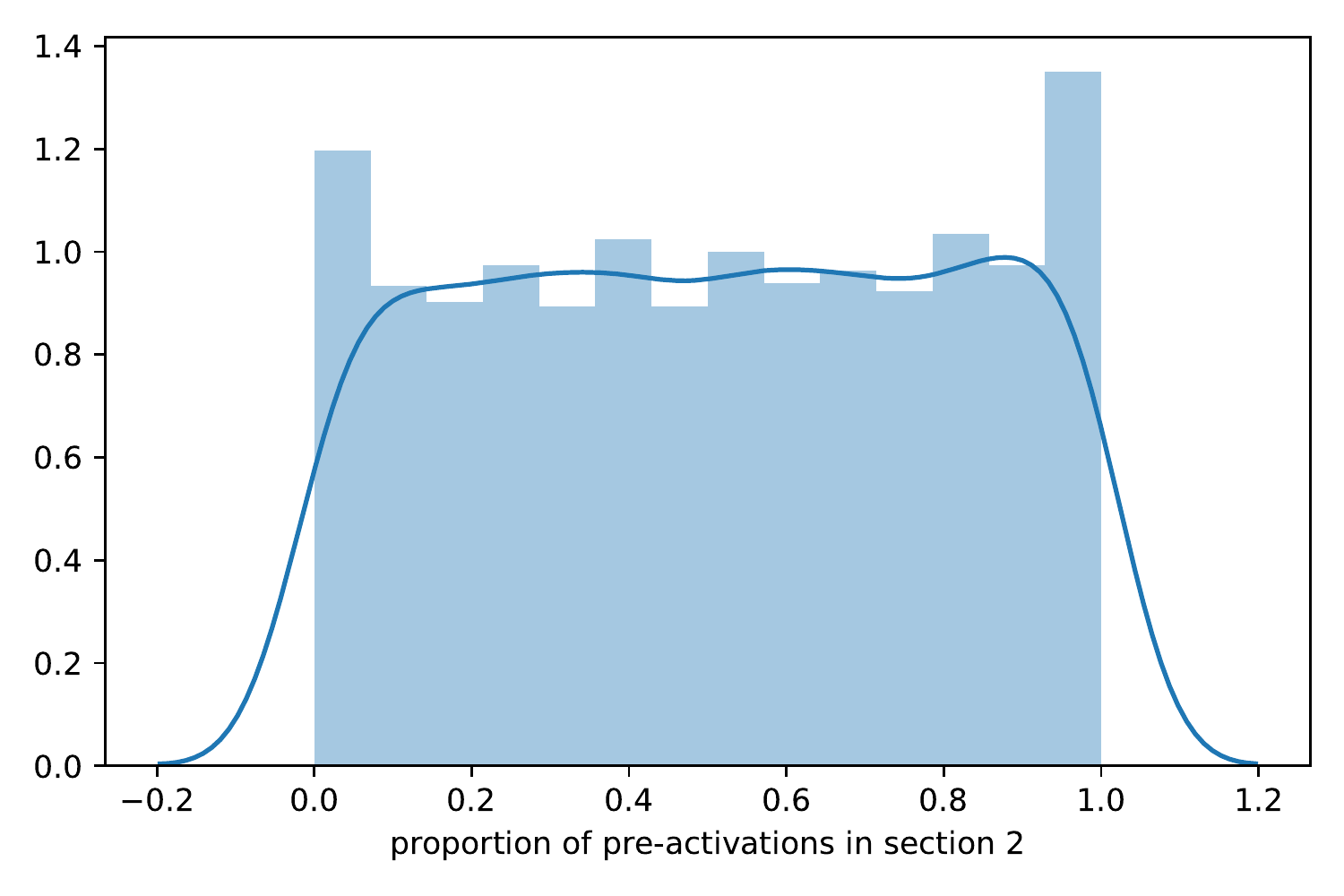}
            \caption{MLP, MNIST data.} 
            \label{fig:probe_agg_data_random_weights_mlp_mnist}
        \end{subfigure}
        \begin{subfigure}[b]{0.236\textwidth}
            \centering
            \includegraphics[width=\textwidth]{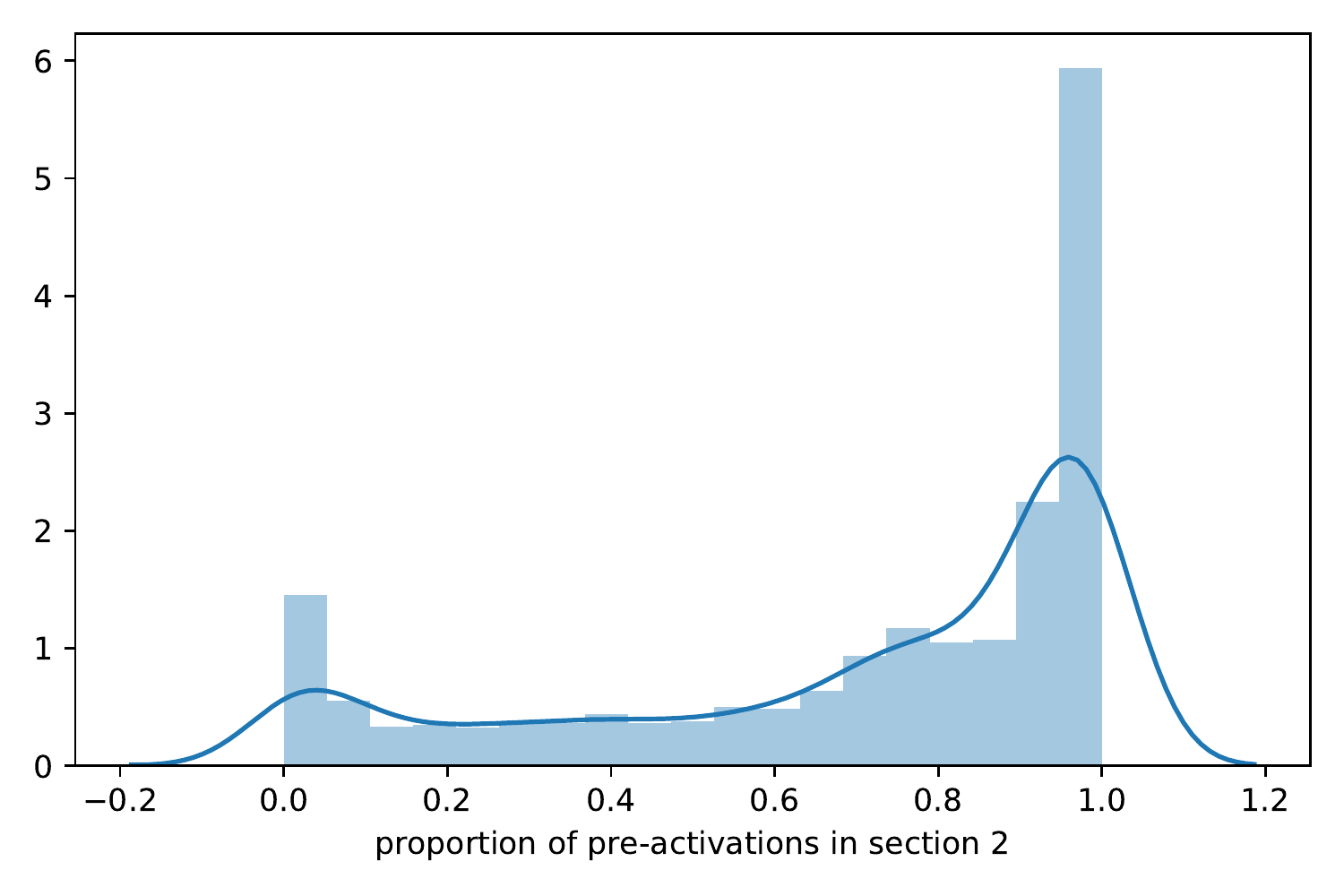}
            \caption{LeNet, MNIST data.} 
            \label{fig:probe_agg_data_random_weights_lenet_mnist}
        \end{subfigure}
        \caption{Experimental distribution of $R_2$ (data averaging; each sample is a single neuron) for random MLP and LeNet \texttt{ReLU} networks, and i.i.d. normal and MNIST data. The blue line is a kernel density estimation fit.} 
        \label{fig:probe_agg_data_random_weights}
    \end{figure*}
  \begin{figure*}[p]
        \centering
        \begin{subfigure}[b]{0.236\textwidth}
            \centering
            \includegraphics[width=\textwidth]{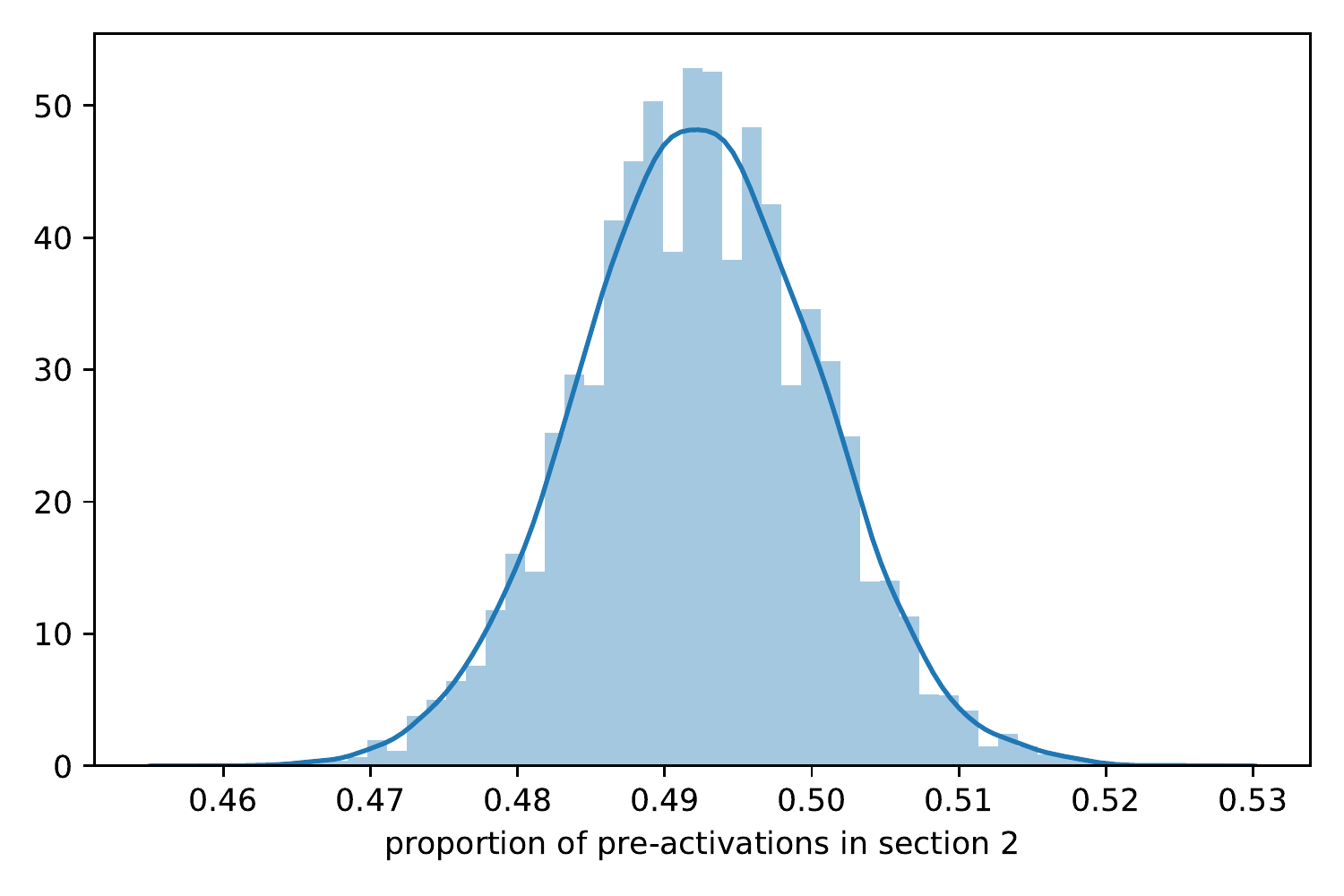}
            \caption{MLP, i.i.d. data.} 
            \label{fig:probe_agg_neuron_random_weights_mlp_iid}
        \end{subfigure}
        \begin{subfigure}[b]{0.236\textwidth}
            \centering
            \includegraphics[width=\textwidth]{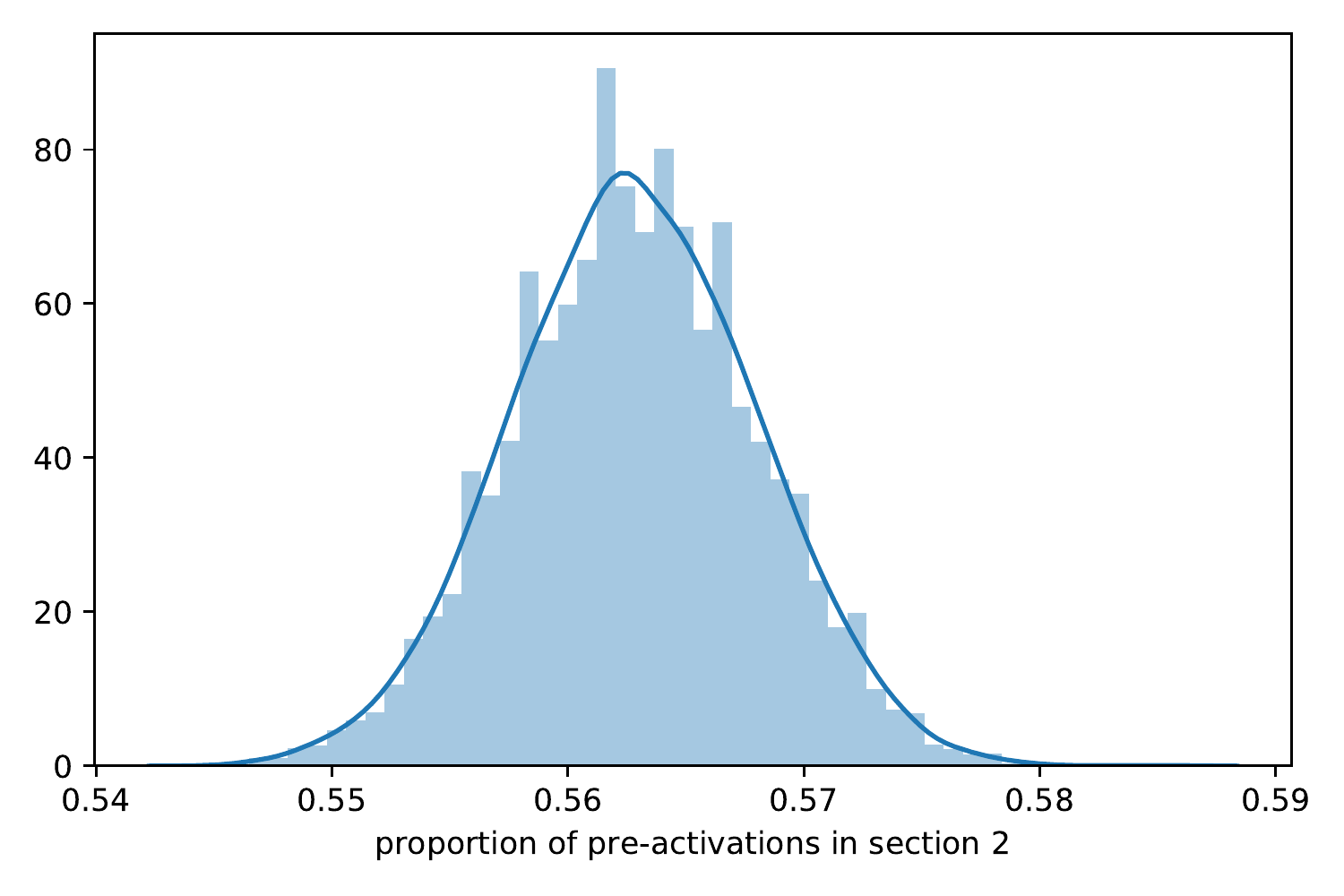}
            \caption{LeNet, i.i.d. data.} 
            \label{fig:probe_agg_neuron_random_weights_lenet_iid}
        \end{subfigure}
        \begin{subfigure}[b]{0.236\textwidth}
            \centering
            \includegraphics[width=\textwidth]{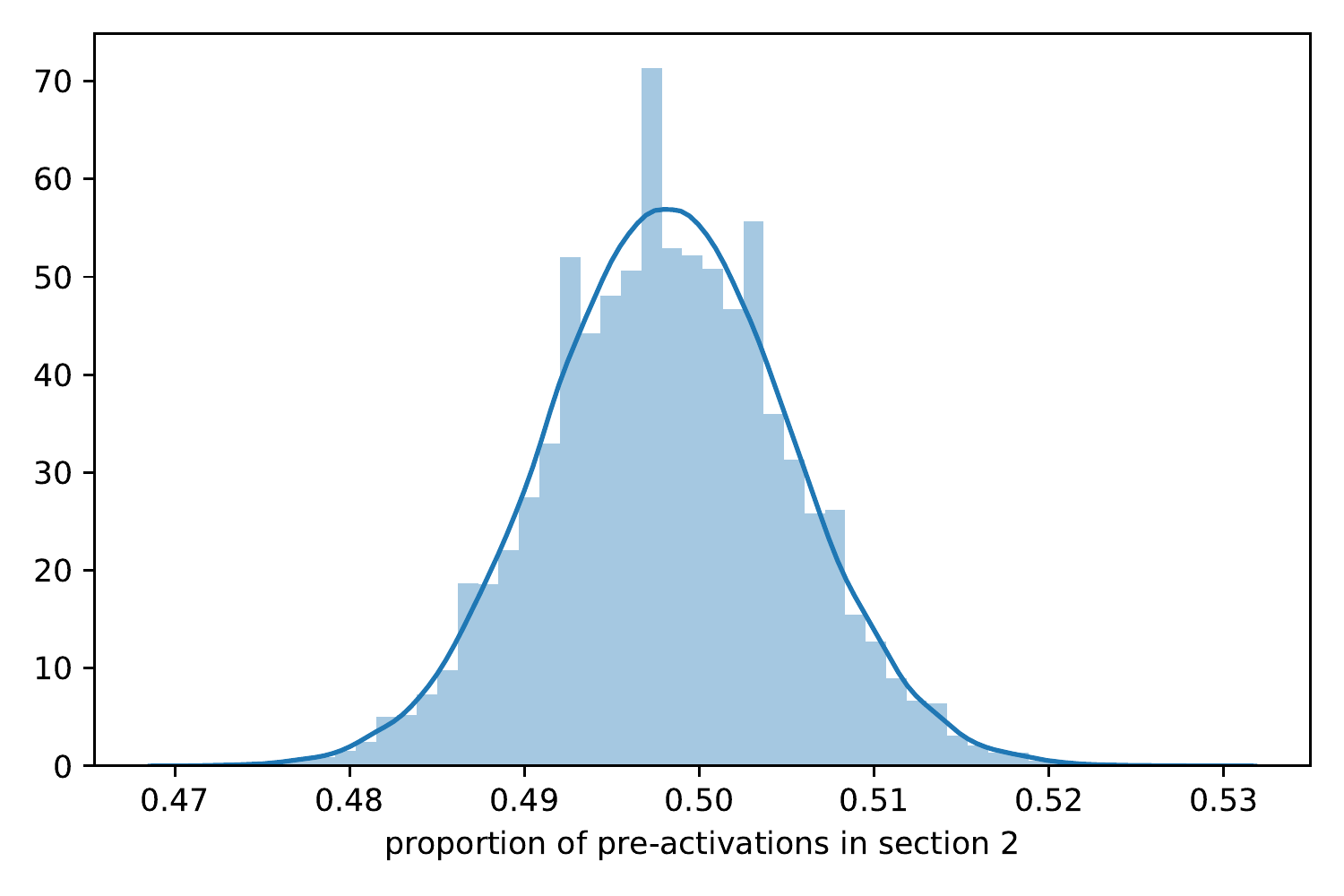}
            \caption{MLP, MNIST data.} 
            \label{fig:probe_agg_neuron_random_weights_mlp_mnist}
        \end{subfigure}
        \begin{subfigure}[b]{0.236\textwidth}
            \centering
            \includegraphics[width=\textwidth]{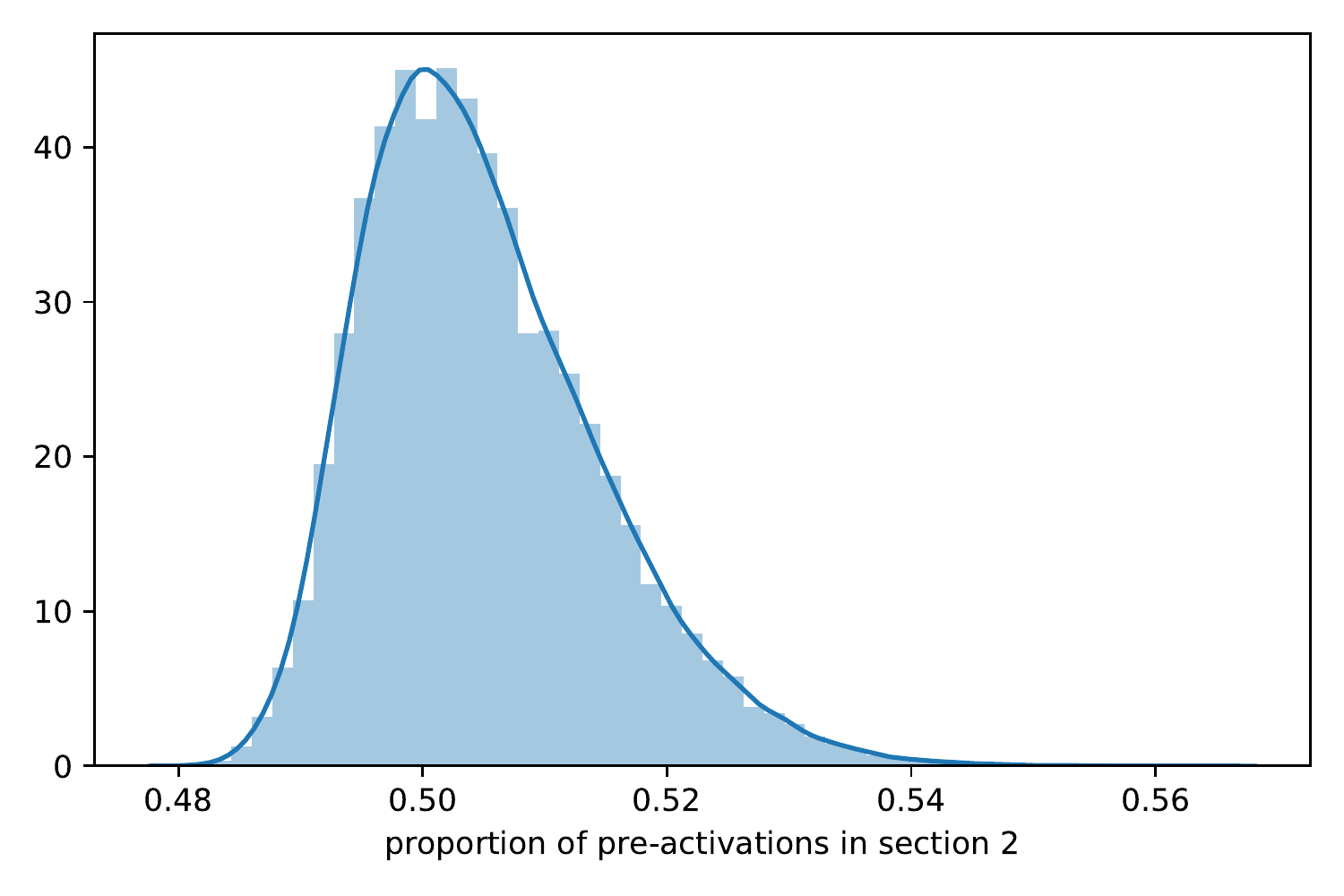}
            \caption{LeNet, MNIST data.} 
            \label{fig:probe_agg_neuron_random_weights_lenet_mnist}
        \end{subfigure}
        \caption{Experimental distribution of $\bar{R}_2$ (neuron averaging; each sample is a single datum) for random MLP and LeNet \texttt{ReLU} networks, and i.i.d. normal and MNIST data. The blue line is a kernel density estimation fit.} 
        \label{fig:probe_agg_neuron_random_weights}
    \end{figure*}
  \begin{figure*}[p]
        \centering
        \begin{subfigure}[b]{0.236\textwidth}
            \centering
            \includegraphics[width=\textwidth]{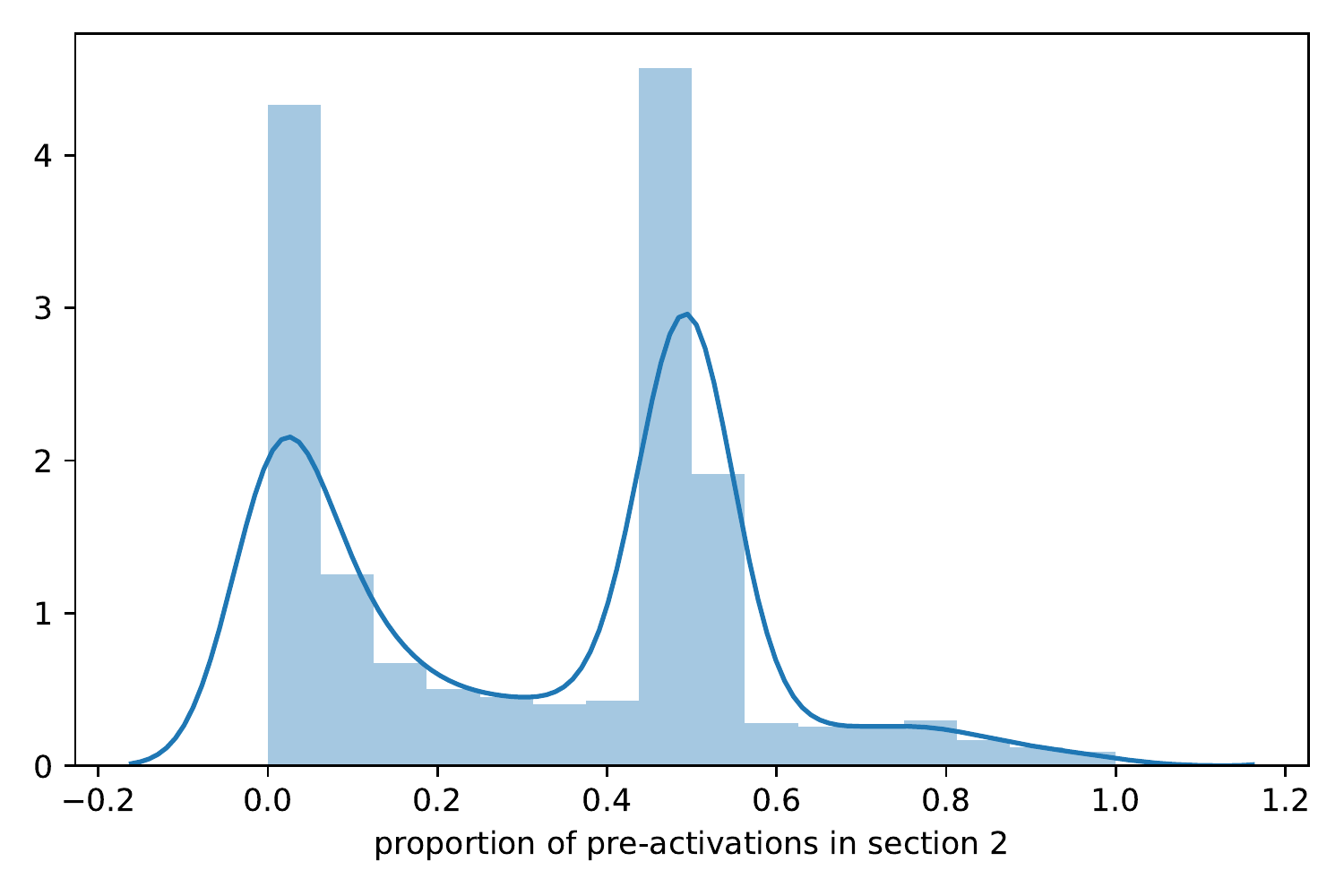}
            \caption{MLP, i.i.d. data.} 
            \label{fig:probe_agg_data_trained_weights_mlp_iid}
        \end{subfigure}
        \begin{subfigure}[b]{0.236\textwidth}
            \centering
            \includegraphics[width=\textwidth]{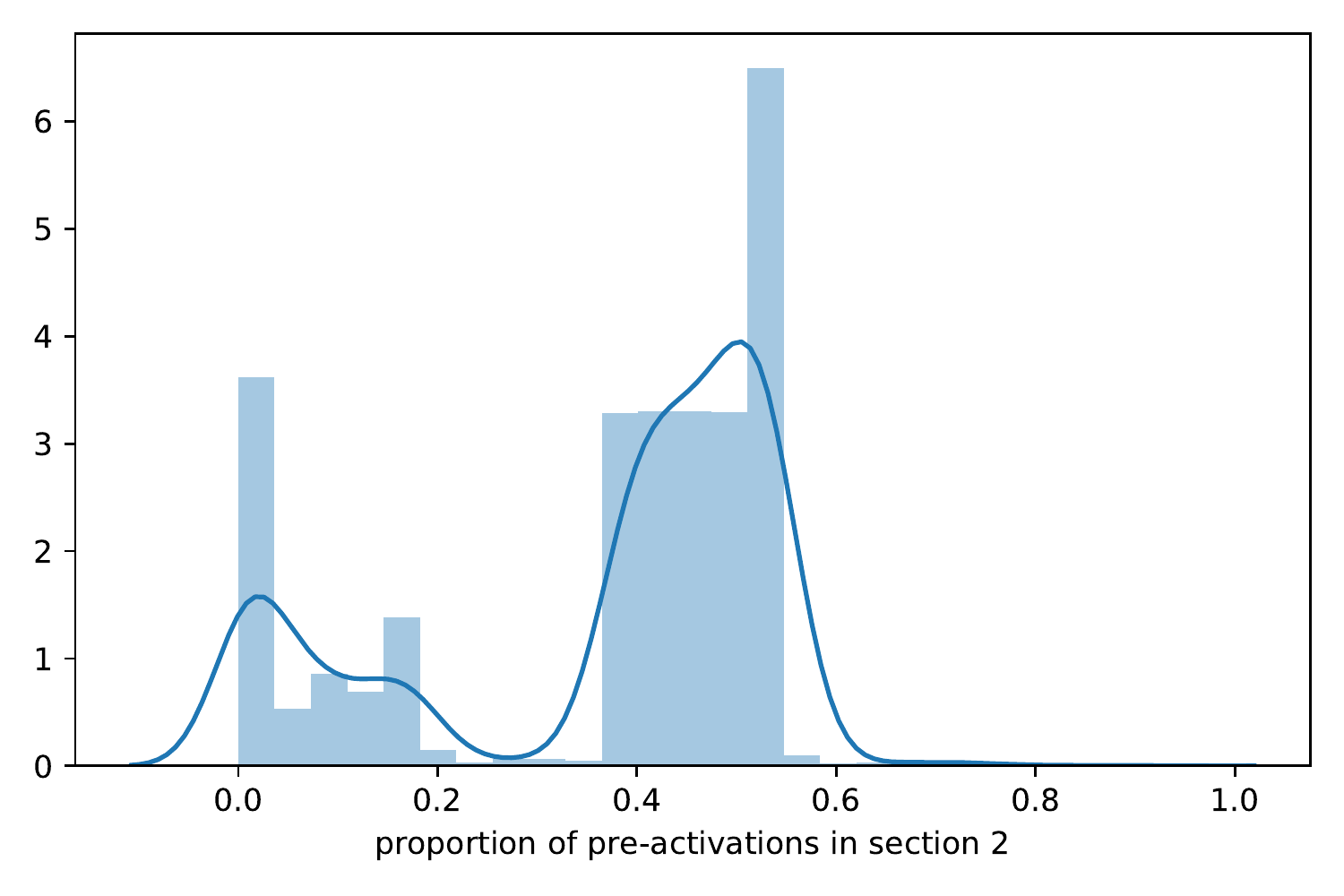}
            \caption{LeNet, i.i.d. data.} 
            \label{fig:probe_agg_data_trained_weights_lenet_iid}
        \end{subfigure}
        \begin{subfigure}[b]{0.236\textwidth}
            \centering
            \includegraphics[width=\textwidth]{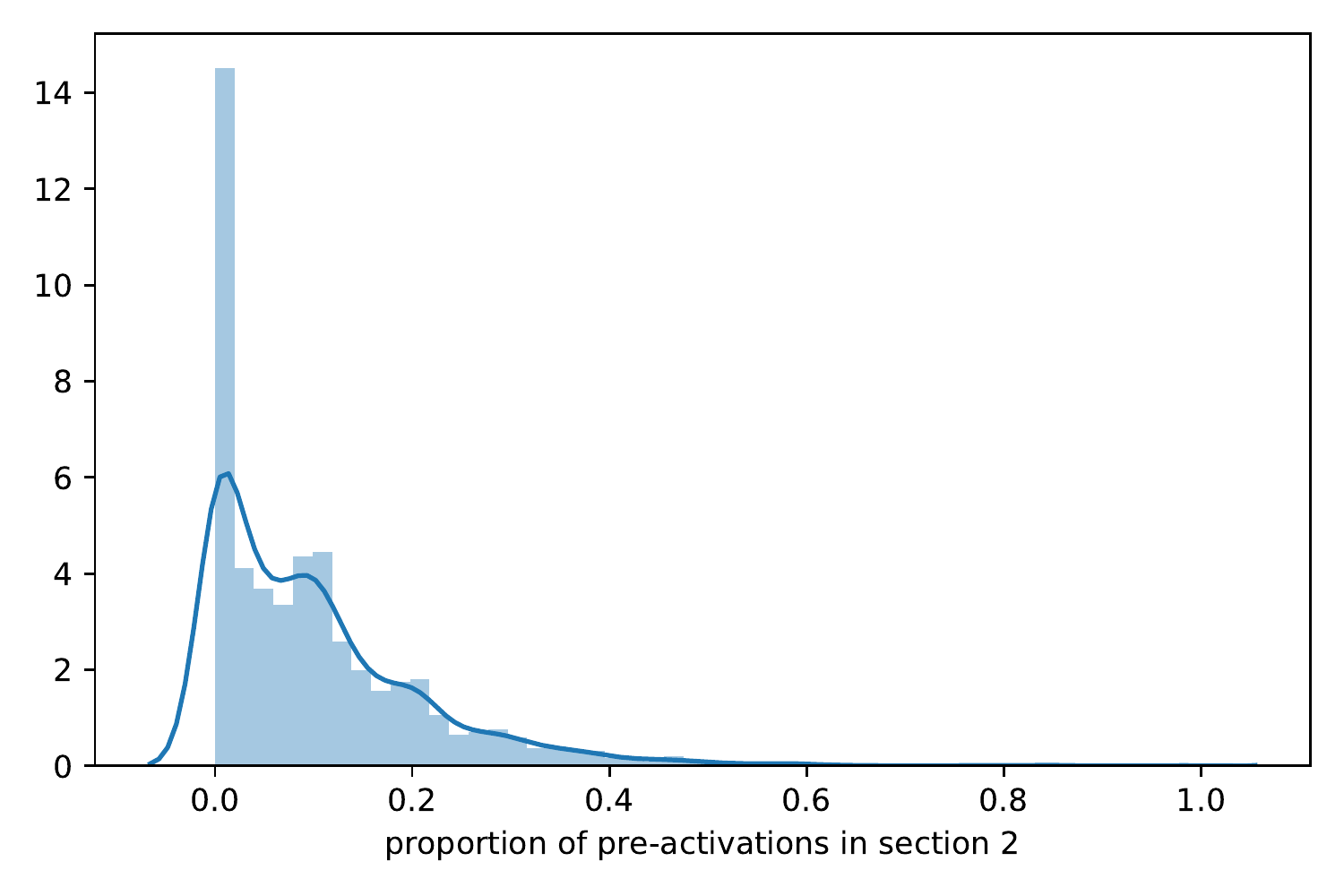}
            \caption{MLP, MNIST data.} 
            \label{fig:probe_agg_data_trained_weights_mlp_mnist}
        \end{subfigure}
        \begin{subfigure}[b]{0.236\textwidth}
            \centering
            \includegraphics[width=\textwidth]{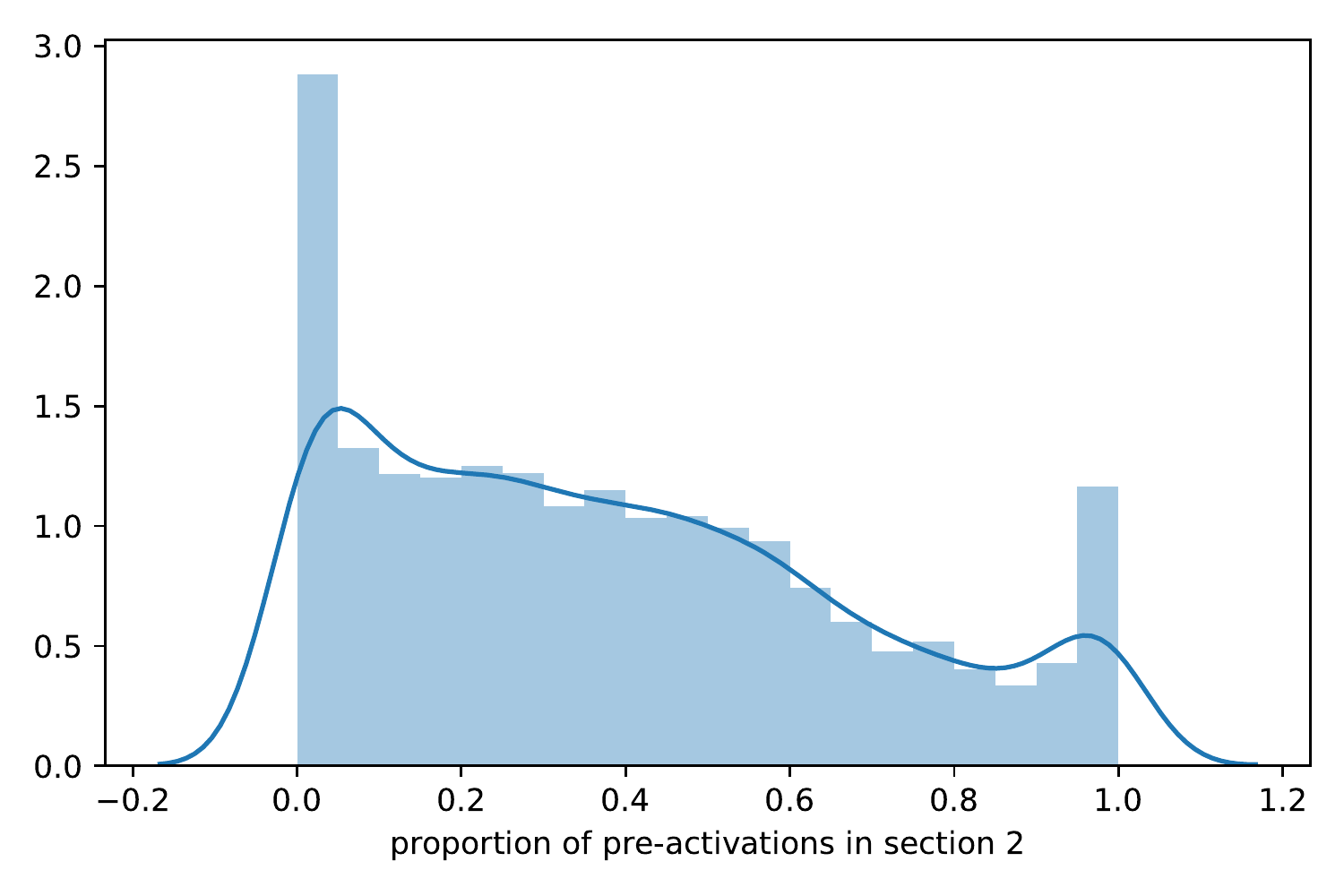}
            \caption{LeNet, MNIST data.} 
            \label{fig:probe_agg_data_trained_weights_lenet_mnist}
        \end{subfigure}
        \caption{Experimental distribution of $R_2$ (data averaging; each sample is a single neuron) for MLP and LeNet \texttt{ReLU} networks trained to high validation accuracy on MNIST, and evaluated on i.i.d. normal and MNIST data. The blue line is a kernel density estimation fit.} 
        \label{fig:probe_agg_data_trained_weights}
    \end{figure*}
  \begin{figure*}[p]
        \centering
        \begin{subfigure}[b]{0.236\textwidth}
            \centering
            \includegraphics[width=\textwidth]{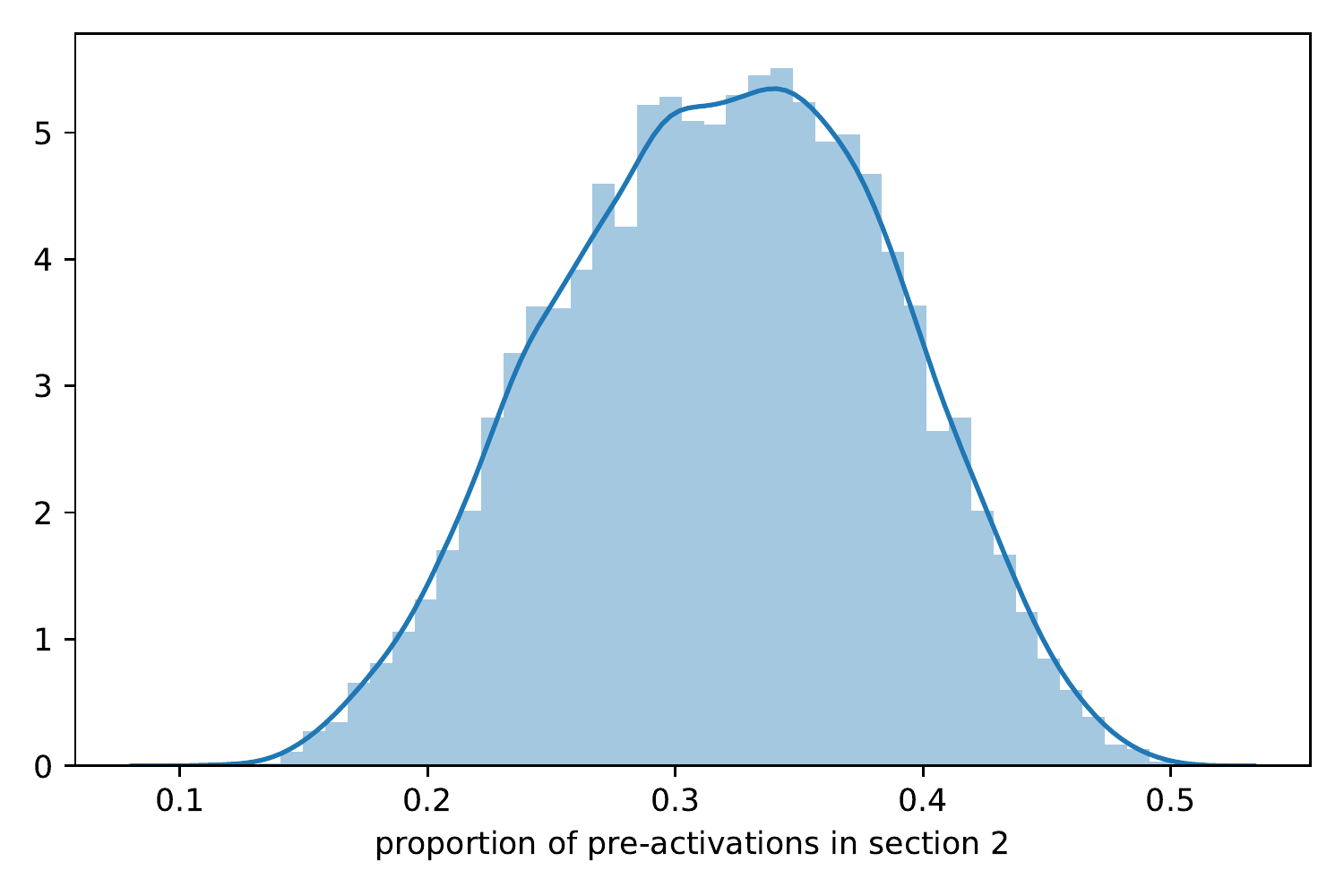}
            \caption{MLP, i.i.d. data.} 
            \label{fig:probe_agg_neuron_trained_weights_mlp_iid}
        \end{subfigure}
        \begin{subfigure}[b]{0.236\textwidth}
            \centering
            \includegraphics[width=\textwidth]{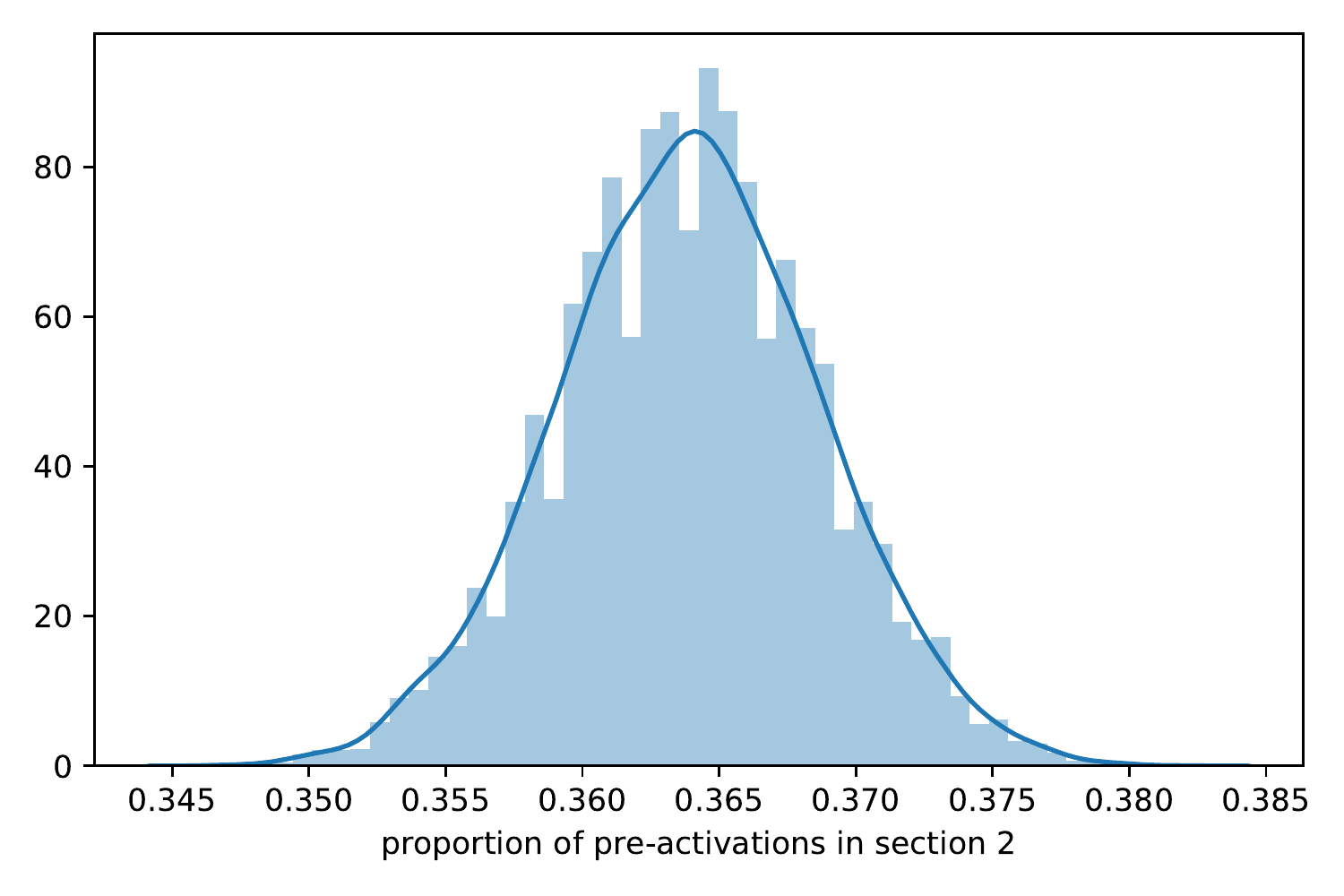}
            \caption{LeNet, i.i.d. data.} 
            \label{fig:probe_agg_neuron_trained_weights_lenet_iid}
        \end{subfigure}
        \begin{subfigure}[b]{0.236\textwidth}
            \centering
            \includegraphics[width=\textwidth]{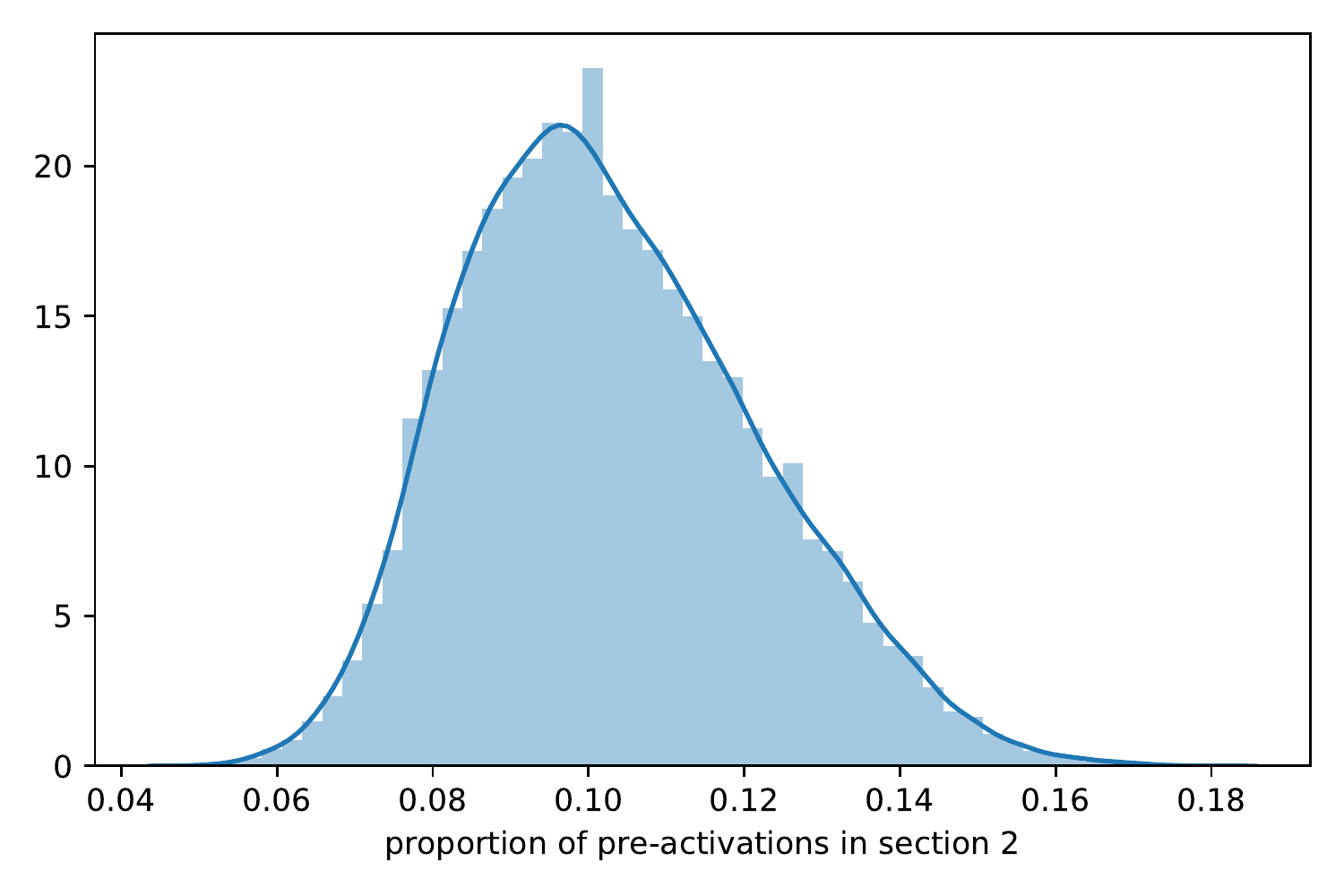}
            \caption{MLP, MNIST data.} 
            \label{fig:probe_agg_neuron_trained_weights_mlp_mnist}
        \end{subfigure}
        \begin{subfigure}[b]{0.236\textwidth}
            \centering
            \includegraphics[width=\textwidth]{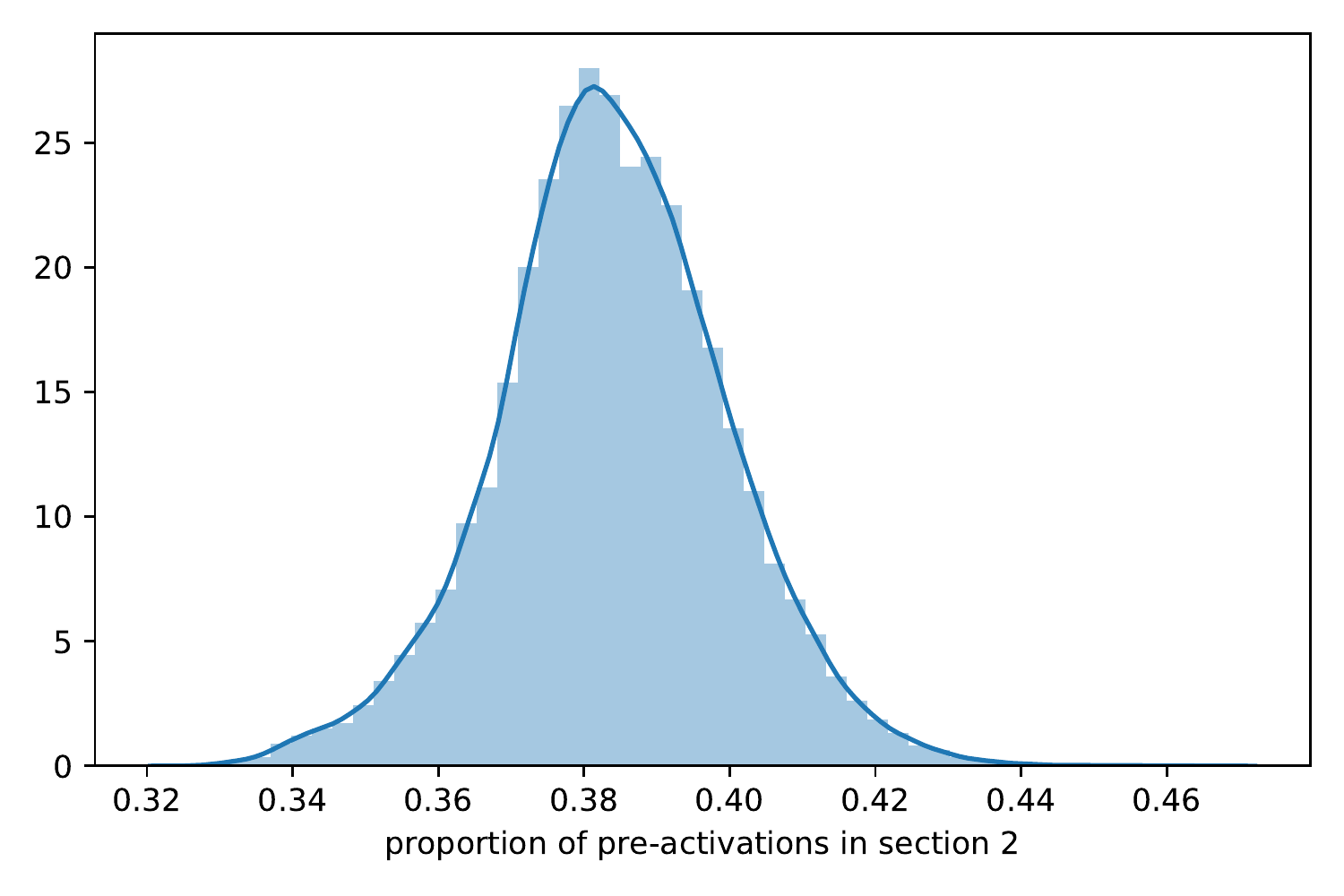}
            \caption{LeNet, MNIST data.} 
            \label{fig:probe_agg_neuron_trained_weights_lenet_mnist}
        \end{subfigure}
        \caption{Experimental distribution of $\bar{R}_2$ (neuron averaging; each sample is a single datum) for MLP and LeNet \texttt{ReLU} networks trained to high validation accuracy on MNIST, and evaluated on i.i.d. normal and MNIST data. The  blue line is a kernel density estimation fit.} 
        \label{fig:probe_agg_neuron_trained_weights}
    \end{figure*}

  \begin{figure*}
        \centering
        \begin{subfigure}[b]{0.236\textwidth}
            \centering
            \includegraphics[width=\textwidth]{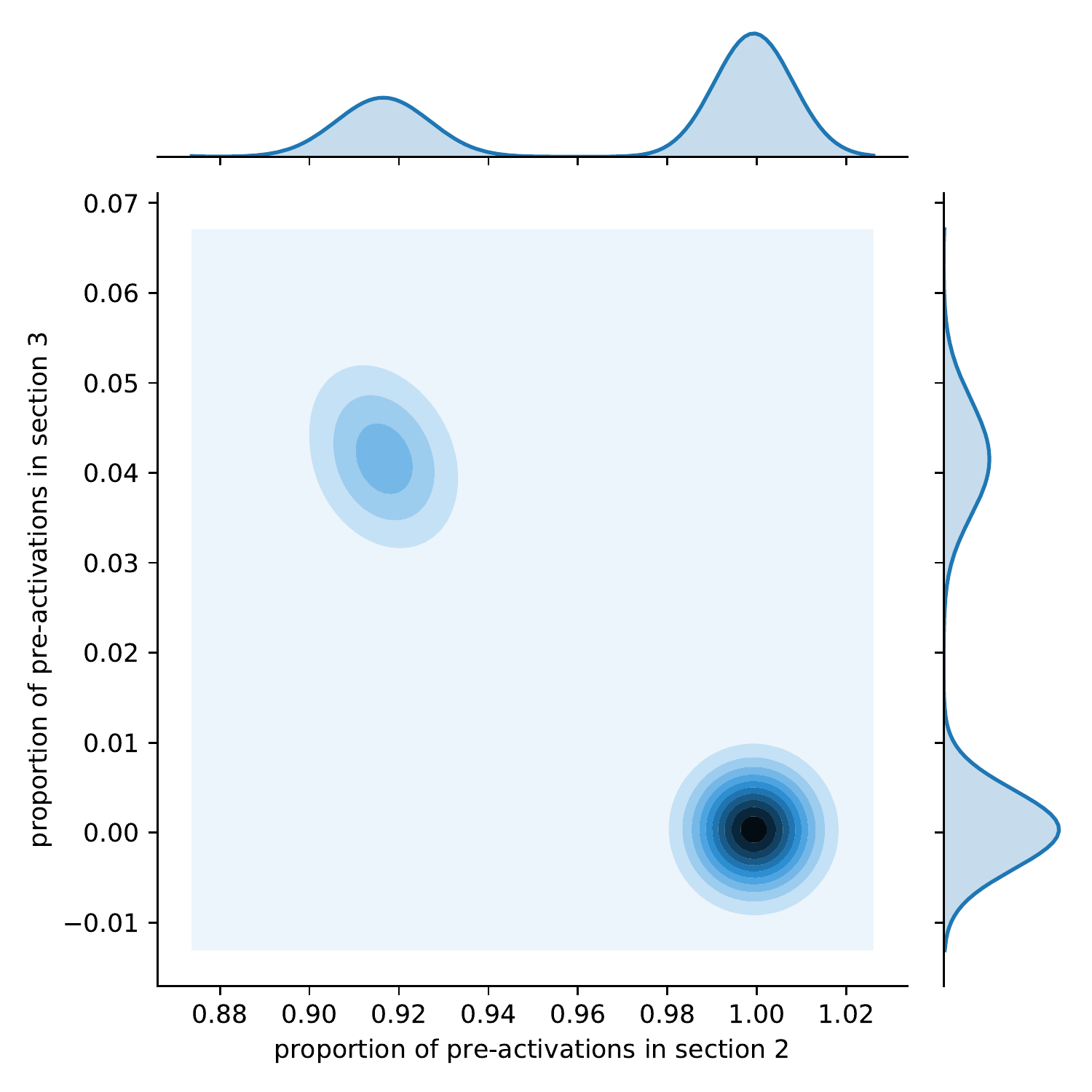}
            \caption{MLP, i.i.d. data.} 
        \end{subfigure}
        \begin{subfigure}[b]{0.236\textwidth}
            \centering
            \includegraphics[width=\textwidth]{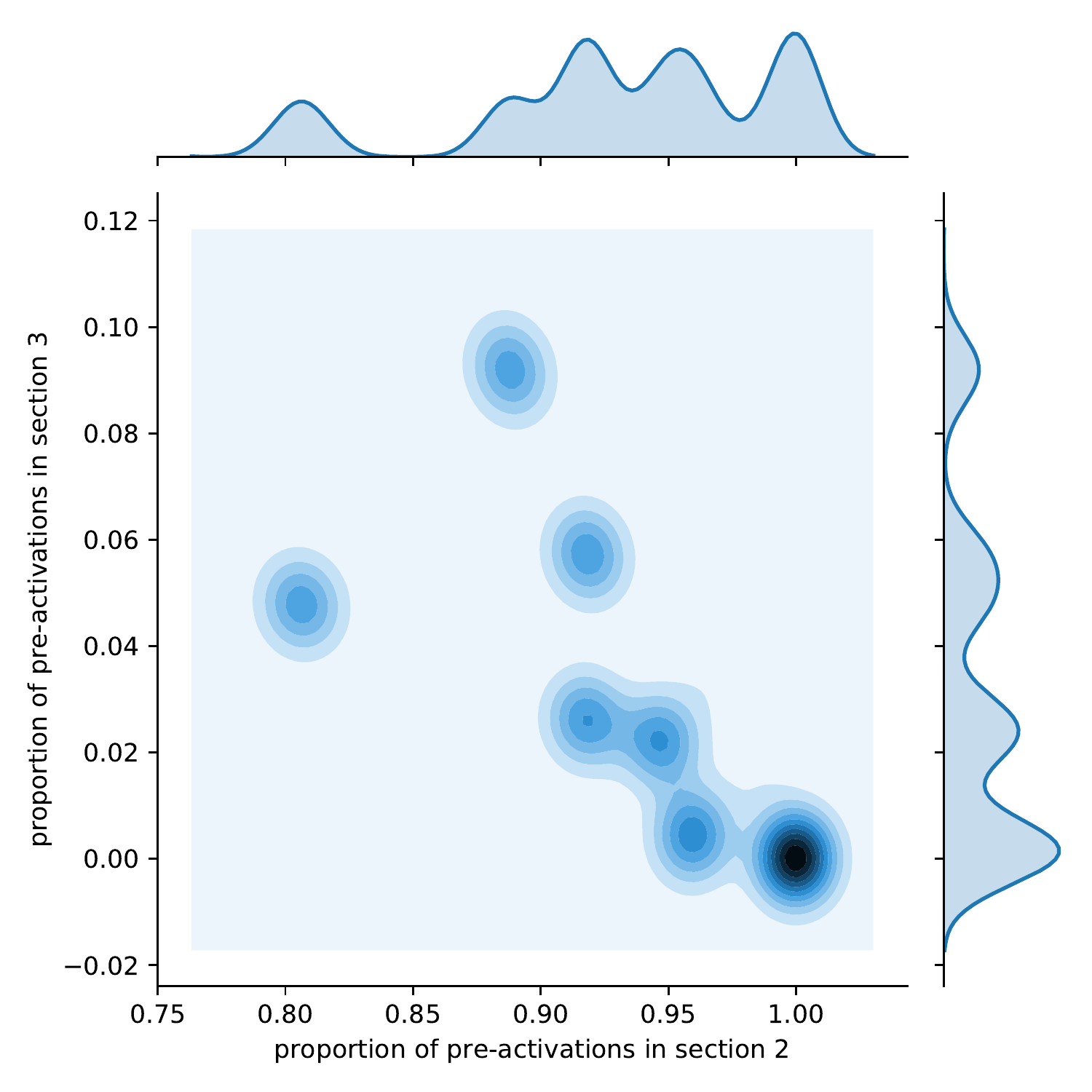}
            \caption{LeNet, i.i.d. data.} 
        \end{subfigure}
        \begin{subfigure}[b]{0.236\textwidth}
            \centering
            \includegraphics[width=\textwidth]{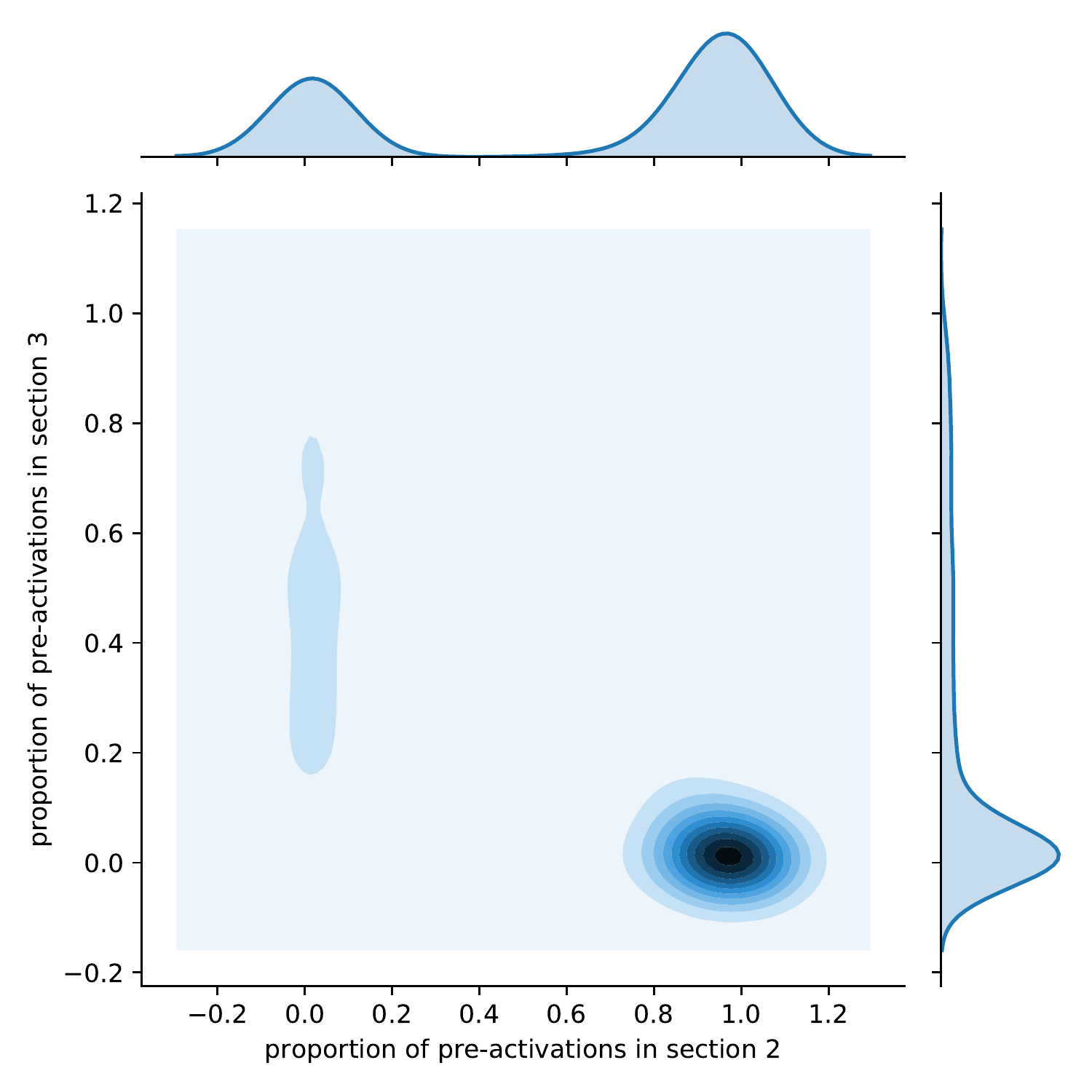}
            \caption{MLP, MNIST data.} 
        \end{subfigure}
        \begin{subfigure}[b]{0.236\textwidth}
            \centering
            \includegraphics[width=\textwidth]{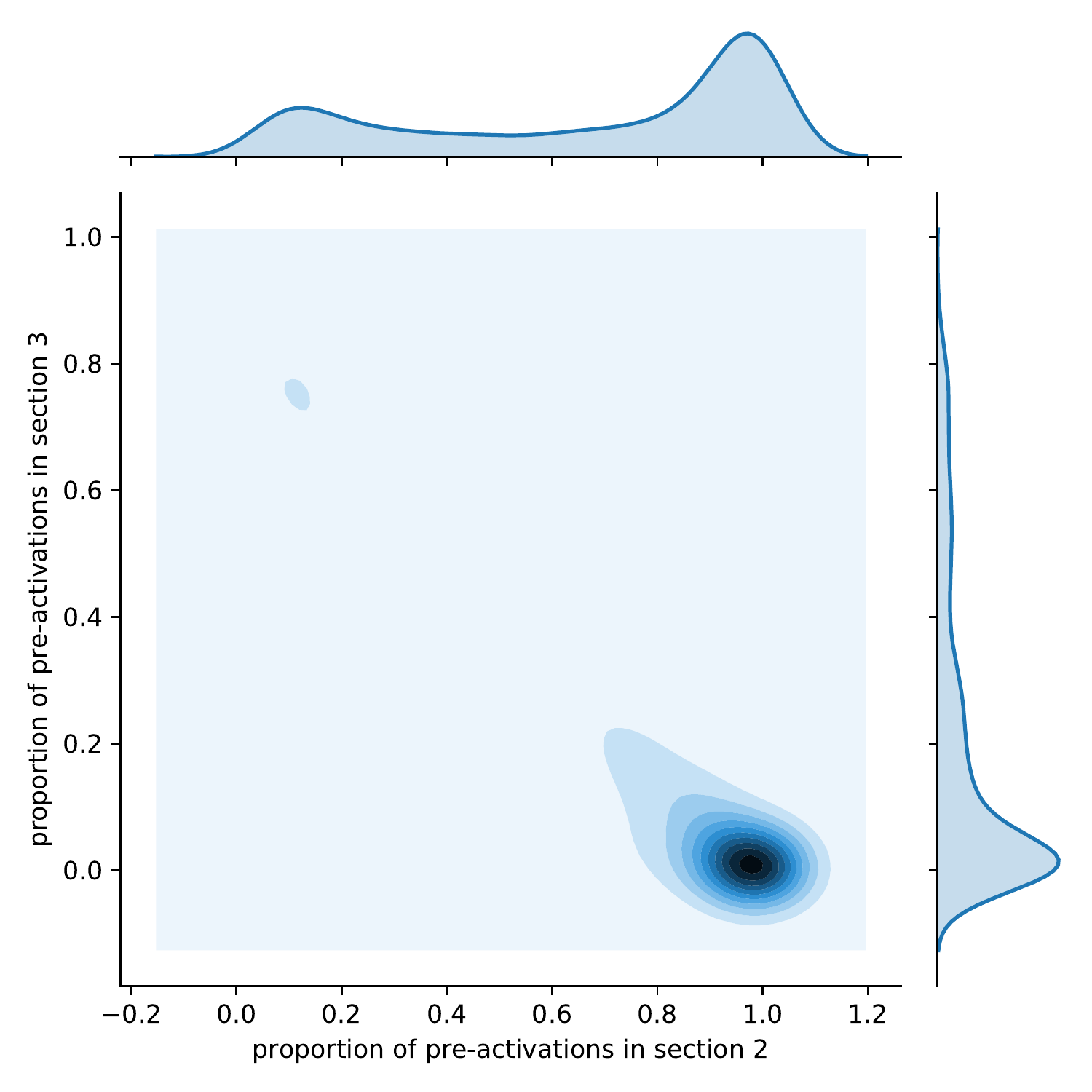}
            \caption{LeNet, MNIST data.} 
        \end{subfigure}
        \caption{Experimental distribution of $(R_2,R_3)$ (data averaging; each sample is a single neuron) for random MLP and LeNet \texttt{HardTanh} networks, and i.i.d. normal and MNIST data. The plots show 2d kernel density estimation fits of the joint and 1d fits of the marginals.} 
                \label{fig:probe_agg_data_random_weights_tanh}
    \end{figure*}
  \begin{figure*}
        \centering
        \begin{subfigure}[b]{0.236\textwidth}
            \centering
            \includegraphics[width=\textwidth]{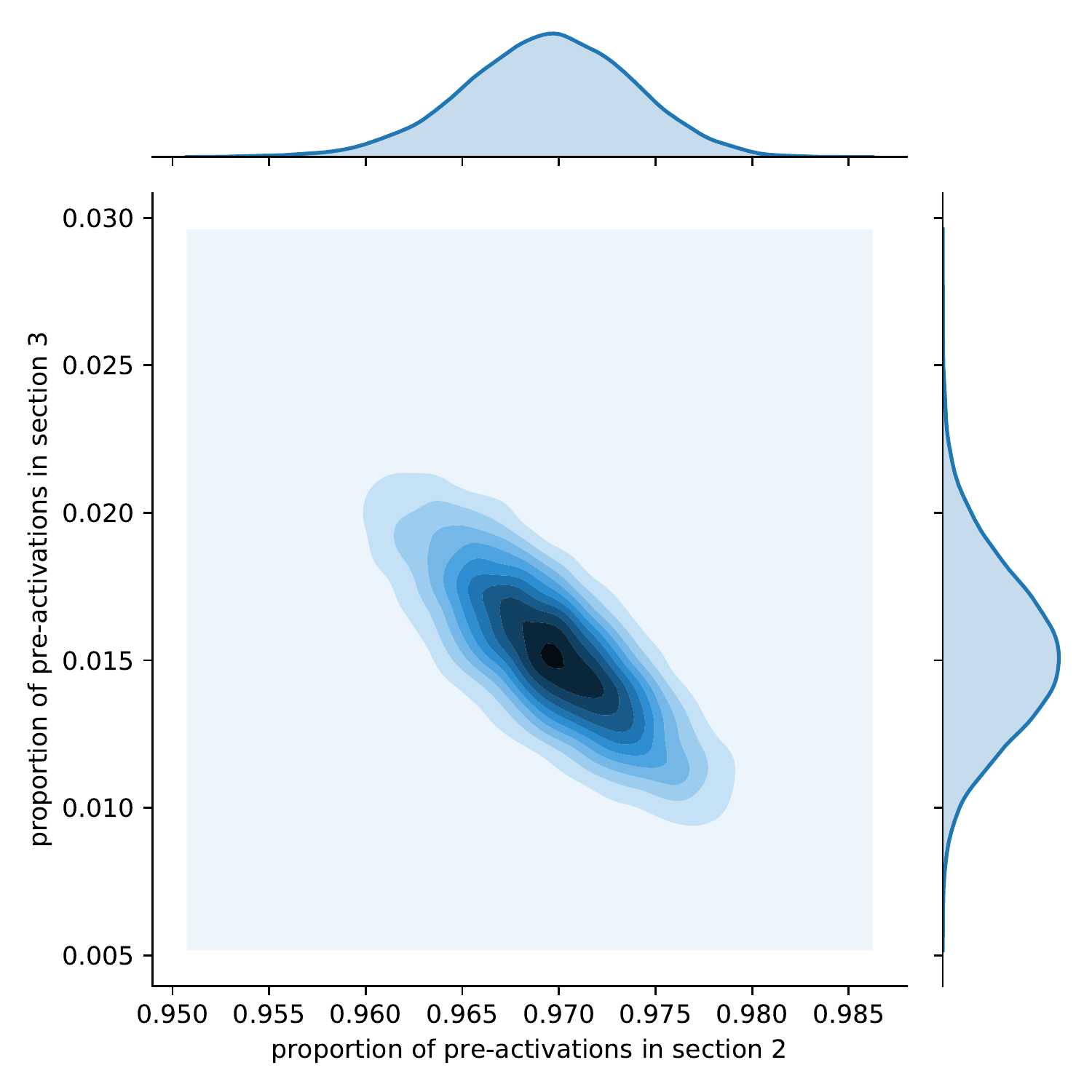}
            \caption{MLP, i.i.d. data.} 
                    \label{fig:probe_agg_neuron_random_weights_tanh_mlp_iid}
        \end{subfigure}
        \begin{subfigure}[b]{0.236\textwidth}
            \centering
            \includegraphics[width=\textwidth]{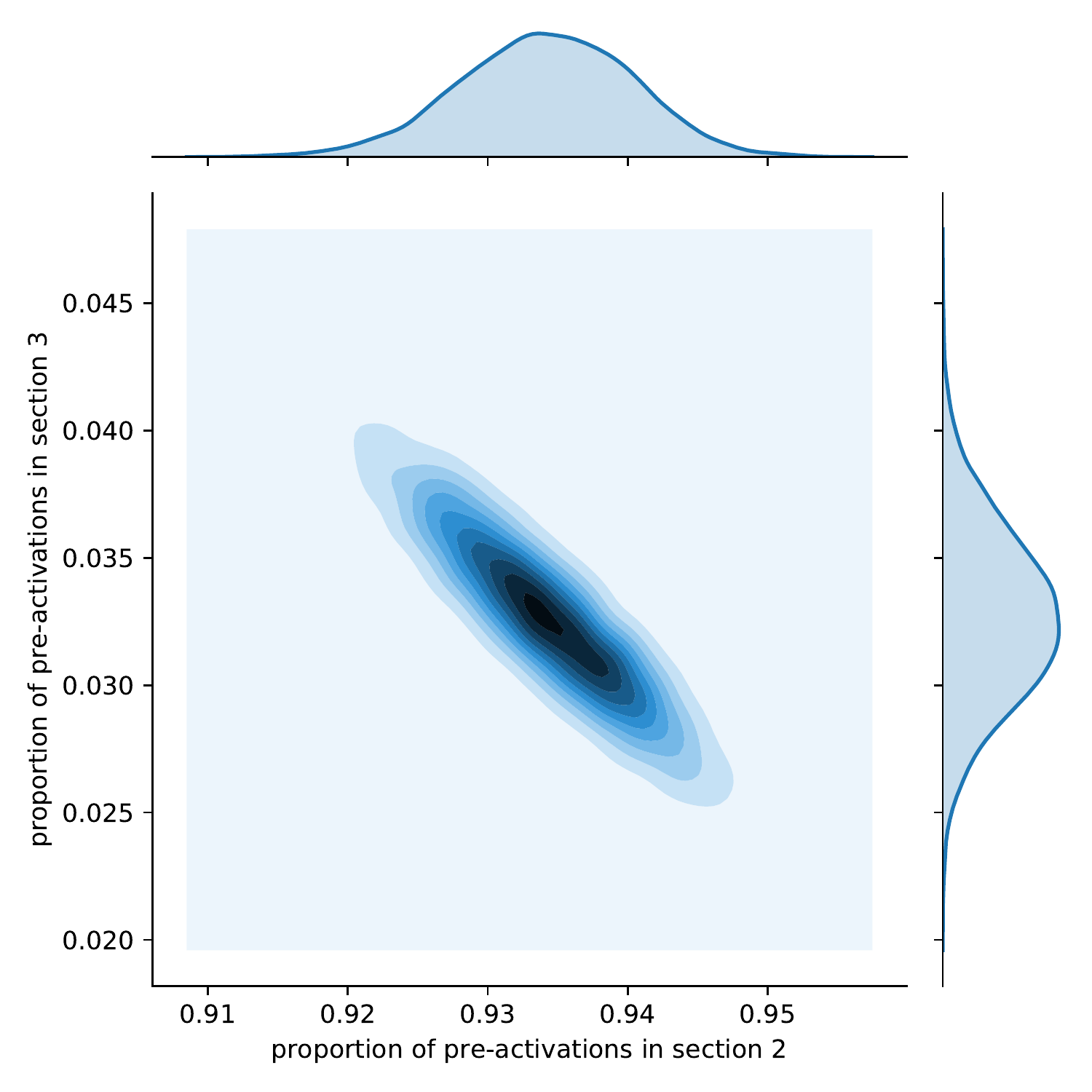}
            \caption{LeNet, i.i.d. data.} 
        \label{fig:probe_agg_neuron_random_weights_tanh_lenet_iid}
        \end{subfigure}
        \begin{subfigure}[b]{0.236\textwidth}
            \centering
            \includegraphics[width=\textwidth]{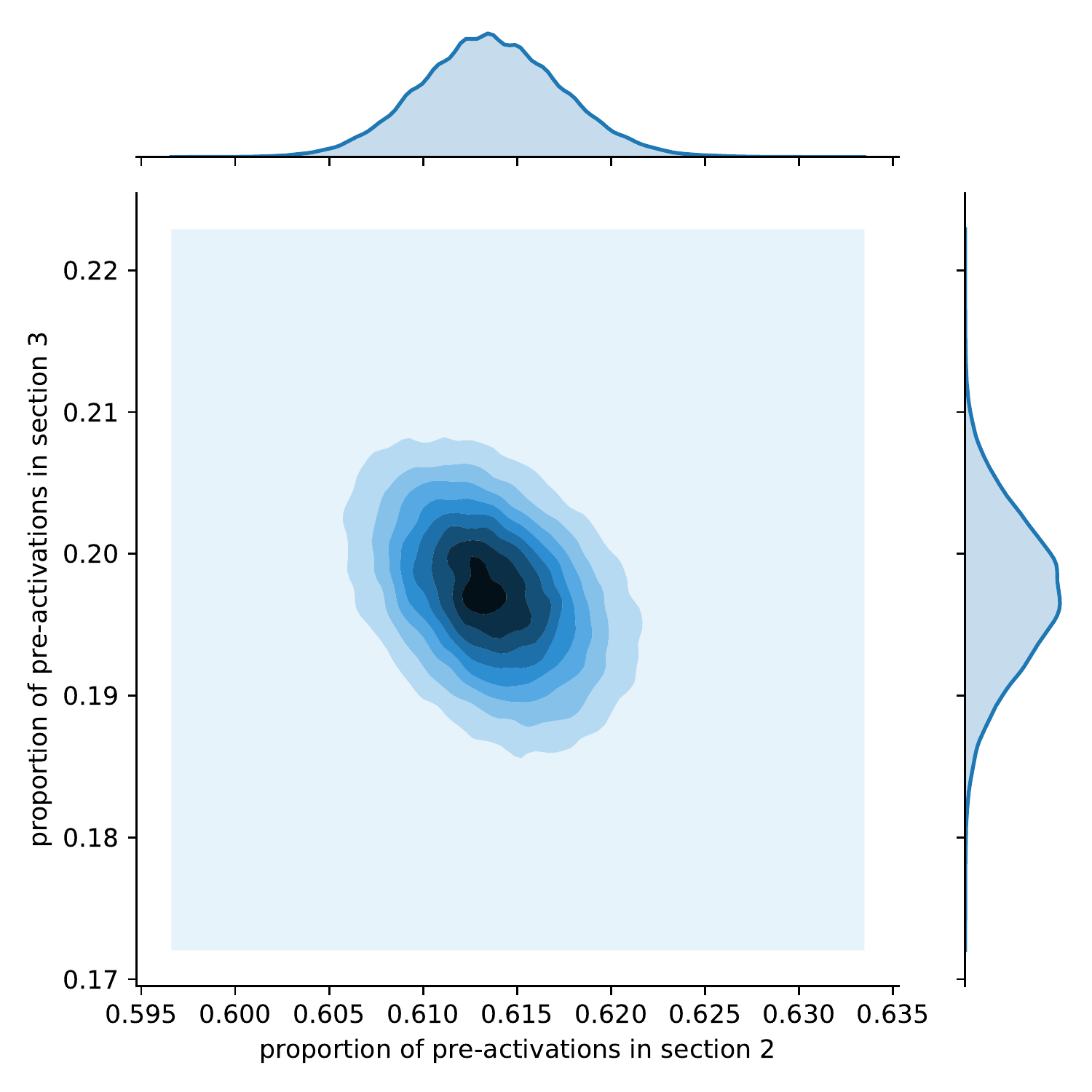}
            \caption{MLP, MNIST data.} 
        \label{fig:probe_agg_neuron_random_weights_tanh_nlp_mnist}
        \end{subfigure}
        \begin{subfigure}[b]{0.236\textwidth}
            \centering
            \includegraphics[width=\textwidth]{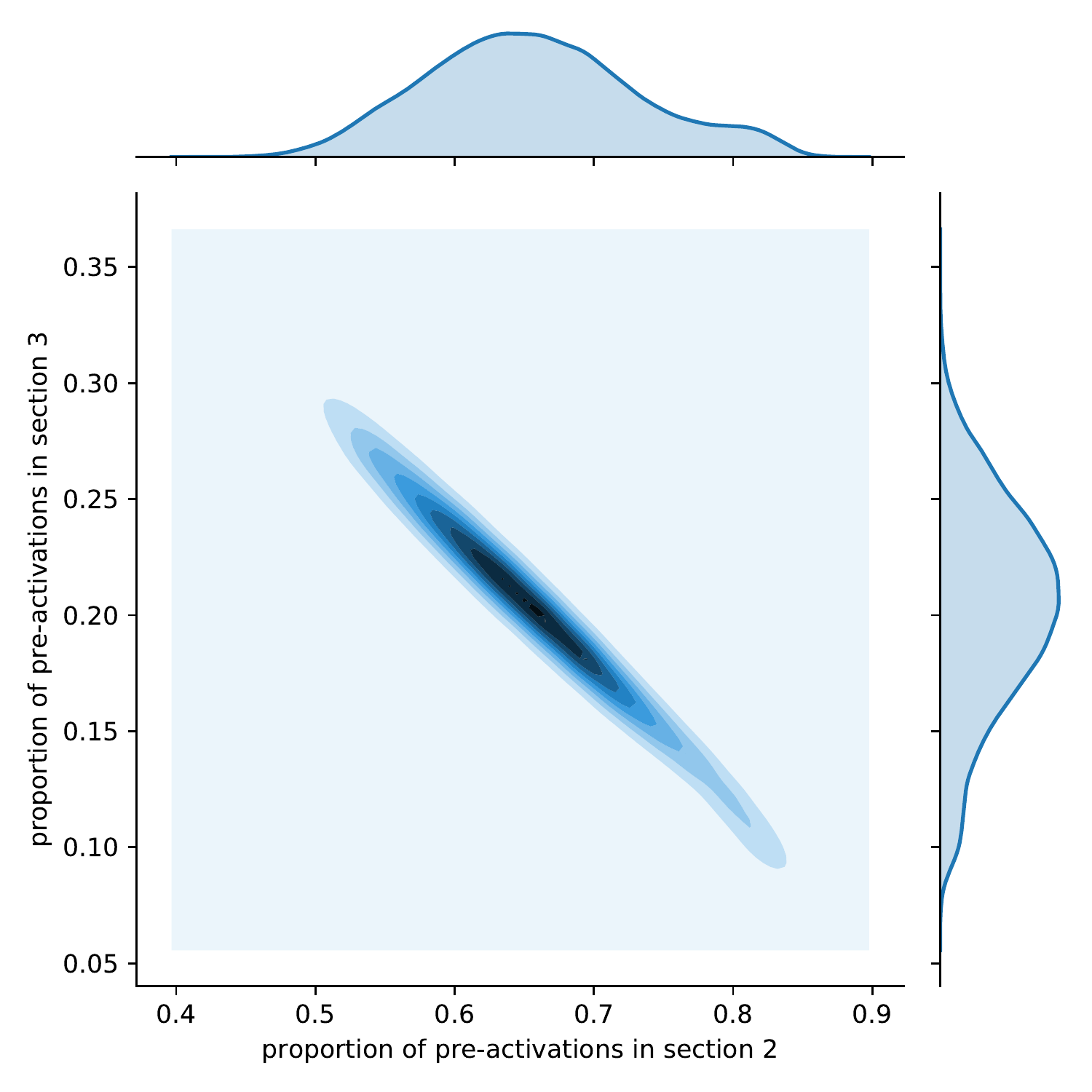}
            \caption{LeNet, MNIST data.} 
        \label{fig:probe_agg_neuron_random_weights_tanh_lenet_mnist}
        \end{subfigure}
        \caption{Experimental distribution of $(\bar{R}_2, \bar{R}_3)$ (neuron averaging; each sample is a single datum) for random \texttt{HardTanh} MLP and LeNet networks, and i.i.d. normal and MNIST data. The plots show 2d kernel density estimation fits of the joint and 1d fits of the marginals.} 
        \label{fig:probe_agg_neuron_random_weights_tanh}
    \end{figure*}
  \begin{figure*}
        \centering
        \begin{subfigure}[b]{0.236\textwidth}
            \centering
            \includegraphics[width=\textwidth]{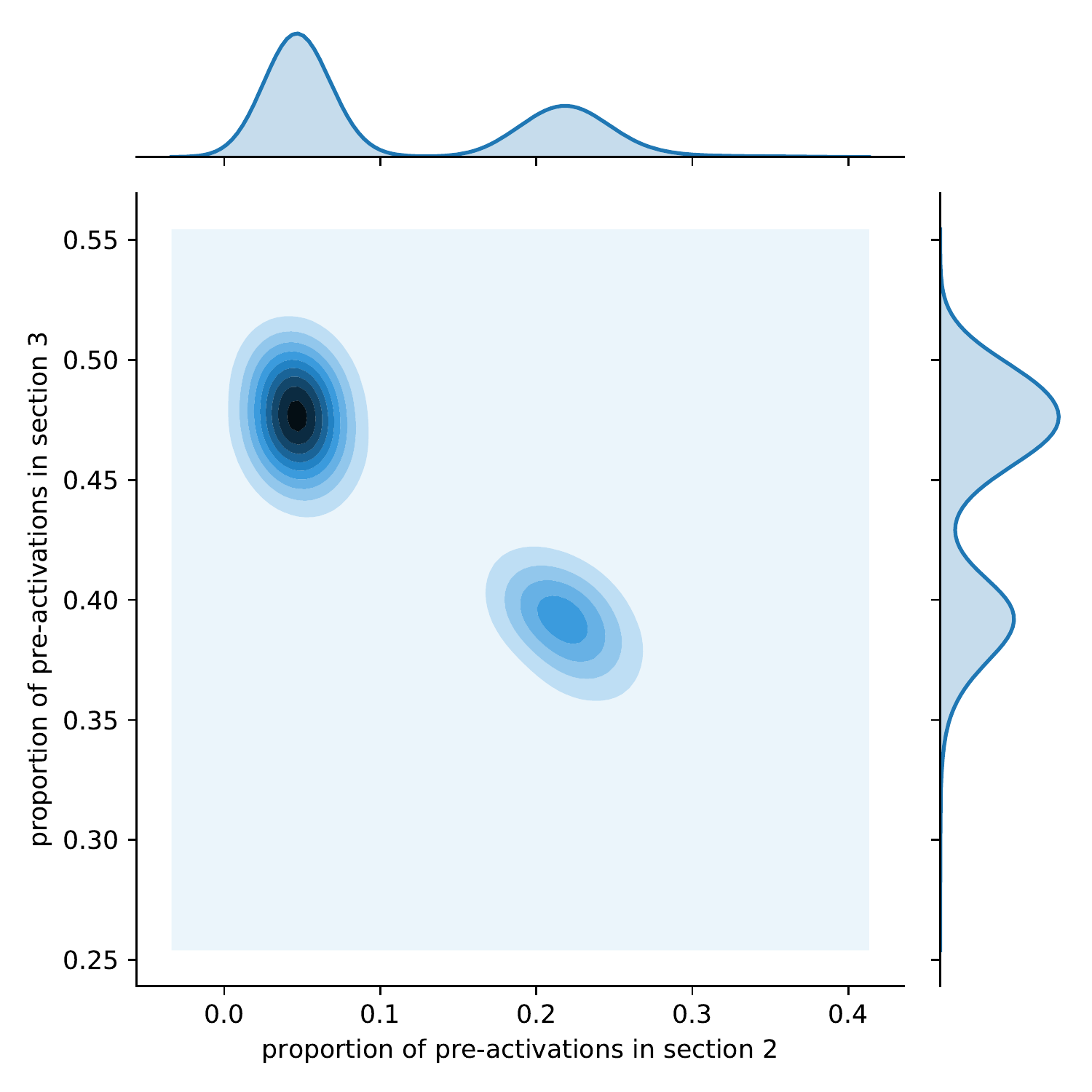}
            \caption{MLP, i.i.d. data.} 
        \end{subfigure}
        \begin{subfigure}[b]{0.236\textwidth}
            \centering
            \includegraphics[width=\textwidth]{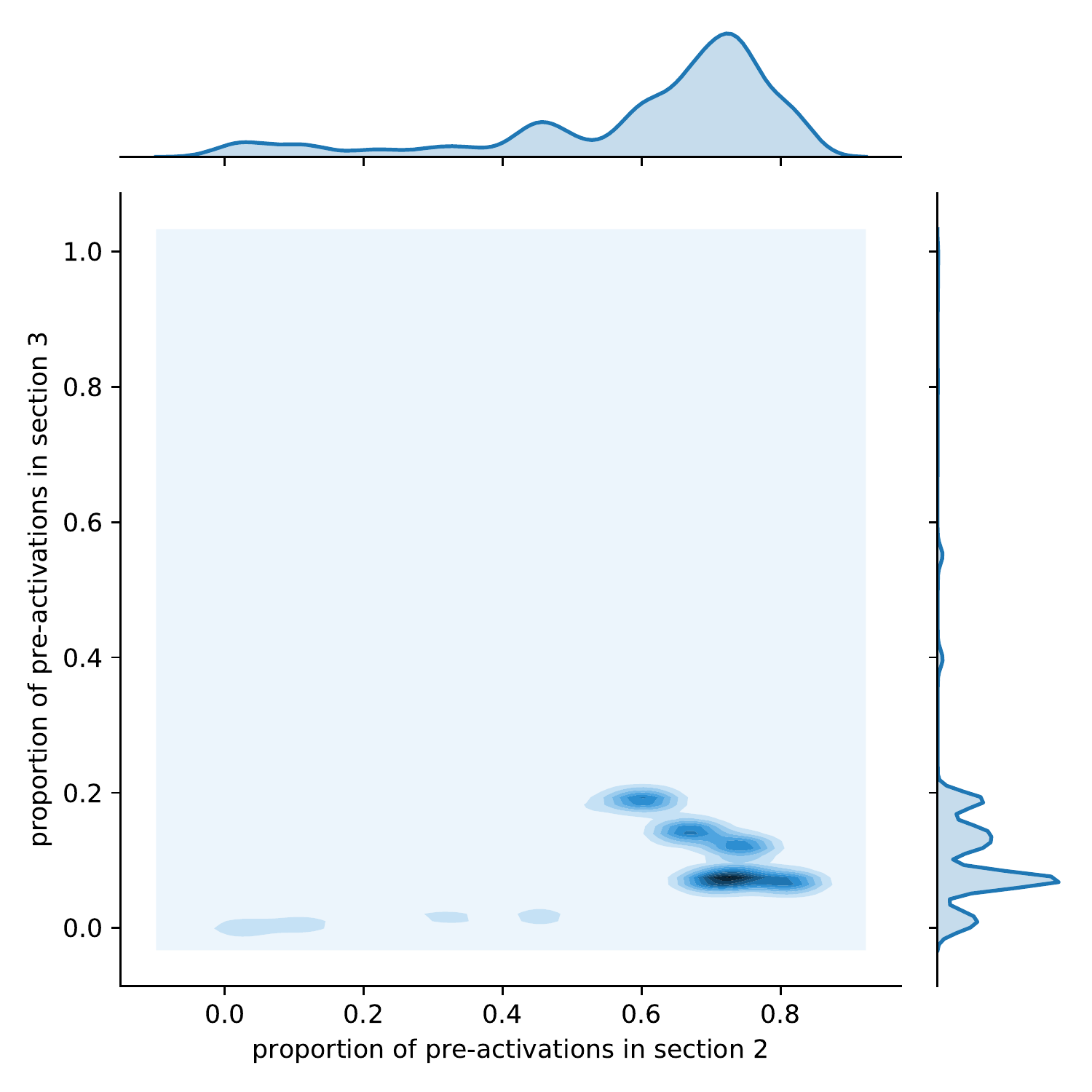}
            \caption{LeNet, i.i.d. data.} 
        \end{subfigure}
        \begin{subfigure}[b]{0.236\textwidth}
            \centering
            \includegraphics[width=\textwidth]{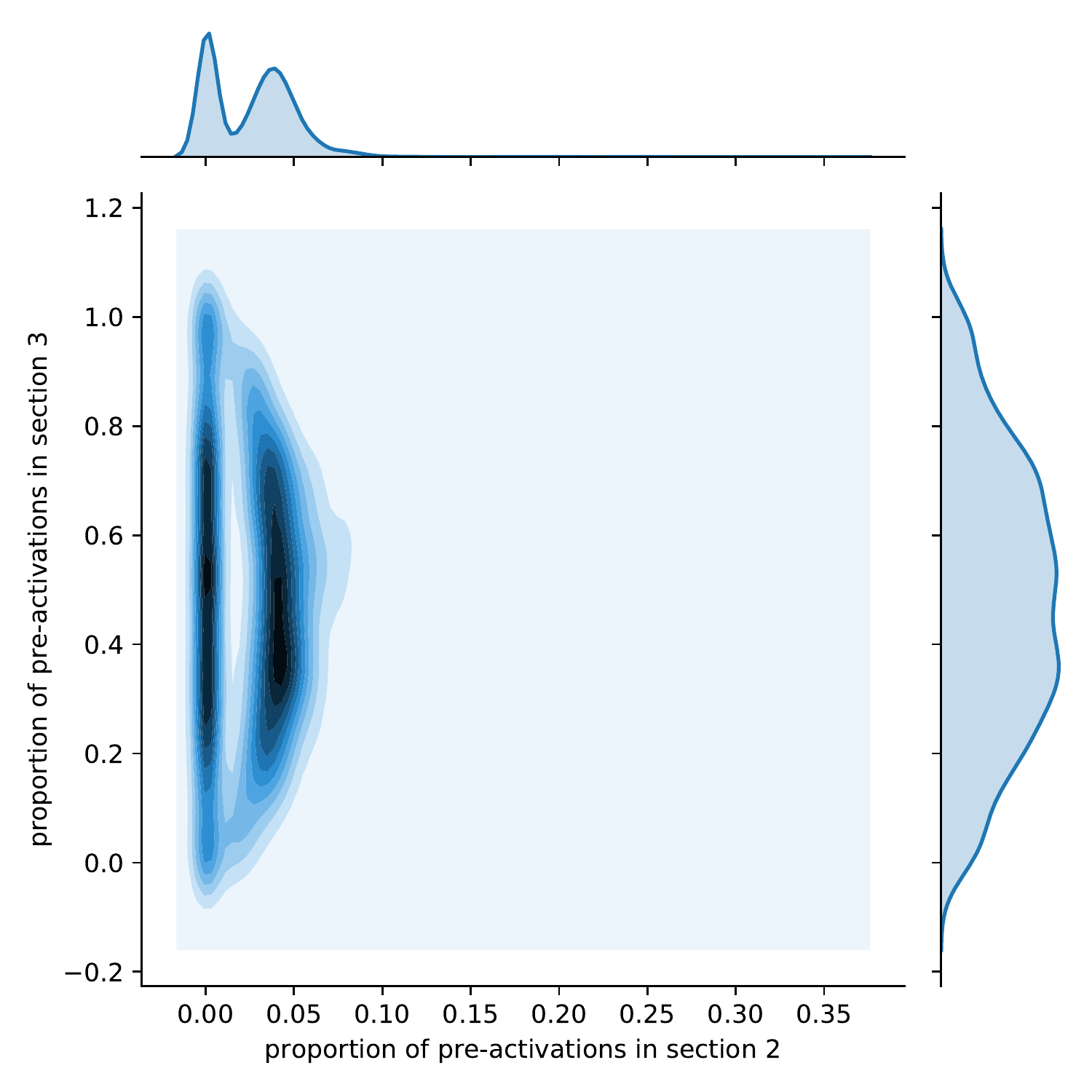}
            \caption{MLP, MNIST data.} 
        \end{subfigure}
        \begin{subfigure}[b]{0.236\textwidth}
            \centering
            \includegraphics[width=\textwidth]{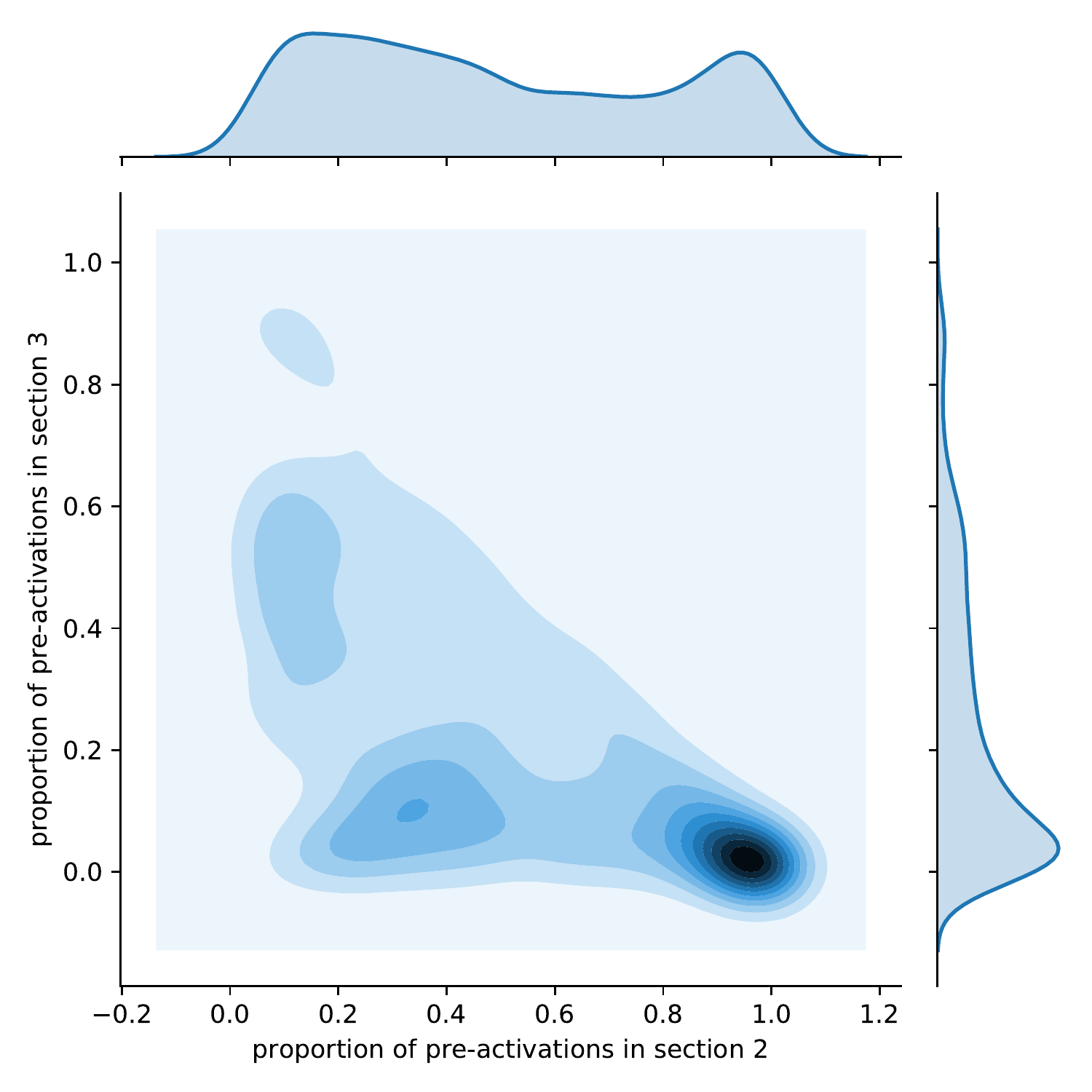}
            \caption{LeNet, MNIST data.} 
        \end{subfigure}
        \caption{Experimental distribution of $(R_2, R_3)$ (data averaging; each sample is a single neuron) for MLP and LeNet \texttt{HardTanh} networks trained to high validation accuracy on MNIST, and evaluated on i.i.d. normal and MNIST data. The plots show 2d kernel density estimation fits of the joint and 1d fits of the marginals.} 
                \label{fig:probe_agg_data_trained_weights_tanh}
    \end{figure*}
  \begin{figure*}
        \centering
        \begin{subfigure}[b]{0.236\textwidth}
            \centering
            \includegraphics[width=\textwidth]{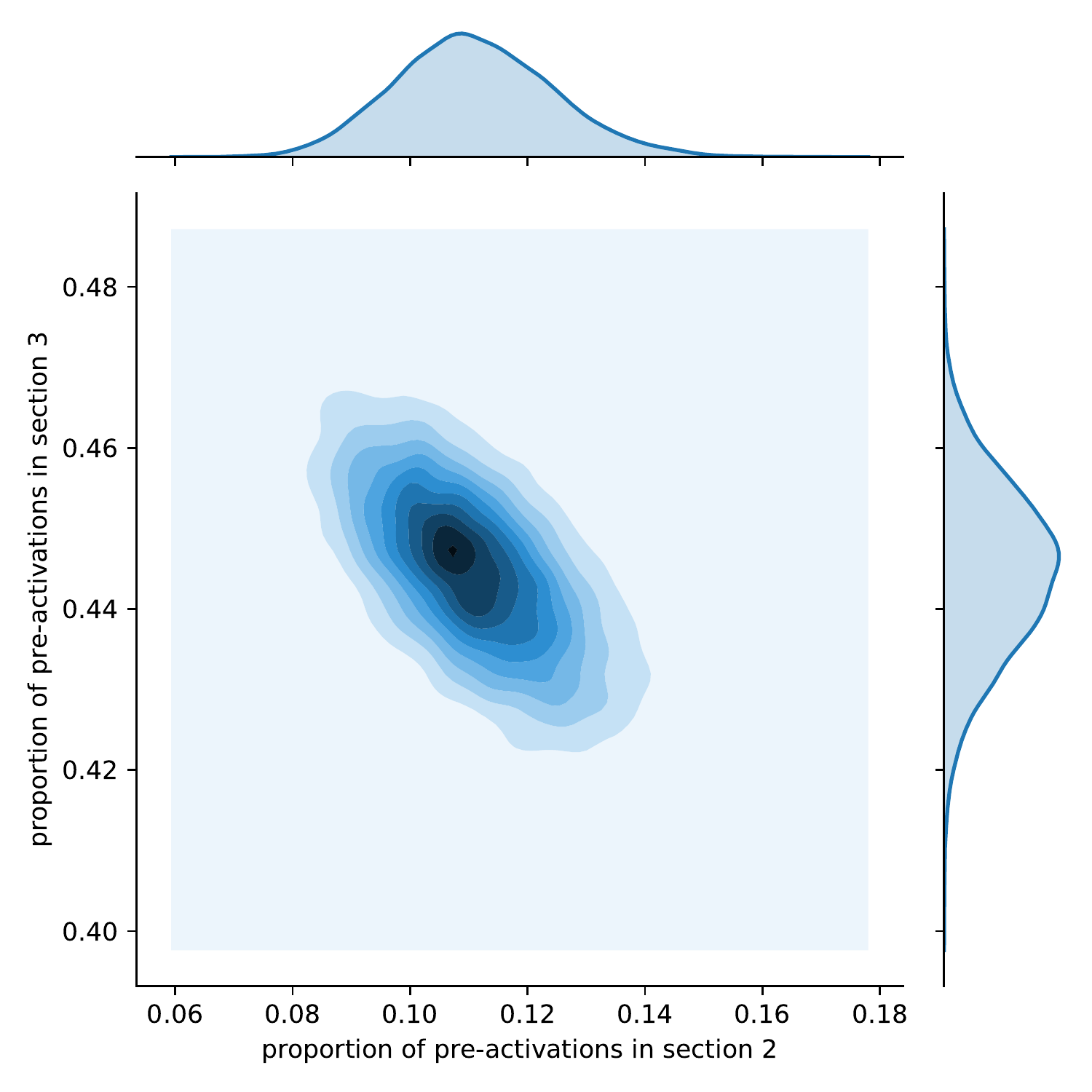}
            \caption{MLP, i.i.d. data.} 
        \end{subfigure}
        \begin{subfigure}[b]{0.236\textwidth}
            \centering
            \includegraphics[width=\textwidth]{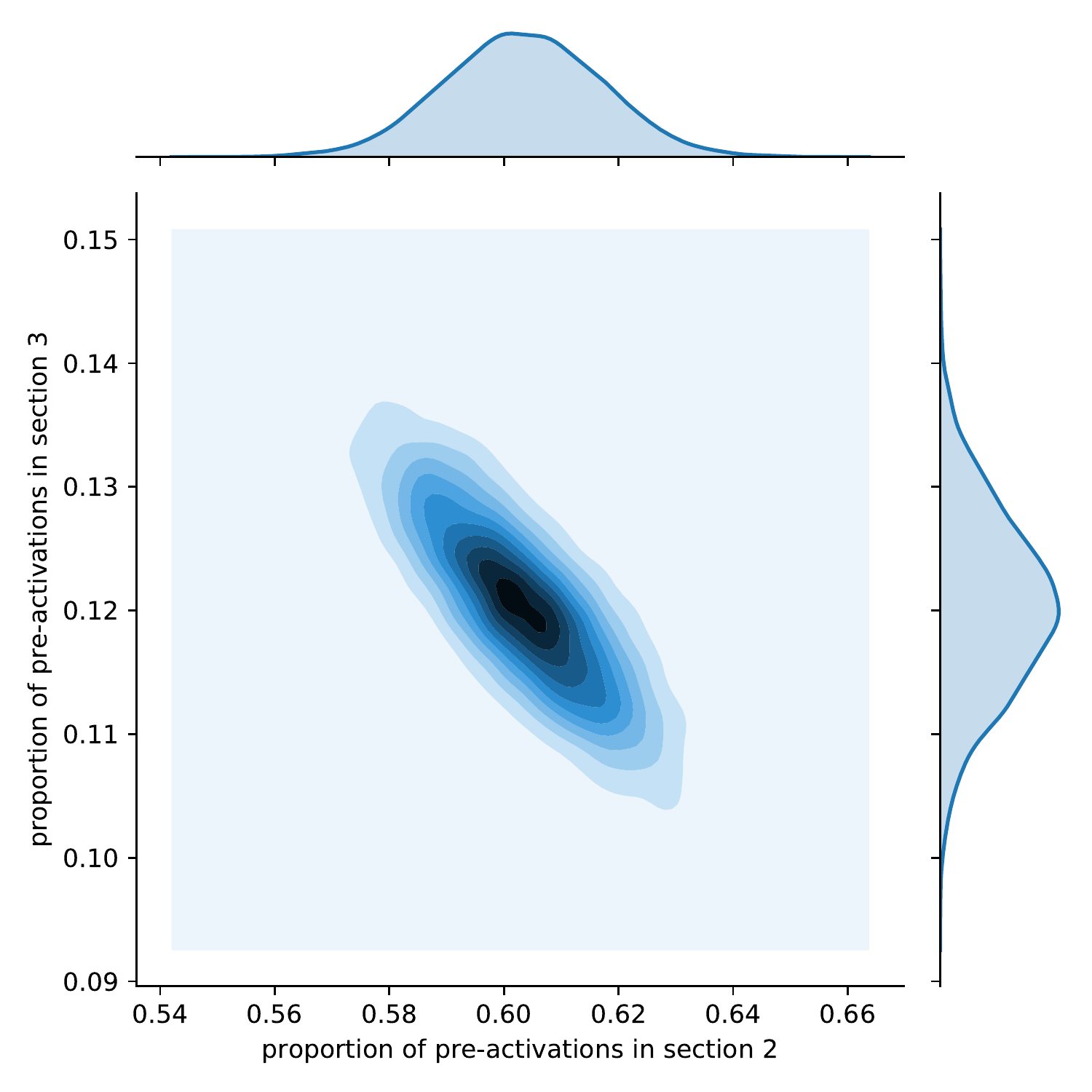}
            \caption{LeNet, i.i.d. data.} 
        \end{subfigure}
        \begin{subfigure}[b]{0.236\textwidth}
            \centering
            \includegraphics[width=\textwidth]{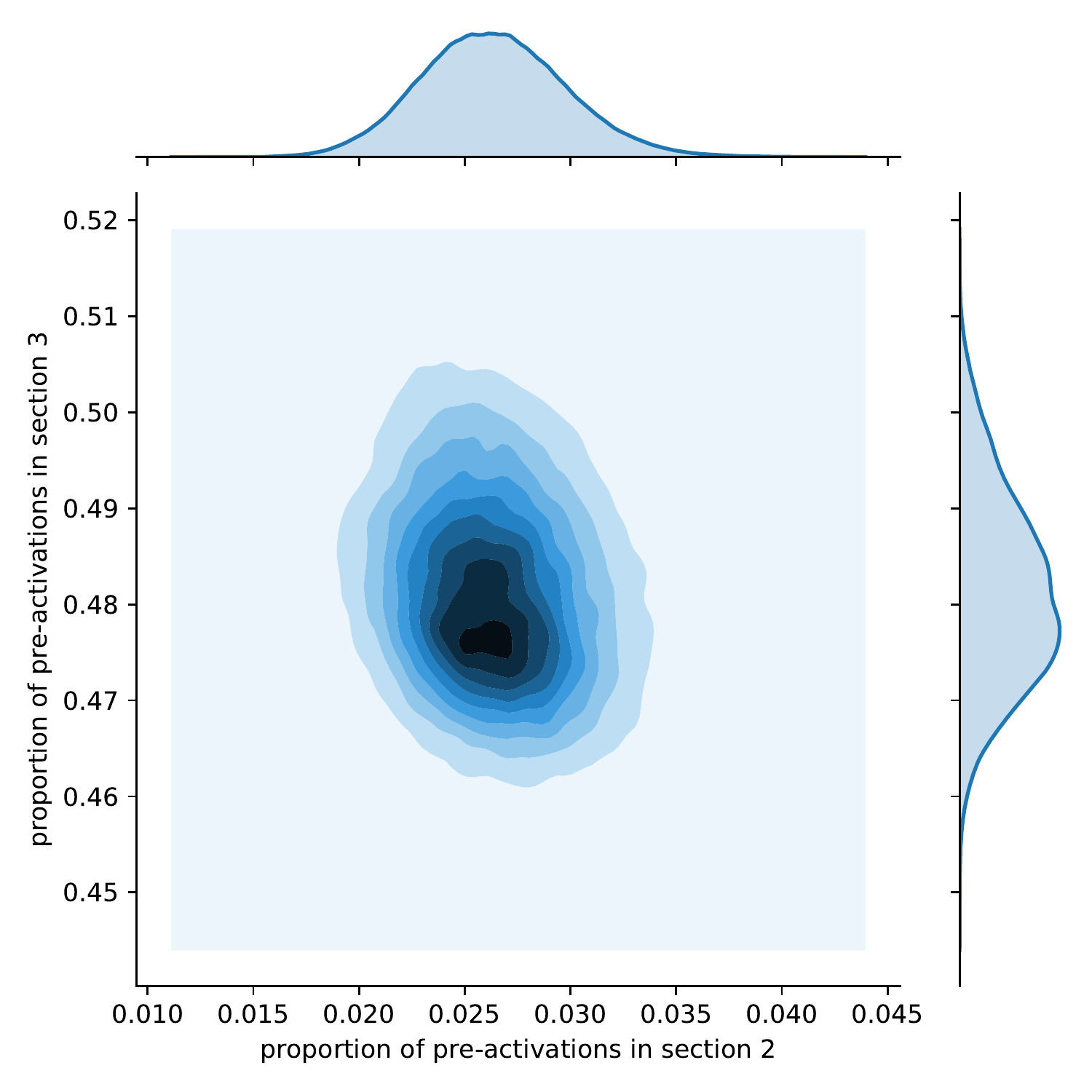}
            \caption{MLP, MNIST data.} 
        \end{subfigure}
        \begin{subfigure}[b]{0.236\textwidth}
            \centering
            \includegraphics[width=\textwidth]{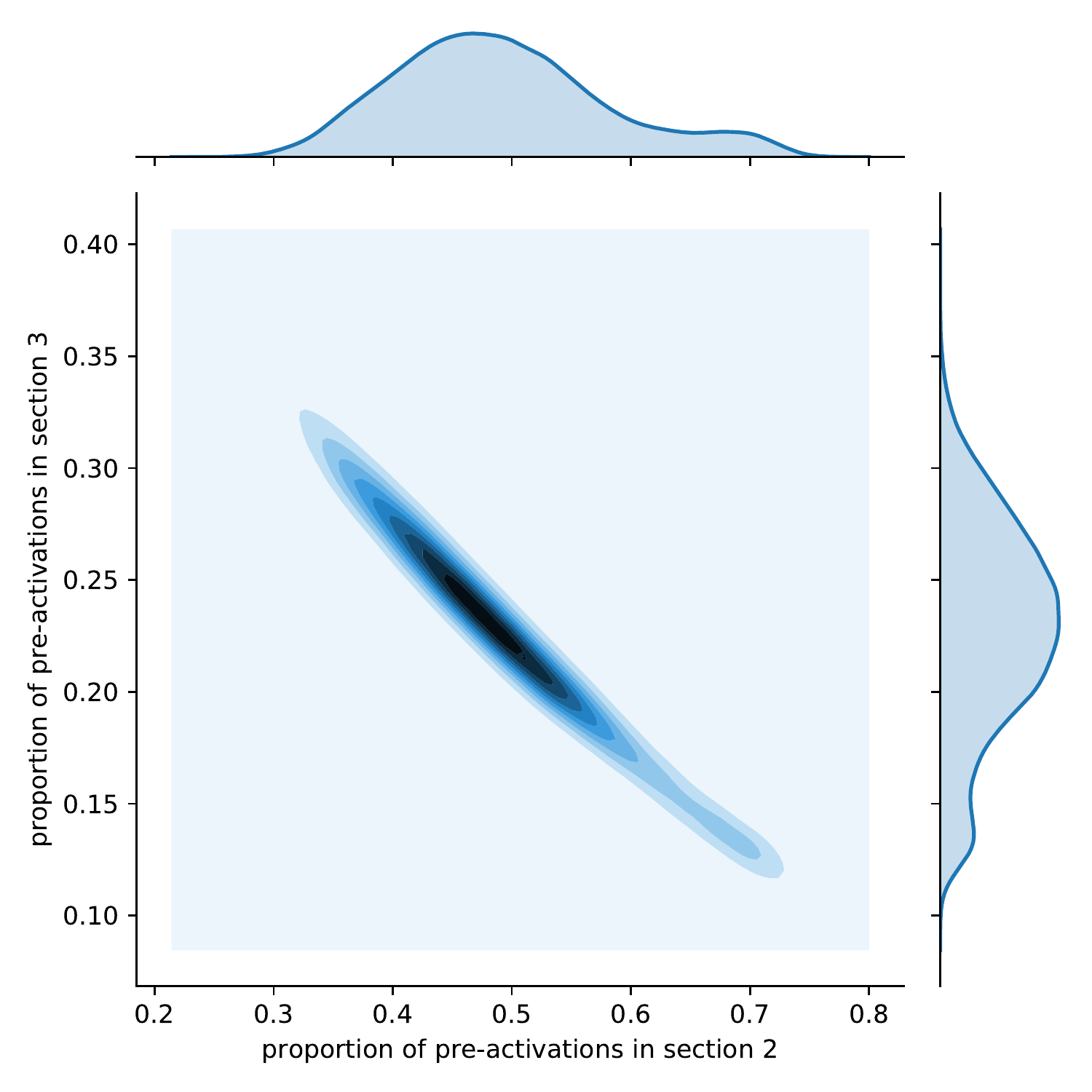}
            \caption{LeNet, MNIST data.} 
        \end{subfigure}
        \caption{Experimental distribution of  $(\bar{R}_2, \bar{R}_3)$ (neuron averaging; each sample is a single datum) for MLP and LeNet \texttt{HardTanh} networks trained to high validation accuracy on MNIST, and evaluated on i.i.d. normal and MNIST data. The plots show 2d kernel density estimation fits of the joint and 1d fits of the marginals.} 
        \label{fig:probe_agg_neuron_trained_weights_tanh}

    \end{figure*}

\begin{enumerate}
    \item The variance of $\bar{R}_2$ is `small' in all cases for \texttt{ReLU} networks except when evaluating MNIST-trained MLP networks on i.i.d. random normal data. This is the least relevant case practically.
    \item For $R_2$, the results are much less convincing, though we do note that, with random weights and i.i.d. data, the MLP network does have quite a strongly peaked distribution. In other cases the variance is undeniably large.
    \item The variance of $\bar{R}_{2,3}$ is `small' in all cases for \texttt{HardTanh} except when evaluating LeNet architectures on MNIST data.
    \item For $R_{3}$ in \texttt{HardTanh} networks, the variance seems to be low when the weights are random, but not when trained.
\end{enumerate}

Overall, we see that in some circumstances, particularly with un-trained weights, the assumption \ref{item: assumption_bernoulli} is not as unreasonable as it first sounds. More importantly for the present work, comparing the three examined activation functions supports the hypothesis that, insofar as modeling the action of the \texttt{ReLU} activation function by independent Bernoulli random variables was valid in \cite{choromanska2015loss}, our analogous modelling of the action of general piece-wise linear functions by independent discrete random variables is also valid. \jstat{Put another way, it does not appear that the assumptions we make here are any stronger than those made in \cite{choromanska2015loss}. We finally note an interesting comparison between, for example, Figures \ref{fig:probe_agg_neuron_random_weights_mlp_iid} and \ref{fig:probe_agg_neuron_random_weights_mlp_mnist}, or equally Figures \ref{fig:probe_agg_neuron_random_weights_tanh_mlp_iid} and \ref{fig:probe_agg_neuron_random_weights_tanh_nlp_mnist}. In both cases, the variance is low for both distributions, and the only difference between the two experiments is the evaluation data, being i.i.d. Gaussian in the one case, and MNIST in the other. These results seem to demonstrate that the assumption of i.i.d. Gaussian data distribution is not trivialising the problem as one might expect a priori.}

\jstat{Taking all of the results of this section together, we see that the case for our extension of \cite{choromanska2015loss} is quite strong, but there are clearly realistic cases where the modelling assumptions applied to activation functions in \cite{choromanska2015loss} are convincingly violated.}

\section{Statement of results}\label{sec:statement_results}
 We shall use \emph{complexity} to refer to any of the following defined quantities which we define precisely as they appear in \cite{auffinger2013random}.

\begin{defn}\label{def:cnk}
For a Borel set $B\subset\mathbb{R}$ and non-negative integer $k$, let \begin{equation}
    \CNk^g(B) = \left|\left\{\vec{w} \in \sqrtsign{N}S^{N-1} ~:~ \grad g(\vec{w})=0, g(\vec{w})\in B, ~ i(\grad^2g)=k\right\}\right|
\end{equation}
where $i(M)$ for a square matrix $M$ is the \emph{index} of $M$, i.e. the number of negative eigenvalues of $M$. We also define the useful generalisation $\ind{x}(M)$ to be the number of eigenvalues of $M$ less than $x$, so $\ind{0}(M) = i(M).$
\end{defn}
\begin{defn}
For a Borel set $B\subset\mathbb{R}$, let \begin{equation}
    \CN^g(B) = \left|\left\{\vec{w} \in \sqrtsign{N}S^{N-1} ~:~ \grad g(\vec{w})=0, g(\vec{w})\in B\right\}\right|.
\end{equation}
\end{defn}

We now state our main identities, which we find simpler to prove by scaling $\vec{w}$ to lie on the hyper-sphere of unit radius: $h(\vec{w}) \defeq N^{-H/2}g(\sqrtsign{N}\vec{w})$.  For convenience, we define \begin{equation}\label{eq:rho_N_redef}
\rho_{\ell}^{(N)} =  \rho_{\ell} N^{-\ell/2}
\end{equation} so that, recalling the form of $g$ in (\ref{eq:g_def}), we obtain \begin{equation}\label{eq:h_def}
    h(\vec{w}) = \sum_{i_1, \ldots, i_H = 1}^{\Lambda} X_{i_1, \ldots, i_H} \prod_{k=1}^H w_{i_k} + \sum_{\ell=1}^H\rho_{\ell}^{(N)} \sum_{i_{\ell + 1}, \ldots, i_H=1}^{\Lambda} \prod_{k=\ell +1}^H w_{i_k}.
\end{equation}Though the complexities have been defined using general Borel sets, as in \cite{auffinger2013random}, we focus on half-infinite intervals $(-\infty, u)$, acknowledging that everything that follows could be repeated instead with general Borel sets \emph{mutatis mutandis}.  We will henceforth be studying the following central quantities (note the minor abuse of notation):

\begin{equation}\label{eq:cnkh_def}
     \CNk^h(\sqrtsign{N}u) = \left|\left\{\vec{w} \in S^{N-1} ~:~ \grad h(\vec{w})=0, h(\vec{w})\in \sqrtsign{N}u, ~ i(\grad^2h)=k\right\}\right|,
\end{equation}
\begin{equation}\label{eq:cnh_def}
     \CN^h(\sqrtsign{N}u) = \left|\left\{\vec{w} \in S^{N-1} ~:~ \grad h(\vec{w})=0, h(\vec{w})\in \sqrtsign{N}u\right\}\right|
\end{equation}
and it will be useful to define a relaxed version of (\ref{eq:cnkh_def}) for $\mathcal{K}\subset \{0,1,\ldots, N\}$:
\begin{equation}\label{eq:cnkh_set_def}
      C_{N, \mathcal{K}}^h(\sqrtsign{N}u) = \left|\left\{\vec{w} \in S^{N-1} ~:~ \grad h(\vec{w})=0, h(\vec{w})\in \sqrtsign{N}u, ~ i(\grad^2h)\in\mathcal{K}\right\}\right|.
\end{equation}

Our main results take the form of two theorems that extend Theorems 2.5 and 2.8 from \cite{auffinger2013random} to our more general spin glass like object $g$, and a third theorem with partially extends Theorem 2.17 of \cite{auffinger2013random}. In the case of Theorem 2.8, we are able to obtain exactly the same result in this generalised setting. For Theorem 2.5, we have been unable to avoid slackening the result slightly, hence the introduction of the quantity $C^h_{N, \mathcal{K}}$ above. In the case of Theorem 2.17, we are only able to perform the calculations of the exact leading order term in one case and obtain a term very similar to that in \cite{auffinger2013random} but with an extra factor dependent on the piece-wise linear approximation to the generalised activation function. This exact term correctly falls-back to the term found in \cite{auffinger2013random} when we take $f=\texttt{ReLU}$.\\

\begin{restatable}{theorem}{auffindk}
\label{thm:auff2.8}%
Recall the definition of $\CN^h$ in (\ref{eq:cnh_def}) and let $\Theta_H$ be defined as in \cite{auffinger2013random}: \begin{align}
    \Theta_H(u) = \begin{cases} \frac{1}{2}\log(H-1) - \frac{H-2}{4(H-1)}u^2 - I_1(u; E_{\infty}) ~~ &\text{if } u\leq -E_{\infty},\\
    \frac{1}{2}\log(H-1) - \frac{H-2}{4(H-1)}u^2 &\text{if } -E_{\infty} \leq u \leq 0,\\
    \frac{1}{2}\log(H-1) &\text{if } 0\geq u,\end{cases}
    \end{align}
    where $E_{\infty} = 2\sqrtsign{\frac{H-1}{H}}$, and $I_1(\cdot; E)$ is defined on $(-\infty, -E]$ as in \cite{auffinger2013random} by \begin{equation}
        I_1(u; E) = \frac{2}{E^2}\int_{u}^{-E} (z^2 - E^2)^{1/2} dz = -\frac{u}{E^2}\sqrtsign{u^2 - E^2} - \log\left(-u + \sqrtsign{u^2 - E^2}\right) + \log E,\label{eq:auffinger_I1_def}
    \end{equation} then \begin{equation}
        \lim_{N\rightarrow\infty} \frac{1}{N}\log\expect C_{N}^{h}(\sqrtsign{N}u) = \Theta_H(u).
    \end{equation}
\end{restatable}

\begin{restatable}{theorem}{auffdepk}
\label{thm:auff2.5}%
Recall the definition of $C_{N, \mathcal{K}}^h$ in (\ref{eq:cnkh_set_def}) and let $\Theta_{H,k}$ be defined as in \cite{auffinger2013random}: \begin{align}
    \Theta_{H,k}(u) = \begin{cases} \frac{1}{2}\log(H-1) - \frac{H-2}{4(H-1)}u^2 - (k+1)I_1(u; E_{\infty}) ~~ &\text{if } u\leq -E_{\infty},\\
     \frac{1}{2}\log(H-1) - \frac{H-2}{H}
     &\text{if } u > -E_{\infty}, \end{cases}
    \end{align}
    then, with $\mathcal{K} = \{k-1, k, k+1\}$ for $k>0$, \begin{equation}
       \Theta_{H,k+1}(u) \leq \lim_{N\rightarrow\infty} \frac{1}{N}\log\expect C_{N,\mathcal{K}}^{h}(\sqrtsign{N}u) \leq \Theta_{H,k-1}(u)
    \end{equation}
        and similarly with $\mathcal{K} = \{0, 1\}$
        \begin{equation}
       \Theta_{H,1}(u) \leq \lim_{N\rightarrow\infty} \frac{1}{N}\log\expect C_{N,\mathcal{K}}^{h}(\sqrtsign{N}u) \leq \Theta_{H,0}(u).
    \end{equation}
    \end{restatable}

\begin{remark}
Note that Theorem \ref{thm:auff2.5} holds for \texttt{ReLU} networks (equivalently, pure multi-spin glass models), as indeed it must. It can be seen as an immediate (weaker) consequence of the Theorem 2.5 in \cite{auffinger2013random} of which it is an analogue in our more general setting.
\end{remark}

\begin{restatable}{theorem}{auffexact}
\label{thm:exact_term}%
Let $u<-E_{\infty}$ and define $v = -\frac{\sqrtsign{2}u}{E_{\infty}}$. Define the function $h$ by (c.f. (7.10) in \cite{auffinger2013random}) \begin{equation}
    h(v) = \left(\frac{|v - \sqrtsign{2}|}{|v + \sqrtsign{2}|}\right)^{1/4} + \left(\frac{|v + \sqrtsign{2}|}{|v - \sqrtsign{2}|}\right)^{1/4},
\end{equation}
\jstat{and the functions
\begin{align}
    q(\theta') =   \frac{1}{2} \sin^2 2\theta' + \frac{1}{4}\left(3+4\cos 4\theta'\right),
\end{align}
 \begin{align}
 j(x, s_1, \theta') = 1 + \frac{1}{\vivacom{2}}s_1\sqrtsign{x^2 - 2}h(x)^2 - s_1^2 q(\theta')|x^2 - 2|h(x)^2,
\end{align}
 \begin{align}
    T(v, s_1) = \frac{2}{\pi}\int_{0}^{\pi/2}j(-v, s_1, \theta')d\theta'.
\end{align}}
The $N-1 \times N-1$ deterministic matrix $S$ is defined subsequently around (\ref{eq:S_def}). $S$ has fixed rank $r=2$ and non-zero eigenvalues \jstat{$\{s_1, N^{-1/2}s_2\}$} where $s_j = \mathcal{O}(1)$. The specific form of $S$ is rather cumbersome and uninformative and so is relegated to Appendix \ref{ap:S_specific}, and the vector $\vec{v}$ is defined in Lemma \ref{lemma:conditional_dist}. Then we have 
\jstat{\begin{align}
     \expect C_{N}^{h}(\sqrtsign{N}u) &\sim \frac{N^{-\frac{\jstat{1}}{2}}}{\sqrtsign{2\pi H}} \jstat{e^{-\frac{\vec{v}^2}{2H}}}\jstat{T(v, s_1)} h(v) e^{N\Theta_H(u)} \frac{e^{I_1(u; E_{\infty}) - \frac{1}{2}u I_1'(u; E_{\infty})}}{\frac{H-2}{2(H-1)}u + I_1'(u; E_{\infty})}.
\end{align}}
\end{restatable}

\begin{figure}
\centering
\begin{subfigure}{0.45\textwidth}
    \centering
    \includegraphics[width=\textwidth]{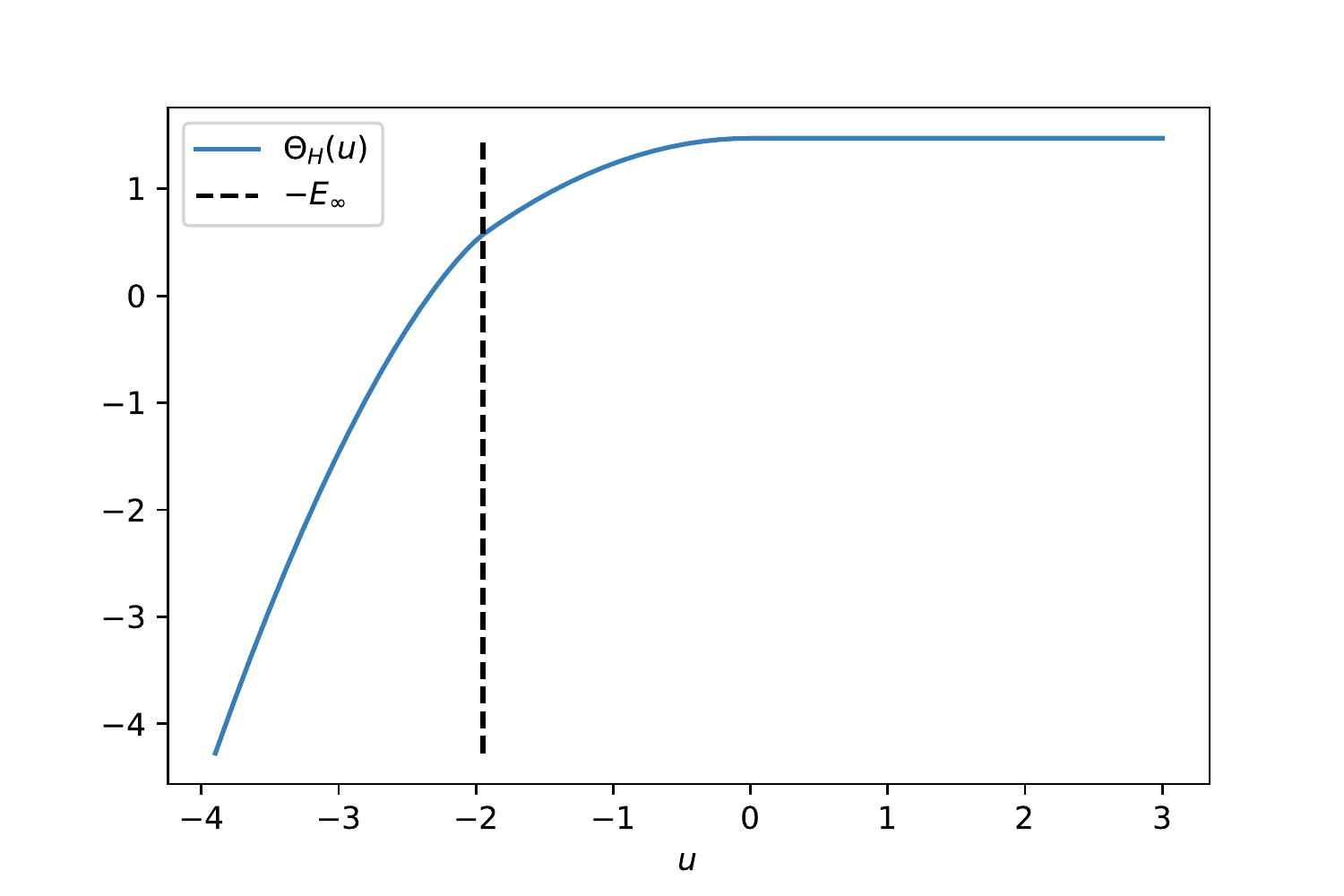}
    \caption{Plot of  $\Theta_H$.}
    \label{fig:theta_h}
\end{subfigure}
\begin{subfigure}{0.45\textwidth}
    \centering
    \includegraphics[width=\textwidth]{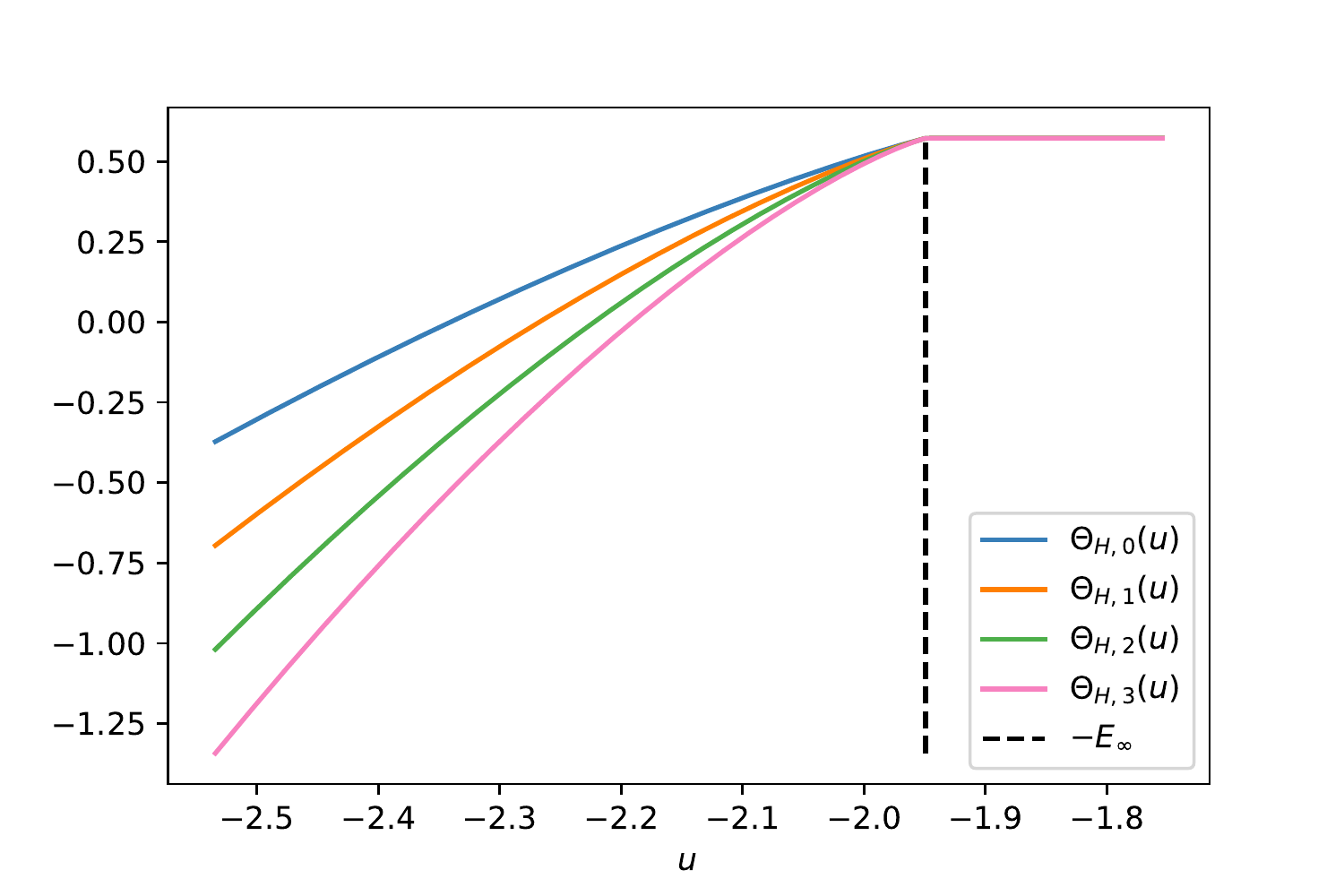}
    \caption{Plot of $\Theta_{H,k}$  $k=0,1,2,3$.}
    \label{fig:theta_hk}
\end{subfigure}
\caption{Plots of the functions $\Theta_H$ and $\Theta_{H,k}$ for $H=20$.}
\end{figure}

We include in Figures \ref{fig:theta_h} and \ref{fig:theta_hk} plots of the functions $\Theta_H$ and $\Theta_{H,k}$ for completeness, though these figures are precisely the same as those appearing in \cite{choromanska2015loss, auffinger2013random}. The critical observation from these plots is that each of the $\Theta_{H,k}$ and $\Theta_H$ are monotonically increasing and that there exist unique $E_0 > E_1 > \ldots > E_{\infty}$ such that $\Theta_{H,k}(-E_k) = 0$ and so the critical values $-E_k$ are the boundaries between regions of exponentially many and `exponentially few' critical points of each respective index.

\jstat{\begin{remark}
It is interesting to compare the expression (\ref{eq:h_def}) to the analogous expression for the model of \cite{ros2019complex}. In that work, when scaled to the unit hypersphere and scaled so that the spin glass term is composed of $\mathcal{O}(1)$ terms, the scale of the deterministic term is $\mathcal{O}(N^{1/2})$, while the corresponding scale in (\ref{eq:h_def}) is $\mathcal{O}(N^{-1/2})$. Based on this, one might well \emph{conjecture} Theorem \ref{thm:auff2.8} and Theorem \ref{thm:auff2.5}, however one would have no means by which to conjecture Theorem \ref{thm:exact_term}, and as far we can see no means to \emph{prove} Theorem \ref{thm:auff2.8} and Theorem \ref{thm:auff2.5}. As mentioned in the introduction, the single fixed distinguished direction in \cite{ros2019complex} is quite a special feature and is not present in (\ref{eq:h_def}). \end{remark}}

\section{GOE expressions for the complexity from Kac-Rice formulae}\label{sec:goe_expressions}
In this section we conduct analysis similar to that in \cite{auffinger2013random, fyodorov2007replica, fyodorov2004complexity} to obtain expressions for the the expected number of critical points of the function $h$ as defined in (\ref{eq:h_def}). We start with an elementary lemma deriving the 2-point covariance function for $h$.

\begin{lemma}\label{lemma:covariance_dot_product}
\jstat{For $\vec{w}\in S^{N-1}$, $h$ is defined as in (\ref{eq:h_def}):  \begin{equation*}
    h(\vec{w}) = \sum_{i_1, \ldots, i_H = 1}^{\Lambda} X_{i_1, \ldots, i_H} \prod_{k=1}^H w_{i_k} + \sum_{\ell=1}^H\rho_{\ell}^{(N)} \sum_{i_{\ell + 1}, \ldots, i_H=1}^{\Lambda} \prod_{k=\ell +1}^H w_{i_k}, ~~~~ X_{i_1, \ldots,, i_H}\overset{\text{i.i.d.}}{\sim}\mathcal{N}(0,1).
\end{equation*}}
For any $\vec{w}, \vec{w}'\in S^{N-1}$ the following holds \begin{equation}
    Cov(h(\vec{w}), h(\vec{w}')) = (\vec{w}\cdot\vec{w}')^H.
\end{equation}
\end{lemma}
\begin{proof}
    Let us begin by writing \begin{equation}
        h(\vec{w}) =  \sum_{i_1, \ldots, i_H = 1}^{N} X_{i_1, \ldots, i_H} \prod_{k=1}^H w_{i_k} + h^{(2)}(\vec{w}) \equiv h^{(1)}(\vec{w}) + h^{(2)}(\vec{w})
    \end{equation}
    where $h^{(2)}$ is deterministic. Then we have \begin{align}
        Cov(h(\vec{w}), h(\vec{w}')) &\equiv \expect\left[h(\vec{w})h(\vec{w}')\right] - \expect h(\vec{w})\expect h(\vec{w}') \notag \\
        &= \expect\left[ h^{(1)}(\vec{w}) h^{(1)}(\vec{w}') - h^{(1)}(\vec{w})h^{(2)}(\vec{w}') - h^{(2)}(\vec{w})h^{(1)}(\vec{w}') + h^{(2)}(\vec{w})h^{(2)}(\vec{w}')\right]\notag\\ & ~~~~- h^{(2)}(\vec{w})h^{(2)}(\vec{w}')\notag \\
        &= \expect \left[ h^{(1)}(\vec{w})h^{(1)}(\vec{w}')\right]\notag \\
        &= \sum_{i_1,\ldots i_H=1}^N \prod_{k=1}^H w_{i_k}w_{i_k}'\notag \\
        &= \prod_{k=1}^H \sum_{i_k=1}^N w_{i_k}w_{i_k}'\notag \\
        &= (\vec{w}\cdot\vec{w}')^H
    \end{align}
    where we have used $\expect h^{(1)} = 0$ in going from the first to the second and the second to the third lines.
\end{proof}

The following lemma calculates the full joint and thence conditional distribution of $h$ and its first and second derivatives. The calculations follow closely those of \cite{auffinger2013random} and the results are required for later use in a Kac-Rice formula.

\begin{lemma}\label{lemma:conditional_dist}
Pick some Cartesian coordinates on $S^{N-1}$ and let $\vec{w}$ be the north-pole of the sphere $\vec{w} = (1,0,0,\ldots)$.  Let $h_i = \partial_i h(\vec{w})$ and $h_{ij} = \partial_i\partial_j h(\vec{w})$ where $\{\partial_i\}_{i=1}^{N-1}$ are the coordinate basis around $\vec{w}$ on the sphere. Then the following results hold.
\begin{enumerate}[label=(\alph*)]
    \item For all $1 \leq i,j,k < N$, $h(\vec{w}), h_i(\vec{w}), h_{jk}(\vec{w})$ are Gaussian random variables whose  distributions are given by 
    \begin{align}
                \expect [h(\vec{w})] &= \sum_{\ell=1}^H \rho_{\ell}^{(N)} \label{eq:derivs_exp_h}\\ 
                Var [h(\vec{w})] &= 1\label{eq:derivs_var_h} \\
               \expect h_i(\vec{w}) &= \jstat{\sum_{\ell=1}^{H-1} \rho_{\ell}^{(N)} \left[(H-\ell) + (H - \ell - 1)\delta_{i1}\right]\equiv v_i\label{eq:derivs_exp_hi}}\\
                \expect [h_{ij}(\vec{w})] &= 
                \jstat{\sum_{\ell=1}^{H-2} \rho_{\ell}^{(N)}\Bigg\{\left[(H-\ell)(H-\ell-1) +1\right] \delta_{i1}\delta_{j1} + (H-\ell - 2)(\delta_{i1} + \delta_{j1}) +1
                \Bigg\} \label{eq:derivs_exp_hij} } \\
                Cov(h(\vec{w}), h_i(\vec{w})) &= 0 \label{eq:derivs_cov_hhi}\\
                Cov(h_i(\vec{w}), h_{jk}(\vec{w})) &= 0 \label{eq:derivs_cov_hihjk}\\
                Cov(h_i(\vec{w}), h_j(\vec{w})) &= H\delta_{ij} \label{eq:derivs_cov_hihj}\\
                 Cov(h(\vec{w}), h_{ij}(\vec{w})) &= -H\delta_{ij} \label{eq:derivs_cov_hhij}\\
                 Cov(h_{ij}(\vec{w}), h_{kl}(\vec{w})) &= H(H-1)(\delta_{ik}\delta{jl} + \delta_{il}\delta_{kl}) + H^2 \delta_{ij}\delta_{kl}.\label{eq:derivs_cov_hijhkl}
     \end{align}
     \vivacom{To reiterate, note that we define the vector $\vec{v}$ in (\ref{eq:derivs_exp_hi}) as \begin{align*}
         v_i =  \sum_{\ell=1}^{H-1} \rho_{\ell}^{(N)} \left[(H-\ell) + (H - \ell - 1)\delta_{i1}\right].
     \end{align*}}
      \item Make the following definitions: \begin{align}
          \xi_0 &= \sum_{\ell=1}^H\rho_{\ell}^{(N)}\\
                    \xi_1 &=  \jstat{\sum_{\ell=1}^{H-2}\rho_{\ell}^{(N)} \left[(H-\ell)(H-\ell -1) +1 \right]}\\
          \xi_2 & =\jstat{\sum_{\ell=1}^{H-2}\rho_{\ell}^{(N)}(H-\ell - 2)} \\
          \xi_3 &= \sum_{\ell=1}^{H-2}\rho_{\ell}^{(N)}
      \end{align}
       Then, conditional on $h(\vec{w}) = x$, for $x\in\mathbb{R}$, the random variables $h_{ij}(\vec{w})$ are independent Gaussians satisfying \begin{align}
           \expect[h_{ij}(\vec{w}) ~|~ h(\vec{w})=x] &=\jstat{ \xi_3 + \xi_2(\delta_{i1} + \delta_{j1}) + \xi_1\delta_{i1}\delta_{j1}} - (x-\xi_0)\delta_{ij} \label{eq:cond_exp_hij} \\
            Var[h_{ij}(\vec{w}) ~|~ h(\vec{w})=x] &= H(H-1)(1+\delta_{ij})\label{eq:cond_var_hij}.
       \end{align}
       Or, equivalently, \begin{align}\label{eq:hij_goe}
           \left(h_{ij}(\vec{w}) ~|~ h(\vec{w})=x\right) 
           \sim  \sqrtsign{2(N-1)H(H-1)} \left( M^{N-1}- \frac{1}{ \sqrtsign{2(N-1)H(H-1)} } H\left(x- \xi_0\right)I + S\right)
       \end{align}
       where $M^{N-1} \sim GOE^{N-1}$ and the matrix $S$ is given by \begin{align}\label{eq:S_def}
         \jstat{  S_{ij} = \frac{1}{ \sqrtsign{2(N-1)H(H-1)} }\left(\xi_3 + \xi_2(\delta_{i1} + \delta_{j1}) + \xi_1\delta_{i1}\delta_{j1}\right).}
       \end{align}
      \jstat{Clearly all entries of $S$ are of order $N^{-1}$, recalling the scale of $\rho_{\ell}^{(N)}$ given in (\ref{eq:rho_N_redef}). Moreover, $S$ is of rank 2 and has eigenvalues $\{s_1, N^{-1/2}s_2\}$ for real $s_i=\mathcal{O}(1)$.}
       \end{enumerate}
       \end{lemma}
    
       
\begin{proof}
\begin{enumerate}[label=(\alph*)]
    \item Becuase the $X_{i_1, \ldots, i_H}$ are centred Gaussians and $\vec{w} = (1,0,0,\ldots, 0)$, we immediately obtain (\ref{eq:derivs_exp_h}). (\ref{eq:derivs_exp_hi})-(\ref{eq:derivs_exp_hij}) can be seen to be true similarly, e.g. (\ref{eq:derivs_exp_hij}) by observing that the stochastic term is again zeroed-out by taking the expectation and the only terms that survive in the non-stochastic part are of the form \begin{equation}
        \frac{\partial^2}{\partial w_i \partial  w_j}w_iw_j w_1^{H-\ell-2} ~ (i,j\neq 1), ~~~  \frac{\partial^2}{\partial w_i \partial w_1 }w_i w_1^{H-\ell-1} ~ (i\neq 1), ~~~  \frac{\partial^2}{\partial w_1^2}  w_1^{H-\ell}.  
    \end{equation}
    
    The remaining results (\ref{eq:derivs_var_h}), (\ref{eq:derivs_cov_hhi})-(\ref{eq:derivs_cov_hijhkl}) all match those in Lemma 3.2 of \cite{auffinger2013random} and follow similarly from Lemma \ref{lemma:covariance_dot_product} and the following (\cite{adler2009random}):\begin{equation}
        Cov\left(\frac{\partial^k\bar{h}(x)}{\partial x_{i_1}\ldots \partial x_{i_k}}, \frac{\partial^l\bar{h}(y)}{\partial y_{j_1}\ldots \partial y_{j_l}}\right) = \frac{\partial^{k+l}Cov(\bar{h}(x),\bar{h}(y))}{\partial x_{i_1}\ldots \partial x_{i_k}\partial y_{j_1}\ldots \partial y_{j_l}}
    \end{equation}
    where $\bar{h}\defeq h\circ \Phi^{-1}$ and $\Phi$ is a coordinate chart around $\vec{w}$.
    
    \item
        (\ref{eq:cond_exp_hij}), (\ref{eq:cond_var_hij}) and the conditional independence result follow from (\ref{eq:derivs_exp_h}), (\ref{eq:derivs_var_h}), (\ref{eq:derivs_exp_hij}), (\ref{eq:derivs_cov_hijhkl}) and the standard result for the conditional distribution of one Gaussian under another (see e.g. \cite{anderson1962introduction} Section 2.5), just as in the proof of Lemma 3.2 in \cite{auffinger2013random}.
        
        To show (\ref{eq:hij_goe}), recall that a $GOE^N$ matrix is a real symmetric random matrix $M$ and whose entries are independent centred Gaussians with with \begin{equation}
            \expect M_{ij}^2 = \frac{1+\delta_{ij}}{2N}.
        \end{equation}
        \jstat{Finally we have to determine the eigenvalues of $S$. With $a = \xi_1 + 2\xi_2 + \xi_3, b=\xi_2 + \xi_3$ and $c=\xi_3$, $S$ has entries \begin{align}
            S = \frac{1}{\sqrtsign{2(N-1)H(H-1)}}\left(\begin{array}{ccccc}
            a & b & b & \ldots & b \\
            b & c & c & \ldots & c \\
            b & c & c & \ldots & c \\
            \vdots & \vdots & \vdots & \vdots & \vdots \\
            b & c & c & \ldots & c \\
            \end{array}\right),
        \end{align}
and so has non-null eigenvectors $(1, u, u, \ldots, u)^T$ with eigenvalues $\left(2(N-1)H(H-1)\right)^{-1/2}\lambda$, where (after some simple manipulation) \begin{align}
    \lambda^2 - (a - c(N-1))\lambda + ca(N-1) - b^2(N-1) = 0, ~~~~~ u = \frac{\lambda - a}{(N-1)b}.
\end{align}
Recalling the scale of $\rho_{\ell}^{(N)} = \mathcal{O}(N^{-\ell/2})$ in (\ref{eq:rho_N_redef}) and the definitions $\xi_1,\xi_2, \xi_3$, we see that $a, b, c=\mathcal{O}(N^{-1/2})$ and so one easily obtains two solutions for $\lambda$, one of order $N^{1/2}$ and another of order $N^{-1/2}$, hence $S$ has two non-zero eigenvalues of order $1$ and $N^{-1/2}$.}
\end{enumerate}
\end{proof}

Our next lemma establishes for use in this context a Kac-Rice fomula that will provide the first step in the computation of $C^h_{N}$ and $C^h_{N, \mathcal{K}}$.

\begin{lemma}\label{lemma:kac_rice}
    Let $\hat{F}$ be a real-valued centred Gaussian field on $S^{N-1}$ that is almost surely (a.s.) $C^2$, $\tilde{F}$ be some non-random, real-valued $C^2$ function on $S^{N-1}$ and let $F \defeq\hat{F} + \tilde{F}$. Let  $\mathcal{A} = \{U_{\alpha}, \Phi_{\alpha}\}_{\alpha\in I}$ be a finite atlas on $S^{N-1}$. Let $h^{\alpha} = h\circ \Phi_{\alpha}^{-1}$, and let $h^{\alpha}_i, h^{\alpha}_{ij}$ denote derivatives of $h$ in the coordinate basis of the chart $(U_{\alpha}, \Phi_{\alpha}).$ Assume that the joint distribution $(F^{\alpha}_i(\vec{x}), F^{\alpha}_{ij}(\vec{x}))$ is non-degenerate for all $\alpha$ and for all $\vec{x}\in S^{N-1}$ and that there exist constants $K_{\alpha}, \beta >0$ such that \begin{equation}\label{eq:var_log_assumption}
        \max_{i,j}\left|Var(\hat{F}_{ij}^{\alpha}(\vec{x})) + Var(\hat{F}_{ij}^{\alpha}(\vec{y})) - 2Cov(\hat{F}_{ij}^{\alpha}(\vec{x}), \hat{F}_{ij}^{\alpha}(\vec{y}))\right| \leq K_{\alpha}\left|\log|x-y|\right|^{-1-\beta}
    \end{equation}Then the following holds \begin{equation}\label{eq:kac_rice}
      \CNk^F(B) = \int_{S^{N-1}} p_{\vec{x}}(0) \mathcal{S}_{N-1}(d\vec{x}) \expect\left[|\det\grad^2 F(\vec{x})|\indic\left\{F(\vec{x})\in B,~ i(\grad^2F(\vec{x})) = k\right\} ~|~ \grad F(\vec{x})=0\right]     \end{equation}

      where $p_{\vec{x}}$ is the density of $\grad F$ at $\vec{x}$ and $\mathcal{S}_{N-1}$ is the usual surface measure on $S^{N-1}$. Similarly,
      \begin{equation}\label{eq:kac_rice_no_k}
      \CN^F(B) = \int_{S^{N-1}} p_{\vec{x}}(0) \mathcal{S}_{N-1}(d\vec{x}) \expect\left[|\det\grad^2 F(\vec{x})|\indic\left\{F(\vec{x})\in B\right\} ~|~ \grad F(\vec{x})=0\right]     \end{equation}
\end{lemma}

The proof of Lemma \ref{lemma:kac_rice} shall rely heavily on the Kac-Rice result Theorem \ref{thm:adler_kac_rice}.

\begin{proof}[Proof of Lemma \ref{lemma:kac_rice}]
Following the proofs of Theorem 12.4.1 in \cite{adler2009random} and Lemma 3.1 in \cite{auffinger2013random}, we will apply Theorem \ref{thm:adler_kac_rice} to the choices \begin{align}
    \phi &\defeq \grad F\notag\\
    \psi &\defeq (F, \grad_{i}\grad_{j}F)\notag\\
    A &\defeq B \times A_k \equiv B \times \{H\in\text{Sym}_{N-1\times N-1} ~|~ i(H)=k\} \subset \mathbb{R}\times \text{Sym}_{N-1\times N-1},\notag\\
    \vec{u} &= 0
\end{align}
Then, if the conditions of Theorem \ref{thm:adler_kac_rice} hold for these choices, we immediately obtain the result. It remains therefore to check the conditions of Theorem \ref{thm:adler_kac_rice}. Firstly, $A$ is indeed an open subset of of $\mathbb{R}\times \text{Sym}_{N-1\times N-1}$ (in turn, isomorphic to some $\mathbb{R}^K$) as can be easily deduced from the continuity of a matrix's eigenvalues in its entries. Condition (a) follows from the assumption of $\hat{F}$ being a.s. $C^2$ and $\tilde{F}$ being $C^2$. Conditions (b)-(f) all follow immediately from the Gaussianity of $\hat{F}$. To establish condition (g), we define $\hat{\omega}(\eta)$ and $\tilde{\omega}(\eta)$ in the obvious way and note that $\tilde{\omega}$ is non-random. Then, because $\tilde{F}$ is continuous, given $\epsilon > 0$ there exists some $\eta_0 >0$ such that for all $\eta < \eta_0$, $\tilde{\omega}(\eta) \leq \epsilon$. Let $\tilde{\omega}_0 \defeq \tilde{\omega}(\eta_0)$ and choose some $\eta_1$ such that for all $\eta < \eta_1$, $\tilde{\omega}(\eta) < \tilde{\omega}_0$. We have $\omega(\eta) \leq \hat{\omega}(\eta) + \tilde{\omega}(\eta)$ and so for $\eta < \eta_1$ \begin{align}
    \mathbb{P}(\omega(\eta) > \epsilon) & \leq \mathbb{P}(\hat{\omega}(\eta) + \tilde{\omega}(\eta) > \epsilon) \notag\\
    &=\mathbb{P}(\hat{\omega}(\eta) > \epsilon - \tilde{\omega}(\eta)) \notag\\
    &\leq \mathbb{P}(\hat{\omega}(\eta) > \epsilon - \tilde{\omega}_0)\label{eq:condition_g_ineq}
\end{align}
and we note that $\epsilon - \tilde{\omega}_0 \geq 0$ by construction. $\hat{\omega}$ is the modulus of continuity for a centred Gaussian field and so the condition (g) follows from (\ref{eq:condition_g_ineq}) and the assumption (\ref{eq:var_log_assumption}) by the Borell-TIS inequality \cite{adler2009random}, just as in the proof of Corollary 11.2.2 in \cite{adler2009random}. (\ref{eq:kac_rice_no_k}) is obtained in precisely the same way but simply dropping the $i(H) = k$ condition.
\end{proof}

\section{Asymptotic evaluation of complexity}\label{sec:asymptotic_evaluation}
In this section we conduct an asymptotic analysis of the GOE expressions for the complexity found in the preceding section. We first consider the case of counting critical points without any condition of the signature of the Hessian, which turns out to be easier. We then introduce the exact signature condition on the Hessian and proceed by presenting the necessary modifications to certain parts of our arguments.

\subsection{Complexity results with no Hessian signature prescription}
\jstat{We need to establish a central lemma, which is a key step towards a generalisation of the results presented in \cite{auffinger2013random} but established by entirely different means, following the supersymmetric calculations of \cite{nock}. Before this main lemma, we require a generalisation of a result from \cite{fyodorov2002characteristic}, whose proof is given at the end of the chapter (Section \ref{ap:low_rank_fyod}).
\begin{restatable}{lemma}{fyodgeneral}
\label{lem:fyod_general}
Given $m$ vectors in $\mathbb{R}^N$ $\vec{x}_1, \ldots, \vec{x}_m$, denote by $Q(\vec{x}_1, \ldots, \vec{x}_m)$ the $m\times m$ matrix whose entries are given by $Q_{ij} = \vec{x}_i^T\vec{x}_j$. Let $F$ be any function of an $m\times m$ matrix such that the integral \begin{equation}
     \int_{\mathbb{R}^N}\ldots\int_{\mathbb{R}^N}d\vec{x}_1\ldots d\vec{x}_m |F(Q)|
\end{equation}
exists, and let $S$ be a real symmetric $N\times N$ matrix of fixed rank $r$ and with non-zero eigenvalues $\{N^{\alpha}s_i\}_{i=1}^r$ for some $\alpha < 1/2$. Define the integral \begin{equation}
     \mathcal{J}_{N, m}(F; S) \defeq \int_{\mathbb{R}^N}\ldots\int_{\mathbb{R}^N}d\vec{x}_1\ldots d\vec{x}_m F(Q) e^{-iN\sum_{i=1}^N \vec{x}_i^T S\vec{x}_i}.
\end{equation} Then as $N\rightarrow\infty$ we have \begin{equation}
    \mathcal{J}_{N,m}(F; S) =\left( 1 + o(1))\right) \frac{\pi^{\frac{m}{2}\left(N - \frac{m-1}{2}\right)}}{\prod_{k=0}^{m-1}\Gamma\left(\frac{N-k}{2}\right)}\int_{\text{Sym}_{\geq 0}(m)}d\hat{Q} \left(\det \hat{Q}\right)^{\frac{N-m-1}{2}}F(\hat{Q})\prod_{i=1}^N\prod_{j=1}^r\left( 1+ \vivacom{2}iN^{\alpha}\hat{Q}_{ii}s_j\right)^{\vivacom{-1/2}}.
\end{equation}
\end{restatable}}

Now we state and prove the main lemma.
\begin{lemma}\label{lemma:nock_deformed}
    Let $S$ be a rank $r$ $N\times N$ symmetric matrix with non-zero eigenvalues $\{s_j\}_{j=1}^r$, where $r=\mathcal{O}(1)$ and $s_j = \mathcal{O}(1)$, and suppose $S$ has all entries of order $\mathcal{O}(N^{-1})$ in a fixed basis. Let $x<0$ and let $M$ denote an $N\times N$ GOE matrix with respect to whose law expectations are understood to be taken. Then
    \begin{align}
                  \expectGOE |\det(M - xI + S)| = K_N\lim_{\epsilon\searrow 0} e^{2N(x^2 - \epsilon^2)}\left(1 + o(1)\right)&\iiint_0^{\pi/2} d\theta d\theta'd\hat{\theta}\iint_0^{\infty}dp_1dp_2 \iint_{\Gamma} dr_1dr_2\notag\\
                  &J_1(p_1, p_2, \theta'; S, N)J_2(r_1,r_2, p_1, p_2)\cos^22\theta \sin2\theta \sin2\hat{\theta}\notag\\
& \exp\Bigg\{-N\Bigg(2\psi^{(+)}_L(r_1; x; \epsilon\cos2\theta\cos2\hat{\theta}) \notag\\ &~~~~~~~~~~+2\psi^{(+)}_U(r_2; x; \epsilon\cos2\theta\cos2\hat{\theta})\notag\\& +\psi^{(-)}_L(p_1; x; \epsilon\cos2\theta')+\psi^{(-)}_U(p_2; x; \epsilon\cos2\theta')\Bigg)\Bigg\}\end{align}
    where \begin{align}
       J_1(p_1, p_2, \theta'; \{s_j\}_{j=1}^r, N) &=\prod_{j=1}^r\left(1 + \vivacom{2}iN^{1/2}s_j(p_1 + p_2) - Ns_j^2\left[\sin^22\theta' (p_1^2 + p_2^2) + \left(3 + 4\cos4\theta'\right)p_1p_2 \right]\right)^{-1/2},\\
        J_2(r_1, r_2, p_1, p_2; \epsilon) &= (r_1 + p_1)(r_2 + p_1)(r_1 + p_2)(r_2 + p_2)|r_1 - r_4|^4 |p_1-p_2| (r_1r_2)^{-2} (p_1p_2)^{-3/2}
    \end{align} 
    and \begin{equation}
        K_N =   \frac{N^{N+3}(-i)^N }{\Gamma\left(\frac{N}{2}\right)\Gamma\left(\frac{N-1}{2}\right) \pi^{3/2}}
    \end{equation}
and the functions $\psi^{\pm}_L, \psi^{(\pm)}_U$ are given by \begin{align}
 \psi^{(\pm)}_L(z; x,\epsilon) &= \frac{1}{2}z^2 \pm i(x+i\epsilon)z - \frac{1}{2}\log z,\\
              \psi^{(\pm)}_U(z;x,\epsilon) &= \frac{1}{2}z^2 \pm i(x-i\epsilon)z - \frac{1}{2}\log z,
\end{align}
and $\Gamma$ is a contour bounded away from zero in $\mathbb{C}$, e.g. that shown in Figure \ref{fig:r1r2_contour}.
\end{lemma}
\begin{proof}
We begin with the useful expression for real symmetric matrices $A$ \cite{fyodorov2005counting, fyodorov2004complexity} \begin{equation}
    |\det A| = \lim_{\epsilon\rightarrow 0} \frac{\det A \det A}{\sqrtsign{\det (A - i\epsilon)}\sqrtsign{\det (A +i\epsilon)}}
\end{equation} where the limit is taken over real $\epsilon$, and WLOG $\epsilon > 0$. We're free to deform the matrices in the numerator for the sake of symmetry in the ensuing calculations, so \begin{align}\label{eq:det_ratio_epsilon}
     |\det A| = \lim_{\epsilon\searrow 0} \frac{\det (A - i\epsilon) \det (A + i\epsilon)}{\sqrtsign{\det (A - i\epsilon)}\sqrtsign{\det (A +i\epsilon)}}.
\end{align}
For convenience of notation we put \begin{equation}
    \Delta_{\epsilon}(M; x, S) = \frac{\det (M - xI + S - i\epsilon) \det (M - xI + S + i\epsilon)}{\sqrtsign{\det (M - xI + S - i\epsilon)}\sqrtsign{\det (M - xI + S +i\epsilon)}}.
\end{equation}

Then we express the determinants and half-integer powers of determinants as Gaussian integrals over anti-commuting and commuting variables respectively as in \cite{nock} and \cite{fyodorov2015random}: \begin{align}
     &\Delta_{\epsilon}(M; x, S) \notag\\
     = &K^{(1)}_N\int d\vec{x}_1 d\vec{x}_2 d\zeta_1 d\zeta_1^{\dagger} d\zeta_2 d\zeta_2^{\dagger} \exp\left\{-i\vec{x}_1^T(M-(x + i\epsilon)I+S)\vec{x}_1 - i\vec{x}_2^T(M-(x-i\epsilon)I + S)\vec{x}_2\right\}\notag \\
      + &\exp\left\{ i \zeta_1^{\dagger}(M-(x+i\epsilon) I+S)\zeta_1
      + i \zeta_2^{\dagger}(M-(x - i\epsilon)I+S)\zeta_2\right\}\label{eq:nock_initial}\end{align}
     where $K^{(1)}_N = (-i)^N \pi^{-N}$, which follows from standard facts about commuting Gaussian integrals and Berezin integration. 
The remainder of the calculation is very similar to that presented in \cite{nock,fyodorov2015random} but we present it in full to keep track of the slight differences. Let \begin{equation}
    A = \vec{x}_1\vec{x}_1^T + \vec{x}_2\vec{x}_2^T + \zeta_1\zeta_1^{\dagger} + \zeta_2\zeta_2^{\dagger}
\end{equation}
and note that, by the cyclicity of the trace, \begin{align}
    \vec{x}_j^T(M-(x \pm i\epsilon)I+S)\vec{x}_j &= \Tr\left((M-(x\pm i\epsilon)I+S)\vec{x}_j\vec{x}_j^T\right)\\
    \zeta_j^{\dagger}(M-(x \pm i \epsilon)I+S)\zeta_j &= -\Tr\left((M-(x\pm i \epsilon)I+S)\zeta_j\zeta_j^{\dagger}\right)
\end{align}
and so we can rewrite (\ref{eq:nock_initial}) as \begin{align}
      \Delta_{\epsilon}(M; x, S)  = K^{(1)}_N\int d\vec{x}_1 d\vec{x}_2 d\zeta_1 d\zeta_1^{\dagger} d\zeta_2 d\zeta_2^{\dagger} &\exp\left\{-i\Tr MA - i\Tr SA + i(x + i\epsilon)\vec{x}_1^T\vec{x}_1 + i(x - i\epsilon)\vec{x}_2^T\vec{x}_2  \right\} \notag\\ &\exp\left\{-i(x + i\epsilon)\zeta_1^{\dagger}\zeta_1 -i (x - i\epsilon)\zeta_2^{\dagger}\zeta_2 \right\}. \label{eq:nock2}
\end{align}
We then define the Bosonic and Fermionic matrices \begin{align}
    Q_B = \left(\begin{array}{cc} \vec{x}_1^T\vec{x}_1 & \vec{x}_1^T\vec{x}_2 \\ \vec{x}_2^T\vec{x}_1 & \vec{x}_2^T\vec{x}_2\end{array}\right), ~~  Q_F =  \left(\begin{array}{cc} \zeta_1^{\dagger}\zeta_1 & \zeta_1^{\dagger}\zeta_2 \\ \zeta_2^{\dagger}\zeta_1 & \zeta_2^{\dagger}\zeta_2\end{array}\right)
\end{align}
and also $B = \vec{x}_1\vec{x}_1^T + \vec{x}_2\vec{x}_2^T$.
Note that (\ref{eq:det_ratio_epsilon}) is true for all real symmetric matrices $A$ and so for \emph{all} real symmetric $M,S$ and real values $x$ we have \begin{equation}
    \lim_{\epsilon\searrow 0} \Delta_{\epsilon}(M; x, S) = |\det\left(M - xI + S\right)|
\end{equation}
and so with respect to the GOE law for $M$ we certainly have \begin{equation}
  \Delta_{\epsilon}(M; x, S) \overset{\text{a.s.}}{\rightarrow} |\det\left(M - xI + S\right)| ~~~ \text{as } \epsilon\searrow 0
\end{equation}
thus meaning that the $\epsilon\searrow 0$ limit can be exchanged with a GOE expectation over $M$. We therefore proceed with fixed $\epsilon>0$ to compute the GOE expectation of $\Delta_{\epsilon}.$

We have the standard Gaussian Fourier transform result for matrices: \begin{align}\label{eq:goe_fourier}
    \expectGOE e^{-i\Tr MA} = \exp\left\{-\frac{1}{8N}\Tr(A + A^T)^2\right\}
\end{align} and from \cite{nock}\footnote{Note that (4.100) in \cite{nock} contains a trivial factor of 4 error that has non-trivial consequences in our calculations.} \begin{equation}
    \Tr(A+A^T)^2 = 4\Tr Q_B^2 - 2\Tr Q_F^2 + 4\zeta_1^T\zeta_2\zeta_2^{\dagger}\zeta_1^* - 8\zeta_1^{\dagger}B\zeta_1 - 8 \zeta_2^{\dagger}B\zeta_2
\end{equation} so we can take the GOE average in (\ref{eq:nock2}) and obtain \begin{align}
    \expectGOE  \Delta_{\epsilon}(M; x, S) = K^{(1)}_N\int &d\vec{x}_1 d\vec{x}_2 d\zeta_1 d\zeta_1^{\dagger} d\zeta_2 d\zeta_2^{\dagger} 
    \exp\left\{ -\frac{1}{2N}\Tr Q_B^2 - i\Tr SB + ix\Tr Q_B + \epsilon \Tr Q_B\sigma \right\}\notag\\
    &\exp\left\{\frac{1}{4N}\Tr Q_F^2 - \frac{1}{2N}\zeta_1^T\zeta_2\zeta_2^{\dagger}\zeta_1^* + \sum_{j=1}^2 \zeta_j^{\dagger}\left(\frac{B}{N} + iS - i(x + i(-1)^{j-1} \epsilon)\right)\zeta_j\right\}.\label{eq:nock3}
\end{align}
where we have defined $$\sigma = \left(\begin{array}{cc} -1 & 0 \\ 0 & 1 \end{array}\right).$$

We can then use the transformation \begin{equation}
    \exp\left\{\frac{1}{4N}\Tr Q_F^2\right\} = \frac{N^2}{\pi Vol(U(2))}\int d\hat{Q}_F \exp\left\{-N\Tr \hat{Q}_F^2 + \Tr Q_F\hat{Q}_F\right\}
\end{equation}
to obtain \begin{align}
        \expectGOE \Delta_{\epsilon}(M; x, S) = K^{(2)}_N&\int d\vec{x}_1 d\vec{x}_2 d\zeta_1 d\zeta_1^{\dagger} d\zeta_2 d\zeta_2^{\dagger} d\hat{Q}_F
    \exp\left\{ -\frac{1}{2N}\Tr Q_B^2 - i\Tr SB + ix\Tr Q_B + \epsilon \Tr Q_B\sigma\right\}\notag\\
    &\exp\left\{-N\Tr \hat{Q}_F^2 + \Tr \hat{Q}_FQ_F -  \frac{1}{2N}\zeta_1^T\zeta_2\zeta_2^{\dagger}\zeta_1^* + \sum_{j=1}^2 \zeta_j^{\dagger}\left(\frac{B}{N} + iS - i(x + i(-1)^{j-1} \epsilon\right)\zeta_j\right\}
\label{eq:nock4}\end{align}
where $K^{(2)}_N = K^{(1)}_N \frac{N^2}{\pi Vol(U(2))}.$
The Fermionic cross-term in (\ref{eq:nock4}) can be dealt with using (see \cite{nock} (4.104)) \begin{equation}
    \exp\left(-\frac{1}{2N}\zeta_1^T\zeta_2\zeta_2^{\dagger}\zeta_1^{*}\right) = \frac{2N}{\pi} \int d^2u \exp\left(-2N\bar{u}u - i\left(u\zeta_1^{\dagger}\zeta_2^{*} + \bar{u}\zeta_2^{\dagger}\zeta_1\right)\right)
\end{equation} where $d^2u = d\Re{u} ~ d\Im{u}$, and so we obtain \begin{align}
     \expectGOE  \Delta_{\epsilon}(M; x, S) = K^{(3)}_N\int &d\vec{x}_1 d\vec{x}_2 d\zeta_1 d\zeta_1^{\dagger} d\zeta_2 d\zeta_2^{\dagger} d\hat{Q}_F d^2u
    \exp\left\{ -\frac{1}{2N}\Tr Q_B^2 - i\Tr SB + ix\Tr Q_B + \epsilon\Tr Q_B \sigma\right\}\notag\\
    &\exp\left\{-N\Tr\hat{Q}_F^2 - 2N u \bar{u}\right\}\notag\\
    &\exp\left\{ \Tr \hat{Q}_FQ_F - i(u\zeta_1^{\dagger}\zeta_2^{*} + \bar{u}\zeta_2^T\zeta_1) + \sum_{j=1}^2 \zeta_j^{\dagger}\left(\frac{B}{N} + iS - i(x + i(-1)^{j-1} \epsilon\right)\zeta_j\right\}\label{eq:nock5}
\end{align}
where $K_N^{(3)} = K_N^{(2)}\frac{2N}{\pi}$.
To simplify the Fermionic component of (\ref{eq:nock5}) and make apparent its form, we introduce $\zeta^T=(\zeta_1^{\dagger}, \zeta_1^T, \zeta_2^{\dagger}, \zeta_2^T)$ and then (\ref{eq:nock5}) reads \begin{align}
        \expectGOE  \Delta_{\epsilon}(M; x, S) = K^{(3)}_N\int &d\vec{x}_1 d\vec{x}_2 d\zeta d\hat{Q}_F d^2u
    \exp\left\{ -\frac{1}{2N}\Tr Q_B^2 - i\Tr SB + ix\Tr Q_B + \epsilon\Tr Q_B \sigma\right\}\notag\\
    &\exp\left\{-N\Tr\hat{Q}_F^2 - 2N u \bar{u}\right\}\notag\\
    &\exp\left\{\frac{1}{2}\zeta^T\mathcal{M}\zeta\right\}\notag\\
     = K^{(3)}_N\int &d\vec{x}_1 d\vec{x}_2 d\hat{Q}_F d^2u
    \exp\left\{ -\frac{1}{2N}\Tr Q_B^2 - i\Tr SB + ix\Tr Q_B + \epsilon\Tr Q_B\sigma\right\}\notag\\
    &\exp\left\{-N\Tr\hat{Q}_F^2 - 2N u \bar{u}\right\}\notag\\
    &\sqrtsign{\det\mathcal{M}}\label{eq:nock6} 
\end{align}
where the matrix $\mathcal{M}$ is given by \begin{equation}
    \mathcal{M} = \left(\begin{array}{cccc}
        0 & A_1 & - iu & q_{12}^* \\
        -A_1 & 0 & - q_{12} & i\bar{u} \\
        iu & q_{12} & 0 & A_2 \\
        -q_{12}^* & -i\bar{u} & -A_2 & 0
    \end{array}\right)\end{equation}
    and, by analogy with (4.107) in \cite{nock}, \begin{equation}
        A_j = q_{jj} - i(x + i(-1)^{j-1}\epsilon)+ \frac{1}{N}B + iS,
    \end{equation}
    where $q_{ij}$ are the entries of $\hat{Q}_F$. To evaluate $\det\mathcal{M}$, we make repeated applications of the well-known result for block $2\times 2$ matrices consisting of  $N\times N$ blocks: $$\det\left(\begin{array}{cc} A & B \\ C & D \end{array}\right) = \det(A - BD^{-1}C)\det(D).$$ This process quickly results in \begin{align}
        \sqrtsign{\det\mathcal{M}} &= \det(A_1A_2 - (u\bar{u} + q_{12}\bar{q}_{12}))\notag\\
        & =  \det\left(\left[\det(\hat{Q}_F - ix- \epsilon\sigma) - \bar{u}u\right]I + \Tr(\hat{Q}_F - ix- \epsilon\sigma)\left(\frac{1}{N}B + iS\right) + \left(\frac{1}{N}B + iS\right)^2\right)\notag\\
        & = \det\left( G_1 + N^{-1}B + iS\right)\det\left( G_2 + N^{-1}B + iS\right)\label{eq:detM_G_factor}
        \end{align}
    where we have chosen $G_1$, $G_2$ to be solutions to \begin{align}
        G_1G_2 &= \det(\hat{Q}_F - ix- \epsilon\sigma) - \bar{u}u\label{eq:G_sim1}\\
        G_1 + G_2 &= \Tr(\hat{Q}_F - ix- \epsilon\sigma)\label{eq:G_sim2}.
    \end{align}
    Recalling the $B$ has rank $2$ we let $O_{B}$ be the $N\times 2$ matrix of the non-null eigenvectors of $B$ and $\lambda^{(B)}_{1,2}$ be its non-null eigenvalues and use the determinantal identity found in equation (3) of \cite{benaych2012large} to write\footnote{Note that we here include explicitly the identity matrix symbols to make plain the dimension of the determinants.}
    \begin{align}
        \det\left(G_j I_N + N^{-1}B + iS\right) &= \det\left(G_jI_N + iS\right)\det\left(I_2 + N^{-1}O_{B}^T\left(G_jI_N + iS\right)^{-1}O_{B}\text{diag}\left(\lambda^{(B)}_1, \lambda^{(B)}_2\right)\right).\label{eq:det_M_expansion_pre_asymp}
    \end{align}

We would now like to apply the integral formula found in Appendix D of \cite{fyodorov2002characteristic} to re-write the integrals over the $N$-dimensional vectors $\vec{x}_1, \vec{x}_2$ as a single integral over a $2\times 2$ symmetric matrix $Q_B$. However, the integrand does not only depend on $\vec{x}_1, \vec{x}_2$ through $Q_B \equiv \left(\begin{array}{cc} \vec{x}_1^T\vec{x}_1 & \vec{x}_1^T\vec{x}_2 \\ \vec{x}_2^T\vec{x}_1 & \vec{x}_2^T\vec{x}_2\end{array}\right)$ thanks to the dependence on the eigenvectors of $B$ in (\ref{eq:det_M_expansion_pre_asymp}) and also in the term $\Tr SB$ in (\ref{eq:nock6}). Before addressing this problem, we will continue to manipulate the $\hat{Q}_F$ and $u$ integrals along the lines of \cite{nock}.\\

First make the change of variables $\hat{Q}_F \leftarrow \hat{Q}_F + ix + \epsilon\sigma$ and $\vec{x}_j \leftarrow \sqrtsign{N}\vec{x}_j$ in (\ref{eq:nock6}) using (\ref{eq:detM_G_factor}) to obtain \begin{align}
        \expectGOE  \Delta_{\epsilon}(M; x, S) = K^{(4)}_N\int &d\vec{x}_1 d\vec{x}_2 d\hat{Q}_F d^2u
    \exp\left\{ -\frac{N}{2}\Tr Q_B^2 - iN\Tr SB + ixN\Tr Q_B + \epsilon N\Tr Q_B\sigma\right\}\notag\\
    &\exp\left\{-N\Tr\hat{Q}_F^2 - 2N \Tr (ix + \epsilon\sigma)\hat{Q}_F - N\Tr(ix + \epsilon\sigma)^2 - 2N u \bar{u}\right\}\notag\\&\prod_{j=1}^2\det\left(G_j + B + iS\right)\label{eq:nock7_prime} 
\end{align}
where $K^{(4)}_N = N^{N}K_N^{(3)}$ and now the terms $G_1, G_2$ are given by the modified versions of (\ref{eq:G_sim1})-(\ref{eq:G_sim2}):\begin{align}
        G_1G_2 &= \det\hat{Q}_F  - \bar{u}u\label{eq:G_sim1_prime}\\
        G_1 + G_2 &= \Tr\hat{Q}_F \label{eq:G_sim2_prime}.
    \end{align}
    
We now diagonalise the Hermitian matrix $\hat{Q}_F = \hat{U}\text{diag}(q_1, q_2)\hat{U}^{\dagger}$ in (\ref{eq:nock7_prime}), but the term $\Tr \sigma\hat{Q}_F$ is not unitarily invariant, so we follow \cite{nock} and introduce an explicit parametrization\footnote{\cite{nock} uses an incorrect parametrization with only two angles. The calculations are are invariant in the extra angles $\alpha,\beta$ and so this detail only matters if one is tracking the multiplicative constants, as we do here.} of the unitary matrix $\hat{U}$ $$\hat{U} =e^{i\hat{\phi}/2} \left(\begin{array}{cc} e^{i\hat{\alpha}/2} & 0 \\ 0 & e^{-i\hat{\alpha}/2}\end{array}\right)\left(\begin{array}{cc} \cos\hat{\theta} & \sin\hat{\theta} \\ -\sin\hat{\theta} & \cos\hat{\theta}\end{array}\right)\left(\begin{array}{cc} e^{i\hat{\beta}/2} & 0 \\ 0 & e^{-i\hat{\beta}/2}\end{array}\right)$$
    where $\hat{\phi},\hat{\alpha}, \hat{\beta}\in [0,2\pi)$, $\hat{\theta}\in [0,\pi/2)$ and elementary calculations give the Jacobian factor $|q_1 - q_2|^2 \sin(2\hat{\theta})$. Further brief elementary calculations give
    \begin{equation}
        \Tr \hat{Q}_F\sigma  = (q_2 - q_1)\cos(2\hat{\theta}).
    \end{equation}
    and so, integrating out $\hat{\phi}, \hat{\alpha}, \hat{\beta}$,
    \begin{align}
        \expectGOE  \Delta_{\epsilon}(M; x, S) = K^{(5)}_N e^{2N(x^2 - \epsilon^2)}\int &d\vec{x}_1 d\vec{x}_2  \iint_{-\infty}^{\infty} dq_1dq_2 \int d^2u\int_{0}^{\pi/2}d\theta \sin2\hat{\theta}\notag\\
    &\exp\left\{ -\frac{N}{2}\Tr Q_B^2 - iN\Tr SB + ixN\Tr Q_B + \epsilon N\Tr Q_B\sigma\right\}\notag\\
    &\exp\left\{-N(q_1^2 + q_2^2)  - 2Nix (q_1 + q_2) - 2N\epsilon(q_2 - q_1)\cos2\hat{\theta} - 2N u \bar{u}\right\}\notag\\
    &\prod_{j=1}^2\det\left(G_j + B + iS\right)|q_1 - q_2|^2\label{eq:nock8_prime} 
\end{align}
with $K^{(5)} = (2\pi)^3 K^{(4)}_N $ and now 
\begin{align}
        G_1G_2 &= q_1q_2 - \bar{u}u\label{eq:G_sim1_prime2}\\
        G_1 + G_2 &= q_1 + q_2 \label{eq:G_sim2_prime2}.
    \end{align}
We form an Hermitian matrix \begin{equation}
        R = \left(\begin{array}{cc} q_1 & \bar{u}\\ u & q_2\end{array}\right)
    \end{equation}
    and so (\ref{eq:nock8_prime}) is rewritten as \begin{align}
              \expectGOE  \Delta_{\epsilon}(M; x, S) = K^{(6)}_Ne^{2N(x^2 - \epsilon^2)}\int &d\vec{x}_1 d\vec{x}_2  \int dR|R_{11} - R_{22}|^2\int_{0}^{\pi/2}d\theta \sin2\hat{\theta}\notag\\
    &\exp\left\{ -\frac{N}{2}\Tr Q_B^2 - iN\Tr SB + ixN\Tr Q_B + \epsilon N\Tr Q_B\sigma\right\}\notag\\
    &\exp\left\{-N\Tr R^2 -2Nix \Tr{R} -2\epsilon N(R_{22} - R_{11})\cos2\hat{\theta} \right\}\notag\\
    &\prod_{j=1}^2\det\left(G_j + B + iS\right)\label{eq:nock9_prime}   
    \end{align}
    with $K_N^{(6)} = \frac{1}{16\pi^2}K_N^{(5)}$ and 
   \begin{align}
        G_1G_2 &= \det R\label{eq:G_sim1_prime22}\\
        G_1 + G_2 &= \Tr R \label{eq:G_sim2_prime22}.
    \end{align}
    The factor of $(16\pi^2)^{-1}$ comes from the change of variables $(q_1, q_2, u, \bar{u}) \mapsto R$. Indeed, clearly $dq_1dq_2dud\bar{u}  = Z^{-1}dR$ for some constant Jacobian factor $Z$. We can most easily determine $Z$ by integrating against a test function:
    \begin{align}
   \frac{4\pi Vol(U(2))}{Z}  =     \frac{1}{Z}\int_{\text{Herm}(2)} dR e^{-\frac{1}{2}\Tr R^2} &= \iint_{-\infty}^{\infty} dq_1 dq_2 \iint_{-\infty}^{\infty} d\Re{u} ~ d\Im{u} e^{-\frac{1}{2}(q_1^2 + q_2^2 + 2u\bar{u})}= 2\pi^2\notag\\
   \implies Z &= \frac{2Vol(U(2))}{\pi} = 16\pi^2.\notag
    \end{align}
    We diagonalise $R = U\text{diag}(r_1, r_2)U^{\dagger}$, but again the integrand in (\ref{eq:nock9_prime}) is not unitarily invariant in $R$ so we repeat the previous procedure using $$U =e^{i\phi/2} \left(\begin{array}{cc} e^{i\alpha/2} & 0 \\ 0 & e^{-i\alpha/2}\end{array}\right)\left(\begin{array}{cc} \cos\theta & \sin\theta \\ -\sin\theta & \cos\theta\end{array}\right)\left(\begin{array}{cc} e^{i\beta/2} & 0 \\ 0 & e^{-i\beta/2}\end{array}\right).$$
 Overall, integrating out $\phi, \alpha, \beta$, (\ref{eq:nock9_prime}) becomes 
   \begin{align}
              \expectGOE  \Delta_{\epsilon}(M; x, S) = K^{(7)}_Ne^{2N(x^2 - \epsilon^2)}\iint_0^{\pi/2} &d\theta d\hat{\theta} \int d\vec{x}_1 d\vec{x}_2  \iint_{-\infty}^{\infty} dr_1dr_2 |r_1 - r_2|^4 \sin2\theta \cos^22\theta\sin2\hat{\theta} \notag\\
    &\exp\left\{ -\frac{N}{2}\Tr Q_B^2 - iN\Tr SB + ixN\Tr Q_B + \epsilon N\Tr Q_B\sigma\right\}\notag\\
    &\exp\{-N(r_1^2 + r_2^2) - 2Ni(x-i\epsilon\cos2\theta\cos2\hat{\theta})r_1\notag\\&~~~ - 2Nix(x + i\epsilon\cos2\theta\cos2\hat{\theta}) \}\notag\\&\prod_{j=1}^2\det\left(G_j + B + iS\right)\label{eq:nock10_prime}   
    \end{align} 
    where $K^{(7)} =(2\pi)^3 K^{(6)} $  and now \begin{align}
                G_1G_2 &= r_1r_2,\label{eq:G_sim1_prime3}\\
        G_1 + G_2 &= r_1 + r_2 \label{eq:G_sim2_prime3}\\
        \iff \{G_1, G_2\} &= \{r_1, r_2\} \label{eq:G_are_r}.
    \end{align}
    We can now clearly take $r_j = G_j$ without loss of generality.
\jstat{The terms $\det(r_j + B + iS)$ and $e^{-iN\Tr SB}$ depend on the eigenvectors of $B$ and prevent an application of the integral formula of \cite{fyodorov2002characteristic} as used by \cite{nock}. In fact, it is possible the adapt this integral formula for use in the presence of the term $e^{-iN \Tr SB}$, as seen in Lemma \ref{lem:fyod_general}.}

\jstat{Since $S$ has all entries of order $N^{-1}$, we can expand the nuisance determinants: \begin{align}
    \det(r_j + B + iS) = \prod_{i=1}^2 (r_j + \lambda^{(B)}_i) (1 + o(1)).\label{eq:simple_det_expan}
\end{align}
For this step to be legitimate in the sense of asymptotic expansions, we must have that the error term is uniformly small in the integration variables $\vec{x}_1, \vec{x}_2, r_1, r_2, \theta, \hat{\theta}$. Note  that the integrand in (\ref{eq:nock10_prime}) is analytic in $r_1, r_2$ and so we can deform the contours of integration from $(-\infty, \infty)$ to $\Gamma$, a contour that, say, runs from $-\infty$ along the real line to $-1$ and then follows the unit semi-circle in the upper half plane to $1$ before continuing to $\infty$ along the real line. We show an example contour in Figure \ref{fig:r1r2_contour}. It is now clear that $r_1, r_2$ are bounded away from $0$ and so the error terms in (\ref{eq:simple_det_expan}) are uniform, so giving
   \begin{align}
              \expectGOE  \Delta_{\epsilon}(M; x, S) = K^{(7)}_Ne^{2N(x^2 - \epsilon^2)}\iint_0^{\pi/2} &d\theta d\hat{\theta} \int d\vec{x}_1 d\vec{x}_2  \iint_{-\infty}^{\infty} dr_1dr_2 |r_1 - r_2|^4 \sin2\theta \cos^22\theta\sin2\hat{\theta} \notag\\
    &\exp\left\{ -\frac{N}{2}\Tr Q_B^2 - iN\Tr SB + ixN\Tr Q_B + \epsilon N\Tr Q_B\sigma\right\}\notag\\
    &\exp\{-N(r_1^2 + r_2^2) - 2Ni(x-i\epsilon\cos2\theta\cos2\hat{\theta})r_1\notag\\
    &~~~~~~- 2Nix(x + i\epsilon\cos2\theta\cos2\hat{\theta}) \}\notag\\&\prod_{i,j=1}^2\det\left(r_j + \lambda^{(B)}_i\right)(1 + o(1))\label{eq:nock10_prime_after_exp}   
    \end{align}}
\begin{figure}
    \centering
    \begin{tikzpicture}
    \draw[step=1cm,gray,very thin] (-1.9,-1.9) grid (5.9,5.9);
    \draw[red,thick] (-1.9,2) -- (1,2);
    \draw[->,red,thick] (1,2) arc (180:0:1cm);
    \draw[red,thick] (3,2) -- (5.9,2);
    \node[green] at (2,2){$\times$};
    \node at (2.3, 1.6) {$r=0$};
    \node at (1.5,3.1){$\Gamma$};
    \end{tikzpicture}
    
    \caption{Example contour $\Gamma$ used for the $r_1, r_2$ integrals to keep away from the origin (denoted by the green cross).}
    \label{fig:r1r2_contour}
\end{figure}
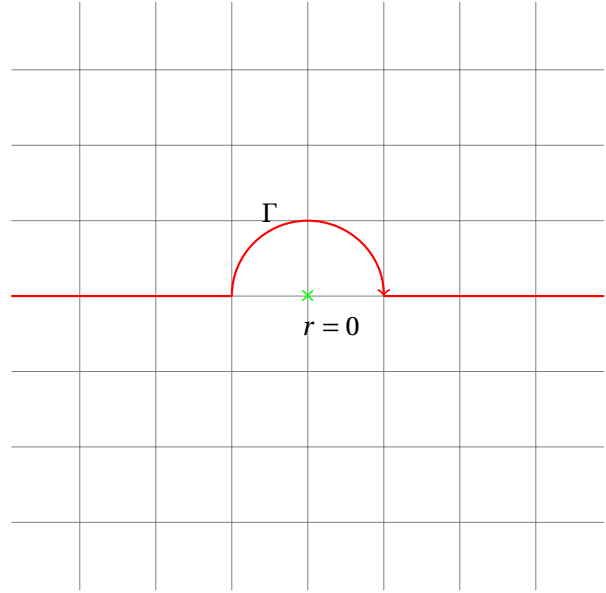

\jstat{Lemma \ref{lem:fyod_general} can now be applied:}
\begin{align}
  \expectGOE  \Delta_{\epsilon}(M; x, S) = K^{(8)}_N e^{2N(x^2 - \epsilon^2)}&\left(1 + o(1)\right)\notag \\
  \iint_0^{\pi/2} &d\theta   d\hat{\theta}\int_{\text{Sym}_{\geq 0}(2)} dQ_B \iint_{\Gamma} dr_1dr_2 \cos^22\theta \sin2\theta   \sin2\hat{\theta}\notag\\
  &  \exp\left\{ -\frac{N}{2}\Tr Q_B^2+ ixN\Tr Q_B + \epsilon N\Tr Q_B\sigma\right\}\notag\\
    &\exp\{-N(r_1^2 + r_2^2) - 2Ni(x-i\epsilon\cos2\theta\cos2\hat{\theta})r_1\notag\\&~~~~~~~~ - 2Nix(x + i\epsilon\cos2\theta\cos2\hat{\theta}) \} \notag\\
    &\jstat{\prod_{j=1}^r\left(1 + \vivacom{2}is_j\Tr Q_B - \vivacom{4}p_{11}p_{22}s_j^2\right)^{-1/2}}\notag \\
    &\prod_{i,j=1}^{2} \left(r_j + \lambda^{(B)}_i\right)|r_1 - r_2|^4(r_1r_2)^{N-2}(\det Q_B)^{\frac{N-3}{2}}, \label{eq:nock12_prime}   
\end{align}

where $p_{ij}$ are the entries of the matrix $Q_B$ and $K^{(8)} = \frac{\pi^N \pi^{-1/2}}{\Gamma\left(\frac{N}{2}\right)\Gamma\left(\frac{N-1}{2}\right)} K^{(7)}_N$.

We now wish to diagonalise $Q_B$ and integrate out its eigenvectors, but as before (around (\ref{eq:nock9_prime})) the integrand is not invariant under the action of the orthogonal group on $Q_B$ and so we instead diagonalise $Q_B = O\text{diag}(p_1, p_2)O^T$ and parametrize $O$ as \begin{equation}
    O = \left(\begin{array}{cc} \cos\theta' &\sin\theta' \\ -\sin\theta' & \cos\theta'\end{array}\right)\label{eq:orthogonal_param}
\end{equation}
but we must be careful to choose domain of integration for $\theta$ and $(p_1, p_2)$ such that the transformation is a bijection. Consider a general positive semi-definite symmetric matrix $$Q_B = \left(\begin{array}{cc} a & c \\ c & b \end{array}\right).$$ Solving for the eigenvalues gives two choices for $(p_1, p_2)$ because of the arbitrary ordering of the eigenvalues. We want a simple product domain for the $(p_1, p_2)$ integrals and both eigenvalues are non-negative, so we choose $(p_1, p_2) \in (\mathbb{R}_{\geq 0})^2$.  One can easily find that \begin{align}
    c &= \frac{p_2 - p_1}{2}\sin2\theta\\
    a &= \frac{p_1 + p_2 + (p_1 - p_2)\cos2\theta}{2}\\
        b &= \frac{p_1 + p_2 + (p_2 - p_1)\cos2\theta}{2}
\end{align}
and so we see immediately that the domain of integration of $\theta$ must be restricted to an interval of length $\pi$ to obtain a bijection. But further, because of the chosen domain for $(p_1, p_2)$ the quantity $(p_1 - p_2)$ takes all values in $\mathbb{R}$ and thus we must in fact restrict $\theta$ to, say, $[0, \pi/2)$ to obtain a bijection.
 The Jacobian of this transformation is $|p_1 - p_2| $ and further \begin{align}
    p_{11}p_{22} & = (p_1\cos^2\theta' + p_2 \sin^2\theta')(p_2\cos^2\theta' + p_1\sin^2\theta')\notag\\
    &=(p_1^2 + p_2^2)(\cos\theta'\sin\theta')^2 + p_1p_2(\cos^4\theta' + \sin^4\theta')\notag\\
    &=  \frac{1}{4}\sin^22\theta' (p_1^2 + p_2^2) + \frac{1}{4}\left(3 + 4\cos4\theta'\right)p_1p_2 
\end{align}
and so we get 

\begin{align}
                  \expectGOE  \Delta_{\epsilon}(M; x, S) = K^{(8)}_N e^{2N(x^2 - \epsilon^2)}&\left(1 + o(1)\right)\iiint_0^{\pi/2} d\theta d\theta'd\hat{\theta}\iint_0^{\infty}dp_1dp_2 \iint_{\Gamma} dr_1dr_2\notag\\
                  &|r_1 - r_2|^4(r_1r_2)^{N-2}(p_1p_2)^{\frac{N-3}{2}}\cos^22\theta \sin2\theta \sin2\hat{\theta}\notag\\
  &  \exp\left\{ -\frac{N}{2}(p_1^2 + p_2^2)  + iN(x-i\epsilon\cos2\theta')p_1 + iN(x+i\epsilon\cos2\theta')p_2 \right\}\notag\\
    &\exp\{-N(r_1^2 + r_2^2) - 2Ni(x-i\epsilon\cos2\theta\cos2\hat{\theta})r_1 \notag\\&~~~~~~- 2Nix(x + i\epsilon\cos2\theta\cos2\hat{\theta}) \} \notag\\
    &\prod_{i,j=1}^{2} \left(r_j + p_i\right)J_1(p_1, p_2, \theta'; \{s_j\}_{j=1}^r, N)
 \label{eq:nock13_prime}   
\end{align}

where
 \jstat{
\begin{equation}
J_1(p_1, p_2, \theta'; \{s_j\}_{j=1}^r, N) =\prod_{j=1}^r\left(1 + \vivacom{2}is_j(p_1 + p_2) - s_j^2\left[\sin^22\theta' (p_1^2 + p_2^2) + \left(3 + 4\cos4\theta'\right)p_1p_2 \right]\right)^{-1/2}.
\end{equation}}

Now let us define the functions \begin{align}
    \psi^{(\pm)}_U(z; x; \epsilon) &= \frac{1}{2}z^2 \pm i(x - i\epsilon)z - \frac{1}{2}\log z\\
      \psi^{(\pm)}_L(z; x; \epsilon) &= \frac{1}{2}z^2 \pm i(x + i\epsilon)z - \frac{1}{2}\log z\\
\end{align}
and also \begin{equation}
    J_2(r_1, r_2, p_1, p_2) = |r_1 - r_2|^4 |p_1 - p_2| (r_1r_2)^{-2} (p_1p_2)^{-\frac{3}{2}} (r_1 + p_1)(r_1 + p_2)(r_2 + p_1)(r_2 + p_2)
\end{equation}
and then we finally rewrite (\ref{eq:nock13_prime}) as 
\begin{align}
                  &\expectGOE  \Delta_{\epsilon}(M; x, S)\notag\\ = K^{(8)}_N e^{2N(x^2 - \epsilon^2)}\left(1 + o(1)\right)\iiint_0^{\pi/2} &d\theta d\theta'd\hat{\theta}\iint_0^{\infty}dp_1dp_2 \iint_{\Gamma} dr_1dr_2\notag\\
                  &J_1(p_1, p_2, \theta'; S, N)J_2(r_1,r_2, p_1, p_2)\cos^22\theta \sin2\theta \sin2\hat{\theta}\notag\\
& \exp\Bigg\{-N\Bigg(2\psi^{(+)}_L(r_1; x; \epsilon\cos2\theta\cos2\hat{\theta}) +2\psi^{(+)}_U(r_2; x; \epsilon\cos2\theta\cos2\hat{\theta})\notag\\& +\psi^{(-)}_L(p_1; x; \epsilon\cos2\theta')+\psi^{(-)}_U(p_2; x; \epsilon\cos2\theta')\Bigg)\Bigg\}.
 \label{eq:nock_lemma1_final}   
\end{align}
\end{proof}

We will need the asymptotic behaviour of the constant $K_N$ defined in Lemma \ref{lemma:nock_deformed}.

\begin{lemma}\label{lemma:K_N_asymp}
    As $N\rightarrow\infty$ \begin{equation}
      K_N\sim \frac{ (-i)^N N^{\frac{9}{2}}}{4\sqrtsign{2}\pi^{\frac{5}{2}}}(2e)^N.
    \end{equation}
\end{lemma}
\begin{proof}
Using Stirling's formula for the Gamma function gives \begin{align}
    K_N &\sim \frac{N^{N+3}(-i)^N}{\pi^{3/2}} N^{-\frac{N}{2} + \frac{1}{2}}(N-1)^{-\frac{N}{2} + 1} 2^{\frac{N}{2} - \frac{1}{2}} 2^{\frac{N}{2} - 1} e^{\frac{N}{2}} e^{\frac{N}{2} - \frac{1}{2}} 
    \left(2\pi\right)^{-1}\notag\\
  &=  \frac{N^{N+3}(-i)^N}{\pi^{3/2}} N^{-N} N^{\frac{3}{2}} 2^{N} 2^{-\frac{5}{2}} e^{N} e^{-\frac{1}{2}}\pi^{-1} \left(\frac{N-1}{N}\right)^{-\frac{N}{2} + 1}\notag\\
  &\sim \frac{ (-i)^N N^{\frac{9}{2}}}{4\sqrtsign{2}\pi^{\frac{5}{2}}}(2e)^N.
\end{align}
\end{proof}
Building on Lemma \ref{lemma:nock_deformed}, we can prove a generalisation of Theorem 2.8 from \cite{auffinger2013random}, namely Theorem \ref{thm:auff2.8}.
\auffindk*
\begin{proof}

Combining Lemmata \ref{lemma:conditional_dist} and \ref{lemma:kac_rice} and observing that the integrand in the Kac-Rice formula of Lemma \ref{lemma:kac_rice} is spherically symmetric, we obtain \begin{align}\label{eq:thm28_1}
\expect C_{N}^{h}(\sqrtsign{N}u) &= \underbrace{\left(2(N-1)(H-1)H\right)^{\frac{N-1}{2}} \omega_N \frac{\jstat{e^{-\frac{\vec{v}^2}{2H}}}}{(2\pi H)^{(N-1)/2}}}_{\defeq \Omega_N} \int_{-\infty}^{u_N} dx ~ \frac{1}{\sqrtsign{2\pi}t} e^{-\frac{x^2}{2t^2}}  \mathbb{E}^{N-1}_{GOE} |\det(M - xI + S)| 
\end{align} 
where $$ u_N = u\sqrtsign{\frac{HN}{2(N-1)(H-1)}}, $$ the variance $t^2 = \frac{H}{2(N-1)(H-1)}$,  $\omega_N = 2\pi^{N/2}/\Gamma(N/2)$ is the surface area of the $N-1$ sphere and $S$ and $\vec{v}$ are defined in Lemma \ref{lemma:conditional_dist}. Note that the first term in $\Omega_N$ comes from the expression (\ref{eq:hij_goe}) and the third term from \jstat{(\ref{eq:derivs_exp_hi}) and} (\ref{eq:derivs_cov_hihj}), i.e. this is the density of $\nabla h$ evaluated at $0$ as appears in Lemma \ref{lemma:kac_rice}. \jstat{The conditions for Lemma \ref{lemma:nock_deformed} are shown to be met in Lemma \ref{lemma:conditional_dist}}, so we obtain 
\begin{align}
    \expect C_{N}^{h}(\sqrtsign{N}u) = &\Omega_N K_{N-1}\sqrtsign{\frac{2(N-1)(H-1)}{H}} \left(1 + o(1)\right)\notag\\&\int_{-\infty}^{u_N}dx ~  \frac{1}{\sqrtsign{2\pi}}  \lim_{\epsilon\searrow 0}\iiint_0^{\pi/2} d\theta d\hat{\theta}d\theta' \iint_0^{\infty}dp_1dp_2 \iint_{\Gamma} dr_1dr_2\notag\\
                  &J_1(p_1, p_2, \theta'; \{s_j\}_{j=1}^r, N-1)J_2(r_1,r_2, p_1, p_2)\cos^22\theta \sin2\theta\sin2\hat{\theta}\notag\\
& \exp\Bigg\{-(N-1)\Bigg(2\psi^{(+)}_L(r_1; x; \epsilon\cos2\theta\cos2\hat{\theta}) +2\psi^{(+)}_U(r_2; x; \epsilon\cos2\theta\cos2\hat{\theta})\notag\\& +\psi^{(-)}_L(p_1; x; \epsilon\cos2\theta')+\psi^{(-)}_U(p_2; x; \epsilon\cos2\theta') - 
\frac{H+1}{H}x^2\Bigg)\Bigg\}\notag \\
 =&c_{N,H}\int_{-\infty}^{u_N}dx ~   \lim_{\epsilon\searrow 0}\iiint_0^{\pi/2} d\theta d\hat{\theta}d\theta'\iint_0^{\infty}dp_1dp_2 \iint_{\Gamma} dr_1dr_2 \notag\\
                  &J_1(p_1, p_2, \theta'; \{s_j\}_{j=1}^r, N-1)J_2(r_1,r_2, p_1, p_2)\cos^22\theta \sin2\theta\sin2\hat{\theta}\notag\\
& \exp\Bigg\{-(N-1)\Bigg(2\psi^{(+)}_L(r_1; x; \epsilon\cos2\theta\cos2\hat{\theta}) +2\psi^{(+)}_U(r_2; x; \epsilon\cos2\theta\cos2\hat{\theta})\notag\\& +\psi^{(-)}_L(p_1; x; \epsilon\cos2\theta')+\psi^{(-)}_U(p_2; x; \epsilon\cos2\theta') - \frac{H+1}{H}x^2\Bigg)\Bigg\}  \label{eq:thm28_mid}
\end{align}

where we have defined the constant \begin{equation}\label{eq:constant_final_defn}
    c_{N,H} = \frac{\Omega_N K_{N-1}\sqrtsign{(H-1)(N-1)}}{\sqrtsign{H\pi}}(1 + o(1)).
\end{equation}
We pause now to derive the asymptotic form of $c_{N,H}$. The vector $\vec{v}$ was defined in Lemma \ref{lemma:conditional_dist} and has entries of order $N^{-1/2}$, so $\vec{v}^2 = \mathcal{O}(1)$. Using Stirling's formula for the Gamma function \begin{align}
    \Omega_N &\sim 2 (N-1)^{\frac{N-1}{2}} (H-1)^{\frac{N-1}{2}}\pi^{1/2} N^{-\frac{N}{2} + \frac{1}{2}} 2^{\frac{N}{2} - \frac{1}{2}} e^{\frac{N}{2}} \left(2\pi\right)^{-1/2}\jstat{e^{-\frac{\vec{v}^2}{2H}}}\notag\\
    &=(H-1)^{\frac{N-1}{2}} (2e)^{\frac{N}{2}} \left(\frac{N-1}{N}\right)^{\frac{N-1}{2}}\jstat{e^{-\frac{\vec{v}^2}{2H}}}\notag\\
    &\sim (H-1)^{\frac{N-1}{2}} (2e)^{\frac{N}{2}} e^{-1/2}\jstat{e^{-\frac{\vec{v}^2}{2H}}}\notag\\ 
    \implies \frac{\Omega_N \sqrtsign{(H-1)(N-1)}}{\sqrtsign{H\pi}} &\sim (H-1)^{\frac{N}{2}} (2e)^{\frac{N}{2}} e^{-1/2} H^{-1/2}\pi^{-1/2} (N-1)^{1/2}\jstat{e^{-\frac{\vec{v}^2}{2H}}}
\end{align}
and so Lemma \ref{lemma:K_N_asymp} gives \begin{align}
    c_{N,H} &\sim \frac{ (-i)^{N-1} (N-1)^{\frac{9}{2}}}{4\sqrtsign{2}\pi^{\frac{5}{2}}}(2e)^{N-1}(H-1)^{\frac{N}{2}} (2e)^{\frac{N}{2}} e^{-1/2} H^{-1/2}\pi^{-1/2} (N-1)^{1/2}\jstat{e^{-\frac{\vec{v}^2}{2H}}}\notag\\
    &\sim \frac{(-i)^{N-1} N^5}{4\pi^3 H^{1/2}} (2e)^{\frac{3}{2}(N-1)} (H-1)^{\frac{N}{2}}\jstat{e^{-\frac{\vec{v}^2}{2H}}}.\label{eq:c_NH_asymp}
\end{align}
In the style of \cite{soton29213}, the multiple integral in (\ref{eq:thm28_mid}) can be written as an expansion over saddle points and saddle points of the integrand restricted to sections of the boundary. Recalling the form of $\psi^{(\pm)}_U$ and $\psi^{(\pm)}_L$, we see that the integrand vanishes on the boundary and so we focus on the interior saddle points. Let us define the exponent function \begin{equation}
        \Phi(r_1, r_2, p_1, p_2, x; S, \epsilon) = 2\psi^{(+)}_L(r_1; x, \epsilon)+ 2\psi^{(+)}_U(r_2; x, \epsilon) + \psi^{(-)}_L(p_1; x, \epsilon) + \psi^{(-)}_U(p_2; x, \epsilon) - \frac{(H+1)}{H} x^2
    \end{equation}
    It is clear that the $\cos\theta, \cos\hat{\theta}$ and $\cos\theta'$ terms in the exponent of (\ref{eq:thm28_mid}) do not affect the saddle point asymptotic analysis, since we take the limit $\epsilon\rightarrow 0$, and $\theta, \hat{\theta}, \theta'\in [0,\pi/2)$ and it is only the signs of the $\mathcal{O}(\epsilon)$ terms that are significant. Therefore, to simplify the exposition, we will suppress these terms.
    The $(r_1,r_2,p_1,p_2)$ components of $\grad\Phi$ are of the form \begin{equation}
        z\mapsto z \pm i(x\pm i\epsilon) - \frac{1}{2z}
    \end{equation}
    and so the only saddle in $\Phi$ restricted to those components is at \begin{align}
        r_1 &= \frac{-i(x+i\epsilon) + (2-(x+i\epsilon)^2)^{1/2}}{2}\defeq z^{(+)}_L\label{eq:saddle_loc_r1}\\
           r_2 &= \frac{-i(x-i\epsilon) + (2-(x-i\epsilon)^2)^{1/2}}{2}\defeq z^{(+)}_U\\
                   p_1 &= \frac{i(x+i\epsilon) + (2-(x+i\epsilon)^2)^{1/2}}{2}\defeq z^{(-)}_L\\
                           p_2 &= \frac{i(x-i\epsilon) + (2-(x-i\epsilon)^2)^{1/2}}{2}\defeq z^{(-)}_U\label{eq:saddle_loc_p2}.
    \end{align}
    To deform the $(r_1,r_2,p_1,p_2)$ contours through this saddle, we are required to choose a branch of the functions in (\ref{eq:saddle_loc_r1} - \ref{eq:saddle_loc_p2}). Each has branch points at  $\pm\sqrtsign{2} + i\epsilon$ or $\pm\sqrtsign{2} -i\epsilon$. Since the initial contour of $x$ integration lies along the real line, we take the following branch cuts in the complex $x$ plane and respective angle ranges (see Figure \ref{fig:branch_cut}) \begin{align}
    [\sqrtsign{2} + i\epsilon, \sqrtsign{2} + i\infty],~~ & [\pi/2, 5\pi/2] \label{eq:branch1}\\
    [\sqrtsign{2} - i\epsilon, \sqrtsign{2} - i\infty],~~ & [-\pi/2, 3\pi/2] \\
    [-\sqrtsign{2} + i\epsilon, -\sqrtsign{2} + i\infty],~~ & [\pi/2, 5\pi/2] \\
    [-\sqrtsign{2} - i\epsilon, -\sqrtsign{2} - i\infty],~~ & [-\pi/2, 3\pi/2]\label{eq:branch4}.    \end{align}

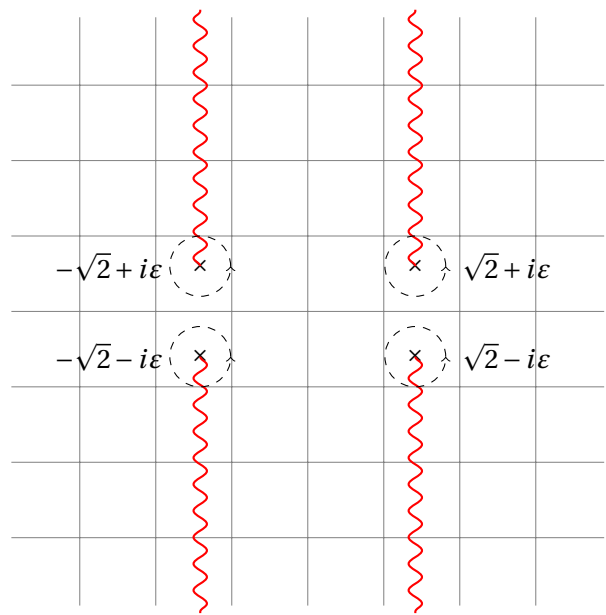
\begin{figure}
    \centering
    \begin{tikzpicture}
    \draw[step=1cm,gray,very thin] (-1.9,-1.9) grid (5.9,5.9);
    \def\R{0.4};
    \def\gapY{1.2};
    \def\gapX{2.83};
    \coordinate (A) at ($(2,2) + (-\gapX/2, \gapY/2)$);
    \coordinate (B) at ($(A) + (\gapX, 0)$);
    \coordinate (C) at ($(A) - (0,\gapY)$);
    \coordinate (D) at ($(B) - (0,\gapY)$);
    \node at (A){$\times$};
    \node at (B){$\times$};
    \node at (C){$\times$};
    \node at (D){$\times$};
    \draw[thick,red,branch cut] (A) to ($(A) + (0,4-\gapY/2)$);
    \draw[thick,red,branch cut] (B) to ($(B) + (0,4-\gapY/2)$);
    \draw[thick,red,branch cut] (C) to ($(C) + (0,-4+\gapY/2)$);
    \draw[thick,red,branch cut] (D) to ($(D) + (0,-4+\gapY/2)$);
    \node at ($(A) + (-1.2,0)$){$-\sqrtsign{2} + i\epsilon$};
    \node at ($(B) + (1.2,0)$){$\sqrtsign{2} + i\epsilon$};
    \node at ($(C) + (-1.2,0)$){$-\sqrtsign{2} - i\epsilon$};
    \node at ($(D) + (1.2,0)$){$\sqrtsign{2} - i\epsilon$};
    \draw[dashed, ->] ($(A) + (\R,0)$) arc (0:360:\R cm);
    \draw[dashed, ->] ($(B) + (\R,0)$) arc (0:360:\R cm);
    \draw[dashed, ->] ($(C) + (\R,0)$) arc (0:360:\R cm);
    \draw[dashed, ->] ($(D) + (\R,0)$) arc (0:360:\R cm);
    \end{tikzpicture}
    \caption{The choice of branch for the $x$ integral in the proof of Theorem \ref{thm:auff2.8}.}
    \label{fig:branch_cut}
\end{figure}

It is simple to compute $\psi_U^{(\pm)}(z^{(\pm)}_U)$ and  $\psi_L^{(\pm)}(z^{(\pm)}_L)$:
\begin{align}
    \psi_L^{(+)}(z^{(+)}_L) = &\frac{1}{4}\left(1+(x+i\epsilon)^2 + \log 2\right) + \frac{1}{4}\log 2 +\frac{1}{4}i(x+i\epsilon)\left(2 - (x+i\epsilon)^2\right)^{1/2}  \notag\\&-\frac{1}{2}\log\left[ -i(x+i\epsilon) + \left(2-(x+i\epsilon)^2\right)^{1/2}\right]\label{eq:psi_L_plus}\\
       \psi_U^{(+)}(z^{(+)}_U) = &\frac{1}{4}\left(1+(x-i\epsilon)^2 + \log 2\right) + \frac{1}{4}\log 2 +\frac{1}{4}i(x-i\epsilon)\left(2 - (x-i\epsilon)^2\right)^{1/2} \notag\\&- \frac{1}{2}\log\left[ -i(x-i\epsilon) + \left(2-(x-i\epsilon)^2\right)^{1/2}\right]\\
           \psi_L^{(-)}(z^{(-)}_L) = &\frac{1}{4}\left(1+(x+i\epsilon)^2 + \log 2\right) + \frac{1}{4}\log 2 -\frac{1}{4}i(x+i\epsilon)\left(2 - (x+i\epsilon)^2\right)^{1/2} \notag\\ &-\frac{1}{2}\log\left[ i(x+i\epsilon) + \left(2-(x+i\epsilon)^2\right)^{1/2}\right]\\
       \psi_U^{(-)}(z^{(-)}_U) = &\frac{1}{4}\left(1+(x-i\epsilon)^2 + \log 2\right) + \frac{1}{4}\log 2 -\frac{1}{4}i(x-i\epsilon)\left(2 - (x-i\epsilon)^2\right)^{1/2} \notag\\&- \frac{1}{2}\log\left[ i(x-i\epsilon) + \left(2-(x-i\epsilon)^2\right)^{1/2}\right]\label{eq:psi_U_minus}.
\end{align}
  Let us consider $x$ still restricted to the real line. We are free to restrict to $\epsilon>0$ and then $x\pm i\epsilon$ lies just above (below) the real line. For $x<-\sqrtsign{2}$ the angle from all four branch points is $\pi$ and so we obtain \begin{align}
      \Phi_{(4)}(x) \defeq \lim_{\epsilon\rightarrow 0} \Phi\left(z_L^{(+)}, z_U^{(+)}, z_L^{(-)}, z_U^{(-)}, x; \epsilon\right) &=              \frac{3}{2}\left(1+x^2+\log 2\right) + \frac{3}{2}\log 2 - \frac{1}{2}x\sqrtsign{x^2 -2}-2\log\left[-ix + i\sqrtsign{x^2 - 2}\right] \notag\\
      &- \log\left[ix + i\sqrtsign{x^2 - 2}\right] - \frac{H+1}{H}x^2 \notag\\
      &=\frac{3}{2}\left(1+\log 2\right) + \frac{H-2}{2H}x^2 + \frac{3}{2}\log 2 - \frac{1}{2}x\sqrtsign{x^2 -2}-\log\left[-ix + i\sqrtsign{x^2 - 2}\right]\notag\\
      & ~~~- \log 2\notag\\
      &=\frac{3}{2}\left(1+\log 2\right)+ \frac{H-2}{2H}x^2 + \frac{1}{2}\log 2 - \frac{1}{2}x\sqrtsign{x^2 -2}-\log\left[-x + \sqrtsign{x^2 - 2}\right]\notag\\ 
      & ~~~ - \log i\notag\\
      &= \frac{3}{2}\left(1+\log 2\right) +\frac{H-2}{2H}x^2+ I_1(x; \sqrtsign{2}) - \log i\label{eq:phi_4_below_bulk}
  \end{align}
 However for $-\sqrtsign{x} < x < \sqrtsign{2}$ the angles about the branch points are $\pi, \pi, 2\pi, 0$ in the order of (\ref{eq:branch1}-\ref{eq:branch4}). It follows that the square root terms in both of $\psi^{(\pm)}_L(z^{(\pm)}_L)$ and both of $\psi^{(\pm)}_U(z^{(\pm)}_U)$ have opposite signs and so \begin{align}
     \Phi_{(4)}(x) &= \frac{3}{2}\left( 1+ \log2\right) + \frac{H-2}{2H}x^2- \frac{3}{2}\log(-2) + \frac{3}{2}\log 2\notag\\
     &= \frac{3}{2}\left( 1 + \log2\right) + \frac{H-2}{2H}x^2
     - \frac{3}{2}\log(-1)\label{eq:phi_4_in_bulk}.
 \end{align}
 Finally, the above reasoning can be trivially extended to $x>\sqrtsign{2}$ to obtain \begin{align}\label{eq:phi_4_above_bulk}
     \Phi_{(4)}(x) = \frac{3}{2}\left(1 + \log{2}\right) + \frac{H-2}{2H}{x^2} + I_1(-x; \sqrtsign{2}) - \log{i}.
 \end{align}
  It is apparent from (\ref{eq:phi_4_below_bulk})\footnote{Note that $I_1(x;\sqrtsign{2})$ is monotonically decreasing on $(-\infty, -\sqrtsign{2}]$.}, (\ref{eq:phi_4_in_bulk}) and (\ref{eq:phi_4_above_bulk}) that the branch choice (\ref{eq:branch1}-\ref{eq:branch4}) and deforming through each of the saddles of in $(r_1, r_2, p_1, p_2)$ gives a contour of steepest descent in $x$ with the critical point being at $x=0$.

We are thus able to write down the leading order asymptotics for (\ref{eq:thm28_mid}) for all real $u$ coming either from the end-point $x=\sqrtsign{2}u/E_{\infty}$ or the critical point $x=0$. We begin with $u< -E_{\infty}$ by using (\ref{eq:phi_4_below_bulk}):
 \begin{align}
    \frac{1}{N}\log\mathbb{E}C^h_{N}(\sqrtsign{N}u) &\sim -\frac{3}{2}\log{2}  -\frac{3}{2} -\frac{H-2}{2H}\frac{Hu^2}{2(H-1)} - I_1(u; E_{\infty}) + \log{i}+ \frac{1}{N}\
    \log c_{N,H}\notag\\
    &\sim \frac{1}{2}\log(H-1) - \frac{H-2}{4(H-1)}u^2 - I_1(u; E_{\infty})
\end{align}
since by (\ref{eq:c_NH_asymp}) \begin{align}
    \log{c_{N,H}} \sim \frac{1}{2}N\log(H-1) + \frac{3}{2}(N-1)(1 + \log{2}) + (N-1)\log(-i).
\end{align}
For $-E_{\infty} \leq u < 0$ we use (\ref{eq:phi_4_in_bulk}):
\begin{align}
    \frac{1}{N}\log\mathbb{E}C^h_{N}(\sqrtsign{N}u) &\sim -\frac{3}{2}\log{2}  -\frac{3}{2} -\frac{H-2}{2H}\frac{Hu^2}{2(H-1)} + \frac{3}{2}\log(-1)+ \frac{1}{N}\
    \log c_{N,H}\notag\\
    &\sim \frac{1}{2}\log(H-1) - \frac{H-2}{4(H-1)}u^2  
\end{align}
since $\frac{3}{2}\log(-1) = \log\left((-1)^{1/2}\right) =\log{i}$. Finally, for $u\geq 0$ the leading contribution comes from the critical point, so 
\begin{align}
    \frac{1}{N}\log\mathbb{E}C^h_{N}(\sqrtsign{N}u) &\sim -\frac{3}{2}\log{2}  -\frac{3}{2} + \frac{3}{2}\log(-1)+ \frac{1}{N}\
    \log c_{N,H}\notag\\
    &\sim \frac{1}{2}\log(H-1).
\end{align}
\end{proof}

We are in-fact able to obtain the exact leading order term in the expansion of $\mathbb{E}C^h_{N}(\sqrtsign{N}u)$ in the case $u<-E_{\infty}$, namely Theorem \ref{thm:exact_term}.
\auffexact*

\begin{proof}
We begin by deriving an alternative form for $h$. For $v>\sqrtsign{2}$ \begin{align}
    h(v)^2 &= \frac{|v - \sqrtsign{2}| + | v + \sqrtsign{2}| + 2|v^2 - 2|^{\frac{1}{2}}}{|v^2 -2|^{\frac{1}{2}}}\notag\\
    &= 2\left( v + |v^2 - 2|^{\frac{1}{2}}\right)|v^2 - 2|^{-\frac{1}{2}}\notag\\
    \implies h(v) &= \sqrtsign{2} \left( v + |v^2 - 2|^{\frac{1}{2}}\right)^{\frac{1}{2}}|v^2 - 2|^{-\frac{1}{4}}\notag\\
   &=2|-v + |v^2 - 2|^{\frac{1}{2}}|^{-\frac{1}{2}}|v^2 - 2|^{-\frac{1}{4}}. \label{eq:alternate_h_form}
\end{align}
This proof now proceeds like that of Theorem \ref{thm:auff2.8} except that we are required to keep track of the exact factors in (\ref{eq:thm28_mid}) and evaluate the $\mathcal{O}(1)$ integrals arising from the saddle point approximation. First note that (using primes to denote $z$ derivatives) \begin{equation}
   { \psi^{(\pm)}_{U,L}}''(z ;x; \epsilon) = 1 + \frac{1}{2z^2}
\end{equation}
and so we abbreviate  ${\psi^{(\pm)}_{U,L}}''= \psi''$. We get the following useful relation (now letting $\epsilon \rightarrow 0$ implicitly for simplicity of exposition) \begin{align}
    \psi''(z^{(\pm)}_{U,L}) &= (z^{(\pm)}_{U,L})^{-2}\left(1 \mp ix z^{(\pm)}_{U,L}\right)\notag \\
    &=\frac{1}{2}(z^{(\pm)}_{U, L})^{-2}\left(2 - x^2  \pm x\sqrtsign{x^2 - 2}\right)\notag \\
    &= i\sqrtsign{x^2 - 2}(z^{(\pm)}_{U, L})^{-1}\label{eq:handy_psi_pp}
\end{align}
where, using our branch choice shown in Figure \ref{fig:branch_cut}, for $x<-\sqrtsign{2}$ the saddle points are \begin{align}
    z_{U,L}^{(\pm)} = \frac{\mp ix + i\sqrtsign{x^2 - 2}}{2}.
\end{align}
We recall the central expression (\ref{eq:thm28_mid}) from the proof of Theorem \ref{thm:auff2.8}:\begin{align}
    \expect C_{N}^{h}(\sqrtsign{N}u) = c_{N,H}\int_{-\infty}^{u_N}dx ~   \lim_{\epsilon\searrow 0}&\iiint_0^{\pi/2} d\theta d\hat{\theta}d\theta'\iint_0^{\infty}dp_1dp_2 \iint_{\Gamma} dr_1dr_2 \notag\\
                  &J_1(p_1, p_2, \theta'; \{s_j\}_{j=1}^r, N-1)J_2(r_1,r_2, p_1, p_2)\cos^22\theta \sin2\theta\sin2\hat{\theta}\notag\\
& \exp\Bigg\{-(N-1)\Bigg(2\psi^{(+)}_L(r_1; x; \epsilon\cos2\theta\cos2\hat{\theta}) +2\psi^{(+)}_U(r_2; x; \epsilon\cos2\theta\cos2\hat{\theta})\notag\\& +\psi^{(-)}_L(p_1; x; \epsilon\cos2\theta')+\psi^{(-)}_U(p_2; x; \epsilon\cos2\theta') - \frac{H+1}{H}x^2\Bigg)\Bigg\}\notag
\end{align}
and we recall the expressions for $J_1, J_2$ from Lemma \ref{lemma:nock_deformed}:
\begin{align}
    J_1(p_1, p_2, \theta'; \{s_j\}_{j=1}^r, N) &=\jstat{\left(1 + iN^{-1/2}s_2(p_1 + p_2) - N^{-1}s_2^2\left[\frac{1}{4}\sin^22\theta' (p_1^2 + p_2^2) + \frac{1}{4}\left(3 + 4\cos4\theta'\right)p_1p_2 \right]\right)^{-1/2}}\notag\\
    & ~~~~~~ \cdot\jstat{ \left(1 + is_1(p_1 + p_2) - s_1^2\left[\frac{1}{4}\sin^22\theta' (p_1^2 + p_2^2) + \frac{1}{4}\left(3 + 4\cos4\theta'\right)p_1p_2 \right]\right)^{-1/2},}\notag\\
        J_2(r_1, r_2, p_1, p_2) &= (r_1 + p_1)(r_2 + p_1)(r_1 + p_2)(r_2 + p_2)|r_1 - r_2|^4 |p_1-p_2| (r_1r_2)^{-2} (p_1p_2)^{-3/2}.\notag
    \end{align} 

We begin by evaluating $J_1$ to leading order at the saddle points:\begin{align}
    \frac{1}{2} \sin^2 2\theta' (z^{(-)})^2 + \frac{1}{4}\left(3+4\cos 4\theta'\right) (z^{(-)})^2 &\equiv q(\theta') (z^{(-)})^2\notag \\
    \implies J_1(z^{(-)}, z^{(-)}, \theta'; \{s_j\}_{j=1}^r, N) &\sim  \jstat{\left(1 + \vivacom{4}iz^{(-)}s_1 - \vivacom{2}q(\theta')\left(z^{(-)}\right)^2s_1^2\right)^{-1/2}\label{eq:J1_at_saddle}.}
\end{align}
\jstat{Recalling \begin{align}
    x + \sqrtsign{x^2 - 2} = \frac{-2}{-x + \sqrtsign{x^2 - 2}} = -\frac{h(x)^2}{2}\sqrtsign{x^2 - 2}, ~~~~~ (z^{(-)})^2 = -\frac{1}{2}\sqrtsign{x^2 - 2}\left( x + \sqrtsign{x^2 - 2}\right)
\end{align}
we obtain \begin{align}
    J_1 \sim 1 + \frac{1}{\vivacom{2}}s_1\sqrtsign{x^2 - 2}h(x)^2 - s_1^2 q(\theta')|x^2 - 2|h(x)^2 \equiv j(x, s_1, \theta').
\end{align}}

We see that $J_2(z^{(+)}, z^{(+)}, z^{(-)}, z^{(-)}) = 0$ and so we are required to expand $J_2$ in the region of $$(r_1, r_2, p_1, p_2) = (z^{(+)}, z^{(+)}, z^{(-)}, z^{(-)}).$$
Following standard steepest descents practice, the integration variables $r_1, r_2, p_1, p_2$ are replaced by scaled variables in the region of the saddle point, i.e. \begin{align}
    r_i &=  z^{(+)} + (N-1)^{-\frac{1}{2}}|{\psi^{(+)}}''(z^{(+)})|^{-\frac{1}{2}}\rho_i\label{eq:exact_term_scaling1}\\
    p_i &=  z^{(-)} + (N-1)^{-\frac{1}{2}}|{\psi^{(-)}}''(z^{(-)})|^{-\frac{1}{2}}\pi_i\label{eq:exact_term_scaling2}
\end{align}
and so \begin{align}
    &J_2(r_1, r_2, p_1, p_2) \notag\\= &(N-1)^{-\frac{5}{2}}|x^2 - 2|^2 (z^{(+)})^{-4}(z^{(-)})^{-3} |{\psi^{(-)}}''(z^{(-)})|^{-\frac{1}{2}}|{\psi^{(+)}}''(z^{(+)})|^{-2}|\rho_1 - \rho_2|^4 |\pi_1 -\pi_2| + o(N^{-\frac{5}{2}}).
\end{align}
Piecing these components together gives \begin{align}
    J_2 J_1 dr_1 dr_2 dp_1dp_2 &= (N-1)^{\jstat{-\frac{9}{2}}} \jstat{j(x, s_1, \theta')} |x^2 - 2|^2 \notag \\
    & ~~~~~~~~~~ |{\psi^{(-)}}''(z^{(-)})|^{-\frac{3}{2}}
    |{\psi^{(+)}}''(z^{(+)})|^{-3}(z^{(+)})^{-4}(z^{(-)})^{-3}\notag \\ & ~~~~~~~~~~ |\rho_1 - \rho_2|^4 |\pi_1 -\pi_2| d\rho_1 d\rho_2 d\pi_1 d\pi_2\notag \\
    &= (N-1)^{-\jstat{\frac{9}{2}}} \jstat{j(x, s_1, \theta')} |x^2 - 2|^{-\frac{1}{4}}  (z^{(+)})^{-1}(z^{(-)})^{-\frac{3}{2}}\notag \\ & ~~~~~~~~~~ |\rho_1 - \rho_2|^4 |\pi_1 -\pi_2| d\rho_1 d\rho_2 d\pi_1 d\pi_2\notag \\
        &= 2(N-1)^{-\jstat{\frac{9}{2}}} \jstat{j(x, s_1, \theta')} |x^2 - 2|^{-\frac{1}{4}}  (z^{(-)})^{-\frac{1}{2}}\notag \\ & ~~~~~~~~~~ |\rho_1 - \rho_2|^4 |\pi_1 -\pi_2| d\rho_1 d\rho_2 d\pi_1 d\pi_2\notag \\
                &= 2^{\jstat{ \frac{3}{2}}}(N-1)^{-\jstat{\frac{9}{2}}} \jstat{j(x, s_1, \theta')} |x^2 - 2|^{-\frac{1}{4}} \left(x + \sqrtsign{x^2 - 2}\right)^{-\jstat{\frac{1}{2}} }\notag \\ & ~~~~~~~~~~ |\rho_1 - \rho_2|^4 |\pi_1 -\pi_2| d\rho_1 d\rho_2 d\pi_1 d\pi_2\label{eq:exact_first_term_jacobian_JJ}.
    \end{align}
Recalling the expression (\ref{eq:alternate_h_form}), we can then write \begin{align}
     J_2 J_1 dr_1 dr_2 dp_1dp_2 &= 2^{\jstat{\frac{3}{2}}}(N-1)^{-\jstat{\frac{9}{2}}}  \jstat{j(x, s_1, \theta') h(-x)} 2^{- \jstat{1}}|\rho_1 - \rho_2|^4 |\pi_1 -\pi_2| d\rho_1 d\rho_2 d\pi_1 d\pi_2\notag \\
     &= 2^{\jstat{\jstat{\frac{1}{2}}}}(N-1)^{-\jstat{\frac{9}{2}}}  \jstat{j(x, s_1, \theta') h(-x)} |\rho_1 - \rho_2|^4 |\pi_1 -\pi_2| d\rho_1 d\rho_2 d\pi_1 d\pi_2
\end{align}
and so using (\ref{eq:c_NH_asymp}), we obtain \begin{align}
      \expect C_{N}^{h}(\sqrtsign{N}u) &\sim \frac{2^{-\jstat{\frac{3}{2}} }N^{\frac{\jstat{1}}{2}}}{\pi^3\sqrtsign{H}}\jstat{e^{-\frac{\vec{v}^2}{2H}}} \frac{Y_{2}^{(4)}}{8} Y_{2}^{(1)} \iint_0^{\pi/2}d\theta d\hat{\theta} ~ \cos^2 2\theta \sin 2\theta\sin2\hat{\theta}\notag\\
      & ~~~\sqrtsign{H-1} \int_{0}^{\pi/2} d\theta' \int_{-\infty}^{\frac{\sqrtsign{2}u}{E_{\infty}}\sqrtsign{\frac{N}{N-1}}} dx ~ h(-x)\jstat{j(x, s_1, \theta')}e^{(N-1)\Theta_H(2^{-\frac{1}{2}}E_{\infty}x)}\label{eq:exact_first_term_midway}
\end{align}
where we have defined the integrals \begin{align}
    Y_{n}^{(\beta)} = \int_{\mathbb{R}^n} d\vec{y} ~ e^{-\frac{1}{2}\vec{y}^2} |\Delta(\vec{y})|^{\beta}
\end{align}
and $\Delta$ is the Vandermonde determinant. Recall that, as in Theorem \ref{thm:auff2.8}, the $x$ integration contour in (\ref{eq:exact_first_term_midway}) is a steepest descent contour and so the leading order term comes from the end point.
Now \begin{align}
   &(N-1) \Theta_H\left(\sqrtsign{\frac{N}{N-1}} u\right)\notag\\ = &(N-1)\frac{1}{2}\log(H-1) - N\frac{H-2}{4(H-1)}u^2 - (N-1)I_1\left(\sqrtsign{\frac{N}{N-1}}u; E_{\infty}\right) \notag\\
   = &(N-1)\frac{1}{2}\log(H-1) - N\frac{H-2}{4(H-1)}u^2 - (N-1)I_1\left(u; E_{\infty}\right) - \frac{N-1}{2N}uI_1'(u; E_{\infty}) + \mathcal{O}(N^{-1}) \notag\\
   = &N\Theta_H(u) - \frac{1}{2}\log(H-1) + I_1(u; E_{\infty}) - \frac{1}{2}uI_1'(u; E_{\infty}) + \mathcal{O}(N^{-1})
\end{align}
and so \begin{align}
     \expect C_{N}^{h}(\sqrtsign{N}u) &\sim \frac{2^{-\jstat{\frac{3}{2}}}N^{\frac{-\jstat{1}}{2}}}{24\pi^3\sqrtsign{H}} \jstat{e^{-\frac{\vec{v}^2}{2H}}} Y_{2}^{(4)} Y_{2}^{(1)} \left(\int_{0}^{\pi/2} d\theta'\jstat{j(-v, s_1, \theta')}\right) h(v) e^{N\Theta_H(u)} \frac{e^{I_1(u; E_{\infty}) - \frac{1}{2}u I_1'(u; E_{\infty})}}{\frac{H-2}{2(H-1)}u + I_1'(u; E_{\infty})}\label{eq:exact_first_term_almost}
\end{align}
where we have defined (c.f. \cite{auffinger2013random} Theorem 2.17) $v = -\sqrtsign{2}uE_{\infty}^{-1}.$
It now remains only to evaluate the various constants in (\ref{eq:exact_first_term_almost}) where possible. Firstly observe \begin{align}
    Y_2^{(1)} &= 2\pi \mathbb{E}_{X_1, X_2\overset{i.i.d.}{\sim}\mathcal{N}(0,1)} |X_1 - X_2| = 2\pi \mathbb{E}_{X\sim \mathcal{N}(0, 2)} |X| = 2\sqrtsign{\pi} \int_0^{\infty} xe^{-\frac{x^2}{4}} = 4\sqrtsign{\pi} 
\end{align}
and similarly \begin{align}
    Y_2^{(4)} &= 2\pi \mathbb{E}_{X_1, X_2\overset{i.i.d.}{\sim}\mathcal{N}(0,1)} (X_1 - X_2)^4 = 2\pi \mathbb{E}_{X\sim \mathcal{N}(0, 2)} X^4 = 24 \pi.
\end{align}
For convenience  we define 
\jstat{\begin{align}
    T(v, s_1) = \frac{2}{\pi}\int_{0}^{\pi/2}j(-v, s_1, \theta')d\theta',
\end{align}}

and then collating our results:
\begin{align}
     \expect C_{N}^{h}(\sqrtsign{N}u) &\sim \frac{N^{-\frac{\jstat{1}}{2}}}{\sqrtsign{2\pi H}} \jstat{e^{-\frac{\vec{v}^2}{2H}}}\jstat{T(v, s_1)} h(v) e^{N\Theta_H(u)} \frac{e^{I_1(u; E_{\infty}) - \frac{1}{2}u I_1'(u; E_{\infty})}}{\frac{H-2}{2(H-1)}u + I_1'(u; E_{\infty})}.\label{eq:exact_first_term_end}
\end{align}

\end{proof}

\begin{remark}\label{rem:exact_k_ind}
Having completed the proof of Theorem \ref{thm:exact_term}, we can now explain why this result generalises only part (a) of the analogous Theorem (2.17) from \cite{auffinger2013random}, namely only the case $u<-E_{\infty}$. Recall that, following standard steepest descent practice, we introduced scaled integration variables in the region of the saddle point (\ref{eq:exact_term_scaling1})-(\ref{eq:exact_term_scaling2}) and so arrived at (\ref{eq:exact_first_term_midway}) with the constant factors $Y_2^{(1)}, Y_2^{(4)}$ resulting from the Laplace approximation integrals over the scaled variables. If we take $-E_{\infty} < u < 0$, say, then $z^{(+)}_U + z^{(-)}_L = 0$ and $z^{(+)}_L + z^{(-)}_U = 0$ and so it is the terms $(r_1 + p_2), (r_2 + p_1)$ that vanish at the saddle point rather than $|r_1 - r_2|^4$ and $|p_1 - p_2|$. It follows that the terms $Y_2^{(1)}, Y_2^{(4)}$ are replaced by the integrals \begin{align}
    \int_{\mathbb{R}} d\pi_1 d\pi_2d\rho_1d\rho_2 ~ e^{-\frac{1}{2}(\pi_1^2 + \pi_2^2 + \rho_1^2 + \rho_2^2)} (\rho_1 + \pi_2)(\rho_2 + \pi_1) = 0.
\end{align} 
It is therefore necessary to keep terms to at least the first sub-leading order in the expansion of $J_1J_2$ around the saddle point, however we cannot do this owing the presence of the $o(1)$ term in the constant $c_{N,H}$ as defined in (\ref{eq:constant_final_defn}) which we cannot evaluate.\\
\end{remark}

\begin{remark}
\jstat{Note that setting all the  $\rho_{\ell}^{(N)}=0$ gives $\vec{v} = 0$, $S=0$, hence $s_1=0$ and so $T = 1$. Consequently (\ref{eq:exact_first_term_end}) recovers the exact spherical $H$-spin glass expression in part (a) of Theorem 2.17 in \cite{auffinger2013random}}. \\
\end{remark}

\begin{remark}
 The function $h(v)$ shows up in \cite{auffinger2013random} in the asymptotic evaluation of Hermite polynomials but arises here by an entirely different route.
\end{remark}

\subsection{Complexity results with  prescribed Hessian signature}
The next theorem again builds on Lemma \ref{lemma:nock_deformed} to prove a generalisation of Theorem 2.5 from \cite{auffinger2013random}. In fact, we will need a modified version of Lemma \ref{lemma:nock_deformed} which we now prove.

\begin{lemma}\label{lemma:nock_deformed_cond_k}
    Let $S$ be a rank $2$ $N\times N$ symmetric matrix with non-zero eigenvalues \jstat{$\{s_j\}_{j=1}^2$}, where and $s_j = \mathcal{O}(1)$.  Let $x<-\sqrtsign{2}$ and let $M$ denote an $N\times N$ GOE matrix with respect to whose law expectations are understood to be taken. Then
    \begin{align}
         &\expectGOE \left[|\det(M - xI + S)|\indic [\ind{x} (M+S)\in\{k-1, k, k+1\}]\right]\notag\\ \leq ~~  &\upsilon_U K_N e^{2Nx^2}\left(1 + o(1)\right)e^{-N(k-1)I_1(x;\sqrtsign{2})}\lim_{\epsilon\searrow 0}\iiint_0^{\pi/2} d\theta d\hat{\theta}d\theta'\iint_0^{\infty}dp_1dp_2 \iint_{\Gamma} dr_1dr_2\notag\\
                  &~~~~~~~~ J_1(p_1, p_2, \theta'; \{s_j\}_{j=1}^r, N)J_2(r_1,r_2, p_1, p_2)\cos^22\theta \sin2\theta\sin2\hat{\theta}\notag\\
& ~~~~~~~~ \exp\Bigg\{-N\Bigg(2\psi^{(+)}_L(r_1; x; \epsilon\cos2\theta\cos2\hat{\theta}) +2\psi^{(+)}_U(r_2; x; \epsilon\cos2\theta\cos2\hat{\theta})\notag\\& ~~~~~~~~~~~~~ +\psi^{(-)}_L(p_1; x; \epsilon\cos2\theta')+\psi^{(-)}_U(p_2; x; \epsilon\cos2\theta')\Bigg)\Bigg\}\end{align}
and 
    \begin{align}
         &\expectGOE \left[|\det(M - xI + S)|\indic [\ind{x} (M+S)\in\{k-1, k, k+1\}]\right]\notag\\ \geq ~~  &\upsilon_L K_N e^{2Nx^2}\left(1 + o(1)\right)e^{-N(k+1)I_1(x;\sqrtsign{2})}\lim_{\epsilon\searrow 0}\iiint_0^{\pi/2} d\theta d\hat{\theta}d\theta'\iint_0^{\infty}dp_1dp_2 \iint_{\Gamma} dr_1dr_2\notag\\
                  &~~~~~~~~ J_1(p_1, p_2, \theta'; \{s_j\}_{j=1}^r, N)J_2(r_1,r_2, p_1, p_2)\cos^22\theta \sin2\theta\sin2\hat{\theta}\notag\\
& ~~~~~~~~ \exp\Bigg\{-N\Bigg(2\psi^{(+)}_L(r_1; x; \epsilon\cos2\theta\cos\hat{\theta}) +2\psi^{(+)}_U(r_2; x; \epsilon\cos2\theta\cos\hat{\theta})\notag\\& ~~~~~~~~~~~~~ +\psi^{(-)}_L(p_1; x; \epsilon\cos2\theta')+\psi^{(-)}_U(p_2; x; \epsilon\cos2\theta')\Bigg)\Bigg\}\end{align}
where the functions $J_1, J_2$, the constant $K_N$ and the functions $\psi^{(\pm)}_{U,L}$ are defined as in Lemma \ref{lemma:nock_deformed}, and the $\upsilon_L, \upsilon_U$ are some constants independent of $N$.
\end{lemma}

\jstat{\begin{remark}
A more general version of this lemma holds with $S$ having any fixed rank $r$. In that case, one considers \begin{equation}
    \expectGOE \left[|\det(M - xI + S)|\indic [\ind{x} (M+S)\in\{k-(r-1),\ldots, k, \ldots, k+(r-1)\}]\right]
\end{equation}
and the statement and proof of the result are immediate extensions of what is given here. We omit this generality, since it is not required here.
\end{remark}}

\begin{proof}
This proof is largely the same as that of Lemma \ref{lemma:nock_deformed}. The first difference arises at (\ref{eq:goe_fourier}), where we are required to compute \begin{equation}\label{eq:goe_fourier_deform}
      \expectGOE \left[e^{-i\Tr MA}\indic[\ind{x}(M+S)=k]\right].
\end{equation} As will become apparent towards the end of this proof, we do not know how to maintain the exact equality constraint\footnote{See Remark \ref{rem:generating_func} below.} on index when $S\neq 0$, hence the slightly relaxed results that we are proving, however we will proceed by performing the calculation for $S=0$ and then show that $S$ can be reintroduced one eigendirection at a time. As in the proof of Theorem A.1 in \cite{auffinger2013random}, we split this expectation by fixing a bound, $R$, for the largest eigenvalue, i.e. \begin{align}
    &\expectGOE \left[e^{-i\Tr MA}\indic[\ind{x}(M)=k]\right] \notag\\ =& \expectGOE \left[e^{-i\Tr MA}\indic[\ind{x}(M)=k, \max\{|\lambda_i(M)|\}_{i=1}^N \leq R]\right]\notag\\ +& \expectGOE \left[e^{-i\Tr MA}\indic[\ind{x}(M)=k, \max\{|\lambda_i(M)|\}_{i=1}^N > R]\right] \label{eq:expect_cond_lemma_split}
\end{align}
We will focus initially on the first expectation on the RHS of (\ref{eq:expect_cond_lemma_split}) and deal with the second term later. Let us abbreviate the notation using $$\mathcal{I}_R(M) = \{\max\{|\lambda_i(M)|\}_{i=1}^N \leq R\}.$$ Recall that $A$ has finite rank and note that $A$ is symmetric without loss of generality, since \begin{equation}
 \Tr M\frac{A + A^T}{2} = \frac{1}{2}\left(\Tr MA + \Tr MA^T \right)= \frac{1}{2}\left( \Tr MA + \Tr AM^T\right) = \Tr MA
\end{equation} and hence $A=\text{diag}(a_1, \ldots, a_{r_A}, 0 \ldots, 0) $ without loss of generality.
We begin by factorising the symmetric matrix $M$ in the GOE integral: \begin{align}\label{eq:expect_cond_lem_1}
   & \expect_M\left[ e^{-i\Tr MA}\indic[\ind{x}(M)=k, \mathcal{I}_R(M)]\right]\notag\\ = &\int \frac{d\mu_E(\Lambda)}{Z_N} \indic[-R \leq \lambda_1 \ldots \leq \lambda_k \leq x\leq \lambda_{k+1} \leq \ldots \lambda_N\leq R] \int d \mu_{Haar}(O) e^{-i\sum_{j=1}^{r_A} a_j \vec{o}_j^T\Lambda\vec{o}_j}
\end{align}
where $\mu_E$ is the un-normalised joint density of ordered GOE eigenvalues, $\mu_{Haar}$ is the Haar measure on the orthogonal group $O(N)$, $\vec{o}_j$ are the rows of the orthogonal matrix $O$ and $Z_N$ is normalisation for the ordered GOE eigenvalues given by the Selberg integral: \begin{align}
    Z_N = \frac{1}{N!} (2\sqrtsign{2})^N N^{-N(N+1)/4} \prod_{i=1}^{N} \Gamma\left(1 + \frac{i}{2}\right).
\end{align}
Much like the proof of Theorem A.1 in \cite{auffinger2013random}, we proceed by
splitting the eigenvalues in (\ref{eq:expect_cond_lem_1}) to enforce the  constraint given by the indicator function:
\begin{align}
      &\expect_M\left[ e^{-i\Tr MA}\indic[\ind{x}(M)=k,\mathcal{I}_R(M)]\right]\notag\\ = &\int d\mu_{Haar}(O)\frac{1}{Z_N}\int_{[-R, x]^k} \prod_{i=1}^k\left( d\lambda_i e^{-N\lambda^2_i/2}\right)\Delta\left(\{\lambda_i\}_{i=1}^k\right)\indic\left[\lambda_1 \leq \ldots \leq \lambda_k\right]  \notag \\
      & \int_{(x,R]^{N-k}} \prod_{i=k+1}^N\left( d\lambda_i e^{-N\lambda^2_i/2}\right)\Delta\left(\{\lambda_i\}_{i=k+1}^N\right)\indic\left[\lambda_{k+1} \leq \ldots \leq \lambda_N\right] \notag\\& e^{-i\sum_{j=1}^{r_A} a_j \vec{o}_j^T\Lambda\vec{o}_j}
       \exp\left(\sum_{j=1}^k\sum_{\ell=k+1}^N \log|\lambda_j - \lambda_{\ell}|\right)\notag\\
        = &\int d\mu_{Haar}(O)\int_{[-R, x]^k} \prod_{i=1}^k\left( d\lambda_i e^{-N\lambda^2_i/2}\right)\Delta\left(\{\lambda_i\}_{i=1}^k\right) \frac{Z_{N-k}}{k!Z_N}  \notag \\
      &\frac{1}{Z_{N-k}(N-k)!} \int_{(x,R]^{N-k}} \prod_{i=k+1}^N\left( d\lambda_i e^{-N\lambda^2_i/2}\right)\Delta\left(\{\lambda_i\}_{i=k+1}^N\right)\notag\\& e^{-i\sum_{j=1}^{r_A} a_j \vec{o}_j^T\Lambda\vec{o}_j}
       \exp\left(\sum_{j=1}^k\sum_{\ell=k+1}^N \log|\lambda_j - \lambda_{\ell}|\right)\notag\\
      =&\int_{[-R_N, x_N]^k}\prod_{i=1}^k\left( d\lambda_i e^{-(N-k)\lambda^2_i/2}\right)\Delta\left(\{\lambda_i\}_{i=1}^k\right) \notag\\
    & \int_{(x_N,R_N]^{N-k}} d\bar{\mu}_E(\Lambda_{N-k}) \int d\mu_{Haar}(O)
    e^{-i\sum_{j=1}^{r_A}\sqrtsign{\frac{N-k}{N}} a_j \vec{o}_j^T\Lambda\vec{o}_j}
    \notag\\
     &\exp\left(\sum_{j=1}^k\sum_{\ell=k+1}^N \log|\lambda_j - \lambda_{\ell}|\right)
        \frac{Z_{N-k}}{k!Z_N} \left(\sqrtsign{\frac{N-k}{N}}\right)^{N + N(N+1)/2}\label{eq:expect_cond_lem_2}
\end{align}
where $x_N\defeq \sqrtsign{\frac{N}{N-k}}x$, $R_N\defeq \sqrtsign{\frac{N}{N-k}}R$ and $\bar{\mu}_E$ is the normalised joint density of un-ordered GOE eigenvalues.

We will first need to deal with the Itzykson-Zuber integral in (\ref{eq:expect_cond_lem_2}) before dealing with the eigenvalue integrals. We follow \cite{guionnet2005fourier}, in particular the proof of Theorem 7 therein. We have the well-known result (Fact 8 in \cite{guionnet2005fourier}) that in the sense of distributions\begin{equation}\label{eq:gram_s}
    (\vec{o}_1, \ldots, \vec{o}_{r_A}) \sim \left(\frac{\tilde{\vec{g}}_1}{||\tilde{\vec{g}}_1||},\ldots, \frac{\tilde{\vec{g}}_{r_A}}{||\tilde{\vec{g}}_{r_A}||}\right)
\end{equation}
where the $(\tilde{\vec{g}}_j)_{j=1}^{r_A}$ are constructed via the Gram-Schmidt process from $(\vec{g}_j)_{j=1}^{r_A}\overset{\text{i.i.d.}}{\sim} \mathcal{N}(\bm{0}, 1)$. (\ref{eq:gram_s}) exactly gives \begin{align}
    \int d\mu_{Haar}(O)
    e^{-i\sum_{j=1}^{r_A}\sqrtsign{\frac{N-k}{N}} a_j \vec{o}_j^T\Lambda\vec{o}_j} = \int \prod_{j=1}^{r_A} \frac{d\vec{g}_j}{\sqrtsign{2\pi}^N} e^{-\frac{\vec{g}_j^2}{2}} \exp\left(-i\sqrtsign{\frac{N-k}{N}}\sum_{j=1}^{r_A} a_j
    \frac{ \tilde{\vec{g}}_j^T\Lambda\tilde{\vec{g}}_j}{ ||\tilde{\vec{g}}_j||^2}\right)
\end{align}
and we will now seek to replace the $\tilde{\vec{g}}_j$ with $\vec{g}_j$ via appropriate approximations. Introduce the event \begin{equation}
    B_N(\upsilon) \defeq\left\{| N^{-1}\langle \vec{g}_i, \vec{g}_j\rangle - \delta_{ij}| \leq N^{-\upsilon}, ~~~ 1\leq i, j \leq r_A\right\}
\end{equation}
and then from \cite{guionnet2005fourier} we immediately conclude that under the i.i.d Gaussian law of the $(\vec{g}_j)_{j=1}^{r_A}$ the complementary event has low probability: \begin{equation}\label{eq:gram_s_bnk}
    \mathbb{P}(B_N(\upsilon)^c) =\mathcal{O}( C(\upsilon) e^{-\alpha N^{1-2\upsilon}})
\end{equation}
where $\alpha, C(\upsilon) > 0$ and we take $0<\upsilon < \frac{1}{2}$ to make this statement meaningful. This enables us to write \begin{align}\label{eq:gram_s_practical}
&\int d\mu_{Haar}(O)
    e^{-i\sum_{j=1}^{r_A}\sqrtsign{\frac{N-k}{N}} a_j \vec{o}_j^T\Lambda\vec{o}_j}\notag\\
    = &\left(1 + \mathcal{O}(e^{-\alpha N^{1-2\upsilon}})\right)\int \prod_{j=1}^{r_A} \frac{d\vec{g}_j}{\sqrtsign{2\pi}^N} e^{-\frac{\vec{g}_j^2}{2}} \exp\left(-i\sqrtsign{\frac{N-k}{N}}\sum_{j=1}^{r_A} a_j
    \frac{ \tilde{\vec{g}}_j^T\Lambda\tilde{\vec{g}}_j}{ ||\tilde{\vec{g}}_j||^2}\right)\indic\{B_N(\upsilon)\}.
\end{align}

Again, directly from \cite{guionnet2005fourier}, given $B_N(\upsilon)$ we have \begin{align}
   ||\tilde{\vec{g}}_j - \vec{g}_j|| \leq N^{\frac{1}{2} - \frac{\upsilon}{2}}
\end{align}
and therefore \begin{align}
    ||\tilde{\vec{g}}_j||^2 = N\left[ 1 + N^{-1}\left(||\tilde{\vec{g}}_j||^2 - ||\vec{g}_j||^2\right) + \left(N^{-1}||\vec{g}_j||^2 - 1\right)\right] = N( 1 + \mathcal{O}(N^{-\upsilon}) )\label{eq:g_tilde_bound_1}
\end{align}
and \begin{align}\label{eq:g_tilde_bound_2}
    \tilde{\vec{g}}_j^T\Lambda \tilde{\vec{g}_j} &=\vec{g}^T\Lambda \vec{g}  + \sum_{i=1}^N (\tilde{g}_i - g_i)^2\lambda_i + 2\sum_{i=1}^N g_i(\tilde{g}_i - g_i)\lambda_i\notag\\
 \implies ~   \Bigg|\frac{\tilde{\vec{g}}_j^T\Lambda \tilde{\vec{g}}_j}{||\tilde{\vec{g}}_j||^2} - \frac{\vec{g}_j^T\Lambda \vec{g}_j}{||\vec{g}_j||^2}\Bigg|  &\lesssim N^{-\frac{\upsilon}{2}}||\Lambda||_{\infty}.
   \end{align}
   We see therefore that, in approximating the $\{\tilde{\vec{g}}_j\}_j$ by $\{\vec{g}_j\}_j$ in (\ref{eq:gram_s_practical}) we introduce an error term in the exponential that is uniformly small in the integration variables $\{\vec{g}_j\}_j.$
Combining (\ref{eq:gram_s_practical}), (\ref{eq:g_tilde_bound_1}) and (\ref{eq:g_tilde_bound_2}) and noting that $||\Lambda||_{\infty} = R_N \sim R$ under the eigenvalue integral in (\ref{eq:expect_cond_lem_2}) gives \begin{align}
        &\int d\mu_{Haar}(O)
    e^{-i\sum_{j=1}^{r_A}\sqrtsign{\frac{N-k}{N}} a_j \vec{o}_j^T\Lambda\vec{o}_j}\notag\\ =  &\left(1 + 
      \mathcal{O}(N^{-\frac{\upsilon}{2}})\right)\int \prod_{j=1}^{r_A} \frac{d\vec{g}_j}{\sqrtsign{2\pi}^N} e^{-\frac{\vec{g}_j^2}{2}} \exp\left(-i\sqrtsign{\frac{N-k}{N}}\sum_{j=1}^{r_A} a_j
    \frac{ \vec{g}_j^T\Lambda\vec{g}_j}{N ( 1 + \mathcal{O}(N^{-\upsilon}))}\right)\notag\\
    =  &\prod_{j=1}^{r_A}\prod_{i=1}^{N}\left(1 + 2iN^{-1}a_j\lambda_i\right)^{-\frac{1}{2}}\left(1 + \mathcal{O}(N^{-\frac{\upsilon}{2}})\right)\notag\\
    =   &\exp\left\{-\frac{N-k}{2}\sum_{j=1}^{r_A} \int d\hat{\mu}_{N-k}(z) \log(1 + 2iN^{-1}a_j z)\right\}\notag\\
    &\exp\left\{-\frac{1}{2} \sum_{j=1}^{r_A}\sum_{i=1}^k\log(1+2iN^{-1}a_j\lambda_i)   \right\}\left(1 + \mathcal{O}(N^{-\frac{\upsilon}{2}})\right) \label{eq:it_zub_split}
\end{align}
where we have defined \begin{equation}
    \hat{\mu}_{N-k} = \frac{1}{N-k}\sum_{i=k+1}^N \delta_{\lambda_i}.
\end{equation}

Following \cite{auffinger2013random}, we now introduce the following function \begin{align}
    \Phi(z, \mu) = -\frac{z^2}{2} + \int d\mu(z') \log|z-z'|
\end{align}
and so and then (\ref{eq:expect_cond_lem_2}) and (\ref{eq:it_zub_split}) can be rewritten as 
\begin{align}
 &\expect_M\left[ e^{-i\Tr MA}\indic[\ind{x}(M)=k,\mathcal{I}_R(M)]\right]\notag\\ = &\int_{[-R_N, x_N]^k}\prod_{i=1}^k d\lambda_i ~ \Delta\left(\{\lambda_j\}_{j=1}^k\right)\exp\left\{-\frac{1}{2} \sum_{j=1}^{r_A}\sum_{i=1}^k\log(1+2iN^{-1}a_j\lambda_i)   \right\}\left(1 + \mathcal{O}(N^{-\frac{\upsilon}{2}})\right) \notag\\
 &\int_{(x_N, R_N]^{N-k}} d\bar{\mu}_E(\Lambda_{N-k}) \notag\exp\left\{-\frac{N-k}{2}\sum_{j=1}^{r_A} \int d\hat{\mu}_{N-k}(z) \log(1 + 2iN^{-1}a_j z)\right\}\\
 &\exp\left((N-k)\sum_{j=1}^k \Phi(\lambda_j, \hat{\mu}_{N-k})\right)    \frac{Z_{N-k}}{k!Z_N} \left(\sqrtsign{\frac{N-k}{N}}\right)^{N + N(N+1)/2}.\label{eq:expect_cond_lem_3}
 \end{align}

We now appeal to the Coulomb gas method \cite{cunden2016shortcut} and in particular the formulation found in \cite{majumdar2011many}. We replace the joint integral of $N-k$ eigenvalues in (\ref{eq:expect_cond_lem_3}) with a functional integral over the continuum eigenvalues density: 
\begin{align}
    &\int_{(x_N, R_N]^{N-k}} d\bar{\mu}_E(\Lambda_{N-K})\exp\left((N-k)\sum_{j=1}^k \Phi(\lambda_j, \hat{\mu}_{N-k})\right)\exp\left\{-\frac{N-k}{2}\sum_{j=1}^{r_A} \int d\hat{\mu}_{N-k}(z)\log\left(1 + \frac{2ia_j z}{N}\right)\right\} \notag\\
    = & \int \mathcal{D}[\mu]e ^{-N^2 \mathcal{S}_x[\mu]} \exp\left((N-k)\sum_{j=1}^k \Phi(\lambda_j, \mu)\right)\exp\left\{-\frac{N-k}{2}\sum_{j=1}^{r_A} \int d\mu(z) \log\left(1 + \frac{2ia_j z}{N}\right)\right\}\label{eq:expect_cond_lem_7}
\end{align}
where the action is defined as \begin{align}
    \mathcal{S}_x[\mu] = &\frac{1}{2}\int dz \mu(z) z^2 - \iint_{z\neq z} dzdz'\mu(z)\mu(z') \log|z-z'| \notag \\
    + &A_1\left(\int dz\theta(R_N - z)\mu(z) - 1\right) + A_2\left(\int dz \mu(z)\theta(z-x) - 1\right) - \Omega
\end{align}
where $\theta$ is the Heaviside step function, $\Omega$ is the constant resulting from the normalisation of the eigenvalue joint density and $A_1,A_2$ are Lagrange multipliers. 

Owing to the $N^2$ rate in (\ref{eq:expect_cond_lem_7}), the integral concentrates around the minimiser of the action. Since $x< -\sqrtsign{2}$ and we have chosen $R>|x|$, it is clear following \cite{majumdar2011many} that the semi-circle law $\mu_{SC}(z) = \pi^{-1}\sqrtsign{2 - z^2}$ minimises this action and further that $\mathcal{S}_x[\mu_{SC}] = 0$, so we have \begin{align}
   & \int \mathcal{D}[\mu]e ^{-N^2 \mathcal{S}_x[\mu]} \exp\left((N-k)\sum_{j=1}^k \Phi(\lambda_j, \mu)\right)\exp\left\{-\frac{N-k}{2}\sum_{j=1}^{r_A} \int d\mu(z) \log(1 + 2iN^{-1}a_j z)\right\}\notag\\
   = &\int_{B_{\delta}(\mu_{SC})} \mathcal{D}[\mu]e ^{-N^2 \mathcal{S}_x[\mu]}\exp\left((N-k)\sum_{j=1}^k \Phi(\lambda_j, \mu)\right)\exp\left\{-\frac{N-k}{2}\sum_{j=1}^{r_A} \int d\mu(z) \log(1 + 2iN^{-1}a_j z)\right\}
 \notag \\ & ~~~~ + e^{-N^2 c_{\delta}}\mathcal{O}(1) \label{eq:expect_cond_lem_8}
\end{align}
where $\delta=\mathcal{O}(N^{-1})$ and $c_{\delta}>0$ is some constant. Performing the usual Laplace method expansion of the action in (\ref{eq:expect_cond_lem_8}) and re-scaling the first non-vanishing derivative to be $\mathcal{O}(1)$, it is clear that the action only contributes a real factor of $\mathcal{O}(1)$ that is independent of the dummy integration variables $\vec{x}_1, \vec{x}_2, \zeta_1, \zeta_1^{\dagger}, \zeta_2, \zeta_2^{\dagger}$ and the other eigenvalues $\lambda_1,\ldots, \lambda_k$ and can therefore be safely summarised as $\mathcal{O}(1)$. Whence   \begin{align}
   & \int \mathcal{D}[\mu]e ^{-N^2 \mathcal{S}_x[\mu]} \exp\left((N-k)\sum_{j=1}^k \Phi(\lambda_j, \mu)\right)\exp\left\{-\frac{N-k}{2}\sum_{j=1}^{r_A} \int d\mu(z) \log(1 + 2iN^{-1}a_j z)\right\}\notag\\
   = &\mathcal{O}(1)\exp\left((N-k)\sum_{j=1}^k \Phi(\lambda_j, \mu_{SC})\right)\exp\left\{-\frac{N-k}{2}\sum_{j=1}^{r_A} \int d\mu_{SC}(z) \log(1 + 2iN^{-1}a_j z)\right\}
 \notag \\ & ~~~~ + e^{-N^2 c_{\delta}}\mathcal{O}(1). \label{eq:expect_cond_lem_9}\end{align}
Now elementary calculations give, noting that the integrand is uniformly convergent in $N$ owing to the compact support of $\mu_{SC}, $\begin{align}
     \int d\mu_{SC}(z) \log(1 + 2iN^{-1}a_j z) &= -\frac{2ia_j}{N}\int d\mu_{SC}(z) z + \frac{2a^2_j}{N^2}\int d\mu_{SC}(z) z^2 + \mathcal{O}(a_j^3N^{-3})\notag\\
     &= \frac{a^2_j}{N^2} ( 1 + \mathcal{O}(a_jN^{-1}))\notag\\
     \implies  \frac{N-k}{2}\sum_{j=1}^{r_A} \int d\mu_{SC}(z) \log(1 + 2iN^{-1}a_j z) &= \frac{\Tr A^2}{2N}  ( 1 + ||A||_{\infty}\mathcal{O}(N^{-1}))\label{eq:expect_cond_lem_10}
\end{align}
where we have implicitly assumed that the spectral radius $||A||_{\infty} \ll N$. This constraint can be introduced by restricting the domains of integration for $\vec{x}_1$ and $\vec{x}_2$ in the anaologue of  (\ref{eq:nock2}) from all of $\mathbb{R}^N$ to balls of radius $o(\sqrtsign{N})$. It is a standard result for Gaussian integrals that this can be achieved at the cost of an exponentially smaller term.
Summarising (\ref{eq:expect_cond_lem_3}), (\ref{eq:expect_cond_lem_7}), (\ref{eq:expect_cond_lem_9}) and (\ref{eq:expect_cond_lem_10}):\begin{align}
 &\expect_M\left[ e^{-i\Tr MA}\indic[\ind{x}(M)=k,\mathcal{I}_R(M)]\right]\notag\\ = &\int_{[-R_N, x_N]^k}\prod_{i=1}^k d\lambda_i ~ \Delta\left(\{\lambda_j\}_{j=1}^k\right)\exp\left\{-\frac{1}{2} \sum_{j=1}^{r_A}\sum_{i=1}^k\log(1+2iN^{-1}a_j\lambda_i)   \right\}\exp\left((N-k)\sum_{j=1}^k \Phi(\lambda_j, \mu_{SC})\right) \notag\\
 &e^{-\frac{\Tr A^2}{2N}}\left(\mathcal{O}(1) + \mathcal{O}(N^{-\frac{\upsilon}{2}}) + \mathcal{O}(N^{-1})||A||_{\infty}\right)   \frac{Z_{N-k}}{k!Z_N} \left(\sqrtsign{\frac{N-k}{N}}\right)^{N + N(N+1)/2}\notag\\
 = &\int_{[-R_N, x_N]^k}\prod_{i=1}^k d\lambda_i ~ \Delta\left(\{\lambda_j\}_{j=1}^k\right)\exp\left((N-k)\sum_{j=1}^k \Phi(\lambda_j, \mu_{SC})\right) \notag\\
 &e^{-\frac{\Tr A^2}{2N}}\left(\mathcal{O}(1) + \mathcal{O}(N^{-\frac{\upsilon}{2}}) + \mathcal{O}(N^{-1})||A||_{\infty}\right)   \frac{Z_{N-k}}{k!Z_N} \left(\sqrtsign{\frac{N-k}{N}}\right)^{N + N(N+1)/2}\notag\\
 = &\int_{[-R_N, x_N]^k}\prod_{i=1}^k d\lambda_i ~ \Delta\left(\{\lambda_j\}_{j=1}^k\right)\exp\left((N-k)\sum_{j=1}^k \Phi(\lambda_j, \mu_{SC})\right) \notag\\
 &e^{-\frac{\Tr A^2}{2N}}\mathcal{O}(1)   \frac{Z_{N-k}}{k!Z_N} \left(\sqrtsign{\frac{N-k}{N}}\right)^{N + N(N+1)/2}\label{eq:expect_cond_lem_11}
 \end{align}
where in the second equality we have Taylor expanded the remaining logarithm and summarised the result with another factor of $(1 + \mathcal{O}(N^{-1})||A||_{\infty})$.

We now wish to follow the proof of Theorem A.1 in \cite{auffinger2013random} and use $\Delta(\{\lambda_j\}_{j=1}^{k}) \leq (2R_N)^k \leq (3R)^k$ for $\lambda_j \in [-R_N, R_N]$ with bound (\ref{eq:expect_cond_lem_11}), however the expectation on the left hand side of (\ref{eq:expect_cond_lem_11}) is not necessarily real. We do however know that the $\mathcal{O}(1)$ term in (\ref{eq:expect_cond_lem_11}) is real to leading order and so we can write  \begin{equation}\label{eq:expect_cond_real_imag}
    \expect_M\left[ e^{-i\Tr MA}\indic[\ind{x}(M)=k,\mathcal{I}_R(M)]\right] = \Re \expect_M\left[ e^{-i\Tr MA}\indic[\ind{x}(M)=k,\mathcal{I}_R(M)]\right] \left( 1 + io(1)\right)
\end{equation}

and thence focus on bounding the real part of the expectation to obtain \begin{align}
     & \Re \expect_M\left[ e^{-i\Tr MA}\indic[\ind{x}(M)=k, \mathcal{I}_R(M)]\right] \notag\\
     \leq &K(3R)^k \frac{Z_{N-k}}{k!Z_N} \left(\sqrtsign{\frac{N-k}{N}}\right)^{N + N(N+1)/2}
    e^{-\frac{\Tr A^2}{2N}} \left(\int_{-R_N}^{x_N} dz e^{(N-k)\Phi(z, \mu)}\right)^k\label{eq:expect_cond_lem_12}
\end{align}
where we have exchanged $\mathcal{O}(1)$ terms for some appropriate constant $K$. Continuing to bound (\ref{eq:expect_cond_lem_12}): \begin{align}
 &\Re \expect_M\left[ e^{-i\Tr MA}\indic[\ind{x}(M)=k, \mathcal{I}_R(M)]\right] \notag\\
     \leq &K(3R)^{2k}\frac{Z_{N-k}}{k!Z_N} \left(\sqrtsign{\frac{N-k}{N}}\right)^{N + N(N+1)/2}
    e^{-\frac{\Tr A^2}{2N}} \exp\left(k(N-k)\sup\limits_{\substack{z\in[-2R, x] \\ \nu\in B_{\delta}(\mu_{SC})}}\Phi(z, \nu)\right)\notag\\
      \leq &K(3R)^{2k} \frac{Z_{N-k}}{k!Z_N} \left(\sqrtsign{\frac{N-k}{N}}\right)^{N + N(N+1)/2}
    e^{-\frac{\Tr A^2}{2N}} e^{-k(N-k)(1/2 + I_1(x; \sqrtsign{2})} \label{eq:expect_cond_lem_upper_bound}
\end{align}
where we have used the same result as used around (A.18) in \cite{auffinger2013random} to take the supremum.

Recalling (\ref{eq:expect_cond_lemma_split}), we can now use (\ref{eq:expect_cond_lem_upper_bound}) and the GOE large deviations principle \cite{arous2001aging} as in \cite{auffinger2013random} to obtain \begin{align}
    \Re\expect_M\left[ e^{-i\Tr MA}\indic[\ind{x}(M)=k]\right] 
      \leq &K''(3R)^k  \frac{Z_{N-k}}{k!Z_N} \left(\sqrtsign{\frac{N-k}{N}}\right)^{N + N(N+1)/2} e^{-k(N-k)(1/2 + I_1(x; \sqrtsign{2}))}
     e^{-\frac{1}{2N}\Tr A^2} + e^{-NR^2}\label{eq:expect_cond_lem_upper_bound_full}
\end{align}
We now seek to obtain a complementary lower bound and again follow \cite{auffinger2013random} in choosing some $y$ and $R'$ such that $y < x < R' < -\sqrtsign{2}$. We then, following a similar procedure as above, find \begin{align}
       \Re\expect_M\left[ e^{-i\Tr MA}\indic[\ind{x}(M)=k]\right] \geq
 & \tilde{K} \frac{Z_{N-k}}{k!Z_N} \left(\sqrtsign{\frac{N-k}{N}}\right)^{N + N(N+1)/2}
     e^{-\frac{1}{2N}\Tr A^2} \exp\left(k(N-k)\sup\limits_{\substack{z\in[y, x] \\ \nu\in B_{\delta}(\mu_{SC})}}\Phi(z, \nu)\right)\end{align}
and taking $y\nearrow x$ we obtain the complement to (\ref{eq:expect_cond_lem_upper_bound_full}): \begin{align}
      \Re\expect_M\left[ e^{-i\Tr MA}\indic[\ind{x}(M)=k]\right] \geq
 & \tilde{K}\frac{Z_{N-k}}{k!Z_N} \left(\sqrtsign{\frac{N-k}{N}}\right)^{N + N(N+1)/2}
    e^{-k(N-k)(1/2 + I_1(x; \sqrtsign{2}))} e^{-\frac{1}{2N}\Tr A^2}.\label{eq:expect_cond_lem_lower_bound}
\end{align}

Next we need the asymptotic beahviour of the Selberg term in  (\ref{eq:expect_cond_lem_upper_bound_full}) and (\ref{eq:expect_cond_lem_lower_bound})

\begin{align}
 T_{N,k} \defeq \frac{Z_{N-k}}{k!Z_N} \left(\sqrtsign{\frac{N-k}{N}}\right)^{N + N(N+1)/2} =   &\underbrace{ \frac{Z_{N-k}(N-k)!}{Z_N N!}\left(\frac{N-k}{N}\right)^{ \frac{(N-k)(N-k+1)}{4} }}_{T_{N,k}'} \notag\\&\frac{N!}{(N-k)!k!}\left(\frac{N-k}{N}\right)^{\frac{N}{2} + \frac{N(N+1)- (N-k)(N-k+1)}{4} }\label{eq:k_theorem_selberg}.
\end{align}
The term $T_{N,k}'$ appears in \cite{auffinger2013random} (defined in A.13) and it is shown there that \begin{equation}
 \lim_{N\rightarrow\infty}   N^{-1}\log T_{N,k}' = \frac{k}{2}.
\end{equation}
Clearly \begin{equation}
   \lim_{N\rightarrow\infty}N^{-1} \log \frac{N!}{(N-k)!k!} = 0
\end{equation}
and it is simple to show that \begin{align}
    \lim_{N\rightarrow\infty} \left(\frac{N-k}{N}\right)^{\frac{N}{2} + \frac{N(N+1)- (N-k)(N-k+1)}{4} } = e^{-\frac{k(k+1)}{2}}
\end{align}
and so we have overall \begin{equation}
    \lim_{N\rightarrow\infty}N^{-1}\log T_{N,k} = \frac{k}{2}.
\end{equation}

So absorbing any $\mathcal{O}(1)$ terms into constants $K_L$ and $K_U$ we have \begin{align}
   K_L e^{-kN(1 + o(1)) I_1(x; \sqrtsign{2})}
     e^{-\frac{1}{2N}\Tr A^2}  \leq  \Re\expect_M\left[ e^{-i\Tr MA}\indic[\ind{x}(M)=k]\right] 
      \leq K_U e^{-kN(1+o(1)) I_1(x; \sqrtsign{2})}
     e^{-\frac{1}{2N}\Tr A^2} \label{eq:expect_cond_lemma_bounds}
\end{align}

Set $S=s_1\vec{e}_1\vec{e}_1^T + s_2\vec{e}_2\vec{e}_2^T$ and $S_1 = s_1\vec{e}_1\vec{e}_1^T$. Suppose $s_1>0$ and $s_2>0$.  By the interlacing property of eigenvalues, we have \begin{equation}\label{eq:interlacing}
    \lambda^{(M)}_1 \leq  \lambda^{(M+S_1)}_1 \leq  \lambda^{(M)}_2 \leq \ldots \leq  \lambda^{(M)}_k \leq  \lambda^{(M+S_1)}_{k} \leq  \lambda^{(M)}_{k+1} \leq \lambda^{(M+S_1)}_{k+1} \leq \ldots \leq \lambda^{(M)}_{N} \leq \lambda^{(M+S_1)}_{N}
\end{equation}

Therefore we have \begin{align}\label{eq:interlace_set}
  \begin{cases}  \{ \ind{x}(M) =k \} \subset \{\ind{x}(M+S_1) \in\{k-1, k\}\} \subset \{\ind{x}(M) \in\{k-1,k, k+1\} \}=\bigsqcup\limits_{j=-1}^1\{\ind{x}(M) = k + j \} ~ &\text{for } k>0,\\
    \{ \ind{x}(M) =k \} \subset \{\ind{x}(M+S_1) =k \} \subset \{\ind{x}(M) \in\{k, k+1\} \}=\bigsqcup\limits_{j=0}^1\{\ind{x}(M) = k + j \} &\text{for } k=0,\\
  \end{cases}
\end{align}
and so (\ref{eq:expect_cond_lemma_bounds}) gives \begin{equation}
\begin{aligned}
      K_L e^{-kN (1+o(1))I_1(x; \sqrtsign{2})}
     e^{-\frac{1}{2N}\Tr A^2}\leq&\Re\expect_M\left[ e^{-i\Tr MA}\indic[\ind{x}(M+S_1)\in\{k-1,k\}]\right]\notag\\ \leq&  3K_U e^{-(k-1)N (1+o(1))I_1(x; \sqrtsign{2})}
     e^{-\frac{1}{2N}\Tr A^2},\\
     e^{-\frac{1}{2N}\Tr A^2}\leq&\Re\expect_M\left[ e^{-i\Tr MA}\indic[\ind{x}(M+S_1)=0]\right] \leq  2K_U
     e^{-\frac{1}{2N}\Tr A^2}.
\end{aligned}    
\end{equation}
We can then extend to $S$ likewise by observing that interlacing gives \begin{align}
       \{ \ind{x}(M+S_1) \in\{k, k+1\} \} &\subset \{\ind{x}(M+S) \in\{k-1, k, k+1\}\} \subset \{\ind{x}(M+S_1) \in\{k-1,k, k+1, k+2\} \}
\end{align} and iterating using (\ref{eq:interlace_set}) yields 
\begin{align}\label{eq:M_S_interlace}
\begin{cases}
       \{ \ind{x}(M) = k+1 \} \subset \{\ind{x}(M+S) \in\{k-1, k, k+1\}\} \subset \bigsqcup\limits_{j=-1}^3\{\ind{x}(M) = k + j \}, ~ &\text{for } k>0\\
      \{ \ind{x}(M) = k+1 \} \subset \{\ind{x}(M+S) \in\{ k, k+1\}\} \subset \bigsqcup\limits_{j=0}^3\{\ind{x}(M) = k + j \}, &\text{for } k=0
\end{cases}
\end{align}
and (\ref{eq:expect_cond_lemma_bounds}) then gives
\begin{equation}\label{eq:expect_cond_lemma_deform_bounds}
\begin{aligned}
    K_L e^{-(k+1)N(1+o(1)) I_1(x; \sqrtsign{2})}
     e^{-\frac{1}{2N}\Tr A^2}&\leq\Re\expect_M\left[ e^{-i\Tr MA}\indic[\ind{x}(M+S)\in\{k-1,k, k+1\}]\right] \\
     &\leq  5K_U e^{-(k-1)N(1+o(1)) I_1(x; \sqrtsign{2})}
     e^{-\frac{1}{2N}\Tr A^2}\\
      K_L e^{-N(1+o(1)) I_1(x; \sqrtsign{2})}
     e^{-\frac{1}{2N}\Tr A^2}&\leq\Re\expect_M\left[ e^{-i\Tr MA}\indic[\ind{x}(M+S)\in\{0, 1\}]\right] 
     \leq  4K_U
     e^{-\frac{1}{2N}\Tr A^2}.
\end{aligned}     
\end{equation}
If instead the \jstat{signs of $s_1, s_2$ are different, then the interlacing will be in the reverse orders}, but the conclusion of (\ref{eq:expect_cond_lemma_deform_bounds}) will be unchanged.
Finally using (\ref{eq:expect_cond_real_imag}) in the analogue of (\ref{eq:nock_initial})

\begin{align}
 &\expect_M[|\det(M - xI + S)|\indic[\ind{x}(M+S)\in\{k-1,k, k+1\}] \notag \\ = &\Re\expect_M[|\det(M - xI + S)|\indic[\ind{x}(M+S)\in\{k-1,k, k+1\}] \notag \\ 
 =&\Re \Bigg\{  K^{(1)}_N\lim_{\epsilon\searrow 0}\int d\vec{x}_1 d\vec{x}_2 d\zeta_1 d\zeta_1^{\dagger} d\zeta_2 d\zeta_2^{\dagger} \exp\left\{-i\vec{x}_1^T(M-(x + i\epsilon)I+S)\vec{x}_1 - i\vec{x}_2^T(M-(x-i\epsilon)I + S)\vec{x}_2\right\}\notag \\
 & ~~~~~~\exp\left\{ i \zeta_1^{\dagger}(M-(x+i\epsilon) I+S)\zeta_1
      + i \zeta_2^{\dagger}(M-(x - i\epsilon)I+S)\zeta_2\right\} \\
  &~~~~~~\expect_M\left[e^{-i\Tr MA}\indic[\ind{x}(M+S)\in\{k-1,k, k+1\}]\right]\Bigg\}\notag \\
 =&\Re \Bigg\{  K^{(1)}_N\lim_{\epsilon\searrow 0}\int d\vec{x}_1 d\vec{x}_2 d\zeta_1 d\zeta_1^{\dagger} d\zeta_2 d\zeta_2^{\dagger} \exp\left\{-i\vec{x}_1^T(M-(x + i\epsilon)I+S)\vec{x}_1 - i\vec{x}_2^T(M-(x-i\epsilon)I + S)\vec{x}_2\right\}\notag \\
 & ~~~~~~\exp\left\{ i \zeta_1^{\dagger}(M-(x+i\epsilon) I+S)\zeta_1
      + i \zeta_2^{\dagger}(M-(x - i\epsilon)I+S)\zeta_2\right\} \\
  &~~~~~~\Re\expect_M\left[e^{-i\Tr MA}\indic[\ind{x}(M+S)\in\{k-1,k, k+1\}]\right](1+io(1))\Bigg\}\notag \\
  =&\Re \Bigg\{  K^{(1)}_N\lim_{\epsilon\searrow 0}\int d\vec{x}_1 d\vec{x}_2 d\zeta_1 d\zeta_1^{\dagger} d\zeta_2 d\zeta_2^{\dagger} \exp\left\{-i\vec{x}_1^T(M-(x + i\epsilon)I+S)\vec{x}_1 - i\vec{x}_2^T(M-(x-i\epsilon)I + S)\vec{x}_2\right\}\notag \\
 & ~~~~~~\exp\left\{ i \zeta_1^{\dagger}(M-(x+i\epsilon) I+S)\zeta_1
      + i \zeta_2^{\dagger}(M-(x - i\epsilon)I+S)\zeta_2\right\} \\
  &~~~~~~\Re\expect_M\left[e^{-i\Tr MA}\indic[\ind{x}(M+S)\in\{k-1,k, k+1\}]\right]\Bigg\}(1+io(1))
\label{eq:expect_cond_lemma_real_bound_expr}
\end{align}

From this point on, the proof proceeds, \emph{mutatis mutandis}, as that for Lemma \ref{lemma:nock_deformed} but applied to the upper and lower bounds on (\ref{eq:expect_cond_lemma_real_bound_expr}) obtained from (\ref{eq:expect_cond_lemma_deform_bounds}). The final range of integration for $p_1$ and $p_2$ will be some intervals $(0, o(1))$ owing to the change of variables used around (\ref{eq:nock7_prime}), but this does not affect the ensuing asymptotics in which the $p_1,p_2$ integration contours are deformed through the saddle point at $z^{(-)}_{U,L}$.
\end{proof}

\begin{remark}\label{rem:generating_func}
We note that if an appropriate generating function for $\indic\left[\ind{x}(M + S)=k\right]$ could be found, that would allow for a straightforward taking of the expectation in (\ref{eq:goe_fourier_deform}), then the calculations of Lemma \ref{lemma:nock_deformed} could be modified to include this extra term and then the desired expectation $\expectGOE \left[|\det(M - xI + S)|\indic [\ind{x} (M+S)=k]\right]$ could be read-off in comparison with the result of Lemma \ref{lemma:nock_deformed}.
\end{remark}

We have established all we need to prove Theorem \ref{thm:auff2.5}.
\auffdepk*
\begin{proof}

First consider $u < -E_{\infty}$. The proof proceeds just as that of Theorem \ref{thm:auff2.8} but applying Lemma \ref{lemma:nock_deformed_cond_k} instead of Lemma \ref{lemma:nock_deformed} and working identically on the upper and lower bounds from Lemma \ref{lemma:nock_deformed_cond_k}.

Now consider $u> -E_{\infty}$. By the interlacing property as used around (\ref{eq:interlacing}), $\ind{x}(M)$ and $\ind{x}(M+S)$ differ by no more than 2. Hence \begin{equation}
    \ind{x}(M+S) \in \mathcal{K} \implies \ind{x}(M) = \mathcal{O}(1)
\end{equation} but for $0 > x > -\sqrtsign{2}$, and $M\sim GOE_N$, the large deviations principle for the GOE \cite{arous1997large} gives \begin{equation}
    \mathbb{P}(\ind{x}(M) = \mathcal{O}(1)) \leq e^{-cN^2}
\end{equation} 
for some constant $c$, hence the $x$ integral analogous to (\ref{eq:thm28_mid}) is exponentially suppressed with quadratic speed in $N$ for $x>-\sqrtsign{2}$. But we have already seen that the integral is only suppressed with linear speed in $N$ for $x < -\sqrtsign{2}$, and further that $\Theta_{H,k}(u)$ is increasing on $(-\infty, -E_{\infty})$ and so, by the Laplace principle, the leading order contribution is from around $x=-\sqrtsign{2}$ and so \begin{equation}
    \lim_{N\rightarrow\infty} \frac{1}{N}\log\expect C_{N,\mathcal{K}}^{h}(\sqrtsign{N}u) = \lim_{N\rightarrow\infty} \frac{1}{N}\log\expect C_{N,\mathcal{K}}^{h}(-\sqrtsign{N}E_{\infty})
\end{equation} for $u > - E_{\infty}$, which completes the proof.

\end{proof}

\begin{remark}
We are clearly unable to provide an exact leading term for $C^h_{N, \mathcal{K}}(\sqrtsign{N}u)$ for any value of $u$ as we did for $C^h_N(\sqrtsign{N}u)$ for $u< -E_{\infty}$ in Theorem \ref{thm:exact_term} because  the presence of $S$ in $\ind{x}(M+S)$ has forced us in Lemma \ref{lemma:nock_deformed_cond_k} to resort to upper and lower bounds on the leading order term. We note that in \cite{auffinger2013random} the authors are also not able to obtain the exact leading term in this case by their rather different methods. Recalling Remark \ref{rem:generating_func}, we conjecture that this term could be obtained by variants of our methods if only a suitable (perhaps approximate) generating function for $\indic[\ind{x}(M+S) = k] $ could be discovered.
\end{remark}

\section{Low rank perturbation of a matrix identity}\label{ap:low_rank_fyod}
\jstat{In this section we establish a modified version of Theorem I from \cite{fyodorov2002characteristic} required in the proof in Lemma \ref{lemma:nock_deformed}. In that Lemma, we are faced with an integral of the form \begin{equation}\label{eq:ap_fyod_deformed}
    \mathcal{I}_N(F; S) = \iint_{\mathbb{R}^N} d\vec{x}_1 d\vec{x}_2 F(Q_B) e^{-iN\Tr SB}
\end{equation}
where the $N\times N$ matrix $B$ is defined as $B=\vec{x}_1\vec{x}_1^T + \vec{x}_2\vec{x}_2^T$, the $2\times 2$ matrix $Q_B$ is given by \begin{equation}
    Q_B = \left(\begin{array}{cc} \vec{x}_1^T\vec{x}_1 & \vec{x}_1^T\vec{x}_2 \\ \vec{x}_2^T\vec{x}_1 & \vec{x}_2^T\vec{x}_2\end{array}\right),
\end{equation}
 $F$ is some suitably nice function and $S$ is some real symmetric matrix of rank $r=\mathcal{O}(1)$ as $N\rightarrow \infty$ and with non-zero eigenvalues $\{N^{-1/2}s_i\}_{i=1}^r$ for $s_i=\mathcal{O}(1)$.
It is sufficient to be able to evaluate a leading order term of $\mathcal{I}_N$ in an expansion for large $N$. \cite{fyodorov2002characteristic} proves the following related result:
\begin{lemma}[\cite{fyodorov2002characteristic}\label{lemma:fyod_lemma} Theorem I]
Given $m$ vectors in $\mathbb{R}^N$ $\vec{x}_1, \ldots, \vec{x}_m$, denote by $Q(\vec{x}_1, \ldots, \vec{x}_m)$ the $m\times m$ matrix whose entries are given by $Q_{ij} = \vec{x}_i^T\vec{x}_j$. Let $F$ be any function of an $m\times m$ matrix such that the integral \begin{equation}
     \int_{\mathbb{R}^N}\ldots\int_{\mathbb{R}^N}d\vec{x}_1\ldots d\vec{x}_m |F(Q)|
\end{equation}
exists and define the integral \begin{equation}
     \mathcal{J}_{N, m}(F) \defeq \int_{\mathbb{R}^N}\ldots\int_{\mathbb{R}^N}d\vec{x}_1\ldots d\vec{x}_m F(Q).
\end{equation} Then we have \begin{equation}
    \mathcal{J}_{N,m}(F) = \frac{\pi^{\frac{m}{2}\left(N - \frac{m-1}{2}\right)}}{\prod_{k=0}^{m-1}\Gamma\left(\frac{N-k}{2}\right)}\int_{\text{Sym}_{\geq 0}(m)}d\hat{Q} \left(\det \hat{Q}\right)^{\frac{N-m-1}{2}}F(\hat{Q}).
\end{equation}
 \end{lemma}
We will prove the following perturbed version of this result and in greater generality than is required in the present work.
\fyodgeneral*}
\begin{proof}
\jstat{The proof of Lemma \ref{lemma:fyod_lemma} presented in Appendix D of \cite{fyodorov2002characteristic} proceeds by induction on $m$ and relies on writing the integration vector $\vec{x}_m$ as $\vec{x}_m = \rho_m O_m\vec{e}_N$ where $\vec{e}_N$ is the $N$-th basis vector in the chosen orthonormal basis, $\rho_m>0$ is a scalar variable and $O_m$ is an orthogonal matrix. The proof proceeds by making a change of variables for the first $m-1$ integration vectors and then finding that the integrand does not depend on $O_m$ and so the integral over $O_m$ with respect to the Haar measure just contributes a volume factor of \begin{equation}
    \frac{2\pi^{N/2}}{\Gamma(N/2)}.
\end{equation}
It is at this point where the $e^{-iN\Tr SB}$ term in (\ref{eq:ap_fyod_deformed}) causes problems because a dependence on $O_m$ remains. Indeed, we have \begin{align}
  \vec{x}_m^T S\vec{x}_m = \rho_m \vec{e}_N^TO^T_m S O_m \vec{e}_N.
\end{align}
Since $S$ is real symmetric we may take, WLOG, $S = N^{\alpha}\text{diag}(s_1, \ldots, s_r, 0, \ldots, 0).$ Then \begin{align}
     e^{-iN\vec{x}_m^T S\vec{x}_m} = e^{-iN^{1 + \alpha}\rho_m \sum_{j=1}^r s_j (o_{Nj})^2}
\end{align}
where $o_{Nj}$ is the $j$-th component of the $N$-th column of $O$. Proceeding with an evaluation of an integral like (\ref{eq:ap_fyod_deformed}) then requires the evaluation of the integral \begin{equation}\label{eq:ap_hciz}
    \int_{O(N)} d\mu_{\text{Haar}}(O_m) e^{-iN^{1 + \alpha}\rho_m \sum_{j=1}^r s_j (o_{Nj})^2}.
\end{equation}
We can now follow \cite{guionnet2005fourier}, in particular the proof of Theorem 7 therein.  We have the well-known result (Fact 8 in \cite{guionnet2005fourier}) that in the sense of distributions\begin{equation}
    (\vec{o}_1, \ldots, \vec{o}_{p}) \sim \left(\frac{\tilde{\vec{g}}_1}{||\tilde{\vec{g}}_1||},\ldots, \frac{\tilde{\vec{g}}_{p}}{||\tilde{\vec{g}}_{p}||}\right)
\end{equation}
for any $p=\mathcal{O}(1)$ and where the $(\tilde{\vec{g}}_j)_{j=1}^{p}$ are constructed via the Gram-Schmidt process from $(\vec{g}_j)_{j=1}^{r_A}\overset{\text{i.i.d.}}{\sim} \mathcal{N}(\bm{0}, 1)$. So in particular \begin{align}\label{eq:ap_gram_s}
    \vec{o}_N \sim \frac{\vec{g}}{||\vec{g}||}, ~~ \vec{g}\sim \mathcal{N}(0,1).
\end{align}}

\jstat{(\ref{eq:ap_gram_s}) then exactly gives 
\begin{align}
    \int_{O(N)} d\mu_{\text{Haar}}(O_m) e^{-iN^{1 + \alpha}\rho_m \sum_{j=1}^r s_j (o_{Nj})^2} &= \int_{\mathbb{R}^N} \frac{d\vec{g}}{(2\pi)^{N/2}} e^{-\frac{\vec{g}^2}{2}} \exp\left(-iN^{1+\alpha}\rho_m \sum_{j=1}^r s_j \frac{g_j^2}{||\vec{g}||^2}\right)
\end{align}}

 \jstat{Introduce the event \begin{equation}
    B_N(\upsilon) \defeq\left\{| N^{-1}\langle \vec{g}, \vec{g}\rangle - 1| \leq N^{-\upsilon} \right\}
\end{equation}
and then from \cite{guionnet2005fourier} we immediately conclude that under the i.i.d Gaussian law of  $\vec{g}$ the complementary event has low probability: \begin{equation}\label{eq:gram_s_bnk2}
    \mathbb{P}(B_N(\upsilon)^c) =\mathcal{O}( C(\upsilon) e^{-\beta N^{1-2\upsilon}})
\end{equation}
where $\beta, C(\upsilon) > 0$ and we take $0<\upsilon < \frac{1}{2}$ to make this statement meaningful. This enables us to write \begin{align}
  \int_{O(N)} d\mu_{\text{Haar}}(O_m) &e^{-iN^{1 + \alpha}\rho_m \sum\limits_{j=1}^r s_j (o_{Nj})^2} \notag\\=  \left(1 + \mathcal{O}(e^{-\beta N^{1-2\upsilon}})\right)\int_{\mathbb{R}^N} &\frac{d\vec{g}}{(2\pi)^{N/2}} e^{-\frac{\vec{g}^2}{2}} \exp\left(-iN^{1+\alpha}\rho_m \sum_{j=1}^r s_j \frac{g_j^2}{||\vec{g}||^2}\right)\indic\{B_N(\upsilon)\}\notag \\
  = \left(1 + \mathcal{O}(e^{-\beta N^{1-2\upsilon}})\right)\int_{\mathbb{R}^N}& \frac{d\vec{g}}{(2\pi)^{N/2}}\indic\{B_N(\upsilon)\}\notag \\ &  e^{-\frac{\vec{g}^2}{2}} \exp\left(-iN^{\alpha}(1+\mathcal{O}(N^{-\upsilon}))\rho_m \sum_{j=1}^r s_j g_j^2\right)\label{eq:ap_gs_practical_fail}
\end{align}
but given $B_N(\upsilon)$ we have $g_j^2 \lesssim N$ for all $j=1,\ldots, N$ and so we do not, as it stands, have uniformly small error terms. We can circumvent this by introducing the following event for $0<\eta<\frac{1}{2}$: \begin{equation}
   E_N^{(r)}(\eta) = \{|g_j|\leq N^{\frac{1}{2} - \eta} ~ \text{ for } j=1,\ldots, r\}.
\end{equation} 
Let us use $\hat{\vec{g}}$ to denote the $N-r$ dimensional vector with components $\left(g_{r+1}, \ldots, g_N\right)$. }

\jstat{Then we have \begin{align}
 \Bigg| |N^{-1}||\hat{\vec{g}}||^2 - 1 | - N^{-1}\sum_{i=1}^r g_j^2  \Bigg| \leq  |N^{-1}||\vec{g}||^2 - 1 | \leq |N^{-1}||\hat{\vec{g}}||^2 - 1 | + N^{-1}\sum_{i=1}^r g_j^2
\end{align}
so if $\eta > \frac{\upsilon}{2}$ then it follows that \begin{equation}
    B_N(\upsilon) ~|~ E_N^{(r)}(\eta) = B_{N-r}(\upsilon).
\end{equation}
But we also have (e.g. \cite{andrews_askey_roy_1999} Appendix C) \begin{align}
    \mathbb{P}( E_N^{(r)}(\eta)) = \left[\text{erf}\left(N^{\frac{1}{2}-\eta}\right)\right]^r = \left[1 - \mathcal{O}(N^{\frac{1}{2} - \eta} e^{-N^{1-2\eta}})\right]^r = 1 - \mathcal{O}(N^{\frac{1}{2} - \eta} e^{-N^{1-2\eta}})
\end{align}
and so  (taking $\eta> \upsilon$, say)\begin{equation}
    \mathbb{P}\left(B_N(\upsilon)\cap E^{(r)}_N(\eta)\right) =     \mathbb{P}\left(B_N(\upsilon) ~|~ E^{(r)}_N(\eta)\right)    \mathbb{P}\left(E^{(r)}_N(\eta)\right) =1 - \mathcal{O}(e^{-\alpha N^{1-2\upsilon}})
\end{equation}
and thus we can replace (\ref{eq:ap_gs_practical_fail}) with 
\begin{align}
  \int_{O(N)} d\mu_{\text{Haar}}(O_m) e^{-iN^{1 + \alpha}\rho_m \sum\limits_{j=1}^r s_j (o_{Nj})^2} = \left(1 + \mathcal{O}(e^{-\beta N^{1-2\upsilon}})\right)\int_{\mathbb{R}^N}& \frac{d\vec{g}}{(2\pi)^{N/2}}\indic\{B_N(\upsilon)\cap E^{(r)}_N(\eta)\}\notag \\ &  e^{-\frac{\vec{g}^2}{2}} \exp\left(-iN^{\alpha}(1+\mathcal{O}(N^{-\upsilon}))\rho_m \sum_{j=1}^r s_j g_j^2\right)\notag\\
  \label{eq:ap_gs_practical_success}
\end{align}
but now $N^{\alpha-\upsilon} g_j^2 \leq N^{\alpha + 1 -\upsilon - 2\eta} \leq N^{\alpha + 1 -3\upsilon} \rightarrow 0 $ as $N\rightarrow \infty$ so long as we choose $\upsilon > \frac{\alpha + 1}{3}$. Given that $\alpha< 1/2$, this choice is always possible for $0< \upsilon < 1/2$. Thus the error term in the exponent of (\ref{eq:ap_gs_practical_success}) is in fact uniformly small in $\vec{g}$ and so we obtain 
\begin{align}
  &\int_{O(N)} d\mu_{\text{Haar}}(O_m) e^{-iN^{1 + \alpha}\rho_m \sum\limits_{j=1}^r s_j (o_{Nj})^2}\notag\\ = &\left(1 + o(1)\right)\int_{\mathbb{R}^N} \frac{d\vec{g}}{(2\pi)^{N/2}}\indic\{B_N(\upsilon)\cap E^{(r)}_N(\eta)\}  \exp\left(-\frac{\vec{g}^2}{2}-iN^{\alpha}\rho_m \sum_{j=1}^r s_j g_j^2\right)\notag\\
   = &\left(1 + o(1)\right)\int_{\mathbb{R}^r} \frac{dg_1\ldots dg_r}{(2\pi)^{r/2}} \exp\left(-\frac{1}{2}\sum_{j=1}^r\left\{1+\vivacom{2}iN^{\alpha}\rho_m  s_j\right\} g_j^2\right)\notag \\
  = &(1 + o(1)) \prod_{j=1}^r \left( 1+ \vivacom{2}iN^{\alpha} \rho_m s_j\right)^{-\frac{1}{2}}.\label{eq:ap_haar_finished}
\end{align}
In the induction step in the proof of \cite{fyodorov2002characteristic}, $\rho_m$ becomes the new diagonal entry of the expanded $\hat{Q}$ matrix. Combining (\ref{eq:ap_haar_finished}) with that proof gives the result
 \begin{align}\label{eq:fyod_lemma_result}
    \mathcal{I}_N(F; S) =(1+ o(1))\frac{\pi^{N - \frac{1}{2}}(1+ o(1))}{\Gamma\left(\frac{N}{2}\right)\Gamma\left(\frac{N-1}{2}\right)}\int_{\text{Sym}_{\geq 0}(m)}d\hat{Q} \left(\det \hat{Q}\right)^{\frac{N-3}{2}}F(\hat{Q}) \prod_{j=1}^r\prod_{i=1}^N \left( 1+ \vivacom{2}iN^{\alpha}\hat{Q}_{ii}s_j\right)^{\vivacom{-1/2}}.
\end{align}}\end{proof}

\vivacom{\begin{remark}
    We note a comparison between Lemma \ref{lem:fyod_general} and the theorem in Appendix C of \cite{fyodorov2019spin}. That result is exact and holds for general functions of projections $\vec{x}_i^T\vec{s}$ onto some arbitrary fixed vector $\vec{s}$, so it is a generalisation of our Lemma \ref{lem:fyod_general} for $r=1$, however it \emph{only} applies to $r=1$. In \cite{fyodorov2019spin}, the function $F(Q_B)$ (in our notation) is replaced by the more general $\mathcal{F}(Q_B; \vec{s}_B)$ where the vector $\vec{s}_B$ has entries $(\vec{s}_B)_i = \vec{s}^T\vec{x}_i$ and $\vec{s}$ is an arbitrary vector. The result analogous Lemma \ref{lem:fyod_general} is 
    \begin{equation}
    \mathcal{J}_{N,m}(\mathcal{F}; \vec{s}) \propto \int_{\text{Sym}_{\geq 0}(m)}d\hat{Q}\int_{\R^m}d\vec{t} \left(\det \hat{Q}\right)^{\frac{N-m-2}{2}}\mathcal{F}(\hat{Q} + \vec{t}\vec{t}^T; \|\vec{s}\|\vec{t}),
\end{equation}
where we omit the constant multiplicative factor since we are content to verify that the functional form agrees with Lemma \ref{lem:fyod_general}.
To use this theorem in the case of Lemma \ref{lem:fyod_general}, the vector $\vec{s}$ is chosen to have norm $\|\vec{s}\|_2 = N^{\alpha/2}s_1^{1/2}$, where $s_1$ is the single non-zero eigenvalue of the rank 1 matrix $S$ and $\mathcal{F}(Q_B; \vec{s}_B) = F(Q_B)e^{-iN\sum_{j=1}^m (\vec{x}_j^T\vec{s})^2}$. With these choices
    \begin{align}
    \mathcal{J}_{N,m}(\mathcal{F}; \vec{s}) &\propto \int_{\text{Sym}_{\geq 0}(m)}d\hat{Q}\int_{\R^m}d\vec{t} \left(\det \hat{Q}\right)^{\frac{N-m-2}{2}}F(\hat{Q} + \vec{t}\vec{t}^T) e^{-iN^{1+\alpha}s_1 \vec{t}^2}\notag\\
 &=  \int_{\R^m}d\vec{t}   \int_{\text{Sym}_{\geq 0}(m)}d\hat{Q} \1\{\vec{t}^T\hat{Q}^{-1}\vec{t} < 1\}\left(\det \hat{Q}- \vec{t}\vec{t}^T\right)^{\frac{N-m-2}{2}}F(\hat{Q}) e^{-iN^{1+\alpha}s_1\vec{t}^2}\notag\\
 &=  \int_{\R^m}d\vec{t}   \int_{\text{Sym}_{\geq 0}(m)}d\hat{Q} \1\{\vec{t}^T\hat{Q}^{-1}\vec{t} < 1\}\det \hat{Q}^{\frac{N-m-2}{2}} (1 - \vec{t}^T\hat{Q}^{-1}\vec{t})^{\frac{N-m-2}{2}}F(\hat{Q}) e^{-iN^{1+\alpha}s_1 \vec{t}^2}\notag\\
 &=  \int_{\|\vec{t}\|_2<1}d\vec{t}   \int_{\text{Sym}_{\geq 0}(m)}d\hat{Q}\det \hat{Q}^{\frac{N-m-1}{2}} (1 - \vec{t}^2)^{\frac{N-m-2}{2}}F(\hat{Q}) e^{-iN^{1+\alpha}s_1\vec{t}^T\hat{Q}\vec{t}}.\label{eq:yan_lemma_remark}
\end{align}
Now \begin{align*}
     \int_{\|\vec{t}\|_2<1}d\vec{t}  (1 - \vec{t}^2)^{\frac{N-m-2}{2}}e^{-iN^{1+\alpha}s_1\vec{t}^T\hat{Q}\vec{t}} &= \int_{\|\vec{t}\|_2<1}d\vec{t} \exp\left\{-N\left(iN^{\alpha}s_1 \vec{t}^T\hat{Q}\vec{t} - \frac{N-m-2}{2N} \log(1 - \vec{t}^2)\right)\right\},
\end{align*}
and so we can evaluate the integral over $\vec{t}$ asymptotically. The saddle point is clearly at $
\vec{t} = 0$, so the leading order contribution as $N\rightarrow\infty$ is from around this point. We proceed by expanding the logarithm and evaluating the integral one coordinate at a time. Also note that $\frac{N-m-2}{2N} \sim \frac{1}{2}$ for large $N$. Thus, writing $\vec{t} = (\vec{t}', t_m)$,
\begin{align*}
     \int_{\|\vec{t}\|_2<1}d\vec{t}  (1 - \vec{t}^2)^{\frac{N-m-2}{2}}e^{-iN^{1+\alpha}s_1\vec{t}^T\hat{Q}\vec{t}} &\sim  \int_{\|\vec{t}\|_2<1}d\vec{t} \exp\left\{-N\left(iN^{\alpha}s \vec{t}^T\hat{Q}\vec{t} + \frac{1}{2}\vec{t}^2 \right)\right\}\\
     &\sim \int_{\|\vec{t}'\|_2<1-\epsilon^2}d\vec{t}'\int_{-\epsilon}^{\epsilon}dt_m \exp\Bigg\{-N\Bigg( 
    \frac{1}{2} t_m^2 (2\hat{Q}_{mm}iN^{\alpha}s_1 + 1)\\
    & ~~~~~~~~~~~ + 2t_m \sum_{j\neq m} \hat{Q}_{mj}t_j'
      + \vec{t}'^T\hat{Q}'\vec{t}' + \frac{1}{2}\vec{t}'^2 \Bigg)\Bigg\}
\end{align*}
where $\hat{Q}'$ is the $m-1 \times m-1$ top left block of $\hat{Q}$ and $\epsilon \ll 1$. Completing the square and applying Laplace's method to the $t_m$ integral gives 
\begin{align*}
     \int_{\|\vec{t}\|_2<1}d\vec{t}  (1 - \vec{t}^2)^{\frac{N-m-2}{2}}e^{-iN^{1+\alpha}s_1\vec{t}^T\hat{Q}\vec{t}}
     &\sim N^{-1/2}\int_{\|\vec{t}'\|_2<1}d\vec{t}' (2\hat{Q}_{mm}iN^{\alpha}s_1 + 1)^{-1/2} \exp\left\{-N\left( \vec{t}'^T\hat{Q}'\vec{t}' + \frac{1}{2}\vec{t}'^2 \right)\right\}
\end{align*}
and so one can clearly iterate to obtain 
\begin{align*}
     \int_{\|\vec{t}\|_2<1}d\vec{t}  (1 - \vec{t}^2)^{\frac{N-m-2}{2}}e^{-iN^{1+\alpha}s_1\vec{t}^T\hat{Q}\vec{t}}
     &\sim N^{-m/2}\prod_{j=1}^m(2\hat{Q}_{jj}iN^{\alpha}s + 1)^{-1/2} 
\end{align*}
and so, recalling (\ref{eq:yan_lemma_remark}), we obtain the same expression as Lemma \ref{lem:fyod_general} (up-to un-tracked constants). 
\end{remark}}

\section{Conclusion}
The interpretation of the results we have presented in this chapter is largely the same as that first given in \cite{choromanska2015loss}. \jstat{Under the chosen modeling assumptions}, the local optima of the the neural network loss surface are arranged so that, above a critical value $-\sqrtsign{N}E_{\infty}$, it is overwhelmingly likely that gradient descent will encounter high-index optima and so `escape' and descend to lower loss.  Below $-\sqrtsign{N}E_{\infty}$, the low-index optima are arranged in a `banded' structure, however, due to the imprecision of Theorem \ref{thm:auff2.5}, the bands are slightly blurred when compared with \cite{choromanska2015loss}. We display the differences in Table \ref{tab:banded}.
\begin{table}[]
    \centering
    \begin{tabular}{c|c|c}
       band  & possible indices \cite{choromanska2015loss} & \textbf{possible indices}  \\
       \hline
        $(-\sqrtsign{N}E_{0},-\sqrtsign{N}E_{1})$  &0&0,1,2\\
        $(-\sqrtsign{N}E_{1},-\sqrtsign{N}E_{2})$  &0,1 &0,1,2,3\\
        $(-\sqrtsign{N}E_{2},-\sqrtsign{N}E_{3})$ & 0,1,2 & 0,1,2,3,4\\
        $(-\sqrtsign{N}E_{3},-\sqrtsign{N}E_{4})$ & 0,1,2,3 & 0,1,2,3,4,5\\
    \end{tabular}
        \caption{Illustration of the banded low-index local optima structure obtained here for neural networks with general activation functions and compared to the analogous results in \cite{choromanska2015loss}.}
    \label{tab:banded}
\end{table}

Our results have plugged a gap in the analysis of \cite{choromanska2015loss} by demonstrating that the specific \texttt{ReLU} activation function required by the technicalities of their derivation is not, in fact, a requirement of the results themselves, which we have shown to hold for any reasonable choice of activation function. At the same time, experimental results imply that a sufficiently precise model for deep neural network loss surfaces should display some non-trivial dependence on the choice of activation function, but we have shown that no dependence at all is seen at the relevant level of logarithmic asymptotic complexity, but is visible in the sharp leading order complexity. In defense of \cite{choromanska2015loss}, we have reduced the scope for their results to be some spurious apparition of an intersection of several unrealistic simplifications. However, with the same result, we have demonstrated an important aspect of neural network architectural design to which the multi-spin glass correspondence is entirely insensitive, so limiting the precision of any statements about real neural networks that can be made using this analysis.

In the pursuit of our aims, we have been forced to approximately reproduce the work of \cite{auffinger2013random} by means of the supersymmetric method of Random Matrix Theory, which we believe is quite novel and have also demonstrated how various steps in these supersymmetric calculations can be adapted to the setting of a GOE matrix deformed by some low-rank fixed matrix including utilising Gaussian approximations to orthogonal matrices in ways we have not previously seen in the literature. \jstat{We believe some of our intermediate results and methods may be of use in other contexts in Random Matrix Theory.}

As highlighted in the main text, there are a few areas for future work that stem immediately from our calculations. We list them here along with other possibilities.
\begin{enumerate}
    \item Constructing an appropriate indicator function (or approximate indicator function) for the index of a matrix so that Theorem \ref{thm:auff2.5} can be precised and to obtain exact leading order terms for $C^h_{N,k}$ that could not be obtained in \cite{auffinger2013random} (see Remark \ref{rem:exact_k_ind}).
    \item The `path-independence' assumption (Section \ref{subsec:modelling_assumptions}, assumption \ref{item: assumption_bernoulli}) is the weakest link in this work (and that of \cite{choromanska2015loss}) and we have shed further light on its validity through experimentation (Section \ref{subsec:discussion_assumptions}). The supersymmetric calculations used here have shown themselves to be powerful and quite adaptable. We therefore suggest that it may be possible to somehow encapsulate the failure of assumption \ref{item: assumption_bernoulli} as a first-order correlation term and repeat the presented analysis in an expansion when this term is small. 
    \item Further, this work and others mentioned in the introduction have shown that studying spin glass like objects in this context is a fruitful area of research and so we would like to study more exotic glassy objects inspired by different neural network architectures and applications and hope to be able to adapt the calculations presented here to such new scenarios. 
\end{enumerate}

%
%
\let\textcircled=\pgftextcircled
\chapter{A spin glass model for generative adversarial networks}
\label{chap:spin_glass_gans}
The content of this chapter was published first as a pre-print in January 2021 (\url{https://arxiv.org/abs/2101.02524}) and later as a journal article: ``A spin glass model for the loss surfaces of generative adversarial networks''. \textbf{Nicholas P
Baskerville}, Jonathan P Keating, Francesco Mezzadri and Joseph Najnudel. \emph{Journal of
Statistical Physics}, 186(2):1-45 2022.
\medskip

\textbf{NPB} suggested the topic, performed most of the calculations and experiments and wrote
the paper. The other authors contributed ideas for possible approaches, provided feedback
on results throughout and made small revisions to the drafts. \text{NPB} and JN collaborated on
the proof of Lemma 4. Jonathan Hodgson helped considerably with the design of Figure 6.
Anonymous reviewers spotted some minor errors, advised on changes of presentation and
extra experiments and provided useful references.

\section{An interacting spin glass model}\label{sec:model}
We use multi-spin glasses in high dimensions as a toy model for neural network loss surfaces without any further justification, beyond that found in \cite{choromanska2015loss} and Chapter \ref{chap:general_activation_functions}. GANs are composed of two networks: \emph{generator} ($G$) and \emph{discriminator} ($D$). $G$ is a map $\mathbb{R}^m\rightarrow\mathbb{R}^d$ and $D$ is a map $\mathbb{R}^d\rightarrow\mathbb{R}$. $G$'s purpose is to generate synthetic data samples by transforming random input noise, while $D$'s is to distinguish between real data samples and those generated by $G$. Given some probability distribution $\mathbb{P}_{data}$ on some $\mathbb{R}^d$, GANs have the following minimax training objective \begin{align}\label{eq:gan}
    \min_{\Theta_G}\max_{\Theta_D}\left\{\mathbb{E}_{\vec{x}\sim \mathbb{P}_{data}} \log D(\vec{x}) + \mathbb{E}_{\vec{z}\sim \mathcal{N}(0, \sigma_z^2)}\log(1 - D(G(\vec{z})))\right\},
\end{align}
where $\Theta_D, \Theta_G$ are the parameters of the discriminator and generator respectively. With $\vec{z}\sim\mathcal{N}(0, \sigma_z^2)$, $G(\vec{z})$ has some probability distribution $\mathbb{P}_{gen}$. When successfully trained, the initially unstructured $\mathbb{P}_{gen}$ examples are easily distinguished by $D$, this in turn drives improvements in $G$, bring $\mathbb{P}_{gen}$ closer to $\mathbb{P}_{data}$. Ultimately, the process successfully terminates when $\mathbb{P}_{gen}$ is very close to $\mathbb{P}_{data}$ and $D$ performs little better than random at the distinguishing task. To construct our model, we introduce two spin glasses:
\begin{align}
    \lD(\vwD) &= \sum_{i_1,\ldots, i_p=1}^{N_D} X_{i_1,\ldots, i_p} \prod_{k=1}^p \wD_{i_k}\end{align}
    \begin{align}
    \lG(\vwD, \vwG) &= \sum_{i_1,\ldots, i_{p+q}=1}^{N_D+N_G}  Z_{i_1,\ldots, i_{p+q}} \prod_{k=1}^{p+q} w_k
\end{align}
where $\vec{w}^T = ({\vwD}^T, {\vwG}^T)$, all the $X_{i_1,\ldots, i_p}$ are i.i.d. $\mathcal{N}(0,1)$ and $Z_{j_1,\ldots, j_{p+q}}$ are similarly i.i.d. $\mathcal{N}(0,1)$. We then define the models for the discriminator and generator losses: \begin{align}
    \LD(\vwD, \vwG) &= \lD(\vwD) - \sigma_z\lG(\vwD, \vwG),\label{eq:ld_def}\\
    \LG(\vwD, \vwG) &= \sigma_z \lG(\vwD,\vwG).\label{eq:lg_def}
\end{align}

$\lD$ plays the role of the loss of the discriminator network when trying to classify genuine examples as such. $\lG$ plays the role of loss of the discriminator when applied to samples produced by the generator, hence the sign difference between $\LD$ and $\LG$. $\vwD$ are the weights of the discriminator, and $\vwG$ the weights of the generator. The $X_{\vec{i}}$ are surrogates for the training data (i.e. samples from $\mathbb{P}_{data}$) and the $Z_{\vec{j}}$ are surrogates for the noise distribution of the generator. For convenience, we have chosen to pull the $\sigma_z$ scale outside of the $Z_{\vec{j}}$ and include it as a constant multiplier in (\ref{eq:ld_def})-(\ref{eq:lg_def}). In reality, we should like to keep $Z_{\vec{j}}$ as i.i.d. $\mathcal{N}(0,1)$ but take $X_{\vec{i}}$ to have some other more interesting distribution, e.g. normally or uniformly distributed on some manifold. Using $[x]$ to denote the integer part of $x$, we take $N_D = [\kappa N], N_G = [\kappa' N]$ for fixed $\kappa\in(0,1)$, $\kappa'=1-\kappa$, and study the regime $N\rightarrow\infty$. Note that there is no need to distinguish between $[\kappa N]$ and $\kappa N$ in the $N\rightarrow\infty$ limit.

\begin{remark} Our model is not supposed to have any direct relationship to GANs. Rather, we have used two spin glasses as models for high-dimensional random surfaces. The spin glasses are related by sharing some of their variables, namely the $\vwD$, just as the two training objectives in GANs share the discriminator weights. In prior work modeling neural network loss surfaces as spin glasses, the number of spins corresponds to the number of layers in the network, therefore we have chosen $p$ spins for $\lD$ and $p+q$ for $\lG$, corresponding to $p$ layers in the discriminator and $q$ layers in the generator, but the generator is only ever seen in the losses composed with the discriminator. One could make other choices of $\lD$ and $\lG$ to couple the two glasses and we consider one such example in the appendix Section \ref{app:bipartite}. \end{remark}

\section{Kac-Rice formulae for complexity}\label{sec:kac}
Training GANs involves jointly minimising the losses of the discriminator and the generator. Therefore, rather than being interested simply in upper-bounding a single spin-glass and counting its stationary points, the complexity of interest comes from jointly upper bounding both $L^{(D)}$ and $L^{(G)}$ and counting points where both are stationary. Using $S^{M}$ to denote the $M$-sphere\footnote{We use the convention of the $M$-sphere being the sphere embedded in $\mathbb{R}^M$.}, we define the complexity

\begin{align}
    C_{N} =\Bigg|\left\{ \vwD\in S^{N_D}, \vwG\in S^{N_G } ~:~ \nabla_D \LD = 0, \nabla_G \LG = 0, \LD\in B_D, \LG\in B_G\right\}\Bigg|
\end{align}
for some Borel sets $B_D, B_G\subset\mathbb{R}$ and where $\nabla_D, \nabla_G$ denote the Riemannian covariate derivatives on the hyperspheres with respect to the discriminator and generator weights respectively.
Note:
\begin{enumerate}
    \item We have chosen to treat the parameters of each network as somewhat separate by placing them on their own hyper-spheres. This reflects the minimax nature of GAN training, where there really are 2 networks being optimised in an adversarial manner rather than one network with some peculiar structure.
    \item We could have taken $\nabla = (\nabla_D, \nabla_G)$ and required $\nabla \LD = \nabla \LG = 0$ but, as in the previous comment, our choice is more in keeping with the adversarial set-up, with each network seeking to optimize separately its own parameters in spite of the other.
    \item We will only be interested in the case $ B_D = (-\infty, \sqrtsign{N} u_D)$ and $B_G= (-\infty, \sqrtsign{N} u_G)$, for $u_D, u_G\in \mathbb{R}$.
\end{enumerate}

So that the finer structure of local minima and saddle points can be probed, we also define the corresponding complexity with Hessian index prescription
\begin{align}
    C_{N, k_D, k_G} =\Bigg|\Bigg\{ \vwD\in S^{N_D  }, \vwG\in S^{N_G } ~:~ &\nabla_D \LD = 0, \nabla_G \LG = 0, \LD\in B_D, \LG\in B_G\notag\\
    &i(\nabla_D^2 L^{(D)}) = k_D, ~i(\nabla_G^2 L^{(G)}) = k_G \Bigg\}\Bigg|,\label{eq:C_Nkk_def}
\end{align}
where $i(M)$ is the index of $M$ (i.e. the number of negative eigenvalues of $M$). We have chosen to consider the indices of the Hessians $\nabla_D^2 \LD$ and $\nabla_G^2\LG$ separately, just as we chose to consider separately vanishing derivatives  $\nabla_D \LD$ and $\nabla_G\LG$. We believe this choice best reflects the standard training loop of GANs, where each iteration updates the discriminator and generator parameters in separate steps.

\medskip
To calculate the complexities, we follow the well-trodden route of Kac-Rice formulae as pioneered by \cite{fyodorov2004complexity,fyodorov2007replica}. For a fully rigorous treatment, we proceed as in \cite{auffinger2013random} and Chapter \ref{chap:general_activation_functions}.

\begin{lemma}\label{lemma:kac-rice-mine}
\begin{align}
    C_N = \int_{S^{N_D}\times S^{N_G}}&d\vwG d\vwD ~~\varphi_{(\nabla_D \LD,
         \nabla_G \LG)}(0)\notag\\
         &\expect\left[  \left|\det \left(\begin{array}{cc} \nabla_D^2 \LD & \nabla_{GD} \LD \\ \nabla_{DG}\LG & \nabla^2_{G} \LG\end{array}\right)\right|  ~\mid~ \nabla_G\LG =0, \nabla_D\LD = 0\right]\mathbbm{1}\left\{\LD\in B_D, \LG\in B_G\right\}\label{eq:kac_rice_first_old}
\end{align}
and therefore
\begin{align}
    C_N = \int_{S^{N_D}\times S^{N_G}}&d\vwG d\vwD ~~\varphi_{(\nabla_D \LD,
         \nabla_G \LG)}(0)\int_{B_D}dx_D \int_{B_G} dx_G ~ \varphi_{\LD}(x_D)\varphi_{\LG}(x_G) \notag\\ &\expect\Bigg[  \left|\det \left(\begin{array}{cc} \nabla_D^2 \LD & \nabla_{GD} \LD \\ \nabla_{DG}\LG  & \nabla^2_{G} \LG\end{array}\right)\right|  ~\mid~ \nabla_G\LG =0, \nabla_D\LD = 0, 
         \LD = x_D, \LG = x_G\Bigg].\label{eq:kac_rice_first}
\end{align}

where $\varphi_{(\nabla_D \LD,\nabla_G \LG)}$ is the joint density of $(\nabla_D \LD,\nabla_G \LG)^T$, $\varphi_{\LD}$ the density of $\LD$, and $\varphi_{\LG}$ the density of $\LG$, all implicitly evaluated at $(\vwD, \vwG)$.
\end{lemma}
\begin{proof}
In the notation of Theorem \ref{thm:adler_kac_rice}, we make the following choices: \begin{align*}
    \phi = \left(\begin{array}{c}
         \nabla_D \LD \\
         \nabla_G \LG
    \end{array}\right), ~~~  \psi = \left(\begin{array}{c}
          \LD \\
         \LG
    \end{array}\right)
\end{align*}
and so \begin{align*}
    A = B_D\times B_G, ~~~ \vec{u} = 0.
\end{align*}
and the manifold $\mathcal{M}$ is taken to be $S^{N_D}\times S^{N_G}$ with the product topology. It is sufficient to check the conditions of Theorem \ref{thm:adler_kac_rice} with the above choices.

\medskip
Conditions (a)-(f) are satisfied due to Gaussianity and the manifestly smooth definition of $L^{(D)}, L^{(G)}$. The moduli of continuity conditions as in (g) are satisfied separately for $L^{(D)}$ and its derivatives on $S^{N_D}$ and  for $L^{(G)}$ and its derivatives on $S^{N_G}$, as seen in the proof of the analogous result for a single spin glass in \cite{auffinger2013random}. But since $\mathcal{M}$ is just a direct product with product topology, it immediately follows that (g) is satisfied, so Theorem \ref{eq:adler_taylor_kac_rice} applies and we obtain (\ref{eq:kac_rice_first_old}). (\ref{eq:kac_rice_first}) follows simply, using the rules of conditional expectation.
\end{proof}

With Lemma \ref{lemma:kac-rice-mine} in place, we can now establish the following Kac-Rice expression specialised to our model:

\begin{lemma}\label{lemma:kac-rice-special}
For $(N-2)\times (N-2)$ GOE matrix $M$ and independent $(N_D - 1)\times (N_D-1)$ GOE matrix $M_1$, define \begin{align}
    H(x, x_1) &\overset{d}{=} bM + b_1\left(\begin{array}{cc} M_1 & 0 \\ 0 & 0 \end{array}\right) - x - x_1\left(\begin{array}{cc} I_{N_D} & 0 \\ 0 & 0 \end{array}\right).
\end{align}
For $u_G, u_D\in \mathbb{R}$, define
\begin{align}
    B =  \left\{(x, x_1)\in\mathbb{R}^2 ~:~ x\leq \frac{1}{\sqrtsign{2}}(p+q)2^{p+q} u_G, ~~ x_1 \geq -(p+q)^{-1} 2^{-(p+q)}p x - \frac{p}{\sqrtsign{2}}u_D\right\}.
 \end{align}
Define the constant 
\begin{align}
    K_N =\omega_{\kappa N}\omega_{\kappa'N} (2(N-2))^{\frac{N-2}{2}} (2\pi)^{-\frac{N-2}{2}}\left(p + \sigma_z^22^{p+1}(p+q)\right)^{-\frac{\kappa N-1}{2}} \left(\sigma_z^2 2^{p+q} (p+q)\right)^{-\frac{\kappa' N-1}{2}}\label{eq:KN_def}
\end{align}
where the variances are 
\begin{align}
    s^2 = \frac{1}{2}\sigma_z^2(p+q)^2 2^{3(p+q)}, ~~~ s_1^2 = \frac{p^2}{2}.
\end{align}
and $\omega_N = \frac{2\pi^{N/2}}{\Gamma(N/2)}$ is the surface area of the $N$ sphere.
The expected complexity $C_N$ is then
\begin{align}
    \mathbb{E} C_N = K_N \int_B    \sqrtsign{\frac{N}{2\pi s^2}}e^{-\frac{N}{2s^2}x^2}dx ~ \sqrtsign{\frac{N}{2\pi s_1^2}} e^{ -\frac{N}{2s_1^2} x_1^2}dx_1 ~\mathbb{E}|\det H(x, x_1)|\label{eq:ecn}.
\end{align}
\end{lemma}
\begin{proof}
Define the matrix \begin{align*}
    \tilde{H} = \left(\begin{array}{cc} \nabla_D^2 \LD & \nabla_{GD} \LD \\ \nabla_{DG}\LG & \nabla^2_{G} \LG\end{array}\right)
\end{align*}
appearing in the expression for $C_N$ in Lemma \ref{lemma:kac-rice-mine}. Note that $\tilde{H}$ takes the place of a Hessian (though it is not symmetric).
We begin with the distribution of
\begin{align*}
    \tilde{H} ~ \mid ~ \{(\lD, \lG) = (x_D, x_G), ~ (\nabla_D\lD, \nabla \lG) = (0, 0)\}.
\end{align*} 
Note that the integrand in (\ref{eq:ecn}) is jointly spherically symmetric in both $\vwD$ and $\vwG$. It is therefore sufficient to consider $\tilde{H}$ in the region of a single point on each sphere. We choose the north poles and coordinate bases on both spheres in the region of their north poles. The remaining calculations are routine Gaussian manipulations, very similar in character to those in the previous chapter, so they are given at the end of this chapter (section \ref{app:gaussian}). One finds \begin{align}
    \tilde{H} \overset{d}{=}
  \sqrtsign{2p(p-1)} \left(\begin{array}{cc}
       \sqrtsign{N_D -1}M^{(D)}_2 & 0 \\
         0 & 0
    \end{array}\right) +&  \sigma_z\sqrtsign{2^{p+q+1}(p+q)(p+q-1)} \left(\begin{array}{cc}
       \sqrtsign{N_D -1}M^{(D)}_1 & -2^{-1/2}G \\
        2^{-1/2} G^T & \sqrtsign{N_G - 1}M^{(G)}
    \end{array}\right) \notag\\
   -& \sigma_z(p+q)x_G2^{p+q} \left(\begin{array}{cc}
        -I_{N_D} & 0  \\
         0 & I_{N_G} 
    \end{array}\right)  - px_D\left(\begin{array}{cc}
        I_{N_D} & 0  \\
         0 & 0
    \end{array}\right)\label{eq:h_tilde_def}
\end{align}
where $M^{(D)}_1, M^{(D)}_2$ are independent $GOE^{N_D - 1}$ matrices, $M^{(G)}$ is an independent $GOE^{N_G - 1}$ matrix and $G$ is an independent $(N_D - 1)\times(N_G - 1)$ Ginibre matrix. Note that the dimensions are $N_D - 1$ and $N_G - 1$ rather $N_D$ and $N_G$. This is simply because the hypersphere $S^{N_D}$ is an $N_D - 1$ dimensional manifold, and similarly $S^{N_G}$.

We can simplify by summing independent Gaussians to obtain
\begin{align}
    \tilde{H} =  \left(\begin{array}{cc}
      \sigma_D\sqrtsign{N_D -1}M^{(D)} & -2^{-1/2}\sigma_GG, \\
      2^{-1/2} \sigma_G G^T & \sigma_G\sqrtsign{N_G - 1}M^{(G)}
    \end{array}\right)
  -\sigma_z(p+q)x_G2^{p+q} \left(\begin{array}{cc}
        -I_{N_D} & 0  \\
         0 & I_{N_G} 
    \end{array}\right)  - px_D\left(\begin{array}{cc}
        I_{N_D} & 0  \\
         0 & 0
    \end{array}\right) \label{eq:H_first}
\end{align}
where \begin{align}\sigma_G&= \sigma_z\sqrtsign{2^{p+q+1}(p+q)(p+q-1)}\\
\sigma_D &= \sqrtsign{\sigma_G^2 + 2p(p-1)}
\end{align}
and $M^{(D)}\sim GOE^{N_D - 1}$ is a GOE matrix independent of $M^{(G)}$ and $G$.

There is an alternative reformulation of $\tilde{H}$ that will also be useful. Indeed, because $M^{(D)}_{1,2} \overset{d}{=} -M^{(D)}_{1,2}$, let us write $\tilde{H}$ as \begin{align}
    \tilde{H} =& \sigma_zJ\left(\sqrtsign{2^{p+q+1}(p+q)(p+q-1)(N_D + N_G - 2)}M_1 - (p+q)x_G2^{p+q}I\right) \notag\\
  &+  \left(\sqrtsign{2p(p-1)(N_D - 1)}\left(\begin{array}{cc} M_2 & 0 \\ 0 & 0 \end{array}\right) - px_D\left(\begin{array}{cc} I_{N_D} & 0 \\ 0 & 0 \end{array}\right)\right)\notag\\
  \overset{d}{=}& J\Bigg[ \sigma_z\sqrtsign{2^{p+q+1}(p+q)(p+q-1)(N_D + N_G - 2)}M_1 - \sigma_z(p+q)x_G2^{p+q}I\notag\\
  &+ \sqrtsign{2p(p-1)(N_D - 1)}\left(\begin{array}{cc} M_2 & 0 \\ 0 & 0 \end{array}\right) + px_D\left(\begin{array}{cc} I_{N_D} & 0 \\ 0 & 0 \end{array}\right)\Bigg]
\end{align}

where $M_1\sim GOE^{N_D + N_G - 2}$ is a GOE matrix of size $N_D + N_G-2$, $M_2\sim GOE^{N_D - 1}$ is an independent GOE matrix of size $N_D-1$ and \begin{align}
    J = \left(\begin{array}{cc} -I_{N_D} & 0 \\ 0 & I_{N_G} \end{array}\right).
\end{align}
If follows that \begin{align}
    |\det \tilde{H}| \overset{d}{=}& \Bigg|\det\Bigg[ \sigma_z\sqrtsign{2^{p+q+1}(p+q)(p+q-1)(N_D + N_G - 2)}M_1 - \sigma_z(p+q)x_G2^{p+q}I\notag\\
  &+ \sqrtsign{2p(p-1)(N_D - 1)}\left(\begin{array}{cc} M_2 & 0 \\ 0 & 0 \end{array}\right) + px_D\left(\begin{array}{cc} I_{N_D} & 0 \\ 0 & 0 \end{array}\right)\Bigg]\Bigg|.\label{eq:H_rewritten}
\end{align}

Now define the constants \begin{align}
b = \sqrtsign{2^{p+q}(p+q)(p+q-1)}\sigma_z, ~~&~~ b_1= \sqrtsign{p(p-1)\kappa}\\
    x = \frac{\sigma_z(p+q)2^{p+q}}{\sqrtsign{N}}x_G, ~~&~~  x_1= -\frac{p}{\sqrtsign{N}}x_D,\label{eq:x_x1_def}
\end{align}  and then we arrive at \begin{align}
    |\det \tilde{H}| \overset{d}{=} \left(2(N-2)\right)^{\frac{N-2}{2}} |\det H(x, x_1)|.
\end{align}
The variances of $\LD$ and $\LG$ are derived from those of $\lG, \lD$ computed in Section \ref{app:gaussian} (see (\ref{eq:ld_var}), (\ref{eq:var_lg})):
\begin{align*}
    Var(\lD) = 1, ~~ Var(\lG) = 2^{p+q}.    
\end{align*}
Similarly the density $\varphi_{(\nabla_D \LD,\nabla_G \LG)}$ is found in (\ref{eq:grad_dens_0}):
\begin{align*}
    \varphi_{\left(\nabla_D L^{(D)}, \nabla_G L^{(G)}\right)}(0) = (2\pi)^{-\frac{N-2}{2}} \left(p + \sigma_z^22^{p+1}(p+q)\right)^{-\frac{N_D - 1}{2}} \left(\sigma_z^2 2^{p+q} (p+q)\right)^{-\frac{N_G-1}{2}}.
\end{align*}

We have now collected all the inputs required for Lemma \ref{lemma:kac-rice-mine}. The domain of integration $B$ arises from the constraints $ L^{(D)} \in (-\infty, \sqrtsign{N} u_D)$ and $L^{(G)} \in (-\infty, \sqrtsign{N} u_G)$ and the re-scaled variables (\ref{eq:x_x1_def}). This completes the proof.
 \end{proof}

 We will need the asymptotic behaviour of the constant $K_N$, which we now record in a small lemma.

\begin{lemma}\label{lemma:kn}
As $N\rightarrow\infty$, \begin{align}
K_N \sim 2^{\frac{N}{2}}\pi^{N/2} \left(\kappa^{\kappa} \kappa'^{\kappa'}\right)^{-N/2} \sqrtsign{\kappa\kappa'}\left(p + \sigma_z^22^{p+1}(p+q)\right)^{-\frac{\kappa N-1}{2}} \left(\sigma_z^2 2^{p+q} (p+q)\right)^{-\frac{\kappa' N-1}{2}}
\end{align}
\end{lemma}

\begin{proof} 
By Stirling's formula \begin{align}
    K_N &\sim 4 \pi^{N} \left(\frac{4\pi}{\kappa N}\right)^{-1/2}\left(\frac{4\pi}{\kappa' N}\right)^{-1/2}\left(\frac{\kappa N}{2 e}\right)^{-\kappa N/2}
    \left(\frac{\kappa' N}{2 e}\right)^{-\kappa' N/2}
    \left(2(N-2)\right)^{\frac{N-2}{2}} \left(2\pi\right)^{-\frac{N-2}{2}}\notag\\
    &~~~~~~ \left(p + \sigma_z^22^{p+1}(p+q)\right)^{-\frac{\kappa N-1}{2}} \left(\sigma_z^2 2^{p+q} (p+q)\right)^{-\frac{\kappa' N-1}{2}}\notag\\
    & \sim 2^{\frac{N}{2}}\pi^{N/2} \left(\kappa^{\kappa} \kappa'^{\kappa'}\right)^{-N/2} \sqrtsign{\kappa\kappa'}\left(p + \sigma_z^22^{p+1}(p+q)\right)^{-\frac{\kappa N-1}{2}} \left(\sigma_z^2 2^{p+q} (p+q)\right)^{-\frac{\kappa' N-1}{2}}
\end{align}
where we have used $(N-2)^{\frac{N-2}{2}} = N^{\frac{N-2}{2}} \left(1- \frac{2}{N}\right)^{\frac{N-2}{2}} \sim N^{\frac{N-2}{2}} e^{-N/2}.$
\end{proof}

\section{Limiting spectral density of the Hessian}\label{sec:lsd}
Our intention now is to compute the the expected complexity $\expect C_N$ via the Coulomb gas method. The first step in this calculation is to obtain the limiting spectral density of the random matrix \begin{align}
    H' = bM + b_1\left(\begin{array}{cc}
     M_1 & 0 \\ 0 & 0\end{array}\right) - x_1\left(\begin{array}{cc}
     I & 0 \\ 0 & 0\end{array}\right),
\end{align}
where, note, $H' = H + xI$ is just a shifted version of $H$ as defined in Lemma \ref{lemma:kac-rice-special}. Here the upper-left block is of dimension $\kappa N$, and the overall dimension is $N$. Let $\mu_{eq}$ be the limiting spectral measure of $H'$ and $\rho_{eq}$ its density. The supersymmetric method provides a way of calculating the expected Stieltjes transforms of $\rho_{eq}$ \cite{Verbaarschot_2004}:\begin{align}
    \langle G(z) \rangle &= \frac{1}{N} \frac{\partial}{\partial J}\Bigg|_{J=0} Z(J)\\
    Z(J) &\coloneqq \mathbb{E}_{H'} \frac{\det(z - H' + J)}{\det(z - H')}.
\end{align}
Recall that a density and its Stieltjes transform are related by the Stieltjes inversion formula  \begin{align}\label{eq:stieljes}
     \rho_{eq}(z) = \frac{1}{\pi}\lim_{\epsilon \rightarrow 0} \Im \langle G(z + i\epsilon)\rangle.
 \end{align}
The function $Z(J)$ can be computed using a supersymmetric representation of the ratio of determinants. Firstly, we recall an elementary result from multivariate calculus, where $M$ is a real matrix: \begin{align}\label{eq:det_complex_id}
    \int \prod_{i=1}^N \frac{d\phi_i d\phi_i^*}{2\pi} e ^{-i\phi^{\dagger}M\phi} = \frac{1}{\det M}.
\end{align}
By introducing Grassmann varibables $\chi_i, \chi_i*$ and a Berezin integral, we obtain a complimentary expression: \begin{align}\label{eq:det_grass_id}
    \int \frac{1}{-i}\prod_{i=1}^N d\chi_i d\chi_i^* e^{-i\chi^{\dagger}M\chi} = \det{M},\end{align}

Using the integral results (\ref{eq:det_complex_id}), (\ref{eq:det_grass_id})  we can then write \begin{align}
 \frac{\det(z - H' + J)}{\det(z - H')} &= \int d\Psi \exp\left\{ -i\phi^{\dagger}(z-H') \phi - i \chi^{\dagger}(z+J - H')\chi\right\}\label{eq:susy_lsd_ratio}
 \end{align}
 where the measure is \begin{align}
     d\Psi = \frac{1}{-i (2\pi)^N}\prod_{t=1}^2 d\phi[t]d\phi^{*}[t] d\chi[t] d\chi^*[t],
 \end{align}
$\phi$ is a vector of $N$ complex commuting variables, $\chi$ and  $\chi^*$ are vectors of $N$ Grassmann variables, and we use the $[t]$ notation to denote the splitting of each of the vectors into the first $\kappa N$ and last $(1-\kappa)N$ components, as seen in \cite{guhr1990isospin}: \begin{align}
    \phi = \left(\begin{array}{c} \phi[1] \\ \phi[2]\end{array}\right).
\end{align}
We then split the quadratic form expressions in (\ref{eq:susy_lsd_ratio}) \begin{align}
    & -\phi^{\dagger}(z-H') \phi -  \chi^{\dagger}(z+J - H')\chi\notag\\
    = & -\phi[1]^{\dagger}(x_1 - b_1M_1) \phi[1]
     -\phi^{\dagger}(z - bM) \phi
    - \chi[1]^{\dagger}(x_1 - b_1 M_1)\chi[1]
        - \chi^{\dagger}(z + J - b M)\chi.
\end{align}
Taking the GOE averages is now simple \cite{Verbaarschot_2004,nock2017characteristic}:\begin{align}
    \mathbb{E}_M \exp\left\{-ib\phi^{\dagger}M \phi - ib\chi^{\dagger}M\chi\right\} &= \exp\left\{- \frac{b^2}{4N}\trg Q^2\right\},\\
  \mathbb{E}_M \exp\left\{-ib_1\phi[1]^{\dagger}M_1 \phi[1] - ib_1\chi[1]^{\dagger}M_1\chi[1]\right\} &= \exp\left\{- \frac{b_1^2}{4\kappa N}\trg Q[1]^2\right\},
\end{align}
where the supersymmetric matrices are given by \begin{align}
    Q = \left(\begin{array}{cc} \phi^{\dagger}\phi & \phi^{\dagger}\chi \\ \chi^{\dagger}\phi & \chi^{\dagger}\chi\end{array}\right), ~~~    Q[1] = \left(\begin{array}{cc} \phi[1]^{\dagger}\phi[1] & \phi[1]^{\dagger}\chi[1] \\ \chi[1]^{\dagger}\phi[1] & \chi[1]^{\dagger}\chi[1]\end{array}\right).
\end{align}
Introducing the tensor notation \begin{align}
    \psi = \phi \otimes \left(\begin{array}{c} 1 \\ 0 \end{array}\right) + \chi \otimes \left(\begin{array}{c} 0 \\ 1 \end{array}\right), ~~     \psi[1] = \phi[1] \otimes \left(\begin{array}{c} 1 \\ 0 \end{array}\right) + \chi[1] \otimes \left(\begin{array}{c} 0 \\ 1 \end{array}\right)
\end{align}
and \begin{align}
    \zeta = \left(\begin{array}{cc} z & 0 \\ 0 & z + J\end{array}\right)
\end{align}we can compactly write \begin{align}
    Z(J) = \int d\Psi \exp\left\{ - \frac{b^2}{4N} \trg Q^2 - \frac{b_1^2}{4\kappa N} \trg Q[1]^2 - i \psi[1]^{\dagger}\psi[1]x_1 - i \psi^{\dagger}\zeta \psi\right\}.
\end{align}
We now perform two Hubbard-Stratonovich transformations \cite{Verbaarschot_2004} \begin{align}
    Z(J) &= \int d\Psi d\sigma d\sigma[1]\exp\left\{ - \frac{N}{b^2} \trg \sigma^2 - \frac{\kappa N}{b_1^2} \trg \sigma[1]^2 - i \psi[1]^{\dagger}(x_1 + \sigma[1])\psi[1] - i \psi^{\dagger}(\sigma + \zeta )\psi\right\},
\end{align}
where $\sigma$ and $\sigma[1]$ inherit their form from $Q, Q[1]$ \begin{align}
    \sigma = \left( \begin{array}{cc} \sigma_{BB}& \sigma_{BF} \\ \sigma_{FB} & i\sigma_{FF}\end{array}\right), ~~~     \sigma[1] = \left( \begin{array}{cc} \sigma_{BB}[1] & \sigma_{BF}[1] \\ \sigma_{FB}[1] & i\sigma_{FF}[1]\end{array}\right)
\end{align}
with $\sigma_{BB}, \sigma_{FF}, \sigma_{BB}[1], \sigma_{FF}[1]$ real commuting variables, and $\sigma_{BF}, \sigma_{FB}, \sigma_{BF}[1], \sigma_{FB}[1]$ Grassmanns; the factor $i$ is introduced to ensure convergence. Integrating out over $d\Psi$ is now a straightforward Gaussian integral in superspace, giving \begin{align}
    Z(J) &= \int d\Psi d\sigma d\sigma[1] \exp\left\{ - \frac{N}{b^2} \trg \sigma^2 - \frac{\kappa N}{b_1^2} \trg \sigma[1]^2 - i \psi[1]^{\dagger}(x_1 + \zeta + \sigma + \sigma[1])\psi[1] - i \psi[2]^{\dagger}(\sigma + \zeta )\psi[2]\right\}\notag\\
    & = \int d\sigma d\sigma[1] \exp\left\{ - \frac{N}{b^2} \trg \sigma^2 - \frac{\kappa N}{b_1^2} \trg \sigma[1]^2 - \kappa N\trg\log(x_1 + \zeta + \sigma + \sigma[1]) -  \kappa' N \trg\log (\sigma + \zeta )\right\}\notag\\
    &=\int d\sigma d\sigma[1] \exp\left\{ - \frac{N}{b^2} \trg (\sigma - \zeta)^2 - \frac{\kappa N}{b_1^2} \trg \sigma[1]^2 - \kappa N\trg\log(x_1  + \sigma + \sigma[1]) -  \kappa' N \trg\log \sigma\right\}.
\end{align}
Recalling the definition of $\zeta$, we have \begin{align}
    \trg(\sigma - \zeta)^2 = (\sigma_{BB} - z)^2 - (i\sigma_{FF} - z- J)^2 
\end{align}
and so one immediately obtains \begin{align}
    \frac{1}{N}\frac{\partial }{\partial J}\Bigg|_{J=0} Z(J) = \frac{2}{b^2}\int d\sigma d\sigma[1] (z - i\sigma_{FF}) \exp\Bigg\{ &-\frac{N}{b^2} \trg (\sigma - z)^2 - \frac{\kappa N}{b_1^2} \trg \sigma[1]^2\notag\\
    &-\kappa N\trg\log(x_1  + \sigma + \sigma[1]) -  \kappa' N \trg\log \sigma\Bigg\}\notag\\
    = \frac{2}{b^2}\int d\sigma d\sigma[1] (z - i\sigma_{FF}) \exp\Bigg\{ &-\frac{N}{b^2} \trg \sigma ^2 - \frac{\kappa N}{b_1^2} \trg \sigma[1]^2 \notag\\
    &-\kappa N\trg\log(x_1  + z + \sigma + \sigma[1]) -  \kappa' N \trg\log ( z+\sigma)\Bigg\}\label{eq:lsd_pre_saddle}
\end{align}
To obtain the limiting spectral density (LSD), or rather its Stieltjes transform, one must find the leading order term in the $N\rightarrow\infty$ expansion for (\ref{eq:lsd_pre_saddle}). This can be done by using the saddle point method on the $\sigma,\sigma[1]$ manifolds. We know that the contents of the exponential must vanish at the saddle point, since the LSD is $\mathcal{O}(1)$, so we in fact need only compute $\sigma_{FF}$ at the saddle point. We can diagonalise $\sigma$ within the integrand of (\ref{eq:lsd_pre_saddle}) and absorb the diagonalising graded $U(1/1)$ matrix into $\sigma[1]$. The resulting saddle point equations for the off-diagonal entries of the new (rotated) $\sigma[1]$ dummy variable are trivial and immediately give that $\sigma[1]$ is also diagonal at the saddle point. The saddle point equations are then \begin{align}
    &\frac{2}{b_1^2}\sigma_{BB}[1] + \frac{1}{\sigma_{BB}[1] + \sigma_{BB} + x_1 + z} = 0\label{eq:boson_saddle1}\\
        &\frac{2}{b^2}\sigma_{BB} + \frac{\kappa}{\sigma_{BB}[1] + \sigma_{BB} + x_1 + z} + \frac{\kappa'}{\sigma_{BB} + x} = 0\label{eq:boson_saddle2}\\
 &\frac{2}{b_1^2}\sigma_{FF}[1] - \frac{1}{\sigma_{FF}[1] + \sigma_{FF} - ix_1 - iz} = 0\label{eq:ferm_saddle1}\\
&\frac{2}{b^2}\sigma_{FF} - \frac{\kappa}{\sigma_{FF}[1] + \sigma_{FF} - ix_1 -iz} - \frac{\kappa'}{\sigma_{FF} - iz} = 0.\label{eq:ferm_saddle2}
\end{align}

\begin{figure}[h]
    \centering
    \begin{subfigure}{0.3\linewidth}
     \centering
     \includegraphics[width=\linewidth]{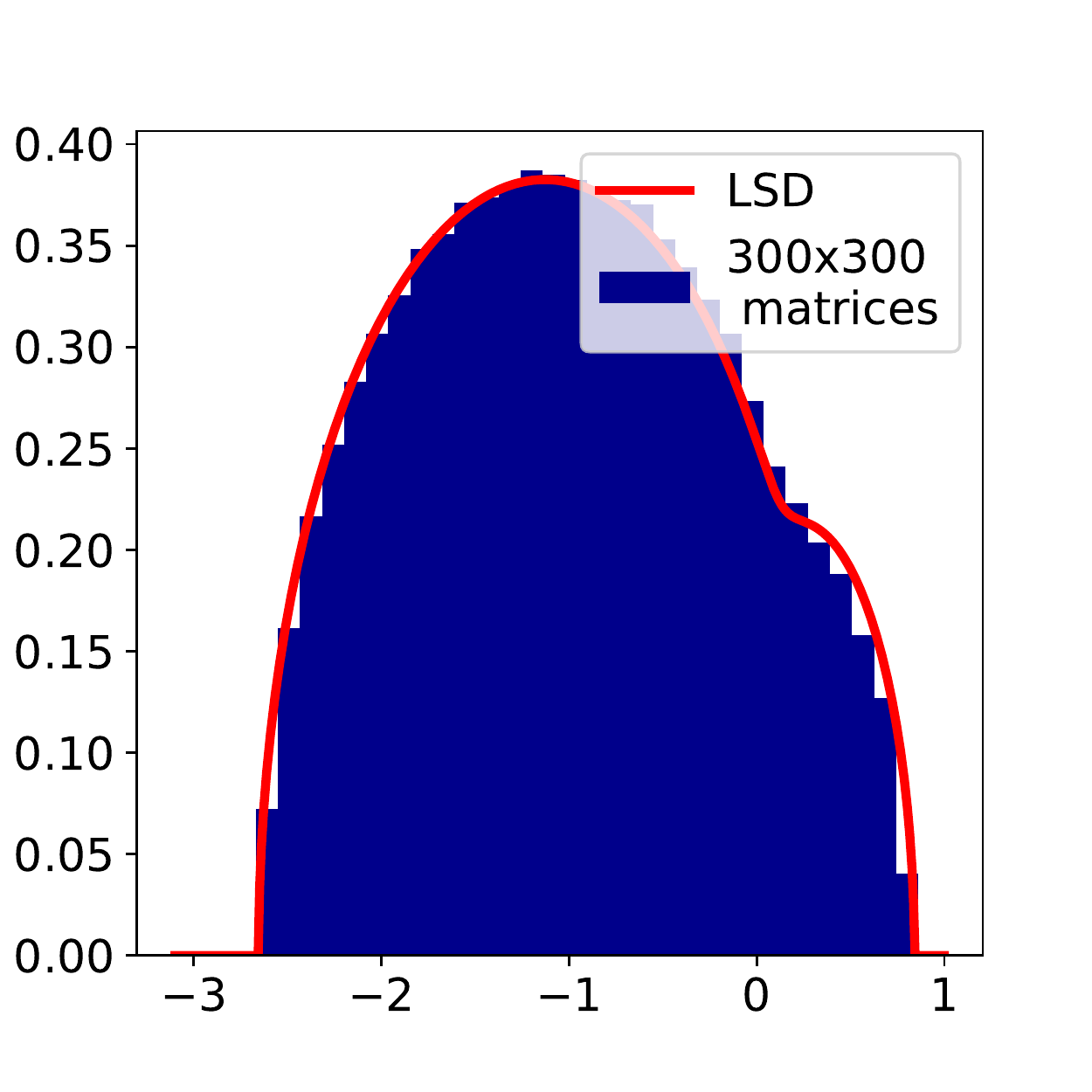}
     \subcaption{Merged}
    \end{subfigure}    
    \begin{subfigure}{0.3\linewidth}
     \centering
     \includegraphics[width=\linewidth]{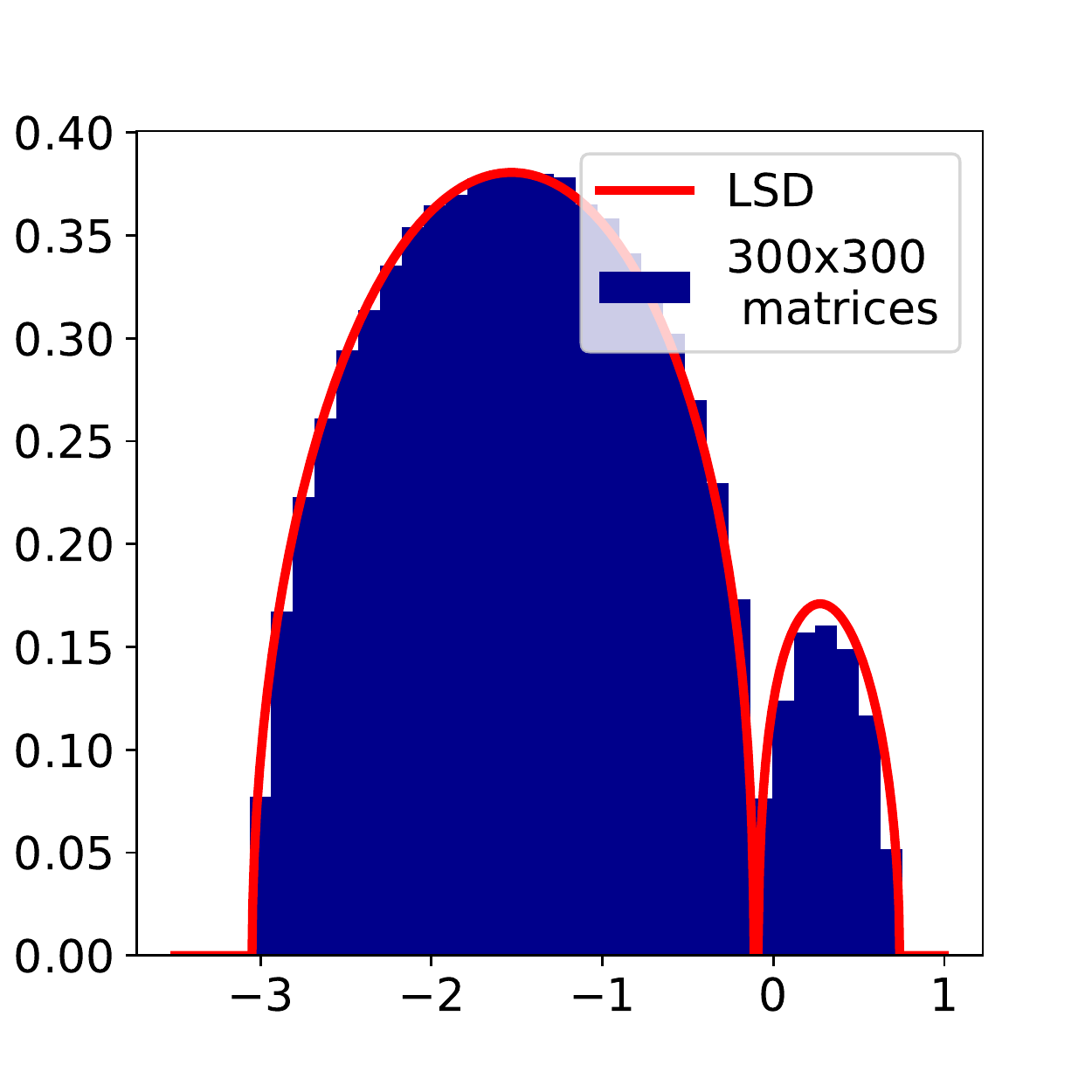}
     \subcaption{Touching}
    \end{subfigure}    
    \begin{subfigure}{0.3\linewidth}
     \centering
     \includegraphics[width=\linewidth]{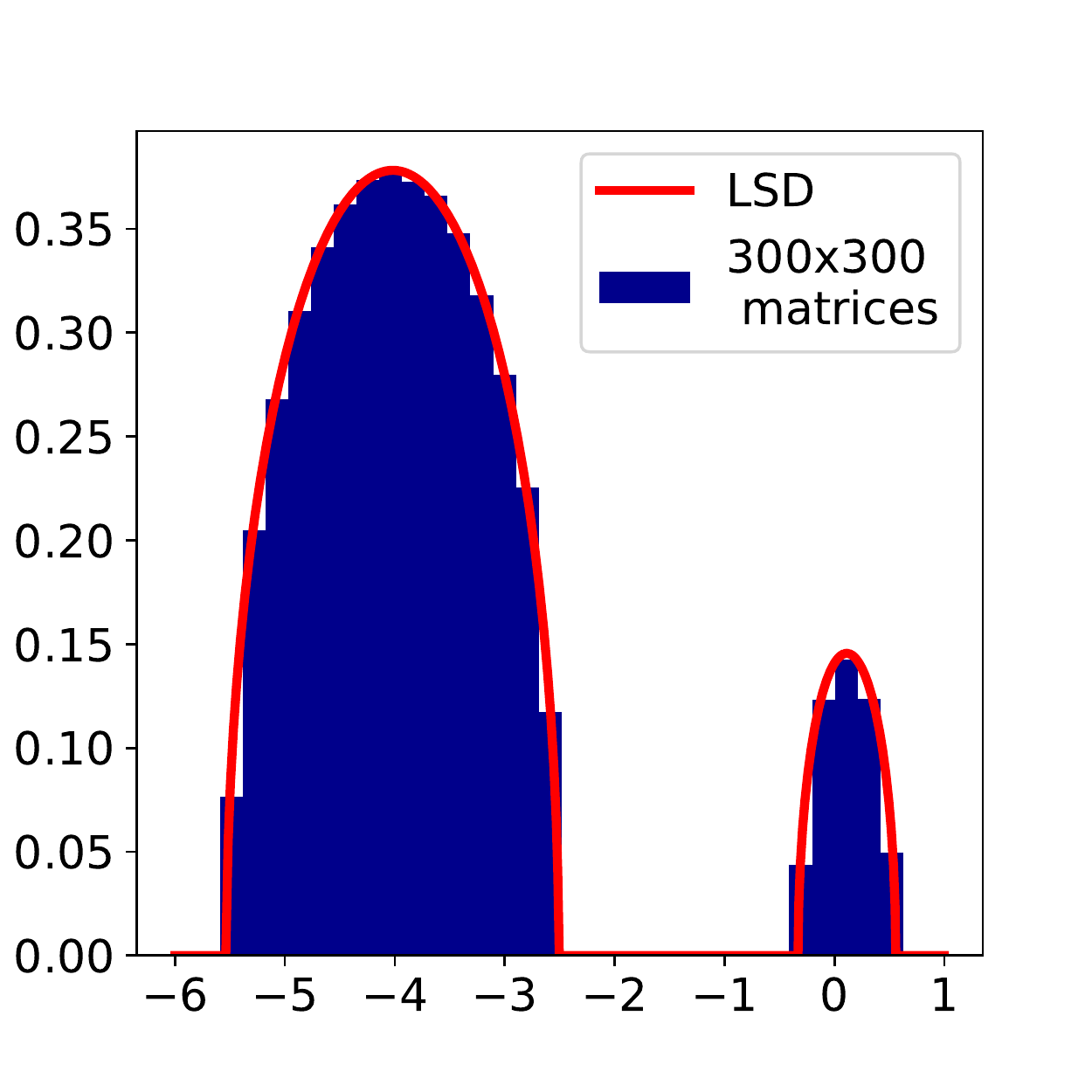}
     \subcaption{Separate}
    \end{subfigure}    
\caption{Example spectra of $H'$ showing empirical spectra from 100 $300\times 300$ matrices and the corresponding LSDs computed from (\ref{eq:master_quartic}). Here $b=b_1=1$, $\kappa=0.9$, $\sigma_z$=1 and $x_1$ is varied to give the three different behaviours.}
\label{fig:spectra}
\end{figure}

(\ref{eq:ferm_saddle1}) and (\ref{eq:ferm_saddle2}) combine to give an explicit expression for $\sigma_{FF}[1]$: \begin{align}
    \sigma_{FF}[1] = \frac{b_1^2}{2\kappa}\left(\frac{2}{b^2}\sigma_{FF} - \kappa'(\sigma_{FF} - iz)^{-1}\right)\label{eq:ferm_saddle3}.
\end{align}
With a view to simplifying the numerical solution of the coming quartic, we define $t = i(\sigma_{FF} - iz)$ and then a line of manipulation with (\ref{eq:ferm_saddle2}) and (\ref{eq:ferm_saddle3}) gives \begin{align}
    \left(t^2 -  zt - \kappa' b^2\right)\left((1 + \kappa^{-1}b^{-2} b_1^2)t^2 - (\kappa^{-1}b_1^2 b^{-2}z - x_1)t - \kappa'\kappa^{-1}b_1^2\right) + b^2\kappa t^2= 0\label{eq:master_quartic}.
\end{align}
 By solving (\ref{eq:master_quartic}) numerically for fixed values of $\kappa, b, b_1, x_1$, we can obtain the four solutions $t_1(z), t_2(z), t_3(z), t_4(z)$. These four solution functions arise from choices of branch for $(z, x_1)\in \mathbb{C}^2$ and determining the correct branch directly is highly non-trivial. However, for any $z\in\mathbb{R}$, at most one of the $t_i$ will lead to a positive LSD, which gives a simple way to compute $\rho_{eq}$ numerically using (\ref{eq:stieljes}) and (\ref{eq:lsd_pre_saddle}): \begin{align}\label{eq:t_lsd}
    \rho_{eq}(z) = \max_i\left\{-\frac{2}{b^2\pi} \Im t_i(z)\right\}.
\end{align}
Plots generated using (\ref{eq:t_lsd}) and eigendecompositions of matrices sampled from the distribution of $H'$ are given in Figure \ref{fig:spectra} and show good agreement between the two. Note the three different forms: single component support, two component support and the transition point between the two, according to the various parameters. In these plots, the larger lobes on the left correspond to the upper left block, which is much larger than the lower-right block (since $\kappa=0.9$ here). One can see this by considering large $x_1$, for which there must be a body of eigenvalues in the region of $-x_1$ owing to the upper left block. Since $x_1$ only features in the upper-left block, not all of the eigenvalues can be located around $-x_1$, and the remainder are found in the other lobe of the density which is around 0 in Figure \ref{fig:spectra}.

\section{The asymptotic complexity}\label{sec:complexity}
\begin{figure}[h]
    \centering
 \includegraphics[width=0.33\textwidth]{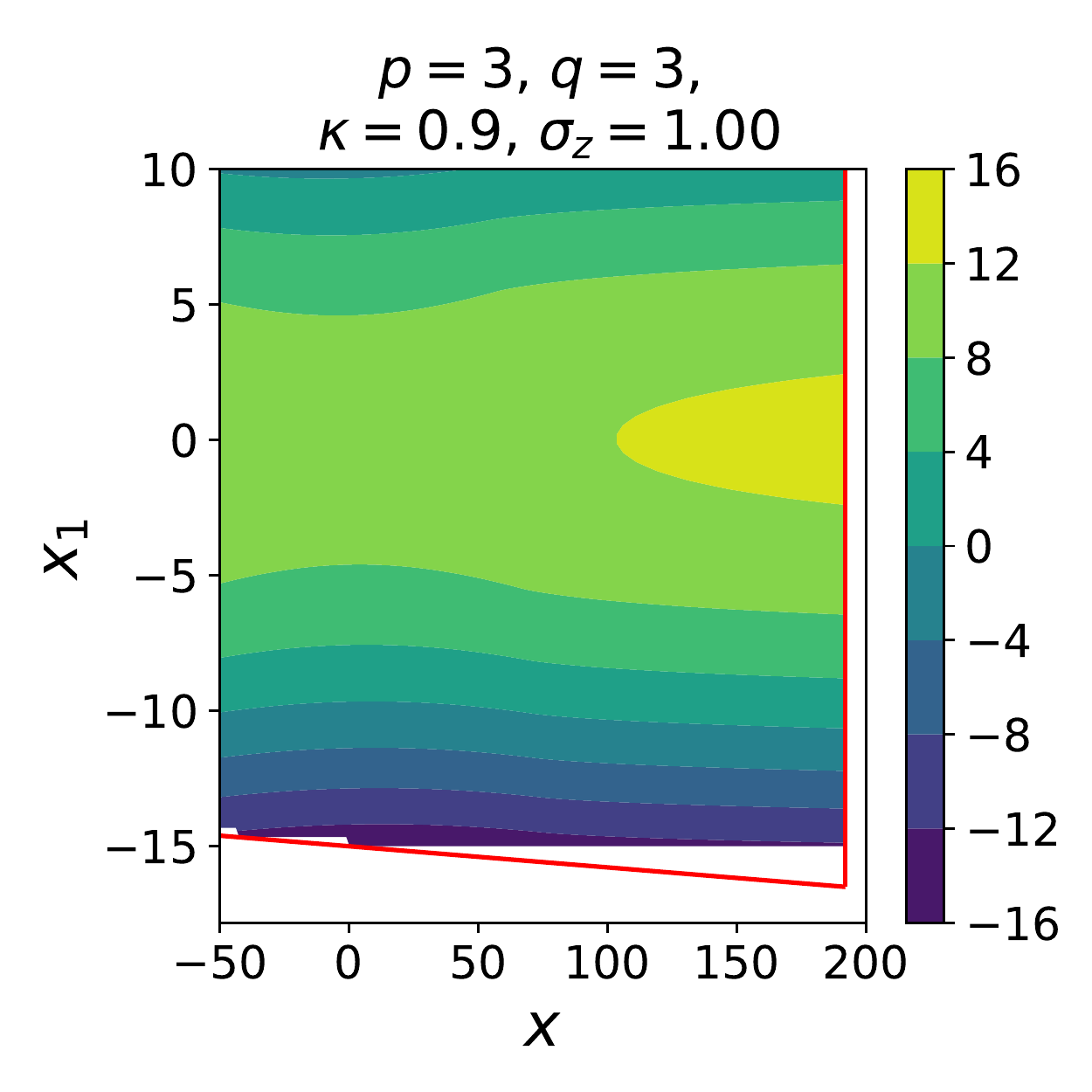}
    \caption{$\Phi$ for  $p=q=3, \sigma_z=1, \kappa=0.9$. Red lines show the boundary of the integration region $B$.}
    \label{fig:phi_indom_latest}
\end{figure}

\begin{figure}[h]
    \centering
    \includegraphics[width=\textwidth]{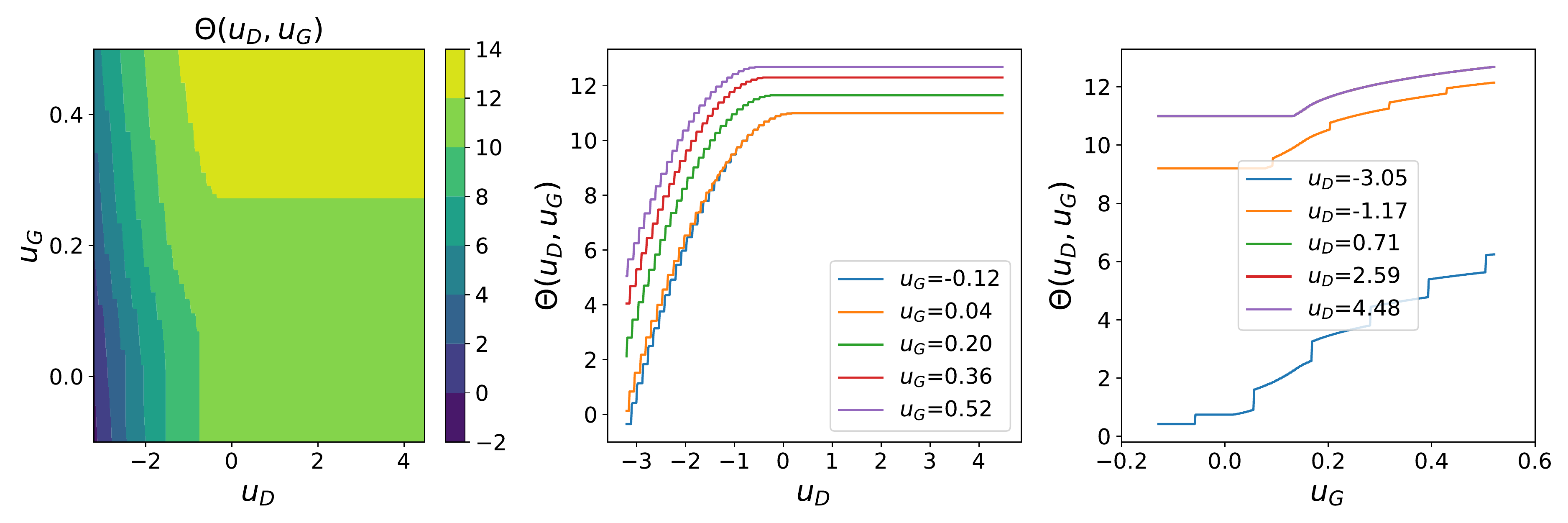}
    \caption{$\Theta$ and its cross-sections, fixing separately $u_D$ and $u_G$. Here $p=q=3, \sigma_z=1, \kappa=0.9$.}
    \label{fig:theta_latest}
\end{figure}

In the previous section, we have found the equilibrium measure, $\mu_{eq}$, of the ensemble of random matrices \begin{align}
    H' = bM + b_1\left(\begin{array}{cc}
     M_1 & 0 \\ 0 & 0\end{array}\right) - x_1\left(\begin{array}{cc}
     I & 0 \\ 0 & 0\end{array}\right), ~~ M\sim GOE^N, ~ M_1\sim GOE^{\kappa N}.
\end{align}
The Coulomb gas approximation gives us a method of computing $\mathbb{E} |\det(H'-x)|$:
\begin{align}
    \mathbb{E} |\det(H'-x)| \approx \exp\left\{N\int \log|z - x| d\mu_{eq}(z)\right\}.\label{eq:quadrature}
\end{align}
We have access to the density of $\mu_{eq}$ pointwise (in $x$ and $x_1$) numerically, and so (\ref{eq:quadrature}) is a matter of one-dimensional quadrature. Recalling (\ref{eq:ecn}), we then have \begin{align}
   \mathbb{E} C_N \approx  K_N'\iint_B dxdx_1 ~ \exp\left\{-(N-2)\left(  \frac{1}{2s^2}x^2 + \frac{1}{2s_1^2} (x_1)^2 - \int\log|z - x| d\mu_{eq}(z) \right)\right\}\equiv K_N'\iint_B dxdx_1 ~ e^{-(N-2)\Phi(x, x_1)}
\end{align}

 where \begin{align}
     K_N' = K_N \sqrtsign{\frac{N-2}{2\pi s_1^2}} \sqrtsign{\frac{N-2}{2\pi s^2}}.
 \end{align}
Due to Lemma \ref{lemma:kn}, the constant term has asymptotic form \begin{align}
    &\frac{1}{N}\log K_N' \notag\\\sim &\frac{1}{2}\log{2} + \frac{1}{2}\log{\pi} - \frac{\kappa}{2}\log\left(p + \sigma_z^22^{p+q}(p+q)\right) - \frac{\kappa'}{2} \log\left(\sigma_z^2(p+q) 2^{p+q}\right) - \frac{\kappa}{2}\log\kappa - \frac{\kappa'}{2}\log\kappa' \notag\\\equiv &K
 \end{align}
We then define the desired $\Theta(u_D, u_G)$ as \begin{align}
    \lim \frac{1}{N} \log\mathbb{E}C_N = \Theta(u_D, u_G)
\end{align}
and we have \begin{align}
    \Theta(u_D, u_G) = K - \min_B \Phi.
\end{align}
Using these numerical methods, we obtain the plot of $\Phi$ in $B$ and a plot of $\Theta$ for some example $p,q,\sigma_z, \kappa$ values, shown in Figures \ref{fig:phi_indom_latest}, \ref{fig:theta_latest}. Numerically obtaining the maximum of $\Phi$ on $B$ is not as onerous as it may appear, since $-\Phi$ grows quadratically in $|x|, |x_1|$ at moderate distances from the origin.

\begin{figure}[h]
    \centering
    \includegraphics[width=\textwidth]{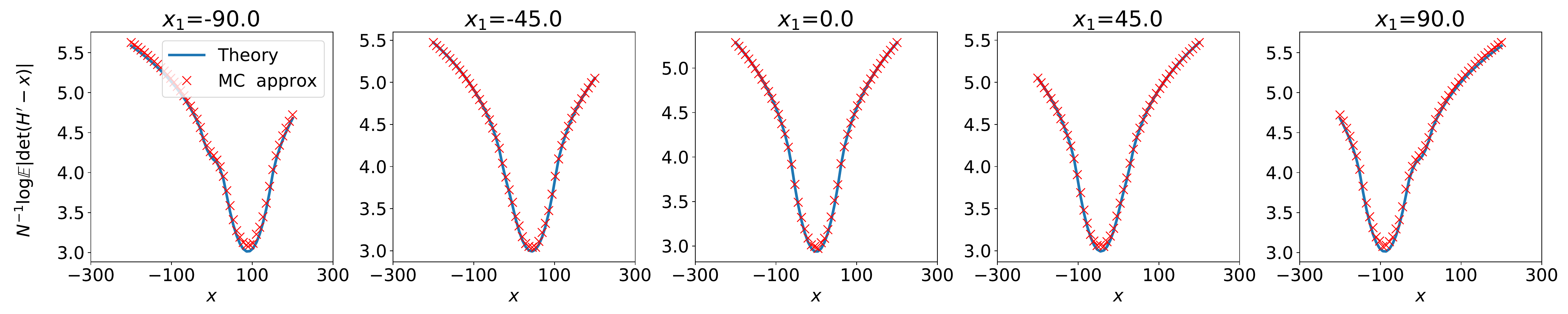}    
    \caption{Comparison of (\ref{eq:quadrature}) and (\ref{eq:mc_det}), verifying the Coulomb gas approximation numerically. Here $p=q=3, \sigma_z=1, \kappa=0.9$. Sampled matrices for MC approximation are dimension $N=50$, and $n=50$ MC samples have been used.}
    \label{fig:mc_coulomb_verify}
\end{figure}

We numerically verify the legitimacy of this Coulomb point approximation with Monte Carlo integration \begin{align}
    \mathbb{E}|\det(H'-x)| \approx \frac{1}{n}\sum_{i=1}^n \prod_{j=1}^N |\lambda_j^{(i)} - x|,\label{eq:mc_det}
\end{align}
where $\lambda^{(i)}_j$ is the $j$-th eigenvalues of the $i$-th i.i.d. sample from the distribution of $H'$. The results, comparing $N^{-1}\log\mathbb{E}|\det(H'-x)|$ at $N=50$ for a variety of $x,x_1$ are show in Figure \ref{fig:mc_coulomb_verify}. Note the strong agreement even at such modest $N$, however to rigorously substantiate the Coulomb gas approximation in (\ref{eq:quadrature}), we must prove a concentration result.

\begin{lemma}\label{lemma:coulomb_det}
Let $(H_N)_{N=1}^{\infty}$ be a sequence of random matrices, where for each $N$ \begin{align}
    H_N \overset{d}{=}  bM + b_1\left(\begin{array}{cc}
     M_1 & 0 \\ 0 & 0\end{array}\right) - x_1\left(\begin{array}{cc}
     I & 0 \\ 0 & 0\end{array}\right)
\end{align}
and $M\sim GOE^N$, $M_1\sim GOE^{\kappa N}$. Let $\mu_N$ be the empirical spectral measure of $H_N$ and say $\mu_N\rightarrow \mu_{eq}$ weakly almost surely. Then for any $(x, x_1)\in\mathbb{R}^2$\begin{align}
\mathbb{E} |\det(H_N-xI)| = \exp\left\{N (1 + o(1))\int \log|z - x| d\mu_{eq}(z)\right\}
\end{align}
 as $N\rightarrow\infty$.
\end{lemma}

\begin{proof}
We begin by establishing an upper bound. Take any $\beta>0$, then 
\begin{align}
    &\int \log|z-x| d\mu_N(z)\notag\\
    = &\int \log|z-x| \indic\{|x-z|\geq e^{\beta}\} d\mu_N(z) +\int \log|z-x| \indic\{\log|x-z| < \beta\} d\mu_N(z)\notag\\
    \leq &\int \log|z-x| \indic\{|x-z|\geq e^{\beta}\} d\mu_N(z) +\int  \min(\log|x-z|, \beta) d\mu_N(z).
\end{align}
Take also any $\alpha>0$, then trivially
\begin{align}
    \int  \min(\log|x-z|, \beta) d\mu_N(z) \leq   \int  \max(-\alpha, \min(\log|x-z|, \beta)) d\mu_N(z).
\end{align}
Overall we have, for any $\alpha, \beta > 0$,
\begin{align}
    &\exp\left\{N\int \log|z-x| d\mu_N(z)
    \right\}\notag\\
    \leq &\exp\left\{ N\int \log|z-x| \indic\{|x-z|\geq e^{\beta}\} d\mu_N(z) \right\}\notag\\
    &\exp\left\{ N \int  \max(-\alpha, \min(\log|x-z|, \beta)) d\mu_N(z) \right\}.
\end{align}
Thence an application of H\"{o}lder's inequality gives
\begin{align}
    \expect |\det (H_N-xI)| &= \expect\left[\exp\left\{N\int \log|z-x| d\mu_N(z)
    \right\}\right] \notag\\
    & \leq \underbrace{\left(\expect\left[\exp\left\{2N\int \max\left(-\alpha, \min\left(\log|x-z|, \beta\right)\right) d\mu_N(z)\right\}\right]\right)^{1/2}}_{A_N}\notag\\ & ~~~~~~~ \underbrace{\left(\expect\left[\exp\left\{2N\int \log|x-z|\indic\{|x-z|\geq e^{\beta}\}d\mu_N(z)\right\}\right]\right)^{1/2}}_{B_N}.
\end{align}
Considering $B_N$, we have \begin{align}
    \log|x-z| \indic\{|x-z| \geq e^{\beta}\} \leq |x-z|^{1/2}\indic\{|x-z| \geq e^{\beta}\} \leq e^{-\beta/2}|x-z|
\end{align}
and so \begin{align}
  \expect\left[\exp\left\{2N\int \log|x-z|\indic\{|x-z|\geq e^{\beta}\}\right\}\right] &\leq  \expect \left[\exp\left\{2N e^{-\beta/2} \frac{\Tr |H_N - xI|}{N}\right\}\right] \notag\\
   &= \expect \left[\exp\left\{2e^{-\beta/2} \Tr |H_N - xI|\right\}\right].
\end{align}
The entries of $H_N$ are Gaussians with variance $\frac{1}{N}b^2, \frac{1}{2N}b^2, \frac{1}{N}(b^2 + b_1^2)$ or $\frac{1}{2N}(b^2+b_1^2)$ and all the diagonal and upper diagonal entries are independent. All of these variances are $\mathcal{O}(N^{-1})$, so
\begin{align}
    |H_N - x|_{ij} \leq |x| + |x_1| + \mathcal{O}(N^{-1/2})|X_{ij}|
\end{align}
where the $X_{ij}$ are i.i.d. standard Gaussians for $i\leq j$. It follows that \begin{align}
       \expect \left[\exp\left\{2e^{-\frac{\beta}{2}} \Tr |H_N- xI|\right\}\right]   &\leq e^{2e^{-\frac{\beta}{2}}N(|x| + |x_1|)}\expect_{X\sim\mathcal{N}(0,1)} e^{2e^{-\frac{\beta}{2}}\mathcal{O}(N^{1/2})|X|}.
\end{align}
Elementary calculations give \begin{align}
    \expect_{X\sim \mathcal{N}(0,1)} e^{c |X|} \leq \frac{1}{2}\left(e^{-c^2} + e^{c^2}\right) \leq e^{c^2}
\end{align}
and so \begin{align}
      \expect \left[\exp\left\{2e^{-\frac{\beta}{2}} \Tr |H_N- xI|\right\}\right]   &\leq e^{2e^{-\frac{\beta}{2}}N(|x| + |x_1|)} e^{4e^{-\beta} \mathcal{O}(N)}\notag\\ &= \exp\left\{2N\left(e^{-\frac{\beta}{2}}(|x| + |x_1|) + e^{-\beta}\mathcal{O}(1)\right)\right\}
\end{align}
thus when we take $\beta\rightarrow \infty$, we have $B_N \leq e^{o(N)}$.

Considering $A_N$, it is sufficient now to show \begin{align}
    \expect \left[\exp\left\{ 2N \int f(z) d\mu_N(z)\right\}\right] = \exp\left\{2N\left(\int f(z) d\mu_{eq}(z) + o(1)\right)\right\}
\end{align}
where $f(z) = 2\max\left(\min(\log|x-z|, \beta), -\alpha\right)$, a continuous and bounded function. For any $\epsilon>0$, we have \begin{align}
    &\expect \left[\exp\left\{2N\int f(z) d\mu_N(z)\right\}\right]\notag\\
    \leq &\exp\left\{2N\left(\int f(z) d\mu_{eq}(z) + \epsilon\right)\right\} + e^{2N||f||_{\infty}}\mathbb{P}\left(\int f(z) d\mu_N(z) \geq \int f(z)d\mu_{eq}(z) + \epsilon\right).
\end{align}
The entries of $H_N$ are Gaussian with $\mathcal{O}(N^{-1})$ variance and so obey a log-Sobolev inequality as required by Theorem 1.5 from \cite{guionnet2000concentration}. The constant, $c$, in the inequality is independent of $N, x, x_1$, so we need not compute it exactly. The theorem from \cite{guionnet2000concentration} then gives
\begin{align}
    \mathbb{P}\left(\int f(z) d\mu_N(z) \geq \int f(z)d\mu_{eq}(z) + \epsilon\right) \leq \exp\left\{-\frac{N^2}{8c}\epsilon^2\right\}.
\end{align}
We have shown \begin{align}
  \expect|\det(H_N - xI)| \leq  A_NB_N &\leq \exp\left\{N(1 + o(1))\left(\int f(z)d\mu_{eq}(z)\right)\right\}\notag\\
  &\leq  \exp\left\{N(1 + o(1))\left(\int \log|x-z|d\mu_{eq}(z)\right)\right\}.\label{eq:lemma_upper_bnd}
\end{align}

We now need to establish a complimentary lower bound to complete the proof. By Jensen's inequality \begin{align}
    \expect |\det(H_N-x)| &\geq \exp\left(N\mathbb{E}\left[\int \log|z-x| d\mu_N(z)\right]\right) \notag\\
    & \geq \exp\left(N\mathbb{E}\left[\int\max\left(-\alpha,  \log|z-x|\right) d\mu_N(z)\right]\right)  \exp\left(N\expect\left[\int \log|z-x| \indic\{|z-x| \leq e^{-\alpha}\}d\mu_N(z)\right]\right)\notag\\
    &\geq \exp\left(N\mathbb{E}\left[\int\min\left(\beta, \max\left(-\alpha,  \log|z-x|\right)\right) d\mu_N(z)\right]\right) \notag\\ &~~~~~\exp\left(N\expect\left[\int \log|z-x| \indic\{|z-x| \leq e^{-\alpha}\}d\mu_N(z)\right]\right)\label{eq:ldp_lower_bound_2_terms}
\end{align}
for any $\alpha, \beta >0$. Convergence in law of $\mu_N$ to $\mu_{eq}$ and the dominated convergence theorem give \begin{align}
    \exp\left(N\mathbb{E}\left[\int\min\left(\beta, \max\left(-\alpha,  \log|z-x|\right)\right) d\mu_N(z)\right]\right) \geq \exp\left\{N \left(\int \log|x-z| d\mu_{eq}(z) + o(1)\right)\right\}
\end{align}
for large enough $\beta$, because $\mu_{eq}$ has compact support. It remains to show that the expectation inside the exponent in the second term of (\ref{eq:ldp_lower_bound_2_terms}) converges to zero uniformly in $N$ in the limit $\alpha \rightarrow \infty$. \\

By (\ref{eq:stieljes}), it is sufficient to consider $\langle G_N(z)\rangle$, which is computed via (\ref{eq:lsd_pre_saddle}). Let us define the function $\Psi$ so that \begin{align}
    \langle G_N(z) \rangle = \frac{2}{b^2} \int d\sigma d\sigma[1] (z-i\sigma_{FF}) e^{-N\Psi(\sigma, \sigma[1])}.
\end{align}
Henceforth, $\sigma_{FF}^*, \sigma_{FF}[1]^*, \sigma_{BB}^*, \sigma_{BB}[1]^*$ are the solution to the saddle point equations (\ref{eq:boson_saddle1}-\ref{eq:ferm_saddle2}) and  $\tilde{\sigma}_{FF}, \tilde{\sigma}_{FF}[1], \tilde{\sigma}_{BB}, \tilde{\sigma}_{BB}[1]$ are integration variables. Around the saddle point \begin{align}
    z - i\sigma_{FF} = z - i\sigma_{FF}^* - iN^{-\frac{1}{r}}\tilde{\sigma}_{FF}
\end{align}
for some $r\geq 2$. We use the notation $\vec{\sigma}$ for $(\sigma_{BB}, \sigma_{BB}[1], \sigma_{FF}, \sigma_{FF}[1])$ and similarly $\vec{\sigma}_{BB}, \vec{\sigma}_{FF}$. A superscript asterisk on $\Psi$ or any of its derivatives is short hand for evaluation at the saddle point. While the Hessian of $\Psi$ may not in general vanish at the saddle point, \begin{align}
    \int d\tilde{\sigma}d\tilde{\sigma}[1] \tilde{\sigma}_{FF} e^{-N \tilde{\vec{\sigma}}^T \nabla^2 \Psi^* \tilde{\vec{\sigma}}} = 0
\end{align}
and so we must go to at least the cubic term in the expansion of $\Psi$ around the saddle point, i.e. \begin{align}
      \langle G_N(z) \rangle = G(z) -  \frac{2i}{b^2 N^{5/3}}\underbrace{\int_{-\infty}^{\infty} d\tilde{\vec{\sigma}}_{BB} d\tilde{\vec{\sigma}}_{FF} \tilde{\sigma}_{FF} e^{-\frac{1}{6} \tilde{\sigma}^i \tilde{\sigma}^j \tilde{\sigma}^k \partial_{ijk}\Psi^*}}_{E(z; x_1)} + \text{ exponentially smaller terms}.
\end{align}
The bosonic (BB) and fermionic (FF) coordinates do not interact, so we can consider derivatives of $\Phi$ as block tensors. Simple differentiation gives \begin{align}
    (\nabla\Psi)_B &= \left(\begin{array}{c}
        \frac{2}{b^2}\sigma_{BB} - \kappa\left(\sigma_{BB} + \sigma_{BB}[1] + z + x_1\right)^{-1} - \kappa'\left(\sigma_{BB} + z\right)^{-1}  \\
        \frac{2}{b_1^2}\sigma_{BB}[1] - \left(\sigma_{BB} + \sigma_{BB}[1] + z + x_1\right)^{-1}\end{array}\right)\notag\\
   \implies  (\nabla^2\Psi)_B &= \left(\begin{array}{cc}
       \kappa\left(\sigma_{BB} + \sigma_{BB}[1] + z + x_1\right)^{-2} + \kappa'\left(\sigma_{BB} + z\right)^{-2}  &
       \kappa\left(\sigma_{BB} + \sigma_{BB}[1] + z + x_1\right)^{-2}\\
     \left(\sigma_{BB} + \sigma_{BB}[1] + z + x_1\right)^{-2} &
      \left(\sigma_{BB} + \sigma_{BB}[1] + z + x_1\right)^{-2}
       \end{array}\right)\\
 \implies (\nabla^3\Psi)_B^* &= \left(\left(\begin{array}{cc}
      A_B\kappa + B_B\kappa' & A_B\kappa \\
      A_B &  A_B
 \end{array}\right), A_B\left(\begin{array}{cc}
      \kappa & \kappa \\
      1 &  1
 \end{array}\right) \right),     
\end{align}
where \begin{align}
    A_B = -\frac{2}{\left(\sigma_{BB}^* + \sigma_{BB}^*[1] + z + x_1\right)^3}, ~~~ B_B= -\frac{2}{\left(\sigma_{BB}^* + z\right)^3}.
\end{align}
$(\nabla^3\Psi)_F^*$ follows similarly with \begin{align}
        A_F = -\frac{2}{\left(\sigma_{FF}^* + \sigma_{FF}^*[1] - iz - ix_1\right)^3}, ~~~ B_F= -\frac{2}{\left(\sigma_{FF}^* - iz\right)^3}.
\end{align}
By the saddle point equations (\ref{eq:boson_saddle1})-(\ref{eq:ferm_saddle2}) we have \begin{align}
    A_B &= 2(\sigma_{BB}[1]^*)^3, ~~ B_B = \frac{2}{(\kappa')^3}\left(\frac{2\kappa}{b_1^2}\sigma_{BB}[1]^* - \frac{2}{b^2} \sigma_{BB}^*\right)^3\label{eq:ab_bb_final}\\
        A_F &= 2(\sigma_{FF}[1]^*)^3, ~~ B_F = \frac{2}{(\kappa')^3}\left(\frac{2\kappa}{b_1^2}\sigma_{FF}[1]^* - \frac{2}{b^2} \sigma_{FF}^*\right)^3.\label{eq:af_bf_final}
\end{align}
Let $\xi_1= \tilde{\sigma}_{BB}, \xi_2 =\tilde{\sigma}_{BB}[1]$. Then \begin{align}
   ( \tilde{\sigma}^i \tilde{\sigma}^j \tilde{\sigma}^k \partial_{ijk}\Phi^*)_B &= \left(A_B\kappa + B_B\kappa'\right)\xi_1^3 + A_B(2\kappa +1 ) \xi_1^2\xi_2[1] + A_B(\kappa +2 ) \xi_1\xi_2^2 + A_B\xi_2^3\notag\\
   &= A_B\left[\xi_2^3 + (2\kappa +1 )\xi_2\xi_1^2 +(2+ \kappa)\xi_1\xi_2^2 + C\xi_1^3\right] + \left(B_B\kappa' + A_B\kappa - CA_B\right)\xi_1^3
\end{align}
for any $C$. Let $\xi_1 = a_1\xi_1'$ and then choose $C = a_1^{-3}$ and $a_1 = (2+\kappa)(2\kappa + 1)^{-1}$ to give \begin{align}
     ( \tilde{\sigma}^i \tilde{\sigma}^j \tilde{\sigma}^k \partial_{ijk}\Phi^*)_B &= A_B(\xi_1' + \xi_2)^3 + (B_B\kappa' + A_B\kappa - CA_B)a_1^3(\xi_1')^3 \equiv A_B\eta^3 + D_B\xi^3
\end{align}
with $\eta = \xi_1' + \xi_2$, $\xi=\xi_1'$, $D_B=B_B\kappa' + A_B\kappa - a_1^{-3}A_B$. The expressions for $ ( \tilde{\sigma}^i \tilde{\sigma}^j \tilde{\sigma}^k \partial_{ijk}\Phi^*)_F$ follow identically. We thus have \begin{align}
    E(z;x_1) \propto \left(\int_0^{\infty} d\xi ~ \xi \int_{\xi}^{\infty}d\eta ~ e^{A_F\eta^3 + D_F\xi^3}\right)\left(\int_0^{\infty} d\xi ~  \int_{\xi}^{\infty}d\eta ~ e^{A_B\eta^3 + D_B\xi^3}\right)
\end{align} 
or perhaps with the the integration ranges reversed depending on the signs of $\Re A_F, \Re A_B, \Re D_F, \Re D_B$. We have \begin{align}
    |E(z;  x_1)| & \leq  \left|\int_0^{\infty} d\xi ~ \xi \int_{\xi}^{\infty}d\eta ~ e^{A_F\eta^3 + D_F\xi^3}\right|\cdot\left|\int_0^{\infty} d\xi ~  \int_{\xi}^{\infty}d\eta ~ e^{A_B\eta^3 + D_B\xi^3}\right|\notag\\
    & \leq  \int_0^{\infty} d\xi ~ \xi \int_{\xi}^{\infty}d\eta ~| e^{A_F\eta^3 + D_F\xi^3}|\cdot\int_0^{\infty} d\xi ~  \int_{\xi}^{\infty}d\eta ~| e^{A_B\eta^3 + D_B\xi^3}|\notag\\
    & \leq  \int_0^{\infty} d\xi ~ \xi \int_{0}^{\infty}d\eta ~| e^{A_F\eta^3 + D_F\xi^3}|\cdot\int_0^{\infty} d\xi ~  \int_{0}^{\infty}d\eta ~| e^{A_B\eta^3 + D_B\xi^3}|\notag\\
    & \leq \left(|\mathfrak{M} D_F|\right)^{-2/3}\left(|\mathfrak{M} A_F|\right)^{-1/3}
    \left(|\mathfrak{M} D_B|\right)^{-1/3}
    \left(|\mathfrak{M} A_B|\right)^{-1/3}\left(\int_0^{\infty} e^{-\xi^3}d\xi\right)^3 \left(\int_0^{\infty}~ \xi e^{-\xi^3}d\xi\right)\label{eq:esd_error_bound}
\end{align}
where we have defined \begin{align}
    \mathfrak{M} y = \begin{cases} \Re y ~~ &\text{if } \Re y \neq 0, \\
    \Im y ~~ &\text{if } \Re y = 0. \\
    \end{cases}
\end{align}
This last bound follows from a standard Cauchy rotation of integration contour if any of $D_F, A_F, D_B, A_B$ has vanishing real part.
(\ref{eq:esd_error_bound}) is valid for $D_B, A_B, D_F, A_F \neq 0$, but if $D_B=0$ and $A_B\neq 0$, then the preceding calculations are simplified and we still obtain an upper bound but proportional to $(|\mathfrak{M} A_B|)^{-1/3}$. Similarly with $A_B=0$ and $D_B\neq 0$ and similarly for $A_F, D_F$. The only remaining cases are $A_B = D_B =0$ or $A_F = D_F =0$. But recall (\ref{eq:af_bf_final}) and (\ref{eq:ferm_saddle1})-(\ref{eq:ferm_saddle2}). We immediately see that $A_F=D_F$ if and only if $\sigma_{FF}=\sigma_{FF}[1]=0$, which occurs for no finite $z, x_1$. Therefore, for \emph{fixed} $(x, x_1)\in\mathbb{R}^2$,  $\alpha > 0$ and any $z\in (x-e^{-\alpha}, x + e^{-\alpha})$ \begin{align}
    |\mathbb{E}\mu_N(z) - \mu_{eq}(z; x_1) | \lesssim N^{-5/3} C(x_1, |x| + e^{-\alpha})
\end{align}
where $C(|x_1|, |x| + e^{-\alpha})$ is positive and is decreasing in $\alpha$. Since $\mu_{eq}$ is bounded, it follows that $\mathbb{E}\mu_N$ is bounded, and therefore \begin{align}
 \mathbb{E}   \int \log|z-x| \indic\{|z-x| \leq e^{-\alpha}\} d\mu_N(z) \rightarrow 0
\end{align}
as $\alpha\rightarrow\infty$ uniformly in $N$, and so the lower bound is completed.
\end{proof}

Equipped with this result, we can now prove the legitimacy of the Coulomb gas approximation in our complexity calculation. The proof will require an elementary intermediate result which has undoubtedly appeared in various places before, but we prove it here anyway for the avoidance of doubt.

\begin{lemma}\label{lem:max_entry_bound}
Let $M_N$ be a random $N\times N$ symmetric real matrix with independent centred Gaussian upper-diagonal and diagonal entries. Suppose that the variances of the entries are bounded above by $cN^{-1}$ for some constant $c>0$. Then there exists some constant $c_e$ such that \begin{align}
    \expect ||M_N||_{\text{max}}^N \lesssim e^{c_eN}.
\end{align}
\end{lemma}
\begin{proof}
Let $\sigma_{ij}^2$ denote the variance of $M_{ij}$. Then
\begin{align}
    \mathbb{E}||M||_{max}^N &\leq \sum_{i,j} \expect|M_{i,j}|^N\notag\\
    &= \sum_{i,j} \expect |\mathcal{N}(0, \sigma_{ij}^2)|^N\notag\\
    &= \sum_{i,j} \sigma_{ij}^N \mathbb{E}|\mathcal{N}(0,1)|^N\notag\\
    &\leq N^2c^{N/2} N^{-N/2} \expect |\mathcal{N}(0,1)|^N.
\end{align}
Simple integration with a change of variables gives \begin{align}
    \expect |\mathcal{N}(0,1)|^N &= 2^{\frac{N+1}{2}}\Gamma\left(\frac{N+1}{2}\right)
\end{align}
and then, for large enough $N$, Stirling's formula gives \begin{align}
      \expect |\mathcal{N}(0,1)|^N &\sim 2^{\frac{N+1}{2}} \sqrtsign{\pi(N+1)} \left(\frac{N+1}{2e}\right)^{\frac{N-1}{2}}\notag\\
     & \sim 2\sqrtsign{\pi} e^{-\frac{N-1}{2}} N^{N/2} \left(\frac{N+1}{N}\right)^{N/2}\notag\\
      &\sim 2\sqrtsign{\pi e} N^{N/2}.
\end{align}
So finally \begin{align}
     \mathbb{E}||M||_{max}^N &\lesssim N^2c^{N/2} = e^{\frac{1}{2}N\log{c}+ 2\log{N}} \leq e^{\left(\frac{1}{2}\log{c} + 2\right)N},
\end{align}
so defining $c_e = \frac{1}{2}\log{2} + 2$ gives the result.
\end{proof}

\begin{theorem}
For any $x_1\in\mathbb{R}$, let $H_N$ be a random $N\times N$ matrix distributed as in the statement of Lemma \ref{lemma:coulomb_det}. Then as $N\rightarrow \infty$
 \begin{align}
&\iint_B dxdx_1 ~ \exp\left\{-N\left(\frac{1}{2s^2}x^2 + \frac{1}{2s_1^2} (x_1)^2\right)\right\}\mathbb{E}|\det(H_N(x_1) - x)|\notag\\
= 
&\iint_B dxdx_1 ~ \exp\left\{-N\left(\frac{1}{2s^2}x^2 + \frac{1}{2s_1^2} (x_1)^2 - \int\log|z - x| d\mu_{eq}(z) + o(1) \right)\right\} +o(1).
 \end{align}
\end{theorem}
\begin{proof}
Let $R > 0$ be some constant, independent of $N$. Introduce the notation $B_{\leq R} = B\cap \{\vec{z}\in\mathbb{R}^2 \mid |z|\leq R\}$, and then \begin{align}
    &\Bigg|\iint_B dxdx_1 ~ \exp\left\{-N\left(\frac{1}{2s^2}x^2 + \frac{1}{2s_1^2} (x_1)^2\right)\right\}\mathbb{E}|\det(H_N(x_1) - x)|\notag\\
    &- \iint_{B_{\leq R}} dxdx_1 ~ \exp\left\{-N\left(\frac{1}{2s^2}x^2 + \frac{1}{2s_1^2} (x_1)^2\right)\right\}\mathbb{E}|\det(H_N(x_1) - x)|\Bigg|\notag\\
    \leq & \iint_{||\vec{x}||\geq R} dxdx_1 ~ \exp\left\{-N\left(\frac{1}{2s^2}x^2 + \frac{1}{2s_1^2} (x_1)^2\right)\right\}\mathbb{E}|\det(H_N(x_1) - x)|.\label{eq:conc_integral_bound1}
\end{align}

We have the upper bound (\ref{eq:lemma_upper_bnd}) of Lemma \ref{lemma:coulomb_det} but this cannot be directly applied to (\ref{eq:conc_integral_bound1}) since the bound relies on uniformity in $x, x_1$ which can only be established for bounded $x, x_1$. We use a much cruder bound instead. First, let \begin{align}
J_N = H_N + x_1 \left(\begin{array}{cc}
     I & 0 \\
     0 & 0
\end{array}\right)
\end{align} and then \begin{align}
    |\det\left(H_N - xI\right)| \leq  ||J_N||_{\text{max}}^N \max\{|x|, |x_1|\}^N =   ||J_N||_{\text{max}}^N \exp\left(N\max\{\log|x|, \log|x_1|\}\right).
\end{align}
$J_N$ has centred Gaussian entries with variance $\mathcal{O}(N^{-1})$, so Lemma \ref{lem:max_entry_bound} applies, and we find 
\begin{align}
    \expect  |\det\left(H_N - xI\right)| \lesssim  \exp\left(N\max\{\log|x|, \log|x_1|\}\right) e^{c_e N}
\end{align}
for some constant $c_e>0$ which is independent of $x, x_1$ and $N$, but we need not compute it.

Now we have \begin{align}
       &\Bigg|\iint_B dxdx_1 ~ \exp\left\{-N\left(\frac{1}{2s^2}x^2 + \frac{1}{2s_1^2} (x_1)^2\right)\right\}\mathbb{E}|\det(H_N(x_1) - x)|\notag\\
    &- \iint_{B_{\leq R}} dxdx_1 ~ \exp\left\{-N\left(\frac{1}{2s^2}x^2 + \frac{1}{2s_1^2} (x_1)^2\right)\right\}\mathbb{E}|\det(H_N(x_1) - x)|\Bigg|\notag\\
    \lesssim & \iint_{||\vec{x}||\geq R} dxdx_1 ~ \exp\left\{-N\left(\frac{1}{2s^2}x^2 + \frac{1}{2s_1^2} (x_1)^2 - \max\{\log|x|, \log|x_1|\} - c_e\right)\right\}.
\end{align}
But, since $\mu_{eq}$ is bounded and has compact support, we can choose $R$ large enough (independent of $N$) so that \begin{align}
  \frac{1}{2s^2}x^2 + \frac{1}{2s_1^2} (x_1)^2 - \max\{\log|x|, \log|x_1|\} - c_e> L > 0\label{eq:theorem_gaussian_bound}
\end{align}
for all $(x, x_1)$ with $\sqrtsign{x^2 + x_1^2} > R$ and for some fixed $L$ independent of $N$.  Whence  \begin{align}
       &\Bigg|\iint_B dxdx_1 ~ \exp\left\{-N\left(\frac{1}{2s^2}x^2 + \frac{1}{2s_1^2} (x_1)^2\right)\right\}\mathbb{E}|\det(H_N(x_1) - x)|\notag\\
    &- \iint_{B_{\leq R}} dxdx_1 ~ \exp\left\{-N\left(\frac{1}{2s^2}x^2 + \frac{1}{2s_1^2} (x_1)^2\right)\right\}\mathbb{E}|\det(H_N(x_1) - x)|\Bigg|\notag\\
    \lesssim & N^{-1}e^{-NL} \rightarrow 0\label{eq:thm_final1}
\end{align}
as $N\rightarrow\infty$. Finally, for $x, x_1$ in $B_{\leq R}$, the result of the Lemma \ref{lemma:coulomb_det} holds uniformly in $x, x_1$, so \begin{align}
    &\iint_{B_{\leq R}} dxdx_1 ~ \exp\left\{-N\left(\frac{1}{2s^2}x^2 + \frac{1}{2s_1^2} (x_1)^2\right)\right\}\mathbb{E}|\det(H_N(x_1) - x)|\notag\\ =  &\iint_{B_{\leq R}} dxdx_1 ~ \exp\left\{-N\left(\frac{1}{2s^2}x^2 + \frac{1}{2s_1^2} (x_1)^2 - \int\log|z-x| d\mu_{eq}(z; x_1)+ o(1)\right)\right\}.\label{eq:thm_final2}
\end{align}
The result follows from (\ref{eq:thm_final1}), (\ref{eq:thm_final2}) and the triangle inequality.
\end{proof}

\subsection{Asymptotic complexity with prescribed Hessian index}
Recall the complexity defined in (\ref{eq:C_Nkk_def}): \begin{align}
    C_{N, k_D, k_G} =\Bigg|\Bigg\{ \vwD\in S^{N_D  }, \vwG\in S^{N_G } ~:~ &\nabla_D \LD = 0, \nabla_G \LG = 0, \LD\in B_D, \LG\in B_G\notag\\
    &i(\nabla_D^2 L^{(D)}) = k_D, ~i(\nabla_G^2 L^{(G)}) = k_G \Bigg\}\Bigg|.\tag{\ref{eq:C_Nkk_def}}
\end{align}

The extra Hessian signature conditions in (\ref{eq:C_Nkk_def}) enforce that both generator and discriminator are at low-index saddle points. Our method for computing the complexity $C_N$ in the previous subsection relies on the Coulomb gas approximation applied to the spectrum of $H'$. However, the Hessian index constraints are formulated in the natural Hessian matrix (\ref{eq:H_first}), but our spectral calculations proceed from the rewritten form (\ref{eq:H_rewritten}). We find however that we can indeed proceed much as in Chapter \ref{chap:general_activation_functions}. Recall the key Hessian matrix $\tilde{H}$ given in (\ref{eq:H_first}) by \begin{align}
   \tilde{H}=  &\left(\begin{array}{cc}
      \sqrtsign{2(N_D - 1)}\sqrtsign{b^2 + b_1^2}M^{(D)} & -bG \\
       b G^T & \sqrtsign{2(N_G-1)}bM^{(G)}
    \end{array}\right)\notag\\
   &-\sqrtsign{N-2}x \left(\begin{array}{cc}
        -I_{N_D} & 0  \\
         0 & I_{N_G} 
    \end{array}\right)  + \sqrtsign{N-2}x_1\left(\begin{array}{cc}
        I_{N_D} & 0  \\
         0 & 0
    \end{array}\right)
\end{align}
where $M^{(D)}\sim GOE^{N_D -1}$, $M^{(G)}\sim GOE^{N_G-1}$, $G$ is $N_D - 1 \times N_G - 1$ Ginibre, and all are independent. Note that we have used (\ref{eq:x_x1_def}) to slightly rewrite (\ref{eq:H_first}).
We must address the problem of computing \begin{align}
    \mathbb{E}|\det \tilde{H}|\mathbbm{1}\left\{ i\left(\sqrtsign{\kappa}(1 + \mathcal{O}(N^{-1}))\sqrtsign{b^2 + b_1^2}M_D + \frac{x+x_1}{\sqrtsign{2}}\right) = k_D, ~ i\left(\sqrtsign{\kappa'}(1 + \mathcal{O}(N^{-1}))bM_G - \frac{x}{\sqrtsign{2}}\right) = k_G\right\}.
\end{align}

Indeed, we introduce integration variables $\vec{y}_1, \vec{y}_2, \zeta_1, \zeta_1^*, \zeta_2,\zeta_2^*$, being $(N-2)$-vectors of commuting and anti-commuting variables respectively. Use $[t]$ notation to split all vectors into the first $\kappa N -1$ and last $\kappa'N-1$ components. Let \begin{align}
    A[t] = \vec{y}_1\vec{y}_1^T + \vec{y}_2\vec{y}_2^T + \zeta_1\zeta_1^{\dagger} + \zeta_2\zeta_2^{\dagger}.
\end{align}

With these definitions, we have (recalling Chapter \ref{chap:general_activation_functions})\begin{align}
    |\det \tilde{H}| = (2(N-2))^{\frac{N-2}{2}} \lim_{\epsilon\searrow 0} \int d\Xi &\exp\Bigg\{-i\sqrtsign{\kappa}(1 + \mathcal{O}(N^{-1})) \sqrtsign{b^2 + b_1^2}\Tr M^{(D)} A[1]\notag\\
    &-i\sqrtsign{\kappa'}(1 + \mathcal{O}(N^{-1})) b \Tr M^{(G)} A[2] \Bigg\}\notag\\
    &\exp\{ \mathcal{O}(\epsilon)\}\exp\{\ldots\}
\end{align}
where $d\Xi$ is the normalised measure of the $\vec{y}_1, \vec{y}_2, \zeta_1, \zeta_1^*, \zeta_2,\zeta_2^*$ and the ellipsis represents terms with no dependence on $M^{(D)}$ or $M^{(G)}$, which we need not write down.
The crux of the matter is that we must compute \begin{align}
   & \mathbb{E}_{M^{(D)}}e^{-i\sqrtsign{\kappa} \sqrtsign{b^2 + b_1^2}\Tr M^{(D)} A[1]}\mathbbm{1}\left\{i\left(M_D + \frac{x + x_1}{\sqrtsign{\kappa}\sqrtsign{b^2 + b_1^2}}(1 + \mathcal{O}(N^{-1}))\right) = k_D\right\},\\
      & \mathbb{E}_{M^{(G)}}e^{-i\sqrtsign{\kappa'} b \Tr M^{(G)} A[2]}\mathbbm{1}\left\{i\left(M_G - \frac{x}{\sqrtsign{\kappa'}b}(1 + \mathcal{O}(N^{-1}))\right) = k_G\right\},
\end{align}
but in Chapter \ref{chap:general_activation_functions} we performed exactly these calculations (see around \vivacom{(\ref{eq:expect_cond_lem_upper_bound_full})}) and so there exist constants $K^{(D)}_U,K^{(D)}_L, K^{(G)}_U,K^{(G)}_L$ such that \begin{align}
  &K^{(D)}_L e^{-Nk_D\kappa(1 + o(1)) I_1(\hat{x}_D; \sqrtsign{2})} e^{-\frac{1}{2N}(b^2 + b_1^2)\Tr A[1]^2}  \notag\\
  \leq  &\Re\mathbb{E}_{M^{(D)}}e^{-i\sqrtsign{\kappa} \sqrtsign{b^2 + b_1^2}\Tr M^{(D)} A[1]}\mathbbm{1}\left\{i\left(M_D + \frac{x + x_1}{\sqrtsign{\kappa}\sqrtsign{b^2 + b_1^2}}(1 + \mathcal{O}(N^{-1}))\right) = k_D\right\} \notag\\
  \leq &K^{(D)}_U e^{-Nk_D\kappa(1 + o(1)) I_1(\hat{x}_D; \sqrtsign{2})} e^{-\frac{1}{2N}(b^2 + b_1^2)\Tr A[1]^2}
\end{align}
and 
\begin{align}
  &K^{(G)}_L e^{-Nk_G\kappa'(1 + o(1)) I_1(\hat{x}_G; \sqrtsign{2})} e^{-\frac{1}{2N}b^2 \Tr A[2]^2}  \notag\\
  \leq  &\Re\mathbb{E}_{M^{(G)}}e^{-i\sqrtsign{\kappa'} b\Tr M^{(G)} A[2]}\mathbbm{1}\left\{i\left(M_G - \frac{x}{\sqrtsign{\kappa'}b}(1 + \mathcal{O}(N^{-1}))\right) = k_G\right\} \notag\\
  \leq &K^{(G)}_U e^{-Nk_G\kappa'(1 + o(1)) I_1(\hat{x}_G; \sqrtsign{2})} e^{-\frac{1}{2N}b^2 \Tr A[2]^2}
\end{align}
where \begin{align}
    \hat{x}_D = -\frac{x + x_1}{\sqrtsign{\kappa}\sqrtsign{b^2 + b_1^2}}, ~~ \hat{x}_G = \frac{x}{\sqrtsign{\kappa'}b}.
\end{align}
Here $I_1$ is the rate function of the largest eigenvalue of the GOE as obtained in \cite{arous2001aging} and used in \cite{auffinger2013random} and Chapter \ref{chap:general_activation_functions}: \begin{align}
    I_1(u; E) = \begin{cases} \frac{2}{E^2}\int_u^{-E} \sqrtsign{z^2 - E^2}dz ~ &\text{ for } u < -E,\\
        \frac{2}{E^2}\int_{E}^u \sqrtsign{z^2 - E^2}dz ~ &\text{ for } u > E,\\
    \infty &\text{ for } |u| < E.
    \end{cases}
\end{align}
Note that for $u< -E$ \begin{align}
    I_1(u; E) = -\frac{u}{E}\sqrtsign{u^2 - E^2} - \log\left(-u + \sqrtsign{u^2 - E^2}\right) + \log{E}
\end{align}
and for $u>E$ we simply have $I_1(u; E) = I_1(-u; E)$. Note also that $I_1(ru; E) = I_1(u, E/r).$

\begin{figure}
    \centering
    \includegraphics[width=\textwidth]{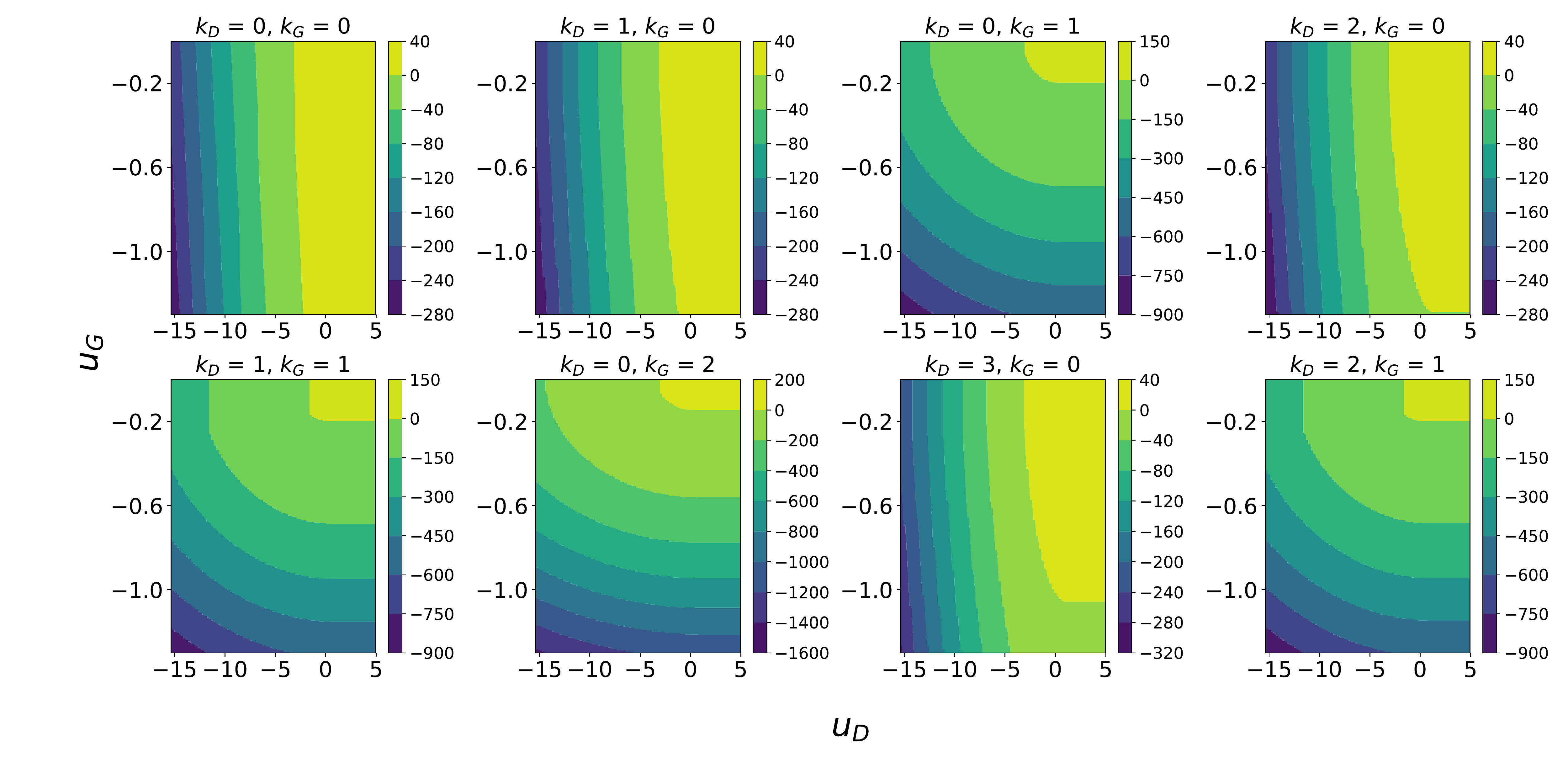}
    \caption{Contour plots of $\Theta_{k_D, k_G}$ for a few values of $k_D, k_G$. Here $p=q=3, \sigma_z=1, \kappa=0.9$.}
    \label{fig:theta_indexs}
\end{figure}

We have successfully dealt with the Hessian index indicators inside the expectation, however we need some way of returning to the form of $\tilde{H}$ in (\ref{eq:H_rewritten}) so the complexity calculations using the Coulomb gas approach can proceed as before. We can achieve this with inverse Fourier transforms:
\begin{align}
    e^{-\frac{1}{2N}(b^2 + b_1^2)\Tr A[1]^2} &= \mathbb{E}_{M_D}e^{-i\sqrtsign{\kappa}\sqrtsign{b^2 + b_1^2}\Tr M_DA[1]}\\
     e^{-\frac{1}{2N}b^2\Tr A[2]^2} &= \mathbb{E}_{M_G}e^{-i\sqrtsign{\kappa'}b\Tr M_GA[2]}   
\end{align}
from which we obtain \begin{align}
&K_Le^{-Nk_D\kappa(1 + o(1)) I_1(\hat{x}_D; \sqrtsign{2})} e^{-Nk_G\kappa'(1 + o(1)) I_1(\hat{x}_G; \sqrtsign{2})}\mathbb{E}|\det \tilde{H}|\notag\\
\leq 
  & \mathbb{E}|\det \tilde{H}|\mathbbm{1}\left\{ i\left(\sqrtsign{\kappa}(1 + \mathcal{O}(N^{-1}))\sqrtsign{b^2 + b_1^2}M_D + \frac{x+x_1}{\sqrtsign{2}}\right) = k_D, ~ i\left(\sqrtsign{\kappa'}(1 + \mathcal{O}(N^{-1}))bM_G - \frac{x}{\sqrtsign{2}}\right) = k_G\right\} \\
  \leq & K_Ue^{-Nk_D\kappa(1 + o(1)) I_1(\hat{x}_D; \sqrtsign{2})} e^{-Nk_G\kappa'(1 + o(1)) I_1(\hat{x}_G; \sqrtsign{2})} \mathbb{E}|\det \tilde{H}|.
  \end{align}
It follows that \begin{align}
&K_N'\iint_B dx dx_1 e^{-(N-2)\left[ \Phi(x, x_1) + k_G\kappa' I_1(x; \sqrtsign{2\kappa'}b) + k_D\kappa I_1\left(( - (x + x_1); \sqrtsign{2\kappa(b^2 + b_1^2)}\right)\right](1 + o(1))} \notag\\
\lesssim
    &C_{N, k_D, k_G} \notag\\
    \lesssim &K_N' \iint_B dx dx_1 e^{-(N-2)\left[ \Phi(x, x_1) + k_G\kappa' I_1(x; \sqrtsign{2\kappa'}b) + k_D\kappa I_1\left(( - (x + x_1); \sqrtsign{2\kappa(b^2 + b_1^2)}\right)\right](1 + o(1))}.
\end{align}
So we see that the relevant exponent in this case is the same as for $C_N$ but with additional GOE eigenvalue large deviation terms, giving the complexity limit \begin{align}
     \lim \frac{1}{N} \log\mathbb{E}C_{N, k_D, k_G} 
&=  \Theta_{k_D, k_G}(u_D, u_G)\notag\\ &= K - \min_B \left\{\Phi +  k_G\kappa' I_1(x; \sqrtsign{2\kappa'}b) + k_D\kappa I_1\left( - (x + x_1); \sqrtsign{2\kappa(b^2 + b_1^2)}\right)\right\}.\label{eq:theta_kd_kg_final}
\end{align}
Plots of $\Theta_{k_D, k_G}$ for a few values of $k_D, k_G$ are shown in Figure \ref{fig:theta_indexs}.

\begin{remark}
Recall that the limiting spectral measure of the Hessian displays a transition as the support splits from one component to two, as shown in Figure \ref{fig:spectra}. Let us comment on the relevance of this feature to the complexity. The spectral measure appears in one place in the above complexity calculations: the Coulomb gas integral $\int d\mu_{eq}(z) \log|z - x|$. The effect of integrating against the measure $\mu_{eq}$ is to smooth out the transition point. In other words, if $\mu_{eq}$ has two components or is at the transition point, one expects to be able to construct another measure $\nu$ supported on a single component such that $\int d\nu(z) \log|z - x| = \int d\mu_{eq}(z) \log|z - x|$. We interpret this to mean that the Coulomb gas integral term does not display any features that can be unambiguously attributed to the transition behaviour of the spectral measure.
\end{remark}

\section{Implications}\label{sec:implications}
\subsection{Structure of low-index critical points}
\begin{figure}[h]
    \centering
    \includegraphics[width=\textwidth]{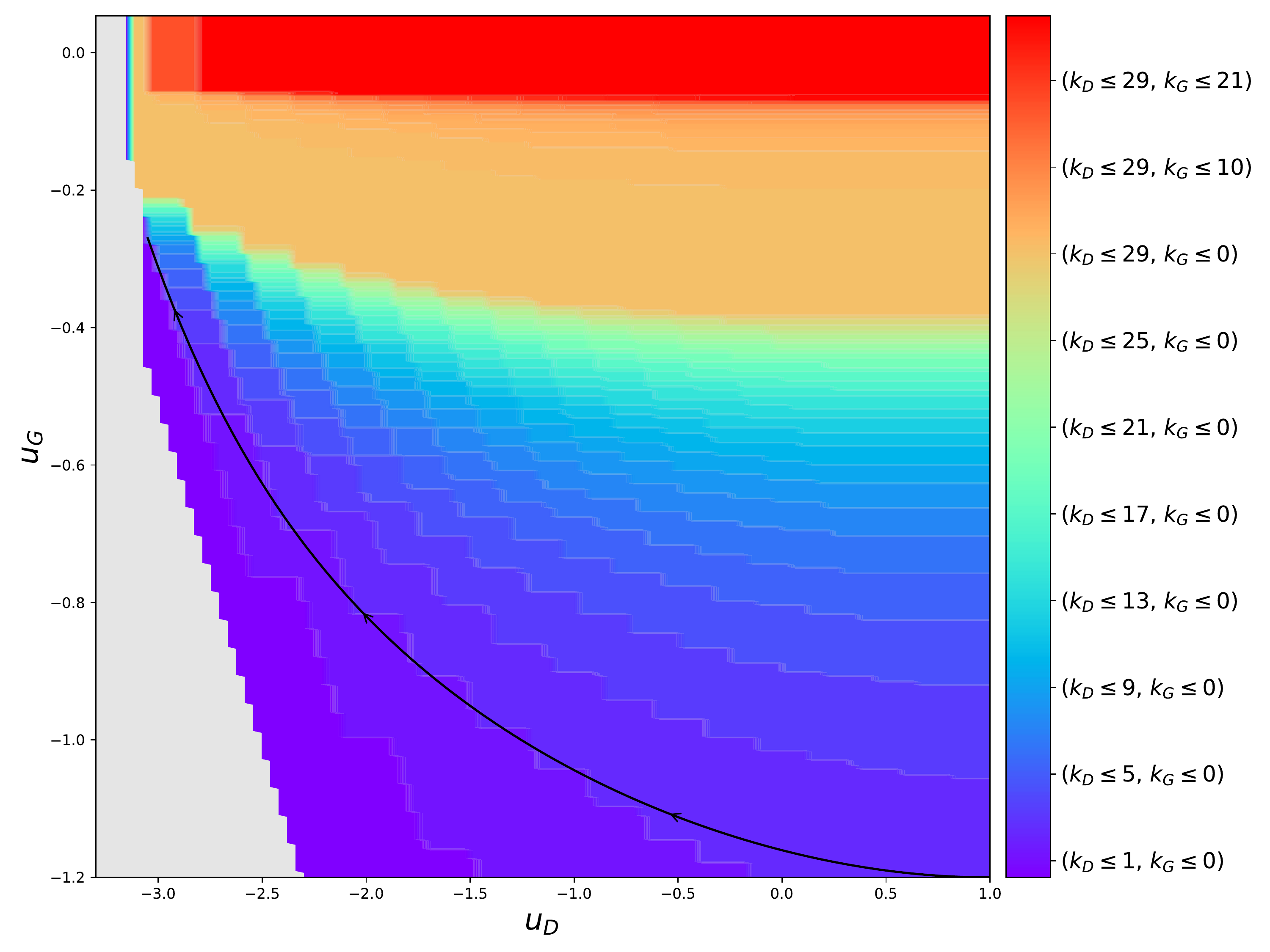}
    \caption{Contours in the $(u_D, u_G)$ plane of the maximum $k_D$ and $k_G$ such that $\Theta_{k_D, k_G}(u_D, u_G)>0$. $k_D$ results shown with a red colour red scheme, and $k_G$ with blue/green. The grey region on the left lies outside the domain of definition of $\Theta_{k_D, k_G}$. Here $p=q=3, \sigma_z=1, \kappa=0.9$. The arrow indicates the approximate location of the contour discussed in the main text.}
    \label{fig:kd_kg_structure}
\end{figure}

We examine the fine structure of the low-index critical points for both spin glasses. \cite{choromanska2015loss} used the `banded structure' of low-index critical points to explain the effectiveness of gradient descent in large multi-layer perceptron neural networks. We undertake to uncover the analogous structure in our dual spin-glass model and thence offer explanations for GAN training dynamics with gradient descent. For a range of $(k_D, k_G)$ values, starting at $(0, 0)$, we compute $\Theta_{k_D, k_G}$ on an appropriate domain. In the $(u_D, u_G)$ plane, we then find the maximum $k_D$, and separately $k_G$, such that $\Theta_{k_D, k_G}(u_D, u_G)>0$. In the large $N$ limit, this procedure reveals the regions in the $(u_D, u_G)$ plane where critical points of each index of the two spin glasses are found. Figure \ref{fig:kd_kg_structure} plots these maximum $k_D, k_G$ values as contours on a shared $(u_D, u_G)$ plane. The grey region in the plot clearly shows the `ground state' boundary beyond which no critical points exist. We use some fixed values of the various parameters: $p=q=3, \sigma_z=1, \kappa=0.9$.

\medskip
These plots reveal, unsurprisingly perhaps, that something resembling the banded structure of \cite{choromanska2015loss} is present, with the higher index critical points being limited to higher loss values for each network. The 2-dimensional analogues of the $E_{\infty}$ boundary of \cite{choromanska2015loss} are evident in the bunching of the $k_D, k_G$ contours at higher values. There is, however further structure not present in the single spin-glass multi-layer perceptron model. Consider the contour of $k_D = 0$ at the bottom of the full contour plot in Figure \ref{fig:kd_kg_structure}. Imagine traversing a path near this contour from right to left (decreasing $u_D$ values); an example path is approximately indicated by a black arrow on the figure. At all points along such a path, the only critical points present are exact local minima for both networks, however the losses range over 
\begin{enumerate}[label=(\roman*)]
    \item low generator loss, high discriminator loss;
    \item some balance between generator and discriminator loss;
    \item high generator loss, low discriminator loss.
\end{enumerate}  These three states correspond qualitatively to known GAN phenomena:

\begin{enumerate}[label=(\roman*)]
    \item discriminator collapses to predicting `real' for all items;
    \item successfully trained model;
    \item generator collapses to producing garbage samples which the discriminator trivially identifies.
\end{enumerate}

\begin{figure}[h]
    \centering
    \includegraphics[width=0.4\textwidth]{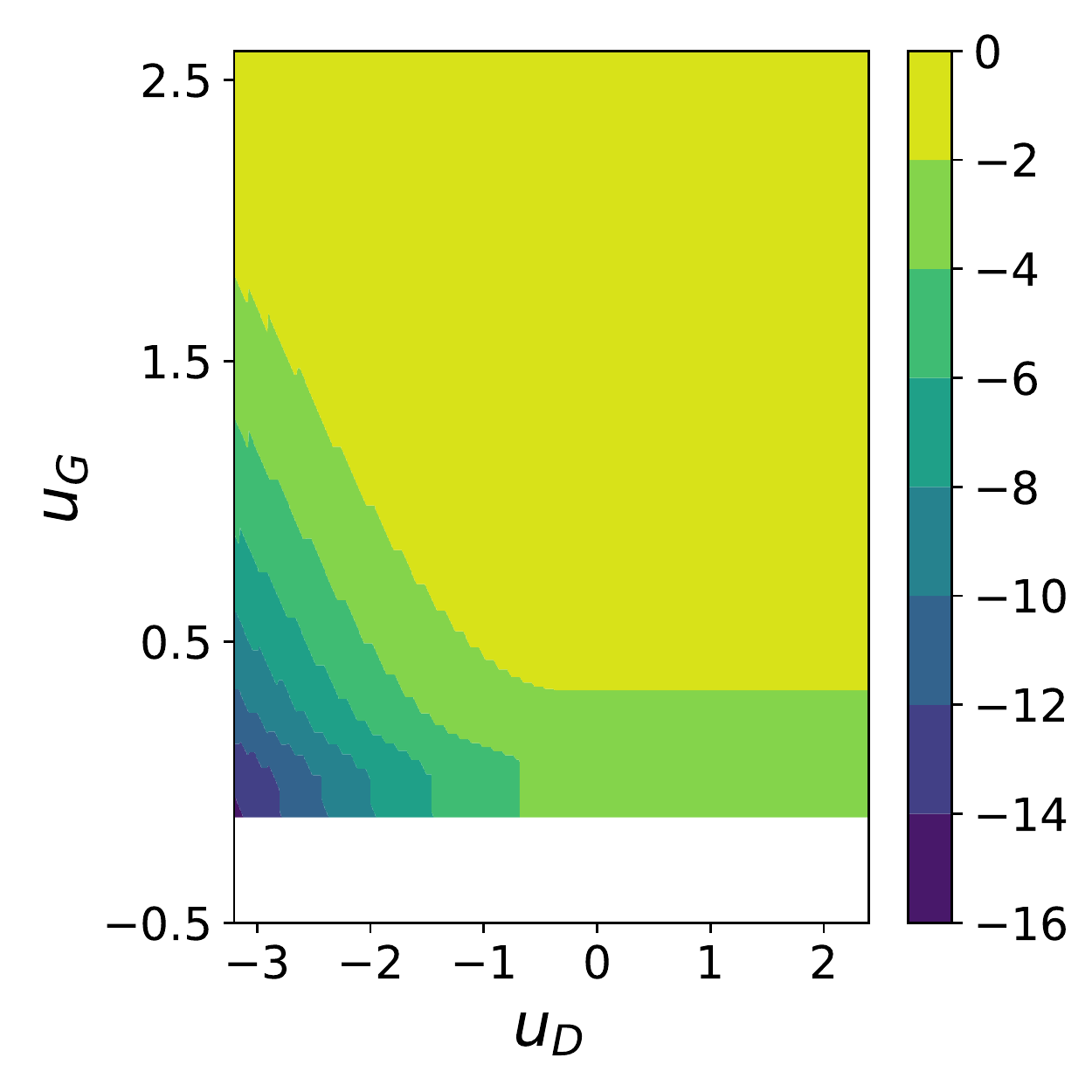}
    \caption{Contour plot of the log ratio quantity given in (\ref{eq:theta_ratio}). This is the approximate proportion of minima with both $L_D \leq u_D$ and $L_G\leq u_G$ out of all points where at at least one of those conditions holds.}
    \label{fig:theta_ratio}
\end{figure}

Overall, the analysis of our model reveals a loss surface that favours convergence to states of low loss for \emph{at least one of the networks}, but not necessarily both. Moreover, our plots of $\Theta$ and $\Theta_{k_D, k_G}$ in Figures \ref{fig:theta_latest}, \ref{fig:theta_indexs} demonstrate clearly the competition between the two networks, with the minimum attainable discriminator loss increasing as the generator loss decreases and vice-versa. We thus have a qualitative similarity between the minimax dynamics of real GANs and our model, but also a new two-dimensional banded critical points structure. We can further illuminate the structure by plotting, for each $(u_D, u_G)$, the approximate proportion of minima with both $L_D \leq u_D$ and $L_G\leq u_G$ out of all points where at at least one of those conditions holds. The expression is \begin{align}\label{eq:theta_ratio}
    \Theta(u_D, u_G) - \max\{\Theta(u_D, \infty), \Theta(\infty, u_G)\}
\end{align} which gives the log of the ratio in units of $N$. We show the plot in Figure \ref{fig:theta_ratio}. Note that, for large $N$, any region of the plot away from a value of zero contains exponentially more bad minima -- where one of the networks has collapsed -- than good minima, with equilibrium between the networks. The model therefore predicts the existence of good local minima (in the bottom left of Figure \ref{fig:theta_ratio}) that are effectively inaccessible due to their being exponentially outnumbered by bad local minima.

The structure revealed by our analysis offers the following explanation of large GAN training dynamics with gradient descent: \begin{enumerate}
    \item As with single feed-forward networks, the loss surface geometry encourages convergence to globally low values of at least one of the network losses.
    \item The same favourable geometry encourages convergence to successful states, where both networks achieve reasonably low loss, but also encourages convergence to failure states, where the generator's samples are too easily distinguished by the discriminator, or the discriminator has entirely failed thus providing no useful training signal to the generator.
\end{enumerate}

\begin{remark}
A natural question in the context of our analysis of low-index critical points is: do such points reflect the points typically reached by gradient descent algorithms used to train real GANs? There has been much discussion in the literature of the analogous question for single networks and spin glasses \cite{choromanska2015loss,baity2019comparing,folena2019rethinking}. It is not clear how to settle this question in our case, but we believe our model and its low-index critical points give a description of the baseline properties to be expected of high-dimensional adversarial optimisation problems late in the optimisation procedure. In addition, the unstructured random noise present in spin glasses may be more appropriate in our model for GANs than it is for single spin-glass models of single networks, as GAN generators do genuinely contain unstructured latent noise, rather than just the highly-structured data distributions seen on real data.
\end{remark}

\begin{remark}
    The issue of meta-stability is also worth mentioning. In single spin glasses, the boundary $E_{\infty}$ between fixed index and unbounded index critical points is meta-stable \cite{crisanti1995thouless,kurchan1993barriers}. From the random matrix theory perspective, the $E_{\infty}$ boundary corresponds to the left edge of the Wigner semi-circle \cite{auffinger2013random}. There are $O(N)$ eigenvalues in any finite interval at the left of the Wigner semi-circle, corresponding to $O(N)$ Hessian eigenvalues in any neighbourhood around zero. The 2D analogue of the $E_{\infty}$ boundary in our double spin-glass model is expected to possess the same meta-stability: the Wigner semi-circle is replaced by the measure studied in Section \ref{sec:lsd}, to which the preceding arguments apply. In the context of deep neural networks, there is a related discussion concerning ``wide and flat local optima'' of the loss surface, i.e. local optima for which many of the Hessian eigenvalues are close to zero. There are strong indications that deep neural networks converge under gradient-based optimisation to such optima \cite{10.1162/neco.1997.9.1.1,Chaudhari_2019,DBLP:conf/iclr/KeskarMNST17,kleinberg2018alternative,baldassi2021unveiling,baldassi2020shaping} and that they are perhaps better for generalisation (i.e. test set loss) than other local optima, however some authors have challenged this view \cite{pmlr-v70-dinh17b,NIPS2017_a5e0ff62,kawaguchi2020generalization,he2019asymmetric,granziol2020flatness}. It is beyond the scope of the present work to analyse the role of meta-stability further, however we note that the indications from machine learning are that it is most significant when considering generalisation, however our work simplifies to the case of a single loss rather than separately considering training and test loss.
    \end{remark}

\subsection{Hyperparameter effects}\label{subsec:hparams}
Our proposed model for GANs includes a few fixed hyperparameters that we expect to control features of the model, namely $\sigma_z$ and $\kappa$. Based on the results of \cite{auffinger2013random,choromanska2015loss} and Chapter \ref{chap:general_activation_functions}, and the form of our analytical results above, we do not expect $p$ and $q$ (the number of layers in the discriminator and generator) to have any interesting effect beyond $p, q \geq 3$; this is clearly a limitation of the model. We would expect there to exist an optimal value of $\sigma_z$ that would result in minimum loss, in some sense. The effect of $\kappa$ is less clear, though we guess that, in the studied $N\rightarrow\infty$ limit, all $\kappa\in(0, 1)$ are effectively equivalent. Intuitively, choosing  $\kappa=0, 1$ corresponds to one network having a negligible number of parameters when compared with the other and we would expect the much larger network to prevail in the minimax game, however our theoretical results above are valid strictly for $\kappa\in (0,1)$. 

\medskip
In the following two subsections we examine effect of $\sigma_z$ and $\kappa$ in our theoretical and in real experiments with a DCGAN \cite{radford2015unsupervised}. Additional supporting plots are given in the appendix.

\subsubsection{Effect of variance ratio}
In the definition of complexity, $u_D$ and $u_G$ are upper bounds on the loss of the discriminator and generator, respectively. We are interested in the region of the $u_D,u_G$ plane such that $\Theta(u_D, u_G)>0$, this being the region where gradient descent algorithms are expected to become trapped. We therefore investigate the minimum loss such that $\Theta > 0$, this being, for a given $\sigma_z$, the theoretical minimum loss attainable by the GAN. We consider two natural notions of loss:
\begin{enumerate}
    \item $\vartheta_D = \min\{u_D\in\mathbb{R} \mid \exists u_G\in\mathbb{R} ~:~ \Theta(u_D, u_G) > 0 \} $;
    \item $\vartheta_G = \min\{u_G\in\mathbb{R} \mid \exists u_D\in\mathbb{R} ~:~ \Theta(u_D, u_G) > 0 \} $.
\end{enumerate}
We vary $\sigma_z$ over a range of values in $(10^{-5}, 10^{2})$ and compute $\vartheta_D, \vartheta_G$.

\begin{figure}[h]
    \centering
    \begin{subfigure}{0.4\linewidth}
     \centering
     \includegraphics[width=\linewidth]{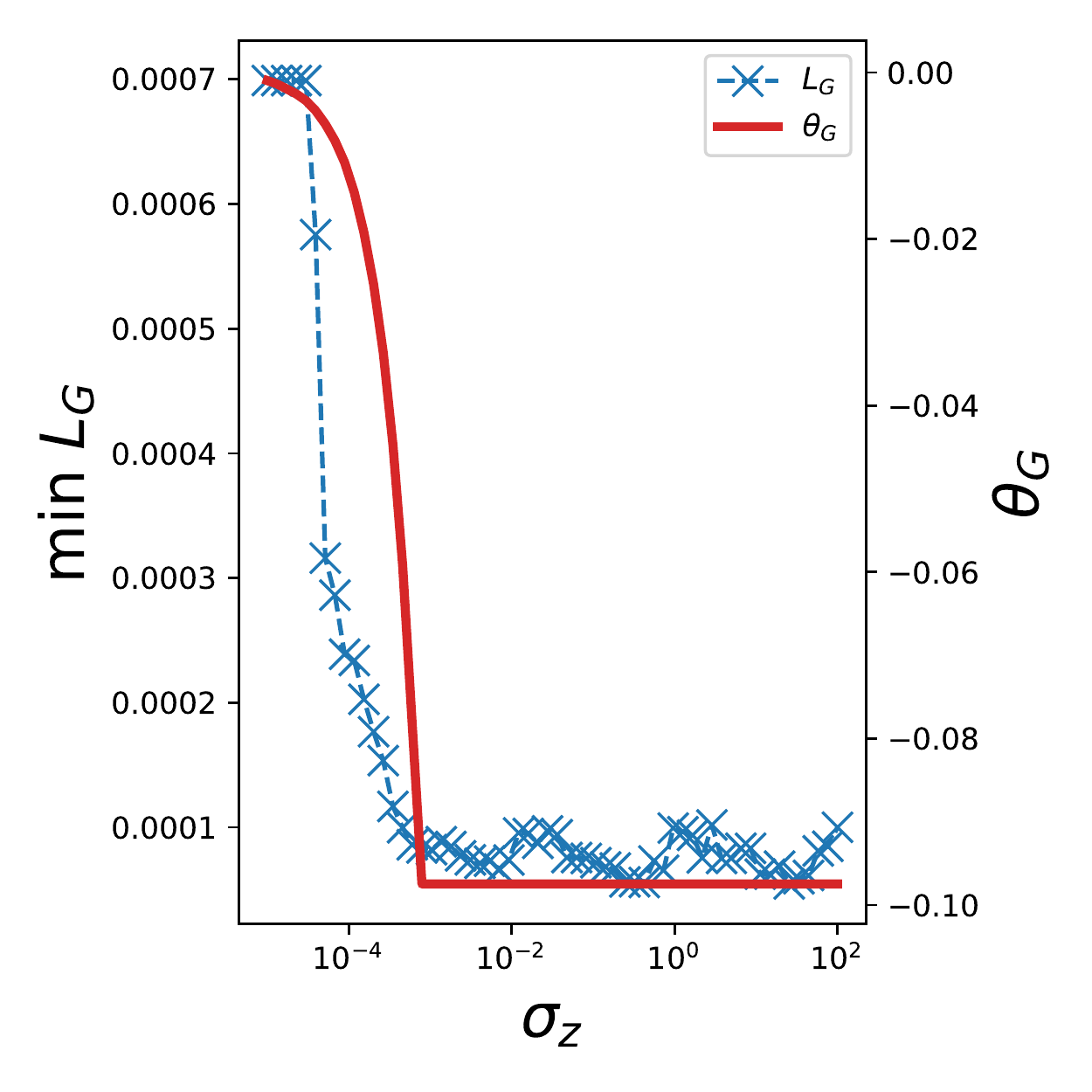}
     \subcaption{Generator}
    \end{subfigure}    
        \begin{subfigure}{0.4\linewidth}
     \centering
     \includegraphics[width=\linewidth]{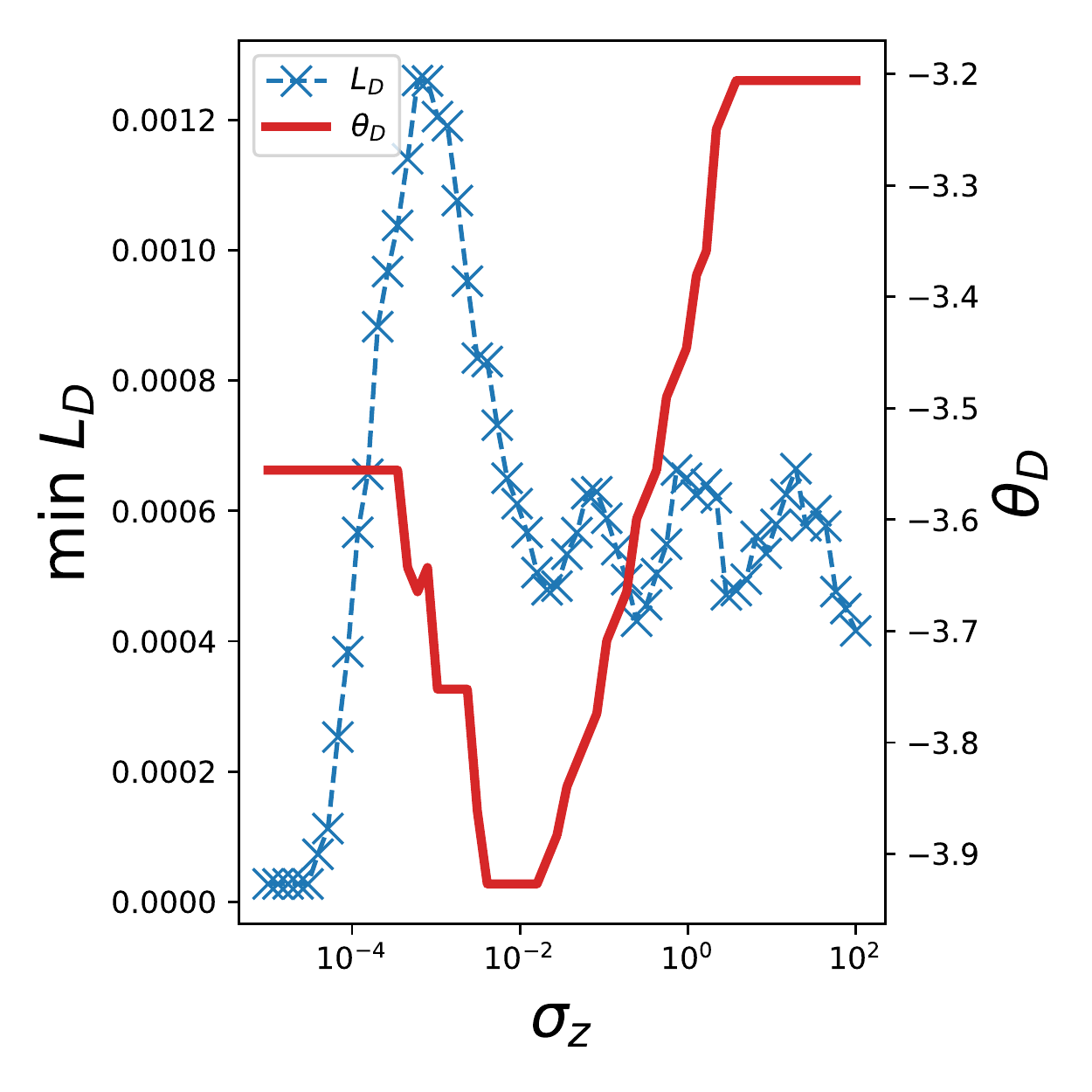}
     \subcaption{Discriminator}
    \end{subfigure}    
    \caption{The effect of $\sigma_z$. Comparison of theoretical predictions of minimum possible discriminator and generator losses to observed minimum losses when training DCGAN on CIFAR10. The blue cross-dashed lines show the experimental DCGAN results, and solid red lines show the theoretical results $\theta_G, \theta_D$. $p=q=5$ and $\kappa=0.5$ are used in the theoretical calculations, to best match the DCGAN architecture. $\sigma_z$ is shown on a log-scale.}
    \label{fig:vary_sigma_results}
\end{figure}

\medskip
To compare the theoretical predictions of the effect of $\sigma_z$ to real GANs, we perform a simple set of experiments. We use a DCGAN architecture \cite{radford2015unsupervised} with 5 layers in each network, using the reference PyTorch implementation from \cite{gan-code}, however we introduce the generator noise scale $\sigma_z$. That is, the latent input noise vector $\vec{z}$ for the generator is sampled from $\mathcal{N}(0, \sigma_z^2I)$. For a given $\sigma_z$, we train the GANs for 10 epochs on CIFAR10 \cite{krizhevsky2009learning} and record the generator and discriminator losses. For each $\sigma_z$, we repeat the experiment 30 times and average the minimum attained generator and discriminator losses to account for random variations between runs with the same $\sigma_z$. We note that the sample variances of the loss were typically very high, despite the PyTorch random seed being fixed across all runs. We plot the sample means, smoothed with rolling averaging over a short window, in the interest of clearly visualising whatever trends are present. The results are shown in Figure \ref{fig:vary_sigma_results}.\\

There is a striking similarity between the generator plots, with a sharp decline between $\sigma_z=10^{-5}$ and around  $10^{-3}$, after which the minimum loss is approximately constant. The picture for the discriminator is less clear. Focusing on the sections $\sigma_z > 10^{-3}$, both plots show a clear minimum, at around $\sigma_z=10^{-1}$ in experiments and $\sigma_z=10^{-2}$ in theory. Note that the scales on the $y$-axes of these plots should not be considered meaningful. Though there is not precise correspondence between the discriminator curves, we claim that both theory and experiment tell the same qualitative story: increasing $\sigma_z$ to at least around $10^{-3}$ gives the lowest theoretical generator loss, and then further increasing to, tentatively, some value in $(10^{-2}, 10^{-1})$ gives the lowest possible discriminator loss at no detriment to the generator. 

We are not aware of $\sigma_z$ tuning being widely used in practice for real GANs, rather it is typically taken to be unity. We have chosen this parameter, as it can be directly paralleled in our spin glass model, therefore allowing for the above experimental comparison. Naturally there are other parameters of real GANs that one might wish to study (such as learning rates and batch sizes) however these are much less readily mirrored in the spin glass model and complexity analysis, precluding comparisons between theory and experiment. Nevertheless, the experimental results in Figure \ref{fig:vary_sigma_results} do demonstrate that tuning $\sigma_z$ in real GANs could be of benefit, as $\sigma_z=1$ does not appear to be the optimal value.

\subsubsection{Effect of size ratio}
\begin{figure}[h]
    \centering
        \begin{subfigure}{0.4\linewidth}
     \centering
     \includegraphics[width=\linewidth]{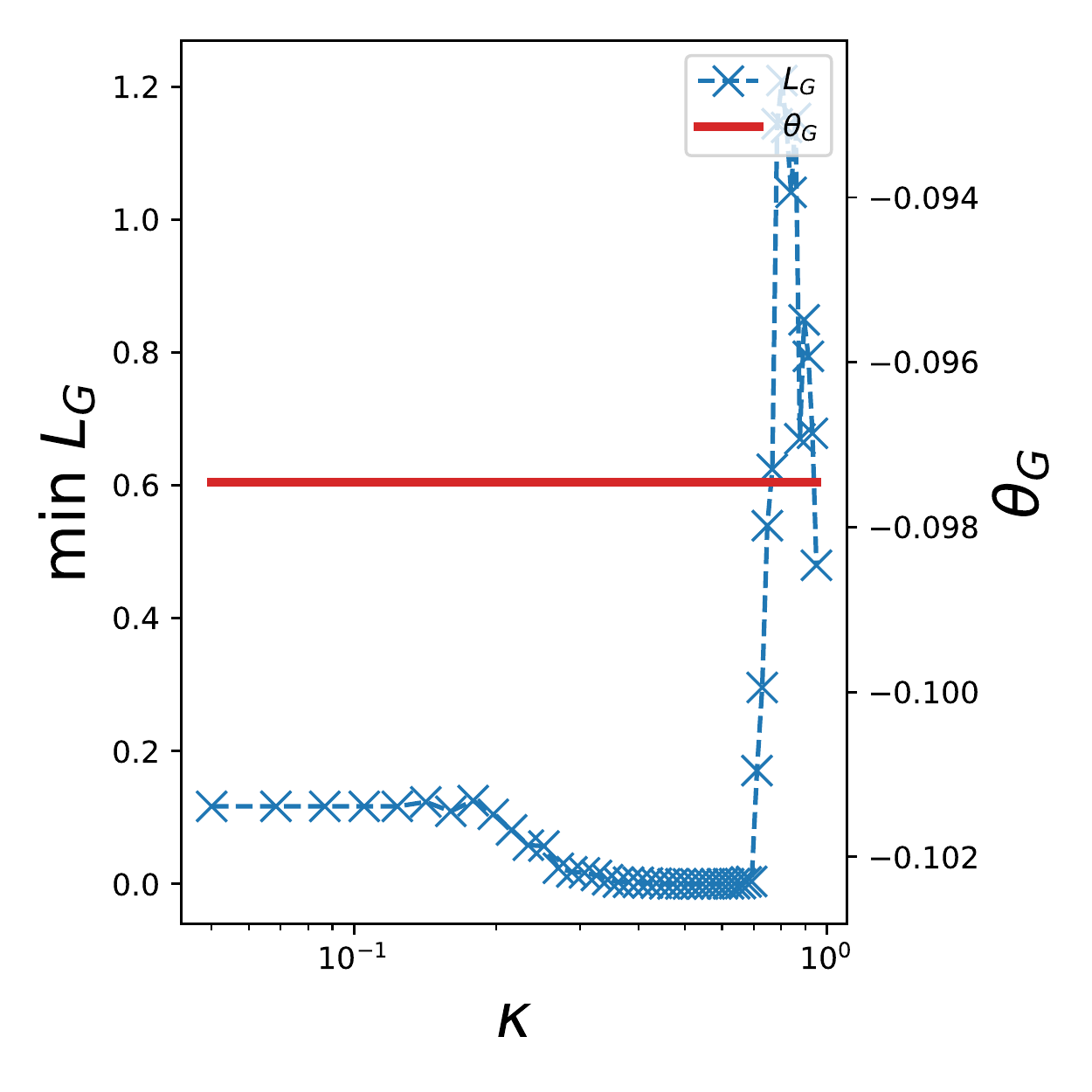}
     \subcaption{Generator}
    \end{subfigure}    
        \begin{subfigure}{0.4\linewidth}
     \centering
     \includegraphics[width=\linewidth]{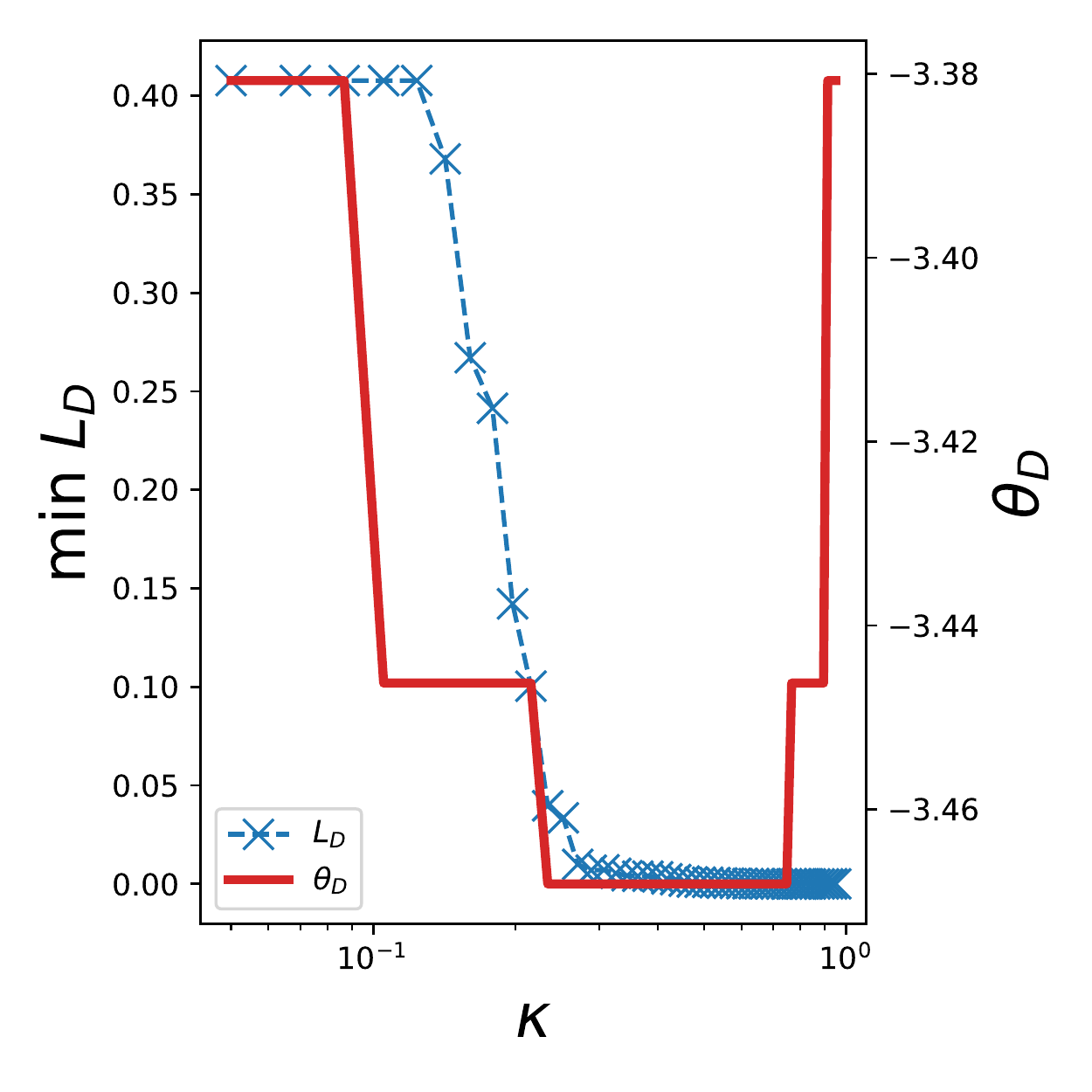}
     \subcaption{Discriminator}
    \end{subfigure}  
    \caption{The effect of $\kappa$. Comparison of theoretical predictions of minimum possible discriminator and generator losses to observed minimum losses when training DCGAN on CIFAR10. The blue cross-dashed lines show the experimental DCGAN results, and the solid red show the theoretical results $\vartheta_G, \vartheta_D$. $p=q=5$ and $\sigma_z=1$ are used in the theoretical calculations, to best match the DCGAN architecture.}
    \label{fig:vary_kappa_results}
\end{figure}

Similarly to the previous section, we can investigate the effect of $\kappa$ using $\vartheta_D, \vartheta_G$ while varying $\kappa$ over $(0,1)$. To achieve this variation in the DCGAN, we vary the number of convolutional filters in each network. The generator and discriminator are essentially mirror images of each other and the number of filters in each intermediate layer are defined as increasing functions\footnote{Number of filters in a layer is either proportional to $n_D$ or $n_D^2$ depending on the layer (and similarly with $n_G$).} of some positive integers $n_G, n_D$.  We fix $n_D + n_G=128$ and vary $n_D$ to obtain a range of $\kappa$ values, with $\kappa = \frac{n_d}{n_d + n_g}$. The results are shown in Figure \ref{fig:vary_kappa_results}.

\medskip
The theoretical model predicts a a broad range of equivalently optimal $\kappa$ values centred on $\kappa=0.5$ from the perspective of the discriminator loss, and no effect of $\kappa$ on the generator loss. The experimental results similarly show a broad range of equivalently optimal $\kappa$ centred around $\kappa=0.5$, however there appear to be deficiencies in our model, particularly for higher $\kappa$ values. The results of the experiments are intuitively sensible: the generator loss deteriorates for $\kappa$ closer to 1, i.e. when the discriminator has very many more parameters than the generator, and vice-versa for small $\kappa$.

\section{Gaussian Hessian calculations}\label{app:gaussian}
In this section we give the full details of the Gaussian calculations for the distribution of the Hessian:
\begin{align}\label{eq:app_cond_hess1}
    \left(\begin{array}{cc} \nabla_D^2 \LD & \nabla_{GD} \LD \\ \nabla_{DG}\LG & \nabla^2_{G} \LG\end{array}\right)  ~\Bigg|~ \nabla_G\LG =0, \nabla_D\LD = 0, \LD\in B_D, \LG\in B_G.
\end{align}

These calculations are routine and consist of repeated application of standard results for conditioning multivariate Gaussians, but the details are nevertheless intricate.

\medskip
Recall the definitions
\begin{align*}
    \LD(\vwD, \vwG) &= \lD(\vwD) - \sigma_z\lG(\vwD, \vwG)\\
    \LG(\vwD, \vwG) &= \sigma_z \lG(\vwD,\vwG)
\end{align*}
and
\begin{align*}
    \lD(\vwD) &= \sum_{i_1,\ldots, i_p=1}^{N_D} X_{i_1,\ldots, i_p} \prod_{k=1}^p \wD_{i_k}\\
    \lG(\vwD, \vwG) &= \sum_{i_1,\ldots, i_{p+q}=1}^{N_D + N_G} Z_{i_1,\ldots, i_{p+q}} \prod_{k=1}^{p+q} w_{i_k}
\end{align*}
for i.i.d. Gaussian $X$ and $Z$, where $\vec{w}^T = ({\vwD}^T, {\vwG}^T)$. As mentioned in the main text, we have spherical symmetry in both $\vwD$ and $\vwG$, so it sufficient to consider the distribution (\ref{eq:app_cond_hess1}) around some fixed specific points on the spheres $S^{N_D}$ and $S^{N_G}$. Following \cite{auffinger2013random}, we choose the north poles. We can select a coordinate basis around both poles, e.g. with \begin{align*}
    \vwD = (\sqrtsign{1 - \vec{u}^2}, \vec{u}), ~~~\vwG = (\sqrtsign{1 - \vec{v}^2}, \vec{v}),
\end{align*}
for $\vec{u}\in\mathbb{R}^{N_D-1}, \vec{v}\in\mathbb{R}^{N_G - 1}$ with $\vec{u}^2 \leq 1, \vec{v}^2 \leq 1$.

We need the joint distributions
\begin{align*}
    \left(\lD, \pD_i \lD, \pD_{jk}\lD\right), ~~ \left(\lG, \pG_i \lG, \pG_{jk}\lG, \pD_l \lG, \pD_{mn}\lG \right)
\end{align*} where the two groups are independent from of each other. \emph{The derivatives $\pD, \pG$ are now Euclidean derivatives with respect to the coordinates $\vec{u}, \vec{v}$.} $\lD$ behaves just like a single spin glass, and so we have \cite{auffinger2013random}:\begin{align}
    Var(\lD) &= 1,\label{eq:ld_var}\\
    Cov(\pD_i \lD, \pD_{jk} \lD) &= 0,\\
    \pD_{ij}\lD ~|~ \{\lD=x_D\} &\sim \sqrtsign{(N_D-1)p(p-1)}GOE^{N_D - 1} - x_DpI.
\end{align}
To find the joint and thence conditional distributions for $\lG$, we first note that $\lG$ is simply a spin glass on a partitioned vector $\vec{w}^T = ({\vwD}^T, {\vwG}^T)$, so \begin{align}
    Cov(\lG(\vwD, \vwG), \lG({\vwD}', {\vwG}')) = \left(\vwD\cdot {\vwD}' + \vwG \cdot {\vwG}'\right)^{p+q}\label{eq:lg_cov_func}
\end{align}
from which, by comparing with \cite{auffinger2013random}, one can obtain the necessary expressions, at the north poles in a coordinate basis. Practically, one writes ${\vwD}^T = (\sqrtsign{1 - \sum_{j}u_j^2}, u_1, \ldots, u_{N_D -1})$, and similarly for $\vwG$. Then one takes derivatives of (\ref{eq:lg_cov_func}) with respect to these new variables around the north poles. Finally, one sets $\vwD={\vwD}'$ and takes $u_j=0 ~ \forall j$, and similarly for $\vwG$. The resulting expressions are largely familiar from the standard spin glass in \cite{auffinger2013random}, except there are extra cross terms between $\vwD$ and $\vwG$:
\begin{align}
    Var(\lG) &= 2^{p+q},\label{eq:var_lg}\\
    Cov(\pG_{ij}\lG, \lG) &= - (p+q)2^{p+q}\delta_{ij},\\
    Cov(\pD_{ij} \lG, \lG) &= - (p+q)2^{p+q}\delta_{ij},\\
    Cov(\pG_{ij}\lG,  \pG_{kl}\lG) &=  2^{p+q}\left[ (p+q)(p+q-1)\left(\delta_{ik}\delta_{jl} + \delta_{il}\delta_{jk}\right) + (p+q)^2 \delta_{ij}\delta_{kl}\right],\\
     Cov(\pG_{ij}\lG,  \pD_{kl}\lG) &=  2^{p+q} (p+q)^2 \delta_{ij}\delta_{kl},\\
      Cov(\pG_i\pD_j\lG,  \pG_{k}\pD_l\lG) &= 2^{p+q} (p+q)(p+q-1) \delta_{ik}\delta_{jl},\label{eq:ginibre}\\
      Cov(\pG_{ij}\lG,  \pG_{k}\pD_l\lG) &= 0\label{eq:gin_goe1}\\
    Cov(\pD_{ij}\lG,  \pD_{k}\pG_l\lG) &= 0\label{eq:gin_goe2},\\
    Cov(\pD_{i}\pG_j \lG, \lG) &= 0.
\end{align}
Also, all first derivatives of $\lG$ are clearly independent of $\lG$ and its second derivatives by the same reasoning as in \cite{auffinger2013random}. Note that \begin{align}
    Cov(\partial^{(D)}_i L^{(D)}, \partial^{(D)}_j L^{(D)})& = (p + \sigma_z^2 2^{p+q}(p+q))\delta_{ij}\\
    Cov(\partial^{(G)}_iL^{(G)}, \partial^{(G)}_j L^{(G)})& = \sigma^2_z 2^{p+q}(p+q)\delta_{ij}\\
     Cov(\partial^{(D)}_iL^{(D)}, \partial^{(G)}_j L^{(G)})& = 0
\end{align}
and so \begin{align}\label{eq:grad_dens_0}
    \varphi_{\left(\nabla_D L^{(D)}, \nabla_G L^{(G)}\right)}(0) = (2\pi)^{-\frac{N-2}{2}} \left(p + \sigma_z^22^{p+1}(p+q)\right)^{-\frac{N_D - 1}{2}} \left(\sigma_z^2 2^{p+q} (p+q)\right)^{-\frac{N_G-1}{2}}.
\end{align}
We need now to calculate the joint distribution of $(\pD_{ij}\lG, \pG_{kl}\lG)$ conditional on $\{\lG = x_G\}$. Denote the covariance matrix for $(\pD_{ij}\lG, \pG_{kl}\lG, \lG)$ by \begin{align}
    \Sigma = \left(\begin{array}{cc}
         \Sigma_{11}&\Sigma_{12}  \\
         \Sigma_{21}&\Sigma_{22} 
    \end{array}\right)
\end{align}
where \begin{align}
    \Sigma_{11} &= 2^{p+q}\left(\begin{array}{cc}
       (p+1)(p+q-1)(1 + \delta_{ij}) + (p+q)^2\delta_{ij}  & (p+q)^2 \delta_{ij}\delta_{kl} \\
             (p+q)^2 \delta_{ij}\delta_{kl} & (p+1)(p+q-1)(1 + \delta_{kl}) + (p+q)^2\delta_{kl}  
    \end{array}\right),\\
    \Sigma_{12} &= -2^{p+q} (p+q)\left(\begin{array}{c}
         \delta_{ij}  \\
          \delta_{kl}
    \end{array}\right),\\
      \Sigma_{21} &= -2^{p+q} (p+q)\left(\begin{array}{cc}
         \delta_{ij}  &\delta_{kl}
    \end{array}\right),\\
    \Sigma_{22} &= 2^{p+q}.
\end{align}
The conditional covariance is then \begin{align}
    \bar{\Sigma} = \Sigma_{11} - \Sigma_{12} \Sigma_{22}^{-1}\Sigma_{21} = 2^{p+q}(p+1)(p+q-1)\left(\begin{array}{cc}
       1 + \delta_{ij} & 0 \\
            0 & 1 + \delta_{kl}
    \end{array}\right).\label{eq:double_goe}
\end{align}

Identical reasoning applied to $(\pG_{ij}\lG, \pG_{kl}\lG, \lG)$ and $(\pD_{ij}\lG, \pD_{kl}\lG, \lG)$ shows that, conditional on $\{\lG = x_G\}$, $\nabla_G^2\lG$ and $\nabla_D^2\lG$ have independent entries up-to symmetry, so \ref{eq:double_goe} demonstrates they are independent GOEs and we have:
\begin{align}
    \left(\begin{array}{cc}
        -\nabla_D^2\lG & -\nabla_G\nabla_D \lG  \\
         \nabla_D\nabla_G \lG & \nabla^2 \lG  
    \end{array}\right) ~|~ \{\lG = x_G\} &\overset{d}{=} 
   \sqrtsign{2^{p+q+1}(p+q)(p+q-1)} \left(\begin{array}{cc}
       \sqrtsign{N_D -1}M^{(D)}_1 & -2^{-1/2}G \\
         2^{-1/2}G^T & \sqrtsign{N_G - 1}M^{(G)}
    \end{array}\right)\notag\\
    &~~~~~~- (p+q)x_G2^{p+1} \left(\begin{array}{cc}
        -I_{N_D} & 0  \\
         0 & I_{N_G} 
    \end{array}\right)
\end{align}
where $M^{(D)}_1\sim GOE^{N_D - 1}$ and $M^{(G)} \sim GOE^{N_G - 1}$ are independent GOEs and $G$ is an independent $N_D - 1 \times N_G - 1$ Ginibre matrix with entries of unit variance.

\section{Conclusion}
We have contributed a novel model for the study of large neural network gradient descent dynamics with statistical physics techniques, namely an interacting spin-glass model for generative adversarial neural networks. We believe this is the first attempt in the literature to incorporate advanced architectural features of modern neural networks, beyond basic single network multi-layer perceptrons, into such statistical physics style models. We have conducted an asymptotic complexity analysis via Kac-Rice formulae and Random Matrix Theory calculations of the energy surface of this model, acting as a proxy for GAN training loss surfaces of large networks. Our analysis has revealed a banded critical point structure as seen previously for simpler models, explaining the surprising success of gradient descent in such complicated loss surfaces, but with added structural features that offer explanations for the greater difficulty of training GANs compared to single networks. We have used our model to study the effect of some elementary GAN hyper-parameters and compared with experiments training real GANs on a standard computer vision dataset. We believe that the interesting features of our model, and their correspondence with real GANs, are yet further compelling evidence for the role of statistical physics effects in deep learning and the value of studying such models as proxies for real deep learning models, and in particular the value of concocting more sophisticated models that reflect aspects of modern neural network design and practice.

\review{Our analysis has focused on the annealed complexity of our spin glass model (i.e. taking the logarithm after the expectation) rather than the quenched complexity (i.e. taking the expectation after the logarithm). Ideally one would compute both, as the quenched complexity is often considered to reflect the typical number of stationary points and is bounded above by the annealed complexity. Computing the quenched complexity is typically more challenging than the annealed and such a calculation for our model could be the subject of a further work requiring considerable technical innovations. Even the elegant and very general methods presented recently in \cite{arous2021exponential} are restricted only to the annealed case. Agreement between annealed and quenched is known only in a few special cases closely related to spherical spin glasses \cite{Sub2017,AufGol2020,BenSubZei2020} and is not expected in general \cite{ros2019complex}. It is conceivable that quenched and annealed complexity agree in the case of our model, as it closely related to spin glasses and possesses no distinguished directions (i.e. spikes) such as are present in \cite{ros2019complex}. Establishing agreement by existing methods requires analysis of pairs of correlated GOE-like matrices. Such an approach for our model may well require analysis of at least 4 correlated matrices (2 per diagonal block), and quite possibly more, including correlations between blocks. We leave this considerable challenge for future work.}

\medskip
From a mathematical perspective, we have extensively studied the limiting spectral density of a novel random matrix ensemble using supersymmetric methods. During the initial explorations for this work, we made considerable efforts to complete the average absolute value determinant calculations directly using a supersymmetric representation, as seen in Chapter \ref{chap:general_activation_functions}, however this was found to be analytically intractable (as expected), but also extremely troublesome numerically (essentially due to analytically intractable and highly complicated Riemann sheet structure in $\mathbb{C}^2$). We were able to sidestep these issues by instead using a Coulomb gas approximation, whose validity we have rigorously proved using a novel combination of concentration arguments and supersymmetric asymptotic expansions. We have verified with numerical simulations our derived mean spectral density for the relevant Random Matrix Theory ensemble and also the accuracy of the Coulomb gas approximation.

\medskip
We hope that future work will be inspired to further study models of neural networks such as we have considered here. Practically, it would be exciting to explore the possibility of using our insights into GAN loss surfaces to devise algorithmic methods of avoiding training failure. Mathematically, the local spectral statistics of our random matrix ensemble may be interesting to study, particularly around the cusp where the two disjoint components of the limiting spectral density merge.

\chapter{Generalised loss surface models and implications}\label{chap:gadam}
The content of this chapter was published first as a pre-print in July 2021 (\url{https://arxiv.org/abs/2003.01247v5}) and was accepted in January 2023 as an article in \emph{Journal of Machine Learning Research}: ``Iterate Averaging in the quest for best test error'', Diego Granziol, \textbf{Nicholas P. Baskerville}, Xingchen Wan, Samuel Albanie and Stephen Roberts. 
\medskip

The experimental ideas behind this paper were conceived and explored by the other authors before \textbf{NPB} joined the project. \textbf{NPB} developed much of the mathematical theory, including constructing all the proofs. In this chapter, we include only the mathematical sections of direct relevance to this thesis, all of which are overwhelmingly \textbf{NPB}'s work.

\section{Introduction}\label{subsec:theory}
The iterate average \cite{polyak1992acceleration} is the arithmetic mean of the model parameters over the optimisation trajectory $\vw_{\mathrm{avg}} = \frac{1}{n}\sum_{i}^{n}\vw_{i}$.
It is a classical variance reducing technique in optimisation and offers optimal asymptotic convergence rates and greater robustness to the choice of learning rate \cite{kushner2003stochastic}.  Indeed, popular regret bounds that form the basis of gradient-based convergence proofs \cite{duchi2011adaptive,reddi2019convergence} often consider convergence for the iterate average~\cite{duchi2018introductory}.  Further, theoretical extensions have shown that the rate of convergence can be improved by a factor of $\log T$ (where $T$ is the iteration number) by \emph{suffix averaging} \cite{rakhlin2011making}, which considers a fraction of the last iterates, \emph{polynomial decay averaging} \cite{shamir2013stochastic} which decays the influence of the previous iterates, or \emph{weighted averaging} \cite{lacoste2012simpler} which weights the iterate by its iteration number. That the final iterate of SGD is sub-optimal in terms of its convergence rate, by this logarithmic factor, has been proved by \cite{harvey2019tight}.
For networks with batch normalisation \cite{ioffe2015batch}, a na\"{i}ve application of IA (in which we simply average the batch normalisation statistics) is known to lead to poor results \cite{defazio2019ineffectiveness}.  However, by computing the batch normalisation statistics for the iterate average using a forward pass of the data at the IA point, \cite{izmailov2018averaging} show that the performance of small-scale image experiments such as CIFAR-10/100 and pretrained ImageNet can be significantly improved. Even for small experiments this computation is expensive, so they further approximate IA by taking the average at the end of each epoch instead of each iteration, referred to as \emph{stochastic weight averaging} (SWA). 

\medskip
In this chapter we examine the variance reducing effect of IA in the context of a quadratic approximation to the true loss combined with additive perturbation models for the batch training loss. The theory we present is high-dimensional (i.e. large number of parameters, $P$) and considers the small batch size (small $B$) regime, which we term the ``deep learning limit''.
Intuitively, any given example from the training set $j \in \mathcal{D}$, will contain \textit{general features}, which hold over the data generating distribution and \textit{instance specific features} (which are relevant only to the training sample in question). For example, for a training image of a dog, we may have that: 
\begin{equation}
	\label{eq:instancespecific}
\overbrace{\underbrace{\nabla L_{\text{sample}}(\vw)}_\textrm{training set example}}^\text{dog $j$}  =\overbrace{ \underbrace{\nabla L_{\text{true}}(\vw)}_\text{general features}}^\text{$4$ legs, snout} + \overbrace{\underbrace{\vepsilon(\vw).}_\text{instance-specific features}}^\text{black pixel in top corner, green grass}
\end{equation}
Under a quadratic approximation to the \emph{true loss}\footnote{The loss under the expectation of the data generating distribution, rather than the loss over the dataset $L_{\text{emp}}(\vw_{k})$.} $L_{\text{true}}(\vw)=\vw^T\mH\vw$, where $\mH = \nabla^{2} L$ is the Hessian of the true loss with respect to the weights and we sample a mini-batch gradient of size $B$ at point $\vw \in \mathbb{R}^{P\times 1}$. The observed gradient is perturbed by $\vepsilon(\vw)$ from the true loss gradient (due to instance specific features).  Under this model the component of the $\vw_{t}$'th iterate along the $j$'th eigenvector $\vphi_{j}$ of the true loss when running SGD with learning rate $\alpha$ can be written:
\begin{equation}\label{eq:weight_updates1}
	\vw_{t}^{T}\vphi_{j} = (1-\alpha\lambda_{j})^{t}\vw_{0}^{T}\vphi_{j} - \alpha(1-\alpha\lambda_{j})^{t-1}\vepsilon(\vw_{1})^{T}\vphi_{j} \cdots ,
\end{equation}
in which $\lambda_j$ are the eigenvalues of $\mH$. The simplest tractable model for the gradient noise $\vepsilon(\vw_t)$ is to assume samples from i.i.d. an isotropic, multivariate Normal. In particular, this assumption removes any dependence on $\vw_t$ and precludes the existence of any distinguished directions in the gradient noise. Using this assumption, we obtain Theorem \ref{theorem:shell} below, which relies on an intermediate result, found in \cite{vershynin2018high}.

\begin{lemma}[\cite{vershynin2018high} Theorem 6.3.2]\label{lem:aniso_bernstein}
    Let $R$ be an $m \times n$ matrix, and let $X=(X_1, \ldots, X_n)\in\mathbb{R}^n$ be a random vector with independent mean-zero unit-variance sub-Gaussian coordinates. Then
    \begin{equation*}
       \mathbb{P}\left( \left|\|RX\|_2 - \|R\|_F\right| > t\right) \leq 2\exp\left(-\frac{ct^2}{K^4 \|R\|^2}\right)
    \end{equation*}
    where $K=\max_i\|X_i\|_{\psi_2}$ and $c>0$ is a constant.
\end{lemma}

\begin{theorem}
	\label{theorem:shell}
	\jmlrreview{
	Assume the quadratic loss model  $L_{\text{true}}(\vw)=\vw^T\mH\vw$, where $\mH$ has eigenvalues $\{\lambda_i\}_{i=1}^P$ and assume the $\{\epsilon_t\}_{t=0}^{n}$ are all i.i.d. Gaussian vectors in $\mathbb{R}^P$ with distribution $\mathcal{N}(0, \sigma^2 B^{-1}I)$ where $B$ is the batch size. 
	Assume the weights are updated according to the rule from (\ref{eq:weight_updates1}) 
	\begin{align}\label{eq:weight_update2}
	    \vw_{t}^{T}\vphi_{j} = (1-\alpha\lambda_{j})^{t}\vw_{0}^{T}\vphi_{j} - \alpha(1-\alpha\lambda_{j})^{t-1}\vepsilon(\vw_{1})^{T}\vphi_{j}.
	\end{align}
	}
	Assume further that $\alpha\lambda_i \ll 1$ for all $i$ and $\lambda_i >0$ for all $i$. Then there exists a constant $c>0$ such that for all $\xi>0$, as $n\rightarrow \infty$
	\begin{equation}\label{eq:thm2_statement}
	    \hspace{-0.4cm}
		\begin{aligned}
			& \mathbb{P}\left(\left|\sqrtsign{\sum_{i}^{P}\left(w_{n,i} - w_{0,i}\mathrm{e}^{-n\alpha\lambda_{i}} ( 1 + o(1))\right)^2}-\sqrt{P\frac{\alpha\sigma^{2}}{B}\bigg\langle\frac{1}{ \lambda(2-\alpha\lambda) }\bigg\rangle} \right|\geq \xi \right) \leq \nu(\xi), \\
			& \mathbb{P}\left(\left|\sqrtsign{\sum_{i}^{P}\left(w_{\mathrm{avg},i} - \frac{w_{0,i}}{\lambda_{i}n\alpha} ( 1 + o(1))\right)^2}-\sqrt{\frac{P\sigma^{2}}{Bn}\bigg\langle\frac{1}{ \lambda }\bigg\rangle}\right| \geq \xi \right) \leq \nu(\xi),
		\end{aligned}
	\normalsize
	\end{equation}
	\normalsize
	where $\nu(\xi) = 2\exp(-c\xi^{2})$.
	\end{theorem}
\begin{proof}
Let $Y = (Y_1, \ldots, Y_P)$ be a random sub-Gaussian vector with independent components. Let \begin{equation*}
    X_i = \frac{Y_i - \mathbb{E}Y_i}{\sqrtsign{\Var Y_i}}, ~~R = \mathrm{diag}(\sqrtsign{\Var Y_1}, \ldots, \sqrtsign{\Var Y_P}).
\end{equation*}
 Lemma \ref{lem:aniso_bernstein} then applies, to give \begin{align*}
   \mathbb{P}\left( \left|\|Y - \mathbb{E}Y\|_2 - \sqrtsign{\sum_{i=1}^P \Var Y_i}\right| > \jmlrreview{\xi}\right) \leq 2\exp\left(-\frac{c\jmlrreview{\xi}^2}{K^4 \|R\|^2}\right).
\end{align*}
We have $K\leq C \max_i \Var Y_i$ for some constant $C>0$ (\cite{vershynin2018high}, exercise 2.5.8), and $\|R\|^2 = (\max_i \sqrtsign{\Var Y_i})^2 = \max_i \Var Y_i$. Hence we obtain \begin{align}\label{eq:aniso_bern_applied}
   \mathbb{P}\left( \left|\|Y - \mathbb{E}Y\|_2 - \sqrtsign{\sum_{i=1}^P \Var Y_i}\right| > \jmlrreview{\xi}\right) \leq 2\exp\left(-\frac{c\jmlrreview{\xi}^2}{(\max_i \Var Y_i)^2}\right)
\end{align}
for some new constant $c>0$. The proof is then completed if we compute the means and variances of $\vw_n$ and $\vw_{\mathrm{avg}}$. To that end, with $\mLambda=\text{diag}\left(\lambda_1, \ldots, \lambda_P\right)$, \jmlrreview{the update rule (\ref{eq:weight_update2}) gives}
 \begin{align}\label{eq:wn_expression_thm1}
    \vw_t = (1-\alpha\mLambda)^n \vw_0 + \alpha\sum_{i=0}^{t-1} (1-\alpha\mLambda)^{t-i-1}\vepsilon_i,
\end{align}
\jmlrreview{for any $1 \leq t\geq n$}.
\jmlrreview{Since $\mLambda$ is diagonal, each component of $\vec{w}_n$ can be treated independently when we sum to obtain $\vec{w}_{avg}$, so for any vector $\vec{v}$ \begin{align}
    \sum_{t=1}^n (1 - \alpha\mLambda)^t\vec{v} = \frac{1 - (1 - \alpha \mLambda)^{t}}{\alpha}\mLambda^{-1} (1-\alpha\mLambda)\vec{v}
\end{align}
So averaging (\ref{eq:wn_expression_thm1}) over $t$ gives}
\begin{align}
    \vw_{avg} &= \frac{1 - (1 - \alpha\mLambda)^n}{\alpha n} \mLambda^{-1}(1-\alpha\mLambda)\vw_0 + \sum_{\jmlrreview{t}=0}^{n-1} \frac{1 - (1-\alpha\mLambda)^{n-t}}{n}\mLambda^{-1}\vepsilon_t \label{eq:wavg_expression_thm1}.
    \end{align}
    \jmlrreview{Since the $\vepsilon_i$ are all i.i.d. centred Gaussians, obtaining the distributions of $\vw_n$ and $\vw_{avg}$ amounts to computing the covariances} \begin{align}
        \jmlrreview{\Cov \left(\alpha \sum_{i=0}^{n-1} (1-\alpha \mLambda)^{n-i-1}\vepsilon_i \right)}
        = &\sigma^2 B^{-1} I \sum_{i=1}^{n-1} \alpha^2(1-\alpha\mLambda)^{2(n-i-1)} \notag \\= &\sigma^2 B^{-1} I \alpha^2(1 - (1 - \alpha\mLambda)^{2n})\left(1 - (1-\alpha\mLambda)^2\right)^{-1}\label{eq:thm2_cov1}
    \end{align}
and similarly \begin{align}
&\jmlrreview{\Cov \left(\sum_{\jmlrreview{t}=0}^{n-1} \frac{1 - (1-\alpha\mLambda)^{n-t}}{n}\mLambda^{-1}\vepsilon_t \right)} \notag \\= 
    &\sum_{t=0}^{n-1} \left(\frac{1 - (1-\alpha\mLambda)^{n-t}}{n}\mLambda^{-1}\right)^2\notag\\ =&\frac{\mLambda^{-2}}{ n^2}\left(n - \frac{2(1 - (1-\alpha\mLambda)^n)}{\alpha}\mLambda^{-1} + \left(1 - (1-\alpha\mLambda)^{2n}\right)\left(1 - (1-\alpha\mLambda)^2\right)^{-1}\right).\label{eq:thm2_cov2}
\end{align}
Now \jmlrreview{using} $\alpha\lambda_i < 1$ for all $i=1,2\ldots, P$, and taking $n\rightarrow\infty$, \jmlrreview{(\ref{eq:wn_expression_thm1}) and (\ref{eq:thm2_cov1}) give
\begin{align}\label{eq:thm2_wn_cov_final}
     \Cov (\vw_n) \sim \sigma^2\alpha^2B^{-1}\left(1 - (1-\alpha\mLambda)^2\right)^{-1} = \sigma^2\alpha B^{-1} \left(2\mLambda - \alpha\mLambda^2\right)^{-1} 
\end{align}}
and similarly \jmlrreview{(\ref{eq:wavg_expression_thm1}) and (\ref{eq:thm2_cov2}) give \begin{align}\label{eq:thm2_wavg_cov_final}
  \Cov(\vw_{avg}) \sim  \frac{1}{ n}\mLambda^{-2}.
\end{align}}
Thus it follows from (\ref{eq:wn_expression_thm1}) and (\ref{eq:thm2_wn_cov_final}) that\begin{align}\label{eq:thm2_wn_sstats}
    \mathbb{E}w_{n, i} = (1-\alpha\lambda_i)^n w_{0, i} \sim e^{-n\alpha\lambda_i}w_{0, i},& ~~ \Var(w_{n, i}) \sim \frac{\sigma^2}{B}\frac{\alpha}{2\lambda_i(1 - \alpha\lambda_i)}
\end{align}
and \jmlrreview{from (\ref{eq:wavg_expression_thm1}) and (\ref{eq:thm2_wavg_cov_final}) it follows}\begin{align}\label{eq:thm2_wavg_sstats}
    \mathbb{E}w_{avg, i} \sim \frac{w_{0,i}}{\lambda_i\alpha n}, ~~~ \Var(w_{avg, i}) = \frac{\sigma^2}{B}\frac{1}{n\lambda_i^2}
\end{align}
where in both cases we have used $\alpha\lambda_i \ll 1 $ to simplify the expected values for large $n$. 
To complete the proof for $\vw_n$, we apply (\ref{eq:aniso_bern_applied}) using (\ref{eq:thm2_wn_sstats}) and \jmlrreview{noting that \begin{align}
    \sqrtsign{\sum_{i=1}^P Var(w_{n,i})} \sim \frac{\sigma^2 P\alpha}{2B} \left\langle\frac{1}{\lambda(1 - \alpha\lambda)}\right\rangle
\end{align}}
\jmlrreview{and $0 < \max_i w_{n,i} < \infty$ since $\lambda_i >0$ and $\alpha\lambda_i < 1$.
The results for $\vw_{avg}$ follows similarly by using (\ref{eq:aniso_bern_applied}) with (\ref{eq:thm2_wavg_sstats}).
This produces two different constants $c>0$ in the statement of (\ref{eq:thm2_statement}), but we can simply take the smaller of the two constants to produce the desired statement.}

\end{proof}
The final iterate attains exponential convergence in the mean of $\vw_{n}$, but does not control the variance term. Whereas for $\vw_{\mathrm{avg}}$, although the convergence in the mean is worse (linear), the variance vanishes asymptotically -- this motivates \textit{tail averaging}, to get the best of both worlds. Another key implication of Theorem \ref{theorem:shell} lies in its dependence on $P$. \jmlrreview{$P$ is a gauge of the model size and appears as a simple linear multiplier of the variances of $\vw_n$ and $\vw_{avg}$, so increasing over-parametrisation implies increasing variance of the final iterate and the IA, however IA provides a counterbalancing variance reduction effect that is entirely absent from the final iterate. This implies that in more complex, over-parameterised models, we expect the benefit of IA over the final iterate to be greater, as IA provides a mechanism to control the weight variance even as it grows with $P$.}

\section{A dependent model for the perturbation} We proceed now to propose a relaxation of the gradient perturbation independence assumption. (\ref{eq:instancespecific}) can be written equivalently as \begin{align}\label{eq:integrated_gradient}
    L_{\text{batch}}(\vw) = L_{\text{true}}(\vw) + \eta(\vw)
\end{align}
where $\eta$ is a scalar field with $\nabla\eta = \vepsilon$. Note that we have neglected an irrelevant arbitrary constant in Equation (\ref{eq:integrated_gradient}) and also that we have $L_{\text{batch}}$ rather than $L_{\text{sample}}$, but this amounts to scaling the per-sample noise variance $\sigma^2$ by the inverse batch size $B^{-1}$. We model $\eta$ as a Gaussian process $\mathcal{GP}(m, k)$, where $k$ is some kernel function $\mathbb{R}^P\times \mathbb{R}^P\rightarrow\mathbb{R}$ and $m$ is some mean function\footnote{It is natural to take $m=0$ in a model for the sample perturbation, however retaining fully general $m$ does not affect our arguments.} $\mathbb{R}^P\rightarrow\mathbb{R}$. As an example, taking $k(\vw, \vw') \propto (\vw^T \vw')^p$ and restricting $\vw$ to a hypersphere results in $\vepsilon$ taking the exact form of a spherical $p$-spin glass, studied previously for DNNs \cite{choromanska2015loss,gardner1988optimal,mezard1987spin,ros2019complex,mannelli2019passed} and in Chapters \ref{chap:general_activation_functions} and \ref{chap:spin_glass_gans} \cite{baskerville2021loss, baskerville2022spin}. \emph{We are not} proposing to model the loss surface (batch or true) as a spin glass (or more generally, a Gaussian process), rather we are modelling the perturbation between the loss surfaces in this way.
We emphasise that this model is a strict generalisation of the i.i.d. assumption above, and presents a rich, but tractable, model of isotropic Gaussian gradient perturbations in which the noise for different iterates is neither independent nor identically distributed.  

Following from our Gaussian process definition, the covariance of gradient perturbations can be computed using a well-known result (see \cite{adler2009random} equation 5.5.4):
 \begin{align} \label{eq:basic_gp_covar}
    \Cov(\epsilon_i(\vw), \epsilon_j(\vw') ) = \partial_{w_i}\partial_{w'_j} k(\vw, \vw').
\end{align}
Further assuming a stationary kernel $k(\vw, \vw') = k\left(-\frac{1}{2}||\vw - \vw'||_2^2\right)$ 
\begin{align}\label{eq:grad_gp_covar}
    & \Cov(\epsilon_i(\vw), \epsilon_j(\vw') ) = (w_i - w'_i)(w'_j - w_j) k''\left(-\frac{1}{2}||\vw - \vw'||_2^2\right)  + \delta_{ij}k'\left(-\frac{1}{2}||\vw - \vw'||_2^2\right).
\end{align}
Thus we have a non-trivial covariance between gradient perturbation at different points in weight-space. This covariance structure can be used to prove the upcoming variance reduction result, but first we require some intermediate lemmas.

\subsection{Intermediate results}
\label{sec:derivations}
In this section we establish some intermediate lemmas that will be required later in the chapter.

\begin{lemma}\label{lemma:incomp_gamma}
Define the function \begin{align}
    r(a; x) = \frac{\gamma(a; x)}{\Gamma(a)},
\end{align}
where $\gamma$ is the lower incomplete gamma function. Assume that \jmlrreview{$x\ll a$}, where $x$ may or may not diverge with $a$, then as $a\rightarrow\infty$, $r(a; x)\rightarrow 0$, and more precisely \begin{align}
    r(a; x) \sim \frac{1}{\sqrtsign{2\pi}} \exp\left(-x + a\log{x} - a - a\log{a} - \frac{1}{2}\log{a}\right).
\end{align}
\end{lemma}
\begin{proof}
    We have $\gamma(a; x) = a^{-1}x^a \,_1F_1(a; 1+a; -x)$, where $\,_1F_1$ is the confluent hypergeometric function of the first kind \cite{andrews_askey_roy_1999}. Then \begin{align}
    r(a; x) &= \frac{a^{-1} x^a \,_1F_1(a; 1+a; -x)}{\Gamma(a)}=\frac{a^{-1} x^a \Gamma(a+1)}{\Gamma(a)^2}\int_0^1 e^{xt} t^{a-1} dt\label{eq:r_int_form}
\end{align}
where we have used a result of \cite{abramowitz1988handbook}. The integral in (\ref{eq:r_int_form}) can be evaluated asymptotically in the limit $x\rightarrow\infty$ with $x \ll a$. Writing the integrand as $e^{xt + (a-1)\log{t}}$ it is plainly seen to have no saddle points in $[0, 1]$ given the condition $x\ll a$. The leading order term therefore originates at the right edge $t=1$. A simple application of Laplace's method leads to \begin{align*}
    r(a; x) & \sim  \frac{a^{-1} x^a \Gamma(a+1) e^{-x}}{\Gamma(a)^2(a- 1 -x)}\\
    & \sim \frac{ x^a  e^{-x}}{a\Gamma(a)}\\
    & \sim \frac{ x^a  e^{-x}}{a \sqrtsign{2\pi a^{-1}} (ae^{-1})^a}\\
    & = \frac{1}{\sqrtsign{2\pi}} \exp\left(-x + a\log{x} - a - a\log{a} - \frac{1}{2}\log{a}\right)
\end{align*}
where the penultimate line makes uses of Stirling's approximation \cite{andrews_askey_roy_1999}. Since $a\gg x$, \begin{align*}
    -x + a\log{x} - a - a\log{a} - \frac{1}{2}\log{a} \sim -a\log{a}\rightarrow-\infty 
\end{align*}
which completes the proof.
\end{proof}

\begin{lemma}\label{lemma:balls_lemma}
Take any $\vx_0,\ldots, \vx_{n-1}\in\mathbb{R}^{P}$ let $\mX\sim\mathcal{N}(\jmlrreview{\bm{\mu}}, \Sigma)$, for any $\bm{\mu}\in\mathbb{R}^P$ and \jmlrreview{ $\Sigma$ such that $\det \Sigma \geq A\sigma^{2P}$ for some constants $A, \sigma>0$.} Consider $P\rightarrow\infty$ with $P\gg \log{n}$ and let $\delta >0$ be $o(P^{\frac{1}{2}})$ (note that $\delta$ and $n$ need not diverge with $P$, but they can). Define \begin{align*}B_i = \{\vx\in\mathbb{R}^P \mid ||\vx-\vx_i|| < \delta \},\end{align*}
then as $P \rightarrow \infty$ \begin{align}\mathbb{P}\left(\mX \in \bigcup_i B_i\right) \rightarrow 0\end{align}
and moreover as $P, n\rightarrow\infty$ \begin{align}\label{eq:balls_lemma_precise}
    n^l\mathbb{P}\left(\mX \in \bigcup_i B_i\right) \rightarrow 0,
\end{align}
\jmlrreview{for any fixed $l>0$.}
\end{lemma}
\begin{proof}
With the Euclidean volume measure, we have \begin{align*}
    Vol\left(\bigcup_i B_i\right) \leq n V_P \delta^P = V_P (\delta n^{1/P})^{P}
\end{align*}
where $V_P$ is the volume of the unit sphere in $P$ dimensions. Therefore \jmlrreview{a sphere of radius $\delta n^{1/P}$ is large enough to enclose all of the $B_i$ and so the probability that $\mX$ lies in any of the $B_i$ is bounded above by the probability that it lies inside the sphere of radius $\delta n^{1/P}$ centred on its mean $\bm{\mu}$.
Note that with $\hat{\sigma}^2 = (\det\Sigma)^{1/P}$, changing variables $\vx = \hat{\sigma}^{-1}\Sigma^{1/2}\vy $ gives \begin{align*}
    \int_{\mathbb{R}^P} d\vx e^{-\frac{\vx^T \Sigma^{-1} \vx}{2}} =    \int_{\mathbb{R}^P}d\vy e^{-\frac{\vy^2}{2 \hat{\sigma}^{2}}}
\end{align*}
since the Jacobian is $1$. Thus we can reduce to a single dimensional Gaussian integral 
} \begin{align}
\mathbb{P}\left(\mX \in \bigcup_i B_i\right) & \leq \frac{1}{(2\pi \jmlrreview{\hat{\sigma}}^2)^{\frac{P}{2}}}\frac{2\pi^{\frac{P}{2}}}{\Gamma(\frac{P}{2})}\int_0^{\delta n^{\frac{1}{P}}}dr ~ e^{-\frac{r^2}{2\jmlrreview{\hat{\sigma}}^2}} r^{P-1}\notag\\
&= \frac{2}{\Gamma(\frac{P}{2})}\int_0^{\frac{\delta n^{\frac{1}{P}}}{\sqrt{2}\jmlrreview{\hat{\sigma}}}} dr ~ e^{-r^2} r^{P-1}\notag\\
& = \frac{1}{\Gamma(\frac{P}{2})} \int_0^{\frac{\delta n^{\frac{2}{P}}}{2\jmlrreview{\hat{\sigma}}^2}} dr ~ e^{-r} r^{\frac{P}{2} - 1}\notag\\
& \leq \frac{1}{\Gamma(\frac{P}{2})} \int_0^{\frac{\delta n^{\frac{2}{P}}}{2\jmlrreview{A^{1/P}}\sigma^2}} dr ~ e^{-r} r^{\frac{P}{2} - 1}\tag{\jmlrreview{using $\hat{\sigma}^2 \geq A^{1/P} \sigma^2$}}\\
& \jmlrreview{\leq \frac{1}{\Gamma(\frac{P}{2})} \int_0^{\frac{\delta n^{\frac{2}{P}}}{2\alpha\sigma^2}} dr ~ e^{-r} r^{\frac{P}{2} - 1}\tag{\jmlrreview{with $\alpha \equiv \inf_P A^{1/P} >0$}}}\\
& \equiv \frac{1}{\Gamma(\frac{P}{2})} \gamma\left(\frac{P}{2}; \frac{n^{\frac{2}{P}}\delta^2}{2\sigma^2\jmlrreview{\alpha}}\right)\label{eq:ball_prob_gamma}
\end{align}
where $\gamma$ is the lower incomplete gamma function. Since $P\gg \log{n}$ and $\delta = o(P^{\frac{1}{2}})$, it follows that \begin{align*}
  x \equiv  \frac{n^{\frac{2}{P}}\delta^2}{2\sigma^2\jmlrreview{\alpha}} = o(P)
\end{align*}
and so Lemma \ref{lemma:incomp_gamma} can be applied to yield the result. 
\jmlrreview{Indeed, recalling that $n\ll e^{P}$, we have
\begin{align*}
n^l \mathbb{P}\left(\mX \in \bigcup_i B_i\right) &\leq e^{lP} r(P/2, x) \sim \frac{1}{\sqrtsign{2\pi}} \exp\left(lP - x + \frac{P}{2}\log{x} - \frac{P}{2} - \frac{P}{2}\log\frac{P}{2} - \frac{1}{2}\log\frac{P}{2}\right)
\end{align*}
for any $l>0$. 
But $x = o(P)$ so for $P$ large enough, the term inside the exponential is negative and diverging with $P$, as required.}
\end{proof}
\medskip
The previous two lemmas are required to prove the next lemma, which will form the foundation of our argument in the next section.

\begin{lemma}\label{lemma:prob_far_apart}
Let $\mX_1,\ldots, \mX_n$ be a sequence of \jmlrreview{jointly multivariate} Gaussian random variables in $\mathbb{R}^P$ such that \jmlrreview{\begin{align*}
    \mX_i ~\mid ~ \{\mX_1,\ldots, \mX_{i-1}\} \sim \mathcal{N}(\bm{\mu}_i, \Sigma_i)
\end{align*} where there exists  a $\sigma>0$ and a constant $A>0$ such that $\det \Sigma_i \geq A\sigma^P$ for all $P$ and $i$.}
Let also $\mX_0$ be any deterministic element of $\mathbb{R}^P.$ \jmlrreview{For $1\leq m \leq n$}, define the events \begin{align*}A_m(\delta) = \{||\mX_i - \mX_j||_2 > \delta \mid 0\leq i < j \leq m\}.\end{align*} 
Consider $P\rightarrow\infty$ with $P\gg \log{n}$ and let $\delta >0$ be $o(P^{\frac{1}{2}})$ (note that $\delta$ and $n$ need not diverge with $P$, but they can). Then $\mathbb{P}(A_n(\delta))\rightarrow 1$ as $P\rightarrow\infty$.
\end{lemma}
\begin{proof}
Let us use the definitions of $B_i$ from Lemma \ref{lemma:balls_lemma}, \jmlrreview{i.e. let \begin{align*}
    B_i = \{\vx\in\mathbb{R}^P \mid ||\vx-\mX_i|| < \delta \}
\end{align*}
for $0 < j < n$.}
Since $A_i(\delta) \subset A_{i-1}(\delta)$ \jmlrreview{for any $i$}, the chain rule of probability gives \begin{align}
    \mathbb{P}(A_n(\delta))& = \mathbb{P}\left(\bigcap_{i\leq n}A_i(\delta)\right) = \mathbb{P}(A_1(\delta))\prod_{i=2}^{n-1} \mathbb{P}(A_i \mid A_{i-1})\notag
\end{align}
but \begin{align*}\mathbb{P}(A_i(\delta) \mid A_{i-1}(\delta))= 1 - \mathbb{P}\left(\mX_i\in\bigcup_{j< i} B_j\right)\end{align*} and so \jmlrreview{
\begin{align}\label{eq:lem3_prob_proof}
 \mathbb{P}(A_n(\delta))& = \mathbb{P}(A_1(\delta))\prod_{i=2}^{n-1} \left(1 -  \mathbb{P}\left(\mX_i\in\bigcup_{j< i} B_j\right)\right)\\ &= \mathbb{P}\left(\mX_1 \in B_0\right) \prod_{i=2}^{n-1} \left(1 -  \mathbb{P}\left(\mX_i\in\bigcup_{j< i} B_j\right)\right).
\end{align}
}
For fixed $n$, the result is now immediate from (\ref{eq:balls_lemma_precise}) in Lemma \ref{lemma:balls_lemma}, since all the probabilities in (\ref{eq:lem3_prob_proof}) converge to $1$ as $P\rightarrow\infty$ and there are only a finite number of terms.

Now consider the case that $n$ also diverges. For any $n$ define 
\begin{align*}
    s_n = \sup_{2\leq i \leq n} \mathbb{P}\left( \mX _i \in \bigcup_{j < i} B_j\right),
\end{align*}
and then \begin{align*}
     \mathbb{P}(A_n(\delta)) \geq \mathbb{P}\left(\mX_1 \in B_0\right) \prod_{i=2}^{n-1} \left(1 -  s_{i-2}\right).
\end{align*}
But, by Lemma \ref{lemma:balls_lemma} we can write $s_n = (n+1)^{-2} f_{n, P}$ where $f_{n, P}\rightarrow 0$ as $P\rightarrow\infty$, say, \vivacom{hence
\begin{align*}
         \mathbb{P}(A_n(\delta)) \geq \mathbb{P}\left(\mX_1 \in B_0\right) \prod_{i=2}^{n-1} \left(1 -  (i-1)^{-2} f_{i-2, P}\right) \geq \mathbb{P}\left(\mX_1 \in B_0\right) \prod_{i=2}^{\infty} \left(1 -  (i-1)^{-2}f_{i-2, P}\right)
\end{align*}
for large $n$, since $|f_{n-2,P}|<1$ and all the extra terms added are strictly between 0 and 1. But 
\begin{align*}
         \log\prod_{i=2}^{\infty} \left(1 -  (i-1)^{-2}f_{i-2, P}\right) \geq -\sum_{i=2}^{\infty} (i-1)^{-2}f_{i-2, P} \geq -\sup_j f_{j-2, P} \sum_{i=2}^{\infty} (i-1)^{-2} = -\frac{\pi^2}{6}\sup_j f_{j-2, P} 
\end{align*}
and so 
\begin{align*}
   \mathbb{P}(A_n(\delta)) \geq e^{-\sup_j f_{j-2, P} \pi^2/6} \mathbb{P}\left(\mX_1 \in B_0\right)
\end{align*}
but $ f_{j-2, P} \rightarrow 0$ for any $j$, so as $P\rightarrow\infty$, $ \mathbb{P}(A_n(\delta))$ is lower bounded by a term converging to $ \mathbb{P}\left(\mX_1 \in B_0\right)$ which, in turn, converges to $1$ by Lemma \ref{lemma:balls_lemma}.}

\end{proof}

\jmlrreview{Recall the Gaussian process covariance structure from above (\ref{eq:grad_gp_covar}):
\begin{align}
    & \Cov(\epsilon_i(\vw), \epsilon_j(\vw') ) = (w_i - w'_i)(w'_j - w_j) k''\left(-\frac{1}{2}||\vw - \vw'||_2^2\right)  + \delta_{ij}k'\left(-\frac{1}{2}||\vw - \vw'||_2^2\right)\tag{\ref{eq:grad_gp_covar}}
\end{align}
}
\begin{lemma}\label{lemma:gp_covar_trace}
Assume the covariance structure (\ref{eq:grad_gp_covar}).
Take any $a_i\in\mathbb{R}$ and define $\bar{\vepsilon} = \sum_{i=1}^n a_i \vepsilon_i$. Then 
\begin{align}\label{eq:gp_trace_cov}
    \Tr~\Cov(\bar{\vepsilon}) &= k'(0)P\sum_{i=1}^na_i^2 + 2P\sum_{1\leq i<j\leq n}a_ia_j\Bigg[k'(-\frac{d_{ij}^2}{2})+ P^{-1}k''(-\frac{d_{ij}^2}{2})d_{ij}^2\Bigg]
\end{align}
where we define $d_{ij} = ||\vw_i - \vw_j||_2$.
\end{lemma}\begin{proof}
Each of the $\vepsilon_i$ is Gaussian distributed with covariance matrix $\Cov(\vepsilon_i)$ given by (\ref{eq:grad_gp_covar}) and the covariance between different gradients $\Cov(\vepsilon_i, \vepsilon_j)$ is similarly given by (\ref{eq:grad_gp_covar}). By standard multivariate Gaussian properties
\begin{align}
   \Cov(\bar{\vepsilon}) &= \sum_{i=1}^na_i^2~\Cov(\vepsilon_i)+ \sum_{i\neq j}a_ia_j \Cov(\vepsilon_i, \vepsilon_j),
\end{align}
then taking the trace
\begin{align}
    \Tr~\Cov(\bar{\vepsilon}) &= \sum_{i=1}^na_i^2\Tr(\Cov(\vepsilon_i))+ 2\sum_{1\leq i<j\leq n}a_ia_j \Tr(\Cov(\vepsilon_i, \vepsilon_j)).
    \end{align}
    Using the covariance structure from (\ref{eq:grad_gp_covar}) gives \begin{align}
\Tr~\Cov(\bar{\vepsilon}) = k'(0)\sum_{i=1}^na_i^2 \Tr{I} + 2\sum_{1\leq i<j\leq n}a_ia_j\Bigg[&k'(-\frac{d_{ij}^2}{2})\Tr{I}\notag\\
&+ k''(-\frac{d_{ij}^2}{2})\Tr(\vw_i - \vw_j)(\vw_j - \vw_i)^T\Bigg]
    \end{align}
from which the result follows.
\end{proof}

\subsection{Main results for dependent noise models}

\begin{theorem}\label{theorem:dependent_noise_shell}
Let $\vw_n$ and $\vw_{avg}$ be defined as in Theorem \ref{theorem:shell} and let the gradient perturbation be given by the covariance structure in (\ref{eq:basic_gp_covar}). Assume that the kernel function $k$ is such that \jmlrreview{$k(-x)$ and its derivatives decay at least as fast as $|x|^M e^{-x}$, for some $M>0$,} as $x\rightarrow\infty$ and define $\sigma^2B^{-1} = k'(0)$. Assume further that $P^{1-\theta}\gg \log n$ \jmlrreview{for some $\theta\in(0,1)$}. Let $\delta=o(P^{1/2})$. Then $\vw_n$ and $\vw_{avg}$ are multivariate Gaussian random variables and, with probability which approaches unity as $P, n\rightarrow\infty$ the iterates $\vw_t$ are all mutually at least $\delta$ apart and 

\small
\begin{alignat}{2}
    &\mathbb{E}w_{n,i} \sim e^{-\alpha\lambda_i n}w_{0,i} ,
    &&\frac{1}{P}\Tr \Cov(\vw_n) \sim \frac{\alpha\sigma^2}{B}\left\langle\frac{1}{\lambda(2-\alpha\lambda)}\right\rangle,\\
    &\mathbb{E}w_{avg, i} \sim \frac{1-\alpha\lambda_i}{\alpha\lambda_i n} w_{0,i},
    &&\frac{1}{P}\Tr \Cov(\vw_{avg}) \leq \frac{\sigma^{2}}{Bn}\left\langle\frac{1}{ \lambda }\right\rangle +  \mathcal{O}(1)\Bigg(k'(-\frac{\delta^2}{2}) + P^{-1}\delta^2k''(-\frac{\delta^2}{2})\Bigg)\label{eq:ia_gp_var}.
\end{alignat}
\end{theorem}
\normalsize

\begin{proof}
We will prove the result in the case $\lambda_i = \lambda ~\forall i$ for the sake of clarity. The same reasoning can be repeated in the more general case; where one gets $P^{-1} f(\lambda) \Tr I$ below, one need only replace it with $\langle f(\lambda)\rangle$, exploiting linearity of the trace. We will also vacuously replace $\sigma^2B^{-1}$ with $\sigma^2$ to save on notation.
For weight iterates $\vw_i$, we have the recurrence \begin{align*}
    \vw_i = (1-\alpha\lambda)\vw_{i-1} + \alpha\vepsilon(\vw_{i-1})
\end{align*}
which leads to \begin{align}\label{eq:wn_expression}
    \vw_n = (1-\alpha\lambda)^n \vw_0 + \alpha\sum_{i=0}^{n-1} (1-\alpha\lambda_i)^{n-i-1}\vepsilon(\vw_i)
\end{align}
and then \begin{align}
    \vw_{avg} &= \frac{1 - (1 - \alpha\lambda)^n}{\alpha\lambda n} (1-\alpha\lambda)\vw_0 + \sum_{i=0}^{n-1}\vepsilon(\vw_i) \frac{1 - (1-\alpha\lambda)^{n-i}}{\lambda n}\label{eq:wavg_expression}.
\end{align}
\jmlrreview{As above, define $\vepsilon_i = \vepsilon(\vw_i)$, for convenience.}
Now define \begin{align*}
    a_i = \alpha(1-\alpha\lambda)^{n-1-i}, ~~~ \bar{a}_i = \frac{1 - (1-\alpha\lambda)^{n-i}}{\lambda n}.
\end{align*}
Next we will apply Lemma \ref{lemma:gp_covar_trace} and utilise Lemma \ref{lemma:prob_far_apart} to bound the variance of $\vw_{avg}$ and $\vw_n$. We first gather the following facts, which were also computed and used in the proof of Theorem $1$:
\begin{align}
    \sum_{i=1}^{n-1} a_i^2 &= \frac{\alpha^2(1 - (1 - \alpha\lambda)^{2n})}{1 - (1-\alpha\lambda)^2}\\
    \sum_{i<j}a_ia_j &= \frac{\alpha}{\lambda}\left(\frac{1 - (1-\alpha\lambda)^n}{\alpha\lambda} - \frac{1 - (1-\alpha\lambda)^{2n}}{1 - (1-\alpha\lambda)^2}\right).
\end{align}
The sum of squares for the $\bar{a}_i$ is simple to obtain similarly \begin{align}
    \sum_{i=0}^{n-1} \bar{a}_i^2 = \frac{1}{\lambda^2 n^2}\left(n - \frac{2(1 - (1-\alpha\lambda)^n)}{\alpha\lambda} + \frac{1 - (1-\alpha\lambda)^{2n}}{1 - (1-\alpha\lambda)^2}\right).
\end{align}
We now use the assumption that $0 < \alpha\lambda < 1$ (required for the convergence of gradient descent) which gives, as $n\rightarrow\infty$, \begin{align}
     \sum_{i=1}^{n-1} a_i^2 & \sim \frac{\alpha^2}{1 - (1 -\alpha\lambda)^2}\label{eq:ai_sq_sum}\\
     \sum_{i<j}a_ia_j & \sim  \frac{\alpha}{\lambda}\left(\frac{1}{\alpha\lambda} - \frac{1}{1 - (1-\alpha\lambda)^2}\right)\label{eq:aiaj_sum}\\
      \sum_{i=1}^{n-1} \bar{a}_i^2 & \sim \frac{1}{\lambda^2 n}\label{eq:abari_sq_sum}
\end{align}
Summing $\sum_{i< j}\bar{a}_i\bar{a}_j$ explicitly is possible but unhelpfully complicated. Instead, some elementary bounds give \begin{align*}
    \sum_{i< j}\bar{a}_i\bar{a}_j &\leq \left(\sum_{i=0}^{n-1}\bar{a}_i\right)^2 = \frac{1}{\lambda^2n^2}\left(n - \frac{1 - (1-\alpha\lambda)^n}{\alpha\lambda}\right)^2 \sim\frac{1}{\lambda^2}
\end{align*}
and \begin{align*}
    \sum_{i< j}\bar{a}_i\bar{a}_j &\geq \sum_{i<j}\left(\frac{1 - (1-\alpha\lambda)^{n-1}}{\lambda n}\right)^2 \sim \frac{1}{2\lambda^2}
\end{align*}
so in particular $\sum_{i < j}\bar{a}_i\bar{a}_j = \mathcal{O}(1)$. Now define the events $A_n(\delta)$ as in Lemma \ref{lemma:prob_far_apart} using $\vepsilon_i$ in place of $\mX_i$. Further, choose $\delta$ large enough so that $k'(-\frac{x^2}{2})$ and $x^2k''(-\frac{x^2}{2})$ are decreasing for $x>\delta$. Define $k'(0) = \sigma^2$. Lemma \ref{lemma:gp_covar_trace} gives \begin{align}
   \frac{1}{P}\Tr \Cov(\vw_n) \mid A_n(\delta) &\leq \sigma^2\sum_{i=1}^n a_i^2 + 2\sum_{i<j} a_ia_j\Bigg(k'(-\frac{\delta^2}{2}) + P^{-1}\delta^2k''(-\frac{\delta^2}{2})\Bigg)\label{eq:trace_wn_bound}
\end{align}
where we note that we have only upper-bounded the second term in (\ref{eq:trace_wn_bound}), so using  (\ref{eq:ai_sq_sum}) and (\ref{eq:aiaj_sum}) and taking $\delta$ large enough we obtain \begin{align}
    \frac{1}{P}\Tr \Cov(\vw_n) \mid A_n(\delta) = \frac{\sigma^2\alpha^2}{1 - (1-\alpha\lambda)^2} + o(1).
\end{align}
Turning now to $\vw_{avg}$ we similarly obtain \begin{align}
    \frac{1}{P}\Tr \Cov(\vw_{avg}) \mid A_n(\delta) \leq \frac{\sigma^2}{n}\frac{1}{\lambda^2}+ \mathcal{O}(1)\Bigg(k'(-\frac{\delta^2}{2}) + P^{-1}\delta^2k''(-\frac{\delta^2}{2})\Bigg)
\end{align}
and, as before, taking $\delta$ large enough we can obtain \begin{align}
    \frac{1}{P}\Tr \Cov(\vw_{avg}) \mid A_n(\delta) = o(1).
\end{align}
Finally recalling (\ref{eq:wn_expression}) and (\ref{eq:wavg_expression}) and writing $(1-\alpha\lambda)^n = \text{e}^{-\alpha\lambda n}+ o(1)$ for large $n$, we obtain the results in the statement of the theorem but conditional on the event $A_n(\delta)$.
\jmlrreview{To complete the proof, we need only to establish that $\mathbb{P}(A_n(\delta)) \rightarrow 1$ $P,n\rightarrow\infty$, which we will do with an application of Lemma \ref{lemma:prob_far_apart}.
Since the loss noise term is a Gaussian process, the $\vepsilon(\vec{w}_i)$ are all jointly Gaussian with the covariance structure (\ref{eq:grad_gp_covar}), but to apply Lemma \ref{lemma:prob_far_apart} we must further establish a lower bound on the covariance of the conditional $\vepsilon_i$. 
Let $\Sigma_n$ be the $P\times P$ covariance matrix of $\vepsilon_n ~\mid~ \{\vepsilon_1, \ldots, \vepsilon_{n-1}\}$, then we are required to show that there exists some $n$-independent $A, \sigma > 0$ such that $\det \Sigma_n > A \sigma^{2P}$ for all $n$ (subject to $\log n \ll P$).
Define $S_n$ to be the $nP \times nP$ covariance matrix of all of the $\{\vepsilon_i\}_{i=1}^n$, i.e. \begin{align*}
    (S_n)_{iP + j, kP + l} = \Cov \left(\epsilon_j(\vw_j), \epsilon_l(\vw_k)\right), ~~ 0\leq i,k < n, ~ 1\leq j,l \leq P,
\end{align*}
and for convenience define $k'(0) = s^2$.
The rules of standard Gaussian conditioning give\begin{align*}
  \Sigma_n = s^2 I - X_n S_{n-1}^{-1} X_n^T,  
\end{align*}
where $X_n$ is the $P \times (n-1)P$ matrix such that $S_n$ has the following block structure
\begin{align}\label{eq:block_sn}
    S_n = \left(\begin{array}{ccc|c}
         & & &  \\
         & S_{n-1} &  & X_n^T\\
         & & & \\ 
         \hline
         & X_n & & s^2 I_P
    \end{array} \right),
\end{align}
so, concretely, from (\ref{eq:grad_gp_covar}) \begin{align}
    (X_n)_{i, Pj + l} = \left( (\vw_n)_i - (\vw_j)_i\right)\left( (\vw_j)_l - (\vw_n)_l\right)k''\left(-\frac{1}{2}d_{jn}^2\right) + \delta_{il} k'\left(-\frac{1}{2}d_{jn}^2\right),
\end{align}
for $1\leq i,l \leq P, ~ 0 \leq j < n - 1$.
We can now Taylor expand the determinant \begin{align*}
    \det \Sigma_n &= s^{2P} \det \left(1 - s^{-2} X_n S_{n-1}^{-1} X_n^T \right)\\
    &= s^{2P} \left(1 - s^{-2} \Tr X_nS_{n-1}^{-1}X_n^T\right) + \ldots
\end{align*}
which is valid provided that the trace term is small compared with 1.
We have \begin{align*}
    |\Tr X_n S_{n-1}^{-1} X_n^T| \leq \Tr X_nX_n^T \|S_{n-1}^{-1}\|_{op} = \|X_n\|_F \|S_{n-1}^{-1}\|_{op}
\end{align*}
where $\|\cdot\|_F, \|\cdot\|_{op}$ are the Frobenius and operator matrix norms respectively.
Hence, it suffices to prove $n,P$-independent bounds $\|S_{n-1}^{-1}\|_{op} < q $ for some $q>0$ and $\|X_n\|_F < c$ for some $0< c < s^2 / 10$, say, valid for all $n$ large enough, to thence obtain $\det \Sigma_n \geq c' s^{2P} $ for some constant $c' > 0$. 
Strictly speaking, one must use a bounded form of the remainder in Taylor's theorem to make precise all of these constants, but in reality we will see that we can make $c$ as small as necessary, so that certainly $c'>0$ exists and the bound $\det \Sigma_n \geq c' s^{2P}$ holds.
Proceeding directly \begin{align*}
   \|X_n\|_F = \Tr X_n X_n^T &= \sum_{i,l=1}^P \sum_{j=0}^{n-2} (X_n)_{i, Pj +l}^2\\
    &= \sum_{j=0}^{n-2} \Bigg\{ P k'\left(- \frac{d_{jn}^2}{2}\right) -2 d_{jn}^2 k'\left(- \frac{d_{jn}^2}{2}\right)k''\left(- \frac{d_{jn}^2}{2}\right) + \left[d_{jn}^2 k''\left(- \frac{d_{jn}^2}{2}\right)\right]^2\Bigg\}\\
    &\leq (n-1)\left(Pk'\left(- \frac{\delta^2}{2}\right) - 2\delta^2 k'\left(- \frac{\delta^2}{2}\right)k''\left(- \frac{\delta^2}{2}\right) + \left[\delta^2 k''\left(- \frac{\delta^2}{2}\right)\right]^2\right),
\end{align*}
but recall that we require $\delta = o(P^{1/2})$, so take for example $\delta = a P^{1/2 - \varphi/2}$ for some $ 0< \varphi < 1$, so 
\begin{align*}
    \|X_n\|_F  \leq (n-1)\left(Pk'\left(- \frac{P^{1-\varphi}}{2}\right) - 2P^{1-\varphi} k'\left(- \frac{P^{1-\varphi}}{2}\right)k''\left(- \frac{P^{1-\varphi}}{2}\right) + \left[P^{1-\varphi} k''\left(- \frac{P^{1-\varphi}}{2}\right)\right]^2\right).
\end{align*}
Now recall that $x k'(-x)$ and $xk''(-x)$ are decaying for large enough $x$, and $\log n \ll P^{1-\theta}$, hence 
\begin{align*}
     \|X_n\|_F  \leq (n-1)\left(2\log^{\frac{1}{1-\theta}} n k'\left(- \frac{\log^{\frac{1-\varphi}{1-\theta}}n}{2}\right)+ \left[\log^{\frac{1-\varphi}{1-\theta}}n ~~k''\left(- \frac{\log^{\frac{1-\varphi}{1-\theta}}n}{2}\right)\right]^2\right).
\end{align*}
Since $\theta > 0$, we can take some $0 < \varphi < \theta$ so that there exists $\chi \in(0,1)$ such that \begin{align}
    \log^{\frac{1 - \varphi}{1 - \theta}} n > \log^{1 + \chi} n
\end{align}
for large enough $n$, and so \begin{align*}
   \|X_n\|_F  \leq (n-1)\left(2\log^{\frac{1}{1-\theta}} n k'\left(- \frac{\log^{1 + \chi}n}{2}\right)+ \left[\log^{1 + \chi}n ~~k''\left(- \frac{\log^{1 + \chi}n}{2}\right)\right]^2\right).
\end{align*}
We assume that $k'(x), k''(x)$ decay at least as fast as $x^M e^{-x}$ for some $M>0$ as $x\rightarrow\infty$, i.e. $k'(x)x^{-M} e^{x} \rightarrow 0$ (and similarly $k''(x)$). Writing $n-1 \leq n = e^{\log n}$, we have 
\begin{align*}
     \|X_n\|_F  \leq 2\log^{\frac{1}{1-\theta}} n k'\left(\log n - \frac{\log^{1 + \chi}n}{2}\right)+ \left[\log^{1 + \chi}n ~~k''\left(\log n- \frac{\log^{1 + \chi}n}{2}\right)\right]^2,
\end{align*}
but for large $n$, $\log^{1+\chi} n \gg \log n$  and so this last expression clearly converges to $0$ as $n\rightarrow\infty$.
Indeed, $e^{-\log^{1+\chi}n /2}$ decays faster than any fixed power of $n$, so the same is true of $\|X_n\|_F$.
Hence we can find the constant $c>0$ such that, for large enough $n>n_0$, say, $\|X_n\|_F< c$, as required.
Now we turn to bounding $\| S_{n-1}^{-1}\|_{op}$, which is done by induction on $n$.
Define the upper bounds  $\| S_{n}^{-1}\|_{op} \leq q_{n}$ for all $n$.
Recalling the block structure (\ref{eq:block_sn}), we get the inverse \begin{align*}
    S_n^{-1} &= \left( \begin{array}{cc}
          (S_{n-1} - s^{-2}X_n^TX_n)^{-1}& 0 \\
         0 & \Sigma_n^{-1}
    \end{array}\right)\left( \begin{array}{cc}
         I & -s^{-2} X_n^T\\
         -s^{-2} X_n & I
    \end{array}\right)\equiv YZ.\end{align*}
    $\|S_n^{-1}\|_{op}$ is bounded above by  $\|X\|_{op}, \|Y\|_{op}$ and so we now bound these norms in turn.
    Since the off diagonals are zero, we have
\begin{align*}
    \|Y\|_{op} \leq \max\{\|\Sigma_n^{-1}\|_{op}, \|(S_{n-1} - s^{-2}X_n^TX_n)^{-1}\|_{op}\}.
\end{align*}
Recalling the expression for $\Sigma_n$ above and expanding the matrix inverse\begin{align*}
    \|\Sigma_n^{-1}\|_{op} &= s^{-2} \| (I - s^{-2}X_n S_{n-1}^{-1}X_n^T)^{-1}\|_{op}\\
    & = s^{-2} \| (I +s^{-2} X_nS_{n-1}^{-1}X_n^T + s^{-4}(X_nS_{n-1}^{-1}X_n^T)^2 + \ldots \|_{op}\\
    & \leq s^{-2} \left(1 + s^{-2} \|X_nS_{n-1}^{-1}X_n^T \|_{op}+ s^{-4} \|_{op}(X_nS_{n-1}^{-1}X_n^T)^2 \|_{op} + \ldots\right)\\
    & \leq s^{-2} \left(1 + s^{-2} \|X_n\|_F\|S_{n-1}^{-1}\|_{op} + s^{-4}\|X_n\|_F^2\|S_{n-1}^{-1}\|_{op}^2 + \ldots\right)\\
    & \leq s^{-2} \left(1 + s^{-2} \|X_n\|_F\|q_{n-1} + s^{-4}\|X_n\|_F^2\|q_{n-1}^2 + \ldots\right)\\
    & \leq s^{-2}(1 + \alpha s^{-2} q_{n-1} \|X_n\|_F)
\end{align*}
for some constant $\alpha>0$, since we have already demonstrated that $\|X_n\|_F\rightarrow 0$ as $n\rightarrow\infty$.
For the other term \begin{align*}
    \|(S_{n-1} - s^{-2}X_n^TX_n)^{-1}\|_{op} &\leq \|S_{n-1}^{-1}\|_{op} \|(I - s^{-2}S_{n-1}^{-1}X_n^TX_n)^{-1}\|_{op}
\end{align*}
from which point, one proceeds just as for $\|\Sigma_n^{-1}\|_{op}$ to obtain \begin{align*}
       \|(S_{n-1} - s^{-2}X_n^TX_n)^{-1}\|_{op} &\leq q_{n-1}(1+\alpha s^{-2} q \|X_n\|_F),
\end{align*}
hence overall \begin{align*}
     \|Y\|_{op} \leq \max\{s^{-2} (1+\alpha s^{-2} q_{n-1} \|X_n\|_F), q_{n-1} (1+\alpha s^{-2} q_{n-1} \|X_n\|_F)\}.
\end{align*}
We can always relax the bound on $\|S_{n-1}^{-1}\|_{op}$ so that $q_{n-1}>s^{-2}$, so we simply have $\|Y\|_{op} \leq q_{n-1} (1 + \alpha s^{-2} q_{n-1} \|X_n\|_F)$.
To bound $\|Z\|_{op}$, we split it into a sum of two matrices \begin{align*}
    \|Z\|_{op} &= \left\|\left( \begin{array}{cc}
         I &  0\\
         0 & I
    \end{array}\right) + \left( \begin{array}{cc}
         0& -s^{-2} X_n^T\\
         -s^{-2} X_n & 0
    \end{array}\right)\right\|_{op}\leq 1 + 2s^{-2}\|X\|_{op} \leq 1 + 2s^{-2} \|X_n\|_F,
\end{align*}
but $\|X_n\|_F \rightarrow 0$ as $n\rightarrow\infty$, so overall we can say \begin{align*}
    \|S_{n}^{-1}\|_{op} \leq q_{n-1}(1 + r_n), ~~ r_n \equiv s^{-2} \|X_n\|_F  (\alpha q_{n-1} + 2 + 2\alpha q_{n-1} \|X_n\|_F),
\end{align*}
which we can simplify to 
\begin{align*}
    \|S_{n}^{-1}\|_{op} \leq q_{n-1}(1 + r_n'), ~~ r_n' \equiv s^{-2} \|X_n\|_F  (\alpha' q_{n-1} + 2)
\end{align*}
and so can say \begin{align*}
    q_n = q_{n-1} + 2s^{-2}\|X_n\|_Fq_{n-1} + s^{-2}\alpha'\|X_n\|_F q_{n-1}^2.
\end{align*}
For large enough $n$, we seek a stability solution to this recurrence, i.e. using the ansatz $q_n = q + h_n$ for $h_n$ small
\begin{align}
    q + h_n = q + h_{n-1} + 2s^{-2}\|X_n\|_F q + 2s^{-2}\|X_n\|_F h_{n-1} + s^{-2}\alpha' \|X_n\|_F (q^2 + 2qh_{n-1} + h_{n-1}^2).
\end{align}
Gathering the leading order terms gives \begin{align*}
    h_n &= h_{n-1} + 2s^{-2}q\|X_n\|_F + s^{-2}\alpha'\|X_n\|_F q^2\\
    \implies h_n &= h_{n_0} + s^{-2}q (2 + q\alpha')\sum_{j=n_0+1}^{n} \|X_j\|_F.
\end{align*}
Recall that $\|X_n\|_F$ decays faster than any fixed power of $n$, so the sum $\sum_{j\geq 2} \|X_j\|_F$ converges, hence for $\varepsilon > 0$ we can take some fixed $n_0$ large enough so that $\sum_{j=n_0 + 1}^n \|X_j\|_F < \varepsilon$ for all $n>n_0$.
We are free to choose $h_{n_0} = 0$ and then for large enough $n_0$, we can guarantee $|h_n| < 1$, say, thus 
\begin{align*}
    q_n \leq\max\left\{ \max_{1\leq m \leq n_0} q_m, q_{n_0} + 1\right\} \equiv q^*.
\end{align*}
Hence we have succeeded in bounding $\|S_{n}^{-1}\|_{op} \leq q^*$ for all $n$.
Combined with the earlier bound on $\|X_n\|_F$, we have now established the bound  $\det \Sigma_n \geq c' s^{2P}$, so we have satisfied the conditions of Lemma \ref{lemma:prob_far_apart} and completed the proof.
}
\end{proof}

Note that Theorem \ref{theorem:dependent_noise_shell} is a generalisation of Theorem \ref{theorem:shell} to the context of our dependent perturbation model. Let us make some clarifying remarks about the theorem and its proof:

\begin{enumerate}
    \item The bound (\ref{eq:ia_gp_var}) in the statement of the theorem relies on \emph{all} iterates being separated by a distance at least $\delta$. Moreover, the bound is only useful if $\delta$ is large enough to ensure the $k'$ and $k''$ terms are small.
    \item Just as in the independent case of Theorem \ref{theorem:shell}, the first term in the bound in (\ref{eq:ia_gp_var}) decays only in the case that the number of iterates $n\rightarrow\infty$.
    \item The remaining conditions on $P, n , \delta$ are required for the high-dimensional probability argument which we use to ensure that all iterates are separated by at least $\delta$.
    \item $P\gg \log{n}$ is a perfectly reasonable condition in the context of deep learning. E.g. for a ResNet-50 with $P\approx 25\times 10^{6}$, violation of this condition would require $n > 10^{10^{7}}$. A typical ResNet schedule on ImageNet has $< 10^{6}$ total steps.
\end{enumerate}

Consequently, our result points to the importance of good separation between weight iterates in IA to retain the independence benefit and variance reduction in a non-independent noise setting, hence one would expect large learning rates to play a crucial role in successful IA. At the same time, our result is particularly adapted to the \emph{deep learning limit} of very many model parameters ($P\rightarrow\infty$), since this is the only regime in which we can argue probabilistically for good separation of weight iterates (otherwise one may simply have to assume such separation). Furthermore, the importance of $P \gg \log n$ indicates that perhaps averaging less frequently than every iteration could be beneficial to generalisation. The following corollary makes this intuition precise.

\begin{corollary}\label{cor:strided}
Let $\vw_{avg}$ now be a strided iterate average with stride $\kappa$, i.e. \begin{align}
    \vw_{avg} = \frac{\kappa}{n}\sum_{i=1}^{\lfloor n/\kappa \rfloor} \vw_i.
\end{align}
Then, under the same conditions as Theorem \ref{theorem:dependent_noise_shell}
\begin{align}
    &\mathbb{E}w_{avg, i} = \frac{\kappa(1-\alpha\lambda_i)^{\kappa}}{n(1 - (1-\alpha\lambda_i)^{\kappa} )} (1 + o(1))w_{0,i},\\
    &\frac{1}{P}\Tr \Cov(\vw_{avg}) \leq \frac{\sigma^2\alpha^2\kappa}{Bn}\left\langle\frac{1}{\left(1 - (1-\alpha\lambda)^{\kappa}\right)^2}\frac{1 - (1-\alpha\lambda)^{2\kappa}}{1 - (1-\alpha\lambda)^{2}}\right\rangle+  \mathcal{O}(1)\Bigg(k'(-\frac{\delta^2}{2}) + P^{-1}\delta^2k''(-\frac{\delta^2}{2})\Bigg)\label{eq:corr_ia_gp_covar}
\end{align}
where the constant $\mathcal{O}(1)$ coefficient of the second term in (\ref{eq:corr_ia_gp_covar}) is independent of $\kappa$.
\end{corollary}
\begin{proof}
The proof is just as in Theorem $2$ (or Theorems $3$ or $4$), differing only in the values of the $\bar{a}_i$. Indeed, a little thought reveals that the generalisation of $\bar{a}_i$ to the case $\kappa > 1$ is \begin{align}
    \bar{a}_i = \frac{\alpha\kappa}{n}(1 - \alpha\lambda)^{\kappa\left(1 + \lfloor\frac{i}{\kappa}\rfloor\right) - 1 - i} \frac{1 - (1-\alpha\lambda)^{\kappa\left(\lfloor\frac{n}{\kappa}\rfloor - \lfloor\frac{i}{\kappa}\rfloor\right)}}{1 - (1-\alpha\lambda)^{\kappa}}.
\end{align}
Note that $\kappa \left\lfloor\frac{i}{\kappa}\right\rfloor - i $ is just the (negative) remainder after division of $i$ by $\kappa$. 
Then for large $n$ \begin{align*}
    \sum_{i} \bar{a}_i^2 &\sim \frac{\alpha^2\kappa^2}{n^2}\frac{(1-\alpha\lambda)^{2(\kappa - 1)}}{\left(1 - (1-\alpha\lambda)^{\kappa}\right)^2} \left\lfloor\frac{n}{\kappa}\right\rfloor \sum_{i=0}^{\kappa-1}(1-\alpha\lambda)^{-2i}\\
     &\leq \frac{\alpha^2\kappa}{n}\frac{(1-\alpha\lambda)^{2(\kappa - 1)}}{\left(1 - (1-\alpha\lambda)^{\kappa}\right)^2} \sum_{i=0}^{\kappa-1}(1-\alpha\lambda)^{-2i}\\
      &= \frac{\alpha^2\kappa}{n}\frac{(1-\alpha\lambda)^{2(\kappa - 1)}}{\left(1 - (1-\alpha\lambda)^{\kappa}\right)^2}\frac{1 - (1-\alpha\lambda)^{-2\kappa}}{1 - (1-\alpha\lambda)^{-2}}\\
    &= \frac{\alpha^2\kappa}{n}\frac{1}{\left(1 - (1-\alpha\lambda)^{\kappa}\right)^2}\frac{1 - (1-\alpha\lambda)^{2\kappa}}{1 - (1-\alpha\lambda)^{2}}.
\end{align*}
and similarly \begin{align}
    \sum_{i< j} \bar{a}_i \bar{a}_j &\sim \frac{\alpha^2\kappa^2}{n^2}\frac{(1-\alpha\lambda)^{2(\kappa-1)}}{(1 - (1-\alpha\lambda)^{\kappa})^2}\sum_{i<j} (1-\alpha\lambda)^{\kappa\lfloor i/\kappa\rfloor - i + \kappa\lfloor j/\kappa\rfloor - j}\\
    &\sim \frac{\alpha^2\kappa^2}{n^2}\frac{(1-\alpha\lambda)^{2(\kappa-1)}}{(1 - (1-\alpha\lambda)^{\kappa})^2}\sum_j (1-\alpha\lambda)^{\kappa\lfloor j/\kappa\rfloor - j}\left\lfloor\frac{j}{\kappa}\right\rfloor\frac{1 - (1-\alpha\lambda)^{-\kappa}}{1 - (1-\alpha\lambda)^{-1}}\\
    &\sim  \frac{\alpha^2\kappa^2}{n^2}\frac{(1-\alpha\lambda)^{2(\kappa-1)}}{(1 - (1-\alpha\lambda)^{\kappa})^2}\left(\frac{1 - (1-\alpha\lambda)^{-\kappa}}{1 - (1-\alpha\lambda)^{-1}}\right)^2\sum_{j=0}^{\lfloor n/\kappa\rfloor}j \\
    &\sim  \frac{\alpha^2}{2}\frac{(1-\alpha\lambda)^{2(\kappa-1)}}{(1 - (1-\alpha\lambda)^{\kappa})^2}\left(\frac{1 - (1-\alpha\lambda)^{-\kappa}}{1 - (1-\alpha\lambda)^{-1}}\right)^2\\
    &=  \frac{\alpha^2}{2}\frac{(1-\alpha\lambda)^{-2}}{(1 - (1-\alpha\lambda)^{-1})^2}.
\end{align}
\end{proof}

 Intuitively, the first term in the covariance  in (\ref{eq:ia_gp_var}) is an ``independence term'', i.e. it is common between Theorems \ref{theorem:shell} and \ref{theorem:dependent_noise_shell} and represents the simple variance reducing effect of averaging. The second variance term in (\ref{eq:ia_gp_var}) comes from dependence between the iterate gradient perturbations. We see from the corollary that an independent model for gradient perturbation would predict an unambiguous inflationary effect of strided IA on variance (the first term in (\ref{eq:corr_ia_gp_covar})). However introducing dependence in the manner that we have predicts a more nuanced picture, where increased distance between weight iterates can counteract the simple ``independent term'' inflationary effect of striding, leaving open the possibility for striding to improve on standard IA for the purposes of generalisation.

\section{Extension of theoretical framework to weight decay and adaptive methods}\label{sec:gp_model_adaptive}
To make a closer connection with the new optimisation algorithms proposed in this work we consider decoupled weight decay (strength $\gamma$) and gradient preconditioning: \begin{align}
    \vw_t = (1 - \alpha\gamma)\vw_{t-1} - \alpha \tilde{\mH}_{t}^{-1}\nabla L_{batch}(\vw_{t-1})
\end{align}
where $\tilde{\mH}_t^{-1}$ is some approximation to the true loss Hessian used at iteration $t$. In the presence of weight decay, we move the true loss minimum away from the origin for the analysis, i.e. $L_{\mathrm{true}}(\vw) = (\vw-\vw^*)^T\mH(\vw-\vw^*)$. The update rule is then
\begin{align}\label{eq:adam_update}
    \vw_t = \left(1-\alpha\gamma - \alpha \tilde{\mH}_t^{-1}\mH\right) \vw_{t-1} + \alpha \mH \vw^* - \alpha \vepsilon(\vw_{t-1}).
\end{align}
We take $\tilde{\mH}_t^{-1}$ to be diagonal in the eigenbasis of $\mH$, with eigenvalues $\tilde{\lambda}_i^{(t)}+\varepsilon$, where $\varepsilon$ is the standard tolerance parameter \cite{kingma2014adam}. One could try to construct the $\tilde{\mH}_t^{-1}$ from the Gaussian process loss model, so making them stochastic and covarying with the gradient noise, however we do not believe this is tractable. Instead, let us heuristically assume that, with high probability, $\tilde{\lambda}_i^{(t)}$ is close to $\lambda_i$, say within a distance $\zeta$, for large enough $t$ and all $i$. If we take a large enough $\zeta$ this is true even for SGD and we expect Adam to better approximate the local curvature matrix than SGD, since this is precisely what it is designed to do. This results in the following theorem.

\begin{theorem}\label{theorem:shell_gadam}
Fix some $\zeta > 0$ and assume that $|\tilde{\lambda}_i^{(t)} - \lambda_i| < \zeta$ for all $t \geq n_0$, for some fixed $n_0(\zeta)$, with high probability. Use the update rule (\ref{eq:adam_update}). Assume that the $\lambda_i$ are bounded away from zero and $\min_i\lambda_i > \zeta$. Further assume $c(\gamma + \varepsilon + \zeta) < 1$, where $c$ is a constant independent of $\varepsilon, \zeta, \gamma$ and is defined in the proof. Let everything else be as in Theorem \ref{theorem:dependent_noise_shell}. Then there exist constants $c_1, c_2, c_3, c_4>0$ such that, with high probability,
\small
\begin{align}
    &|\mathbb{E}w_{n,i}-w^*_i| \leq e^{-\alpha(1 +\gamma -c(\varepsilon+\zeta))n}w_{0,i} + c_1(\varepsilon +\zeta + \gamma)\label{eq:thm4_1} \\
    &\left|\frac{1}{P}\Tr \Cov(\vw_n)- \frac{\alpha\sigma^2}{B(2-\alpha)}\right| \leq   c_2(\varepsilon + \zeta + \gamma) + o(1),\label{eq:thm4_2}\\
    &|\mathbb{E}w_{avg, i} - w^*_i| \leq \frac{1-\alpha(1 +\gamma - c(\varepsilon + \zeta ))}{\alpha(1 + \gamma - c(\varepsilon+\zeta)) n} (1 + o(1))w_{0,i} + c_3(\varepsilon + \zeta + \gamma)\label{eq:thm4_3}\\
    &\left|\frac{1}{P}\Tr \Cov(\vw_{avg}) - \frac{\sigma^{2}}{Bn} -  \mathcal{O}(1)\Bigg(k'(-\frac{\delta^2}{2}) + P^{-1}\delta^2k''(-\frac{\delta^2}{2})\Bigg)\right| \leq c_4 (\gamma, + \zeta + \epsilon).\label{eq:thm4_4}
\end{align}
\end{theorem}
\begin{proof}
We begin with the equivalent of (\ref{eq:wn_expression}) for update rule (\ref{eq:adam_update}): \begin{align}
    \vw_n = \prod_{i=0}^{n-1}\left(1-\alpha\gamma -\alpha \tilde{\mH}_i^{-1}\Lambda\right)\vw_0 &+ \sum_{i=}^{n-1}\alpha \tilde{\mH}_i^{-1}\Lambda \prod_{j=i+1}^{n-1} \left(1 - \alpha\gamma -\alpha\tilde{\mH}_j^{-1}\Lambda \right)\vw^* \notag\\
    & - \sum_{i=}^{n-1}\alpha \tilde{\mH}_i^{-1}\Lambda \left[\prod_{j=i+1}^{n-1} \left(1 - \alpha\gamma -\alpha\tilde{\mH}_j^{-1}\Lambda \right)\right]\vepsilon(\vw_i).
\end{align}
To make progress, we need the following bounds valid for all $t\geq n_0$
 \begin{align*}
     \frac{\lambda_i}{\tilde{\lambda}_i^{(t)}+ \varepsilon} = \frac{\lambda_i}{\lambda_i + \tilde{\lambda}_i^{(t)}  -\lambda_i + \varepsilon}<\frac{\lambda_i}{\lambda_i + \varepsilon - \zeta} < 1 + |\varepsilon-\zeta|\lambda_i^{-1}
\end{align*}

and \begin{align*}
     \frac{\lambda_i}{\tilde{\lambda}_i^{(t)}+ \varepsilon} = \frac{\lambda_i}{\lambda_i + \tilde{\lambda}_i^{(t)}  -\lambda_i + \varepsilon}>\frac{\lambda_i}{\lambda_i + \varepsilon + \zeta} > 1 - (\varepsilon + \zeta)\lambda_i^{-1}
\end{align*}
where the final inequality in each case can be derived from Taylor's theorem with Lagrange's form of the remainder \cite{shirali2014introduction}.
Since the $\lambda_i$ are bounded away from zero, we have established \begin{align}\label{eq:precondition_success_bound}
      \left|\frac{\lambda_i}{\tilde{\lambda}_i^{(t)}+ \varepsilon} - 1\right| < c(\varepsilon + \zeta)
\end{align}
where the constant $c= 1+(\min_j\{\lambda_j\})^{-1}$, say.
From this bound we can in turn obtain \begin{align}
 &1 - \alpha(\gamma + 1 + c(\varepsilon + \zeta))<   1 - \alpha(\gamma + \tilde{(\lambda}_i^{(t)} + \varepsilon)^{-1}\lambda_i) < 1 - \alpha(\gamma + 1 - c(\varepsilon + \zeta))\notag\\
 \implies &1 - \alpha( 1 + c(\varepsilon + \zeta + \gamma))<   1 - \alpha(\gamma + \tilde{(\lambda}_i^{(t)} + \varepsilon)^{-1}\lambda_i) < 1 - \alpha(1 - c(\varepsilon + \zeta + \gamma))
\end{align}
where the second line exploits the assumption $c(\gamma + \varepsilon + \zeta) < 1$ and our choice $c > 1$. Thus \begin{align}
    \sum_{t=0}^{n-1} \alpha \frac{\lambda_k}{\tilde{\lambda}_k^{(t)}} \prod_{j=t+1}^{n-1}\left(1-\alpha\gamma -\alpha(\tilde{\lambda}_k^{(j)} + \varepsilon)\lambda_k\right) &< \sum_{t=0}^{n-1} \alpha (1 + c(\varepsilon + \zeta))\left( 1 - \alpha(\gamma + 1 - c(\varepsilon + \zeta))\right)^{n-1-t}\notag\\
    &< 1 + c_1(\zeta + \varepsilon + \gamma)
\end{align}
where the second inequality follows, for large $n$, by summing the geometric series and again using Lagrange's form of the remainder in Taylor's theorem. $c_1$ is some constant, derived from $c$ that we need not determine explicitly. A complementary lower bound is obtained similarly (for large $n$). We have thus shown that \begin{align}
    |\mathbb{E}w_{n,i} - w^*_i| < c_1(\varepsilon + \zeta + \gamma) + \prod_{t=0}^{n-1}\left(1-\alpha\gamma -\alpha (\tilde{\lambda}_i^{(t)} + 
    \varepsilon)^{-1}\lambda_i\right)w_{0,i}.
\end{align}
Reusing the bound (\ref{eq:precondition_success_bound}) then yields (\ref{eq:thm4_1}). The remaining results, (\ref{eq:thm4_2})-(\ref{eq:thm4_4}) follow similarly using the same bounds and ideas as above, but applied to the corresponding steps from the proof of Theorem $2$.
\end{proof}

Theorem \ref{theorem:shell_gadam} demonstrates the same IA variance reduction as seen previously, but in the more general context of weight decay and adaptive optimisation. As expected, improved estimation of the true Hessian eigenvalues (i.e. smaller $\zeta$) reduces the error in recovery of $\vw^*$. Moreover, increasing the weight decay strength $\gamma$ decreases the leading order error bounds in (\ref{eq:thm4_1}) and (\ref{eq:thm4_3}), but only up to a point, as the other error terms are valid and small only if $\gamma$ is not too large.

\section{Conclusion}
We have proposed a Gaussian Process perturbation between the batch and true risk surfaces and derive the phenomenon of improved generalisation for large learning rates and larger weight decay when combined with iterate averaging observed in practice. We have extended this formalism to include adaptive methods and showed that we expect further improvement when using adaptive algorithms.
\chapter{A random matrix approach to damping in deep learning}\label{chap:damp}

The content of this chapter was published first as a pre-print in March 2022 (\url{https://arxiv.org/abs/2011.08181v5}) and later as a journal article: ``A random matrix theory approach to damping in deep learning''. Diego Granziol, \textbf{Nicholas P
Baskerville}. \emph{Journal
of Physics: Complexity}, 3.2 (2022): 024001.
\medskip

DG conceived of the main idea behind this work and published it as a pre-print, along with other collaborators, before \textbf{NPB} joined the project. \textbf{NPB} introduced the random matrix model and derived the adaptive damping algorithm. \textbf{NPB} also overhauled the existing mathematical content, only some of which is included in this chapter. All the experiments in this chapter were actually executed by DG but \textbf{NPB} contributed equally to their design, analysis and write-up.

\section{The Spiked Model for the Hessian of the Loss}\label{sec:spiked_model}
We conjecture that a key driver of the adaptive generalisation gap is the fact that adaptive methods fail to account for the greater levels of noise associated with their estimates of flat directions in the loss landscape. The fundamental principle underpinning this conjecture, that sharp directions contain information from the underlying process and that flat directions are largely dominated by noise, is theoretically motivated from the spiked covariance model~\cite{baik2004eigenvalues}. This model has been successfully applied in Principal Component Analysis (PCA), covariance matrix estimation and finance \cite{bloemendal2016principal, everson2000inferring,bun2017cleaning,bun2016my}.  We revisit this idea in the context of deep neural network optimisation.

In particular, we consider a spiked additive signal-plus-noise random matrix model for the batch Hessian of deep neural network loss surfaces. In this model, results from random matrix theory suggest several practical implications for adaptive optimisation. We use linear shrinkage theory \cite{bun2016my,bun2016rotational,bun2017cleaning} to illuminate the role of damping in adaptive optimisers and use our insights to construct an adaptive damping scheme that greatly accelerates optimisation. We further demonstrate that typical hyper-parameter settings for adaptive methods produce a systematic bias in favour flat directions in the loss landscape and that the adaptive generalisation gap can be closed by redressing the balance in favour of sharp directions. To track the bias towards flat vs sharp directions we define
\begin{equation}
    \reff := \frac{\alpha_{\text{flat}}}{\alpha_{\text{sharp}}},
\end{equation}
where $\alpha_{\mathrm{flat}}$ and $\alpha_{\mathrm{sharp}}$ are the learning rates along the flat and sharp directions,  respectively and this ratio encapsulates the noise-to-signal ratio as motivated by our conjecture (the terms \textit{flat} and \textit{sharp} are defined more precisely below). 

\subsection{Sharp directions from the true loss surface survive, others wash out}

We can rewrite the (random) batch hessian $\mH_{\text{batch}}$ as the combination  of the (deterministic) true hessian $\mH_{\text{true}}$ plus some fluctuations matrix:
\begin{equation}\label{eq:additive_noise}
    \mH_{\text{batch}}(\vw) = \mH_{\text{true}}(\vw) + \mX(\vw).
\end{equation}
In \cite{granziol2020learning} the authors consider the difference between the batch and empirical Hessian, although this is not of interest for generalisation, the framework can be extended to consider the true Hessian. The authors further show, under the assumptions of Lipschitz loss continuity, almost everywhere double differentiable loss and that the data are drawn i.i.d from the data generating distribution that the elements of $\mX(\vw)$ converge to normal random variables\footnote{Note that although a given batch Hessian is a fixed deterministic property, we are interested in generic properties of batches drawn at random from the data generating distribution for which we make statements and can hence model the fluctuations matrix as a random matrix.}. Under the assumptions of limited dependence between and limited variation in the variance of the elements of the fluctuations matrix, the spectrum of the fluctuations matrix converges to the Wigner semi-circle law \cite{granziol2020learning,wigner1993characteristic}, i.e. weakly almost surely
\begin{align}
    \frac{1}{P}\sum_{i=1}^P \delta_{\lambda_i(\mX)} \rightarrow \mu_{SC},
\end{align}
where the $\lambda_i(\mX)$ are the eigenvalues of $\mX$ and $d\mu_{SC}(x) \propto \sqrt{2P^2 - x^2}dx$.
The key intuition in this chapter is that sharp directions of the true loss surfaces, that is directions in which the true Hessian has its largest eigenvalues, are more reliably estimated by the batch loss than are the flat directions (those with small Hessian eigenvalues).
This intuition is natural in random matrix theory and is supported by results such as the following.
\begin{theorem}
	\label{theorem:overlap}
Let $\{\vtheta_i\}_{i=1}^P$,  $\{\vphi\}_{i=1}^P$ be the orthonormal eigenbasis of the true Hessian $\nabla^2 L_{\mathrm{true}}$ and batch Hessian $\nabla^2 L_{\mathrm{batch}}$ respectively. Let also $\nu \geq \ldots \geq \nu_P$ be the eigenvalues of $\nabla^2 L_{\mathrm{true}}$. Assume that $\nu_i = 0$ for all $i > r$, for some fixed $r$. Assume that $\mX$ is a generalised Wigner matrix. Then as $P\rightarrow\infty$ the following limit holds almost surely	\begin{equation}\label{eq:overlap}
		|\vtheta_{i}^{T}\vphi_{i}|^{2} \rightarrow \begin{cases} 1-\frac{P\sigma^{2}}{B\nu{i}^{2}} &\mbox{if } |\nu_{i}| > \sqrt{\frac{P}{B}}\sigma,\\
			0 & \mbox{otherwise}, \end{cases}
	\end{equation}
where $\sigma$ is the sampling noise per Hessian element.
\end{theorem}

\begin{proof}
    This is a direct application of a result of \cite{capitaine2016spectrum} which is given more explicitly in the case of GOE Wigner matrices by \cite{benaych2011eigenvalues}. In particular, we use a scaling of $\mX$ such that the right edge of the support of its spectral semi-circle is roughly at $P^{1/2}B^{-1/2}\sigma$. The expression in Section 3.1 of \cite{benaych2011eigenvalues} can then be applied to $P^{-1/2}\mH_{\text{batch}}$ and re-scaled in $\sqrt{P}$ to give the result. Note that the substantiation of the expression from \cite{benaych2011eigenvalues} in the case of quite general Wigner matrices is given by Theorem 16 of \cite{capitaine2016spectrum}. 
\end{proof}

Results like Theorem \ref{theorem:overlap} are available for matrix models other than Wigner, such as rotationally invariant models \cite{belinschi2017outliers}, and are conjectured to hold for quite general\footnote{Roughly speaking, models for which a local law can be established \cite{erdHos2017dynamical}.} models \cite{benaych2011eigenvalues}. Convergence of the spectral measure of $P^{-1/2}\mX$ to the semi-circle is necessary to obtain (\ref{eq:overlap}), but not sufficient. The technicalities to rigorously prove Theorem \ref{theorem:overlap} without assuming a Wigner matrix for $\mX$ are out of scope for the present work, requiring as they would something like an optimal local semi-circle law for $\mX$ \cite{erdHos2017dynamical}. We require only the general heuristic principle from random matrix theory encoded in (\ref{eq:overlap}), namely that \emph{only sharp directions retain information from the true loss surface}. It is expected that this principle will hold for a much wider class of random matrices than those for which it has been rigorously proven. This is acutely important for adaptive methods which rely on curvature estimation, either explicitly for stochastic second order methods or implicitly for adaptive gradient methods.

\begin{figure}[h!]
	\centering
	\begin{subfigure}[b]{0.36\linewidth}
		\includegraphics[trim=0cm 0 0 0,clip,width=\textwidth]{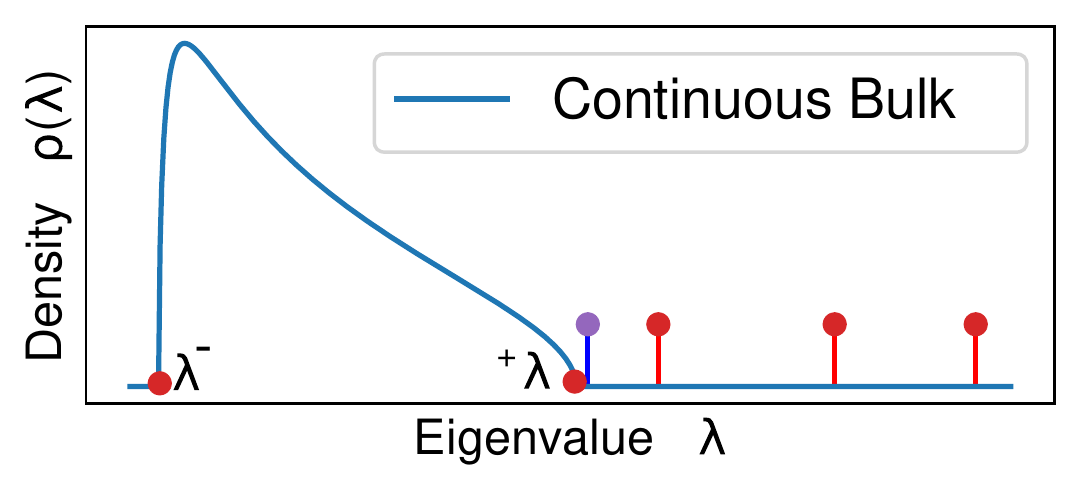}
		\vspace{-15pt}
		\caption{Hypothetical $\rho(\lambda)$}
		\label{subfig:standardmpoutlier}
	\end{subfigure}
	\begin{subfigure}[b]{0.31\linewidth}
		\includegraphics[trim=0cm 0 0 0,clip,width=\textwidth]{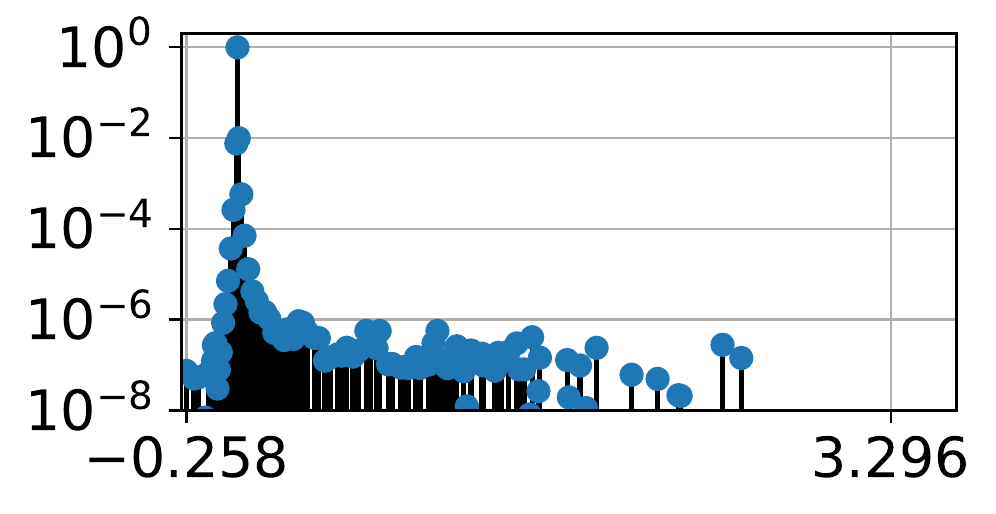}
		\vspace{-15pt}
		\caption{Val Acc$ = 94.3$, SGD}
		\label{subfig:sgdc10}
	\end{subfigure}
	\begin{subfigure}[b]{0.31\linewidth}
		\includegraphics[trim=0cm 0 0 0,clip, width=\textwidth]{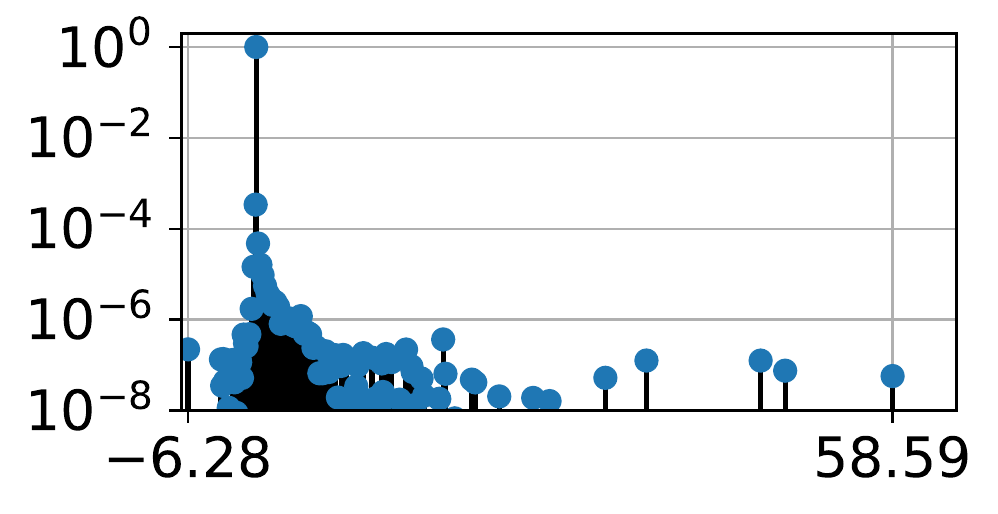}
		\vspace{-15pt}
		\caption{Val Acc$= 95.1$, Adam}
		\label{subfig:gadamc10}
	\end{subfigure}
	\caption{(a) Hypothetical spectral density plot with a sharply supported continuous bulk region, a finite size fluctuation {\color{blue}shown in blue} corresponding to the Tracy-Widom region and three well-separated outliers {\color{red}shown in red}. (b,c) VGG-$16$ Hessian on the CIFAR-$10$ dataset at epoch $300$ for SGD and Adam respectively. Note the "sharper" solution has better validation accuracy.}
	\label{tab:whatissharp}
	\vspace{-10pt}
\end{figure}

\medskip
The spectrum of the noise matrix occupies a continuous region that is sharp in the asymptotic limit \cite{bun2017cleaning} known as \textit{bulk} supported between $[\lambda_{-},\lambda_{+}]$ \cite{bun2017cleaning,bun2016my,bun2016rotational} and observed in DNNs \cite{granziol2019mlrg,papyan2018full,sagun2017empirical}. Within this bulk
eigenvectors are uniformly distributed on the unit sphere \cite{benaych2011eigenvalues} and 
all information about the original eigenvalue/eigenvector pairs is lost \cite{baik2005phase}. Hence from a theoretical perspective it makes no sense to estimate these directions and move along them accordingly. 
An eigenvalue, $\lambda_i$, corresponds to a \textit{flat} direction if $\lambda_{i} \leq \lambda_{+}$. For finite-size samples and network size, there exists a region beyond the predicted asymptotic support of the noise matrix, called the Tracy--Widom region \cite{tracy1994level,el2007tracy}, where there may be isolated eigenvalues which are part of the noise matrix spectrum (also shown in Figure \ref{subfig:standardmpoutlier}). The width of the Tracy--Widom region is very much less than that of the bulk. Anything beyond the Tracy--Widom region $\lambda_{i} \gg \lambda_{+}$, $\lambda_{i} \ll \lambda_{-}$ is considered an outlier and corresponds to a \textit{sharp} direction. \emph{Such directions represent underlying structure from the data}. The eigenvectors corresponding to these eigenvalues can be shown to lie in a cone around their true values \cite{benaych2011eigenvalues} (see Theorem \ref{theorem:overlap}).
In Figure \ref{subfig:sgdc10}, we show the Hessian of a VGG-$16$ network at the $300\textsuperscript{th}$ epoch on CIFAR-100. Here, similar to our hypothetical example, we see a continuous region, followed by a number of eigenvalues which are close to (but not within) the bulk, and finally, several clear outliers.

\section{Detailed experimental investigation of Hessian directions}\label{sec:logisticexp}
\label{subsec:mnist}
In this section we seek to validate our conjecture that movements in the sharp direction of the loss landscape are inherently vital to generalisation by studying a convex non-stochastic example. For such a landscape there is only a single global minimum and hence discussions of bad minima are not pertinent.
We implement a second-order optimiser based on the Lanczos iterative algorithm \cite{meurant2006lanczos} (LanczosOPT) against a gradient descent (GD) baseline. 

\medskip
\paragraph{Note on Lanczos} The Lanczos algorithm is an iterative algorithm for learning approximations to the eigenvalues/eigenvectors of any Hermitian matrix, requiring only matrix--vector products. The values and vectors learned by Lanczos are known as Ritz values/vectors, which are related to the eigenvalue/eigenvector pairs of the matrix. For example, when using a random vector in the matrix vector product, the Ritz values with a weight given by the first element squared of the corresponding Ritz vector, can be shown to give a moment matched approximation to the spectral density of the underlying matrix. In the same way that the power iteration algorithm converges to the largest eigenvalue (with a rate of convergence depending on the size of the spectral gap $\frac{\lambda_{1}-\lambda_{2}}{\lambda_{1}}$) the Lanczos Ritz values converge to well separated outliers\footnote{Intuitively once the largest outlier has been learned, since Lanczos maintains an orthogonal search space, it converges to the next largest outlier}. 
Similar to the power iteration algorithm, this convergence is irrespective of the original seed vector as long as it is not orthogonal to the associated eigenvectors.

\medskip
We employ a training set of $1$K MNIST \cite{lecun1998mnist} examples using logistic regression and validate on a held out test set of $10$K examples. Each optimiser is run for $500$ epochs.   Since the number of well-separated outliers from the spectral bulk is at most the number of classes \cite{papyan2018full} (which is $n_{c}=10$ for this dataset), we expect the Lanczos algorithm to pick out these well-separated outliers when the number of iterations $k \gg n_{c}$ \cite{granziol2019mlrg,meurant2006lanczos} and therefore use $k=50$. 
To investigate the impact of scaling steps in the Krylov subspace given by the sharpest directions, we consider the update $\vw_{k+1}$ of the form:
\begin{equation}
\label{eq:lanczosopt}        
\vw_{k} -\alpha\bigg(\frac{1}{\eta}\sum_{i=1}^{k}\frac{1}{\lambda_{i}+\delta}\vphi_{i}\vphi_{i}^{T}\nabla L(\vw_{k})+\sum_{i=k+1}^{P}\frac{1}{\delta}\vphi_{i}\vphi_{i}^{T}\nabla L(\vw_{k})\bigg)
\end{equation}
where $P=7850$ (the number of model parameters) and hence the vast majority of flat directions remain unperturbed. 
Note that in the case that $k=P=7850$ we would have a fully second order method, whereas in the case where $k=0$, by resolution of the identity, we would have gradient descent with learning rate $\frac{\alpha}{\delta}$. Hence equation \ref{eq:lanczosopt} can be seen as scaling the $k$ Ritz eigenvectors by their respective Ritz values, whilst leaving the remaining directions (which by the previous argument are typically the "flatter" directions) unchanged from their gradient descent counterpart. Whilst Equation \ref{eq:lanczosopt} would naively require $\mathcal{O}(P^{3})$ operations, i.e a full eigendecomposition, it can in fact equivalently be implemented in the following manner
\begin{equation}
\vw_{k} -\alpha\bigg(\frac{1}{\eta}\sum_{i=1}^{k}\bigg[\frac{1}{\lambda_{i}+\delta}-\frac{1}{\delta}\bigg]\vphi_{i}\vphi_{i}^{T}\nabla L(\vw_{k})+\frac{1}{\delta}\nabla L(\vw_{k})\bigg),
\end{equation}
which requires only $k$ Hessian vector products and hence is of computational complexity $\mathcal{O}(kP)$.

\begin{figure}[h!]
	\centering
	\begin{subfigure}[b]{0.59\textwidth}
		\includegraphics[width=\textwidth,trim={0cm 0.2cm 0 0},clip]{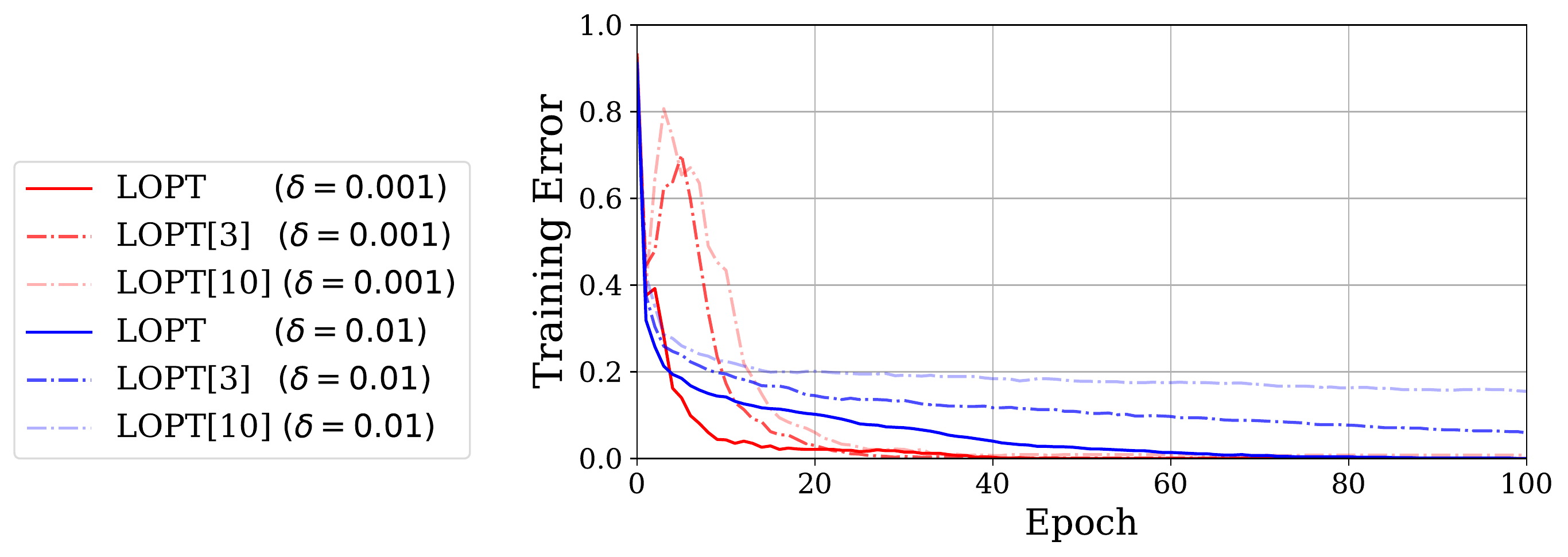}
	\end{subfigure}
	\begin{subfigure}[b]{0.4\textwidth}
		\includegraphics[width=\textwidth,trim={0cm 0.25cm 0 0},clip]{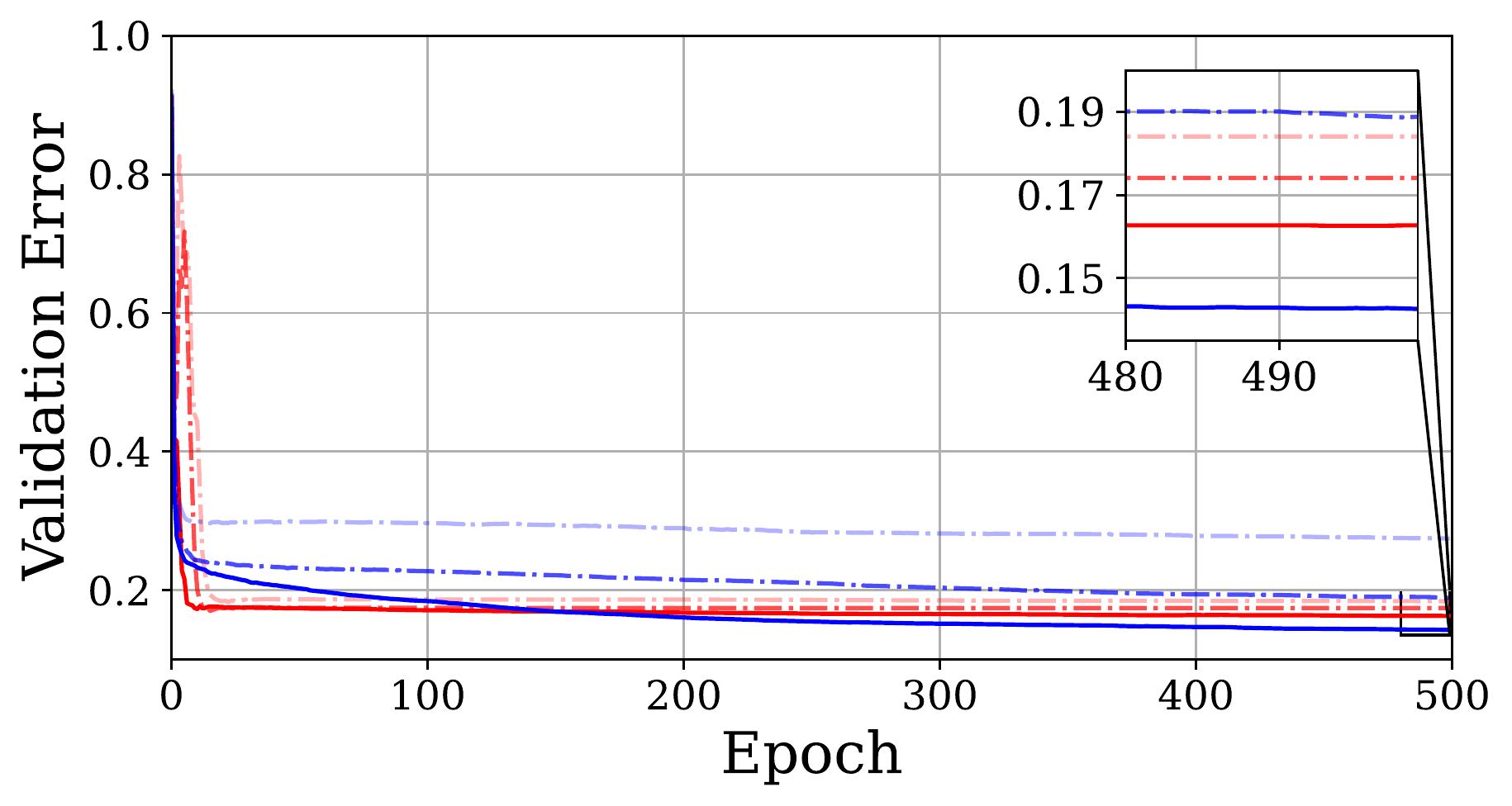}
	\end{subfigure}
	\vspace{-0.3cm}
	\caption{Training/test error of LanczosOPT (LOPT) optimisers for logistic regression on the MNIST dataset with fixed learning rate $\alpha=0.01$ across different damping values, $\delta$. LOPT$[\eta]$ denotes a modification to the LOPT algorithm that perturbs a subset of update directions by a factor of $\eta$. Best viewed in colour. }
	\label{fig:logisticlrandeps1}
\end{figure}

\begin{figure}[h!]
	\centering
	\begin{subfigure}[b]{0.59\textwidth}
		\includegraphics[width=\textwidth,trim={0cm 0.2cm 0 0},clip]{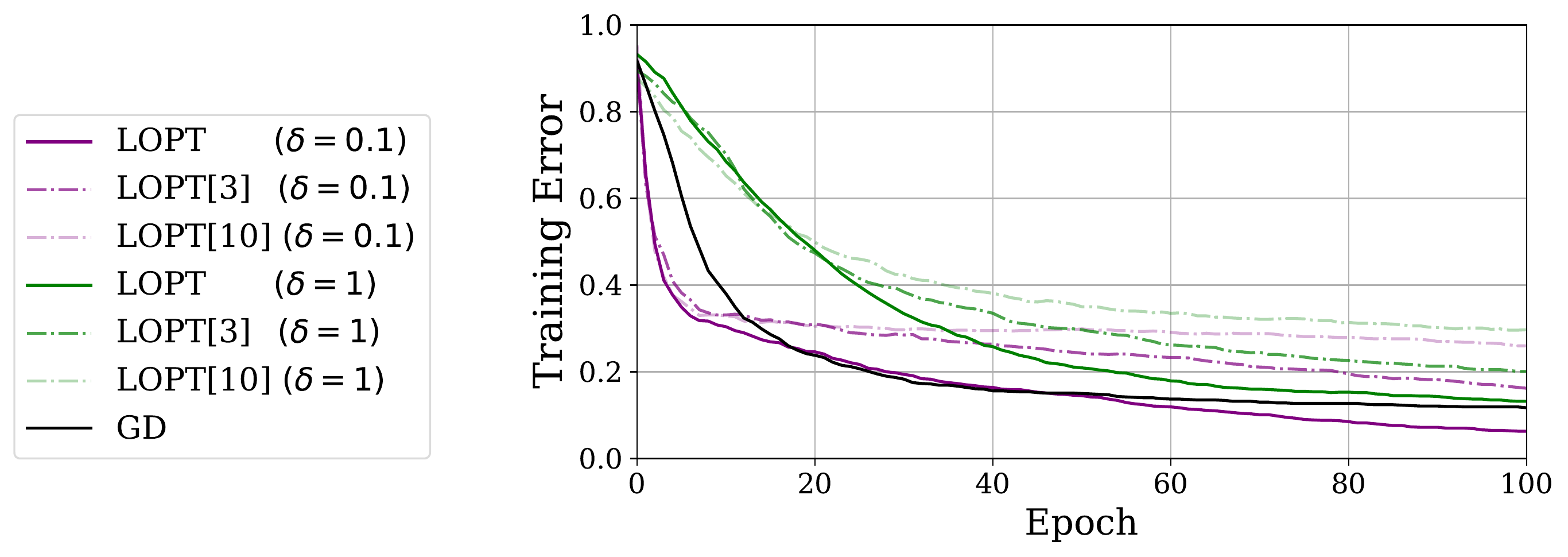}
	\end{subfigure}
	\begin{subfigure}[b]{0.4\textwidth}
		\includegraphics[width=\textwidth,trim={0cm 0.25cm 0 0},clip]{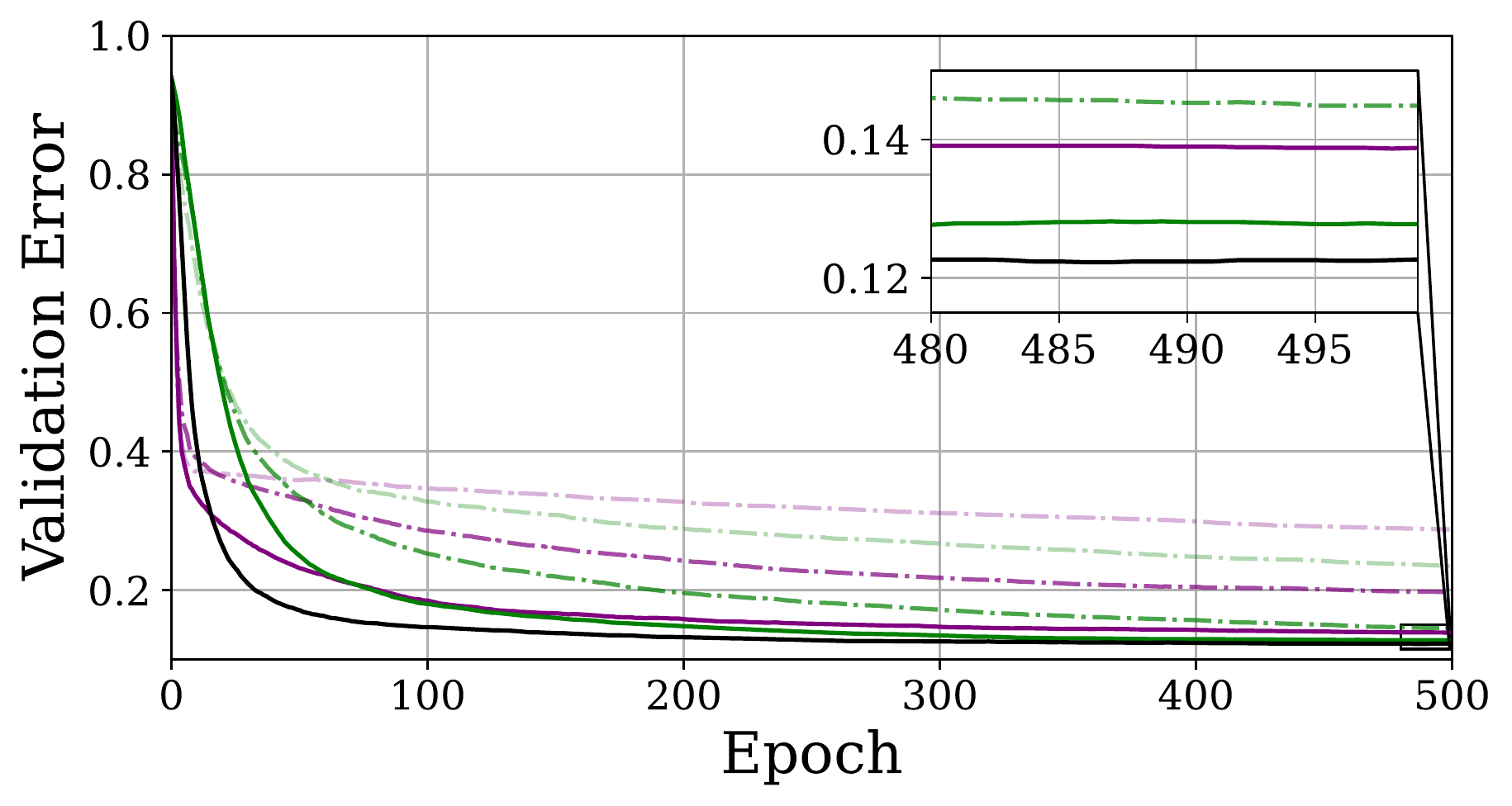}
	\end{subfigure}
	\vspace{-0.3cm}
	\caption{Training/test error of LanczosOPT/Gradient Descent (LOPT/GD) optimisers for logistic regression on the MNIST dataset with fixed learning rate $\alpha=0.01$ across different damping values, $\delta$. LOPT$[\eta]$ denotes a modification to the LOPT algorithm that perturbs a subset of update directions by a factor of $\eta$. Best viewed in colour. }
	\label{fig:logisticlrandeps2}
\end{figure}

To explore the effect of the sharp directions explicitly 
as opposed to implicitly,
we have introduced perturbations to the optimiser (denoted LOPT$[\eta]$), in which we reduce the first term in the parenthesis of Equation~\ref{eq:lanczosopt} by a factor of $\eta$ (we explore scaling factors of $3$ and $10$). This reduces movement in sharp directions, consequently increases reliance on flat directions 
during the optimisation trajectory (we increase $\reff$). This differs from simply increasing $\delta$, which while reducing the movement in all directions, actually relatively increases movement in the sharper directions (decreases $\reff$). To see this consider the case where $\lambda_{i} \gg \delta$, in such an instance, increasing $\delta$ does not appreciably change movement in the sharp directions, whereas it massively decreases movement in flat directions.
For a fixed $\alpha$, $\delta$ controls the $\reff$. 
\label{subsec:logisticexp}
\paragraph{Experimental Results}
We show the training and validation curves for various values of damping $\delta$ and specific sharpness reduction factor $\eta$ in Figures \ref{fig:logisticlrandeps1} and \ref{fig:logisticlrandeps2}. For ease of exposition we only show curves of adjacent values of damping and in order to focus on the speed of convergence we only show the first $100$ epochs of training. We have the full $500$ epochs of training, along with all curves colour coded on the same graph in \ref{app:moremnist}. 
We use colours to distinguish $\delta$ values and dashing/opacity to indicate $\eta$ values (dashed is larger than solid, and dashed with lower opacity is larger still).
Note that as given by our central hypothesis, increasing $\delta$ increases generalisation (we decrease $\reff$), whereas increasing $\eta$ decreases generalisation (we increase $\reff$).
\newline

\begin{figure}
	\begin{subfigure}[b]{0.48\linewidth}
		\includegraphics[width=\textwidth]{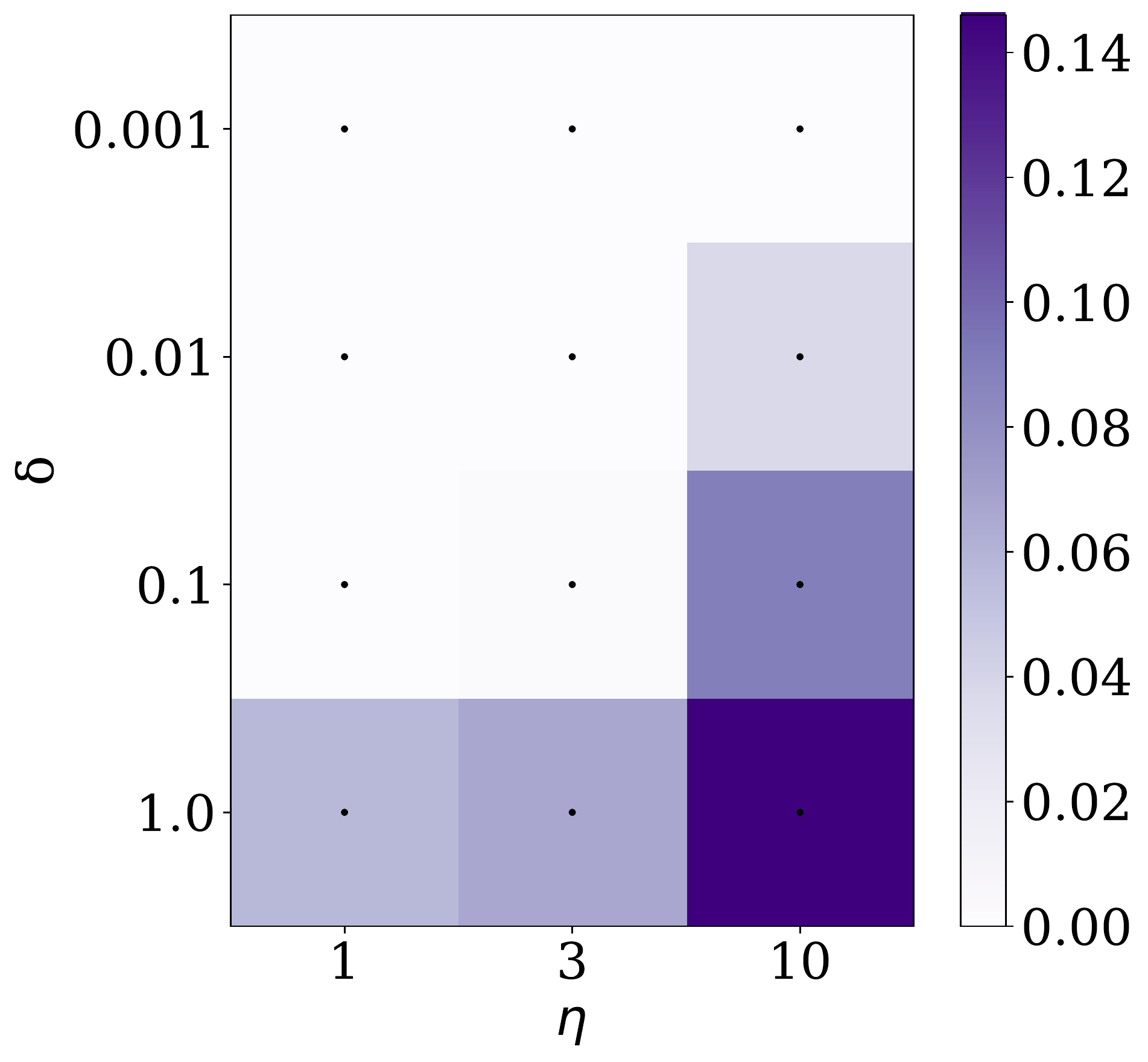}
		\caption{$\Delta(\delta,\eta)$ Training}
	\end{subfigure}
	\begin{subfigure}[b]{0.48\linewidth}
		\includegraphics[width=\textwidth]{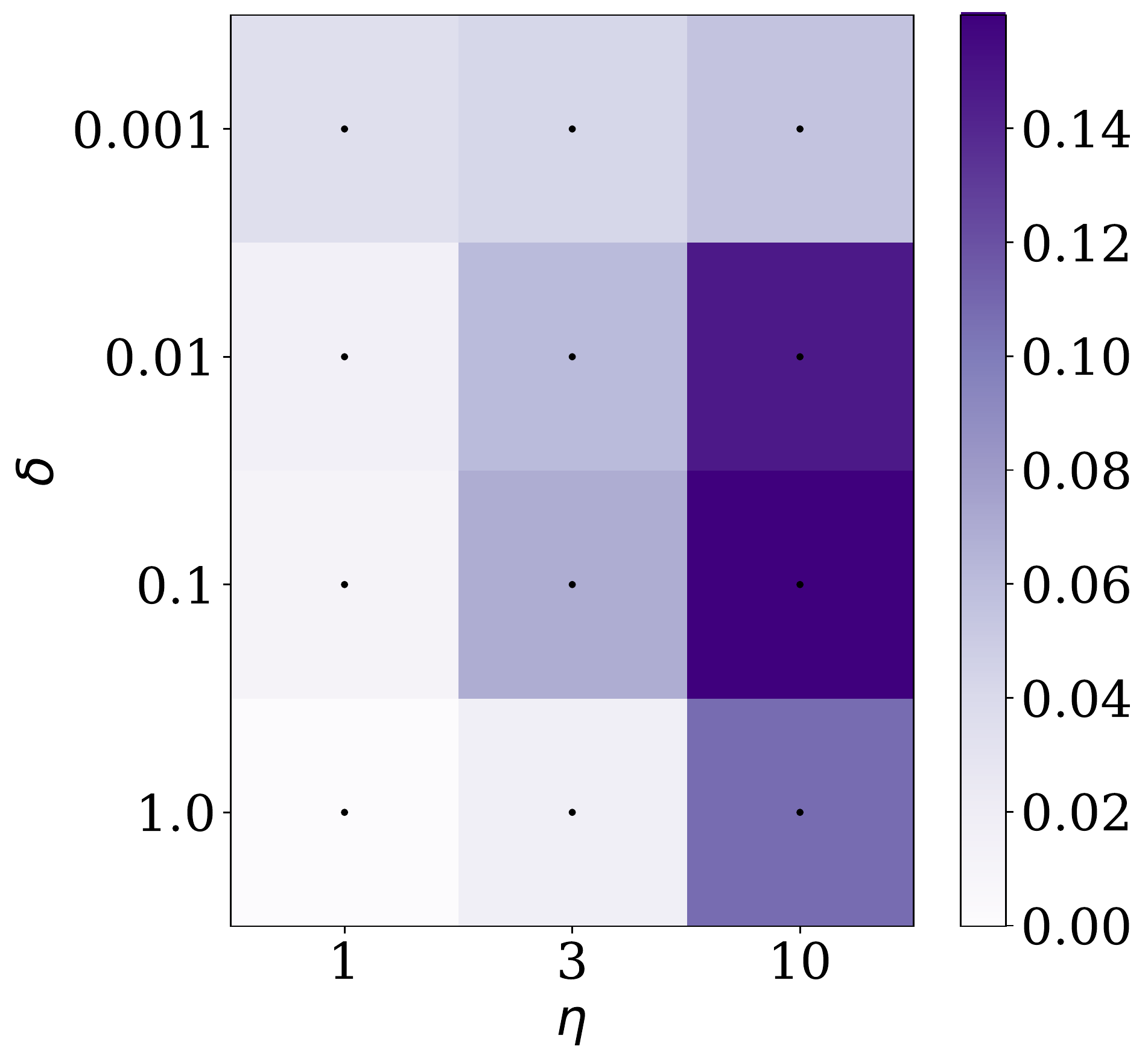}
		\caption{$\Delta(\delta,\eta)$  Testing}
	\end{subfigure}
	\captionof{figure}{Error change with damping/sharp direction perturbation $\delta, \eta$ in LanczosOPT, relative to the single best run. Darker regions indicate higher error. The lowest attained training error ($0$) and validation error ($0.13$) are used as reference points of zero.}
	\label{fig:logisticheatmap}
\end{figure}

We see in Figure \ref{fig:logisticlrandeps1} that despite an initial instability in training for $\eta=3,10$, the red lines with lowest value of damping $\delta=0.001$, all converge quickly to $0$ training error (See \ref{app:moremnist}). However the generalisation as measured by the validation error decreases as we increase $\eta$. This can be seen as the lighter dashed lines (denoting a decrease in movement in the sharpest directions only) increase in validation error. For the blue lines with $\delta=0.01$, whilst increasing $\delta$ decreases the rate of convergence, $\eta=3$ attains a final training error of $0$, yet differs markedly in validation error for its $\eta=1$ counterpart. Similarly so the change in validation error for $\eta=10$ from $\eta=3$ is much larger than the change in training error.
For larger values of $\delta$ as shown in Figure \ref{fig:logisticlrandeps2}, whilst we see an effect on both training and validation, the effect on validation is much more stark. To show this in an intuitive way, in Figure \ref{fig:logisticheatmap}, we use a heat map to show the difference from the best training and testing error as a function of $\delta$ and $\eta$.
The best training error was $0$ and attained at $\eta=1, \delta=0.001$, whereas the best testing error was $0.13$ and attained at $\eta=1, \delta=1.0$.
It is the difference from these values that is shown in Figure \ref{fig:logisticheatmap} (so the top left square is $0$ for training and similarly the bottom left for testing).
As we increase $\reff$ (by decreasing the value of $\delta$ for a fixed $\alpha$ value of $0.01$), the generalisation of the model suffers correspondingly. 
For each fixed value of $\delta$, we see clearly that perturbations of greater magnitude cause greater harm to generalisation than training. We also note that for larger values of $\delta$ the perturbed optimisers suffer more gravely in terms of the effect on both training and validation. It is of course possible that for such large values of $\delta$ we have not converged even after $500$ epochs. We show the full training curves in Figure \ref{fig:logisticlrandeps}.
We observe that the generalisation of all algorithms is worsened by explicit limitation of movement in the sharp directions (and an increase of $\reff$), however for extremely low damping measures (which are typical in adaptive optimiser settings) there is no or very minimal impact in training performance (upper region of Figure \ref{fig:logisticheatmap}(a). 
A consequence of this which is already employed in practical machine learning is the use of $\delta$ tuning. Essentially using larger than default values of $\delta$ (decreasing $\reff$) so as to not simply avoid problems of numerical stability but also generalise better.

\paragraph{Fashion MNIST:}
We repeat the experimental procedure for the FashionMNIST dataset ~\cite{xiao2017fashion}, which paints an identical picture (at slightly higher testing error) The full training curves are given in Figure \ref{fig:fashionlogisticlrandeps}.


\section{The role of damping}\label{sec:nnexperiments}
Consider a general iterative optimiser that seeks to minimise the scalar loss $L(\vw)$ for a set of model parameters $\vw \in \mathbb{R}^P$. Recall the $k+1$-th iteration of such an optimiser can be written\footnote{Ignoring additional features such as momentum and explicit regularisations.} as follows:
\begin{equation}
\vw_{k+1} \leftarrow \vw_{k} - \alpha_k \mB^{-1} \nabla L_{\mathrm{batch}}(\vw_{k})
\end{equation}
where $\alpha_k$ is the global learning rate. For SGD, $\mB = \mI$ whereas for adaptive methods, $\mB$ typically
comprises some form of approximation to the Hessian i.e. $\mB \approx \nabla^{2}L_{\mathrm{batch}}(\vw_{k})$. 
Writing this update in the eigenbasis of the Hessian\footnote{We assume this to be positive definite or that we are working with a positive definite approximation thereof.} $\nabla^{2}L_{\mathrm{batch}}(\vw_{k}) = \sum_{i}^{P}\lambda_{i}\vphi_{i}\vphi_{i}^{T} \in \mathbb{R}^{P\times P}$, where $\lambda_1\geq \lambda_2\geq \dots \geq \lambda_{P} \geq 0$ represent the ordered scalar eigenvalues, the parameter step takes the form:
\begin{equation}
\label{eq:secondorderopt}
\begin{aligned}
\vw_{k+1} = \vw_{k} - \sum_{i=1}^{P}\frac{\alpha}{\lambda_{i}+\delta}\vphi_{i}\vphi_{i}^{T}\nabla L_{\mathrm{batch}}(\vw_{k}).
\end{aligned}
\end{equation}
Here, $\delta$ is a damping (or numerical stability) term. This damping term (which is typically grid searched \cite{dauphin2014identifying} or adapted during training \cite{martens2015optimizing}) can be interpreted as a trust region \cite{dauphin2014identifying} that is required to stop the optimiser moving too far in directions deemed flat ($\lambda_{i} \approx 0$), known to dominate the spectrum in practice \cite{granziol2020learning,papyan2018full,ghorbani2019investigation}, and hence diverging. In the common adaptive optimiser Adam \cite{kingma2014adam}, it is set to $10^{-8}$. For small values of $\delta$, $\alpha$ must also be small to avoid optimisation instability, hence global learning rates and damping are coupled in adaptive optimisers.
\subsection{Adaptive updates and damping} \label{subsec:dampingadaptive}
The learning rate in the flattest ($\lambda \approx 0$) directions is approximately $\frac{\alpha}{\delta}$, which is larger than the learning rate in the sharpest ($\lambda_{i} \gg \delta$) directions  $\frac{\alpha}{\delta+\lambda_{i}}$. 
This difference in per direction effective learning rate makes the best possible (damped) training loss reduction under the assumption that the loss function can be effectively modelled by a quadratic \cite{Martens2016}. Crucially, however, it is agnostic to how accurately each eigenvector component of the update estimates the true underlying loss surface, which is described in Theorem \ref{theorem:overlap}. Assuming that the smallest eigenvalue $\lambda_{P} \ll \delta$,  we see that $\reff = 1+ \frac{\lambda_{1}-\lambda_{P}}{\delta}$. This is in contrast to SGD where 
$\vw_{k+1} = \vw_{k} - \sum_{i=1}^{P}\alpha\vphi_{i}\vphi_{i}^{T}\nabla L_{\mathrm{batch}}(\vw_{k})$ and hence $\reff = 1$. Note that we can ignore the effect of the overlap between the gradient and the eigenvectors of the batch Hessian because we can rewrite the SGD update in the basis of the batch Hessian eigenvectors and hence reduce the problem to one of the relative learning rates.

The crucial point to note here is that the difference in $\reff$ is primarily controlled by the damping parameter: smaller values yield a larger $\reff$, skewing the parameter updates towards flatter directions.


To further explore our central conjecture for modern deep learning architectures (where a large number of matrix--vector products is infeasible) we employ the KFAC \cite{martens2015optimizing} and Adam \cite{kingma2014adam} optimisers on the VGG-$16$~\cite{simonyan2014very} network on the CIFAR-$100$~\cite{krizhevsky2009learning} dataset. The VGG-16 allows us to isolate the effect of $\reff$, as opposed to the effect of different regularisation implementations for adaptive and non-adaptive methods as discussed by \cite{loshchilov2018decoupled,zhang2018three}.
\subsection{VGG16: a laboratory for adaptive optimisation}
The deep learning literature contains very many architectural variants of deep neural networks and a large number of engineering ``tricks'' which are employed to obtain state of the art results on a great variety of different tasks. The theory supporting the efficacy of such tricks and architectural designs is often wanting and sometimes entirely absent. Our primary objective in this work is to illuminate some theoretical aspects of adaptive optimisers such as appropriate damping and Hessian estimation, so we require a simple and clean experimental environment free from, where possible, interference from as many different competing effects. To this end, the VGG architecture \cite{simonyan2014very} for computer vision is particularly appropriate. With 16 layers, the VGG has over $16$ million parameters and is capable of achieving competitive test error on a variety of standard computer vision datasets while being trained without batch normalisation \cite{ioffe2015batch} or weight decay. Indeed, features such as weight decay and batch normalisation obscure the effect of learning rate and damping, meaning that even quite poor choices can ultimately give reasonable results given sufficient training iterations\cite{granziol2020learning}. In contrast the VGG clearly exposes the effects of learning rate and damping, with training being liable to fail completely or diverge if inappropriate values are used. Furthermore as shown in \cite{granziol2020learning} the VGG is highly unstable if too large a learning rate is used. This allows us to very explicitly test whether amendments provided by theory are helpful in certain contexts, such as training stability, as unstable training very quickly leads to divergence.

\paragraph{Learning Rate Schedule} For all experiments unless specified,  we use the following learning rate schedule for the learning rate at the $t$-th epoch:
\begin{equation}
	\alpha_t = 
	\begin{cases}
		\alpha_0, & \text{if}\ \frac{t}{T} \leq 0.5 \\
		\alpha_0[1 - \frac{(1 - r)(\frac{t}{T} - 0.5)}{0.4}] & \text{if } 0.5 < \frac{t}{T} \leq 0.9 \\
		\alpha_0r, & \text{otherwise}
	\end{cases}
\end{equation}
where $\alpha_0$ is the initial learning rate. $T$ is the total number of epochs budgeted for all CIFAR experiments. We set $r = 0.01$ for all experiments.

\begin{table*}[ht]
	\vspace{-0.1cm}
	\begin{minipage}[b]{0.98\linewidth}\centering
		\vspace{15pt}
		\hspace{-1.5cm}
		\begin{minipage}[b]{0.40\linewidth}\centering
			\centering
			\setlength\tabcolsep{3.5pt} 
			\begin{tabular}{@{}lrr@{}}
				\toprule
				& \multicolumn{2}{c}{$\alpha$} \\
				\cmidrule(lr){2-3}
				\multicolumn{1}{c}{$\delta$} &  \multicolumn{1}{c}{0.0004} &  \multicolumn{1}{c}{0.001}  \\
				\midrule
				$1$e-$7$ &  \textbf{53.1}(62.9) & \textbf{}  \\
				$4$e-$4$ &  \textbf{21.1}(64.5) & \textbf{}  \\
				$1$e-$3$ & \textbf{9.9}(63.5) & \textbf{20.8}(64.4)  \\
				$5$e-$3$ & \textbf{} & \textbf{9.1}(66.2)  \\
				$8$e-$3$ & \textbf{} & \textbf{2.4}(65.8) \\
				\bottomrule
			\end{tabular}
			\vspace{6pt}
		\end{minipage}
		\hspace{0.1cm}
		\begin{minipage}[b]{0.40\linewidth}
			\centering
			\setlength\tabcolsep{3pt} 
			\begin{tabular}{@{}lrrr@{}}
				\toprule
				& \multicolumn{3}{c}{$\alpha$} \\
				\cmidrule(lr){2-4}
				\multicolumn{1}{c}{$\delta$} & \multicolumn{1}{c}{0.1} & \multicolumn{1}{c}{0.001} & \multicolumn{1}{c}{0.0001} \\
				\midrule
				$1$e-$1$  & \textbf{6.7}(65.0) & \textbf{} & \\
				$1$e-$2$  & & \textbf{20.8}(64.8) & \textbf{} \\
				$1$e-$3$  & & \textbf{} &\textbf{48.2}(62.2) \\
				$3$e-$4$  & & \textbf{} &  \textbf{527.2}(60.2) \\
				$1$e-$4$  & & \textbf{} &  \textbf{711.3}(56.0) \\ \bottomrule
			\end{tabular}
			\vspace{6pt}
		\end{minipage}
		\caption{\textbf{Spectral norms and generalisation}. We report the spectral norm $\lambda_{1}$ at the end of training in \textbf{bold}, with corresponding validation accuracy (in parentheses) for learning rate/damping $\alpha,\delta$ using Adam (left) and KFAC (right) to train a VGG-$16$ network on CIFAR-$100$.}
		\label{tab:tableofspectralnorm}
	\end{minipage}
\end{table*}

\subsection{KFAC with VGG-16 on CIFAR-100:} 

\begin{figure*}[h!]
	\centering
	\begin{subfigure}[b]{0.60\textwidth}
	    \vspace{-0.1cm}
		\includegraphics[width=\textwidth]{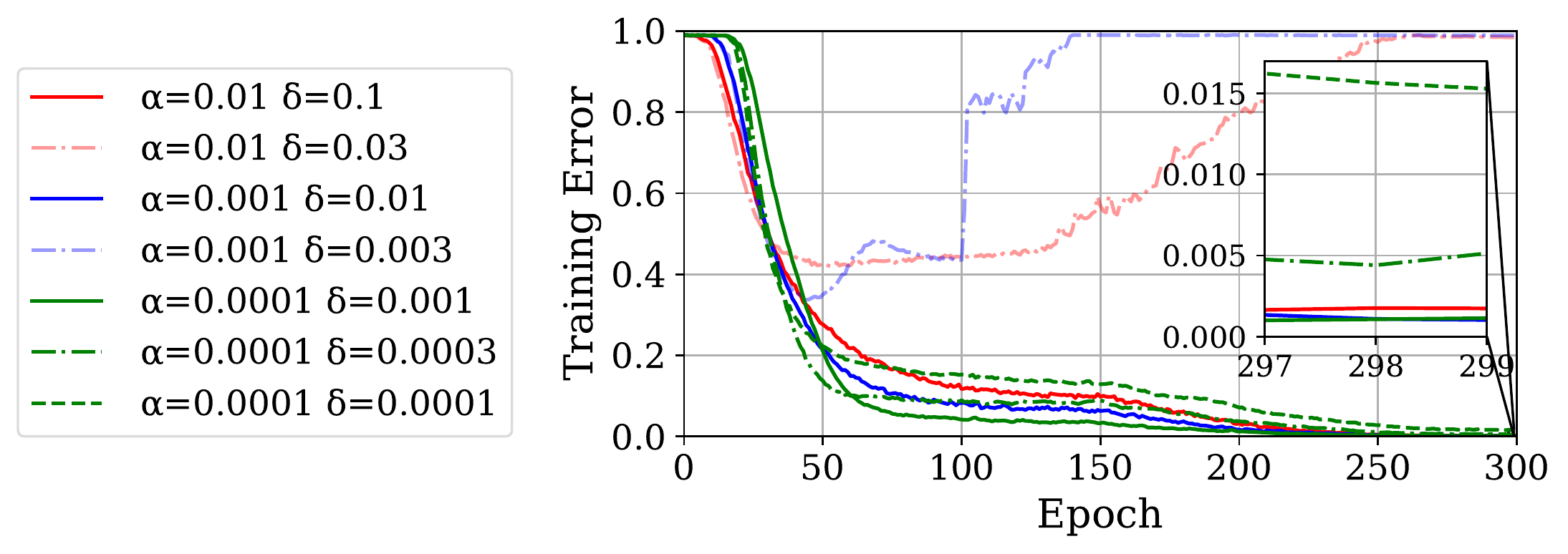}
	\end{subfigure}
	\begin{subfigure}[b]{0.39\textwidth}
		\includegraphics[width=\textwidth]{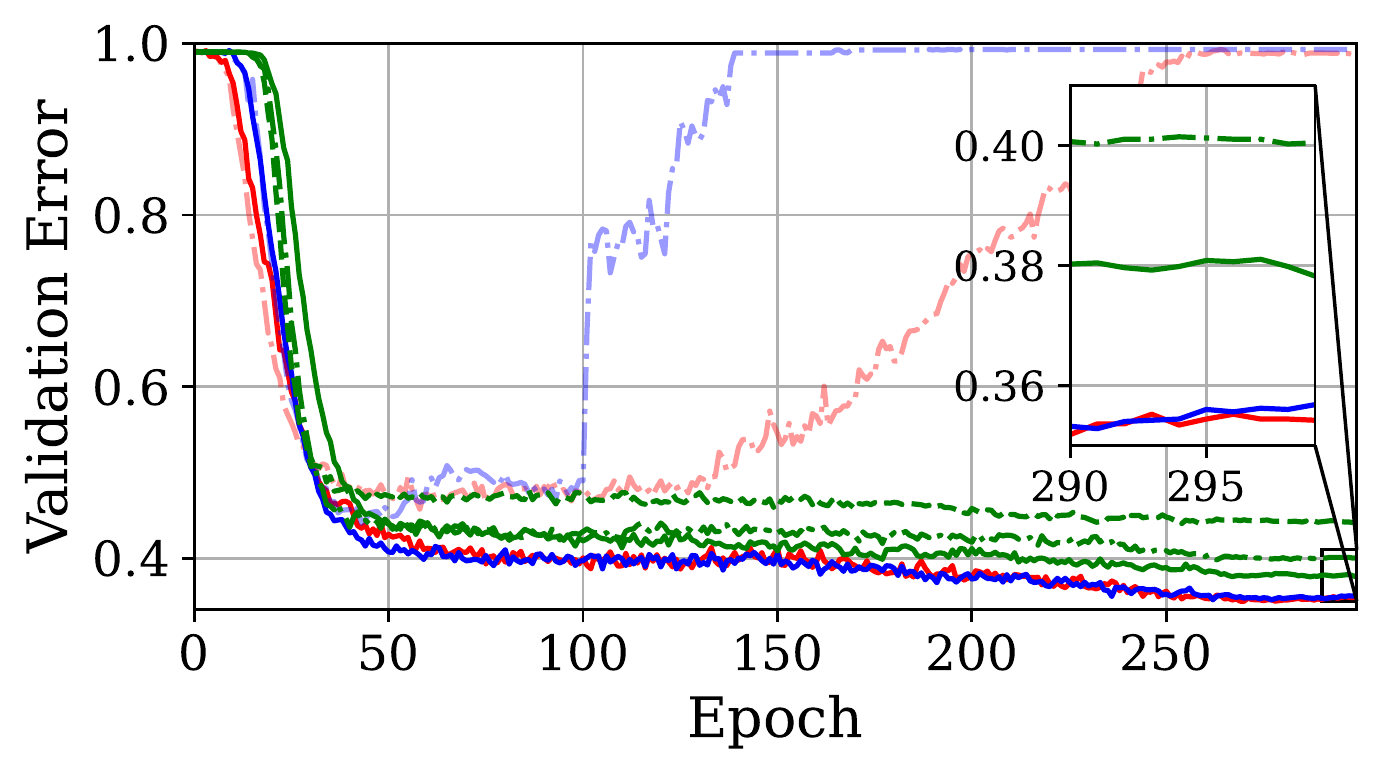}
	\end{subfigure}
	\vspace{-12pt}
	\caption{Training/validation error of the KFAC optimiser for VGG-$16$ on the CIFAR-$100$ dataset with various learning rates $\alpha$ and damping values, $\delta$. The same colour denotes the same learning rate, increasing levels of dashedness denote an ever decreasing damping value.}
	\label{fig:kfaclrandeps}
\end{figure*}

By decreasing the global learning rate $\alpha$ whilst keeping the damping-to-learning-rate ratio $\kappa = \frac{\delta}{\alpha}$ constant, we increase the $\reff$, $\reff$, which is determined by $\frac{\lambda_{i}}{\kappa \alpha}+1$. As shown in Tab.~\ref{tab:tableofspectralnorm} and in Figure \ref{fig:kfaclrandeps} we observe that as we increase $\reff$ the training performance is effectively unchanged, but generalisation suffers ($35\%\rightarrow37.8\%$). 
Whilst decreasing the damping results in poor training for large learning rates, for very low learning rates the network efficiently trains with a lower damping coefficient. Such regimes further increase $\reff$ and we observe that they generalise more poorly. For $\alpha = 0.0001$ dropping the damping coefficient $\delta$ from $0.0003$ to $0.0001$ drops the generalisation further to $60.2\%$ and then $56\%$ respectively. Similar to logistic regression, for both cases the drop in generalisation is significantly larger than the drop in training accuracy. 

\begin{figure*}[h!]
	\centering
	\begin{subfigure}[b]{0.57\textwidth}
		\includegraphics[width=\textwidth]{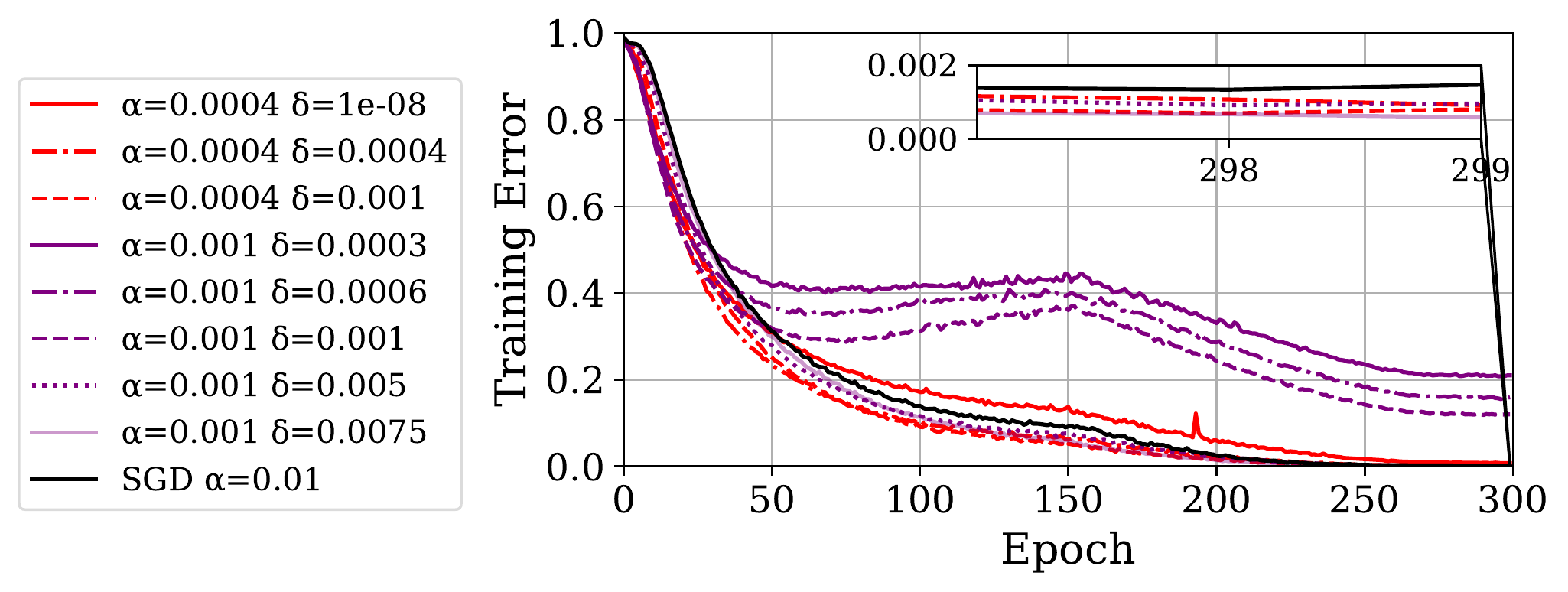}
	\end{subfigure}
	\begin{subfigure}[b]{0.40\textwidth}
		\includegraphics[width=\textwidth]{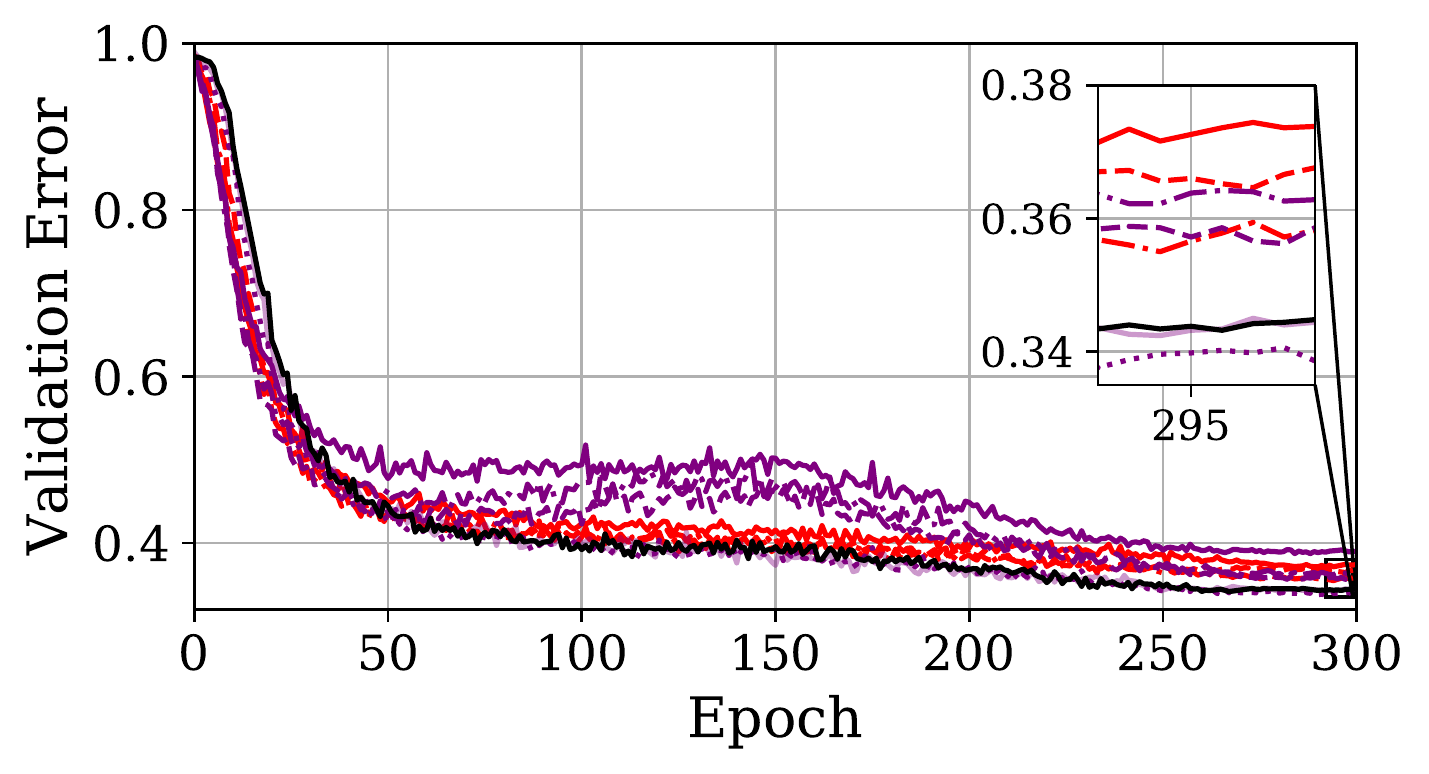}
	\end{subfigure}
	\vspace{-5pt}
	\caption{Training/validation error of the Adam optimiser for VGG-$16$ on the CIFAR-$100$ dataset with various learning rates $\alpha$ and damping values, $\delta$. The same colour denotes the same learning rate, increasing levels of dashedness denote an ever increasing damping value.}
	\label{fig:adamlrandeps}
\end{figure*}

\paragraph{Adam with VGG-16 on CIFAR-100:}
\label{subsec:adamexp}
We employ Adam with a variety of learning rate and damping coefficients with results as shown in Tab.~\ref{tab:tableofspectralnorm} and in Figure \ref{fig:adamlrandeps} and compare against a baseline SGD with $\alpha = 0.01$ (corresponding to optimal performance). For the largest learning rate with which Adam trains ($\alpha = 0.0004$) with the standard damping coefficient $\delta = 10^{-8}$, we see that Adam under-performs SGD, but that this gap is reduced by simply increasing the damping coefficient without harming performance. Over-damping decreases the performance.  For larger global learning rates enabled by a significantly larger than default damping parameter, when the damping is set too low, the training is unstable (corresponding to the dotted lines). Nevertheless, many of these curves with poor training out-perform the traditional setting on testing. We find that for larger damping coefficients $\delta = 0.005, 0.0075$ Adam is able to match or even beat the SGD baseline, whilst converging faster. We show that this effect is statistically significant in Tab.~\ref{tab:seeds}. This provides further evidence that for real problems of interest, adaptive methods are not worse than their non-adaptive counterparts as argued by \cite{wilson2017marginal}. We note as shown in Tab.~\ref{tab:tableofspectralnorm}, that whilst increasing $\delta$ always leads to smaller spectral norm, this does not always coincide with better generalisation performance. We extend this experimental setup to include both batch normalisation \cite{ioffe2015batch} and decoupled weight decay \cite{loshchilov2018decoupled}. We use a learning rate of $0.001$ and a decoupled weight decay of $[0,0.25]$. For this experiment  using a larger damping constant slightly assists training and improves generalisation, both with and without weight decay.

\label{sec:bnexp}
\begin{figure*}[h!]
	\centering
	\begin{subfigure}[b]{0.61\textwidth}
		\includegraphics[width=\textwidth]{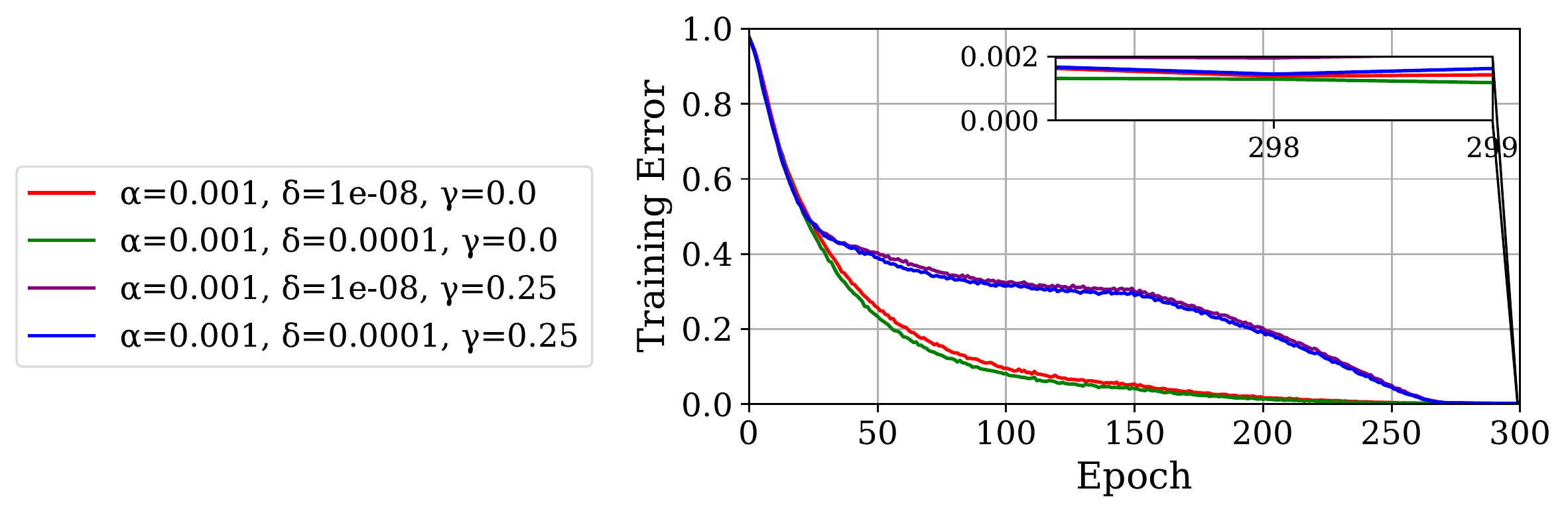}
	\end{subfigure}
	\begin{subfigure}[b]{0.37\textwidth}
		\includegraphics[width=\textwidth]{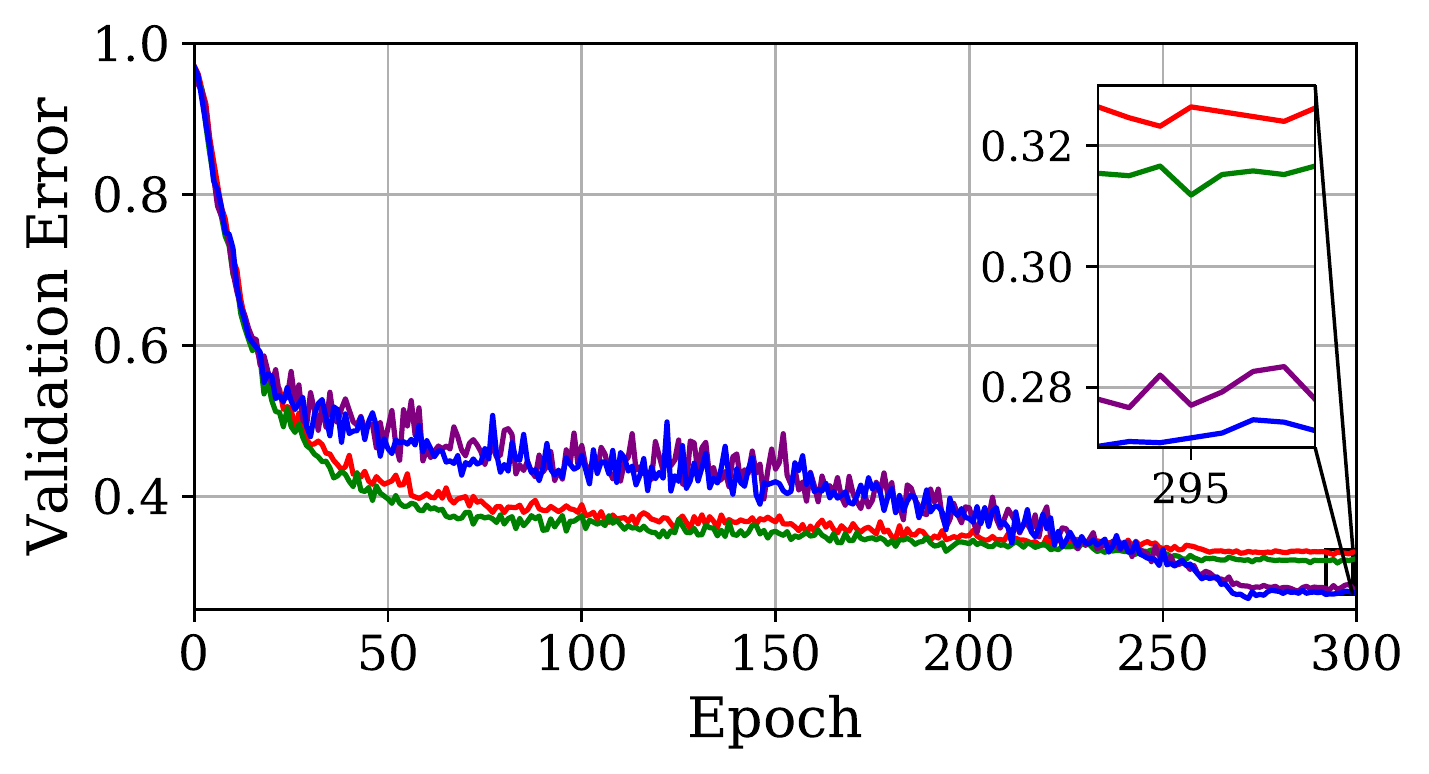}
	\end{subfigure}
	\caption{Training/validation error of the Adam optimiser for VGG-$16$BN using Batch Normalisation and Decoupled Weight Decay on the CIFAR-$100$ dataset with various learning rates $\alpha$ and damping values, $\delta$.}
	\label{fig:adambnlrandeps}
\end{figure*}

\paragraph{ResNet-$50$ ImageNet.}

\begin{figure}
\begin{minipage}[t]{1\textwidth}
		\centering
		\begin{subfigure}[b]{0.49\textwidth}
			\includegraphics[width=\textwidth]{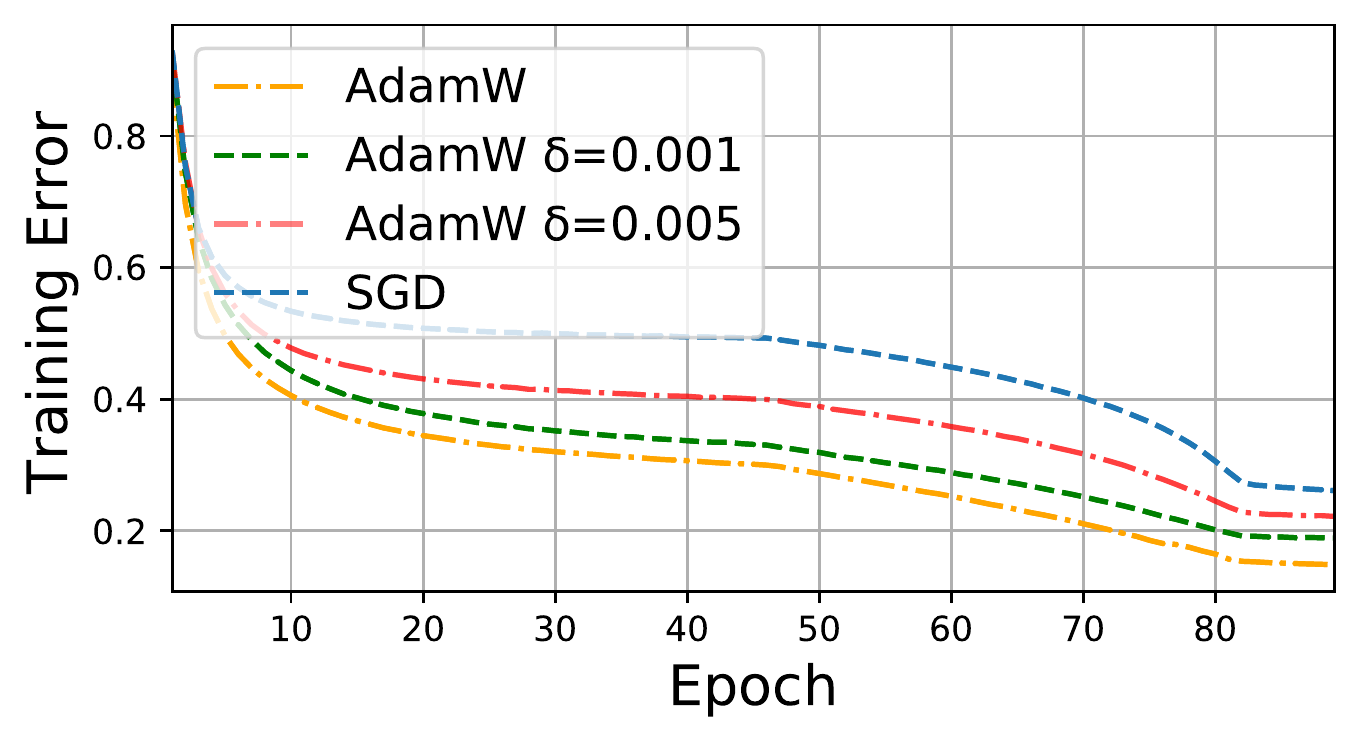}
			\caption{ResNet-$50$ Training Error}
			\label{subfig:r50train}
		\end{subfigure}
		\begin{subfigure}[b]{0.49\textwidth}
			\includegraphics[width=\textwidth]{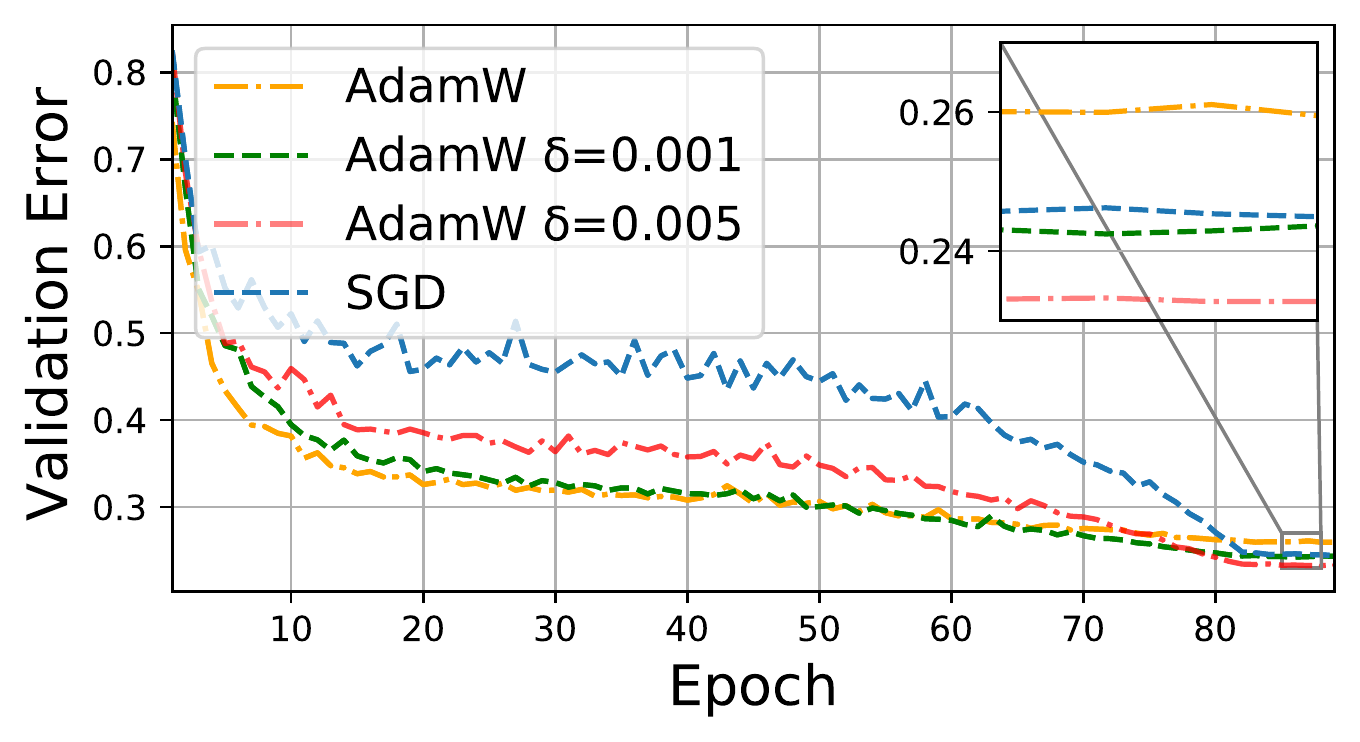}
			\caption{ResNet-$50$ Testing Error}
			\label{subfig:r50test}
		\end{subfigure}
\end{minipage}
\caption{(a-b) The influence of $\delta$ on the generalisation gap. Train/Val curves for ResNet-$50$ on ImageNet. The generalisation gap is completely closed with an appropriate choice of $\delta$.}
\label{fig:res50adamw}
\end{figure}

As shown in Figure \ref{subfig:r50train},\ref{subfig:r50test}, these procedures have practical impact on large scale problems. Here we show that under a typical $90$ epoch ImageNet setup~\cite{he2016deep}, with decoupled weight decay $0.01$ for AdamW and $0.0001$ for SGD, that by increasing the numerical stability constant $\delta$ the generalisation performance can match and even surpass that of SGD, which is considered state-of-the-art and beats AdamW without $\delta$ tuning by a significant margin.

\begin{table}[h]
		\begin{tabular}{@{}llllll@{}}
		\toprule
		\textbf{Dataset} & \textbf{Classes}& \textbf{Model Architecture} & \textbf{SGD} & \textbf{Adam-D} & \textbf{Adam} \\ \midrule
		CIFAR-100 & 100 & VGG16  & 65.3 $\pm$ 0.6 & 65.5 $\pm$ 0.7 & 61.9 $\pm$ 0.4 \\
		\midrule
		
		ImageNet & 1000& ResNet50 & 75.7 $\pm$ 0.1 & 76.6 $\pm$ 0.1 & 74.04* \\ \bottomrule
	\end{tabular}
	\vspace{15pt}
	\caption{Statistical Significance. Comparison of test accuracy across CIFAR 100 (5 seeds) and ImageNet (3 seeds). \textbf{Adam-D} denotes Adam with increased damping ($\delta=5e^{-3}$ for CIFAR-100, $\delta=1e^{-4}$ for ImageNet). *Since it is well established that Vanilla Adam does not generalise well for ImageNet, we do not run this experiment for multiple seeds, we simply report a single seed result for completeness. A more complete discussion for Adam and its generalisation in vanilla form can be found in \cite{granziol2020iterate}.}
	\label{tab:seeds}
\end{table}

\section{Optimal adaptive damping from random matrix theory} \label{sec:adaptive}

Recall the scaling applied in the direction of the $i^{\mathrm{th}}$ eigenvector in (\ref{eq:secondorderopt}). We make the following observation
\begin{equation}
	\begin{aligned}
		\label{eq:diegosderivation}
		& \frac{1}{\lambda_{i}+\delta} = \frac{1}{\beta\lambda_{i}+(1-\beta)}\cdot\frac{1}{\kappa} \\
	\end{aligned}
\end{equation}
where $\kappa = \beta^{-1}, \beta = (1+\delta)^{-1}$. Hence, using a damping $\delta$ is formally equivalent to applying linear shrinkage with factor $\beta=(1+\delta)^{-1}$ to the estimated Hessian and using a learning rate of $\alpha\beta$.
Shrinkage estimators are widely used in finance and data science, with linear shrinkage being a common simple method applied to improve covariance matrix estimation \cite{ledoit2004well}.
The practice of shrinking the eigenvalues while leaving the eigenvectors unchanged is well-established in the fields of sparse component analysis and finance \cite{bun2017cleaning}. 
In the shrinkage literature, the typically considered models are additive and multiplicative \cite{potters2020first}, i.e.
\begin{align*}
    \mE = \mC + \mX, ~~~ \mE = \mC^{1/2}\mX \mC^{1/2}
\end{align*}
where $\mE$ is the observed matrix, $\mC$ is the non-corrupted (or signal) matrix, and $\mX$ is the noise matrix.
White Wishart $\mX$ is the simplest example in the multiplicative case, and Wigner matrices are the simplest choice in the additive case.
In generality, shrinkage estimators are estimators of $\mC$ given $\mE$, and it is common to consider rotationally invariant (or, more precisely, equivariant) estimators which reduce the problem to computing the eigenvalues and eigenvectors of $\mE$ and then correcting, or \emph{shrinking}, the eigenvalues \vivacom{while} keeping the eigenvectors fixed to obtain improved estimation of $\mC$.
Optimal\footnote{Optimality is commonly defined in terms of Frobenuis norm, but some authors have considered the \emph{minimum variance} loss \cite{ledoit2020analytical}.} estimators are constructed in \cite{bun2016rotational, ledoit2011eigenvectors, ledoit2012nonlinear} and most recently \cite{ledoit2020analytical}.
We note in passing that such estimators are only possible in the large matrix limit, where functions of the inaccessible matrix $\mC$ can be replaced by equivalent quantities depending only on $\mE$.
The optimal shrinkage estimators are generally non-linear functions of the eigenvalues of $\mE$ and depend on integral transforms of the limiting spectral measure of $\mE$ and also on the noise matrix $\mX$.
In some very special cases, the optimal shrinkage estimators simplify greatly, for example, in the multiplicative case, if $\mC$ is an inverse Wishart matrix, the linear shrinkage estimator $\Tilde{\mH} = \beta \mH + (1-\beta)\mI = \argmin_{\mH^*}||\mH^*-\mH_{\mathrm{true}}||_2$ is optimal and an explicit expression for the optimal $\beta$ is found depending only on the dimensionality of the model and the noise variance \cite{ledoit2004well,bun2016rotational}.

In our optimisation context, the additive noise model is perhaps the most natural with $\mC$ being the true loss Hessian and $\mE$ the batch loss Hessian, however we cannot expect any special forms on $\mX$ or $\mC$ that will produce closed form expressions for the optimal rotational invariant estimator and the linear shrinkage estimator is almost certainly not optimal.
We suggest that there is no particular reason to break with rotational invariance in this work, as intuitively any distinguished directions of $H_{\text{batch}}$ are those of $H_{\text{true}}$.
However linear shrinkage has the great advantage of being simple to integrate into existing adaptive optimisers and it acts intuitively to reduce the movement of the optimiser in pure-noise directions.
In fact, it is known that general non-linear shrinkage estimators retain the property of increasing the smallest eigenvalues and decreasing the largest \cite{ledoit2012nonlinear}.
Our interpretation reveals that the damping parameter should not be viewed as a mere numerical convenience to mollify the effect of very small estimate eigenvalues, but rather that an optimal $\delta$ should be expected, representing the best linear approximation to the true Hessian and an optimal balancing of variance (the empirical Hessian) and bias (the identity matrix). This optimal choice of $\delta$ will produce an optimiser that more accurately descends the directions of the true loss.

The linear shrinkage interpretation given by (\ref{eq:diegosderivation}) is an elementary algebraic relation but does not by itself establish any meaningful link between damping of adaptive optimisers and linear shrinkage estimators. To that end, we return to the random matrix model (\ref{eq:additive_noise}) for the estimated Hessian:
Let us write the Hessian as \begin{align*}
    \mH_{\text{batch}} = \mH_{\text{true}} + \mX
\end{align*}
where $\mX$ is a random matrix with $\mathbb{E}\mX = 0$. Note that this model is entirely general, we have simply defined $\mX = \mH_{\text{batch}}  - \mathbb{E}\mH_{\text{batch}} $ and $\mathbb{E}\mH_{\text{batch}}  = \mH_{\text{true}}$. We then seek a linear shrinkage estimator $\tilde{\mH}(\beta) = \beta \mH_{\text{batch}} + (1-\beta)\mI$ such that $E(\beta)= P^{-1}\Tr (\tilde{\mH} - \mH_{\text{true}})^2$ is minimised. Note that this is the same objective optimised by \cite{bun2016rotational} to obtain optimal estimators for various models. In this context, we are not finding the optimal estimator for $\mH_{\text{true}}$ but rather the optimal \emph{linear shrinkage} estimator. We have \begin{align*}
    E(\beta) = \frac{1}{P}\Tr\left[(\beta-1) \mH_{\text{true}} + \beta\mX + (1-\beta)\mI\right]^2 \equiv \frac{1}{P}\Tr \left[  (\beta - 1)\mH_{\text{true}}+ \mY_{\beta}\right]^2
\end{align*}
where $\mY_{\beta} = \beta\mX + (1-\beta)\mI$.

A natural assumption in the case of deep learning is that $\mH_{\text{true}}$ is low-rank, i.e. for $P\rightarrow\infty$ either $\text{rank}(\mH_{\text{true}}) = r$ is fixed or  $\text{rank}(\mH_{\text{true}}) = o(P)$. Empirical evidence for this assumption is found in \cite{granziol2020learning,sagun2016eigenvalues,sagun2017empirical,papyan2018full,ghorbani2019investigation}. In this case the bulk of the spectrum of $\mY_{\beta}$ is the same as that of $(\beta-1) \mH_{\text{true}} + \mY_{\beta}$ \cite{benaych2011eigenvalues,capitaine2016spectrum,belinschi2017outliers}. We will also assume that $\mX$ admits a deterministic limiting spectral measure $\mu_X$ such that \begin{align}
    \frac{1}{P}\sum_{j=1}^P \delta_{\lambda(X)_i} \rightarrow \mu
\end{align}
weakly almost surely. Say $\omega_X(x) dx = d\mu(x)$. Then $\mY_{\beta}$ has limiting spectral density \begin{align*}
    \omega_Y(y) = \beta^{-1}\omega_X(\beta^{-1}(y - 1 + \beta)).
\end{align*}
Then for large $P$ 
\begin{align*}
    E(\beta) \approx \beta^{-1}\int y^2 \omega_X(\beta^{-1}(y - 1 + \beta)) ~dy &=  \int (\beta x + 1 - \beta)^2 \omega_X(x) ~ dx \\
    &=  \beta^2 \mu_X(x^2) + (1-\beta)^2
\end{align*}
as the centred assumption on $\mX$ means that $\int x\omega_X(x) ~dx = 0$. $\mu_X(x^2)$ is shorthand for $\int x^2\omega_X(x)~ dx$. $E(\beta)$ is thus minimised to leading order at $\beta = (1 + \mu_X(x^2))^{-1}$. Recalling that $\beta^{-1} = (1+\delta)^{-1}$, this yields $\delta = \mu_X(x^2)$ i.e. the optimal level of damping at large finite $P$ is approximately
\begin{align}\label{eq:finaloptdamp}
\delta=P^{-1}\Tr \mX^2.
\end{align}
Note that the value (\ref{eq:finaloptdamp}) is a very natural measure of the Hessian noise variance. Therefore if the random matrix model described above is appropriate and the linear shrinkage interpretation (\ref{eq:diegosderivation}) is meaningful we should expect it to result in close to optimal performance of a given adaptive optimiser. The purpose of adaptive optimisers is to accelerate training, in part by allowing for larger stable learning rates. As discussed throughout this chapter, such optimisation speed often comes at the cost of degraded generalisation. In this context, `optimal performance' of adaptive optimisers should be taken to mean fast training and good generalisation. 
As we have discussed above, very large values of $\delta$ recover simple non-adaptive SGD, so using (\ref{eq:finaloptdamp}) we should be able to obtain generalisation performance at least as good as SGD and faster optimisation than any choice of $\delta$ including the default very small values often used and the larger values considered in Section \ref{sec:nnexperiments}. 

The value of (\ref{eq:finaloptdamp}) can be easily learned by estimating the variance of the Hessian. The Hessian itself cannot be computed exactly, as it is far too large for $P \geq O(10^7)$, however one can compute $\mH \vv$ (and hence $\mH^2 \vv$) for any vector $\vv$, using $\nabla^2 L \vv = \nabla (\vv^T\nabla L)$. The full approach is given in Algorithm \ref{alg:hessvar}.
\begin{algorithm}[H]
	\begin{algorithmic}[1]
		\STATE {\bfseries Input:} Sample Hessians $\mH_{i}\in \mathbb{R}^{P\times P}$, $1\leq i < N$
		\STATE {\bfseries Output:} Hessian Variance $ \sigma^{2}$
		\STATE $\vv \in \mathbb{R}^{1\times P} \sim \mathcal{N}(\boldsymbol{0}, \mI)$
		\STATE Initialise $\sigma^{2}=0, i = 0$, $\vv \leftarrow \vv/||\vv||$
		\FOR{$i < N$}
		\STATE $\sigma^{2} \leftarrow \sigma^{2} + \vv^{T}\mH_{i}^{2}\vv$
		\STATE $i \leftarrow i + 1$
		\ENDFOR
		\STATE $\sigma^{2} \leftarrow \sigma^{2} - [\vv^{T}(1/N\sum_{j=1}^{N}\mH_{j})\vv]^{2}$
	\end{algorithmic}
	\caption{Algorithm to estimate the Hessian variance}
	\label{alg:hessvar}
\end{algorithm}

\paragraph{Extension to non-linear shrinkage.} If, as we demonstrate below, our interpretation of damping as linear shrinkage is meaningful, it is natural to ask if we can replace linear shrinkage with more general non-linear shrinkage, effectively defining new adaptive optimisers that replace $\lambda_i + \delta$ in (\ref{eq:secondorderopt}) by $f(\lambda_i)$ for some non-linear $f$.
Indeed, non-linear shrinkage is known to outperform linear shrinkage in general \cite{ledoit2012nonlinear, ledoit2020analytical}, so we should expect to see further improvements beyond our optimal damping approach, but there are substantial obstacles to progress in this direction.
Absent the strongly simplifying assumptions that lead to linear shrinkage, one must handle integral transforms of the spectral density of $\mH_{\text{batch}}$ to compute general non-linear shrinkage estimators.
There are various approaches sin the literature that make use of parametric and kernel estimation fits to these transforms or the spectral density itself \cite{potters2020first,ledoit2012nonlinear,ledoit2020analytical} and there are simpler approaches that use cross-validation to construct improved estimators of the true eigenvalues-eigenvector pairs \cite{abadir2014design}.
It is, however, observed by Ledoit and Wolf \cite{ledoit2020analytical} that these methods are infeasible for matrices larger than around $1000\times 1000$.
Ledoit and Wolf \cite{ledoit2020analytical} propose a new, non-parametric non-linear shrinkage estimator that is quite conceptually simple to implement and can scale to larger matrices, but careful inspection reveals that the required computation time for each shrinkage evaluation is nevertheless $O(P^2)$, where in our case $P$ is on the order of $10^7$, so even this approach is infeasible.

\subsection{Experimental Design and Implementation Details}

In order to test our hypothesis for the derived optimal $\delta$ (\ref{eq:finaloptdamp}), we run the classical VGG network \cite{simonyan2014very} with $16$ layers on the CIFAR-$100$ dataset, without weight decay or batch normalisation. This gives us maximal sensitivity to the choice of learning rate and appropriate damping. 

\medskip
Now in practice the damping coefficient is typically grid searched over several runs \cite{dauphin2014identifying} or there are heuristics such as the Levenberg–Marquardt to adapt the damping coefficient \cite{martens2015optimizing}, which however we find does not give stable training for the VGG. We hence compare against a fixed set damping value $\delta$ and a learned damping value as given by our equation (\ref{eq:finaloptdamp}). We find that the variance of the Hessian (\ref{eq:finaloptdamp}) at a random point in weight space (such as at initialisation) or once network divergence has occurred is zero, hence the initial starting value cannot be learned as, with a damping of near zero, the network entirely fails to train (no change in training loss from random). This is to be expected, as in this case the local quadratic approximation to the loss inherent in adaptive methods breaks down. Hence we initialise the learning algorithm with some starting value $\delta^{*}$, which is then updated every $100$ training iterations using equation (\ref{eq:finaloptdamp}). Strictly speaking we should update every iteration, but the value of $100$ is chosen arbitrarily as a computational efficiency. Since we are using the variance of the Hessian, which is expensive to compute compared to a simple gradient calculation, we do not want to compute this quantity too often if it can be helped. We run our experiments on a logarithmic grid search in near factors of $3$. So learning rates and damping rates, either flat or learned are on the grid of $0.0001,0.0003,0.001...$.

We find under this setup that the time taken per epoch against the flat damping schedule is only doubled. We get identical results for using a damping gap of $10$ and so do not consider this to be a very relevant hyper-parameter. We further calculate the variance of the Hessian over a sub-sample of $10000$ examples and do not calculate the variance sample by sample, but over batches of $128$ to speed up the implementation. Under the assumption that the data is drawn i.i.d from the dataset the variance is simply reduced by a factor $(\frac{1}{B}-\frac{1}{N}) \approx \frac{1}{B}$ for a small batch size. We do not consider the impact of using only a sub-sample of the data for estimation, but we expect similar results to hold compared to the entire dataset as long as the sub-sample size $S\gg B$. This should allow such a method to be used even for very large datasets, such as ImageNet (with $1$-million images), for which a pass of the entire dataset is extremely costly. In theory the sub-sample size and mini-batch size for Hessian variance estimation could be two hyper-parameters which are tuned by considering the effect of reduction on training set or validation set loss metrics with the trade off for computational cost. We do not conduct such analysis here.

We also incorporate an exponential moving average into the learned damping with a co-efficient of $0.7$\footnote{This value is not tuned and in fact from our plots it may be advisable to consider higher values for greater stability} to increase the stability of the learned damping.

\subsection{Experiment on CIFAR-100 using KFAC to validate the optimal linear shrinkage}

\begin{figure}[h]
	\centering
	\begin{subfigure}[b]{0.49\linewidth}
		\includegraphics[trim=0cm 0 0 0,clip,width=\textwidth]{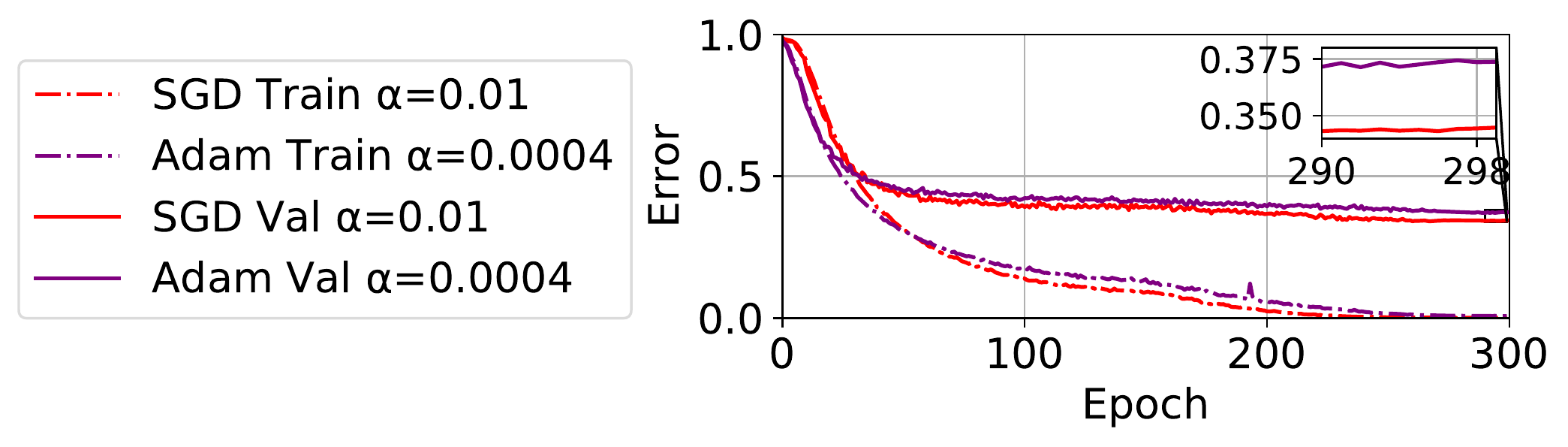}
		\vspace{-15pt}
		\caption{SGD quickly outgeneralises Adam}
		\label{subfig:adamvssgd}
	\end{subfigure}
	\begin{subfigure}[b]{0.49\linewidth}
		\includegraphics[trim=0cm 0 0 0,clip, width=\textwidth]{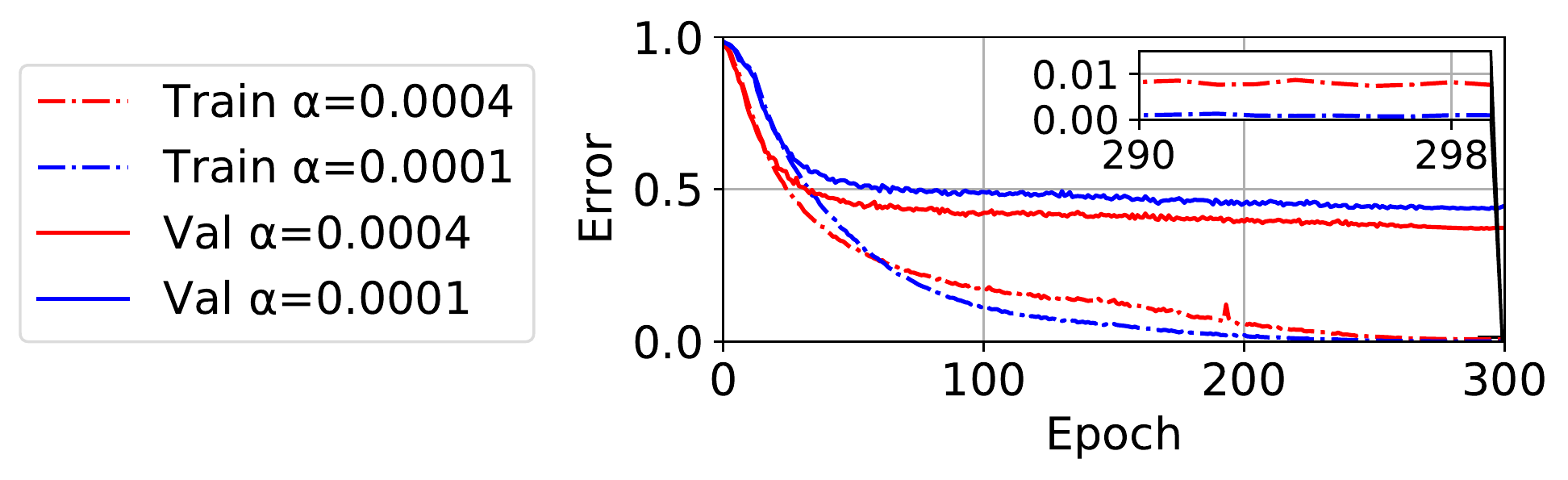}
		\vspace{-15pt}
		\caption{Adam Train/Val for Learning Rates $\{\alpha_{i}\}$}
		\label{subfig:adamlr}
	\end{subfigure}
	\vspace{0pt}
	\caption{\textbf{Adaptive Generalisation Gap and its extent are clearly visible without regularisation.} Train/Val Error on CIFAR-$100$ using VGG-$16$ without batch normalisation and weight decay.}
	\label{fig:adamsgd}
\end{figure}

\medskip
For large damping values $\delta$ we simply revert to SGD with learning rate $\alpha/\delta$, so we follow the typical practice of second order methods and use a small learning rate and correspondingly small damping coefficient.  However as shown in Figure \ref{fig:adamsgd} the generalisation and optimisation are heavily dependent on the global learning rate, with larger learning rates often optimising less well but generalising better and vice versa for smaller learning rates. We hence investigate the impact of our damping learner on learning rates one order of magnitude apart. Where in the very low learning rate regime, we show that our method achieves significantly improved training stability with low starting damping and fast convergence and for the large learning rate regime that we even exceed the SGD validation set result.

\section{Previous Work}\label{sec:motivation}

\begin{figure}[!h]
	\begin{subfigure}[b]{0.66\textwidth}
		\includegraphics[width=\textwidth]{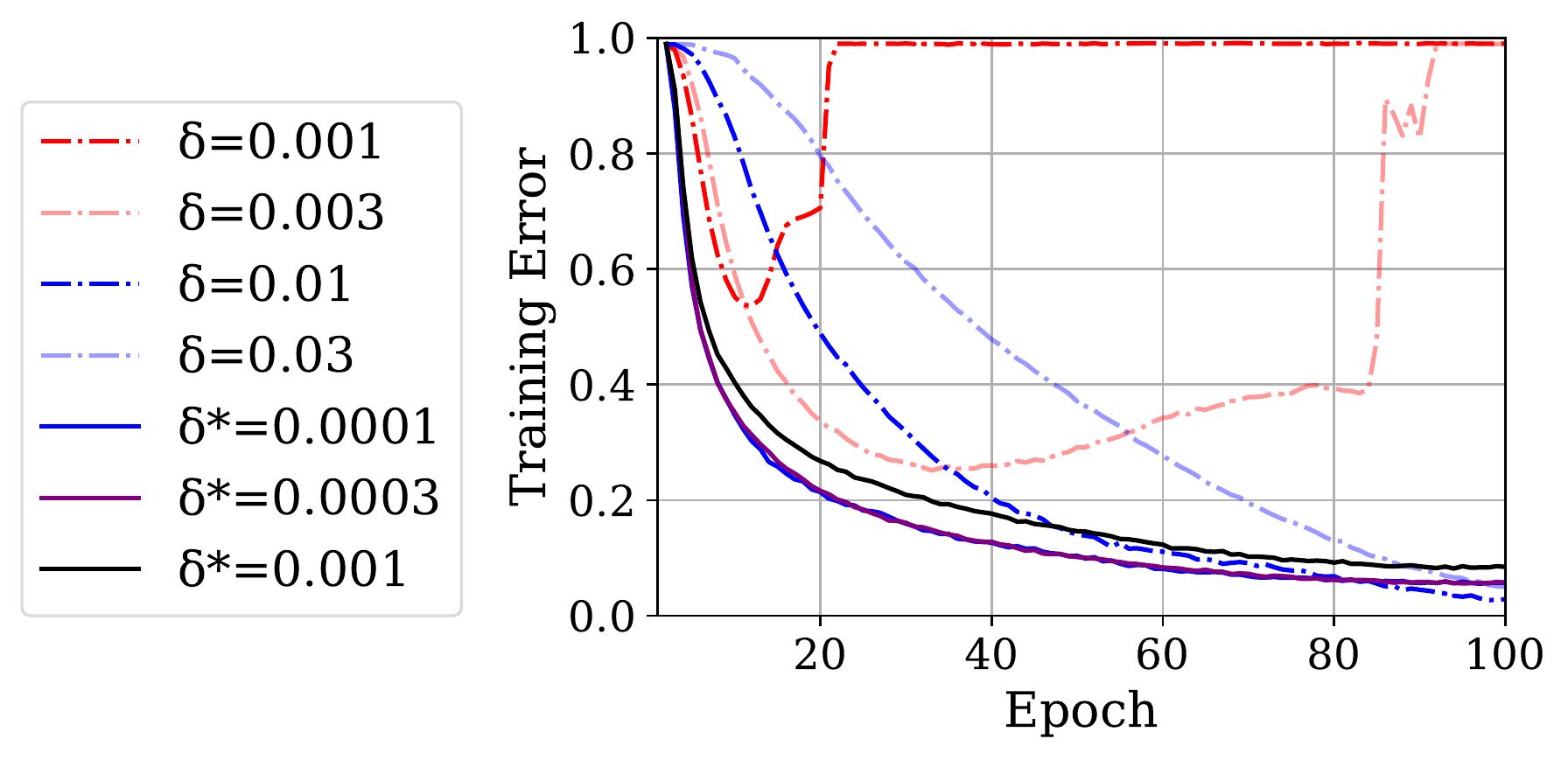}
		\caption{Training Error as a function of Epoch}
		\label{subfig:damperr}
	\end{subfigure}
	\begin{subfigure}[b]{0.33\textwidth}
		\includegraphics[width=\textwidth]{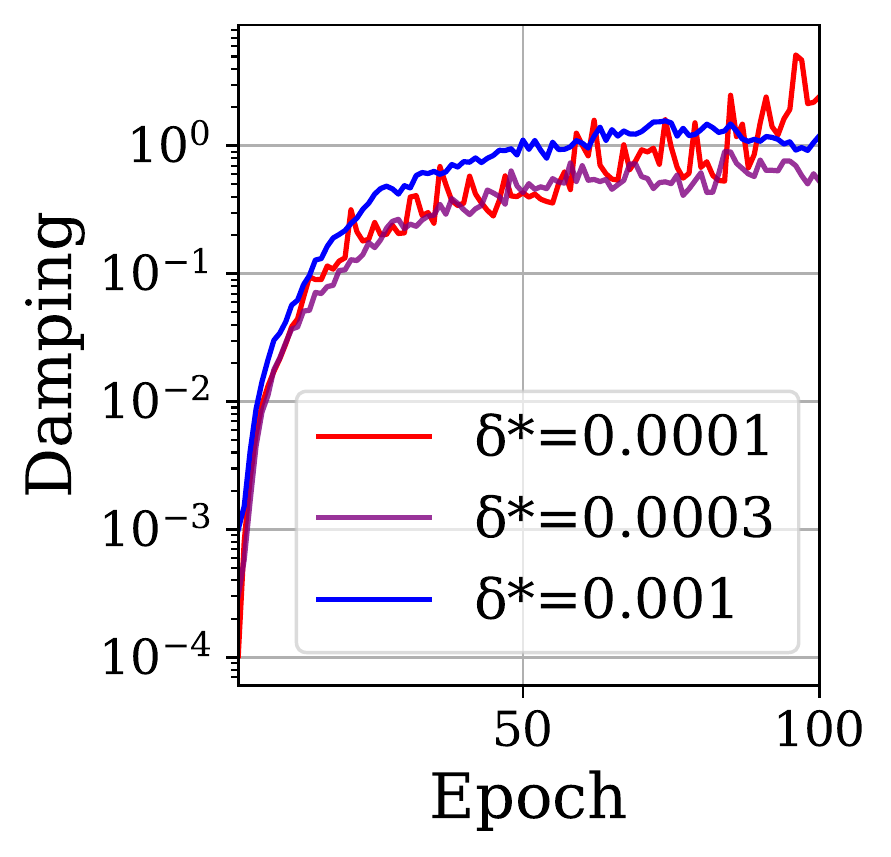}
		\caption{Damping per Epoch}
		\label{subfig:damping}
	\end{subfigure}
	\caption{VGG-$16$ on CIFAR-$100$ dataset using the KFAC optimiser with $\gamma=0$ (no weight decay) for a learning rate of $\alpha=0.0001$, batch size $B=128$ and damping set by $\delta$. For adaptive damping methods the damping is given an initial floor value of $\delta^{*}$ and is then updated using the variance of the Hessian every $100$ steps.}
\end{figure}

\paragraph{Training KFAC with Auto-Damping:}
We show the results for a global learning rate of $0.0001$ in Figure \ref{subfig:damperr}. We see that for the flat damping methods with low values of damping, that training becomes unstable and diverges, despite an initially fast start. Higher damped methods converge, but slowly. In stark contrast, our adaptive damping method is relatively insensitive to their chosen initial values. We show here $\delta^{*} = \alpha,3\alpha,10\alpha$ and all converge and moreover significantly faster than all flat damping methods. The smaller initial damping coefficients $\delta = \alpha,3\alpha$ converge faster than the larger and, interestingly, follow very similar damping trajectories throughout until the very end of training, as shown in Figure \ref{subfig:damping}.

\begin{figure}[!h]
	\begin{subfigure}[b]{0.58\textwidth}
		\includegraphics[width=\textwidth]{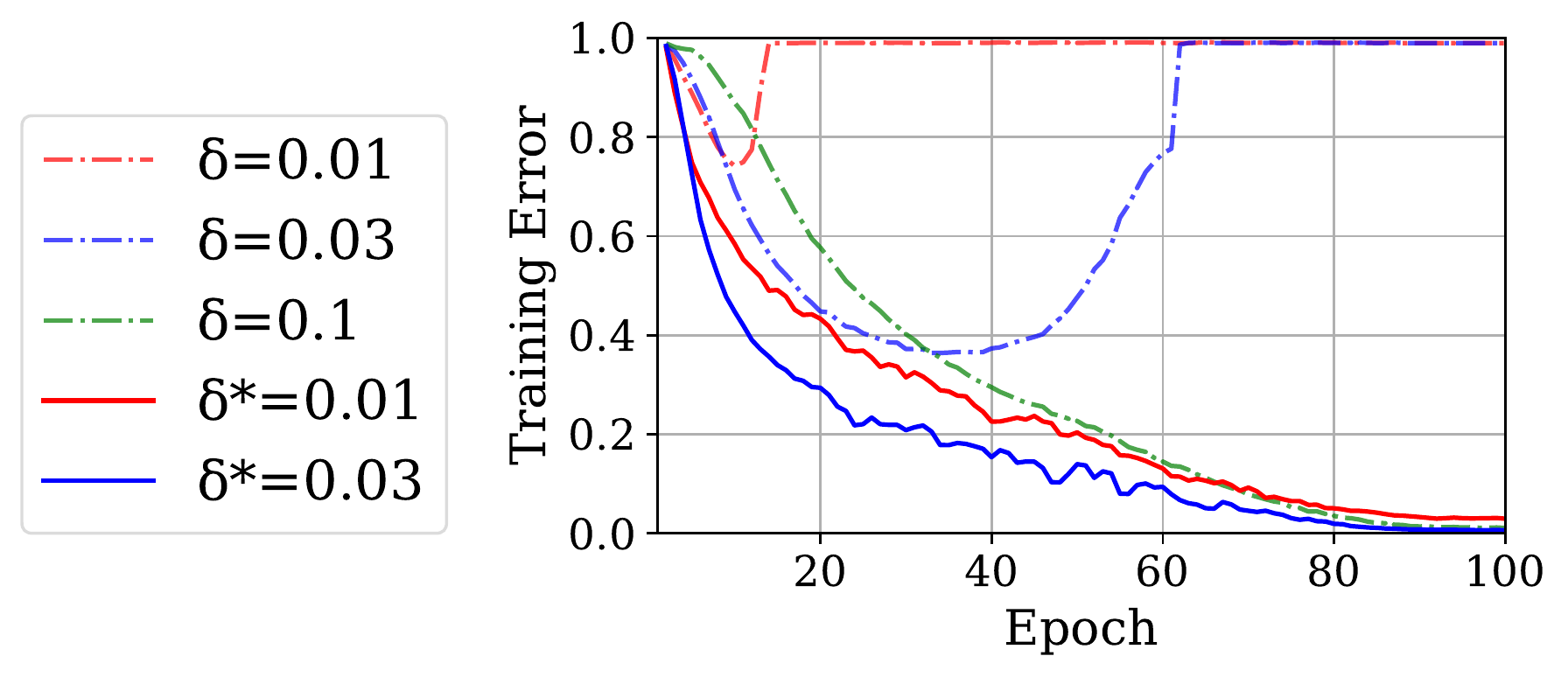}
		\caption{Training Error}
		\label{subfig:trainkfac}
	\end{subfigure}
	\begin{subfigure}[b]{0.41\textwidth}
		\includegraphics[width=\textwidth]{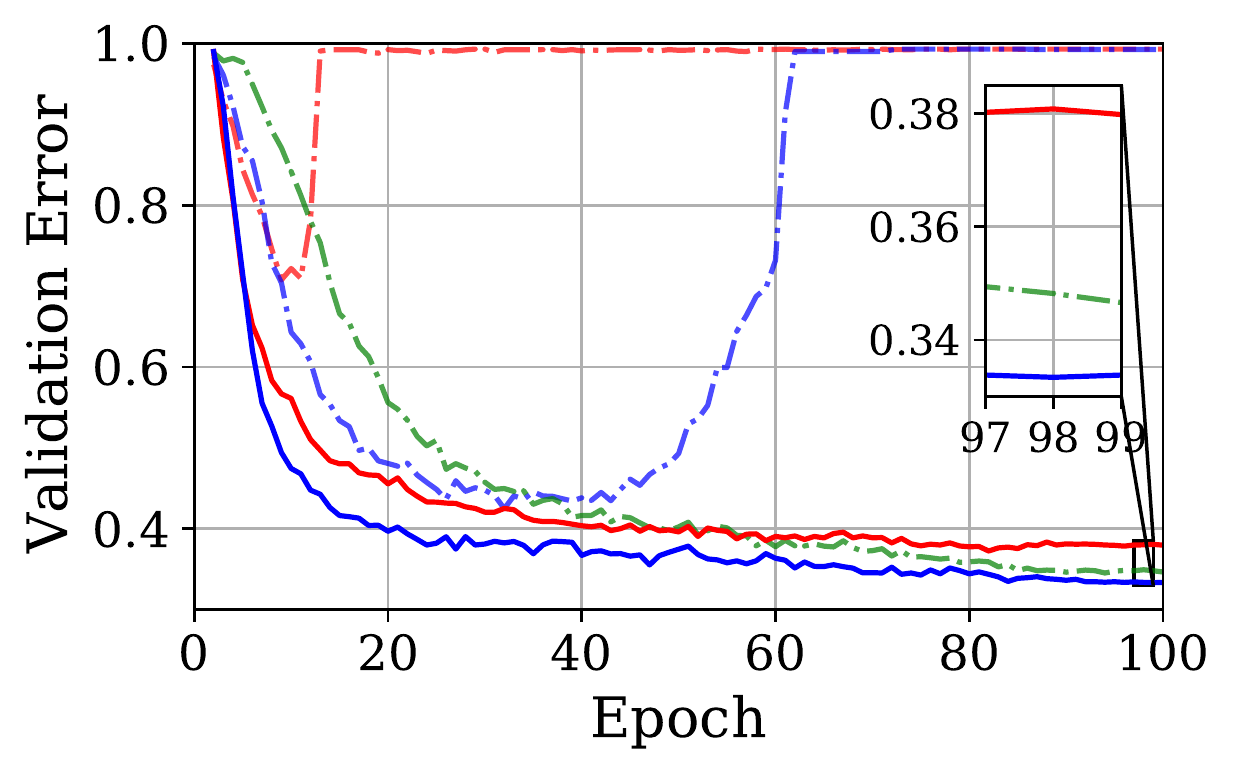}
		\caption{Val Error}
		\label{subfig:valkfac}
	\end{subfigure}
	\caption{VGG-$16$ on CIFAR-$100$ dataset using the KFAC optimiser with $\gamma=0$ (no weight decay) for a learning rate of $\alpha=0.001$, batch size $B=128$ and damping set by $\delta$. For adaptive damping methods the damping is given an initial floor value of $\delta^{*}$ and is then updated using the variance of the Hessian every $100$ steps.}
\end{figure}

\paragraph{Getting Great Generalisation with KFAC and Auto-Damping:}
We similarly train KFAC on the VGG-$16$ with a larger learning rate of $0.001$, in order to achieve better generalisation. Here we see in Figure \ref{subfig:trainkfac} that relatively low values of flat damping such as $0.01$ and  $0.03$ very quickly diverge, whereas a large value of $0.1$ converges slowly to a reasonable test error. The corresponding learned damping curves of $0.01$ and $0.03$ however converge quickly and the $0.03$ initialised damping curve even beats the generalisation performance of the large flat damped version and the test result of SGD on $3$x as many training epochs.
\paragraph{A further look at the value of adaptive damping}
\begin{figure}[!h]
	\begin{subfigure}[b]{0.46\textwidth}
		\includegraphics[width=\textwidth]{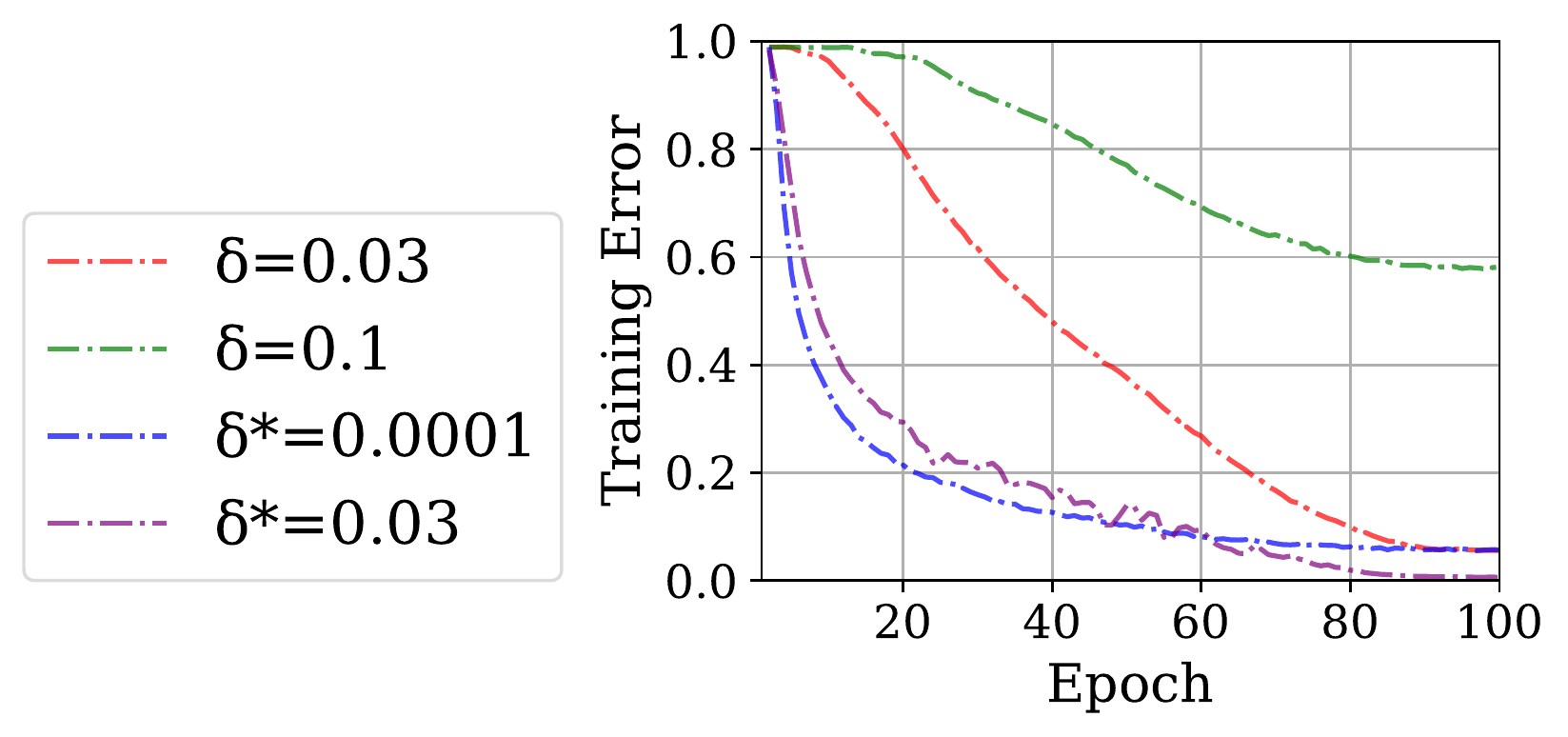}
		\caption{Training Error}
		\label{subfig:trainkfacclean}
	\end{subfigure}
	\begin{subfigure}[b]{0.29\textwidth}
		\includegraphics[width=\textwidth]{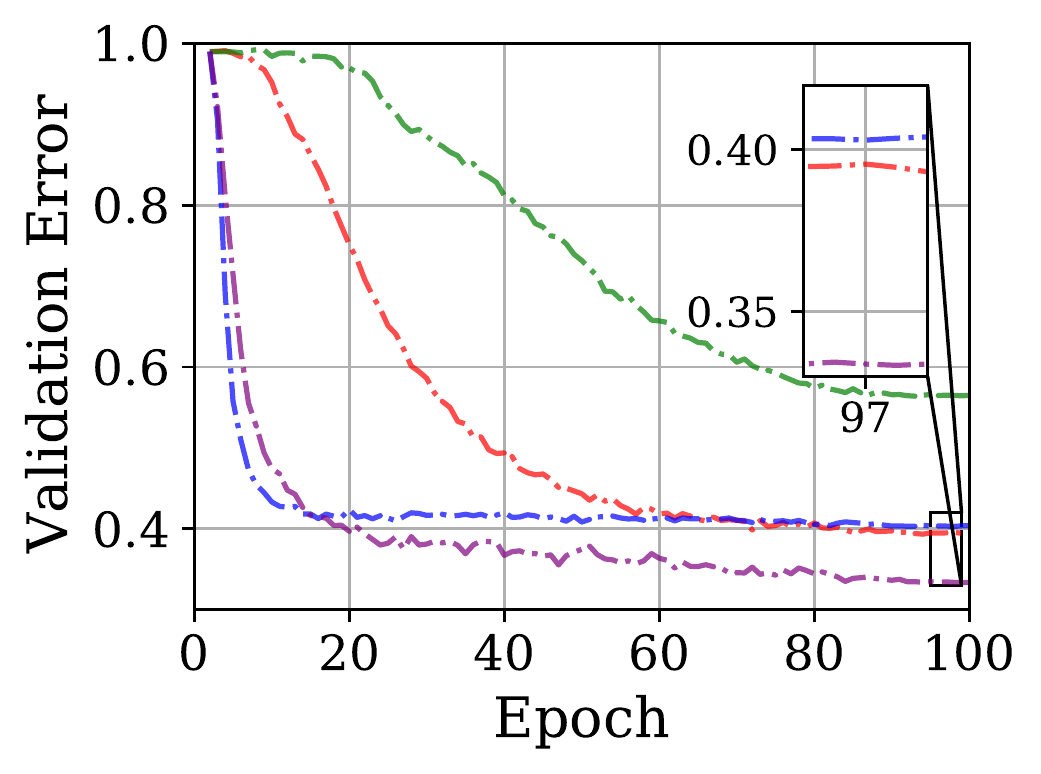}
		\caption{Val Error}
		\label{subfig:valkfacclean}
	\end{subfigure}
	\begin{subfigure}[b]{0.235\textwidth}
		\includegraphics[width=\textwidth]{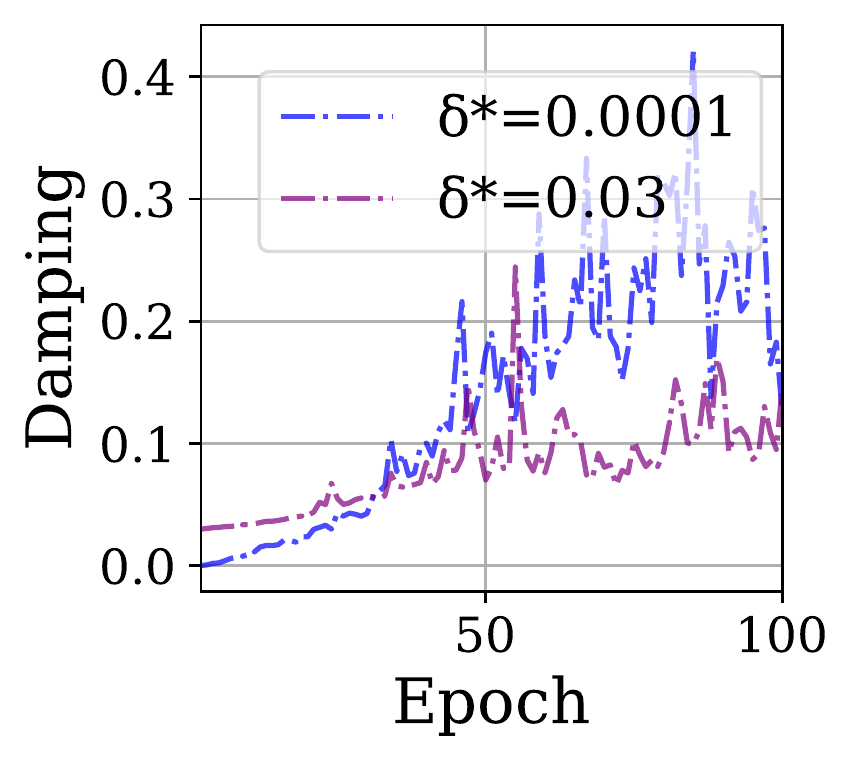}
		\caption{Damping}
		\label{subfig:dampkfac}
	\end{subfigure}
	\caption{VGG-$16$ on CIFAR-$100$ dataset using the KFAC optimiser with $\gamma=0$ (no weight decay) for a learning rate of $\alpha=0.0001$, batch size $B=128$ and damping set by $\delta$. For adaptive damping methods the damping is given an initial floor value of $\delta^{*}$ and is then updated using the variance of the Hessian every $100$ steps.}
	\label{fig:kfacclean}
\end{figure}
To elucidate the impact and workings of the adaptive damping further, we consider a select set of curves the learning rate of $\alpha = 0.0001$, shown in Figure \ref{fig:kfacclean}. here we see that starting with an initial damping of $\delta=\alpha$, the adaptive method reaches a comparable generalisation score to the flat damping of $\delta=0.03$ but at a much faster convergence rate. The initial damping of $\delta=0.03$ converges not quite as quickly but trains and generalises better than its lower starting damping counterpart. Note from Figure \ref{subfig:dampkfac} that even though the damping of this curve reaches $\approx 0.1$ that starting with a flat damping of $0.1$ never achieves a comparable generalisation (or even trains well). This implies as expected that it is important to adjust damping during training.

\subsection{Adam with Auto-Damping}
\begin{figure}[h!]
	\begin{subfigure}[b]{0.62\textwidth}
		\includegraphics[width=\textwidth]{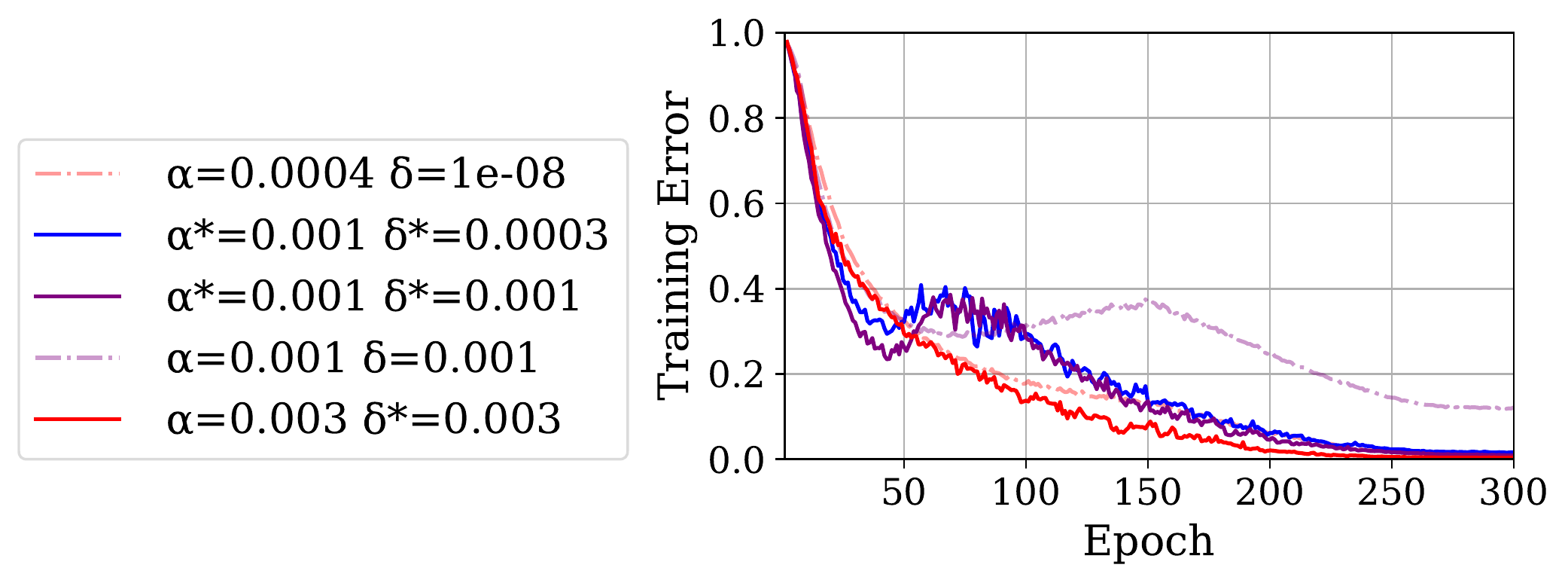}
		\caption{Training Error}
		\label{subfig:trainadam}
	\end{subfigure}
	\begin{subfigure}[b]{0.37\textwidth}
		\includegraphics[width=\textwidth]{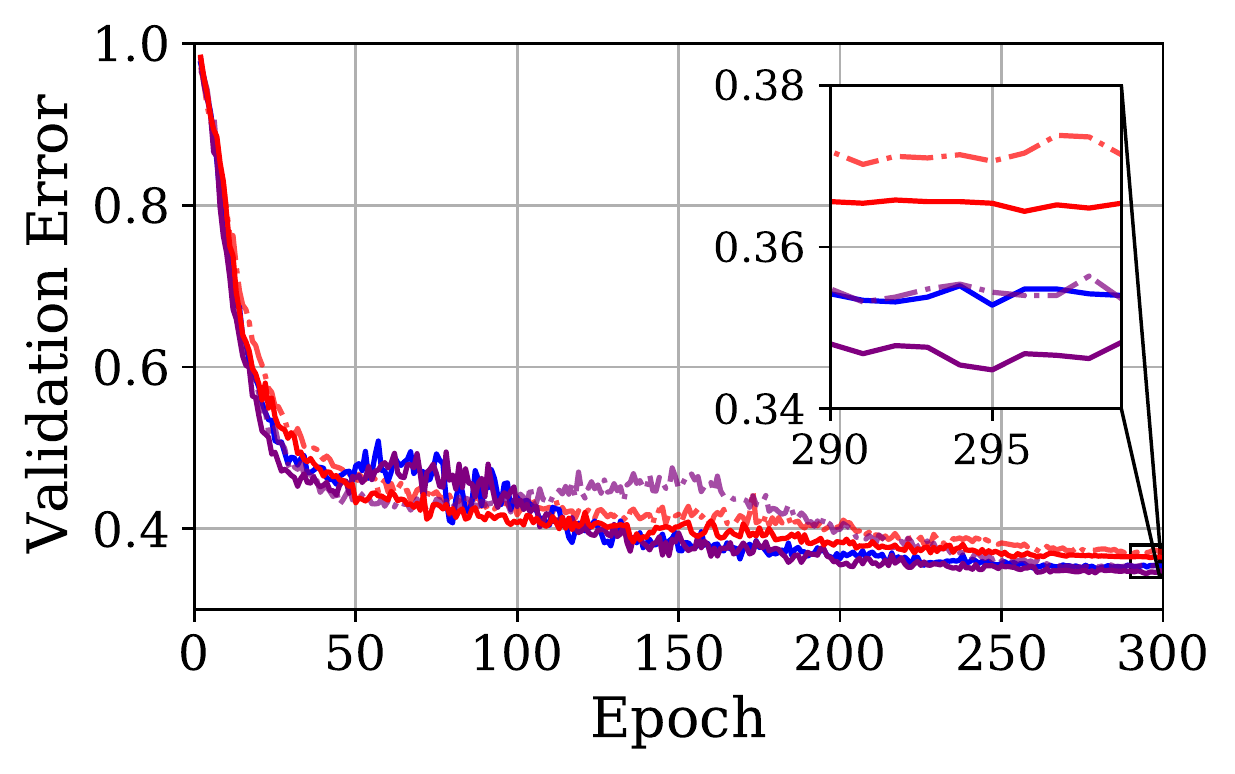}
		\caption{Val Error}
		\label{subfig:valadam}
	\end{subfigure}
	\caption{VGG-$16$ on CIFAR-$100$ dataset using the Adam optimiser with $\gamma=0$ (no weight decay) for a learning rate of $\alpha$, batch size $B=128$ and damping set by $\delta$. For adaptive damping methods the damping is given an initial floor value of $\delta^{*}$ and is then updated using the variance of the Hessian every $100$ steps. $\alpha^{*}$ refers to the use of an alternative ramp up schedule where the base learning rate is increased by a factor of $5$ at the start of training before being decreased.}
	\label{fig:adamautodamp}
\end{figure}

Given that Adam does not employ an obvious curvature matrix, it is curious to consider whether our learned damping estimator can be of practical value for this optimiser. As discussed in the previous section, Adam's implied curvature can be considered a diagonal approximation to the square root of the gradient covariance. The covariance of the gradients has been investigated to have similarities to the Hessian \cite{jastrzebski2020the}. However the nature of the square root, derived from the regret bound in \cite{duchi2011adaptive} presents an interesting dilemma. In the case of very very small eigenvalues of $\mB$, the square root actually reduces their impact on the optimisation trajectory, hence it is very plausible that the learned damping could be too harsh (as it is expected to work optimally for the eigenvalues of $\mH$ and not $\sqrt{\mH})$.  This is actually exactly what we see in Figure \ref{fig:adamautodamp}.
Whilst an increase in learning rate and damping, along with auto-damping improves both the convergence and validation result over the standard baseline (where the damping is kept at the default value and maximal learning rate is found which stably trains) the improvements are small and do not make up the gap with SGD. More specifically they are not better than just using a larger learning rate in combination with a larger flat damping, defeating the purpose of learning the damping factor online. 

To alleviate the effect of overly harsh damping, we consider an alternate learning rate schedule where the base learning rate is increased by a factor of $5$ early in training and then subsequently decreased. The constant $5$ is not tuned but simply a place-holder to consider a more aggressive learning rate schedule to counter-act the effect of the damping learner. These curves are marked with $\alpha^{*}$ in Figure \ref{fig:adamautodamp}. 

\paragraph{Warm up Learning Rate Schedule} For all experiments unless specified,  we use the following learning rate schedule for the learning rate at the $t$-th epoch:
\begin{equation}
	\alpha_t = 
	\begin{cases}
		\alpha_0, & \text{if}\ \frac{t}{T} \leq 0.1 \\
		\alpha_0[1+\frac{(\kappa-1)(\frac{t}{T}-0.1)}{0.2} , & \text{if}\ \frac{t}{T} \leq 0.3 \\
		\alpha_0[\kappa - \frac{(\kappa - r)(\frac{t}{T} - 0.3)}{0.6}] & \text{if } 0.3 < \frac{t}{T} \leq 0.9 \\
		\alpha_0r, & \text{otherwise}
	\end{cases}
\end{equation}
where $\alpha_0$ is the initial learning rate. $T$ is the total number of epochs budgeted for all CIFAR experiments. We set $r = 0.01$ and $\kappa = 5$.

While this introduces some slight training instability early in training, which could potentially be managed by altering the schedule, we find that such a schedule boosts the validation performance, particularly so for auto-damped methods, as shown by the blue curve in Figure \ref{subfig:valadam}, which surpasses the generalisation of SGD (shown in Figure \ref{fig:adamsgd}).

\begin{figure}[h!]
	\begin{subfigure}[b]{0.48\textwidth}
		\includegraphics[width=\textwidth]{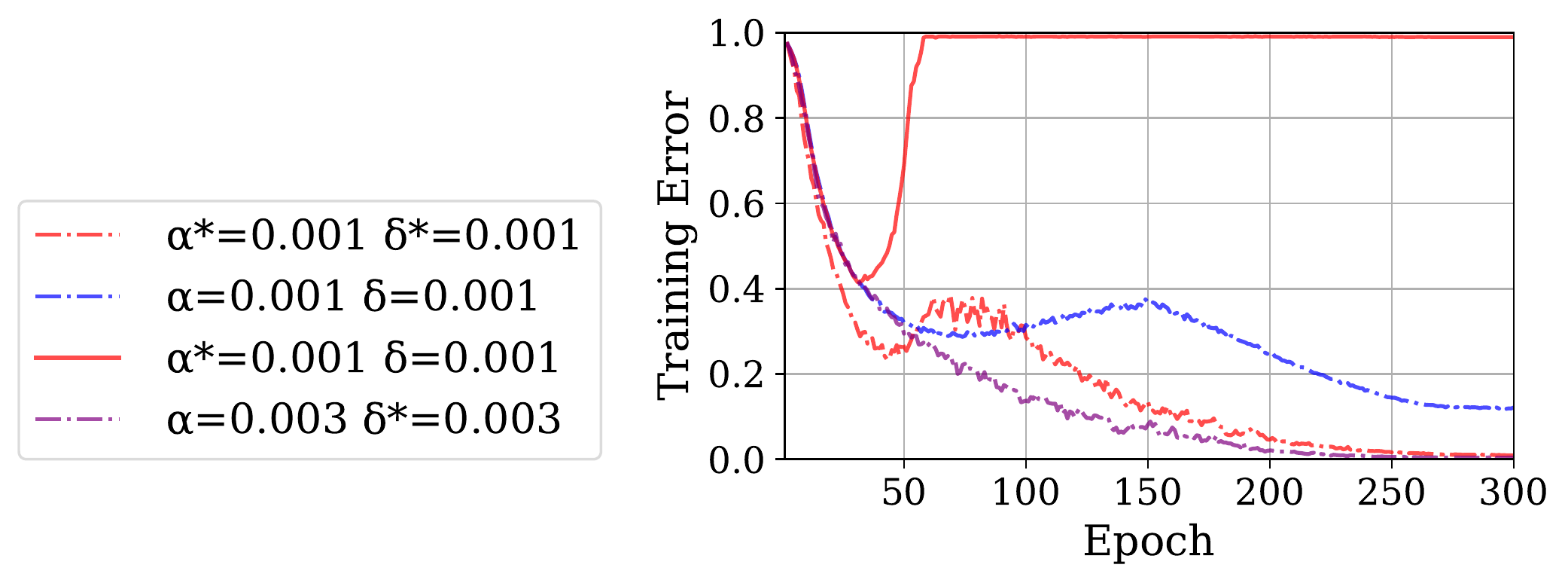}
		\caption{Training Error}
		\label{subfig:closetrainadam}
	\end{subfigure}
	\begin{subfigure}[b]{0.29\textwidth}
		\includegraphics[width=\textwidth]{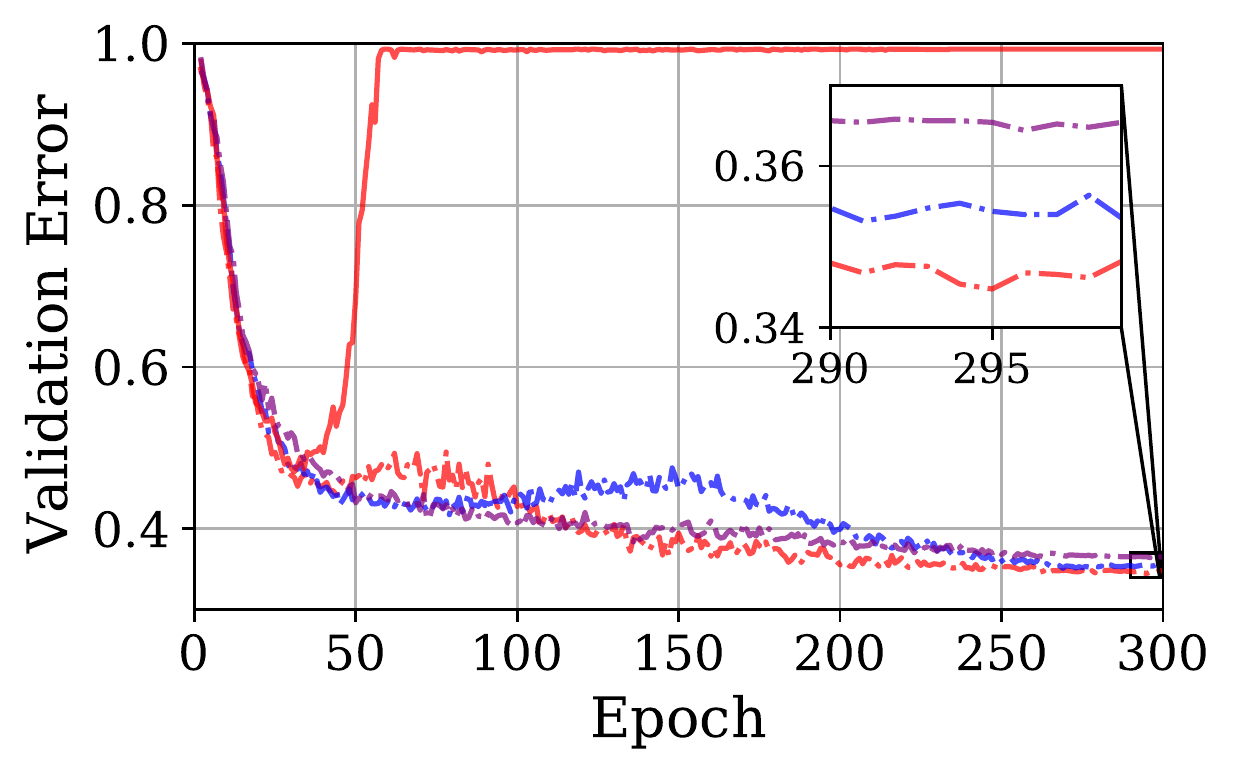}
		\caption{Val Error}
		\label{subfig:closevaladam}
	\end{subfigure}
	\begin{subfigure}[b]{0.205\textwidth}
		\includegraphics[width=\textwidth]{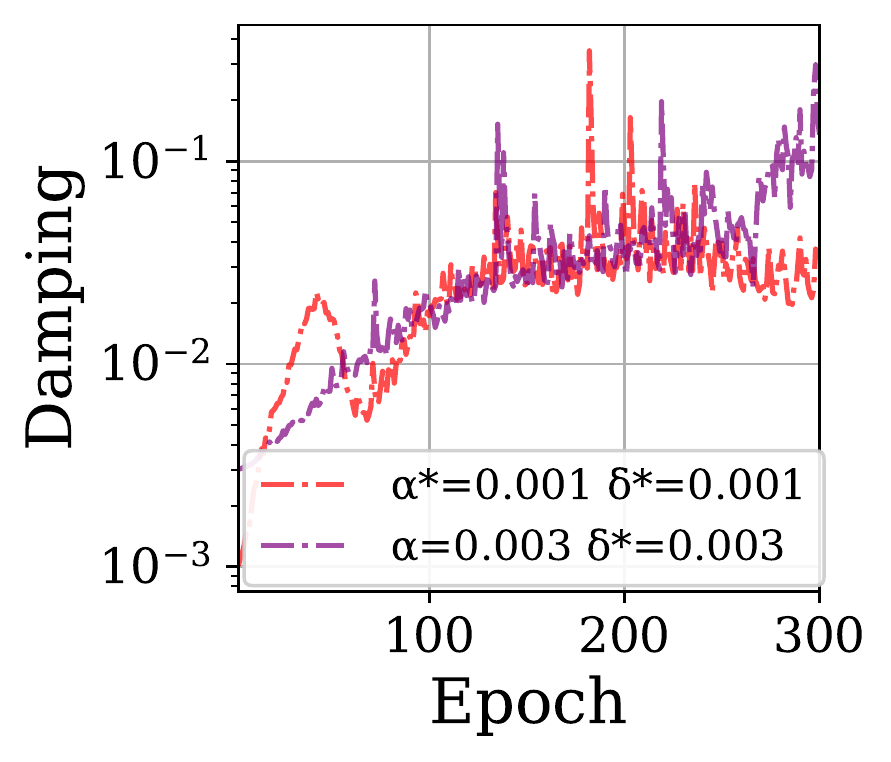}
		\caption{Damping}
		\label{subfig:closeadamdamp}
	\end{subfigure}
	\caption{VGG-$16$ on CIFAR-$100$ dataset using the Adam optimiser with $\gamma=0$ (no weight decay) for a learning rate of $\alpha$, batch size $B=128$ and damping set by $\delta$. For adaptive damping methods the damping is given an initial floor value of $\delta^{*}$ and is then updated using the variance of the Hessian every $100$ steps. $\alpha^{*}$ refers to the use of an alternative ramp up schedule where the base learning rate is increased by a factor of $5$ at the start of training before being decreased.}
	\label{fig:adamautodampclean}
\end{figure}

To more clearly expose the combined impact of adaptive damping and this alternative learning schedule we consider the variations in Figure \ref{fig:adamautodampclean} for a learning rate and damping both equal to $0.0001$. Here we see that the aggressive learning rate schedule with flat damping diverges, whereas the autodamping stabilises training, allowing for convergence to a solution with excellent generalisation. We see here in Figure \ref{subfig:closeadamdamp} that the damping coefficient reacts to this large learning rate increase by increasing its rate of damping early, stabilising training. We also show for reference that the typical linear decay schedule, with a larger learning rate and initial damping does not supersede the validation result of smaller learning rate and flat damping counter-part (it does however train better). This demonstrates the necessity of an alternative learning rate schedule to bring out the value of the adaptive damping. We remark however that optimal results in deep learning almost always require some degree of hand-crafted tuning of the learning rate. Our adaptive damping method is not proposed as a panacea, but just an optimal method of setting the damping coefficient. Since changing the damping coefficient effectively changes to geometry of the loss surface, it is entirely reasonable that the learning rate may have to be tweaked to give best results.

\section{Conclusion}
In this chapter we have showed using a spiked random matrix model for the batch loss of deep neural networks that we expect sharp directions of loss surface to retain more information about the true loss surface compared to flatter directions. For adaptive methods, which attempt to minimise an implicit local quadratic of the sampled loss surface, this leads to sub-optimal steps with worse generalisation performance.
We further investigate the effect of damping on the solution sharpness and find that increasing damping always decreases the solution sharpness, linking to prior work in this area. We find that for large neural networks an increase in damping both assists training and is even able to best the SGD test baseline. An interesting consequence of this finding is that it suggests that damping should be considered an essential hyper-parameter in adaptive gradient methods as it already is in stochastic second order methods. Moreover, our random matrix theory model motivates a novel interpretation of damping as linear shrinkage estimation of the Hessian. We establish the validity of this interpretation by using shrinkage estimation theory to derive an optimal adaptive damping scheme which we show experimentally to dramatically improve optimisation speed with adaptive methods \emph{and} closes the adaptive generalisation gap.

\medskip
Our work leaves open several directions for further investigation and extension. Mathematically, there is the considerable challenge of determining optimal assumptions on the network, loss function and data distribution such that the key outlier overlap result in Theorem \ref{theorem:overlap}, or sufficiently similar analogues thereof, can be obtained. On the experimental side, we have restricted ourselves to computer vision datasets and a small number of appropriate standard network architectures. These choices helped to maintain clarity on the key points of investigation, however they are clearly limiting. In particular, it would be natural to reconsider our investigations in situations for which adaptive optimisers typically obtain state of the art results, such as modern natural language processing \cite{devlin2018bert}. Practically speaking, we have proposed a novel, theoretically motivated and effective adaptive damping method, but it is reliant on relatively expensive Hessian variance estimates throughout training. Future work could focus on cheaper methods of obtaining the required variance estimates. 
\chapter{Appearance of local random matrix statistics}\label{chap:spacings}
The content of this chapter was published first as a pre-print in February 2021 (\url{https://arxiv.org/abs/2102.06740}) and later as a journal article: ``Appearance of random matrix theory in deep learning''. \textbf{Nicholas P Baskerville}, Diego
Granziol and Jonathan P Keating. \emph{Physica A: Statistical Mechanics and its Applications},
590:126742, 2022.
\medskip

\textbf{NPB} performed the calculations, designed, coded and ran most of the experiments and
wrote most of the random matrix theory aspects of the paper. DG assisted with writing
code, ran the training of a few of the neural networks and wrote some of the more machine
learning oriented sections of the paper. JPK proposed the research idea, advised throughout
and contributed several sections to the paper. Anonymous reviewers spotted some minor
errors, advised on changes of presentation and extra experiments and provided useful
references.

\section{Preliminaries}
\label{sec:prelim}
Consider a neural network with weights $\vec{w}\in\mathbb{R}^P$ and a dataset with distribution $\mathbb{P}_{\mathrm{data}}$. For the purposes of our discussion, a neural network, $f_{\vec{w}}$ say, is just a non-linear function from some $\mathbb{R}^d$ to some $\mathbb{R}^c$, parametrised by $\vec{w}$. Neural networks can be defined in many different ways in terms of their weights (the architecture of the network), but these details will not play role in our discussion. What will be important is that the number of weights $P$ will be large, i.e. approaching 10,000 even in the simplest of cases. Let $L(\vec{w}, \vec{x})$ be the loss of the network for a single datum $\vec{x}$ and let $\mathcal{D}$ denote any finite sample of data points from $\mathbb{P}_{\mathrm{data}}$. A simple example of $L$ is the squared error $L(\vec{w}, (\vec{x}, \vec{y})) = ||f_{\vec{w}}(\vec{x}) - \vec{y}||_2^2$, where $\mathbb{P}_{\mathrm{data}}$ is a distribution on tuples of features $\vec{x}$ and labels $\vec{y}$. The \emph{true loss} is given by \begin{align}
    \mathcal{L}_{true}(\vec{w}) = \mathbb{E}_{\vec{x}\sim\mathbb{P}_{\mathrm{data}}} L(\vec{w}, \vec{x})
\end{align}
and the \emph{empirical loss} (or training loss) is given by \begin{align}
        \mathcal{L}_{emp}(\vec{w}, \mathcal{D}) = \frac{1}{|\mathcal{D}|}\sum_{\vec{x}\in\mathcal{D}} L(\vec{w}, \vec{x}).
\end{align}
Where $\mathcal{D}$ denotes the dataset. The true loss is a deterministic function of the weights, while the empirical loss is a random function with the randomness coming from the random sampling of the finite dataset $\mathcal{D}$. The empirical Hessian $\mH_{emp}(\vw) = \nabla^{2} \mathcal{L}_{emp}(\vw)$, describes the loss curvature at the point $\vw$ in weight space. By the spectral theorem, the Hessian can be written in terms of its eigenvalue/eigenvector pairs $\mH_{emp} = \sum_{i}^{P}\lambda_{i}\vphi_{i}\vphi_{i}^{T}$, where the dependence on $\vw$ has been dropped to keep the notation simple. The eigenvalues of the Hessian are particularly important, being explicitly required in second-order optimisation methods, and characterising the stationary points of the loss as local minima, local maxima or generally saddle points of some other index.

For a matrix drawn from a probability distribution, its eigenvalues are random variables. The eigenvalue distribution is described by the joint probability density function (j.p.d.f) $p(\lambda_1, \lambda_2, \ldots, \lambda_P)$, also known as the $P$-point correlation function.
The simplest example is the \emph{empirical spectral density (ESD)}, $\rho^{(P)}(\lambda) = \frac{1}{P}\sum_{i}^{P}\delta(\lambda-\lambda_{i})$.  Integrating $\rho^{(P)}(\lambda)$ over an interval with respect to $\lambda$ gives the fraction of the eigenvalues in that interval.   
Taking an expectation over the random matrix ensemble, we obtain the \emph{mean spectral density} $\mathbb{E}\rho^{(P)}(\lambda)$, which is a deterministic probability distribution on $\mathbb{R}$. Alternatively, taking the $P\rightarrow\infty$ limit, assuming it exists, gives the \emph{limiting spectral density (LSD)} $\rho$, another deterministic probability distribution on $\mathbb{R}$. A key feature of many random matrix ensembles is \emph{self-averaging} or \emph{ergodicity}, meaning that the leading order term (for large $P$) in $\mathbb{E}\rho^{(P)}$ agrees with $\rho$. Given the j.p.d.f, one can obtain the mean spectral density, known as the $1$-point correlation function (or any other $k$-point correlation function) by marginalisation \begin{equation}
    \mathbb{E}\rho^{(P)}(\lambda) = \int p(\lambda, \lambda_2, \ldots, \lambda_P) d\lambda_2\ldots d\lambda_P.
\end{equation}A GOE matrix is an example of a \emph{Wigner random matrix}, namely a real-symmetric (or complex-Hermitian) matrix with otherwise i.i.d. entries and off-diagonal variance $\sigma^2$.\footnote{The GOE corresponds to taking the independent matrix entries to be normal random variables.} The mean spectral density for Wigner matrices is known to be Wigner's semicircle \cite{mehta2004random} \begin{equation}
    \rho_{SC}(\lambda) = \frac{1}{2\pi \sigma^2 P}\sqrt{4P\sigma^2 - \lambda^2}\indic_{|\lambda|\leq 2\sigma\sqrtsign{P}}.
\end{equation} The radius of the semicircle\footnote{Using the Frobenius norm identity $\sum_{i}^{P}\lambda_{i}^{2} = P^{2}\sigma^{2}$} is proportional to $\sqrtsign{P}\sigma$, hence scaling Wigner matrices by $1/\sqrt{P}$ leads to a limit distribution when $P\rightarrow\infty$.  This is the LSD. 
With this scaling, there are, on average, $\mathcal{O}(P)$ eigenvalues in any open subset of the compact spectral support. In this sense, the mean (or limiting) spectral density is \emph{macroscopic}, meaning that, as $P\rightarrow\infty$, one ceases to see individual eigenvalues, but rather a continuum with some given density.

\section{Motivation: Microscopic Universality}

Random Matrix Theory was first developed in physics to explain the statistical properties of nuclear energy levels, and later used to describe the spectral statistics in atomic spectra, condensed matter systems, quantum chaotic systems etc; see, for example \cite{weidenmuller2008random, beenakker1997random, berry1987quantum, bohigas1991random}. \emph{None of these physical systems exhibits a semicircular empirical spectral density}. However they all generically show agreement with RMT at the level of the mean eigenvalue spacing when local spectral statistics are compared.  Our point is that while neither multi-layer perceptron (MLP) nor Softmax Regression Hessians are described by the Wigner semicircle law which holds for GOE matrices (c.f. Figure 1a) -- their spectra contain outliers, large peaks near the origin and the remaining components of the histogram also do not match the semicircle -- nevertheless Random Matrix Theory can still (and we shall demonstrate does) describe spectral fluctuations on the scale of their mean eigenvalue spacing.

\medskip
It is worth noting in passing that possibilities other than random-matrix statistics exist and occur. For example, in systems that are classically integrable, one finds instead Poisson statistics \cite{berry1977level, berry1987quantum}; similarly, Poisson statistics also occur in disordered systems in the regime of strong Anderson localisation \cite{efetov1999supersymmetry}; and for systems close to integrable one finds a superposition of random-matrix and Poisson statistics \cite{berry1984semiclassical}. So showing that Random Matrix Theory applies is far from being a trivial observation.  Indeed it remains one of the outstanding challenges of mathematical physics to prove that the spectral statistics of any individual Hamiltonian system are described by it in the semiclassical limit.

\medskip
Physics RMT calculations re-scale the eigenvalues to have a mean level spacing of $1$ and then typically look at the \emph{nearest neighbour spacings distribution} (NNSD), i.e. the distribution of the distances between adjacent pairs of eigenvalues.  One theoretical motivation for considering the NNSD is that it is independent of the Gaussianity assumption and reflects the symmetry of the underlying system. It is the NNSD that is universal (for systems of the same symmetry class) and not the average spectral density, which is best viewed as a parameter of the system. The aforementioned transformation to give mean spacing $1$ is done precisely to remove the effect of the average spectral density on the pair correlations leaving behind only the universal correlations. To the best of our knowledge no prior work has evaluated the NNSD of artificial neural networks and this is a central focus of this chapter.

\begin{figure}[h]
    \centering
    \includegraphics[width=0.8\textwidth]{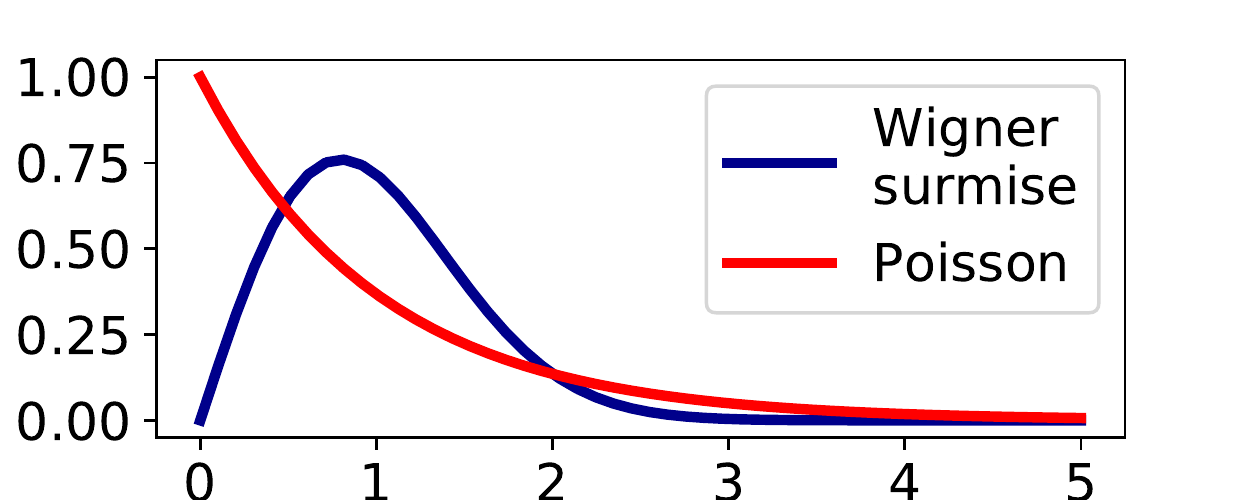}
    \caption{The density of the Wigner surmise.}
    \label{fig:wigner}
\end{figure}

In contrast to the LSD, other $k$-point correlation functions are also normalised such that the mean spacing between adjacent eigenvalues is unity. At this \emph{microscopic} scale, the LSD is locally constant and equal to 1 meaning that its effect on the eigenvalues' distribution has been removed and only microscopic correlations remain. In the case of Wigner random matrices, for which the LSD varies slowly across the support of the eigenvalue distribution, this corresponds to scaling by $\sqrt{P}$. On this scale the limiting eigenvalue correlations when $P\to\infty$ are {\em universal}; that is, they are the same for wide classes of random matrices, depending only on symmetry \cite{guhr1998random}.  For example, this universality is exhibited by the NNSD. Consider a $2\times 2$ GOE matrix, in which case the j.p.d.f has a simple form: \begin{equation}
    p(\lambda_1, \lambda_2) \propto |\lambda_1 - \lambda_2| e^{-\frac{1}{2}(\lambda_1^2 + \lambda_2^2)}.
\end{equation}
Making the change of variables $\nu_1 = \lambda_1 - \lambda_2, \nu_2 = \lambda_1 + \lambda_2$, integrating out $\nu_2$ and setting $s = |\nu_1|$ results in a density $\rho_{Wigner}(s) = \frac{\pi s}{2}e^{-\frac{\pi}{4}s^2}$, known as the \emph{Wigner surmise} (see Figure \ref{fig:wigner}). For larger matrices, the j.p.d.f must include an indicator function $\indic\{\lambda_1\leq \lambda_2\leq \ldots\lambda_P\}$ before marginalisation so that one is studying pairs of \emph{adjacent} eigenvalues. While the Wigner surmise can only be proved exactly, as above, for the $2\times 2$ GOE, it holds to high accuracy for the NNSD of GOE matrices of any size provided that the eigenvalues have been scaled to give mean spacing 1.\footnote{An exact formula for the NNSD of GOE matrices of any size, and one that holds in the large $P$ limit, can be found in \cite{mehta2004random}.} The Wigner surmise density vanishes at $0$, capturing `repulsion' between eigenvalues that is characteristic of RMT statistics, in contrast to the distribution of entirely independent eigenvalues given by the \emph{Poisson law} $\rho_{Poisson}(s) = e^{-s}$. The Wigner surmise is universal in that the same density formula applies to all real-symmetric random matrices, not just the GOE or Wigner random matrices. 

\section{Methodology}\label{sec:methods}
Prior work \cite{granziol2019mlrg,papyan2018full,ghorbani2019investigation} focusing on the Hessian empirical spectral density has utilised fast Hessian vector products \cite{pearlmutter1994fast} in conjunction with Lanczos \cite{meurant2006lanczos} methods. 
However, these methods approximate only macroscopic quantities like the spectral density, not microscopic statistics such as nearest neighbour spectral spacings. 
For modern neural networks, the $\mathcal{O}(P^{3})$ Hessian eigendecomposition cost will be prohibitive,  e.g. for a Residual Network (Resnet) \cite{he2016deep} with $34$ layers $P=10^{7}$. Hence, We restrict to models small enough to perform exact full Hessian computation and eigendecomposition. 

\medskip
We consider single layer neural networks for classification (softmax regression), 2-hidden-layer MLPs\footnote{Hidden layer widths: 10, 100.} and 3 hidden-layer MLPs\footnote{Hidden layer widths: 10, 100, 100.}. On MNIST \cite{deng2012mnist}, the Hessians are of size $7850\times 7850$ for logistic regression, $9860\times 9860$ for the small MLP and $20060\times 20060$ for the larger 3 hidden-layer MLP, so can be computed exactly by simply applying automatic differentiation twice, and the eigenvalues can be computed exactly in a reasonable amount of time.
We also consider a single layer applied to CIFAR-$10$  \cite{krizhevsky2009learning} classification with pre-trained Resnet-$34$ embedding features \cite{he2016deep, resnet-torch}. While we cannot at present study the full Hessian of, for example, a Resnet-$34$, we can study the common transfer learning use-case of training only the final layer on some particular task \cite{sharif2014cnn}. The Hessians can be computed at any data point or over any collection of data points. We consider Hessians computed over the entire datasets in question, and over batches of size 64. We separately consider test and train sets.

\medskip
In order to extend the relevance of our analysis to beyond logistic regression and MLP, we consider one of the simplest convolutional neural networks (CNN) of the form of LeNet \cite{lecun1998mnist} on CIFAR-$10$. Compared to the standard LeNet (which has over $50000$ parameters) we reduce the number of neurons in the first fully connected layer from $120$ to $35$ and the second from $84$ to $50$. Note that the resulting architecture contains a bottleneck in the intermediate layer, in contrast to the ``hour-glass'' shapes that are necessary to maintain manageable parameter numbers with full MLP architectures. Despite reducing the total number of parameters by a factor of $3$ we find the total validation accuracy drop to be no more than $2\%$. The total validation accuracy of $69\%$ is significantly below state of the art $\approx 95\%$, but we are clearly in the regime where significant learning can and does take place, which we consider sufficient for the purposes of this manuscript. We also extend our experiments beyond the cross entropy loss function, by considering a regression problem ($L_{2}$ loss) and beyond the high-dimensional feature setting of computer vision with the Bike dataset\footnote{\href{https://archive.ics.uci.edu/ml/datasets/Bike+Sharing+Dataset}{https://archive.ics.uci.edu/ml/datasets/Bike+Sharing+Dataset} (accessed 14/10/21)} which has only 13-dimensional feature vectors and a single-dimensional regressand (see Appendix \ref{app:preproc} for details of our data pre-processing). The architecture in this case widens considerably in the first layer (from 13 inputs to 100 neurons) and that gradually tapers to the single output. The final test loss (i.e. mean squared error) of the trained model is 0.044 which is competitive with baseline results \cite{wang2019exact}\footnote{\cite{wang2019exact} report an RMSE of 0.220 on Bike (which corresponds to 0.048 mean squared error) using a Gaussian process regression model with exact inference.}

\paragraph{Training details:}All networks were trained using SGD for 300 epochs with initial learning rate 0.003, linear learning rate decay to 0.00003 between epoch 150 and 270, momentum 0.9 and weight decay $5\times 10^{-4}$. We use a PyTorch \cite{paszke2017automatic} implementation. Full code to reproduce our results is made available \footnote{\url{https://github.com/npbaskerville/dnn-rmt-spacings}}. Full descriptions of all network architectures are given in the Appendix \ref{app:architectures}.

\section{Spectral spacing statistics in RMT}
\label{sec:spacings}
Consider a random $P\times P$ matrix $M_P$ with ordered $\lambda_1 \leq \lambda_2 \leq \ldots \leq \lambda_P$. Let $I_{ave}$ be the mean spectral cumulative density function for the random matrix ensemble from which $M_P$ is drawn. The \emph{unfolded spectrum} is defined as \begin{align}
    l_i = I_{ave}(\lambda_i).
\end{align}
The unfolded spacings are then defined as \begin{align}
    s_i = l_i - l_{i-1}, ~~~ i=2, \ldots, P.
\end{align}

With this definition, the mean of the $s_i$ is unity, which means that this transformation has brought the eigenvalues on to the microscopic scale on which universal spectral spacing statistics emerge. We are investigating the presence of Random Matrix Theory statistics in neural networks by considering the nearest neighbour spectral spacings of their Hessians. Within the Random Matrix Theory literature, it has been repeatedly observed \cite{bohigas1991random, berry1987quantum} that the unfolded spacings of a matrix with RMT pair correlations follow universal distributions determined only by the symmetry class of the $M_P$. Hessians are real symmetric, so the relevant universality class is GOE and therefore the unfolded neural network spacings should be compared to the Wigner surmise
\begin{equation}\label{eq:wigner}
    \rho_{Wigner}(s) = \frac{\pi s}{2}e^{-\frac{\pi}{4}s^2}.
\end{equation}
A collection of unfolded spacings $s_2,\ldots, s_P$ from a matrix with GOE spacing statistics should look like a sample of i.i.d. draws from the Wigner surmise density (\ref{eq:wigner}).
For some known random matrix distributions, $I_{ave}$ may be available explicitly, or at least via highly accurate quadrature methods from a known mean spectral density. For example, for the $P\times P$ GOE  \cite{abuelenin2012effect} $I_{ave}^{GOE}(\lambda) $ is given by:
\begin{align}
   P\left[\frac{1}{2} + \frac{\lambda}{2\pi P}\sqrt{2P - \lambda^2} + \frac{1}{\pi}\arctan\left(\frac{\lambda}{\sqrt{2 P - \lambda^2}}\right)\right].
\end{align}
However, when dealing with experimental data where the mean spectral density is unknown, one must resort to using an approximation to $I_{ave}$. Various approaches are used in the literature, including polynomial spline interpolation \cite{abuelenin2012effect}. The approach of \cite{scholak2014spectral, unfoldr} is most appropriate in our case, since computing Hessians over many mini-batches of data results in a large pool of spectra which can be used to accurately approximate $I_{ave}$ simply by the empirical cumulative density.
Suppose that we have $m$ samples $(M^{(i)}_P)_{i=1}^m$ from a random matrix distribution over symmetric $P\times P$ matrices. Fix some integers $m_1, m_2 > 0$ such that $m_1 + m_2 = m$. The spectra of the matrices $(M^{(i)}_P)_{i=1}^{m_1}$ can then be used to construct an approximation to $I_{ave}$.  More precisely, let $\Lambda_1$ be the set of all eigenvalues of the $(M^{(i)}_P)_{i=1}^{m_1}$, then we define \begin{align}
    \tilde{I}_{ave}(\lambda) = \frac{1}{|\Lambda_1|} |\{\lambda' \in \Lambda_1 \mid \lambda' < \lambda\}|.
\end{align}
For each of the matrices  $(M^{(i)}_P)_{i=m_1 + 1}^m$, one can then use $\tilde{I}_{ave}$ to construct their unfolded spacings. When the matrix size $P$ is small, one can only study the spectral spacing distribution by looking over multiple matrix samples. However, the same spacing distribution is also present for a single matrix in the large $P$ limit. A clear disadvantage of studying unfolded nearest neighbour spectral spacings with the above methods is the need for a reasonably large number of independent matrix samples. This rules-out studying the unfolded spacings of a single large matrix. Another obvious disadvantage is the introduction of error by the approximation of $I_{ave}$, giving the opportunity for local spectral statistics to be distorted or destroyed. An alternative statistic is the consecutive spacing ratio of \cite{atas2013distribution}. In the above notation, the ratios for a single $P\times P$ matrix are defined as \begin{align}
   r_i= \frac{\lambda_{i} - \lambda_{i-1}}{\lambda_{i-1} - \lambda_{i-2}}, ~~ 2 \leq i \leq P.
\end{align}
\cite{atas2013distribution} proved a `Wigner-like surmise' for the spacing ratios, which for the GOE is \begin{align}
P(r) = \frac{27(r + r^2)}{8(1 + r + r^2)^{5/2}}.
\end{align}

In our experiments, we can compute the spacing ratios for Hessians computed over entire datasets or over batches, whereas the unfolded spacing ratios can only be computed in the batch setting, in which case a random $\frac{2}{3}$ of the batch Hessians are reserved for computing $\tilde{I}_{ave}$ and the remaining $\frac{1}{3}$ are unfolded and analysed. This split is essentially arbitrary, except that  we err on the side of using more to compute $\tilde{I}_{ave}$ since even a single properly unfolded spectrum can demonstrate universal local statistics.

\section{Results}
We display results as histograms of data along with a plot of the Wigner (or the Wigner-like) surmise density. We make a few practical adjustments to the plots.  Spacing ratios are truncated above some value, as the presence of a few extreme outliers makes visualisation difficult. We choose a cut-off at 10. Note that around 0.985 of the mass of the Wigner-like surmise is below 10, so this is a reasonable adjustment. The hessians have degenerate spectra. The Wigner surmise is not a good fit to the observed unfolded spectra if the zero eigenvalues are retained. Imposing a lower cut-off of $10^{-20}$ in magnitude is sufficient to obtain agreement with Wigner.\footnote{For example, in the case of the 3-hidden-layer MLP on MNIST shown in Figure \ref{fig:deep_mlp_spacings}, among 157 batch-wise spectra the proportion of eigenvalues below the cut-off was between $0.29$ and $0.40$.} This is below the machine precision, so these omitted eigenvalues are indistinguishable from 0.

\begin{figure}[h!]
\centering
    \begin{subfigure}[b]{0.35\textwidth}
        \centering
        \includegraphics[width=\textwidth]{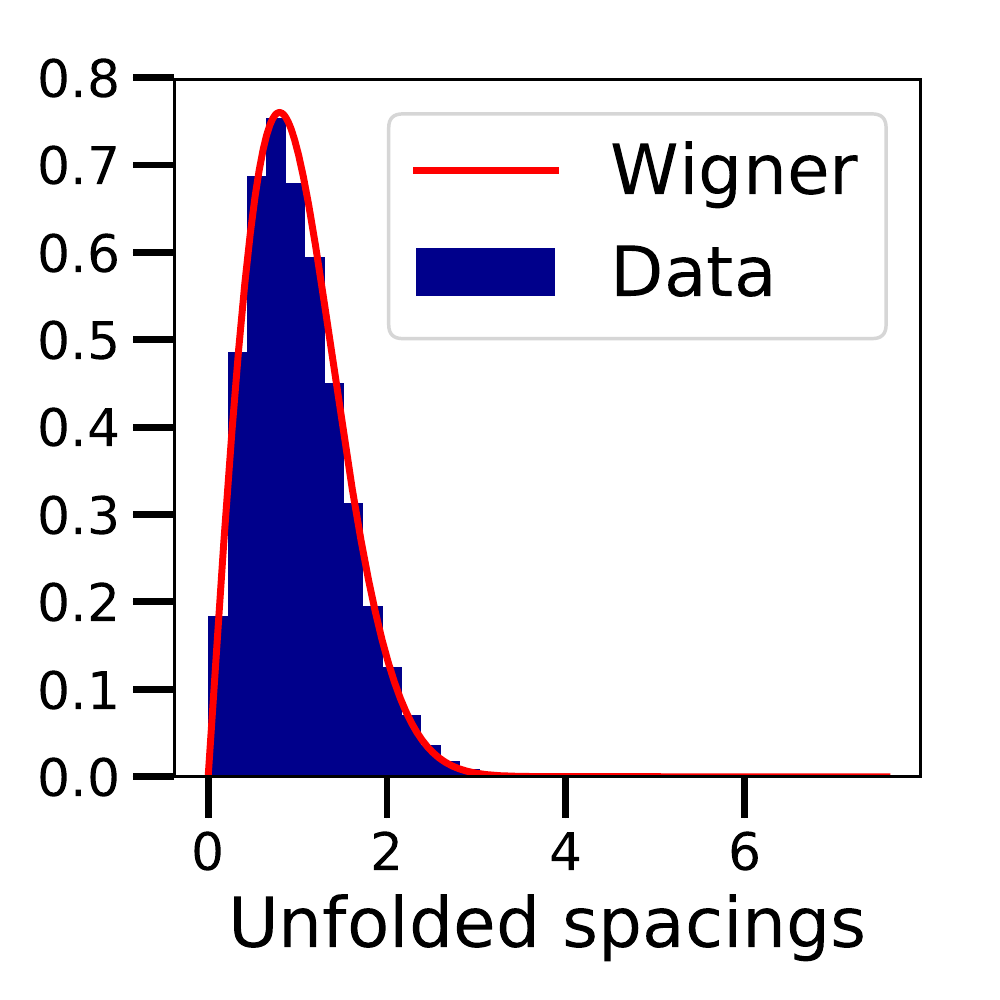}
        \caption{Unfolded spacings. Batch size 64.}
    \end{subfigure}
    \begin{subfigure}[b]{0.35\textwidth}
        \centering
        \includegraphics[width=\textwidth]{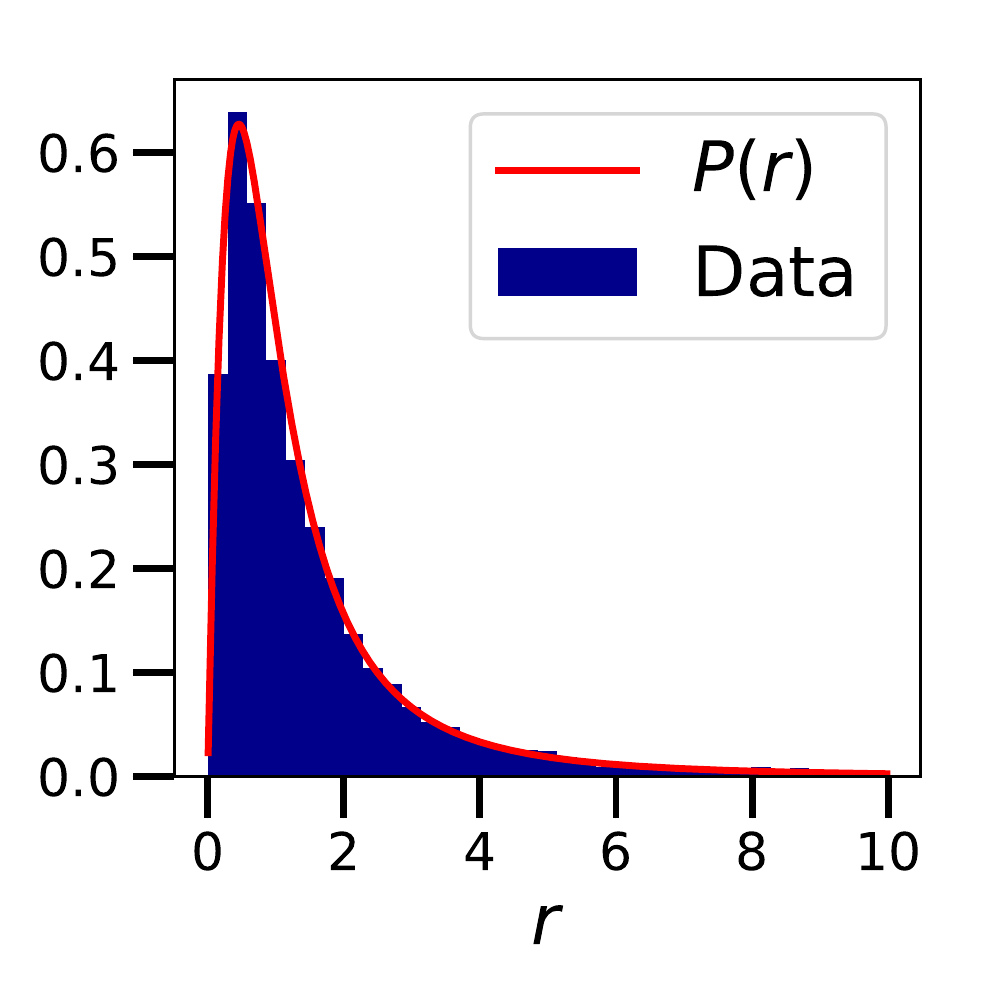}
        \caption{Spacing ratios. Entire dataset.}
    \end{subfigure}
    \caption{Spacing distributions for the Hessian of a logistic regression trained Resnet-$34$ embeddings of CIFAR10. Hessians computed over the test set.}
    \label{fig:cifar_resnet_spacings}
\end{figure}

\begin{figure}[h!]
\centering
    \begin{subfigure}[b]{0.35\textwidth}
        \centering
        \includegraphics[width=\textwidth]{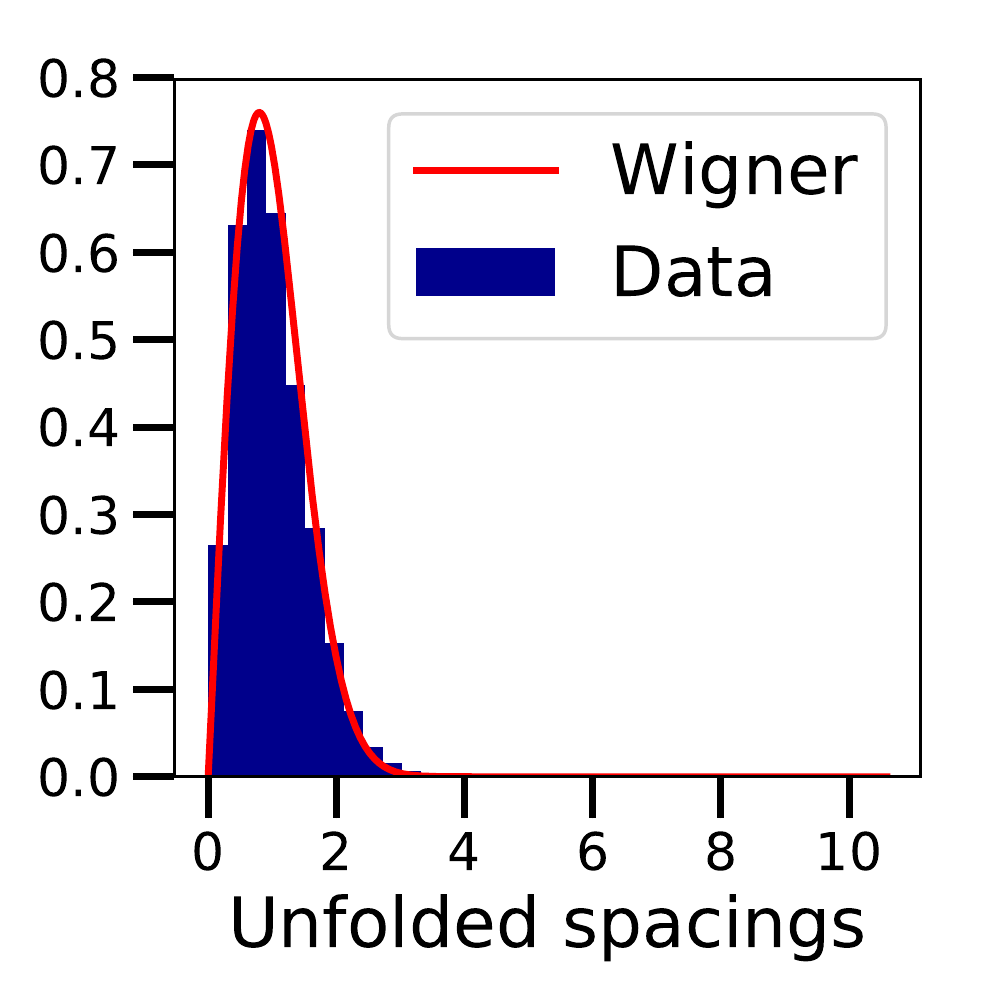}
        \caption{Unfolded spacings. Batch size 64.}
    \end{subfigure}
    \begin{subfigure}[b]{0.35\textwidth}
        \centering
        \includegraphics[width=\textwidth]{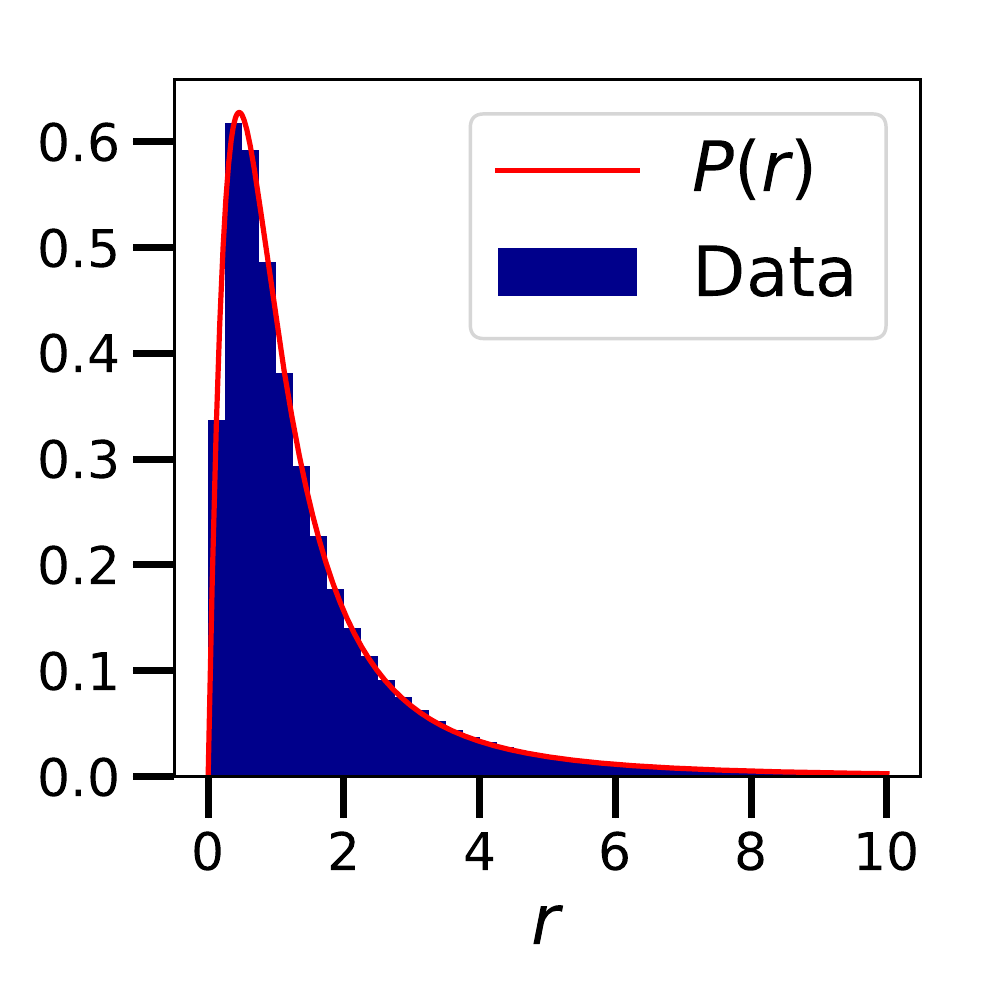}
        \caption{Spacing ratios. Batch size 64.}
    \end{subfigure}

    \caption{Spacing distributions for the Hessian of a 3-hidden-layer MLP trained on MNIST. Hessians computed over the test set.}
    \label{fig:deep_mlp_spacings}
\end{figure}

\subsection{MNIST and MLPs}
We show results in Figures \ref{fig:cifar_resnet_spacings} and \ref{fig:deep_mlp_spacings}, with further plots in the Appendix. We also considered randomly initialised networks and we evaluated the Hessians over train and test datasets separately in all cases. Unfolded spacings were computed only for Hessians evaluated on batches of 64 data points, while spacing ratios were computed in batches and over the entire dataset.
We observe a striking level of agreement between the observed spectra and the GOE. There was no discernible difference between the train and test conditions, nor between batch and full dataset conditions, nor between trained and untrained models. Note that the presence of GOE statistics for the untrained models is not a foregone conclusion. Of course, the weights of the model are indeed random Gaussian, but the Hessian is still a function of the data set, so it is not the case the Hessian eigenvalue statistics are bound to be GOE a priori. Overall, the very close agreement between Random Matrix Theory predictions and our observations for several different architectures, model sizes and datasets demonstrates a clear presence of RMT statistics in neural networks.

Our results indicate that models for the loss surfaces of large neural networks should include assumptions of GOE local statistics of the Hessian,  but ideally avoid such assumptions on the global statistics. To further illustrate this point, consider a Gaussian process $\mathcal{L}_{emp}\sim \mathcal{GP}(0, k)$ where $k$ is some kernel function. Following from our Gaussian process definition, the covariance of derivatives of the empirical loss can be computed using a well-known result (see \cite{adler2009random} equation 5.5.4), e.g.
 \begin{align} 
 \nonumber
    Cov(\partial_i \mathcal{L}_{emp}(\vec{w}), \partial_j\mathcal{L}_{emp}(\vec{w}') ) = \partial_{w_i}\partial_{w'_j} k(\vec{w}, \vec{w}')
\end{align}
and further, assuming a stationary kernel $k(\vec{w}, \vec{w}') = k\left(-\frac{1}{2}||\vec{w} - \vec{w}'||_2^2\right)$ (note abuse of notation) \begin{align}\label{eq:grad_gp_covar_2}
\nonumber
    & Cov(\partial_i \mathcal{L}_{emp}(\vec{w}), \partial_j\mathcal{L}_{emp}(\vec{w}') )  \\ 
    & = (w_i - w'_i)(w'_j - w_j) k''\left(-\frac{1}{2}||\vec{w} - \vec{w}'||_2^2\right)  + \delta_{ij}k'\left(-\frac{1}{2}||\vec{w} - \vec{w}'||_2^2\right). 
\end{align}

\begin{figure}
    \centering
    \begin{subfigure}[b]{0.3\textwidth}
        \centering
        \includegraphics[width=\textwidth]{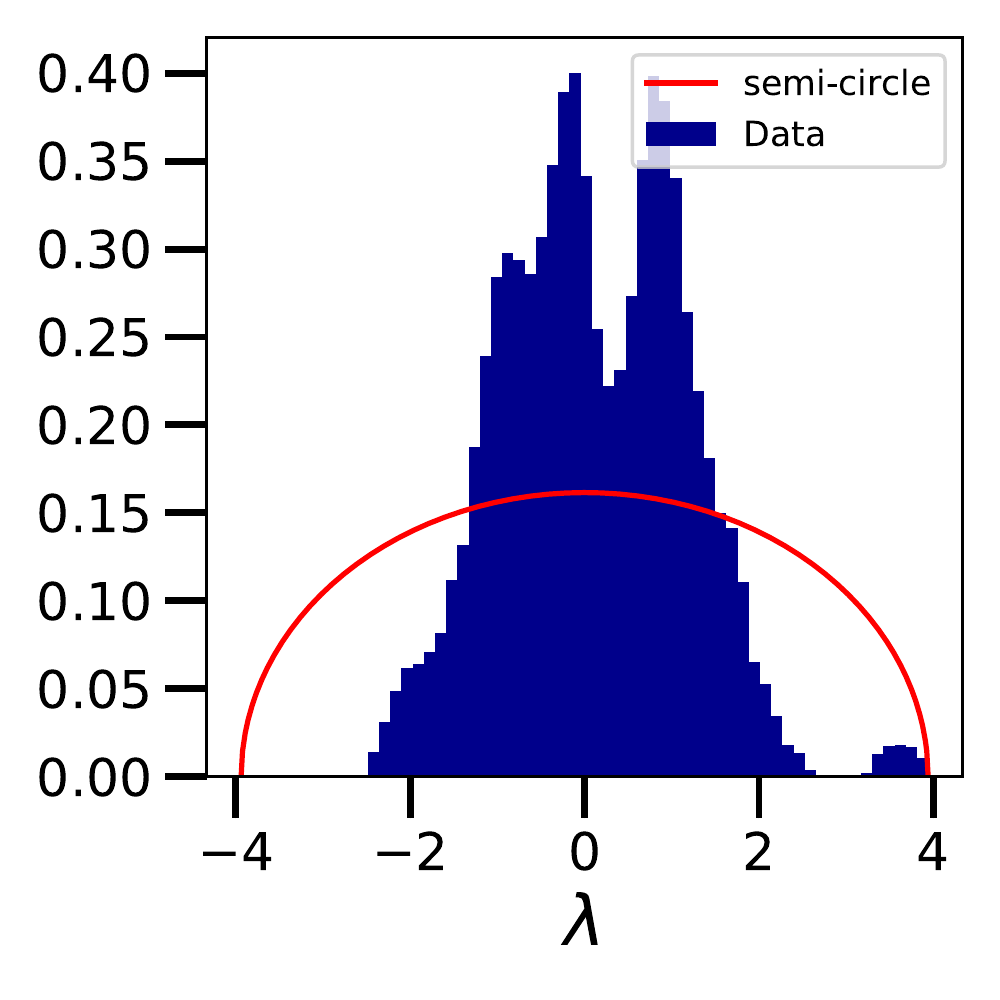}
        \caption{$k''(0)=10^{-4}$}
    \end{subfigure}
    \begin{subfigure}[b]{0.3\textwidth}
        \centering
        \includegraphics[width=\textwidth]{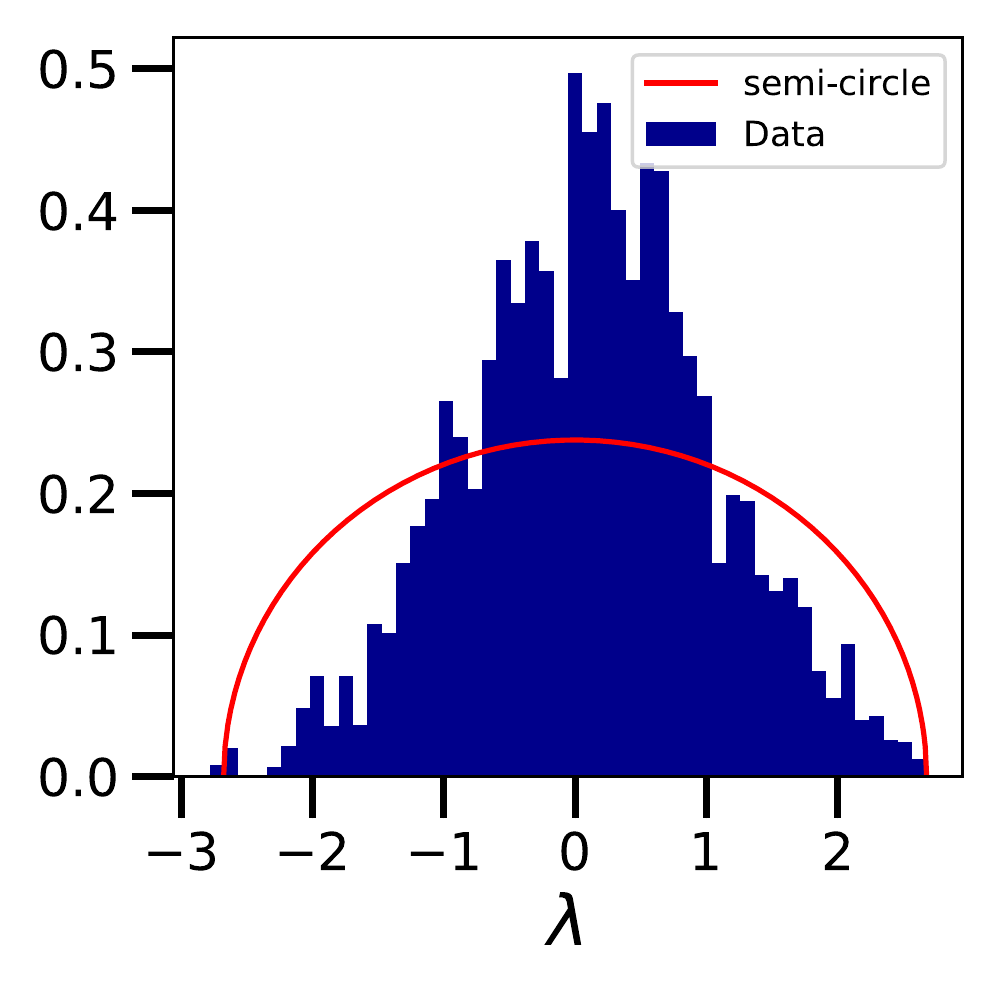}
        \caption{$k''(0)=10^{-3}$}
    \end{subfigure}
    \begin{subfigure}[b]{0.3\textwidth}
        \centering
        \includegraphics[width=\textwidth]{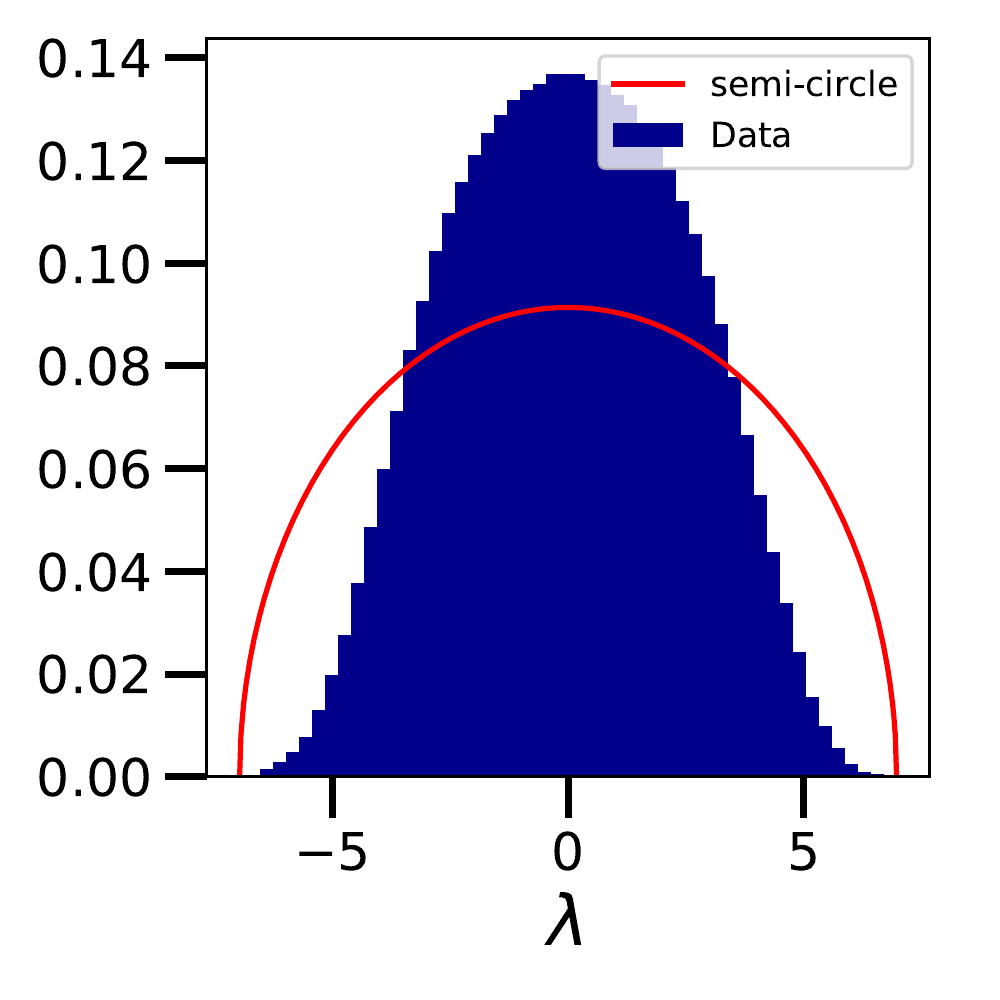}
        \caption{$k''(0)=10^{-1}$}
    \end{subfigure}
    
    \begin{subfigure}[b]{0.3\textwidth}
        \centering
        \caption{$k''(0)=10$}
    \end{subfigure}
    \begin{subfigure}[b]{0.3\textwidth}
        \centering
        \includegraphics[width=\textwidth]{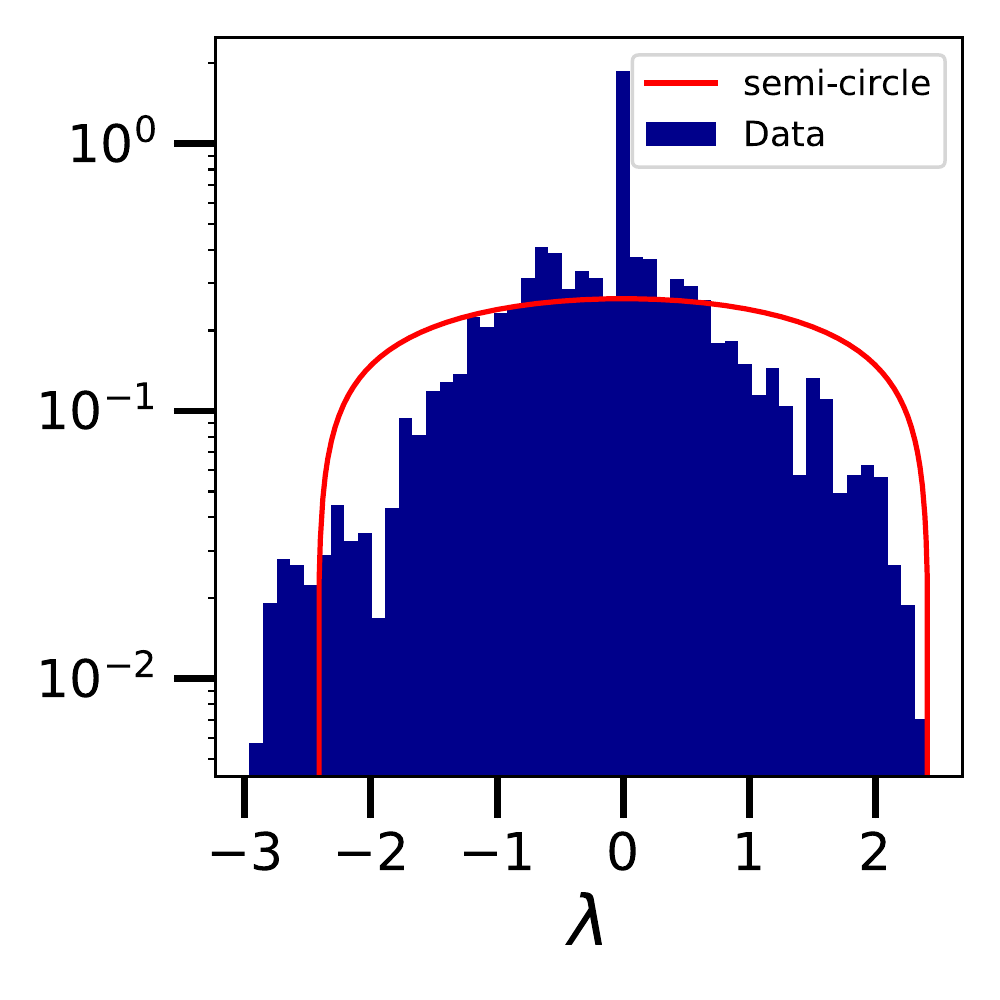}
        \caption{$k''(0)=10^{-3}*$}
    \end{subfigure}
    \begin{subfigure}[b]{0.3\textwidth}
        \centering
        \includegraphics[width=\textwidth]{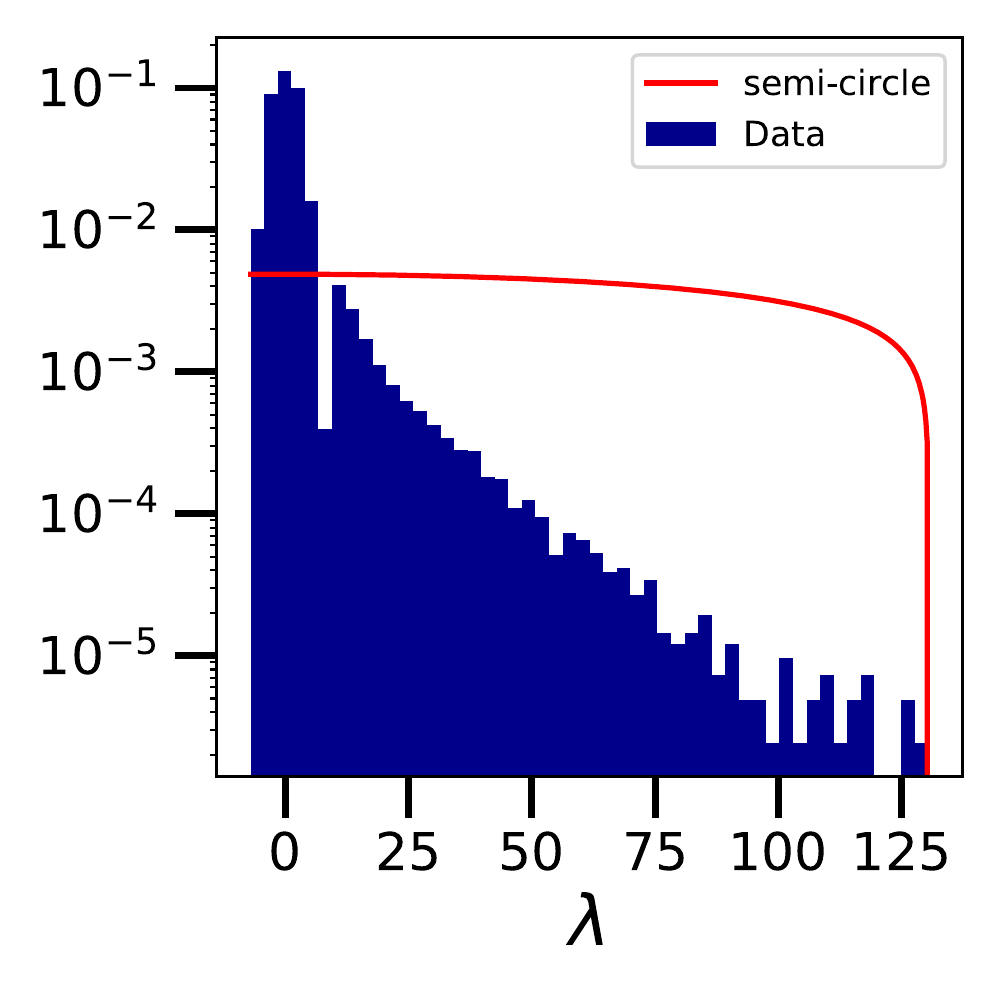}
        \caption{$k''(0)=0.0001\dagger$}
    \end{subfigure}

    \caption{Spectral densities of Gaussian process Hessians with various kernel choices. All use $k'(0)=1$. The dimension is $300$ in all cases except (d), in which the Hessian is padded to 400 dimensions with zeros. All histograms are produced with 100 independent Hessian samples. $* = 100$ degenerate directions. $\dagger = 20$ outliers}
\label{fig:gp_kernel_densities}
\end{figure}

\begin{figure}
    \centering
    \begin{subfigure}[b]{0.3\textwidth}
        \centering
        \includegraphics[width=\textwidth]{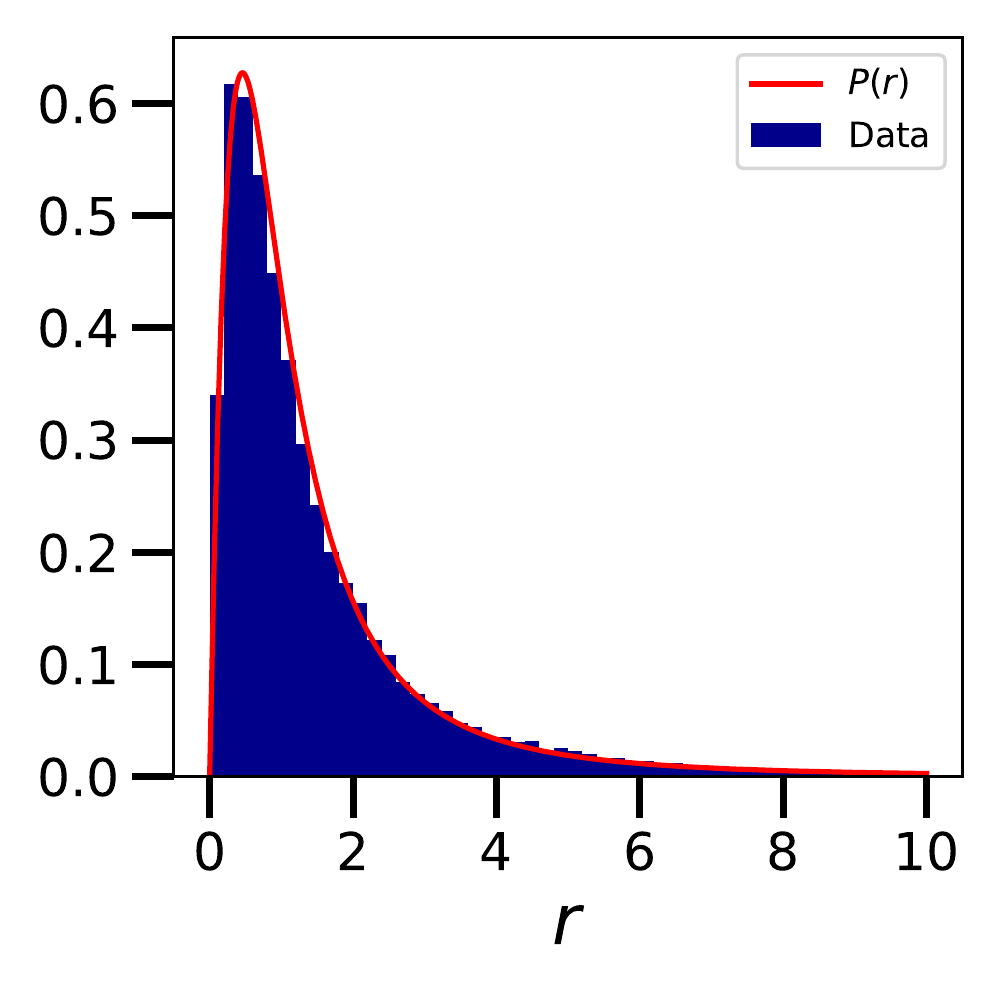}
        \caption{$k''(0)=10^{-4}$}
    \end{subfigure}
    \begin{subfigure}[b]{0.3\textwidth}
        \centering
        \includegraphics[width=\textwidth]{509ed5e9664bc848.pdf}
        \caption{$k''(0)=10^{-3}$}
    \end{subfigure}
    \begin{subfigure}[b]{0.3\textwidth}
        \centering
        \includegraphics[width=\textwidth]{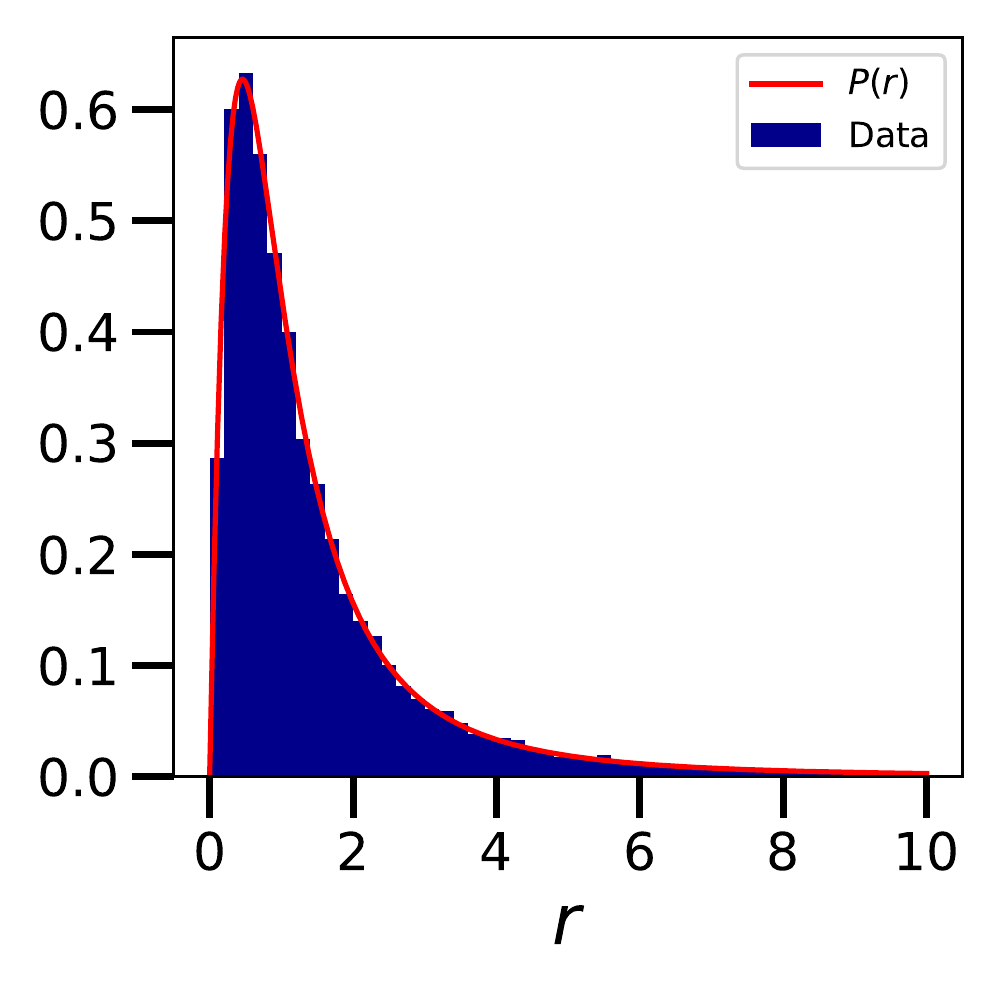}
        \caption{$k''(0)=10^{-1}$}
    \end{subfigure}
    
    \begin{subfigure}[b]{0.3\textwidth}
        \centering
        \caption{$k''(0)=10$}
    \end{subfigure}
    \begin{subfigure}[b]{0.3\textwidth}
        \centering
        \includegraphics[width=\textwidth]{509ed5e9664bc848.pdf}
        \caption{$k''(0)=10^{-3}*$}
    \end{subfigure}
    \begin{subfigure}[b]{0.3\textwidth}
        \centering
        \includegraphics[width=\textwidth]{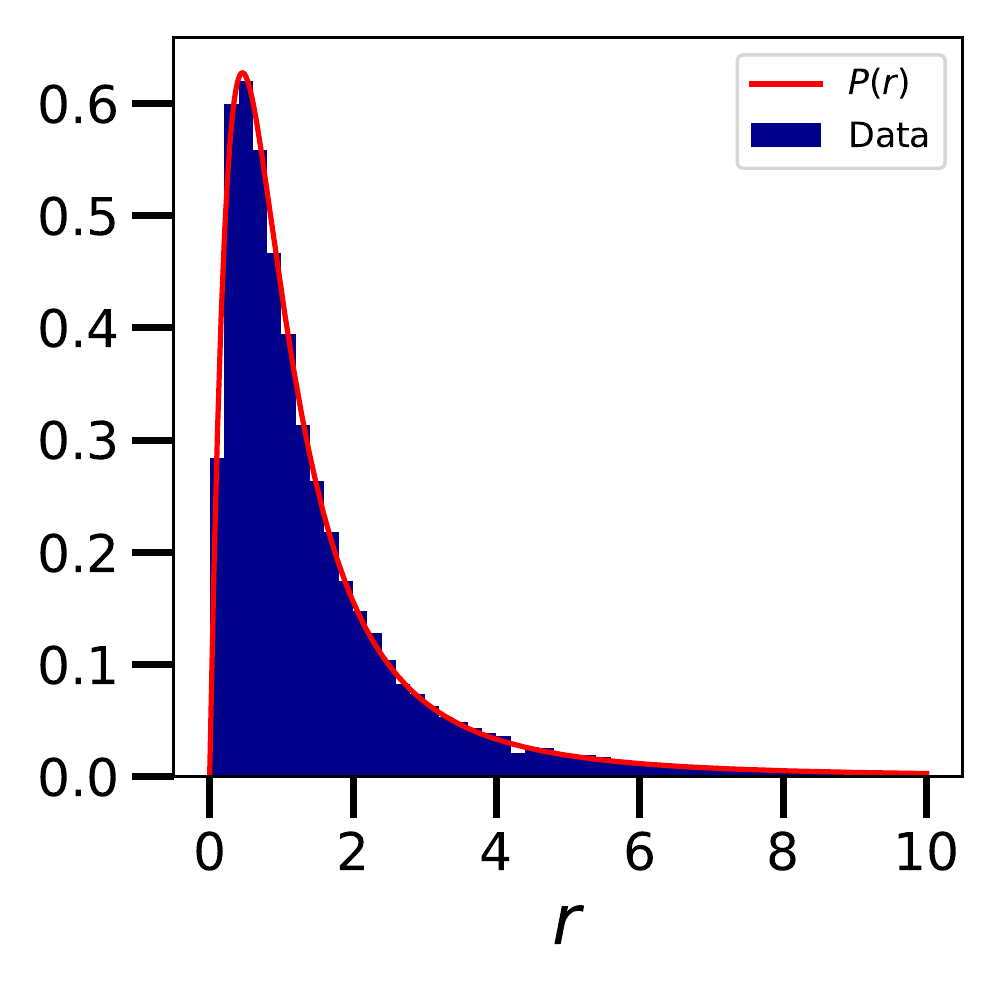}
        \caption{$k''(0)=0.0001\dagger$}
    \end{subfigure}

    \caption{Consecutive spacing ratios of Gaussian process Hessians with various kernel choices. All use $k'(0)=1$. The dimension is $300$ in all cases except (d), in which the Hessian is padded to 400 dimensions with zeros. $* = 100$ degenerate directions. $\dagger = 20$ outliers.}
    \label{fig:gp_kernel_ratios}
\end{figure}

Differentiating (\ref{eq:grad_gp_covar_2}) further, we obtain 
\begin{align}
   & Cov(\partial_{ij}\mathcal{L}_{emp}(\vec{w}), \partial_{kl} \mathcal{L}_{emp}(\vec{w}))  = k''(0)\left(\delta_{ik}\delta_{jl} + \delta_{il}\delta_{jk}\right) + k'(0)^2 \delta_{ij}\delta_{kl}
\end{align}
The Hessian $\mH_{emp}$ has Gaussian entries with mean zero, so the distribution of $\mH_{emp}$ is determined entirely by $k'(0)$ and $k''(0)$. Neglecting to choose $k$ explicitly, we vary the values of $k'(0)$ and $k''(0)$ to produce nearest neighbour spectral spacings ratios and spectral densities. The histograms for spectral spacing ratios are indistinguishable and agree very well with the GOE, as shown in Figure \ref{fig:gp_kernel_ratios}. The spectral densities are shown in Figure \ref{fig:gp_kernel_densities}, including examples with rank degeneracy, introduced by defining $k$ only on a lower-dimensional subspace of the input space, and outliers, introduced by adding a fixed diagonal matrix to the Hessian. Figure \ref{fig:gp_kernel_densities} shows varying levels of agreement with the semi-circle law, depending on the choice of $k'(0), k''(0)$.

\vivacom{
\begin{remark}
    The covariance structure in (\ref{eq:grad_gp_covar_2}) is very close to that of a GOE matrix. If $k'(0) = 0$ and $k''(0)\neq =0$, then the covariance would be exactly that of a GOE matrix. With general values $k'(0)\neq 0 \neq k''(0)$, we see that the second term is non-zero on, and only on, the diagonals of the Hessian, however it does induce dependence between all diagonal elements. We have been unable to compute the limiting spectral density exactly but we suspect it may well be possible.
\end{remark}}

\subsection{Beyond the MLP}
\begin{figure}[h!]
\centering
    \begin{subfigure}[b]{0.35\textwidth}
        \centering
        \includegraphics[width=\textwidth]{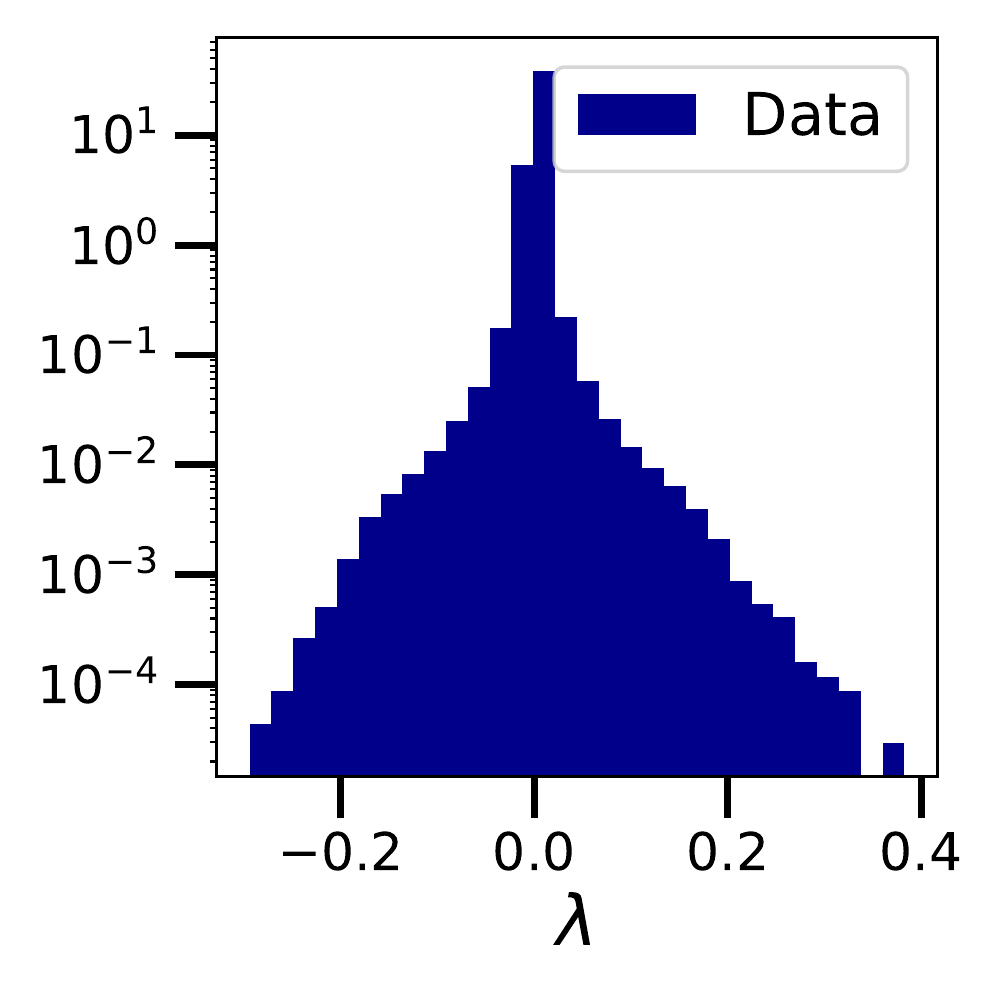}
        \caption{Mean spectral density.}
    \end{subfigure}
    \begin{subfigure}[b]{0.35\textwidth}
        \centering
        \includegraphics[width=\textwidth]{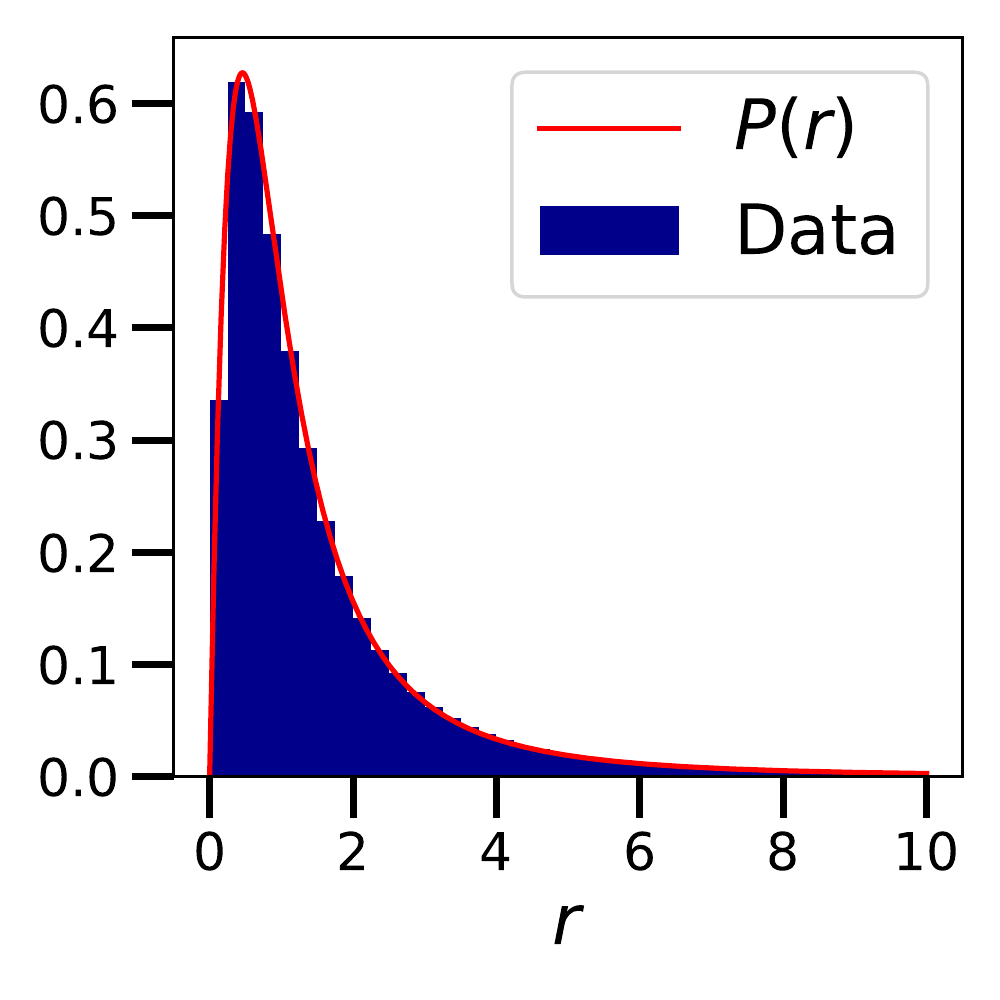}
        \caption{Spacing ratios.}
    \end{subfigure}
    \caption{Spectral statistics for the Hessian of a CNN trained on CIFAR10. Hessians computed over batches of size 64 on the test set.}
    \label{fig:cifar_lenet}
\end{figure}

Figure \ref{fig:cifar_lenet} shows the mean spectral density and adjacent spacing ratios for the Hessian of a CNN trained on CIFAR10. As with the MLP networks and MNIST data considered above, we see an obviously non-semicircular mean level density but the adjacent spacing ratios are nevertheless described by the universal GOE law.

\subsection{Beyond image classification}
\begin{figure}[h!]
\centering
    \begin{subfigure}[b]{0.35\textwidth}
        \centering
        \includegraphics[width=\textwidth]{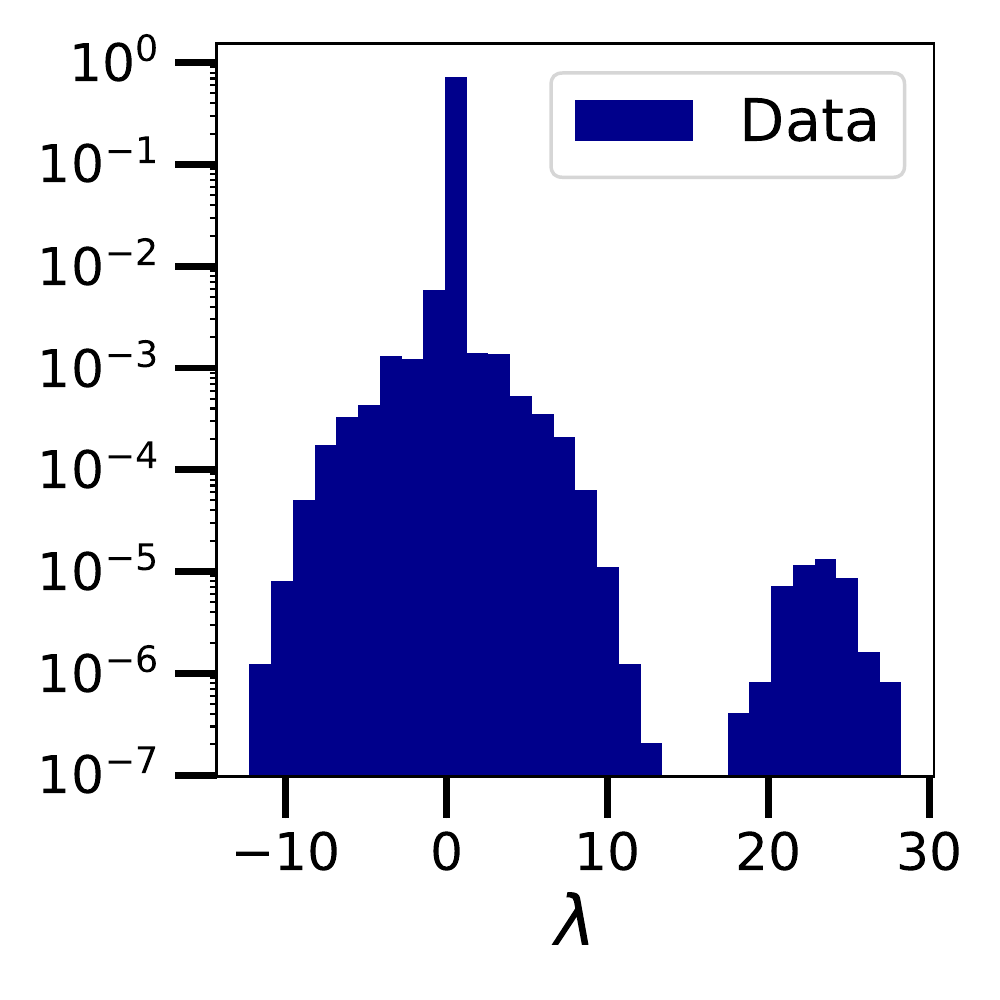}
        \caption{Mean spectral density.}
    \end{subfigure}
    \begin{subfigure}[b]{0.35\textwidth}
        \centering
        \includegraphics[width=\textwidth]{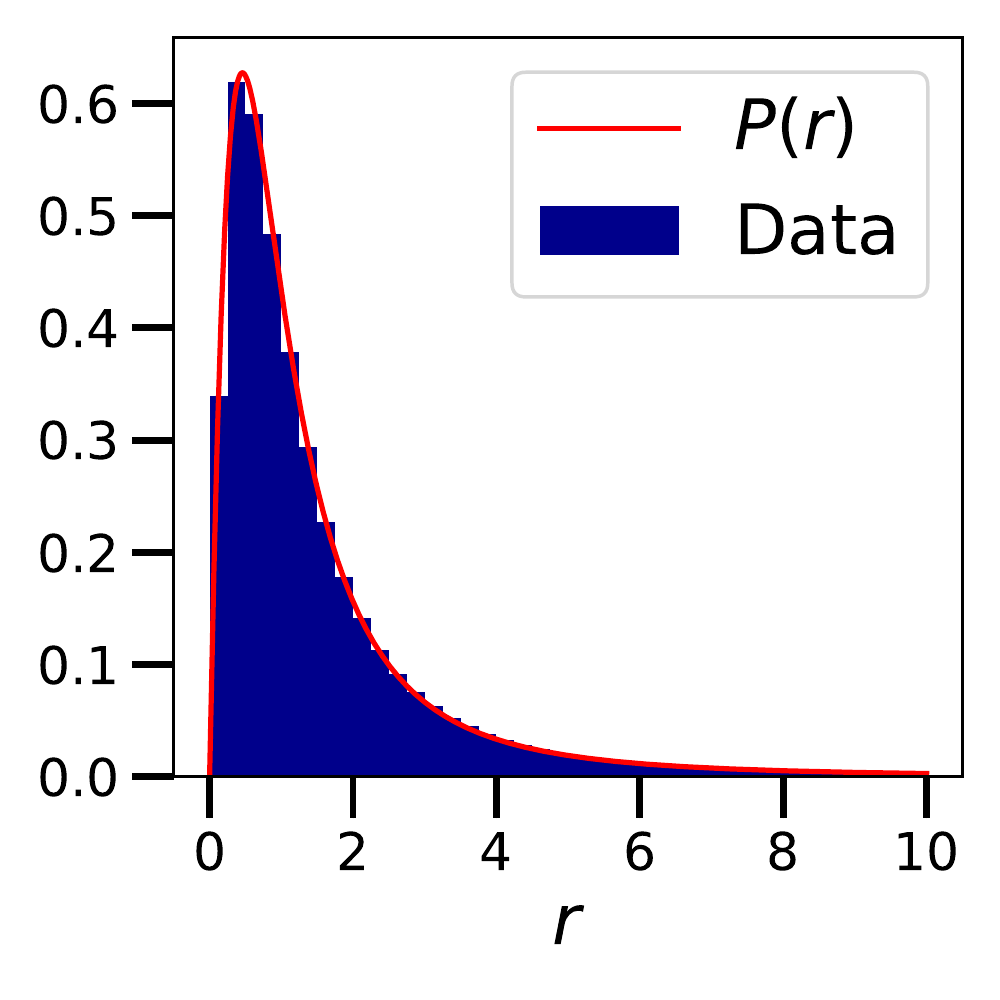}
        \caption{Spacing ratios.}
    \end{subfigure}
    \caption{Spectral statistics for the Hessian of an MLP trained on the Bike dataset. Hessians computed over batches of size 64 on the test set.}
    \label{fig:mlp_bike_hessian}
\end{figure}

Figure \ref{fig:mlp_bike_hessian} shows the mean spectral density and adjacent spacing ratios for the Hessian of an MLP trained on the Bike dataset. Once again we see an obviously non-semicircular mean level density but the adjacent spacing ratios are nevertheless described by the universal GOE law. This serves to demonstrate that there is nothing special about image data or, more importantly, high input feature dimension, since the Bike dataset has only 13 input features.

\subsection{Beyond the Hessian}
\begin{figure}[h!]
\centering
    \begin{subfigure}[b]{0.35\textwidth}
        \centering
        \includegraphics[width=\textwidth]{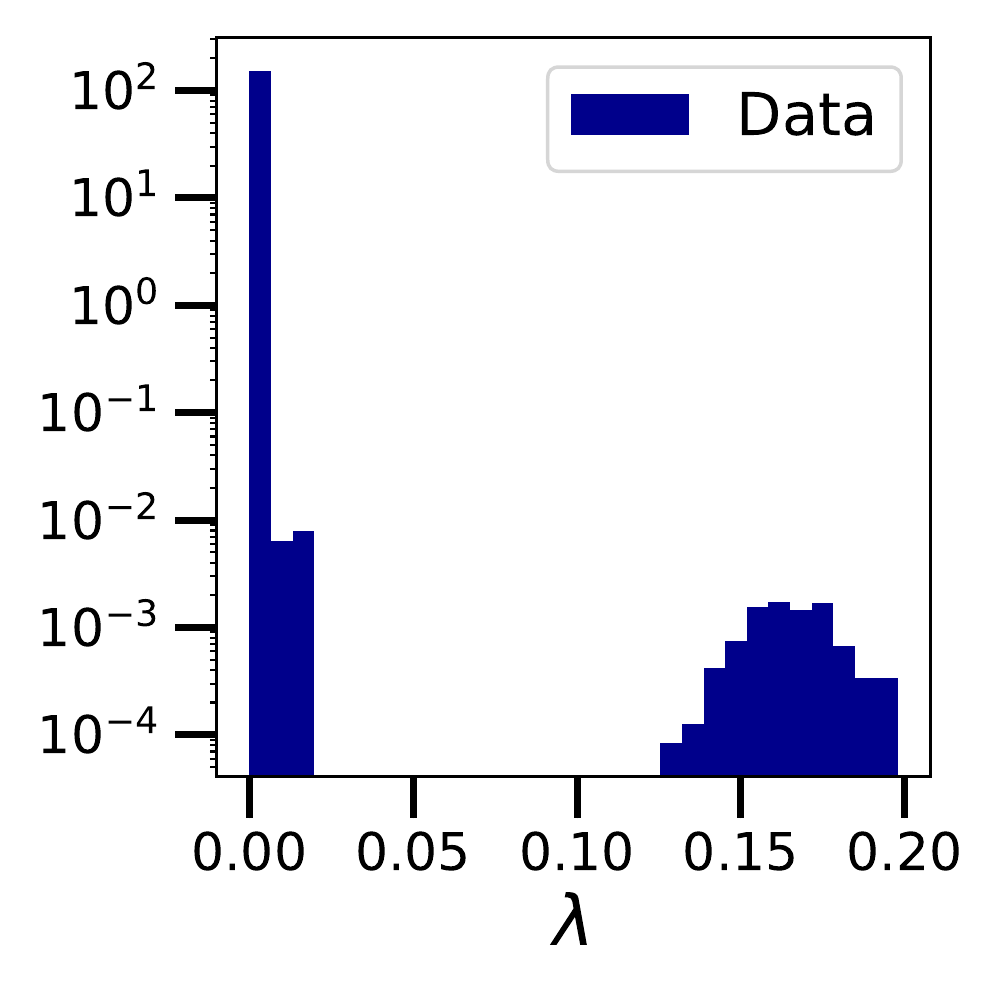}
        \caption{Mean spectral density.}
    \end{subfigure}
    \begin{subfigure}[b]{0.35\textwidth}
        \centering
        \includegraphics[width=\textwidth]{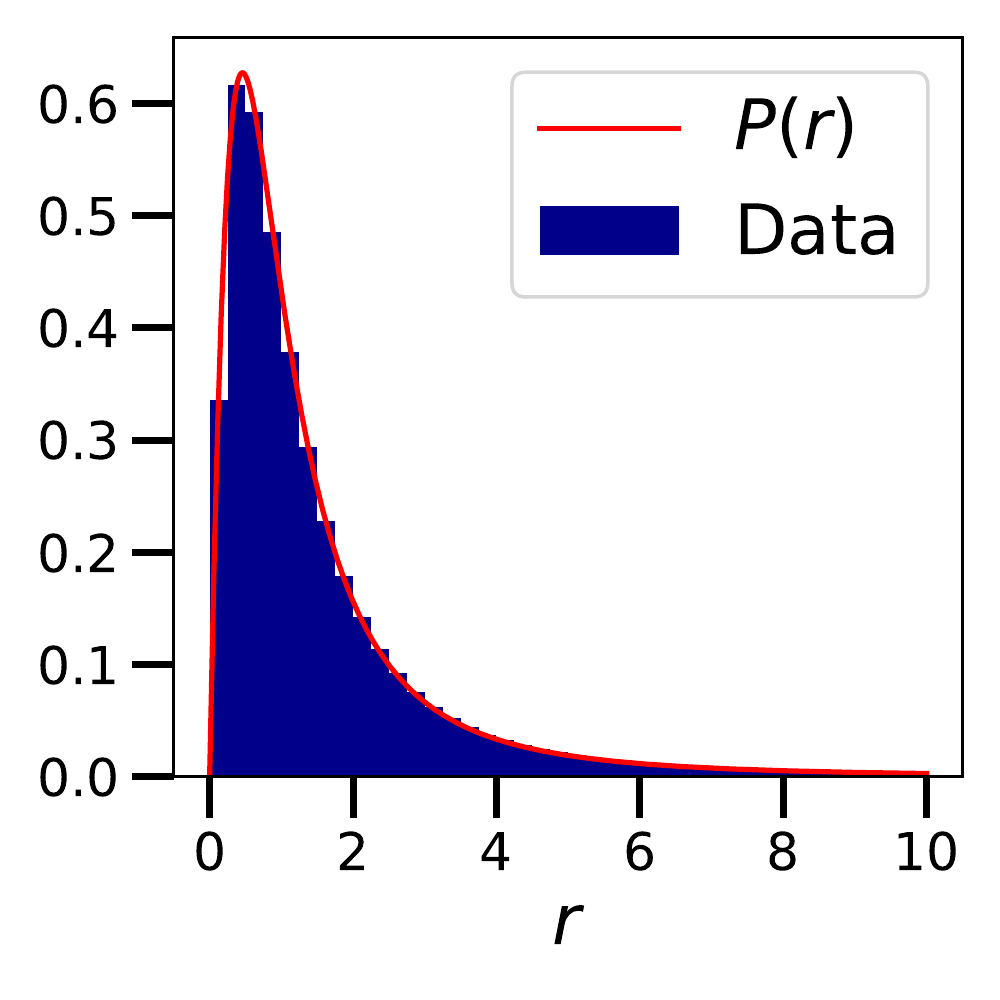}
        \caption{Spacing ratios.}
    \end{subfigure}
    \caption{Spectral statistics for the Gauss-Newton matrix of an MLP trained on the Bike dataset. Matrices computed over batches of size 64 on the test set.}
    \label{fig:mlp_bike_gn}
\end{figure}

Given that the Hessian is not the only matrix of interest in Machine Learning, it is pertinent to study whether our empirical results hold more generally. There have been lots of investigations for the Gauss-Newton \cite{loke2002comparison,pennington2017geometry}, or generalised Gauss-Newton (which is the analogue of the Gauss-Newton when using the cross entropy instead of square loss) matrices, particularly in the fields of optimisation \cite{dauphin2014identifying,martens2012training,martens2015optimizing,martens2014new}. We consider the Gauss-Newton of the network trained on the Bike dataset with square loss. In this case the Gauss Newton $\vec{G} = \vec{J^{T}}\vec{J}$ shares the same non-null subspace as the Neural Tangent Kernel (NTK) \cite{jacot2018neural,cai2019gram}, where $\vec{J}$ denotes the Jacobian, i.e the derivative of the output with respect to the weights, which in this case is simply a vector. The NTK is used for the analysis of trajectories of gradient descent and is particularly interesting for large width networks, where it can be analytically shown that weights remain close to their initialisation and the network is well approximated by its linearisation.
Figure \ref{fig:mlp_bike_gn} shows the mean spectral density and adjacent spacing ratios for the Gauss-Newton matrix of an MLP trained on the Bike dataset. The results are just as for the Hessians above: universal GOE spacings, but the mean density is very much not semicircular. This is an interesting result because even for a different matrix employed in a different context we still see the same universal RMT spacings.

\section{Conclusion}
We have demonstrated experimentally the existence of random matrix statistics in small neural networks on the scale of the mean eigenvalue separation. This provides the first  direct evidence of universal RMT statistics present in neural networks trained on real datasets. Hitherto the role of random matrix theory in deep learning has been unclear. Prior work has studied theoretical models with specific assumptions leading to specific random matrix ensembles. Though certainly insightful, it is not clear to what extent any of these studies are applicable to real neural networks. This work aims to shift the focus by demonstrating the clear presence of universal random matrix behaviour in real neural networks. We expect that future theoretical studies will start from this robust supposition. 

When working with a neural network on some dataset, one has information a priori about its Hessian. Its distribution and correlation structure may well be entirely inaccessible, but correlations between Hessian eigenvalues on the local scale can be assumed to be universal and overall the matrix can be rightly viewed as a random matrix possessing universal local statistics.

We focus on small neural networks where Hessian eigendecomposition is feasible. Future research that our work motivates could develop methods to approximate the level spacing distribution of large deep neural networks for which exact Hessian spectra cannot be computed. If the same RMT statistics are found, this would constitute a profound universal property of neural networks models; conversely, a break-down in these RMT statistics would be an indication of some fundamental separation between different network sizes or architectures.

A few recent works \cite{loureiro2021capturing, goldt2020gaussian, adlam2020neural} considered and used the idea of \emph{Gaussian equivalence} to make theoretical progress in neural network models with fewer assumptions than previously required (e.g. on the data distribution). The principle is that complicated random matrix distributions on non-linear functions of random matrices can be replaced in  calculations training and test loss by their Gaussian equivalents, i.e. Gaussian matrices with matching first and second moments. This idea reflects a form of universality and can drastically increase the tractability of calculations. The random matrix universality we have here demonstrated in neural networks may be related, and should be considered as a possible source of other analogous universality simplifications that can render realistic but intractable models tractable.

One intriguing possible avenue is the relation to chaotic systems. Quantum systems with chaotic classical limits are know to display RMT spectral pairwise correlations, whereas Poisson statistics correspond to integrable systems. We suggest that the presence of GOE pairwise correlations in neural network Hessians, as opposed to Poisson, indicates that neural network training dynamics cannot be reduced to some simpler, smaller set of dynamical equations.
\chapter{Universal characteristics of loss surfaces}\label{chap:univ}
The content of this chapter was published first as a pre-print in May 2022 (\url{https://arxiv.org/abs/2205.08601}) and later as a journal article in December 2022: ``Universal characteristics of neural network loss surfaces from random matrix theory''. \textbf{Nicholas P Baskerville}, Jonathan P Keating, Francesco Mezzadri, Joseph Najnudel and Diego Granziol. \emph{Journal of Physics A: Mathematical and Theoretical}.
\medskip

\textbf{NPB} proposed all three of the main ideas, performed all calculations, proved most of the results, did the vast majority of the write-up and did all analysis of experimental results. DG conducted the neural network training, extracted the empirical Hessian data and contributed to the write-up of sections pertaining to experiments. JN provided several important ideas for the proof in Appendix A. Anonymous reviewers provided helpful feedback on the presentation and spotted several typos.

\section{General random matrix model for loss surface Hessians}\label{sec:hess_model}

\subsection{The model}\label{subsec:hess_model}
Given a loss function $\mathcal{L}: \mathscr{Y}\times \mathscr{Y}\rightarrow\R$, a data generating distribution $\P_{\text{data}}$ supported on $\mathscr{X}\times \mathscr{Y}$ and a neural network $f_{\vec{w}}: \mathscr{X}\rightarrow\mathscr{Y}$ parametrised by $\vec{w}\in\mathbb{R}^N$, its batch Hessian is given by
\begin{align}\label{eq:batch_hessian_def}
    H_{\text{batch}} = \frac{1}{b}\sum_{i=1}^b \frac{\partial^2}{\partial \vec{w}^2} \mathcal{L}(f_{\vec{w}}(\vec{x_i}), y_i), ~~ (\vec{x}_i, y_i)\overset{\text{i.i.d.}}{\sim}\mathbb{P}_{\text{data}}
\end{align}
and its true Hessian is given by
\begin{align}
    H_{\text{true}} = \E_{(\vec{x}, y)\sim\mathbb{P}_{\text{data}}} \frac{\partial^2}{\partial \vec{w}^2}  \mathcal{L}(f_{\vec{w}}(\vec{x}), y).
\end{align}
Both  $H_{\text{batch}}$ and $H_{\text{true}}$ are $N\times N$ matrix functions of $\vec{w}$; $H_{\text{batch}}$ is random but $H_{\text{true}}$ is deterministic. Only in very specific cases and under strong simplifying assumptions can one hope to obtain the distribution of $H_{\text{batch}}$ or the value of $H_{\text{true}}$ from $\mathcal{L}, \P_{\text{data}}$ and $f_{\vec{w}}$. Inspired by the success of many random matrix theory applications, e.g. in Physics, we will instead seek to capture the essential features of deep neural network Hessians in a sufficiently general random matrix model.

\medskip
We introduce the following objects:
\begin{itemize}
    \item A sequence (in $N$) of random real symmetric $N\times N$ matrices $X$. $X$ possesses a limiting spectral probability measure $\mu$, i.e. if $\lambda_1,\ldots,\lambda_N$ are the eigenvalues of $X$ then
    \begin{align}
        \frac{1}{N}\sum_{i=1}^N \delta_{\lambda_i} \rightarrow \mu
    \end{align}
    weakly almost surely. We further assume that $\mu$ has compact support and admits \review{a} smooth density with respect to Lebesgue measure.
    
    \item A sequence (in $N$) of deterministic real symmetric $N\times N$ matrices $A$ with eigenvalues
    \begin{align}
        \theta_1,\ldots,\theta_p,\xi_1,\ldots \xi_{N-p-q},\theta_1',\ldots,\theta_q'
    \end{align}
    for fixed integers $p, q$. We assume the existence of limiting measure $\nu$ such that, weakly, \begin{align}
        \frac{1}{N-p-q}\sum_{i=1}^{N-p-q} \delta_{\xi_i} \rightarrow \nu
    \end{align}
    where $\nu$ is a compactly supported probability measure. The remaining eigenvalues satisfy 
    \begin{align}
        \theta_1 >\ldots > \theta_p > \rmes(\nu), ~ \theta_1' < \ldots < \theta_q' < \lmes(\nu).
    \end{align}
    $\nu$ is also assumed to be of the form $\nu = \epsilon\eta + (1-\review{\epsilon})\delta_0$ where $\eta$ is a compactly supported probability measure which admits a density with respect to Lebesgue measure.
    \item A decreasing function $\sbplain: \N \rightarrow (0, 1)$.
\end{itemize}
With these definitions, we construct the following model for the Hessian:
\begin{align}\label{eq:hessian_def}
    H_{\text{batch}}\equiv H = \sbf X + A
\end{align}
where $b$ is the batch size. We have dropped the subscript on $H_{\text{batch}}$ for brevity. Note that $H$ takes the place of the batch Hessian and $A$ taken the place of the true Hessian. $\sbf X$ takes the place of the random noise introduced by sampling a finite batch at which to evaluate the Hessian. $\sbf$ is an overall scaling induced in $X$ by the batch-wise averaging.

\medskip
This model is almost completely general. Note that we allow the distribution of $X$ and the value of $A$ to depend on the position in weight space $\vec{w}$. The only restrictions imposed by the model are
\begin{enumerate}
    \item the existence of $\nu$;
    \item the position of $\theta_i, \theta_j'$ relative to the support of $\nu$;
    \item $\nu$ may only possess an atom at $0$;
    \item the fixed number of  $\theta_i, \theta_j'$;
    \item the existence of $\mu$;
    \item the existence of the scaling $\sbf$ in batch size.
\end{enumerate}
All of the above restrictions are discussed later in the section. Finally, we must introduce some properties of the noise model $X$ in order to make any progress. We introduce the assumption that the eigenvectors of $X$ obey \emph{quantum unique ergodicity} (QUE)\review{ \cite{bourgade2017eigenvector}}. The precise meaning of this assumption and a thorough justification and motivation is given later in this section. For now it suffices to say that QUE roughly means that the eigenvectors of $X$ are \emph{delocalised} or that they behave roughly like the rows (or columns) of a uniform random $N\times N$ orthogonal matrix (i.e. a matrix with Haar measure). QUE is known to hold for standard ensembles in random matrix theory, such as quite general Wigner matrices, Wishart matrices, adjacency matrices of certain random graphs etc. Moreover, as discussed further section \ref{subsec:que_just} below, QUE can be thought of as a property of quite general random matrix models.

\subsection{Quantum unique ergodicity}
Quantum unique ergodicity was introduced in Chapter \ref{chap:maths} but for convenience we recall some details here.
It is well known that the eigenvectors of quite general random matrices display a universal property of \emph{delocalisation}, namely
\begin{align}
    |u_k|^2 \sim \frac{1}{N}
\end{align}
for any component $u_k$ of an eigenvector $\vec{u}$.
Universal delocalisation was conjectured by Wigner along with the Wigner surmise for adjacent eigenvalue spacing.
Both of these properties, and the more familiar phenomenon of universal correlation functions on the microscopic scale have since been rigorously established for quite a variety of matrix models e.g. \cite{erdos2017dynamical, erdHos2012universality,erdHos2019random}.
\cite{bourgade2017eigenvector} show that the eigenvectors of generalised Wigner matrices obey \emph{Quantum unique ergodicity}, a particular form of delocalisiation, stronger than the above statement. Specifically, they are shown to be approximately Gaussian in the following sense (\cite{bourgade2017eigenvector} Theorem 1.2):
\begin{align}\label{eq:que_def}
    \sup_{||\vec{q}|| = 1}\sup_{\substack{I\subset [N],\\ |I| = n}} \left|\E P\left(\left(N|\vec{q}^T\vec{u}_k|^2\right)_{k\in I}\right) - \E P\left(\left(|\mathcal{N}_j|^2\right)_{j=1}^{\review{n}}
    \right)\right| \leq N^{-\epsilon},
\end{align}
for large enough $N$, where $\mathcal{N}_j$ are i.i.d. standard normal random variables, $(\vec{u}_k)_{k=1}^N$ are the normalised eigenvectors, $P$ is any polynomial in $n$ variables and $\epsilon > 0$. \review{Note that the set $I$ in this statement is a subset of $[N]$ of \emph{fixed size} $n$; $n$ is not permitted to depend on $N$.}

\subsection{Batch Hessian outliers}\label{sec:interlude}
Let $\{\lambda_i\}$ be the eigenvalue\review{s} of $H$. To set the context of our results, let us first simplify and suppose momentarily that $\sbplain = 1$ and, instead of mere QUE, $X$ has eigenvectors distributed with Haar measure, and $A$ is fixed rank, i.e. $\xi_i=0 ~\forall i$, then the results of \cite{benaych2011eigenvalues} would apply and give \begin{align}
    \lambda_j \overset{a.s.}{\rightarrow} \begin{cases}
                    g_{\mu}^{-1}(1/\theta_j) ~~&\text{if } \theta_j > 1/g_{\mu}(\rmes(\mu)),\\
                    \rmes(\mu) &\text{otherwise},
                \end{cases}\label{eq:bgn_outlier_1}
\end{align}
for $j=1,\ldots, p$, and 
\begin{align}
    \lambda_{N-j+1} \overset{a.s.}{\rightarrow} \begin{cases}
                    g_{\mu}^{-1}(1/\theta_j') ~~&\text{if } \theta_j' < 1/g_{\mu}(\lmes(\mu)),\\
                    \lmes(\mu) &\text{otherwise},
                \end{cases}\label{eq:bgn_outlier_2}                
\end{align}
for $j=1,\ldots, q$.
\review{
What follows is our main results for the outliers of $H$ under the general conditions described above.
\begin{theorem}\label{thm:outlier}
Let $H$ be the Hessian matrix model defined in (\ref{eq:hessian_def}) and meeting all the conditions in Section \ref{sec:hess_model}. Then there exist $U_{\epsilon}, L_{\epsilon}\in\mathbb{R}$ such that, for $j=1, \ldots, p$,
\begin{align}
    \lambda_j  = \begin{cases} \omega^{-1}(\theta_j) & \text{ if } \omega^{-1}(\theta_j)> U_{\epsilon},\\
    U_{\epsilon} & \text{ otherwise}.
    \end{cases}\label{eq:top_outlier_thm}
\end{align}
and for  $j=1, \ldots, q$,
\begin{align}
    \lambda_{N-j+1}  = \begin{cases} \omega^{-1}(\theta_j') & \text{ if } \omega^{-1}(\theta_j) < L_{\epsilon},\\
    L_{\epsilon} & \text{ otherwise},
    \end{cases}\label{eq:btm_outlier_thm}
\end{align}
and
\begin{align}
\omega^{-1}(\theta)  = \theta + \sbf R_{\mu}(\sbf\theta^{-1}) + \epsilon \sbf^2 d_{\eta}(\theta)R_{\mu}'(\sbf\theta^{-1}) + \mathcal{O}(\epsilon^2)
\end{align}
\end{theorem}
where we define $d_{\eta}(z) = g_{\eta}(\theta_j) - \theta_j^{-1}$.
}

\paragraph{\textbf{An interlude on prior outlier results}} {\em It was conjectured in \cite{benaych2011eigenvalues} that (\ref{eq:bgn_outlier_1})-(\ref{eq:bgn_outlier_2}) still hold when $X$ has delocalised eigenvectors in some sense, rather than strictly Haar. Indeed, a careful consideration of the proof in that work does reveal that something weaker than Haar would suffice, for example QUE. See in particular the proof of the critical Lemma 9.2 therein which can clearly be repeated using QUE. There is a considerable subtlety here, however, which is revealed best by considering more recent results on deformations of general Wigner matrices. \cite{knowles2017anisotropic} shows that very general deterministic deformations of general Wigner matrices possess an optimal anisotropic local law, i.e. $Y + B$ for Wigner $Y$ and deterministic symmetric $B$. It is expected therefore that $Y+B$ has delocalised eigenvectors in the bulk. Consider the case where $B$ is diagonal, and say that $B$ has a fixed number of ``spike'' eigenvalues $\varphi_1>\ldots>\varphi_r$ and remaining eigenvalues $\zeta_1,\ldots, \zeta_{N-r}$ where the empirical measure of the $\zeta_i$ converges to some measure $\tau$ and $\varphi_r > \rmes(\tau)$. We can then split $B = B_{i} + B_o$ where $B_i$ contains only the $\zeta_j$ and $B_o$ only the $\varphi_j$. The previously mentioned results applies to $Y + B_i$ and then we might expect the generalised result of \cite{benaych2011eigenvalues} to apply to give outliers $g_{\mu_{SC}\boxplus \tau}^{-1}(1/\varphi_i)$ of $Y + B$. This contradicts, however, another result concerning precisely the the outliers of such generally deformed Wigner matrices. It was shown in \cite{capitaine2016spectrum} that the outliers of $Y+B$ are $\omega^{-1}(\varphi_j)$ where $\omega$ is the subordination function such that $g_{\mu_{SC} \boxplus \tau}(z) = g_{\tau}(\omega(z))$. These two expressions coincide when}
\begin{align}
    &\omega^{-1}(z) = g_{\mu_{SC}\boxplus\tau}^{-1}(z^{-1})\notag\\
    \iff &\omega^{-1}(z) = \omega^{-1}(g_{\tau}^{-1}(z^{-1}))\notag\\
    \iff &g_{\tau}^{-1}(z^{-1}) = z\notag\\
    \iff &g_{\tau}(z) = z^{-1}\notag\\
    \iff &\tau = \delta_0,
\end{align}
{\em i.e. only when $B$ is in fact of negligible rank as $N\rightarrow\infty$. This apparent contradiction is resolved by the observation that the proof in \cite{benaych2011eigenvalues} in fact relies implicitly on an \emph{isotropic local law}. Note in particular section 4.1, which translated to our context, would require $\vec{v}^TG_{Y + B_i}(z)\vec{v}\approx g_{\mu_{SC}\boxplus\tau}(z)$ with high probability for general unit vectors $\vec{v}$. Such a result holds if and only if $Y + B_i$ obeys an isotropic local law and is violated if its local law is instead anistropic, as indeed it is, thanks to the deformation.}

\medskip
\begin{proof}[Proof of Theorem \ref{thm:outlier}]
 The conditions on $X$ required to invoke Theorem \ref{thm:nearly_free_addn} \review{from Section \ref{sec:que}} are satisfied, so we conclude that
\begin{align}
    \hat{g}_{H}(z) = g_{\mu_b \boxplus \nu}(z) + o(1) = g_{\nu}(\omega(z)) + o(1) = \hat{g}_{A}(\omega(z)) + o(1)
\end{align}
where $\omega$ is the subordination function such that $g_{\mu_b\boxplus\nu}(z) = g_{\nu}(\omega(z))$ and $\mu_b$ is the limiting spectral measure of $\sbf X$. The reasoning found in \cite{capitaine2016spectrum} then applies regarding the outliers of $H$. Indeed, suppose that $\lambda$ is an outlier of $H$, i.e. $\lambda$ is an eigenvalue of $H$ contained in $\R \backslash \text{supp}(\mu\boxplus\nu)$. Necessarily $\hat{g}_H$ possesses a singularity at $\lambda$, and so $\hat{g}_A$ must have a singularity at $\omega(\lambda)$. For this singularity to persist for all $N$, $\omega(\lambda)$ must coincide with one of the outliers of $A$ which, unlike the bulk eigenvalues $\xi_j$, remain fixed for all $N$. Therefore we have the following expressions for the outliers of $H$:
\begin{align}\label{eq:outlier_sets}
    \{\omega^{-1}(\theta_j) \mid \omega^{-1}(\theta_j)\in \R \backslash\text{supp}(\mu_b\boxplus\nu)\}\cup \{\omega^{-1}(\theta_j') \mid \omega^{-1}(\theta_j')\in \R \backslash\text{supp}(\mu_b\boxplus\nu)\}.
\end{align}

We now consider $\epsilon$ to be small and analyse these outlier locations as a perturbation in $\epsilon$. Firstly note that 
\begin{align}
    g_{\mu_b}(z) = \int \frac{d\mu_b(x)}{z - x} = \int \frac{d\mu(x/\sbf)}{z - x} = \sbf\int \frac{d\mu(x)}{z - \sbf x} = g_{\mu}(z/\sbf).
\end{align}
Also
\begin{align}
    \omega^{-1}(z) &= g_{\mu_b \boxplus \nu}^{-1}(g_{\nu}(z))\label{eq:omega_inv_def}\\
    &= R_{\mu_b}(g_{\nu}(z)) + g_{\nu}^{-1}(g_{\nu}(z))\notag\\
    &= R_{\mu_b}(g_{\nu}(z)) + z.\label{eq:omega_prog1}
\end{align}
We now must take care in computing $R_{\mu_b}$ from $g_{\mu_b}$.
Recall that the $R$-transform of a measure is defined as a formal power series \cite{anderson2010introduction}
\begin{align}
   R(z) = \sum_{n=0}^{\infty} k_{n+1} z^n
\end{align}
where $k_n$ is the $n$-th cumulant of the measure.
It is known \cite{anderson2010introduction} that $k_n=C_n$ where the functional inverse of the Stieljtes transform of the measure is given by the formal power series
\begin{align}
    K(z) = \frac{1}{z} + \sum_{n=1} C_n z^{n-1}.
\end{align}
Now let $m_n$ be the $n$-th moment of $\mu$ and similarly let $m_n^{(b)}$ be the $n$-th moment of $\mu_b$, so formally
\begin{align}
    g_{\mu}(z) = \sum_{n\geq 0} m_n z^{-(n+1)}, ~~ g_{\mu_b}(z) = \sum_{n\geq 0} m_n^{(b)} z^{-(n+1)}.
\end{align}
Also let $k_n$ be the $n$-th cumulant of $\mu$ and $k_n^{(b)}$ be the $n$-th cumulant of $\mu_b$.
Referring to the proof of Lemma 5.3.24 in \cite{anderson2010introduction} we find the relations \begin{align}
    m_n &= \sum_{r=1}^n \sum_{\substack{0\leq i_1,\ldots, i_r\leq n-r \\ i_1+\ldots + i_r = n-r}} k_r m_{i_1}\ldots m_{i_r},\\
    m_n^{(b)} &= \sum_{r=1}^n \sum_{\substack{0\leq i_1,\ldots, i_r\leq n-r \\ i_1+\ldots + i_r = n-r}} k_r^{(b)} m_{i_1}^{(b)}\ldots m_{i_r}^{(b)}.
\end{align}
\review{Note, in particular, that $m_1 = k_1$.}
But clearly the moments of $\mu_b$ have a simple scaling in $\sbf$, namely $m_n^{(b)} = \sbf^{n} m_n$, hence
\begin{align}
    m_n = \sbf^{-n}\sum_{r=1}^n \sum_{\substack{0\leq i_1,\ldots, i_r\leq n-r \\ i_1+\ldots + i_r = n-r}} k_r^{(b)} m_{i_1}\ldots m_{i_r} \sbf^{n-r}
\end{align}
from which we deduce $k_n^{(b)} = \sbf^n k_n$, which establishes that $R_{\mu_b}(z) = \sbf R_{\mu}(\sbf z)$.
Recalling (\ref{eq:omega_prog1}) we find
\begin{align}
    \omega^{-1}(z) = \sbf R_{\mu} (\sbf g_{\nu}(z)) + z.
\end{align}
The form of $\nu$ gives \begin{align}
    g_{\nu}(z) &= \review{(1-\epsilon)\int \frac{dt}{z-t} \delta_0(t) + \epsilon \int \frac{d\eta(t)}{t-z}=} \frac{1-\epsilon}{z} + \epsilon g_{\eta}(z) = \frac{1}{z} + \epsilon\left(g_{\eta}(z) - \frac{1}{z}\right)
\end{align}
and so we can expand to give 
\begin{align}
    \omega^{-1}(\theta_j) &= \theta_j + \sbf R_{\mu}(\sbf\theta_j^{-1}) + \epsilon \sbf^2 \left(g_{\eta}(\theta_j) - \theta_j^{-1}\right)R_{\mu}'(\sbf\theta_j^{-1}) + \mathcal{O}(\epsilon^2)\notag\\
    &= \theta_j + \sbf R_{\mu}(\sbf\theta_j^{-1}) + \epsilon \sbf^2 d_{\eta}(\theta_j)R_{\mu}'(\sbf\theta_j^{-1}) + \mathcal{O}(\epsilon^2)\label{eq:omega_inv_used_form}
\end{align}
where we have defined $d_{\eta}(z) = g_{\eta}(\theta_j) - \theta_j^{-1}$.
\review{The argument with the lower outliers $\{\theta_j'\}_{j=1}^q$ is identical.}

\medskip
The problem of determining the support of $\mu_b\boxplus\nu$ is difficult and almost certainly analytically intractable, with \cite{bao2020support} containing the most advanced results in that direction.
However overall, we have a model for deep neural network Hessians with a spectrum consisting, with high-probability, of a compactly supported bulk $\mu_b\boxplus\nu$ and a set of outliers given by (\ref{eq:omega_inv_used_form}) (and similarly for $\theta_j'$) subject to (\ref{eq:outlier_sets}).
\review{The constants $L_{\epsilon}, U_{\epsilon}$ in the statement (\ref{eq:top_outlier_thm})-(\ref{eq:btm_outlier_thm}) of the theorem are simply the lower and upper edges of the support of $\supp(\mu_b \boxplus \nu)$.}
\end{proof}

Note that (\ref{eq:omega_inv_used_form}) reduces to outliers of the form $\theta_j + \sbf^2 R_{\mu}(\theta_j^{-1})$ if $\epsilon=0$ or $d_{\eta} = 0$, as expected from \cite{benaych2011eigenvalues}\footnote{Note that $d_{\eta}=0 \iff \eta = \delta_0$ which is clearly equivalent (in terms of $\nu$) to $\epsilon=0$.}.

\medskip
(\ref{eq:omega_inv_used_form}) is a generalised form of the result used in \cite{granziol2020learning}. We have the power series
\begin{align}
    R_{\mu}(\sbf \theta_j^{-1}) &= k_1^{(\mu)} + \frac{k_2^{(\mu)}\sbf}{\theta_j} + \frac{k_3^{(\mu)}\sbf^2}{\theta_j^2} + \ldots,\\
    d_{\eta}(\theta_j) &= \frac{m_1^{(\eta)}}{\theta_j^2} + \frac{m_2^{(\eta)}}{\theta_j^3} + \ldots
\end{align}
where $m_n^{(\eta)}$ are the moments of $\eta$ and $k_n^{(\mu)}$ are the cumulants of $\mu$. In the case that the spikes $\theta_j$ are large enough, we approximate by truncating these power series to give 
\begin{align}
    \omega^{-1}(\theta_j) &\approx \theta_j + \sbf m_1^{(\mu)} + \sbf^2k_2^{(\mu)}\left(\frac{1}{\theta_j} + \frac{\epsilon m_1^{(\eta)}}{\theta_j^2}\right)
\end{align}
where the approximation is more precise for larger $b$ and smaller $\epsilon$ and we have used the fact that the first cumulant of any measure matches the first moment.
One could consider for instance a power law for $\sbf$, i.e.
\begin{align}
    \omega^{-1}(\theta_j) &\approx \theta_j +  \frac{k_1^{(\mu)}}{b^{\upsilon}} + \frac{k_2^{(\mu)}}{b^{2\upsilon}}\left(\frac{1}{\theta_j} + \frac{\epsilon m_1^{(\eta)}}{\theta_j^2}\right) =\theta_j +  \frac{m_1^{(\mu)}}{b^{\upsilon}} + \frac{k_2^{(\mu)}}{b^{2\upsilon}}\left(\frac{1}{\theta_j} + \frac{\epsilon m_1^{(\eta)}}{\theta_j^2}\right) \label{eq:omega_inv_approx_form}
\end{align}
for some $\upsilon > 0$.
In the case that $\mu$ is a semicircle, then all cumulants apart from the second vanish, so setting $\epsilon = 0$ recovers \emph{exactly} \begin{align}
    \omega^{-1}(\theta_j) = \theta_j + \frac{\sigma^2}{4b^{2\upsilon}\theta_j}
\end{align}
where $\sigma$ is the radius of the semicircle.
To make the link with \cite{granziol2020learning} obvious, we can take $\upsilon = 1/2$ and $\mu$ to be the semicircle, so giving
\begin{align}
    \omega^{-1}(\theta_j) \approx \theta_j + \frac{\sigma^2}{4b\theta_j}
\end{align}
where we have truncated $\mathcal{O}(\epsilon)$ term.
We present an argument in favour of the $\upsilon=1/2$ power law below, but we allow for general $\upsilon$ when comparing to experimental data.

\begin{remark}
It is quite possible for $\mu$'s density to have a sharp spike at the origin, or even for $\mu$ to contain a $\delta$ atom at $0$, as observed empirically in the spectra of deep neural network Hessians.
\end{remark}

\subsection{Experimental results} 
The random matrix Hessian model introduced above is quite general and abstract.
Necessarily the measures $\mu$ and $\eta$ must be allowed to be quite general as it is well established experimentally \cite{papyan2018full, granziol2020beyond, baskerville2022appearance} that real-world deep neural network Hessians have spectral bulks that are not familiar as being any standard canonical examples from random matrix theory.
That being said, the approximate form in (\ref{eq:omega_inv_approx_form}) gives quite a specific form for the Hessian outliers.
In particular, the constants $m_1^{(\mu)}, m_1^{(\eta)}$ and $m_2^{(\mu)}, \epsilon > 0$ are shared between all outliers at all batch sizes.
If the form of the Hessian outliers seen in (\ref{eq:omega_inv_approx_form}) is not observed experimentally, it would suggest at least one of the following does not hold:
\begin{enumerate}
    \item batch sampling induces a simple multiplicative scaling on the Hessian noise (\ref{eq:hessian_def});
    \item the true Hessian is approximately low-rank (as measured by $\epsilon$) and has a finite number of outliers;
    \item the Hessian noise model $X$ has QUE.
\end{enumerate}
In view of this third point, agreement with (\ref{eq:omega_inv_approx_form}) provides an indirect test for the presence of universal random matrix statistics in deep neural network Hessians. 

\medskip
We can use Lanczos power methods \cite{meurant2006lanczos} to compute good approximations to the top few outliers in the batch Hessian spectra of deep neural networks \cite{granziol2020learning}.
Indeed the so-called Pearlmutter trick \cite{pearlmutter1994fast} enables efficient numerical computation of Hessian-vector products, which is all that one requires for power methods.
Over a range of batch sizes, we compute the top 5 outliers of the batch Hessian for 10 different batch seeds.
We repeat this procedure at every 25 epochs throughout the training of \review{two standard deep neural networks for computer vision tasks,} VGG16 and WideResNet$28\times10$, on the CIFAR100 dataset \cite{krizhevsky2009learning} and at every epoch during the training of a simple multi-layer perceptron network on the MNIST dataset \cite{lecun1998gradient}.
By the end of training each of the models have high test accuracy, specifically the VGG$16$ architecture which does not use batch normalisation, has a test accuracy of $\approx 75\%$, whereas the WideResNet$28\times10$ has a test accuracy of $\approx 80 \%$. The MLP has a test set accuracy of $\approx 95\%$. \review{Full experimental details are given in Appendix \ref{app:outlier_exp_details}.}

\begin{figure}[h!]
    \centering
    \begin{subfigure}{0.3\linewidth}
     \centering
     \includegraphics[width=\linewidth]{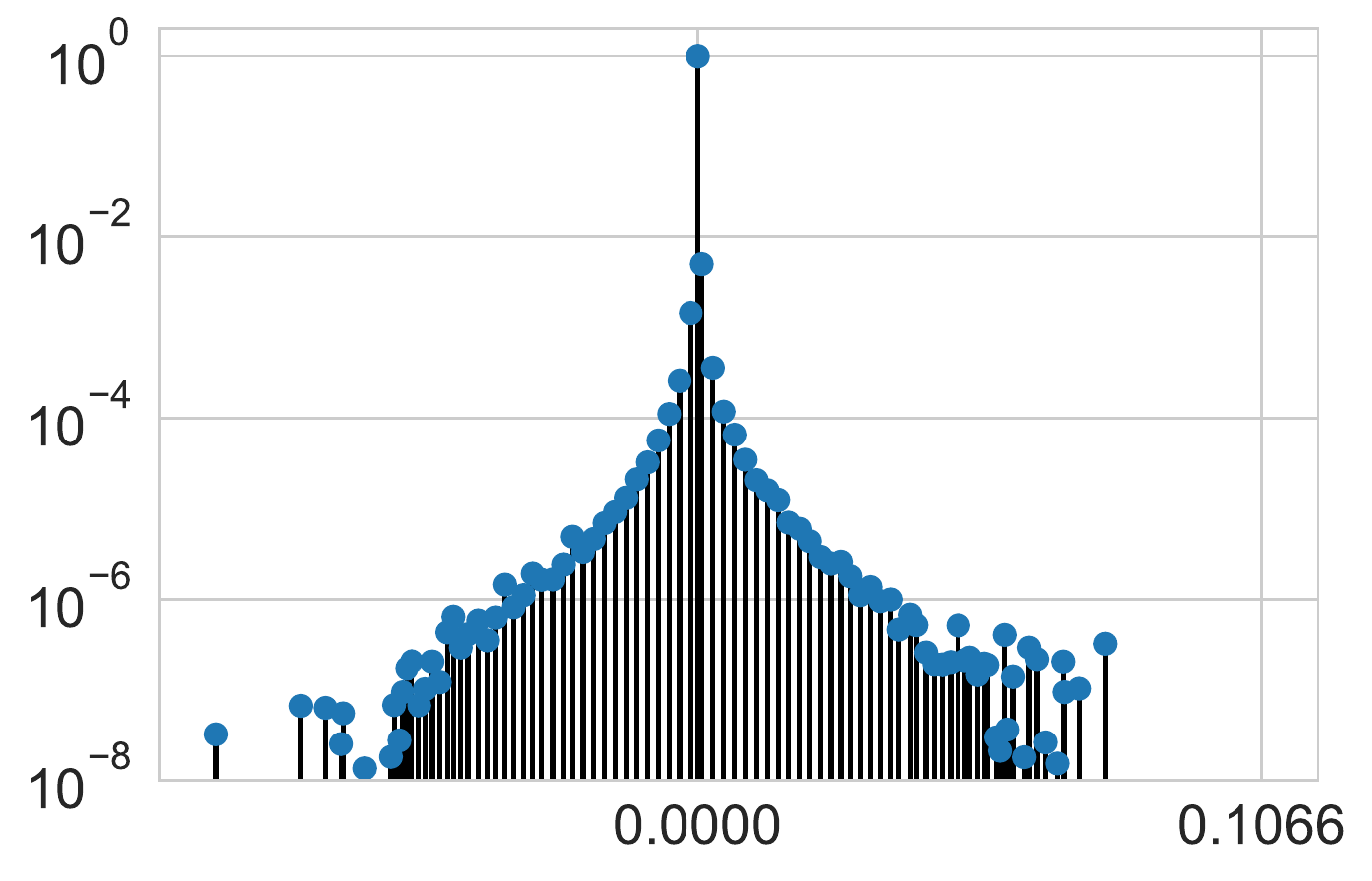}
     \subcaption{Epoch 0}

    \end{subfigure}
    \begin{subfigure}{0.3\linewidth}
     \centering
     \includegraphics[width=\linewidth]{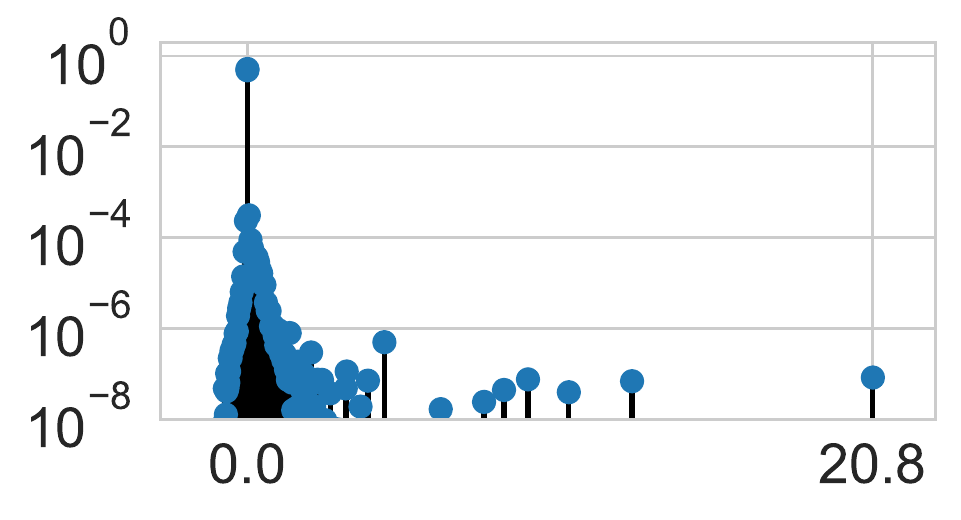}
     \subcaption{Epoch 25}

    \end{subfigure}
    \begin{subfigure}{0.3\linewidth}
     \centering
     \includegraphics[width=\linewidth]{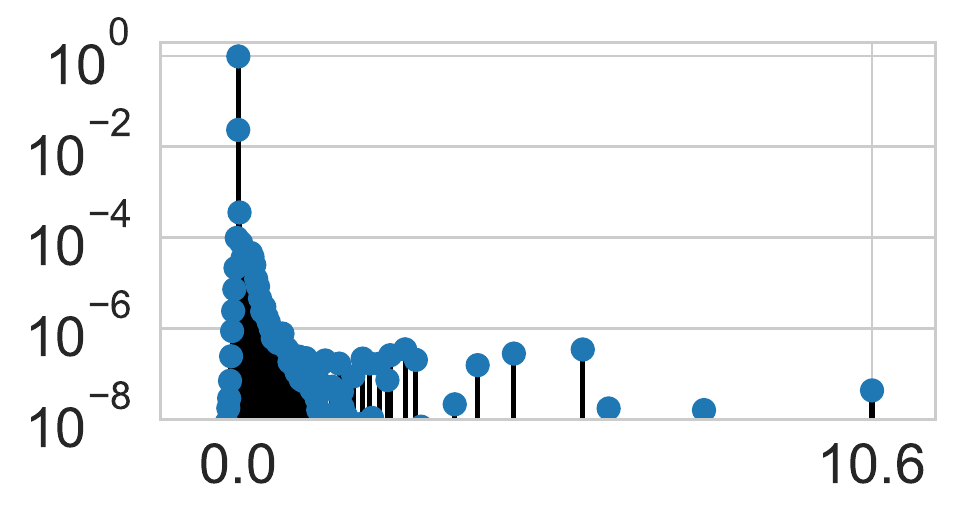}
     \subcaption{Epoch 100}

    \end{subfigure}
    
    \begin{subfigure}{0.3\linewidth}
     \centering
     \includegraphics[width=\linewidth]{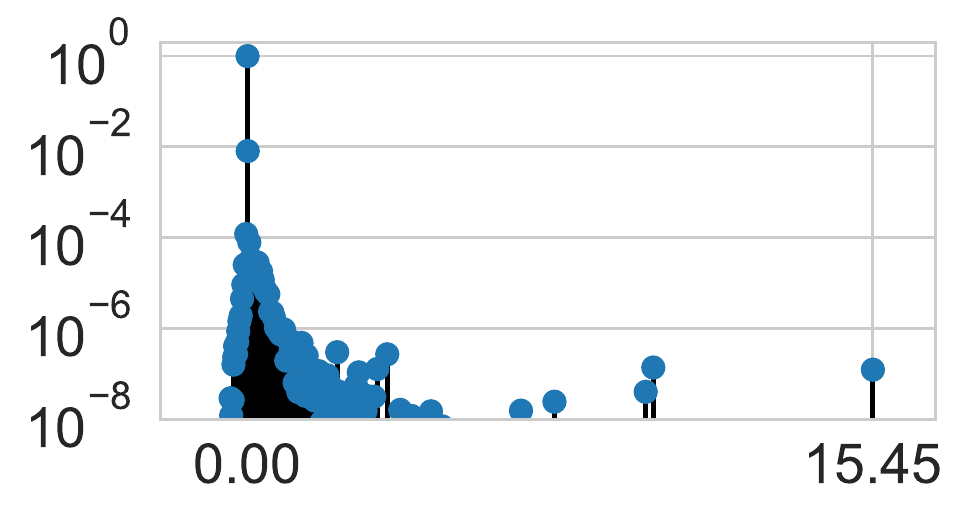}
     \subcaption{Epoch 150}

    \end{subfigure}
    \begin{subfigure}{0.3\linewidth}
     \centering
     \includegraphics[width=\linewidth]{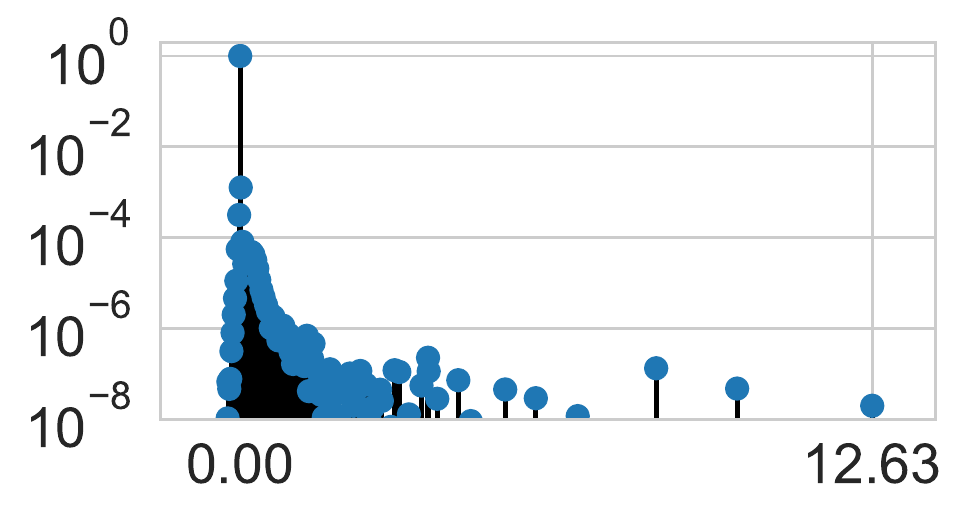}
     \subcaption{Epoch 200}

    \end{subfigure}
    \begin{subfigure}{0.3\linewidth}
     \centering
     \includegraphics[width=\linewidth]{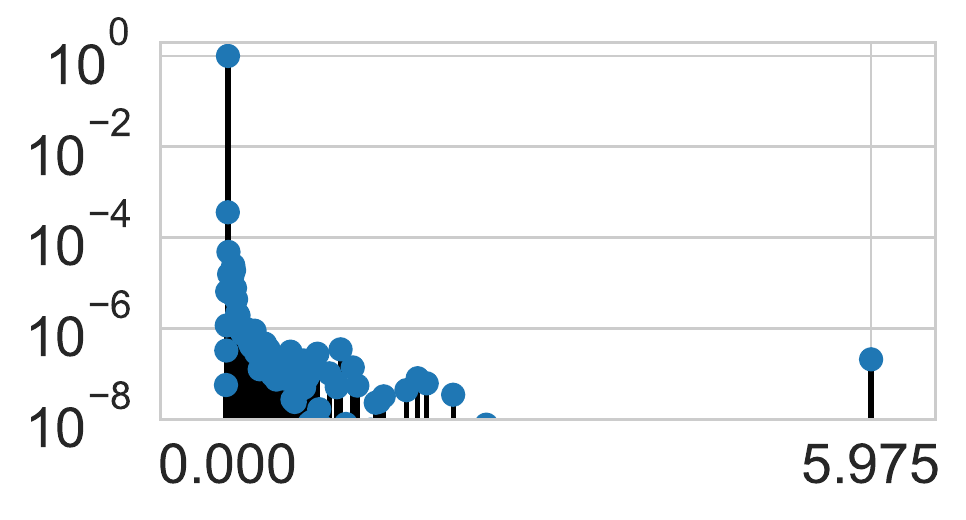}
     \subcaption{Epoch 300}

    \end{subfigure}

    \centering
    \caption{Approximate empirical spectral densities of the Hessian of the VGG16 network trained on MNIST at various stages from initialisation to the end of training. Note the clear presence of large outliers present already at epoch 25.}
    \label{fig:outlier_histos}
\end{figure}

\begin{remark}
There is a subtlety with regard to obtaining the top outliers using the Lanczos power method.
Indeed, since Lanczos provides, in some sense, an approximation to the whole spectrum of a matrix, truncating at $m$ iterations for a $N\times N$ matrix cannot produce good approximations to all of the $m$ top eigenvalues.
In reality, experimental results \cite{papyan2018full,granziol2019deep} show that, for deep neural networks, and using sufficiently many iterations ($m$), the top $r$ eigenvalues may be recovered, for $r\ll m$.
We display some spectral plots of the full Lanczos results in the Figure \ref{fig:outlier_histos} which demonstrate clearly a large number of outliers, and clearly more than $5$. These are not intended to be exhaustive and we recommend references such as \cite{papyan2018full} for detailed discussion of spectral densities like these.
As a result, we can have confidence that our numerical procedure is indeed recovering approximations to the top few eigenvalues required for our experiments.
\end{remark}

\medskip
Let $\lambda^{(i, j, e)}_b$ be the top $i$-th empirical outlier (so $i=1$ is the top outlier) for the $j$-th batch seed and a batch size of $b$ for the model at epoch $e$.
To compare the experimental results to our theoretical model, we propose the following form:
\begin{align}\label{eq:omega_fit_form}
    \lambda^{(i, j, e)}_b \approx \theta^{(i, e)} + \frac{\alpha^{(e)}}{b^{\upsilon}} + \frac{\beta^{(e)}}{b^{2\upsilon}}\left(\frac{1}{\theta^{(i, e)}} + \frac{\gamma^{(e)}}{(\theta^{(i, e)})^2}\right)
\end{align}
where $\beta^{(e)} > 0$ (as the second cumulant of a any measure of non-negative) and $\theta^{(i, e)} > \theta^{(i+1, e)} > 0$ for all $i,e$.
The parameters $\alpha^{(e)}, \beta^{(e)}, \gamma^{(e)}$ and $\theta^{(i, e)}$ need to be fit to the data, which could be done with standard black-box optimisation to minimise squared error in (\ref{eq:omega_fit_form}), however we propose an alternative approach which reduces the number of free parameters and hence should regularise the optimisation problem.
Observe that (\ref{eq:omega_fit_form}) is linear in the parameters $\alpha^{(e)}, \beta^{(e)}, \gamma^{(e)}$ so, neglecting the positivity  constraint on $\beta^{(e)}$, we can in fact solve exactly for optimal values.
Firstly let us define $\barlam$ to be the empirical mean of $\lam$ over the batch seed index $j$.
Each epoch will be treated entirely separately, so let us drop the $e$ superscripts to streamline the notation.
We are then seeking to optimise $\alpha, \beta, \gamma, \theta^{(i)}$ to minimise \begin{align}\label{eq:opt_straight_form}
    E = \sum_{i,b}\left(\barlamnoe - \theta^{(i)} - \frac{\alpha}{b^{\upsilon}} - \frac{\beta}{\theta^{(i)}b^{2\upsilon}} - \frac{\beta\gamma}{b^{2\upsilon} (\theta^{(i)})^2}\right)^2.
\end{align}
Now make the following definitions
\begin{align}
    y_{ib} = \barlamnoe - \theta^{(i)}, ~ \vec{x}_{ib} = \left(\begin{array}{c} b^{-\upsilon} \\ (\theta^{(i)} b)^{-2\upsilon} \\ (b^{2\upsilon} (\theta^{(i)})^2)^{-1}\end{array}\right), ~ \vec{w} = \left(\begin{array}{c} \alpha \\ \beta \\ \beta\gamma\end{array}\right),
\end{align}
so that 
\begin{align}\label{eq:lin_reg_form}
    E = \sum_{i,b} (y_{ib} - \vec{w}^T \vec{x}_{ib})^2.
\end{align}
Finally we can define the $n$-dimensional \review{vector} $\vec{Y}$ by flattening the matrix $(y_{ib})_{ib}$, and the $3\times n$ matrix $\vec{X}$ by stacking the vectors $\vec{x}_{ib}$ and then flattening of the $i,b$ indices.
That done, we have have a standard linear regression problem with design matrix $\vec{X}$ and parameters $\vec{w}$.
For fixed $\theta$, the global minimum of $E$ is then attained at parameters
\begin{align}
    \vec{w}^*(\vec{\theta}) = (\vec{X}\vec{X}^T)^{-1}\vec{X}\vec{Y}
\end{align}
where the dependence on the parameters $\vec{\theta}$ is through $\vec{Y}$ and $\vec{X}$ as above.
We thus have $$\alpha = w^*_1, \beta = w^*_2, \gamma = w^*_3/w^*_2$$
and can plug these values back in to (\ref{eq:opt_straight_form}) to obtain an optimisation problem only over the $\theta^{(i)}$.
There is no closed form solution for the optimal $\theta^{(i)}$ for this problem, so we fit them using gradient descent.
The various settings and hyperparameters of this optimisation were tuned by hand to give convergence and are detailed in \ref{sec:impl_fitting}.
To address the real constraint $\beta > 0$, we add a penalty term to the loss (\ref{eq:opt_straight_form}) which penalises values of $\theta^{(i)}$ leading to negative values of $\beta$.
The constraint $\theta^{(i)} > \theta^{(i+1)} > 0$ is implemented using a simple differentiable transformation detailed in \ref{sec:impl_constraints}..
Finally, the exponent $\upsilon$ is selected by fitting the parameters for each $\upsilon$ in $\{-0.1, -0.2, \ldots, -0.9\}$ and taking the value with the minimum mean squared error $E$.

\begin{figure}[h!]
    \centering
    \begin{subfigure}{0.3\linewidth}
     \centering
     \includegraphics[width=\linewidth]{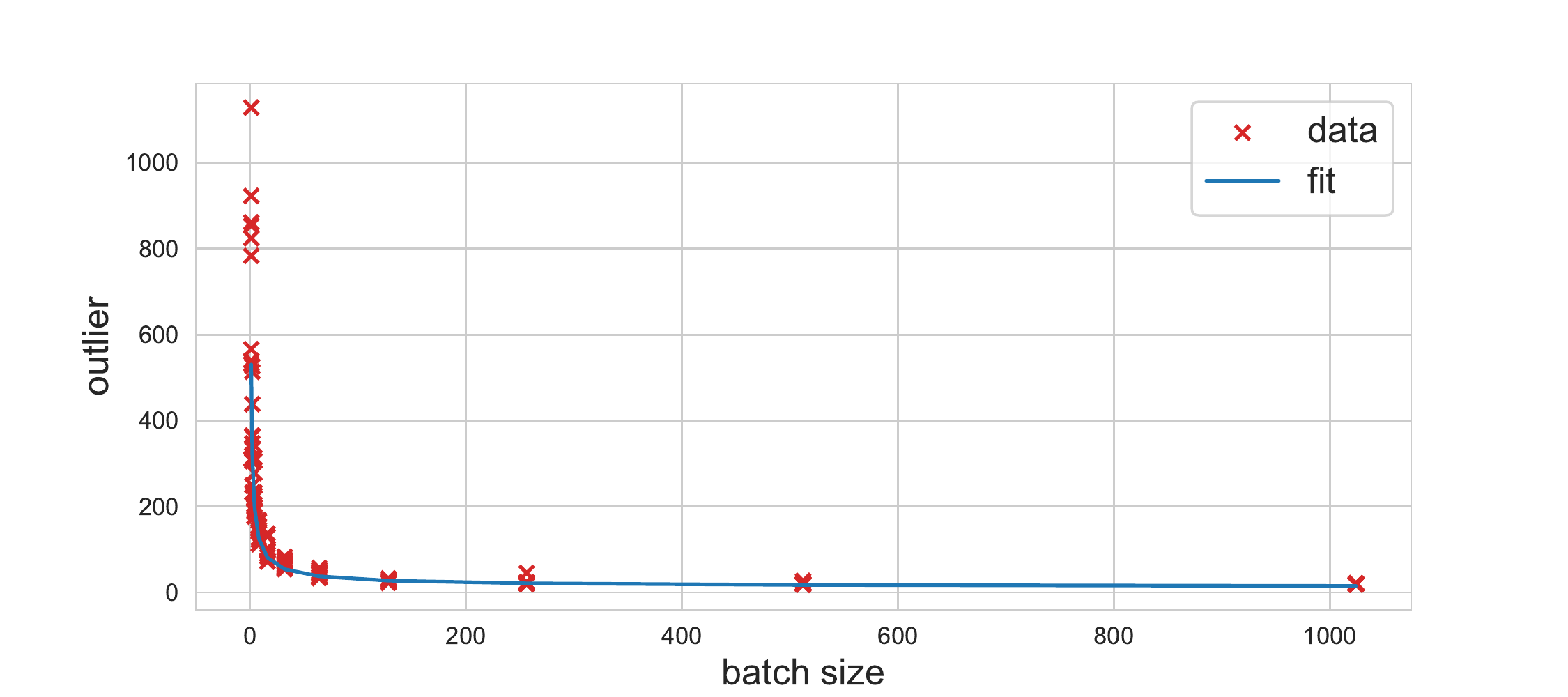}
     \subcaption{Outlier 1, epoch 0}
    \end{subfigure}
    \begin{subfigure}{0.3\linewidth}
     \centering
     \includegraphics[width=\linewidth]{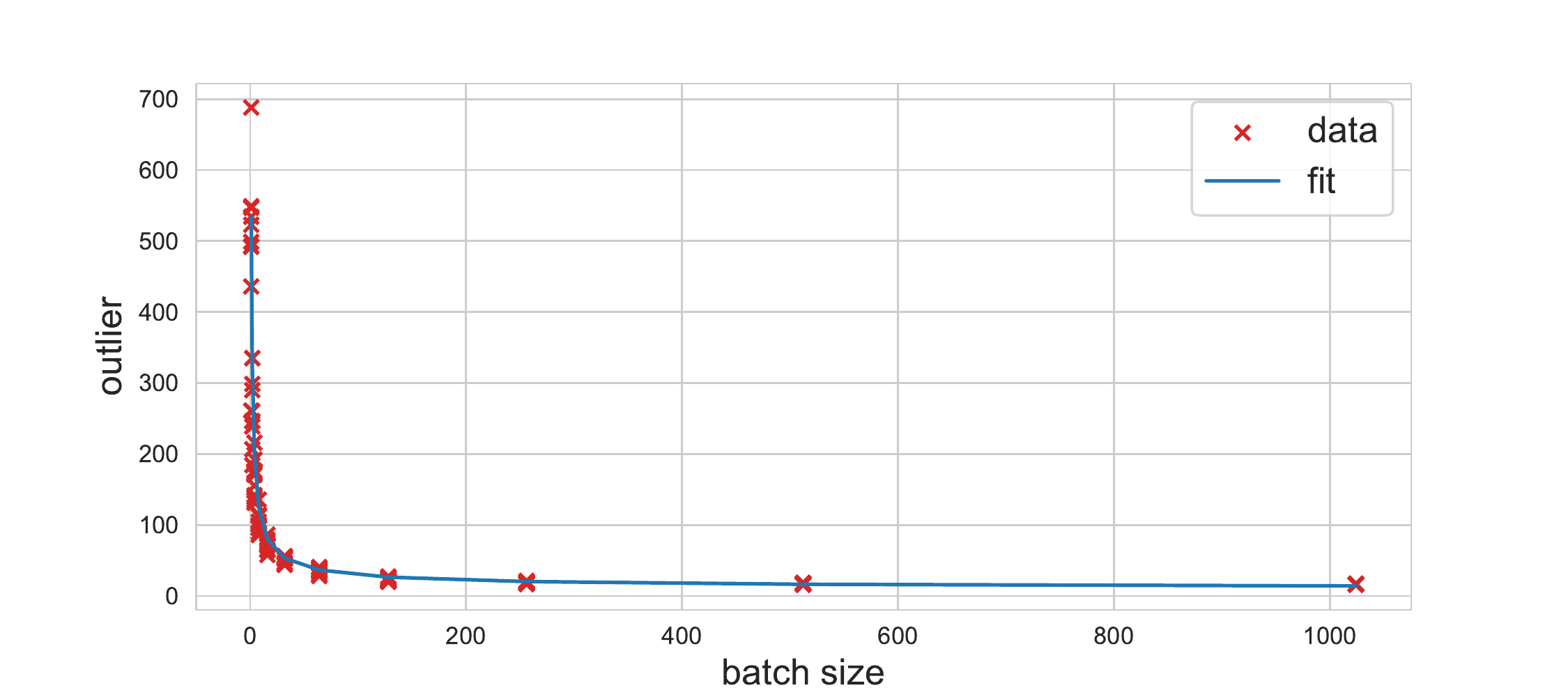}
     \subcaption{Outlier 3, epoch 0}
    \end{subfigure}
    \begin{subfigure}{0.3\linewidth}
     \centering
     \includegraphics[width=\linewidth]{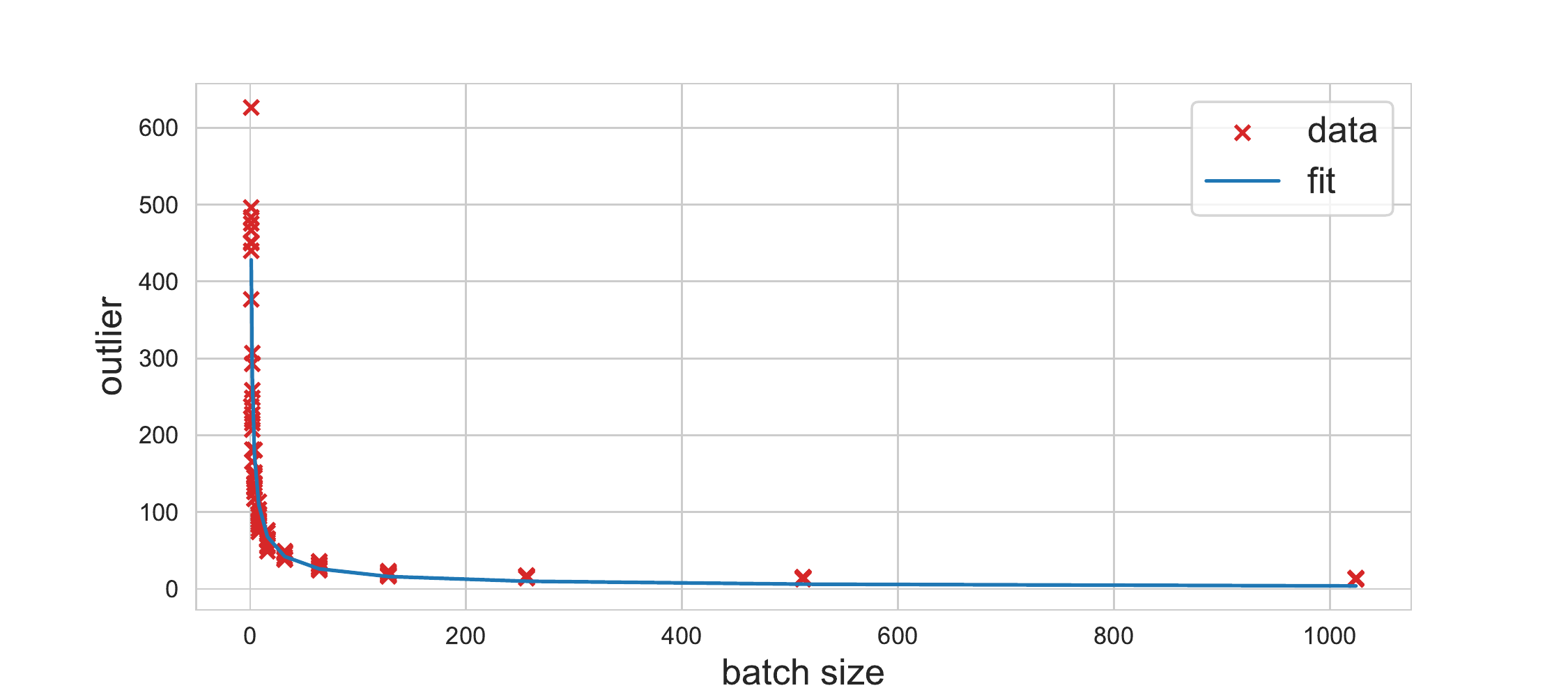}
     \subcaption{Outlier 5, epoch 0}
    \end{subfigure}
    
    \begin{subfigure}{0.3\linewidth}
     \centering
     \includegraphics[width=\linewidth]{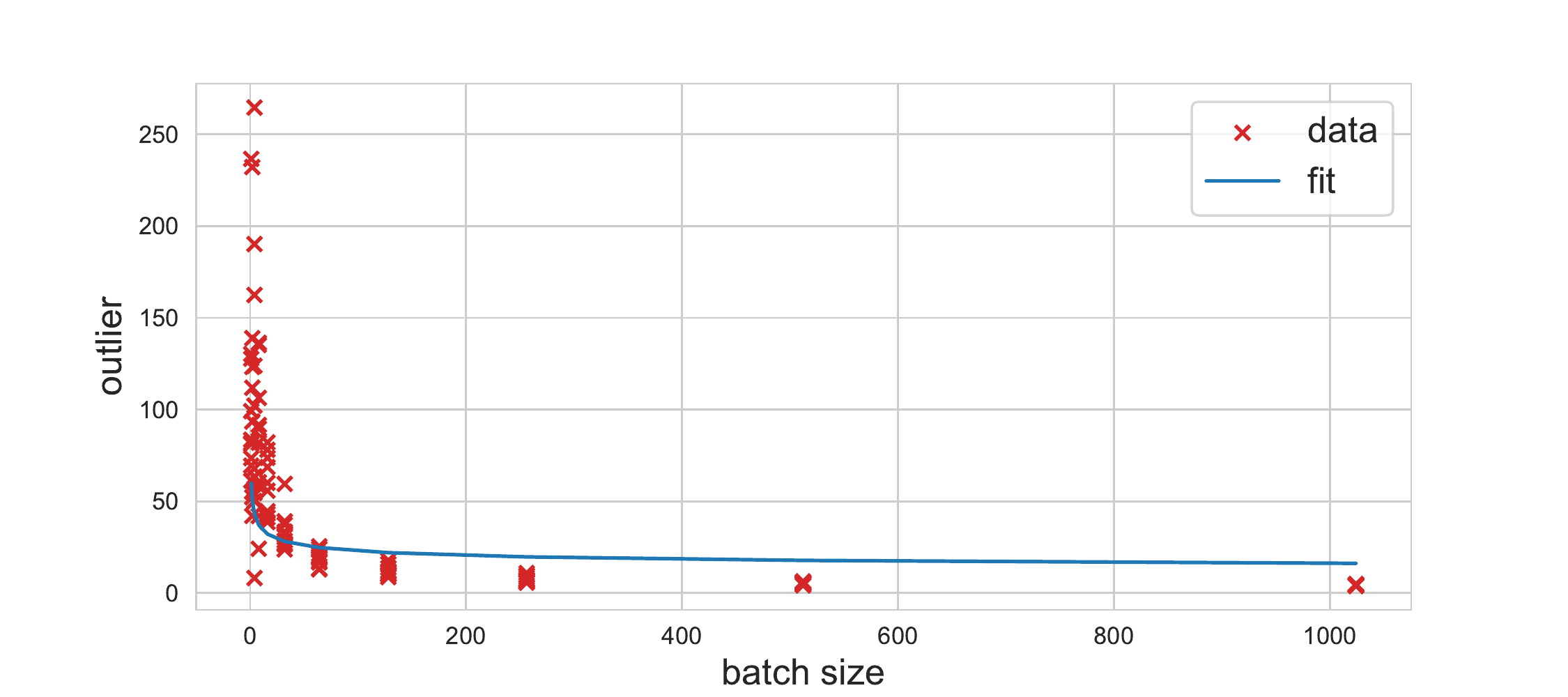}
     \subcaption{Outlier 1, epoch 25}
    \end{subfigure}
    \begin{subfigure}{0.3\linewidth}
     \centering
     \includegraphics[width=\linewidth]{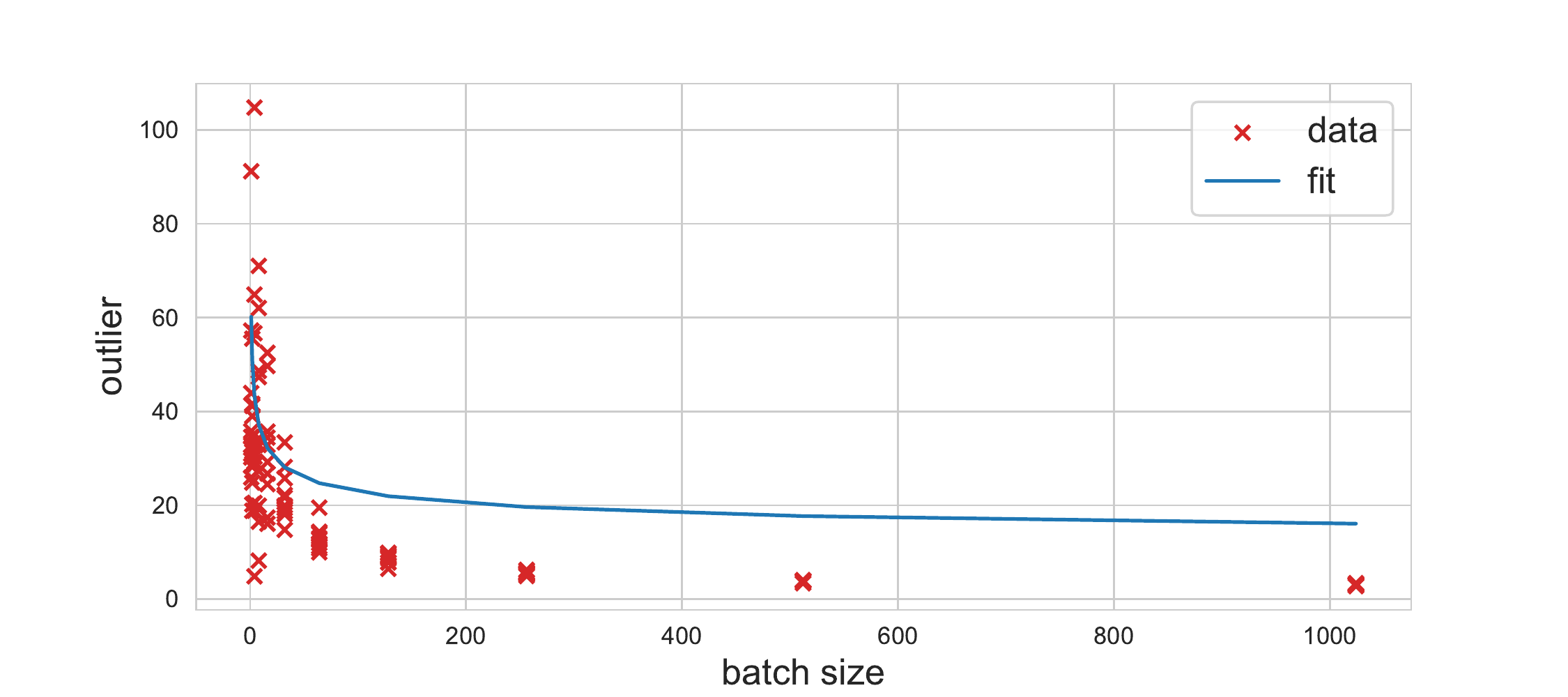}
     \subcaption{Outlier 3, epoch 25}
    \end{subfigure}
    \begin{subfigure}{0.3\linewidth}
     \centering
     \includegraphics[width=\linewidth]{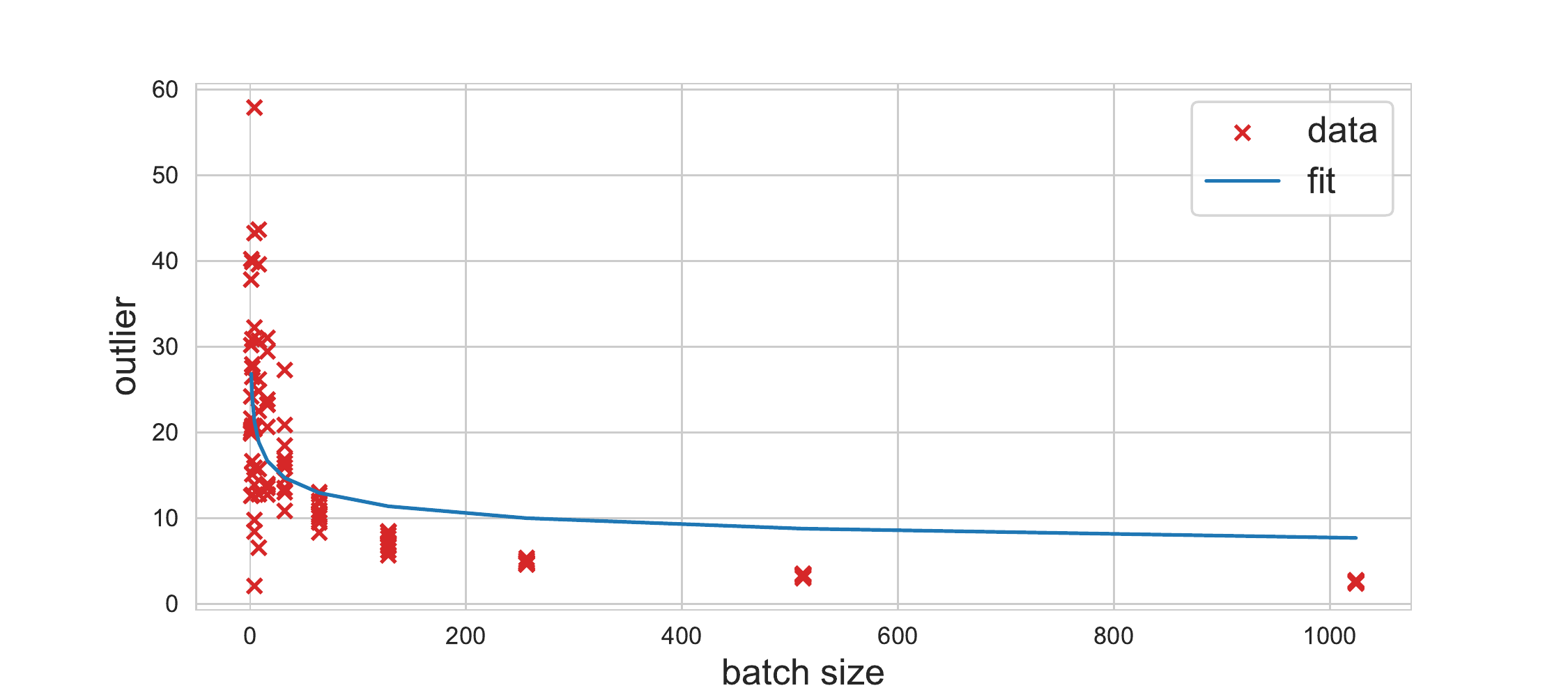}
     \subcaption{Outlier 5, epoch 25}
    \end{subfigure}
    
    \begin{subfigure}{0.3\linewidth}
     \centering
     \includegraphics[width=\linewidth]{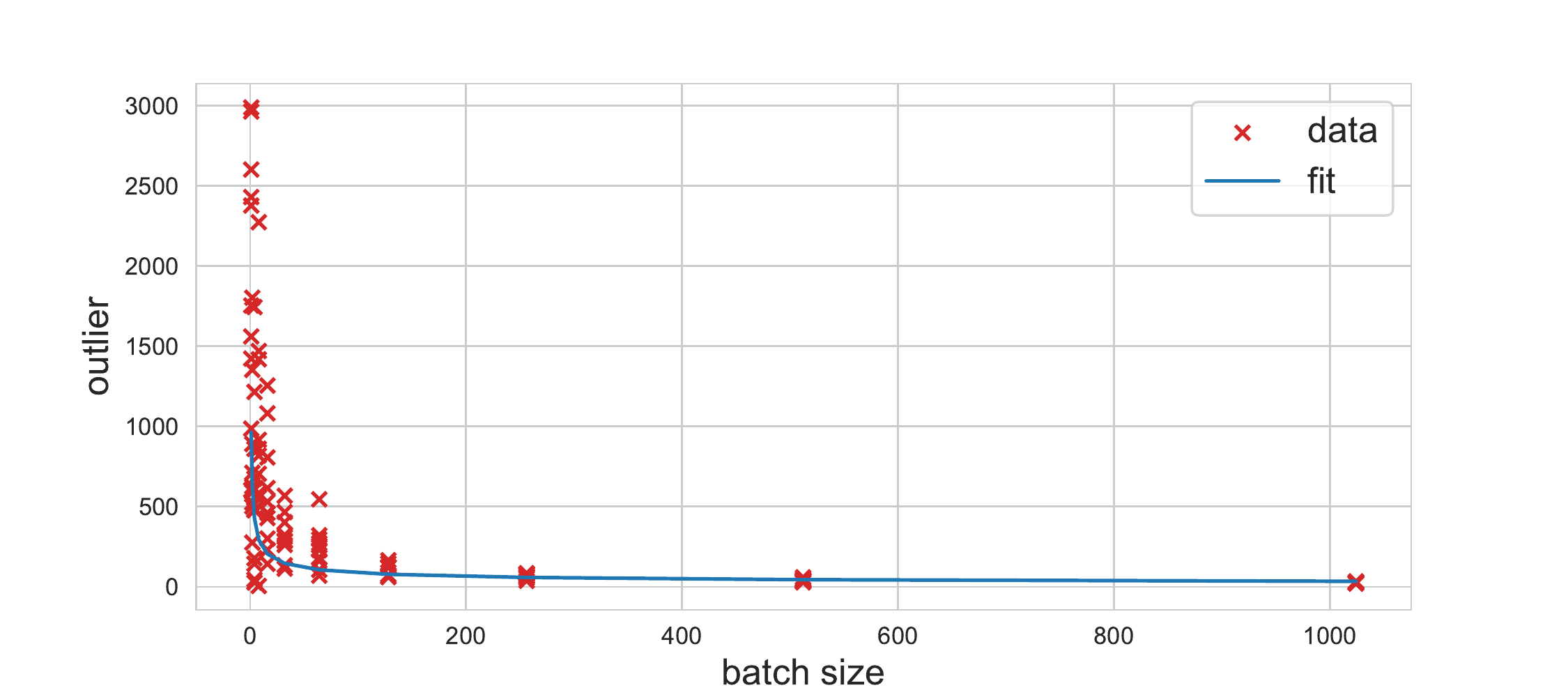}
     \subcaption{Outlier 1, epoch 250}
    \end{subfigure}
    \begin{subfigure}{0.3\linewidth}
     \centering
     \includegraphics[width=\linewidth]{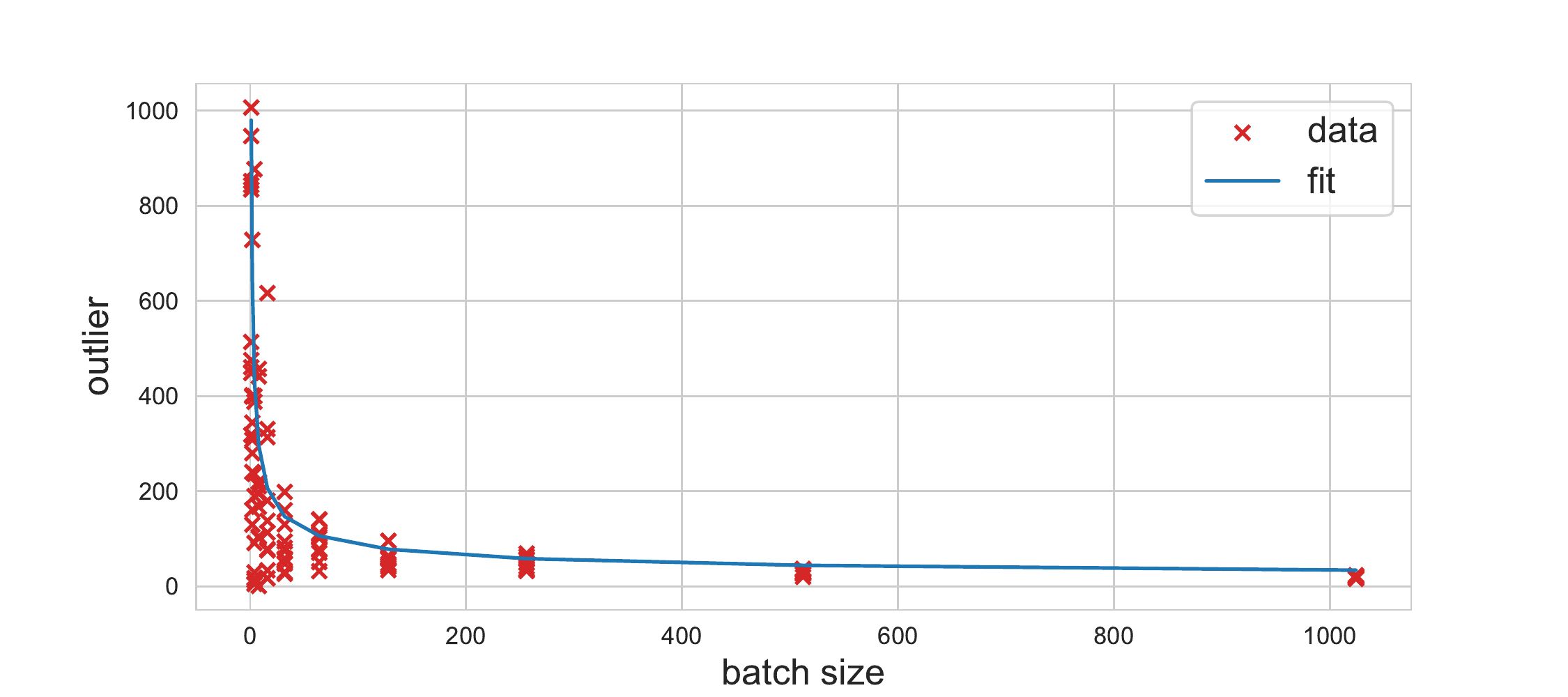}
     \subcaption{Outlier 3, epoch 250}
    \end{subfigure}
    \begin{subfigure}{0.3\linewidth}
     \centering
     \includegraphics[width=\linewidth]{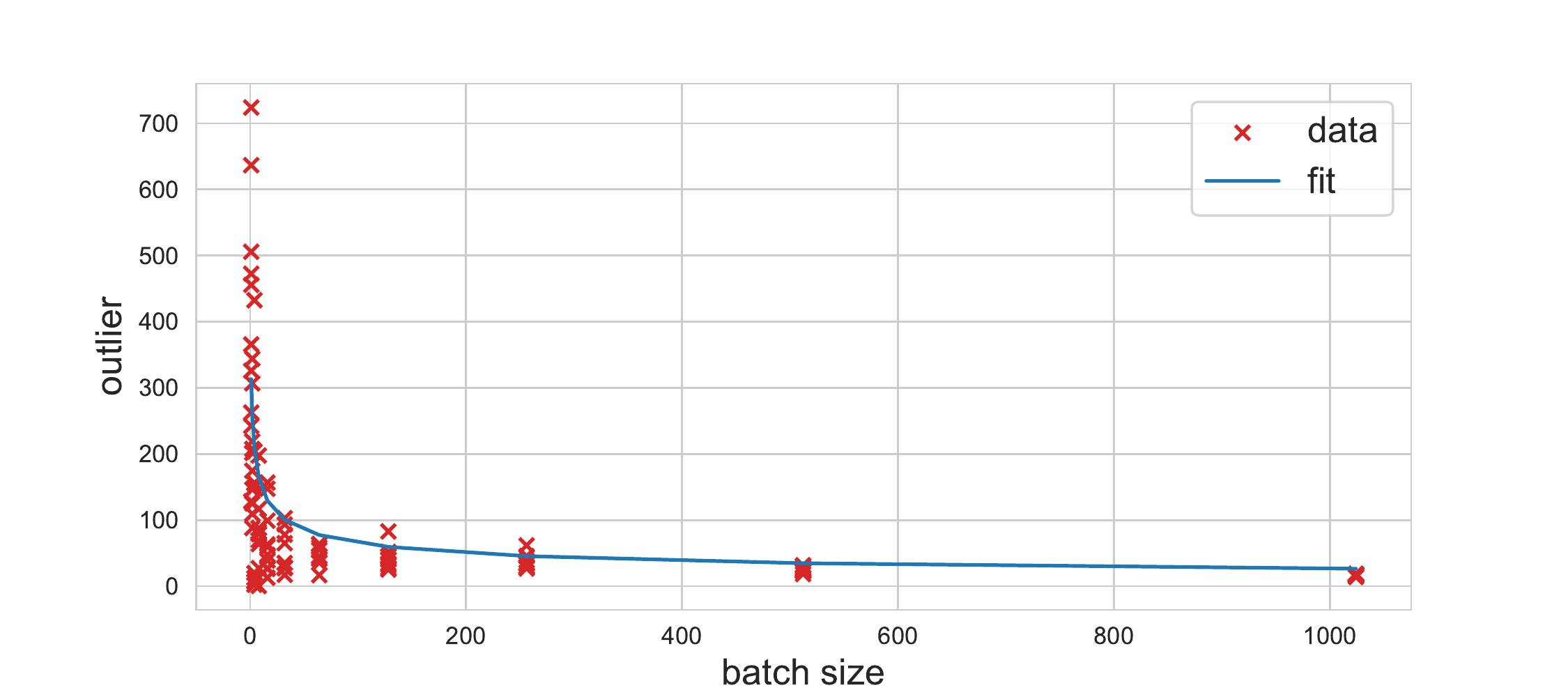}
     \subcaption{Outlier 5, epoch 250}
    \end{subfigure}
    
    \begin{subfigure}{0.3\linewidth}
     \centering
     \includegraphics[width=\linewidth]{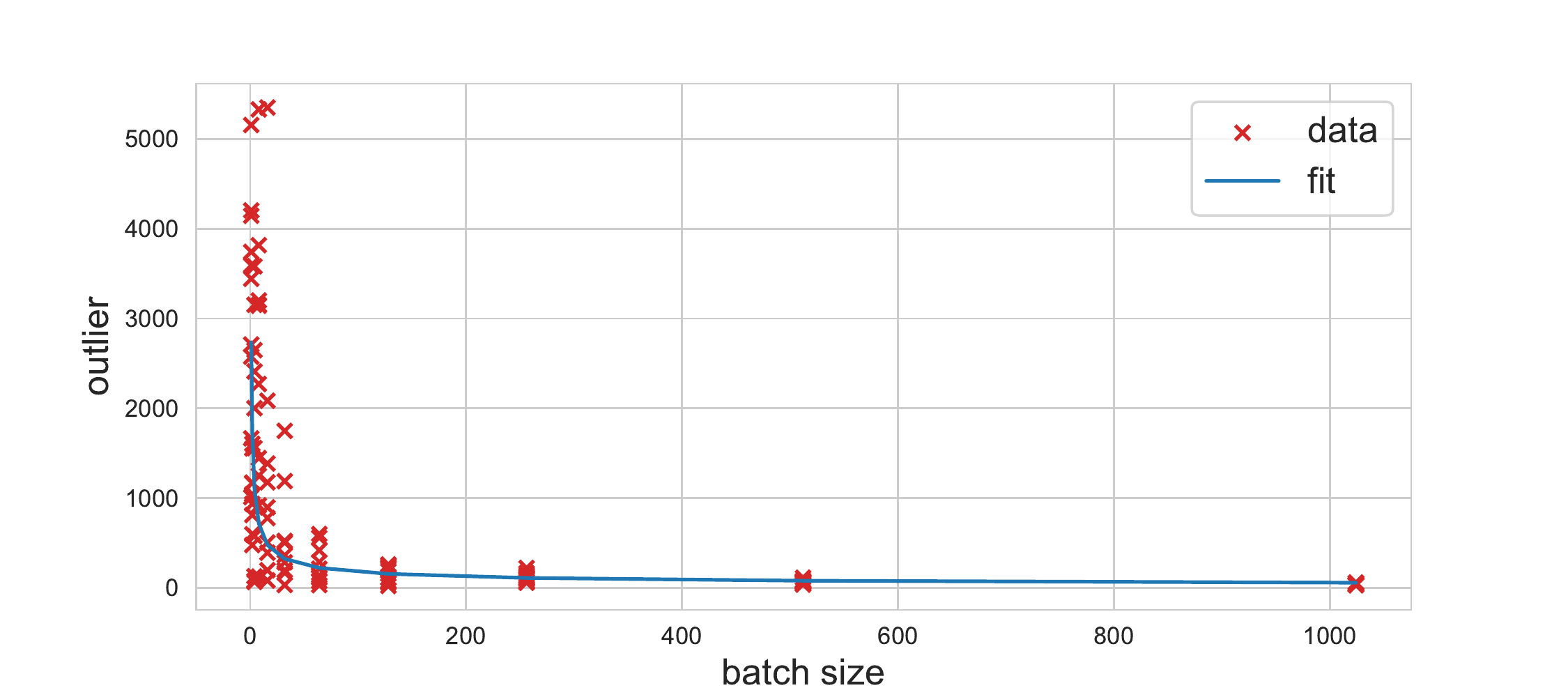}
     \subcaption{Outlier 1, epoch 300}
    \end{subfigure}
    \begin{subfigure}{0.3\linewidth}
     \centering
     \includegraphics[width=\linewidth]{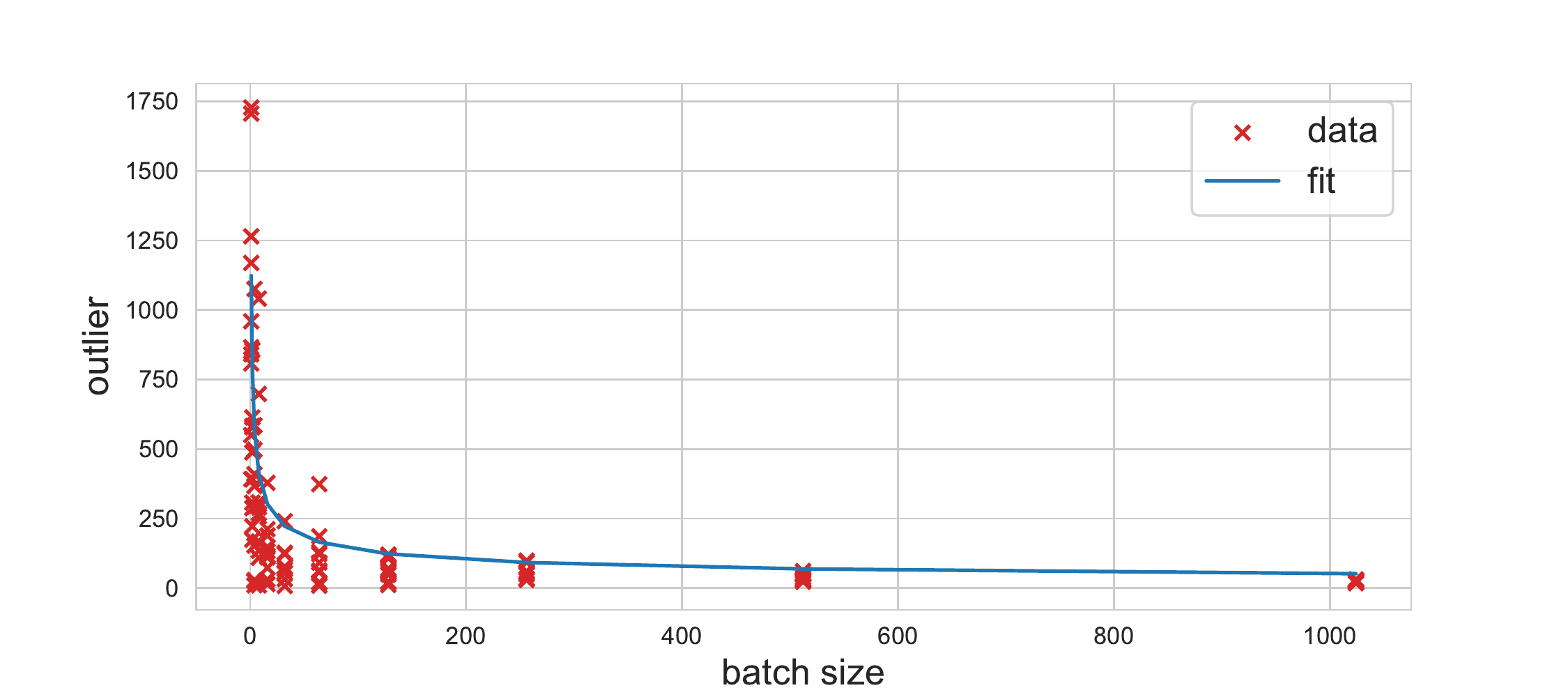}
     \subcaption{Outlier 3, epoch 300}
    \end{subfigure}
    \begin{subfigure}{0.3\linewidth}
     \centering
     \includegraphics[width=\linewidth]{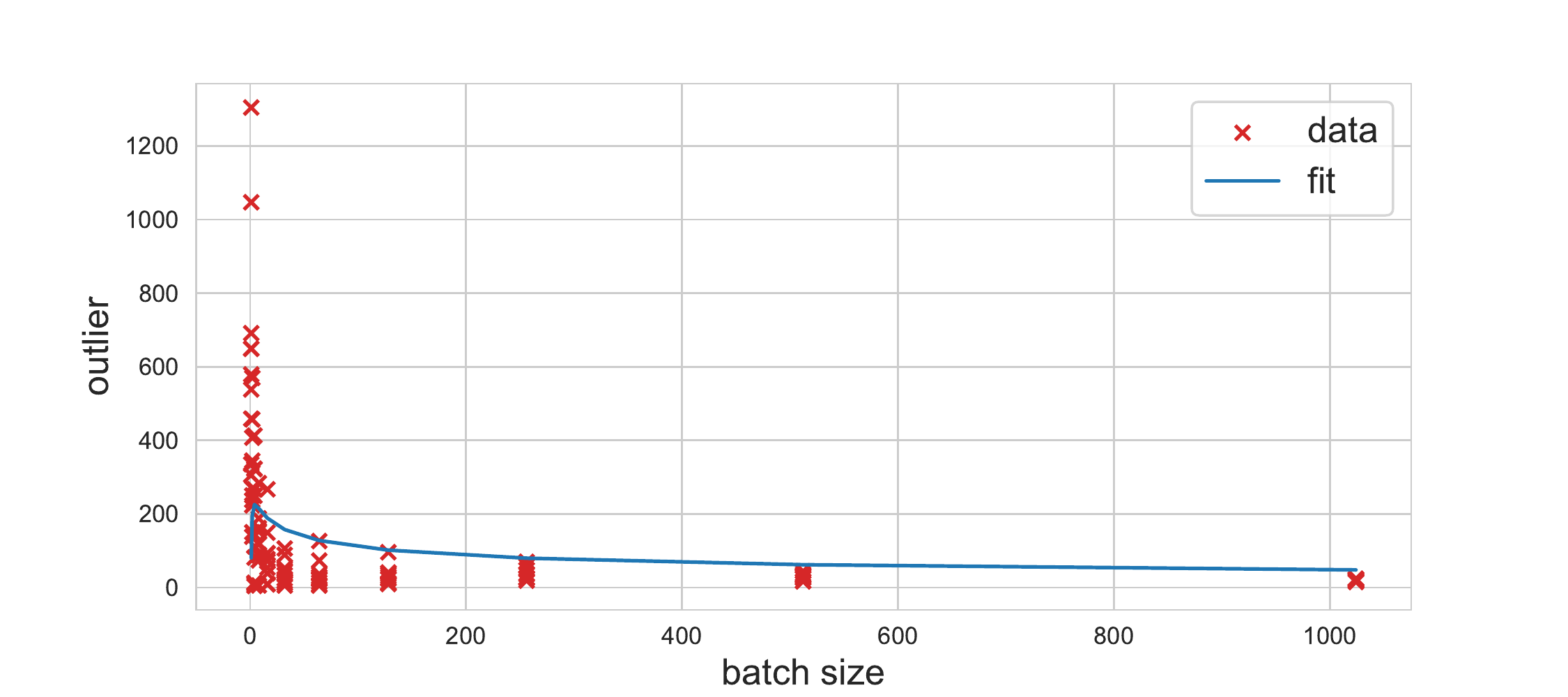}
     \subcaption{Outlier 5, epoch 300}
    \end{subfigure}
    \centering
    \caption{The batch-size scaling of the outliers in the spectra of the Hessians of the Resnet loss on CIFAR100. Training epochs increase top-to-bottom from initialisation to final trained model. Left-to-right the outlier index varies (outlier 1 being the largest). Red cross show results from Lanczos approximations over 10 samples (different batches) for each batch size. The blue lines are parametric power law fits of the form (\ref{eq:omega_fit_form}).}
    \label{fig:outlier_fit_resnet}
\end{figure}

\medskip
The above process results in 12 fits for VGG and Resnet and 10 for MLP (one per epoch).
For each of these, we have a theoretical fit for each of the $5$ top outliers as a function of batch size which can be compared graphically to the data, resulting in $(2\times 12 + 10)\times 5 = 170$ plots.
Rather than try to display them all, we will select a small subset that illustrates the key features.
Figure \ref{fig:outlier_fit_resnet} shows results for the Resnet at epochs 0 (initialisation), 25, 250 and 300 (end of training) and outliers 1, 3 and 5.
Between the three models, the Resnet shows consistently the best agreement between the data and the parametric form (\ref{eq:omega_fit_form}).
The agreement is excellent at epoch 0 but quickly degrades to that seen in the second row of Figure \ref{fig:outlier_fit_resnet}, which is representative of the early and middle epochs for the Resnet.
Towards the end of training the Resnet returns to good agreement between theory and data, as demonstrated in the third and fourth rows of Figure \ref{fig:outlier_fit_resnet} at epochs 250 and 300 respectively.

\begin{figure}[h!]
    \centering
    \begin{subfigure}{0.3\linewidth}
     \centering
     \includegraphics[width=\linewidth]{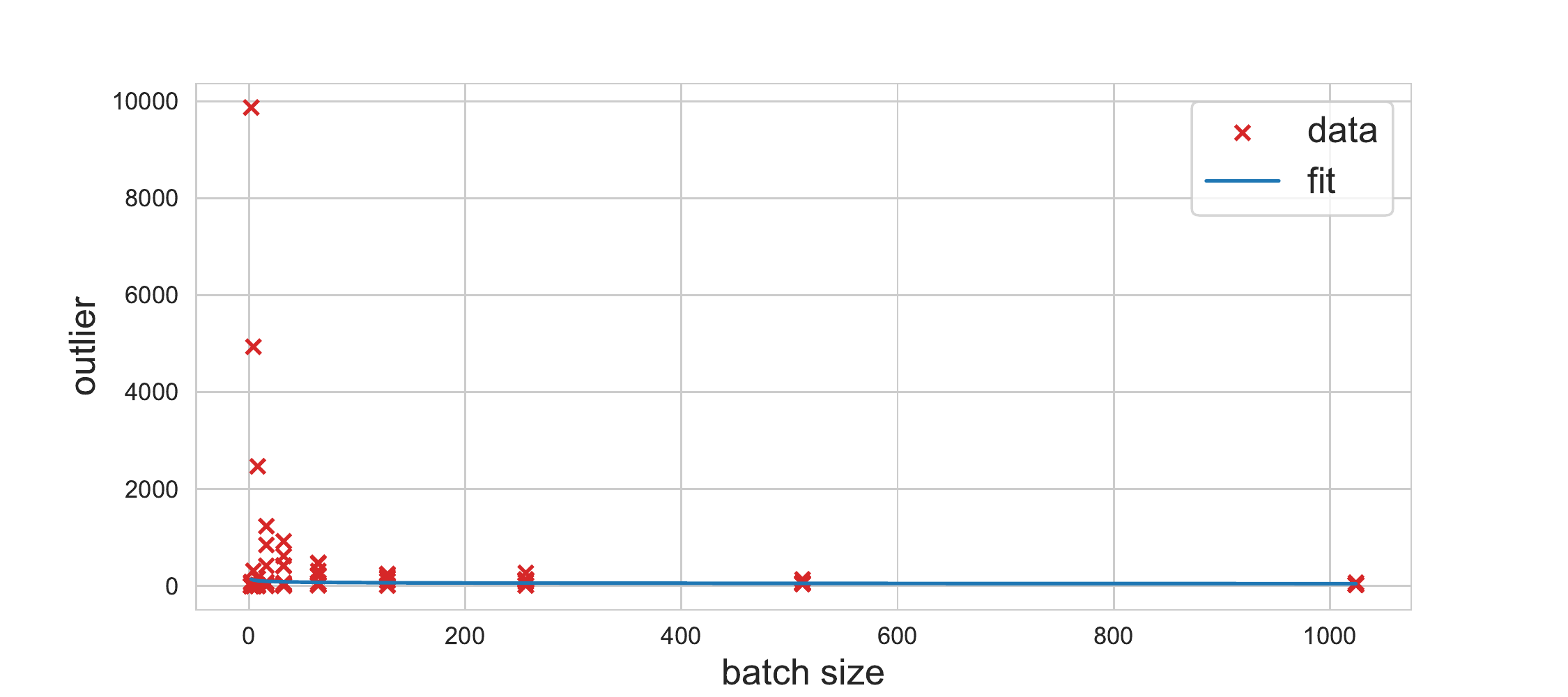}
     \subcaption{VGG16, epoch 300, outlier 1}
    \end{subfigure}
    \begin{subfigure}{0.3\linewidth}
     \centering
     \includegraphics[width=\linewidth]{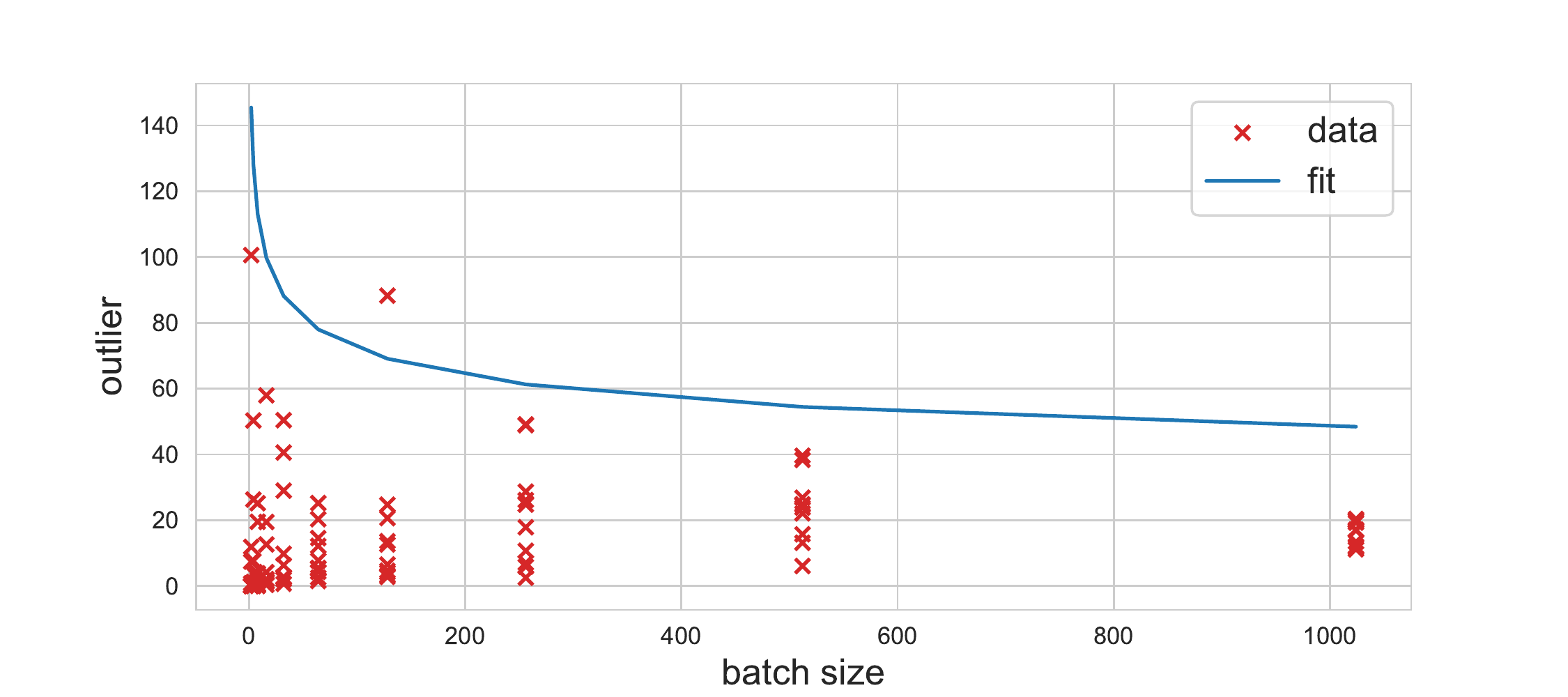}
     \subcaption{VGG16, epoch 300, outlier 3}
    \end{subfigure}
    \begin{subfigure}{0.3\linewidth}
     \centering
     \includegraphics[width=\linewidth]{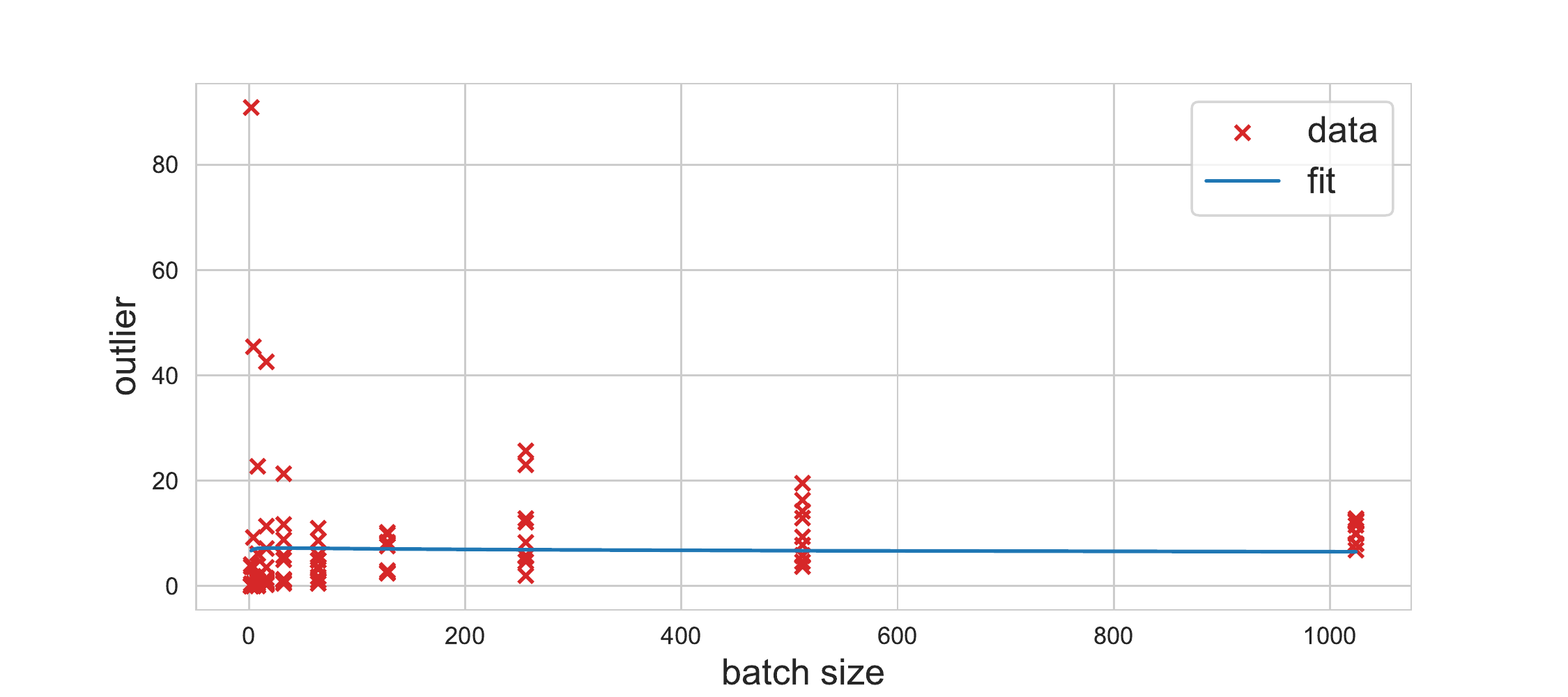}
     \subcaption{VGG16, epoch 300, outlier 5}
    \end{subfigure}
    
    \begin{subfigure}{0.3\linewidth}
     \centering
     \includegraphics[width=\linewidth]{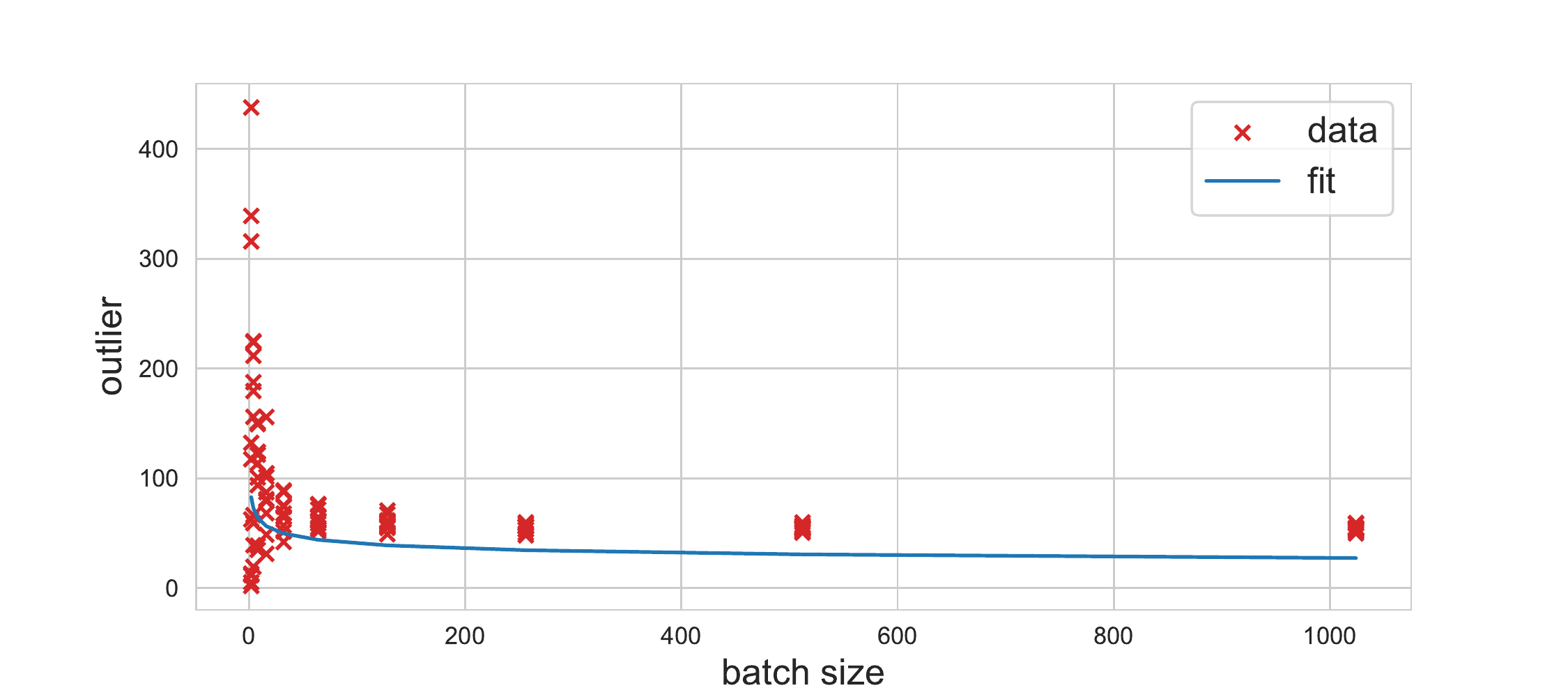}
     \subcaption{MLP, epoch 1, outlier 1}
    \end{subfigure}
    \begin{subfigure}{0.3\linewidth}
     \centering
    \includegraphics[width=\linewidth]{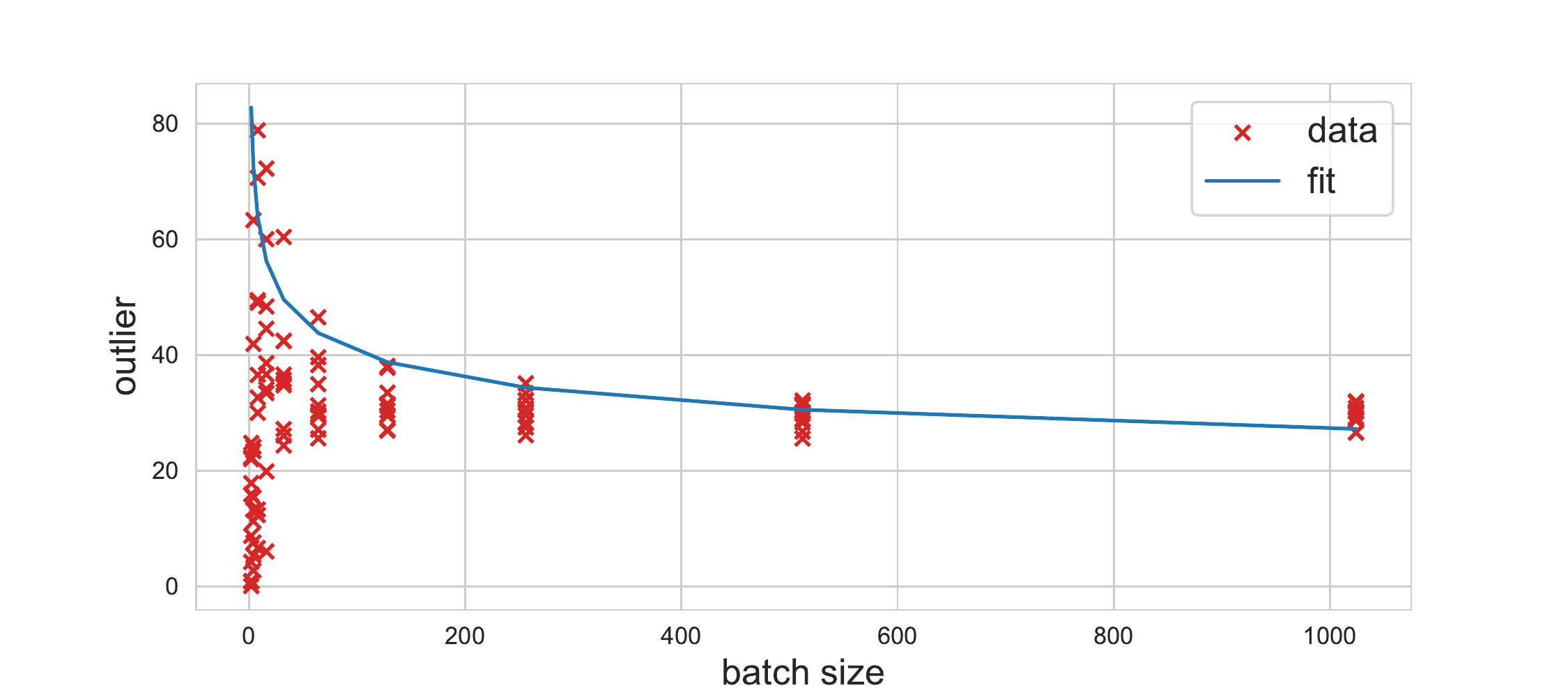}
     \subcaption{MLP, epoch 1, outlier 3}
    \end{subfigure}
    \begin{subfigure}{0.3\linewidth}
     \centering
    \includegraphics[width=\linewidth]{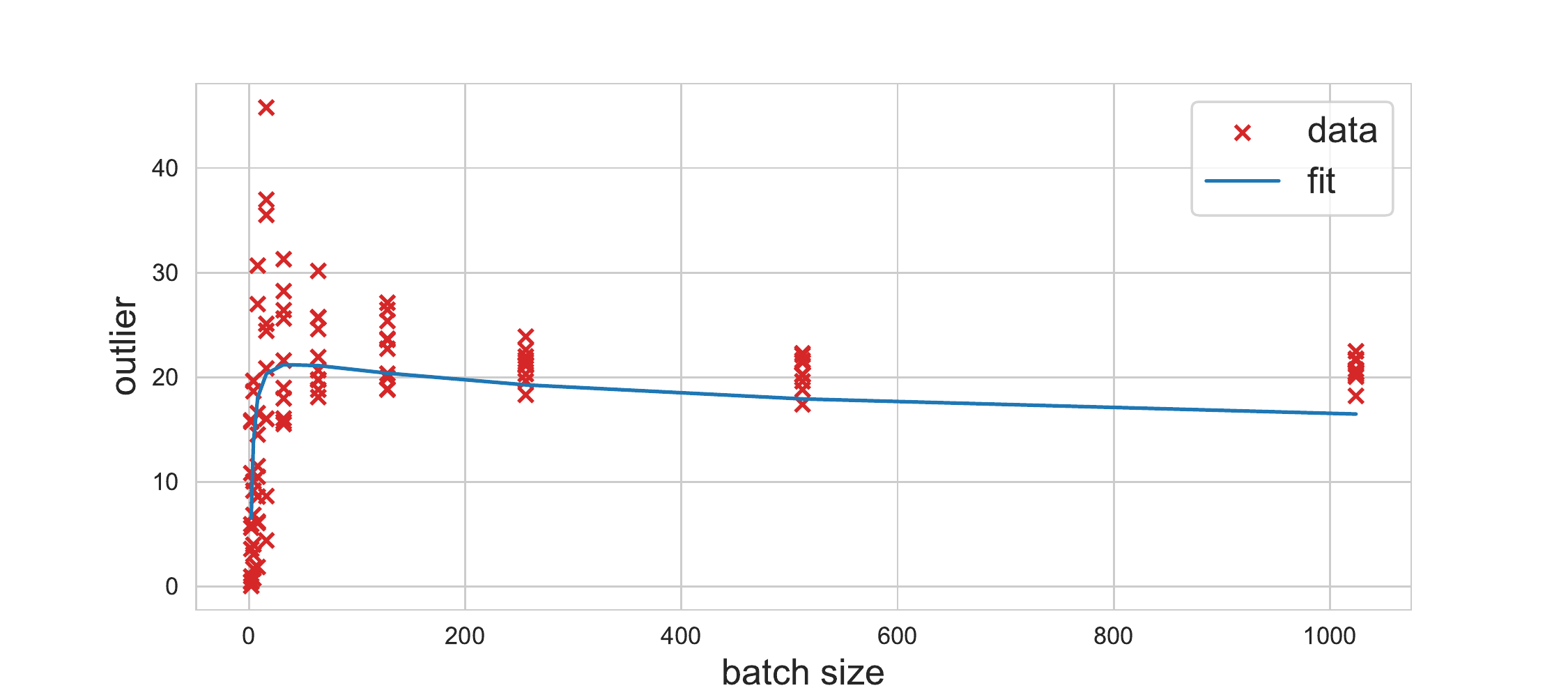}
     \subcaption{MLP, epoch 1, outlier 5}
    \end{subfigure}
    \centering
    \caption{Left-to-right the outlier index varies (outlier 1 being the largest). Red cross show results from Lanczos approximations over 10 samples (different batches) for each batch size. The blue lines are parametric power law fits of the form (\ref{eq:omega_fit_form}). This plot show the final epoch (300) for the VGG16 on CIFAR100 and the first epoch for the MLP on MNIST, both being examples of the parametric fit failing to match the data.}
    \label{fig:outlier_fit_ugly_ducklings}
\end{figure}

The VGG16 also has excellent agreement between theory and data at epoch 0, and thereafter is similar to the early epochs of the Resnet, i.e. reasonable, but not excellent, until around epoch 225 where the agreement starts to degrade significantly until the almost complete failure at epoch 300 shown in the first row of Figure \ref{fig:outlier_fit_ugly_ducklings}.
The MLP has the worst agreement between theory and data, having again excellent agreement at epoch 0, but really quite poor agreement even by epoch 1, as shown in the second row of Figure \ref{fig:outlier_fit_ugly_ducklings}.

The experimental results show an ordering Resnet $>$ VGG $>$ MLP, in terms of how well the random matrix theory loss surface predictions explain the Hessian outliers.
We conjecture that this relates to the difficulty of the loss surfaces.
Resnets are generally believed to have smoother, simpler loss surfaces \cite{li2018visualizing} and be easier to train than other architectures, indeed the residual connections were originally introduced for precisely this reason.
The VGG is generally more sensitive to training set-up, requiring well-tuned hyperparameters to avoid unstable or unsuccessful training (see Chapter \ref{chap:damp} \cite{granziol22damping}).
The MLP is perhaps too small to benefit from high-dimensional highly over-parametrised effects.

\begin{figure}[h!]
    \centering
    \begin{subfigure}{0.3\linewidth}
     \centering
     \includegraphics[width=\linewidth]{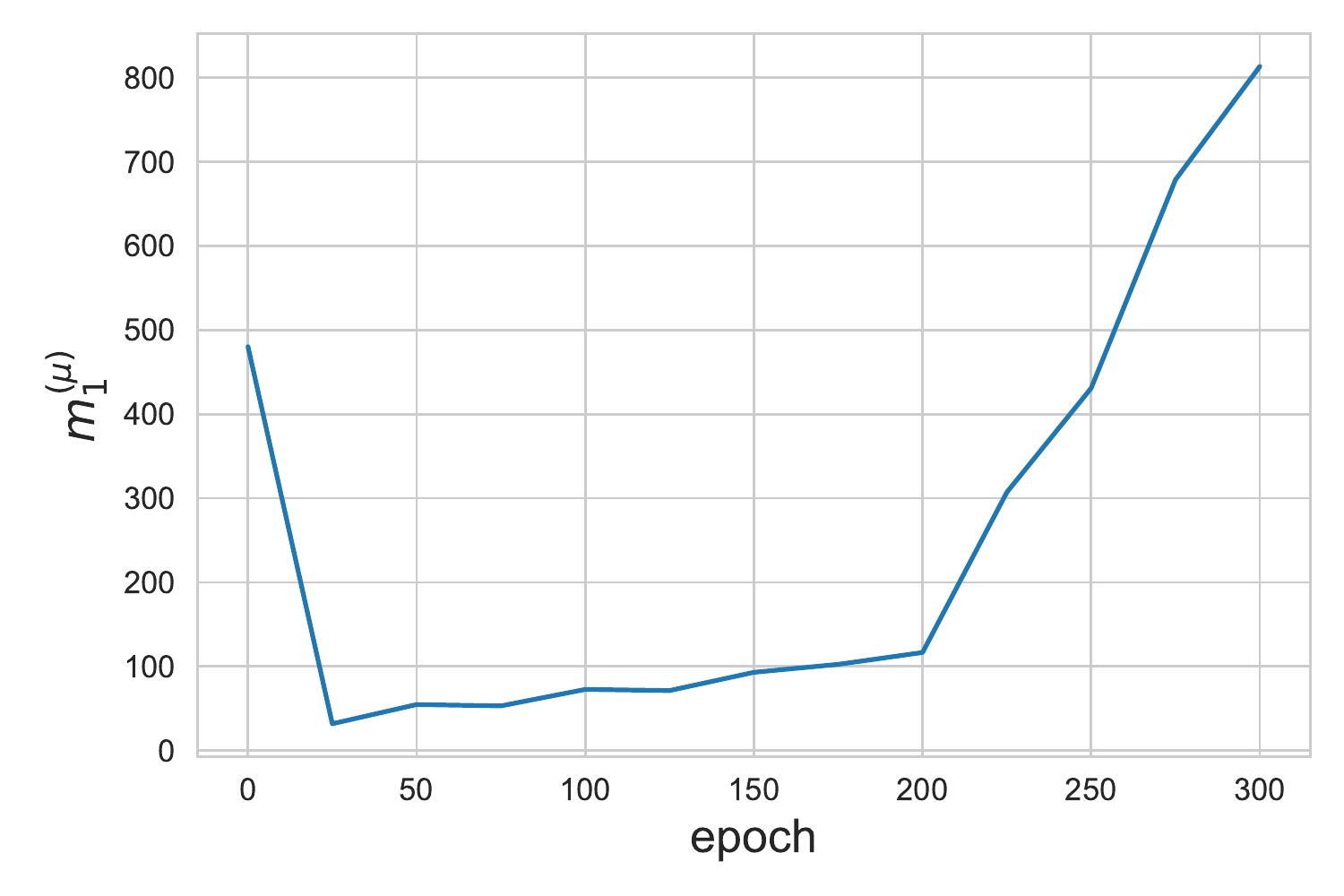}
     \subcaption{$m_1^{(\mu)}$, Resnet on CIFAR100}
          \label{fig:resnet_m1}

    \end{subfigure}
    \begin{subfigure}{0.3\linewidth}
     \centering
     \includegraphics[width=\linewidth]{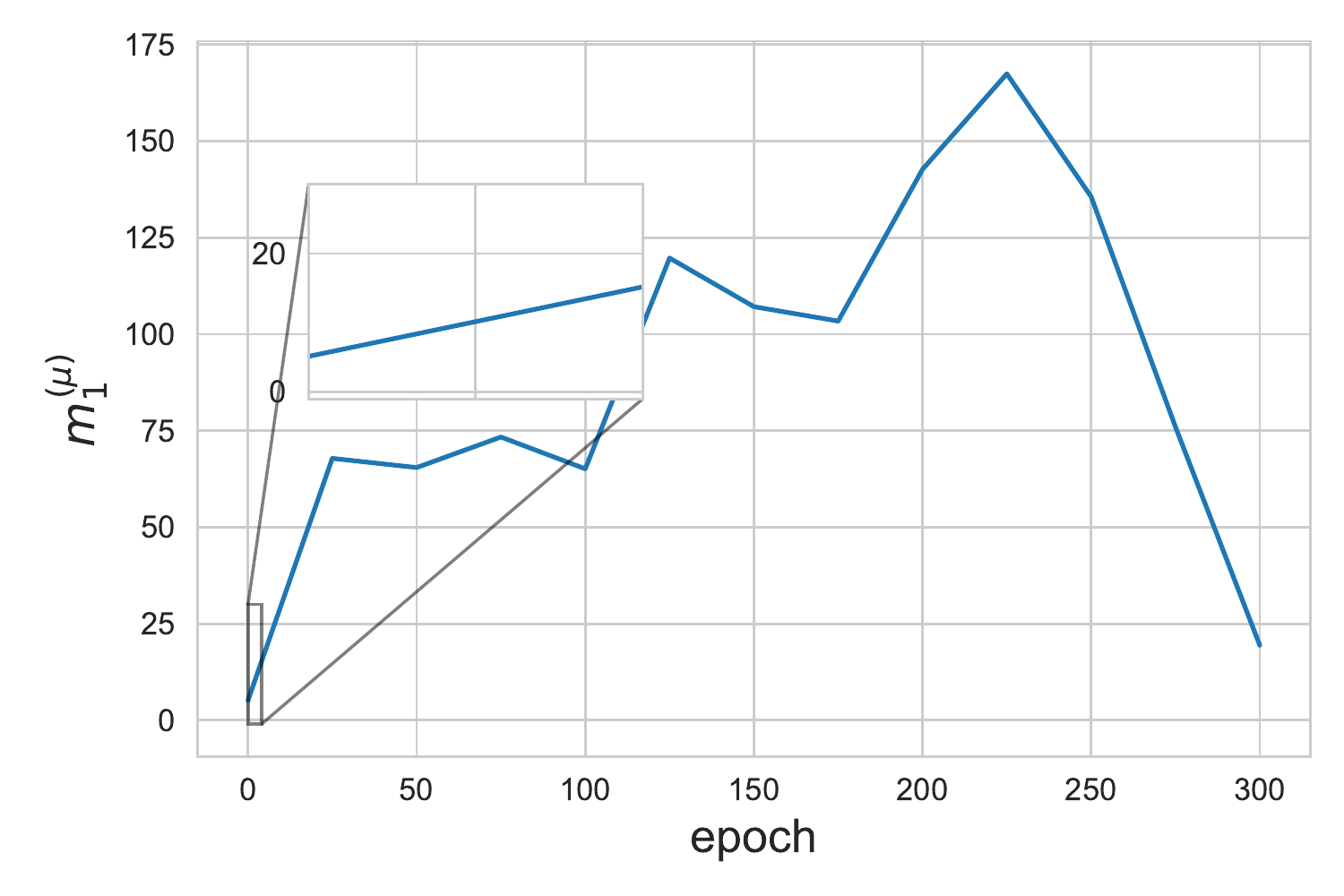}
     \subcaption{$m_1^{(\mu)}$, VGG16 on CIFAR100}
     \label{fig:vgg_m1}
    \end{subfigure}
    \begin{subfigure}{0.3\linewidth}
     \centering
     \includegraphics[width=\linewidth]{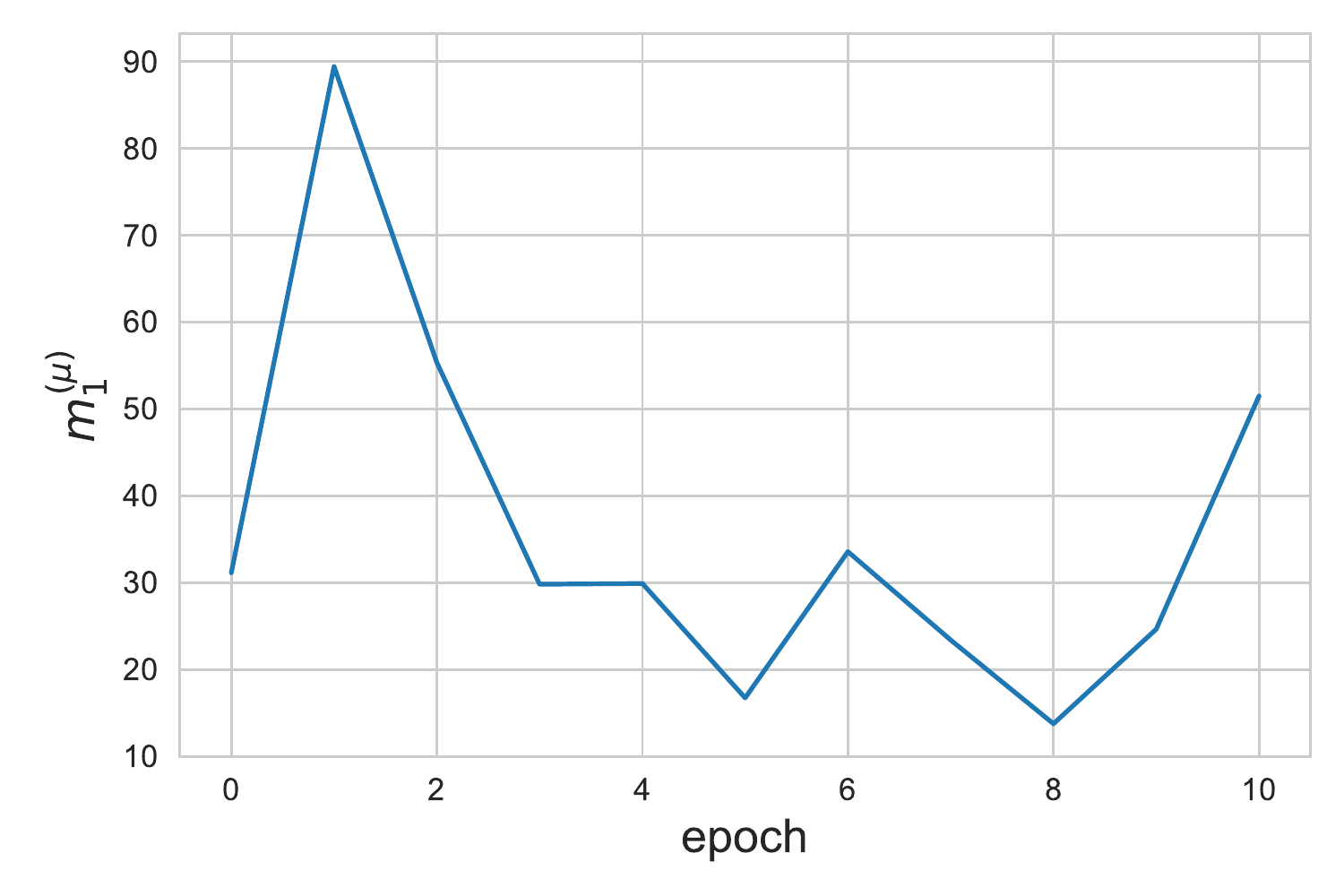}
     \subcaption{$m_1^{(\mu)}$, MLP on MNIST}
    \end{subfigure}
    
    \begin{subfigure}{0.3\linewidth}
     \centering
     \includegraphics[width=\linewidth]{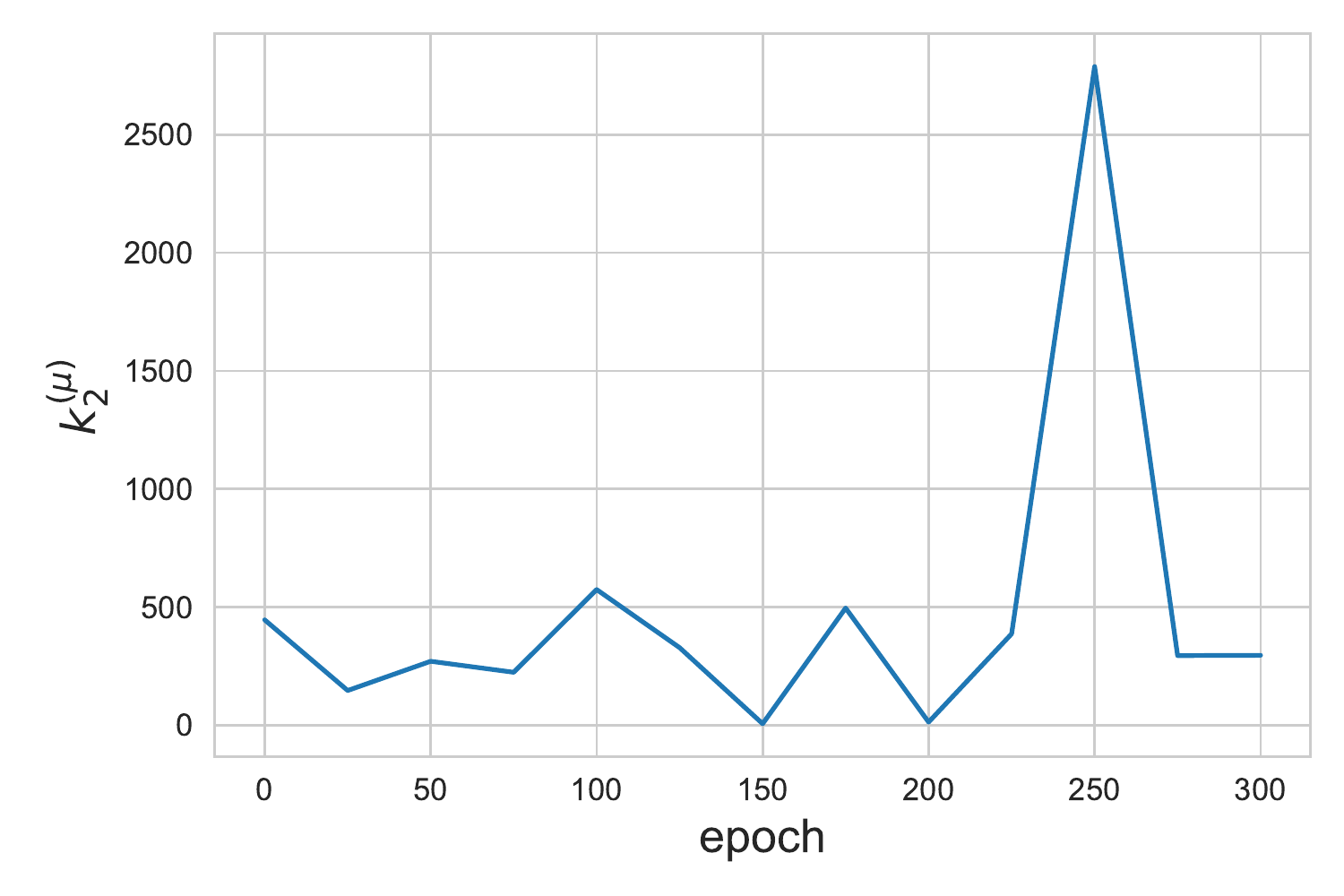}
     \subcaption{$m_2^{(\mu)}$, Resnet on CIFAR100}
    \end{subfigure}
    \begin{subfigure}{0.3\linewidth}
     \centering
     \includegraphics[width=\linewidth]{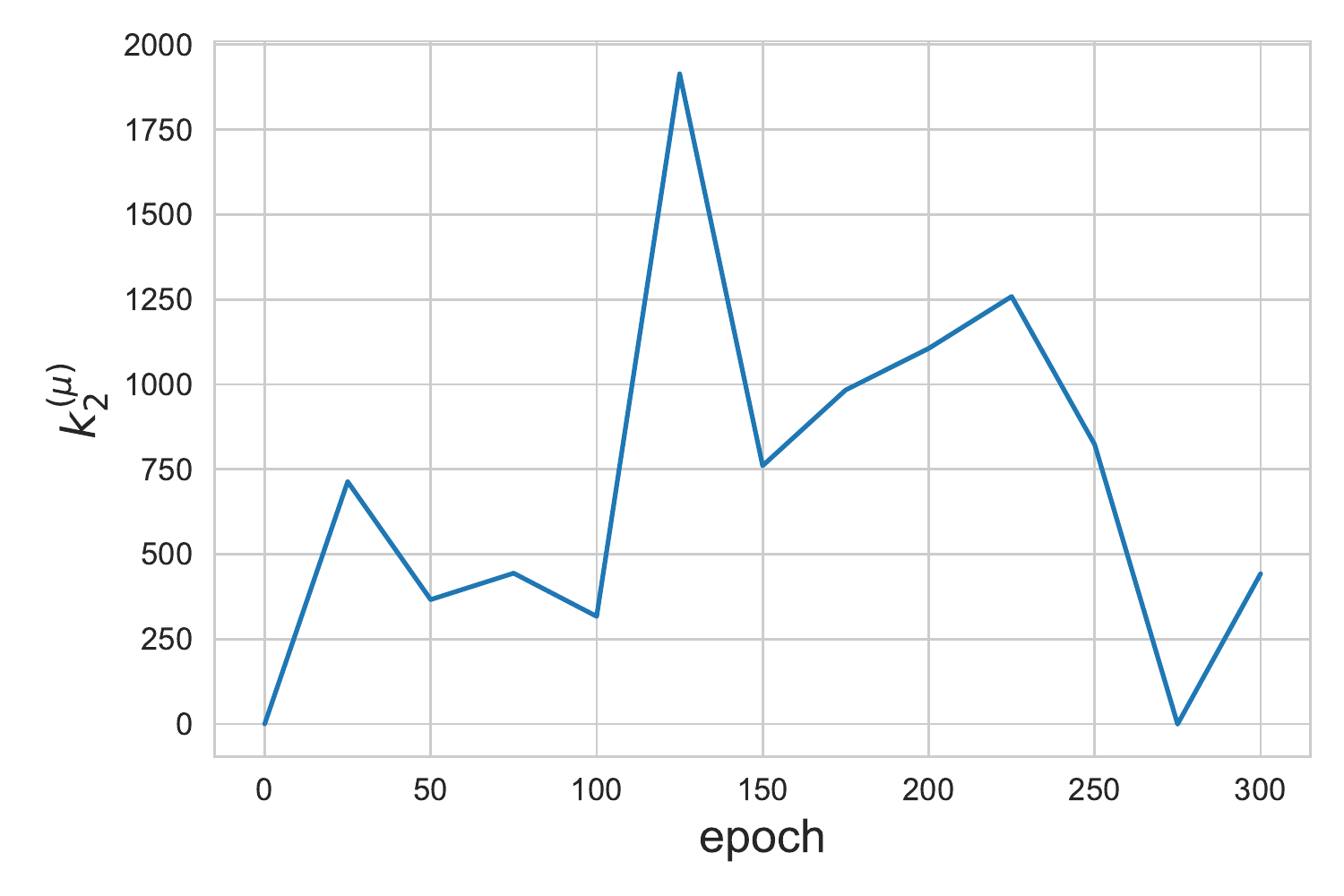}
     \subcaption{$m_2^{(\mu)}$, VGG16 on CIFAR100}
    \end{subfigure}
    \begin{subfigure}{0.3\linewidth}
     \centering
     \includegraphics[width=\linewidth]{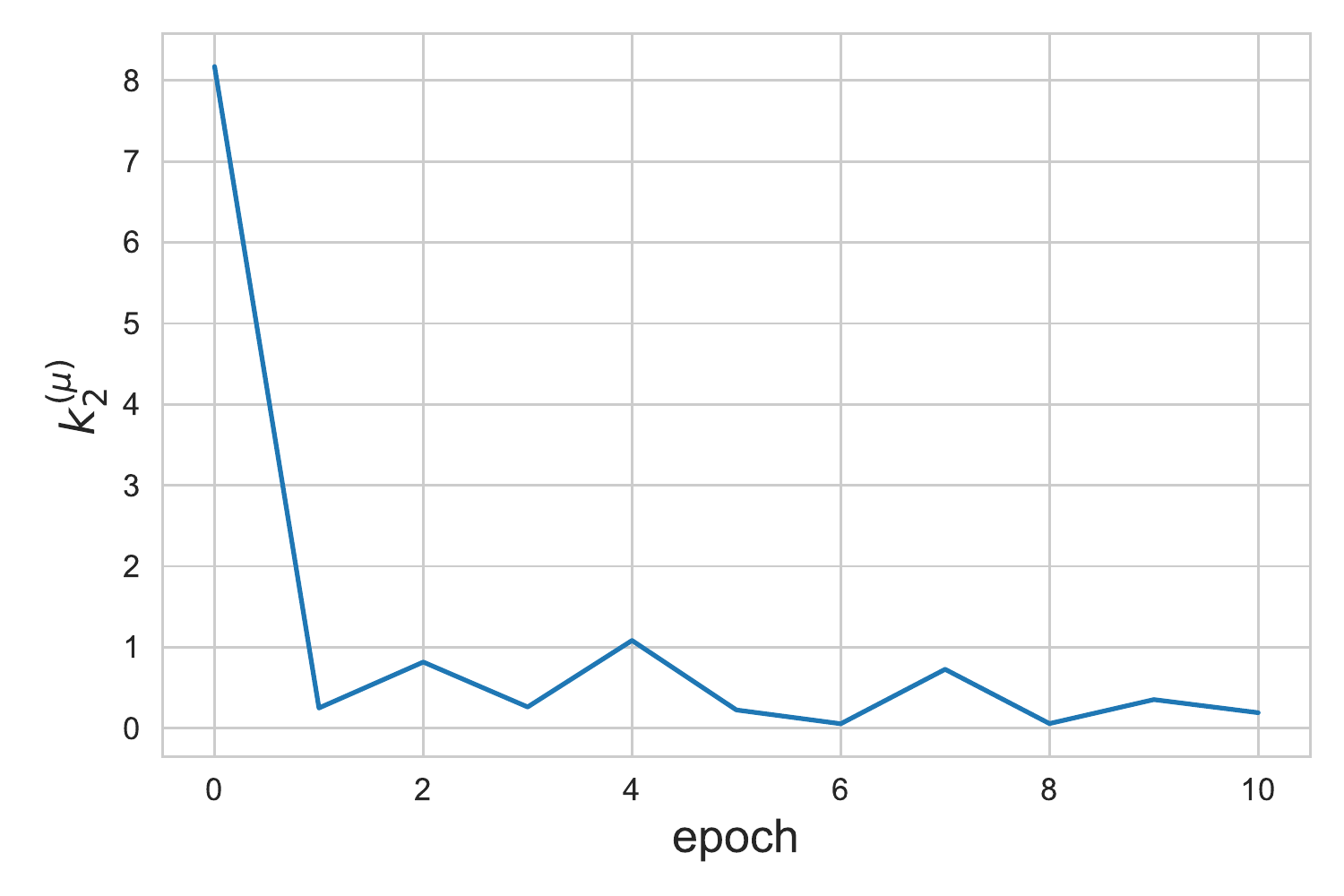}
     \subcaption{$m_2^{(\mu)}$, MLP on MNIST}
    \end{subfigure}
    
    \begin{subfigure}{0.3\linewidth}
     \centering
     \includegraphics[width=\linewidth]{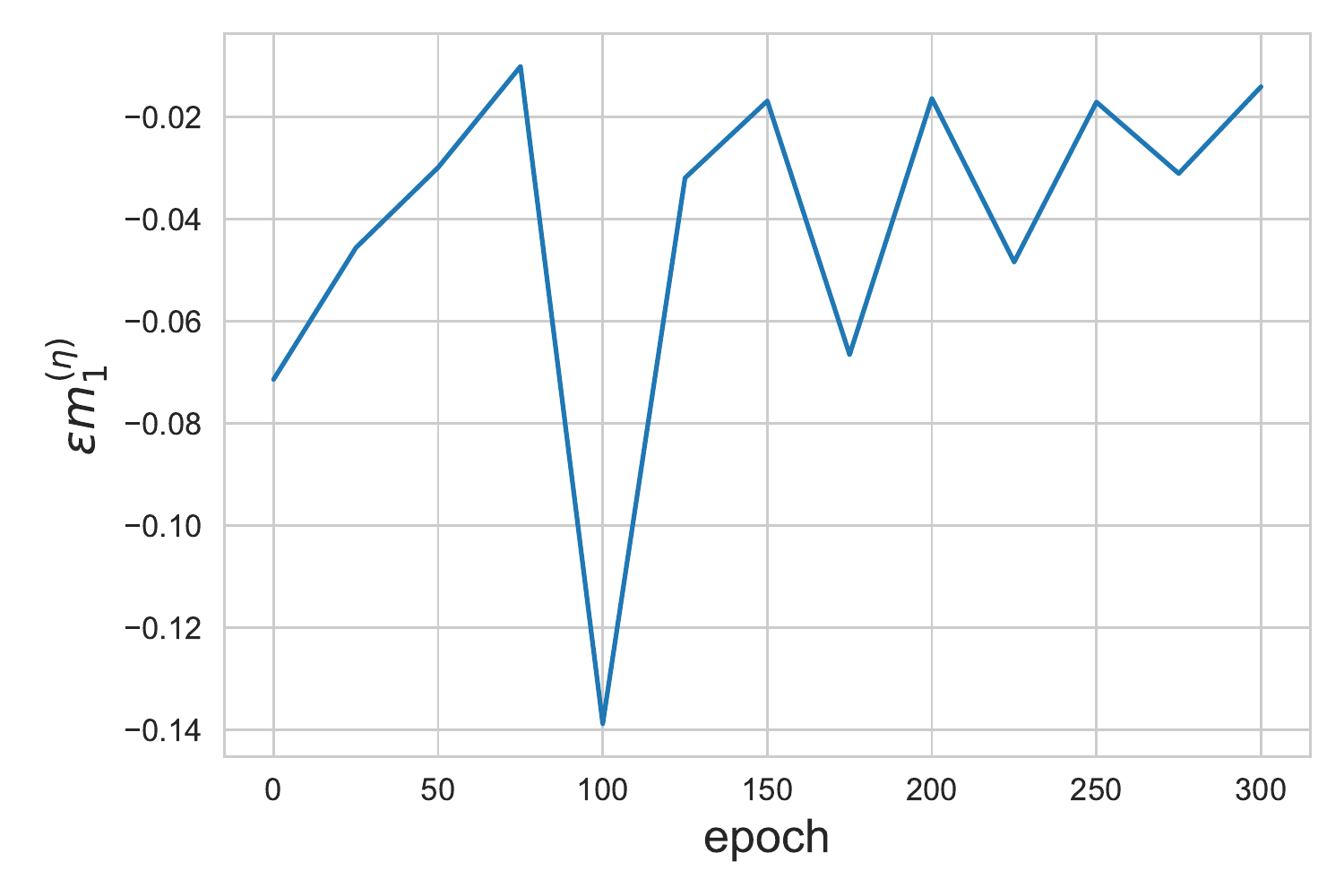}
     \subcaption{$\epsilon m_1^{(\eta)}$, Resnet on CIFAR100}
    \end{subfigure}
    \begin{subfigure}{0.3\linewidth}
     \centering
     \includegraphics[width=\linewidth]{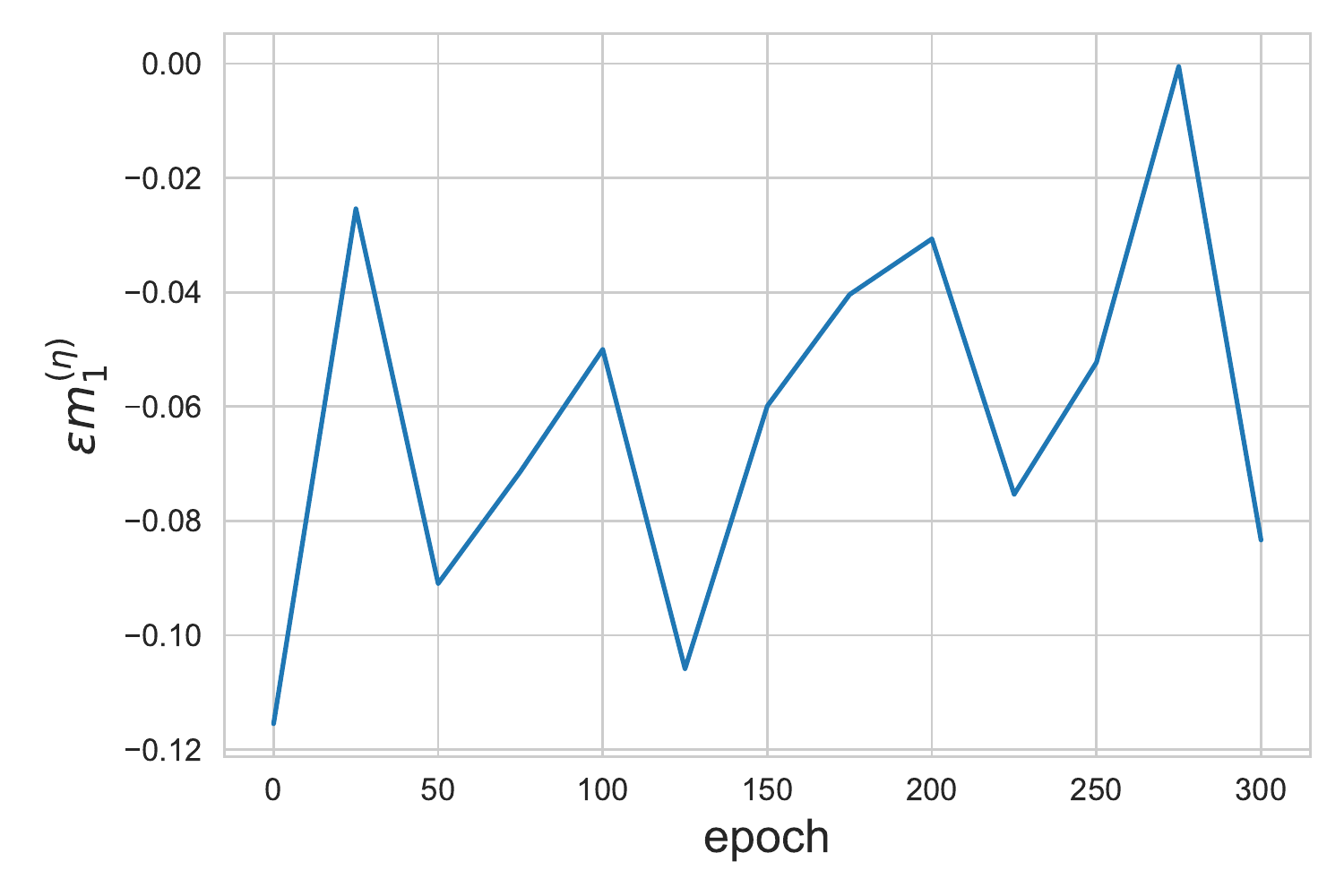}
     \subcaption{$\epsilon m_1^{(\eta)}$, VGG16 on CIFAR100}
    \end{subfigure}
    \begin{subfigure}{0.3\linewidth}
     \centering
     \includegraphics[width=\linewidth]{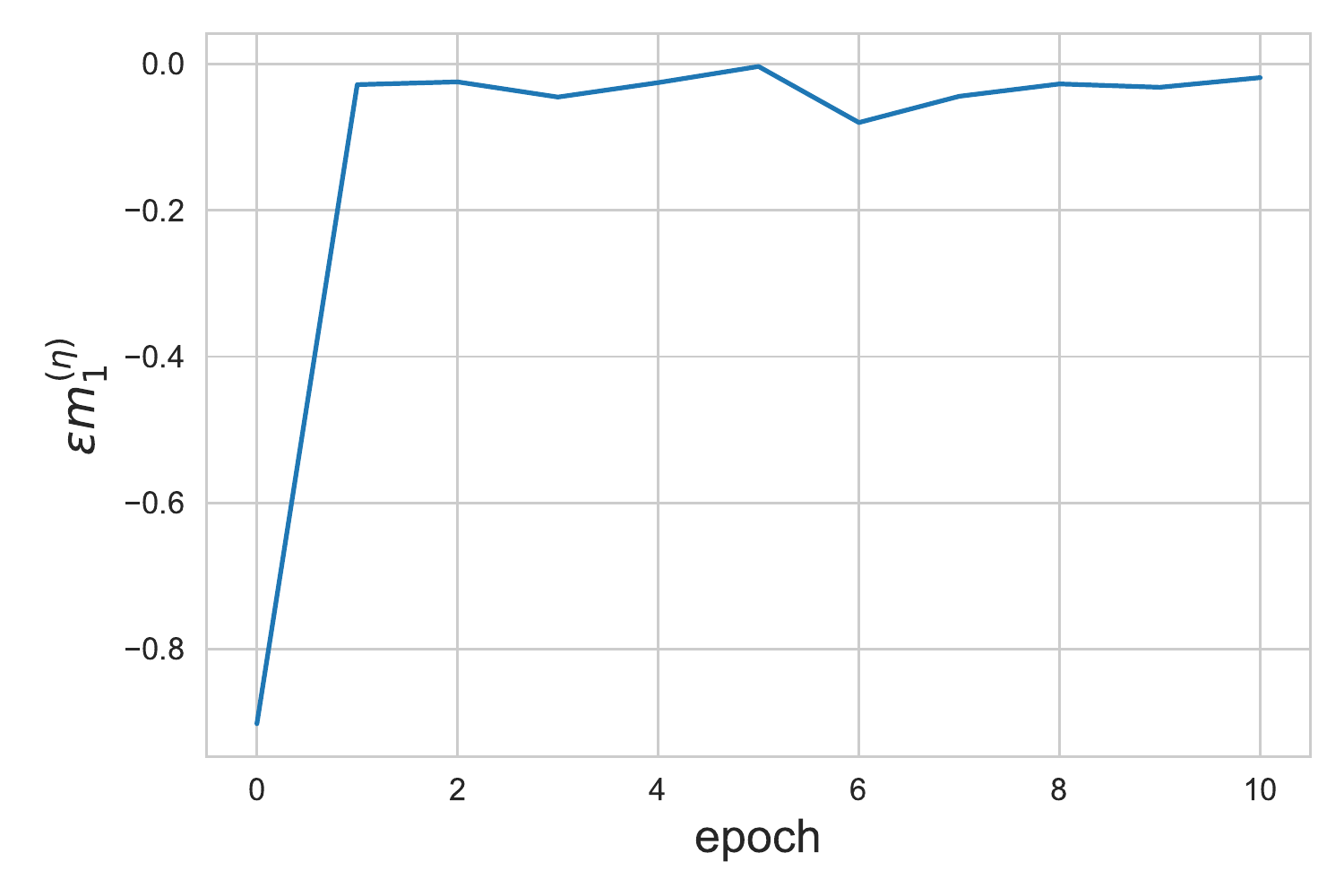}
     \subcaption{$\epsilon m_1^{(\eta)}$, MLP on MNIST}
    \end{subfigure}
    
    \begin{subfigure}{0.3\linewidth}
     \centering
     \includegraphics[width=\linewidth]{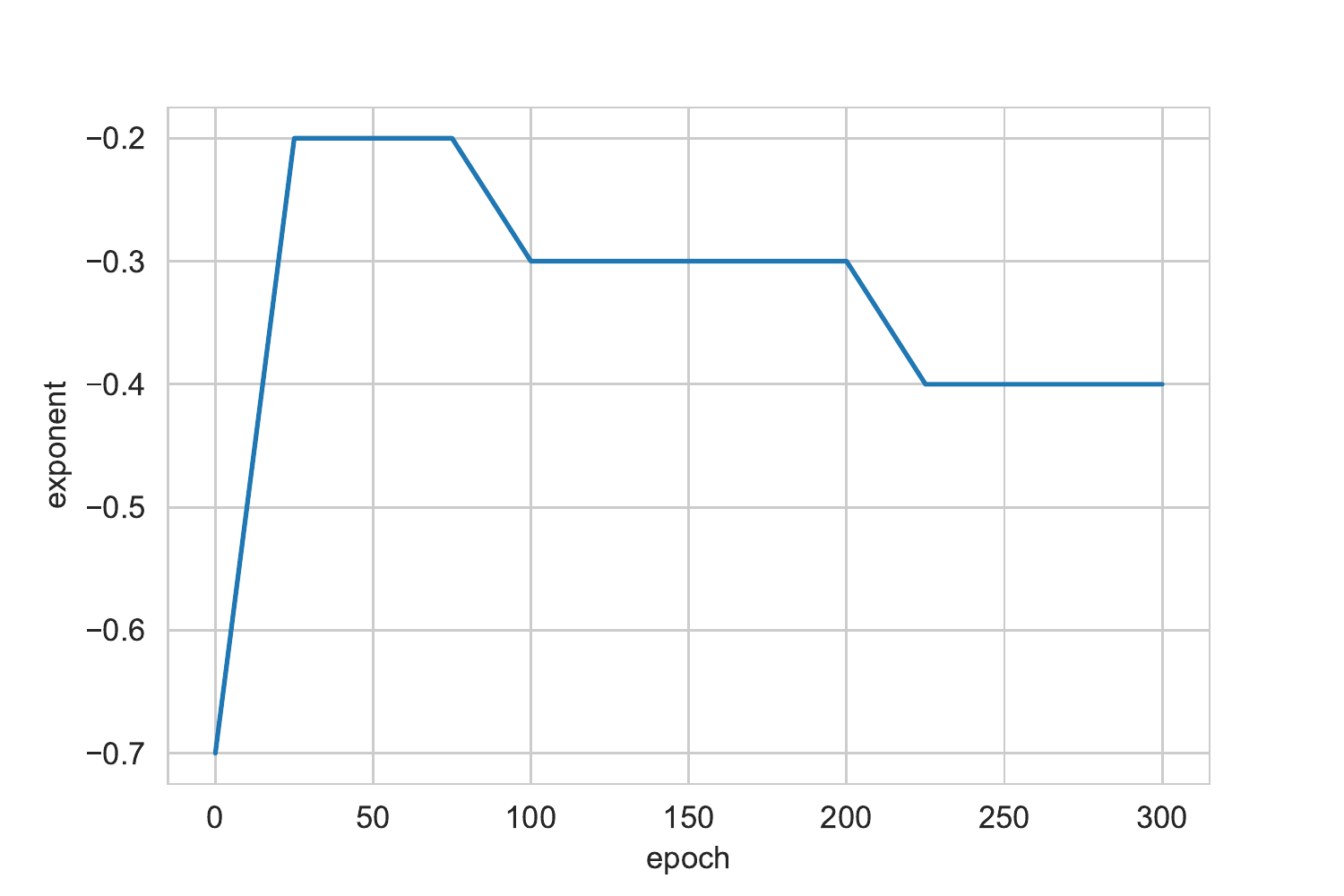}
     \subcaption{Exponent $\upsilon$, Resnet on CIFAR100}
               \label{fig:resnet_q}

    \end{subfigure}
    \begin{subfigure}{0.3\linewidth}
     \centering
     \includegraphics[width=\linewidth]{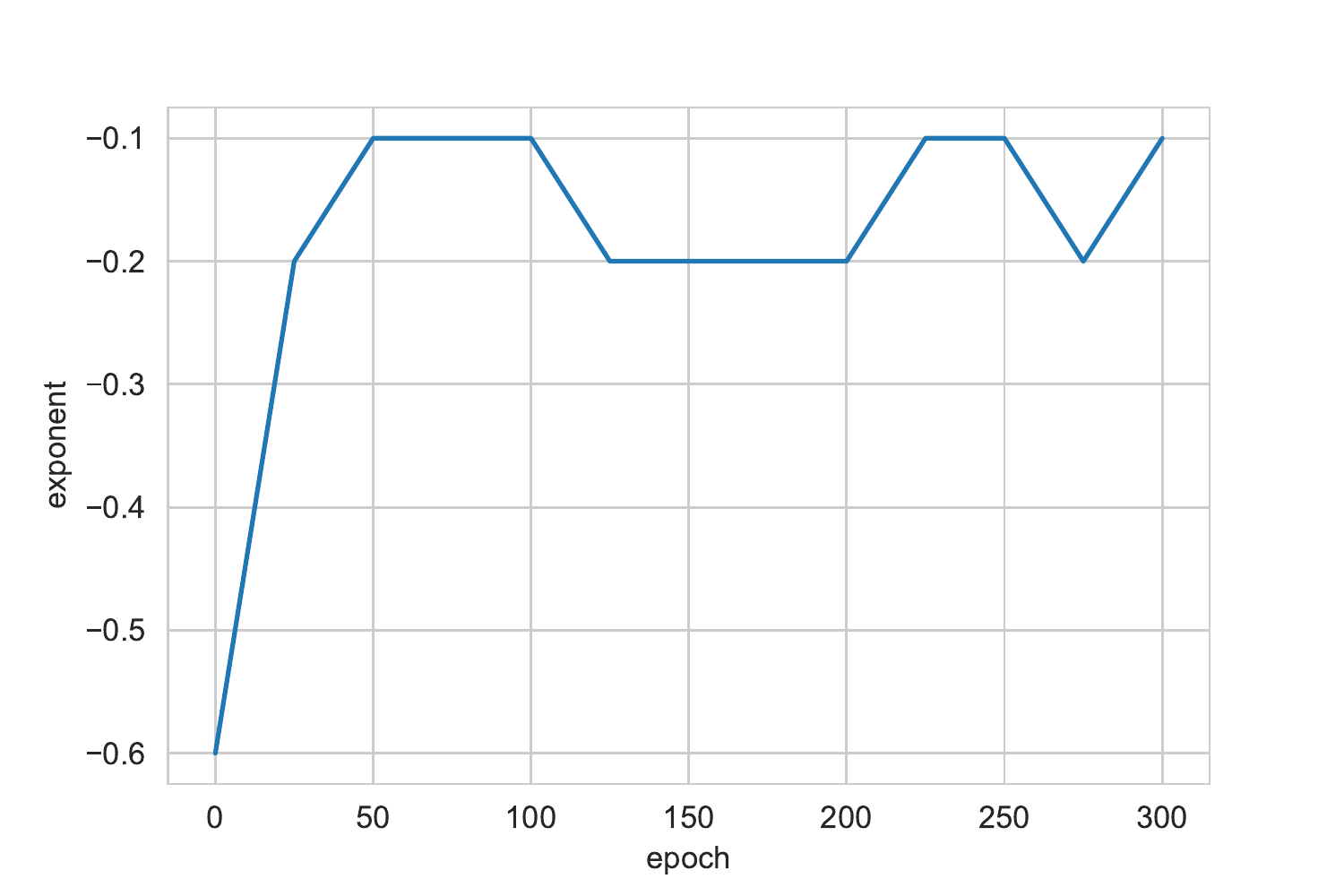}
     \subcaption{Exponent $\upsilon$, VGG16 on CIFAR100}
                    \label{fig:vgg_q}

    \end{subfigure}
    \begin{subfigure}{0.3\linewidth}
     \centering
     \includegraphics[width=\linewidth]{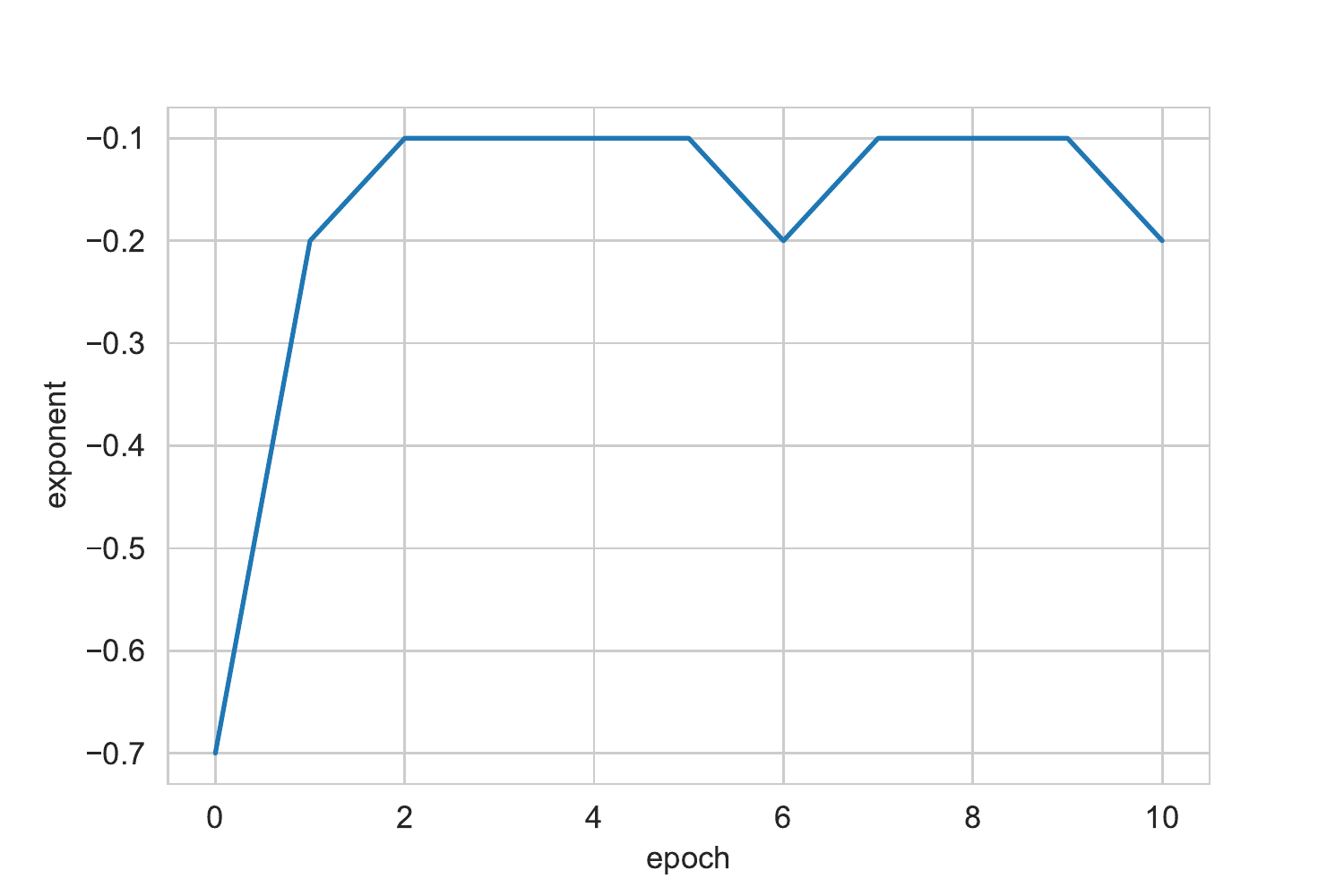}
     \subcaption{Exponent $\upsilon$, MLP on MNIST}
                    \label{fig:mlp_q}

    \end{subfigure}    
    
    \begin{subfigure}{0.3\linewidth}
     \centering
     \includegraphics[width=\linewidth]{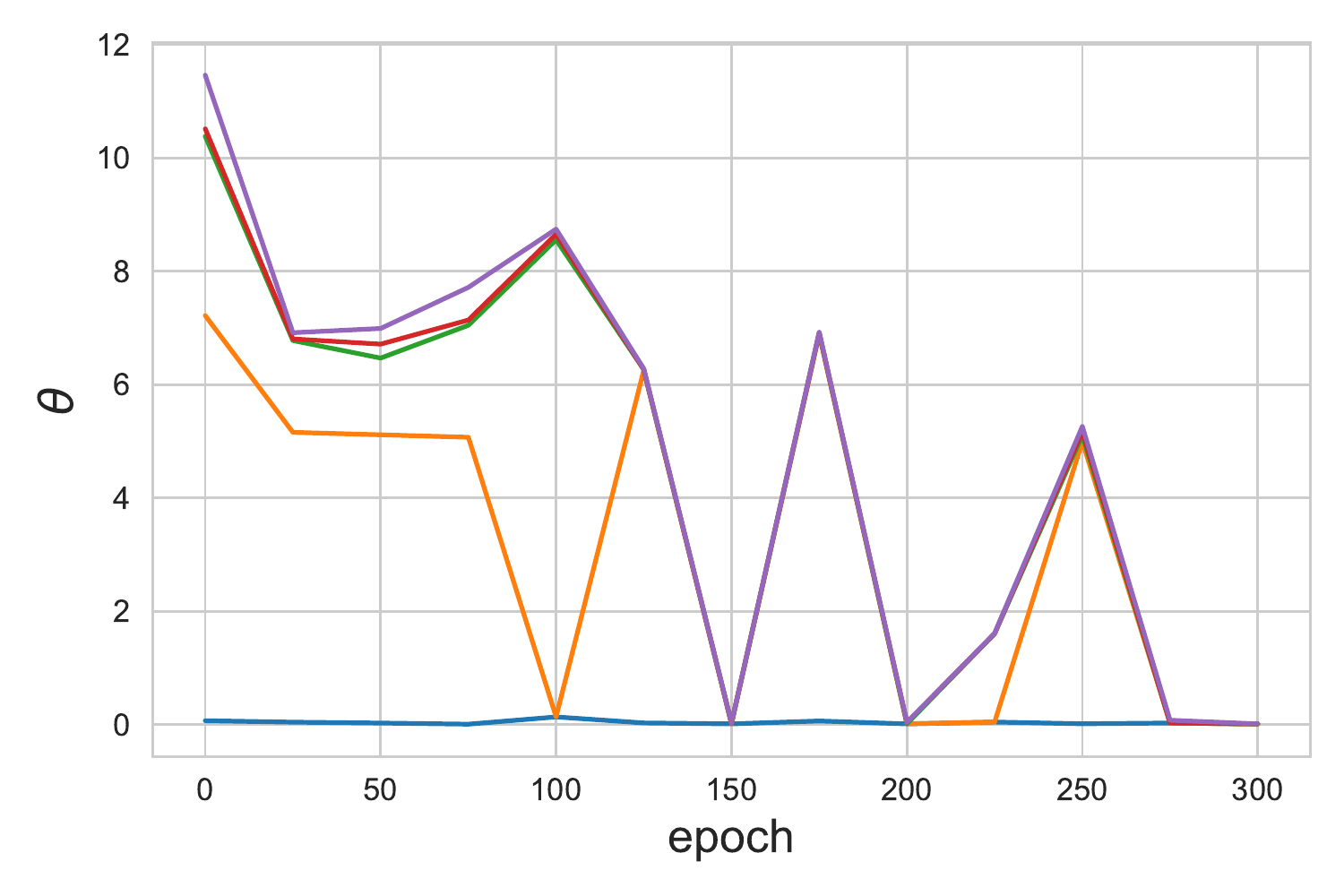}
     \subcaption{$\theta_1,\ldots, \theta_5$, Resnet on CIFAR100}
    \end{subfigure}
    \begin{subfigure}{0.3\linewidth}
     \centering
     \includegraphics[width=\linewidth]{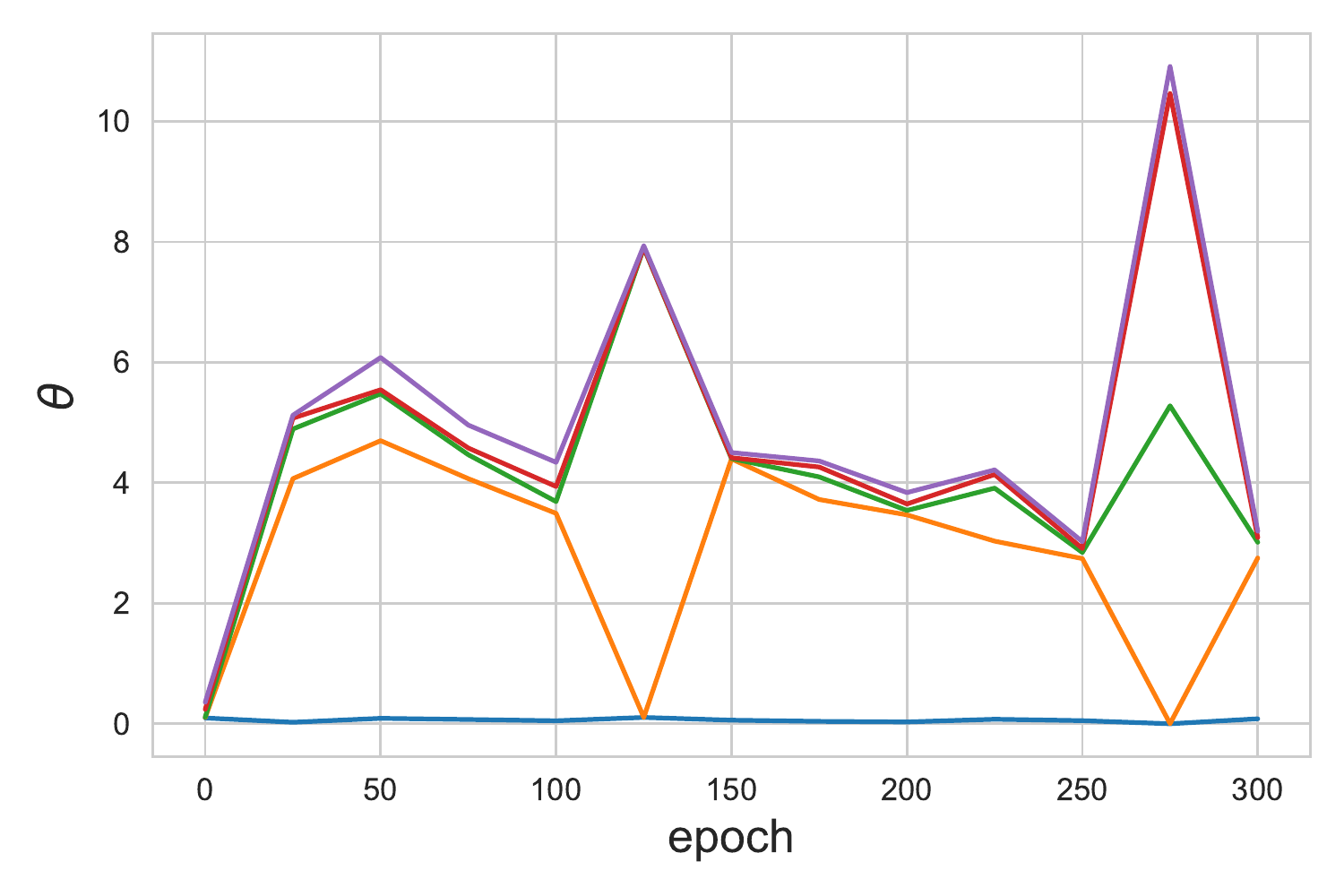}
     \subcaption{$\theta_1,\ldots, \theta_5$, VGG16 on CIFAR100}
    \end{subfigure}
    \begin{subfigure}{0.3\linewidth}
     \centering
     \includegraphics[width=\linewidth]{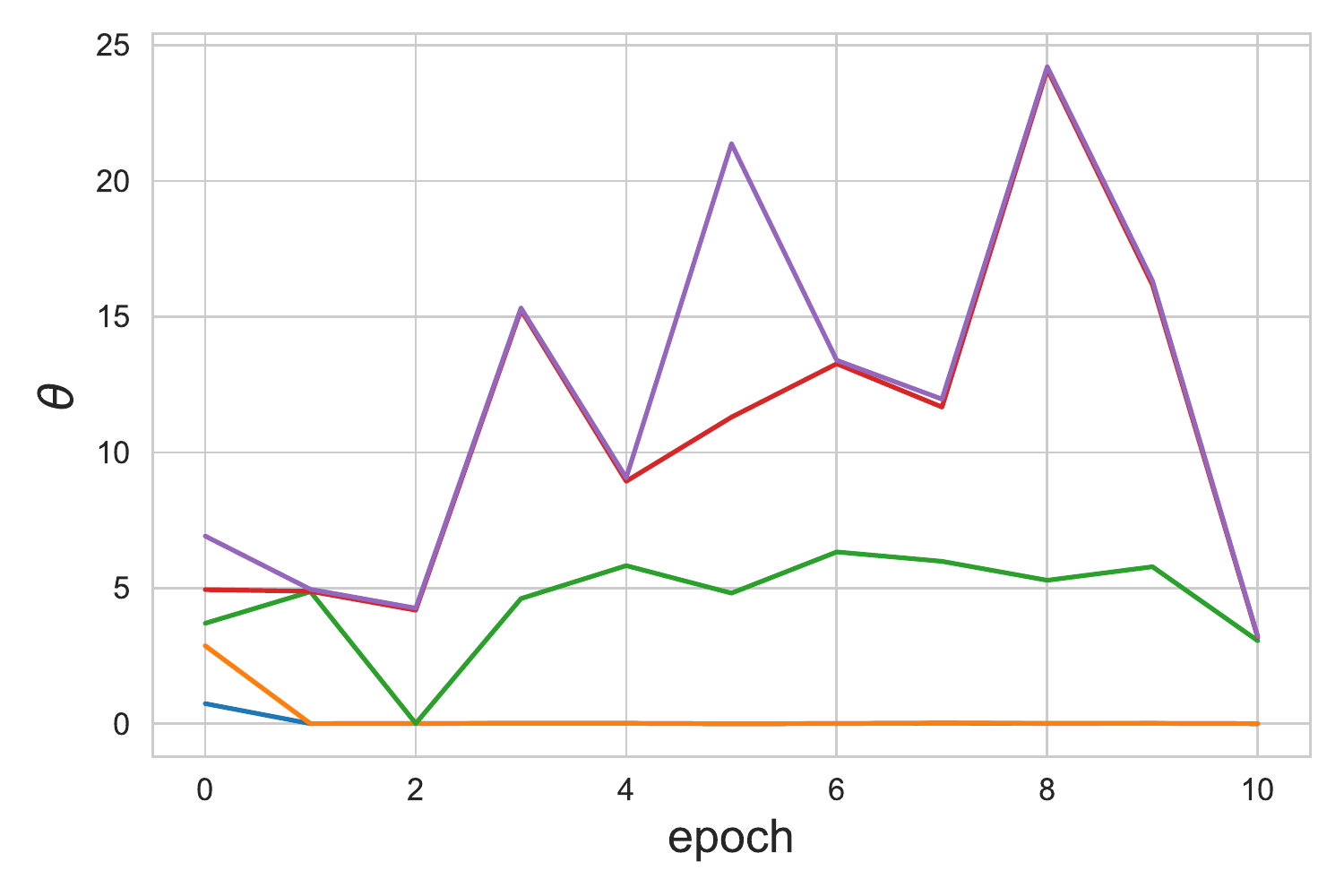}
     \subcaption{$\theta_1,\ldots, \theta_5$, MLP on MNIST}
    \end{subfigure}
    \centering
    \caption{The parameter values produced when fitting experimental neural network Hessian outlier data to (\ref{eq:omega_fit_form}).}
    \label{fig:outlier_fit_summary_params}
\end{figure}

\medskip
The parameter values obtained for all models over all epochs are shown in Figure \ref{fig:outlier_fit_summary_params}, with a column for each model.
There are several interesting features to draw out of these plots, however note that we cannot meaningfully interpret the parameters for the MLP beyond epoch 0, as the agreement with (\ref{eq:omega_fit_form}) is so poor. 
Firstly consider the parameter $m_1^{(\mu)}$, which is interpreted as the first moment (i.e. mean) of the spectral density of the noise matrix $X$.
$m_1^{(\mu)}=0$ is significant, as it is seen in the case of the a symmetric measure $\mu$, such as the Wigner semicircle used by \cite{granziol2020learning}.
For the VGG, $m_1^{(\mu)}$ starts close to 0 (Figure  \ref{fig:vgg_m1}) and generally grows with training epochs (note that the right hand side of this plot is not trustworthy, as we have observed that the agreement with (\ref{eq:omega_fit_form}) does not survive to the end of training).
For the Resnet, we see a similar upwards trend (Figure  \ref{fig:resnet_m1}), with the notable exception that of initialisation (epoch 0). 
These two observations together, suggest that training encourages a skew in the spectrum of $X$ away from symmetry around 0, however for some structural reason the Resnet is highly skewed at initialisation.

Note that for all models this parameter starts close to 0 and generally grows with training epochs, noting that the right hand side of Figure \ref{fig:vgg_m1} at the higher epochs should be ignored owing to the bad fit discussed above.

It is interesting also to observe that $\epsilon m_1^{(\eta)}$ remains small for all epochs particularly compared to $m_1^{(\mu)}, k_2^{(\mu)}$.
This is consistent with the derivation of (\ref{eq:omega_fit_form}), which relies on $\epsilon$ being small, however we emphasise that \emph{this was not imposed as a numerical constraint} but arises naturally from the data.
Recall that the magnitude of $\epsilon m_1^{(\eta)}$ measures the extent of the deviation of $A$ from being exactly low rank, so its small but non-zero values suggest that it is indeed important to allow for the true Hessian to have non-zero rank in the $N\rightarrow\infty$ limit.
Finally, we comment that the best exponent is generally not $\upsilon=1/2$.
Again, the results from the Resnet are the most reliable and they appear to show that the batch scaling, as characterised by $\upsilon$, is not constant throughout training, particularly comparing epoch 0 and epoch 300, say.


\subsection{Justification and motivation of QUE}\label{subsec:que_just}

We recall the various types of local law first introduced in section \ref{sec:universal_intro}.
All provide high probability control on the error between the (random) matrix Green's function $G(z) = (z - X)^{-1}$ and certain deterministic equivalents.
In all cases we use the set 
\begin{align}
    \vec{S} = \left\{E + i\eta \in \C \mid |E| \leq \omega^{-1}, ~ N^{-1 + \omega} \leq \eta \leq \omega^{-1}\right\}
\end{align}
for $\omega\in(0, 1)$ and the local law statements holds for all (large) $D>0$ and (small) $\xi > 0$ and for all large enough $N$.
The \emph{averaged local law} states:
\begin{align}
   \sup_{z\in\vec{S}} \P\left(\left|\frac{1}{N}\Tr G(z) - g_{\mu}(z)\right| > N^{\xi}\left(\frac{1}{N\eta} + \sqrtsign{\frac{\Im g_{\mu}(z)}{N\eta}}\right)\right) \leq N^{-D}.
\end{align}
The \emph{isotropic local law} states:
\begin{align}\label{eq:iso_local_law}
    \sup_{\|\vec{u}\|,\|\vec{v}\|  = 1, z\in\vec{S}}\P\left( |\vec{u}^TG(z)\vec{v} - g_{\mu}(z)| > N^{\xi}\left(\frac{1}{N\eta} + \sqrtsign{\frac{\Im g_{\mu}(z)}{N\eta}}\right)\right) \leq N^{-D}.
\end{align}
The \emph{anisotropic local law} states:
\begin{align}\label{eq:aniso_local_law}
    \sup_{\|\vec{u}\|,\|\vec{v}\|  = 1, z\in\vec{S}}\P\left( |\vec{u}^TG(z)\vec{v} - \vec{u}^T\Pi(z)\vec{v}| > N^{\xi}\left(\frac{1}{N\eta} + \sqrtsign{\frac{\Im g_{\mu}(z)}{N\eta}}\right)\right) \leq N^{-D}
\end{align}
where $\Pi(\cdot)$ is an $N\times N$ deterministic matrix function on $\mathbb{C}$.
The \emph{entrywise local law} states:
\begin{align}\label{eq:entry_local_law}
    \sup_{z\in\vec{S}, 1\leq i,j\leq N}\P\left( |G_{ij}(z) - \Pi_{ij}(z)| > N^{\xi}\left(\frac{1}{N\eta} + \sqrtsign{\frac{\Im g_{\mu}(z)}{N\eta}}\right)\right) \leq N^{-D}.
\end{align}
As mentioned above, quantum unique ergodicity was proved for general Wigner matrices in \cite{bourgade2017eigenvector}.
It appears that the key ingredient in the proof of QUE (\ref{eq:que_def}) in \cite{bourgade2017eigenvector} is the isotropic local semicircle law (\ref{eq:iso_local_law}) for general Wigner matrices.
Indeed, all the intermediate results in Sections 4 of \cite{bourgade2017eigenvector} take only (\ref{eq:iso_local_law}) and general facts about the Dyson Brownian Motion eigenvector flow given by \begin{align}
    d\lambda_k &= \frac{dB_{kk}}{\sqrtsign{N}} + \left(\frac{1}{N}\sum_{\ell\neq k} \frac{1}{\lambda_k - \lambda_{\ell}}\right)dt,\\
    du_k &= \frac{1}{\sqrtsign{N}}\sum_{\ell\neq k}\frac{dB_{kl}}{\lambda_k - \lambda_{\ell}} u_{\ell} - \frac{1}{2N} \sum_{\ell\neq k}\frac{dt}{(\lambda_k - \lambda_{\ell})^2} u_{k}.
\end{align}
This can be generalised to 
\begin{align}
    d\lambda_k &= \frac{dB_{kk}}{\sqrtsign{N}} + \left(-V(\lambda_i) + \frac{1}{N}\sum_{\ell\neq k} \frac{1}{\lambda_k - \lambda_{\ell}}\right)dt,\\
    du_k &= \frac{1}{\sqrtsign{N}}\sum_{\ell\neq k}\frac{dB_{kl}}{\lambda_k - \lambda_{\ell}} u_{\ell} - \frac{1}{2N} \sum_{\ell\neq k}\frac{dt}{(\lambda_k - \lambda_{\ell})^2} u_{k}.
\end{align}
where $V$ is a potential function.
Note that the eigenvector dynamics are unaffected by the presence of the potential $V$, so we expect to be able to generalise the proof of \cite{knowles2017anisotropic} to any random matrix ensemble with an isotropic local law by defining the potential $V$ so that the invariant ensemble with distribution $Z^{-1}e^{-N \Tr V(X)}dX$ has equilibrium measure $\mu$ ($Z$ is a normalisation constant).
We show how to construct such a $V$ from $\mu$ in Section \ref{sec:inv}.

The arguments so far suffice to justify a generalisation of the ``dynamical step'' in the arguments of \cite{bourgade2017eigenvector}, so it remains to consider the ``comparison step''. The dynamical step establishes QUE for the matrix ensemble with a small Gaussian perturbation, but in the comparison step one must establish that the perturbation can be removed without breaking QUE.
To our knowledge no such argument has been articulated beyond generalized Wigner matrices, with the independence of entries and comparable scale of variances being critical to the arguments given by \cite{bourgade2017eigenvector}.
Our guiding intuition is that QUE of the form (\ref{eq:que_def}) is a general property of random matrices and can reasonably be expected to hold in most, if not all, cases in which there is a local law and universal local eigenvalue statistics are observed.
At present, we are not able to state a precise result establishing QUE in sufficient generality to be relevant for this work, so we shall take it as an assumption.
\begin{assump}\label{thm:que_gaussian}
    Let $X$ be an ensemble of $N\times N$ real symmetric random matrices. Assume that $X$ admits a limiting spectral measure is $\mu$ with Stieljtes transform $m$. Suppose that the isotropic local law (\ref{eq:iso_local_law}) holds for $X$ with $\mu$.
    Then there is some set $\mathbb{T}_N \subset [N]$ with $|\mathbb{T}_N^c| = o(N)$ such that with $|I|=n$, for any polynomial $P$ in $n$ indeterminates, there exists some $\epsilon(P) > 0$ such that for large enough $N$ we have
    \begin{align}
        \sup_{\substack{I \subset \mathbb{T}_N, |I|=n,\\ \|\vec{q}\| = 1}} \left| \E\left(P\left(\left(N(\vec{q}^T\vec{u}_k)^2\right)_{k\in I}\right)\right) - \E\left(P\left(\left(|\mathcal{N}_j|^2\right)_{k\in I}\right)\right)\right| \leq N^{-\epsilon}.
    \end{align}
\end{assump}
Note that the isotropic local law in Assumption \ref{thm:que_gaussian} can be obtained from the weaker entrywise law (\ref{eq:entry_local_law}) as in Theorem 2.14 of \cite{alex2014isotropic} provided there exists a $C>0$ such that $\E|X_{ij}|^2 \leq CN^{-1}$ for all $i,j$ and there exists $C_p > 0$ such that $\E |\sqrtsign{N}X_{ij}|^p \leq C_p$ for all $i,j$ and integer $p>0$.

\begin{remark}
In \cite{bourgade2017eigenvector} the restriction $I\subset\TN$ is given for the explicit set 
\begin{align}
\mathbb{T}_N = [N] \backslash \{(N^{1/4}, N^{1-\delta}) \cup (N - N^{1-\delta}, N - N^{1/4})\}
\end{align}
for some $0< \delta < 1$.
In the case of generalised Wigner matrices, this restriction on the indices has since been shown to be unnecessary \cite{benigni2022optimal, benigni2020eigenvectors,benigni2021fluctuations}.
In our context, we could simply take as an assumption all results holds with $\mathbb{T}_N = [N]$, however our results can in fact be proved using only the above assumption that $|\mathbb{T}_N^c| = o(N)$, so we shall retain this weaker form of the assumptions.
\end{remark}

This section is not intended to prove QUE from explicit known properties of deep neural network Hessians, but rather to provide justification for it as a reasonable modeling assumption in the noise model for Hessians defined in section \ref{subsec:hess_model}.
We have shown how QUE can be obtained from an isotropic (or entrywise) local law beyond the Wigner case.
It is important to go beyond Wigner or any other standard random matrix ensemble, as we have observed above that the standard macroscopic spectral densities of random matrix theory such as the semicircle law are not observed in practice.
That said, we are not aware of any results establishing QUE in the more general case of anistropic local laws, and this appears to be a very significant technical challenge.
We must finally address why a local law assumption, isotropic or otherwise, may be reasonable for the noise matrix $X$ in our Hessian model.
Over the last decade or so, universal local statistics of random matrices in the form of $k$-point correlation functions on the appropriate microscopic scale have been established for a litany of random matrix ensembles.
An immediate consequence of such results is that, on the scale of unit mean eigenvalue spacing, Wigner's surmise holds \review{to a very good approximation}, depending only on the symmetry class (orthogonal, unitary or symplecitic).
\review{Such universality results are rather older for invariant ensembles \cite{deift1999orthogonal,erdos2017dynamical} and can be established with orthogonal polynomial techniques, however the recent progress focusing on non-invariant ensembles, beginning with Wigner matrices \cite{erdHos2012bulk} and proceeding to much more general ensembles \cite{erdHos2019random}, is built on a very general ``three step strategy'' (though see \cite{erdHos2012universality} for connections between universality in invariant and non-invariant ensembles).} 
As with the QUE proof discussed above, the key ingredient in these proofs, as part of the three step strategy \cite{erdos2017dynamical}, is establishing a local law.
The theoretical picture that has emerged is that, \review{for very general random matrices,} when universal local eigenvalue statistics are observed in random matrices, it is due to the mechanism of short time scale relaxation of local statistics under Dyson Brownian Motion made possible by a local law.

In Chapter \ref{chap:spacings} \cite{baskerville2022appearance} we observed that universal local eigenvalue statistics do indeed appear to be present in the Hessian of real, albeit quite small, deep neural networks.
Given all of this context, we propose that a local law assumption of some kind is reasonable  for deep neural network Hessians and not particularly restrictive.
As we have shown, if we are willing to make the genuinely restrictive assumption of an isotropic local law for the Hessian noise model, then QUE follows.
However an anistropic local law is arguably more plausible as we expect deep neural networks Hessians to contain a good deal of dependence between entries, and such correlations are know to generically lead to anisotropic local laws \cite{erdHos2019random}.

\subsection{Motivation of true Hessian structure}
In this section we revisit and motivate the assumptions made about the Hessian in Section \ref{subsec:hess_model}.
Firstly note that one can always define $A = \E H_{\text{batch}}$ and it is natural then to associate $A$ with the true Hessian $H_{\text{true}}$.
In light of (\ref{eq:batch_hessian_def}), it is natural to expect some fixed form of the law for $H_{\text{batch}} - A$ for any batch size, but with an overall scaling $\sbf$, which must naturally be decreasing in $b$ \review{as experimental results show that the overall spectral width of the batch Hessians of neural networks decreases with increasing batch size.}
Next we address the assumptions made about the spectrum of $A$.
The first assumption one might think to make is that $A$ has fixed rank relative to $N$, with spectrum consisting only of the spikes $\theta_i, \theta_j'$.
\review{Indeed, it has been repeatedly observed, in our own experiments and others \cite{papyan2018full,granziol2020learning}, that neural network Hessians contain a number of spectral outliers separated from the spectral bulk. 
It is natural to conjecture that such outliers arise from some outliers in an underlying structured deterministic matrix of which the batch Hessian is a noisy version, as in the case of BBP style phase transitions in random matrix theory.
The outliers in neural network Hessians have been associated with inter-class separation in the case of classification models \cite{papyan2019measurements} and it can be observed that spectra lack (or have smaller and fewer) outliers at the start of training, or if they are intentionally trained to give poor (i.e. random) predictive performance.
That being said, in almost any experiment with sensibly trained neural networks, spectral outliers are observed, and over a range of batch sizes (and hence noise levels) suggesting that some of the spike eigenvalues in the true Hessian are above the phase transition threshold.
}

Behind such an assumption is the intuition that the data distribution does not depend on $N$ and so, in the over-parametrised limit $N\rightarrow\infty$, the overwhelming majority of directions in weight space are unimportant.
The form we take for $A$ in the above is a strict generalisation of the fixed rank assumption; $A$ still has a fixed number of spiked directions, but the parameter $\epsilon$ controls the rank of $A$.
Since any experimental investigation is necessarily limited to $N<\infty$, the generalisation to $\review{\epsilon}>0$ is particularly important.
Compact support of the measures $\mu$ and $\eta$ is consistent with experimental observations of deep neural network Hessian spectra.

\subsection{The batch size scaling}
Our experimental results considered $\sbf = b^{-\upsilon}$ and $\upsilon=1/2$ is the value required to give agreement with \cite{granziol2020learning}, a choice which we now justify.
From (\ref{eq:batch_hessian_def}) we have \begin{align}
    H_{\text{batch}} = \frac{1}{b}\sum_{i=1}^b\left(H_{\text{true}} + X^{(i)}\right)
\end{align}
where $X^{(i)}$ are i.i.d. samples from the law of $X$.
Suppose that the entries $X_{ij}$ were Gaussian, with $\text{Cov}(X_{ij}, X_{kl}) = \Sigma_{ij,kl}$.
Then $Z = X^{(p)}_{ij} + X^{(q)}_{ij}$ has 
\begin{align}
    \text{Cov}(Z_{ij}, Z_{kl}) = \E X_{ij}^{(p)}X_{kl}^{(p)} + \E X_{ij}^{(q)}X_{kl}^{(q)} -  \E X_{ij}^{(p)}\E X_{kl}^{(p)} - \E X_{ij}^{(q)}\E X_{kl}^{(q)}  = 2\Sigma_{ij, kl}.
\end{align}
In the case of centred $X$, one then obtains\begin{align}
    \frac{1}{b}\sum_{i=1}^b X^{(i)} \overset{d}{=} b^{-1/2}X.
\end{align}
Note that this does not quite match the case described in Section \ref{subsec:hess_model}, since we do not assume \review{there that} $\E X = 0$, however we take this a rough justification for $\sbf = b^{-1/2}$ as an ansatz.
Moreover, numerical experimentation with $\sbf = b^{-\upsilon}$ for values of $\upsilon>0$ shows that $q = 1/2$ gives a reasonable fit to the data (\review{note that the values shown in Figures \ref{fig:resnet_q}, \ref{fig:vgg_q}, \ref{fig:mlp_q} are those producing the best fit, but $\upsilon=1/2$ was seen to be not much inferior).}
\FloatBarrier

\section{Spectral free addition from QUE}\label{sec:que}

\subsection{Intermediate results on QUE}\label{sec:fourier}
This section establishes some intermediate results that follow from assuming QUE for the eigenvectors of a matrix. They will be crucial for our application in the following section. 

\begin{lemma}\label{lem:que_preserved_under_rotation}
    Consider a real orthogonal $N\times N$ matrix $U$ with rows $\{\vec{u}_i^T\}_{i=1}^N$. Assume that $\{\vec{u}_i\}_{i=1}^N$ are the eigenvectors of a real random symmetric matrix with QUE. Let $P$ be a fixed $N\times N$ real orthogonal matrix. Let $V = UP$ and denote the rows of $V$ by $\{\vec{v}_i^T\}_{i=1}^N$. Then $\{\vec{v}_i\}_{i=1}^N$ also satisfy QUE.
\end{lemma}
\begin{proof}
   Take any unit vector $\vec{q}$, then for any $k=1,\ldots, N$
   \begin{align*}
       \vec{q}^T \vec{v}_k = \sum_{j}q_jV_{kj} &= \sum_{j, l}q_j U_{kl}P_{lj} = (P\vec{q})^T \vec{u}_k.
   \end{align*}
   But $\|P\vec{q}\|_2=\|\vec{q}\|_2=1$ since $P$ is orthogonal, so the statement of QUE for $\{\vec{u}_i\}_{i=1}^N$   transfers directly to $\{\vec{v}_i\}_{i=1}^N$ thanks to the supremum of all unit $\vec{q}$.
\end{proof}

\begin{lemma}\label{lem:weaker_col_que}
    Consider a real orthogonal $N\times N$ matrix $U$ with rows $\{\vec{u}_i^T\}_{i=1}^N$. Assume that $\{\vec{u}_i\}_{i=1}^N$ are the eigenvectors of a real random symmetric matrix with QUE. Let $\ell_0(\vec{q}) = \sum_i \1\{q_i \neq 0\}$ count the non-zero elements of a vector with respect to a fixed orthonormal basis $\{\vec{e}_i\}_{i=1}^N$. 
    For any fixed integer $s>0$, define the set \begin{align}
            \Vs = \left\{\vec{q}\in \R^N \mid \|\vec{q}\|=1, ~ \ell_0(\vec{q})=s, ~ q_i=0 ~\forall i\in \TN^c\right\}
    \end{align}
    \review{where, recall the definition
    \begin{align*}
\mathbb{T}_N = [N] \backslash \{(N^{1/4}, N^{1-\delta}) \cup (N - N^{1-\delta}, N - N^{1/4})\}.
\end{align*}}
    Then the columns $\{\vec{u}_i'\}_{i=1}^N$ of $U$ satisfy a weaker form of QUE (for any fixed $n, s>0$):
    \begin{align}\label{eq:que_weak_def}
    \sup_{\substack{\vec{q}\in\Vs \\ }}\sup_{\substack{I \subset \TN\\ |I| = n}} \left|\E P\left(\left(N|\vec{q}^T\vec{u}_k|^2\right)_{k\in I}\right) - \E P\left(\left(|\mathcal{N}_j|^2\right)_{j=1}^m
    \right)\right| \leq N^{-\epsilon}.
\end{align}
We will denote this form of QUE as $\hQUE$.
\end{lemma}
\begin{proof}
   Take some $\vec{q}\in \Vs$. Then there exists some $J\subset\TN$ with $|J|=s$ and non-zero $\{q_k\}_{k\in J}$ such that\begin{align*}
       \vec{q}^T\vec{u}_k' = \sum_{j\in J} q_j \vec{e}_j^T \vec{u}_k'.
   \end{align*}
   Take $\{\vec{e}_i\}_{i=1}^N$ to be a standard basis with $(\vec{e}_i)_j = \delta_{ij}$, then $\vec{e}_j^T \vec{u}_k' = U_{jk} = \vec{e}_k^T\vec{u}_j$ so \begin{align*}
       \vec{q}^T\vec{u}_k' = \sum_{j\in J} q_j \vec{e}_k^T\vec{u}_j
   \end{align*}
   but then the coefficients $q_j$ can be absorbed into the definition of the general polynomial in the statement (\ref{eq:que_def}) of QUE for $\{\vec{u}_i\}_{i=1}^N$, which completes the proof, noting that the sum only includes indices contained in $\TN$ owing to the definition of $\Vs$.
\end{proof}

\begin{lemma}\label{lem:explicit_error}
    Fix some real numbers $\{y_i\}_{i=1}^r$. Fix also a diagonal matrix $\Lambda$ and an orthonormal set of vectors $\{\vec{v}_i\}_{i=1}^N$ that satisfies $\hQUE$. Then there exists an $\epsilon>0$ and $\vec{\eta}_i\in\C^N$ with
      \begin{align}
       \eta_{ij}^2 &\in [-1, 1] ~ \forall j\in\TN,\\
       \eta_{ij}^2 &\in [-N^{\epsilon}, N^{\epsilon}] ~ \forall j\in\TN^c.
   \end{align}
 such that for any integer $l>0$\begin{align}\label{eq:explicit_error_lem1}
        \mathbb{E} \left(\sum_{i=1}^r y_i \vec{v}_i^T\Lambda \vec{v}_i\right)^l - \mathbb{E}\left(\sum_{i=1}^r y_i \frac{1}{N}\vec{g}_i^T\Lambda \vec{g}_i\right)^l = N^{-(1+\epsilon)l} \left( \sum_{i=1}^r y_i \vec{\eta}_i^T\Lambda \vec{\eta}_i\right)^l
    \end{align}
    where the $\vec{g}_i$ are i.i.d. Gaussians $N(0, I_N)$. 
    
\end{lemma}
\begin{proof}
   Let $\{\vec{e}_i\}_{i=1}^N$ be the standard orthonormal basis from above. Then 
   \begin{align}
       \E \left(\sum_{i=1}^r y_i \vec{v}_i^T\Lambda \vec{v}_i\right)^l &= \E \sum_{i_1,\ldots, i_l=1}^r\prod_{k=1}^{l} y_{i_k} \vec{v}_{i_k}^T\Lambda\vec{v}_{i_k}\notag\\
       &= \E \sum_{i_1,\ldots, i_l=1}^r\sum_{j_1,\ldots, j_l=1}^N\prod_{k=1}^{l} y_{i_k} \lambda_{j_k} (\vec{e}_{j_k}^T\vec{v}_{i_k})^2\label{eq:que_error_bound1}\\
       \implies \E \left(\sum_{i=1}^r y_i \vec{v}_i^T\Lambda \vec{v}_i\right)^l-\mathbb{E}\left(\sum_{i=1}^r y_i \frac{1}{N}\vec{g}_i^T\Lambda \vec{g}_i\right)^l &= N^{-l}\sum_{i_1,\ldots, i_l=1}^r\sum_{j_1,\ldots, j_l=1}^N\prod_{k=1}^{l} y_{i_k} \lambda_{j_k} \left[N\E(\vec{e}_{j_k}^T\vec{v}_{i_k})^2 - \E (\vec{e}_{j_k}^T\vec{g}_{i_k})^2\right]\notag\\
       &=N^{-l}\sum_{i_1,\ldots, i_l=1}^r\sum_{j_1,\ldots, j_l\in\TN}\prod_{k=1}^{l} y_{i_k} \lambda_{j_k} \left[N\E(\vec{e}_{j_k}^T\vec{v}_{i_k})^2 - \E (\vec{e}_{j_k}^T\vec{g}_{i_k})^2\right]\notag\\
       &+ N^{-l}\sum_{i_1,\ldots, i_l=1}^r\sum_{\substack{j_1\in\TN^c,\\ j_2,\ldots, j_l\in\TN}}\prod_{k=1}^{l} y_{i_k} \lambda_{j_k} \left[N\E(\vec{e}_{j_k}^T\vec{v}_{i_k})^2 - \E (\vec{e}_{j_k}^T\vec{g}_{i_k})^2\right]\notag\\
       &+ \ldots
   \end{align}
   The ellipsis represents the similar terms where further of the $j_1, \ldots, j_r$ are in $\TN^c$.
   For $j\in\TN^c$ the terms \begin{align}
       \left[N\E(\vec{e}_{j_k}^T\vec{v}_{i_k})^2 -\E (\vec{e}_{j_k}^T\vec{g}_{i_k})^2\right]
   \end{align}
   are excluded from the statement of $\hQUE$, however we can still bound them crudely.
   Indeed \begin{align}
       \sum_{j\in\TN^c} N(\vec{e}_j^T\vec{v}_i)^2 = \sum_{j=1}^N N(\vec{e}_j^T\vec{v}_i)^2 - \sum_{j\in\TN}N(\vec{e}_j^T\vec{v}_i)^2 = N - \sum_{j\in\TN}N(\vec{e}_j^T\vec{v}_i)^2
   \end{align}
   but since the bound of $\hQUE$ applies for $j\in\TN$ \begin{align}
       N\E (\vec{e}_j^T\vec{v}_i)^2 = \E(\vec{e}_j^T\vec{g})^2 + o(1) = 1 + o(1) ~~ \forall j\in\TN,
   \end{align}
   then \begin{align}
       \sum_{j\in\TN^c} N(\vec{e}_j^T\vec{v}_i)^2 = N - N(1 + o(1)) = o(N) ~ \implies ~  \E(\vec{e}_j^T\vec{v}_i)^2 = o(1) ~ \forall j \in\TN^c.
   \end{align}
   Note that this error term is surely far from optimal, but is sufficient here.
   Overall we can now say \begin{align}
      \left|\left[N\E(\vec{e}_{j}^T\vec{v}_{i})^2 -\E (\vec{e}_{j}^T\vec{g}_{i})^2\right]\right| \leq 1 + o(1) \leq 2 ~ \forall j \in\TN^c.
   \end{align}

   We can apply $\hQUE$ to the terms in square parentheses to give $\epsilon_{1}, \ldots, \epsilon_{r}>0$ such that 
   \begin{align}
       |N\E(\vec{e}_{j_k}^T\vec{v}_{i_k})^2 - \E (\vec{e}_{j_k}^T\vec{g}_{i_k})^2| \leq N^{-\epsilon_{i_k}} ~~~ \forall j_k\in\TN ~\forall i_k=1,\ldots, r.
   \end{align}
   We can obtain a single error bound by setting $\epsilon = \min_i \epsilon_i$, where clearly $\epsilon > 0$ and then write
   \begin{align}\label{eq:eta_defn}
    N\E(\vec{e}_{j_k}^T\vec{v}_{i_k})^2 - \E (\vec{e}_{j_k}^T\vec{g}_{i_k})^2 = \eta_{i_kj_k}^2 N^{-\epsilon}    
   \end{align}
   where $ \eta_{i_kj_k}^2 \in [-1, 1]$. To further include the indices $j\in\TN^c$, we extend the expression (\ref{eq:eta_defn}) to all $j_k$ by saying
   \begin{align}
       \eta_{i_kj_k}^2 &\in [-1, 1] ~ \forall j_k\in\TN,\\
       \eta_{i_kj_k}^2 &\in [-N^{\epsilon}, N^{\epsilon}] ~ \forall j_k\in\TN^c.
   \end{align}
Overall we have \begin{align}
     \E \left(\sum_{i=1}^r y_i \vec{v}_i^T\Lambda \vec{v}_i\right)^l-\mathbb{E}\left(\sum_{i=1}^r y_i \frac{1}{N}\vec{g}_i^T\Lambda \vec{g}_i\right)^l = N^{-l(1+\epsilon)}\sum_{i_1,\ldots, i_l=1}^r\sum_{j_1,\ldots, j_l=1}^N\prod_{k=1}^{l} y_{i_k} \lambda_{j_k} \eta_{i_kj_k}^2
\end{align}
but by comparing with (\ref{eq:que_error_bound1}) we can rewrite as
\begin{align}
    \E \left(\sum_{i=1}^r y_i \vec{v}_i^T\Lambda \vec{v}_i\right)^l-\mathbb{E}\left(\sum_{i=1}^r y_i \frac{1}{N}\vec{g}_i^T\Lambda \vec{g}_i\right)^l = \left( \sum_{i=1}^r N^{-(1+\epsilon)}y_i \vec{\eta}_i^T\Lambda \vec{\eta}_i\right)^l
\end{align}
where $\vec{\eta}_i^T = (\eta_{i1},\ldots, \eta_{iN})$.

\end{proof}

\subsection{Main result}
\begin{theorem}\label{thm:nearly_free_addn}
    Let $X$ be an $N\times N$ real symmetric random matrix and let $D$ be an $N\times N$ symmetric matrix (deterministic or random). Let $\hat{\mu}_X, \hat{\mu}_{D}$ be the empirical spectral measures of the sequence of matrices $X, D$ and assume there exist deterministic limit measures $\mu_X, \mu_D$. Assume that $X$ has QUE, i.e. \ref{thm:que_gaussian}.
    Assume also the $\hat{\mu}_X$ concentrates in the sense that 
    \begin{align}\label{eq:concentration_condition}
        \P(W_1(\hat{\mu}_X, \mu_X) > \delta) \lesssim e^{-N^{\tau} f(\delta)}
    \end{align}
    where $\tau>0$ and $f$ is some positive increasing function.
    Then $H = X + D$ has a limiting spectral measure and it is given by the free convolution $\mu_X \boxplus \mu_D$.
\end{theorem}
\begin{remark}
A condition like (\ref{eq:concentration_condition}) is required so that the Laplace method can be applied to the empirical measure $\hat{\mu}_X$. There are of course other ways to formulate such a condition.
Consider for example the conditions used in Theorems 1.2 and 4.1 of \cite{arous2021exponential}.
There it is assumed the existence of a sequence of deterministic measures $(\mu_N)_{N \geq 1}$ and a constant $\kappa>0$ such that for large enough $N$
\begin{align}\label{eq:concentration_condition_deterministic}
    W_1(\E \hat{\mu_X}, \mu_N) \leq N^{-\kappa}, ~~ W_1(\mu_N, \mu_X) \leq N^{-\kappa},
\end{align}
which is of course just a deterministic version of (\ref{eq:concentration_condition}).
\cite{arous2021exponential} introduce the extra condition around concentration of Lipschitz traces:
\begin{align}\label{eq:lipschitz_traces}
    \P\left( \left|\frac{1}{N} \Tr f(H_N) - \frac{1}{N}\E \Tr f(H_N)\right| > \delta \right) \leq \exp\left(-\frac{c_{\zeta}}{N^{\zeta}} \min\left\{\left(\frac{N\delta}{\|f\|_{Lip}}\right)^2, \left(\frac{N\delta}{\|f\|_{Lip}}\right)^{1+\epsilon_0} \right\}\right),
\end{align}
for all $\delta>0$, Lipschitz $f$ and $N$ large enough, where $\zeta, c_{\zeta}>0$ are some constants.
As shown in the proof of Theorem 1.2, this condition is sufficient to obtain \begin{align}\label{eq:concetration_integral}
    \P \left( \left|\int |\lambda| d\hat{\mu}_X(\lambda) - \int |\lambda| d\E\hat{\mu}_X(\lambda)\right| \leq t \right) \leq \exp\left(-\frac{c_{\zeta}}{N^{\zeta}} \min\left\{ (2Nt\eta)^2, (2Nt\eta)^{1+\epsilon_0}\right\}\right)
\end{align}
for any $t>0$ and for large enough $N$.
Note that \cite{arous2021exponential} prove this instead for integration against a regularised version of $\log|\lambda|$, but the proof relies only the integrand's being Lipschitz, so it goes through just the same here.
(\ref{eq:concetration_integral}) and (\ref{eq:concentration_condition_deterministic}) clearly combine to give (\ref{eq:concentration_condition}).
The reader may ignore this remark if they are content to take (\ref{eq:concentration_condition}) as an assumption.
Alternatively, as we have shown, (\ref{eq:concentration_condition}) can be replaced by (\ref{eq:concentration_condition_deterministic}) and (\ref{eq:lipschitz_traces}), conditions which have already been used for quite general results in the random matrix theory literature.
\end{remark}
\begin{proof}
We shall denote use the notation \begin{align}
    G_H(z) = \frac{1}{N}\Tr (z - H)^{-1}.
\end{align}
Recall the supersymmetric approach to calculating the expected trace of the resolvent of a random matrix ensemble:
\begin{align}
    \E_H G_H(z) = \frac{1}{N}\frac{\partial}{\partial \j}\Bigg|_{\j=0} \mathbb{E}_H Z_H(\j)
\end{align}
where \begin{align}
    Z_H(\j) &= \frac{\det(z + \j - H)}{\det(z - H)} = \int d\Psi e^{-i\Tr AH} e^{i\Tr \Psi\Psi^{\dagger}J},\\
    A &= \phi\phi^{\dagger} + \chi\chi^{\dagger},\\
    J &= I_N \otimes \left(\begin{array}{cc} z & 0 \\ 0 & \j + z\end{array}\right),\\
    d\Psi &= \frac{d\phi d\phi^* d\chi d\chi^*}{-(2\pi)^N i},\\
    \Psi &= \left(\begin{array}{c} \phi \\ \chi \end{array}\right)
\end{align}
with $\phi\in\C^N$ and $\chi,\chi^*$ being $N$-long vectors of anti-commuting variables.
Independence of $X$ and $D$ gives
\begin{align}
    \E_H Z_H(\j) &= \int d\Psi e^{i\Tr \Psi\Psi^{\dagger}J} \E_{X,D} e^{-i\Tr A(X + D)} \notag\\
    &= \int d\Psi e^{i\Tr \Psi\Psi^{\dagger}J} \E_D e^{-i\Tr AD} \E_X e^{-i\Tr AX}.
\end{align}
$\E_D$ simply means integration against a delta-function density if $D$ is deterministic. 

\medskip
Let us introduce some notation: for $N\times N$ matrices $K$, $\Phi_X(K) = \E_X e^{-i\Tr XK}$, and similarly $\Phi_D$. We also define a new matrix ensemble $\bar{X} \overset{d}{=} O^T \Lambda O$, where $\Lambda=\text{diag}(\lambda_1, \ldots, \lambda_N)$ are equal in distribution to the eigenvalues of $X$ and $O$ is an entirely independent Haar-distributed orthogonal matrix.

Now
\begin{align}
   &\E_H Z_H(\j) = \int d\Psi e^{i\Tr \Psi\Psi^{\dagger}J} \Phi_{\bX}(K)\Phi_D(K) + \int d\Psi e^{i\Tr \Psi\Psi^{\dagger}J} (\Phi_X(A) - \Phi_{\bX}(A))\Phi_D(A)\notag\\
   \implies &\E G_{D+X}(z) = \E G_{D+\bX}(z) + \frac{1}{N}\frac{\partial}{\partial \j}\Bigg|_{\j = 0} \int d\Psi e^{i\Tr \Psi\Psi^{\dagger}J} (\Phi_X(A) - \Phi_{\bX}(A))\Phi_D(A) \equiv   \E G_{D+\bX}(z) + E(z)
\end{align}
and so we need to analyse the error term $E(z)$.\\

Now consider $X = U^T\Lambda U$ where the rows of $U$ are the eigenvectors $\{\vec{u}_i\}_i$ of $X$. Say also that $K = Q^TYQ$ for diagonal $Y = (y_1, \ldots, y_r, 0, \ldots, 0)$, where we note that $K$ has fixed rank, by construction. Then \begin{align*}
    \Tr XK = Y (UQ^T)^T\Lambda (UQ^T)
\end{align*}
but Lemma \ref{lem:que_preserved_under_rotation} establishes that the rows of $UQ^T$ obey QUE, since the rows of $U$ do. Further, Lemma \ref{lem:weaker_col_que} then establishes that the columns of $UQ^T$ obey $\hQUE$ as required by Lemma \ref{lem:explicit_error}. Let $\{\vec{v}_i\}$ be those columns, then we have \begin{align}\label{eq:trace_as_vsum}
    \Tr XK = \sum_{i=1}^r y_i \vec{v}_i^T \Lambda \vec{v}_i.
\end{align}

The expectation over $X$ can be split into eigenvalues and conditional eigenvectors 
\begin{align}\label{eq:phix_expansion}
    \Phi_X(K) = \E_{\Lambda}\E_{U\mid \Lambda}\sum_{l=0}^{\infty} \frac{1}{l!} (-i)^l\left(\Tr U^T \Lambda UK\right)^l.
\end{align}
We can simply bound \begin{align}
   \left| \sum_{l=0}^{n} \frac{1}{l!} (-i)^l\left(\Tr U^T \Lambda U\right)^l\right| \leq e^{|\Tr U^T\Lambda UK\review{|}}
\end{align}
for any $n$, but clearly \begin{align}
    \E_{U\mid \Lambda}e^{|\Tr U^T\Lambda UK|} < \infty
\end{align}
since, whatever the distribution of $U\mid\Lambda$, the integral is over a compact group (the orthogonal group $O(N)$) and the integrand has no singularities. 
Therefore, by the dominated convergence theorem
\begin{align}
     \Phi_X(K) = \E_{\Lambda}\sum_{l=0}^{\infty} \frac{1}{l!} (-i)^l\E_{U\mid \Lambda}\left(\Tr U^T \Lambda UK\right)^l
\end{align}
and in precisely the same way
\begin{align}\label{eq:phibx_expansion}
      \Phi_X(K) = \E_{\Lambda}\sum_{l=0}^{\infty} \frac{1}{l!} (-i)^l\E_{O\sim \mu_{Haar}}\left(\Tr O^T \Lambda OK\right)^l.
\end{align}
Recalling (\ref{eq:trace_as_vsum}) we now have
\begin{align}\label{eq:phix_conditional_expansion}
    \Phi_X(K) = \E_{\Lambda}\sum_{l=0}^{\infty} \frac{1}{l!} (-i)^l \E_{U\mid \Lambda}\left(\sum_{i=1}^r y_i \vec{v}_i^T \Lambda \vec{v}_i\right)^l.
\end{align}
and similarly 
\begin{align}\label{eq:phixbar_conditional_expansion}
    \Phi_{\bX}(K) = \E_{\Lambda}\sum_{l=0}^{\infty} \frac{1}{l!} (-i)^l \E_{U\mid \Lambda}\left(\sum_{i=1}^r y_i \bar{\vec{v}}_i^T \Lambda \bar{\vec{v}}_i\right)^l.
\end{align}
where the $\bar{\vec{v}}_i$ are defined in the obvious way from $\bX$.
We would now apply $\hQUE$, but to do so we must insist that $\E_{\Lambda}$ is taken over the \emph{ordered} eigenvalues of $X$.
Having fixed that convention, \review{L}emma \ref{lem:explicit_error} can be applied to the terms 
\begin{align}
    \E_{U\mid \Lambda}\left(\sum_{i=1}^r y_i \vec{v}_i^T \Lambda \vec{v}_i\right)^l
\end{align}
in (\ref{eq:phix_conditional_expansion}).
The terms in $\Phi_{\bX}$ can be treated similarly.
This results in \begin{align}
    \Phi_X(K) - \Phi_{\bX}(K) &= \E_{\Lambda}\Bigg[\sum_{l=0}^{\infty} \frac{i^l}{l!} \left\{\E_{\{\vec{g}_i\}_{i=1}^r}\left(\sum_{i=1}^r y_i \frac{1}{N}\vec{g}_i^T\Lambda \vec{g}_i\right)^l + \left(\sum_{i=1}^r N^{-(1+\epsilon)} y_i \vec{\eta}_i^T\Lambda \vec{\eta}_i\right)^l \right\}\notag\\
    & ~~~~~~~~-\sum_{l=0}^{\infty} \frac{i^l}{l!} \left\{\E_{\{\vec{g}_i\}_{i=1}^r}\left(\sum_{i=1}^r y_i \frac{1}{N}\vec{g}_i^T\Lambda \vec{g}_i\right)^l + \left(\sum_{i=1}^r N^{-(1+\epsilon)} y_i \bar{\vec{\eta}}_i^T\Lambda \bar{\vec{\eta}}_i\right)^l \right\}\Bigg]
\end{align}
The exponential has infinite radius of convergence, so we may re-order the terms in the sums to give cancellation
\begin{align*}
    \Phi_X(K) - \Phi_{\bX}(K) = \E_{\Lambda}\sum_{l=1}^{\infty}\frac{1}{l!} N^{-(1+\epsilon)l}(-i)^l\left(\sum_{i=1}^r y_i \vec{\eta}_i^T\Lambda \vec{\eta}_i\right)^l -\E_{\Lambda}\sum_{l=1}^{\infty}\frac{1}{l!} N^{-(1+\epsilon)l}(-i)^l\left(\sum_{i=1}^r y_i \bar{\vec{\eta}}_i^T\Lambda \bar{\vec{\eta}}_i\right)^l.
\end{align*}
Here $\epsilon>0$ and $\vec{\eta}_i, \tilde{\vec{\eta}}_i\in \C^N$ with 
\begin{align}
   -1\leq [(\vec{\eta}_i)_j]^2 , [(\bar{\vec{\eta}}_i)_j]^2 \leq 1 ~&~ \forall i=1,\ldots, r, ~  \forall j\in\TN,\\
   -N^{\epsilon}\leq [(\vec{\eta}_i)_j]^2 , [(\bar{\vec{\eta}}_i)_j]^2 \leq N^{\epsilon} ~&~ \forall i=1,\ldots, r, ~  \forall j\in\TN.
\end{align}

Simplifying, we obtain \begin{align}
    \Phi_X(K) - \Phi_{\bX}(K) = \E_{\Lambda}\exp\left(-iN^{-(1+\epsilon)}\sum_{i=1}^r y_i \vec{\eta}_i^T\Lambda \vec{\eta}_i\right) - \E_{\Lambda}\exp\left(-iN^{-(1+\epsilon)}\sum_{i=1}^r y_i \tilde{\vec{\eta}}_i^T\Lambda \tilde{\vec{\eta}}_i\right).
\end{align}
Since $|\TN^c| \leq 2N^{1-\delta}$ we have \begin{align}
    \sum_{j\in\TN^c} |\lambda_j| \leq \mathcal{O}(N^{1-d} N^{-1}) \Tr |\Lambda|
\end{align}
and so \begin{align}
    |\vec{\eta}_i^T\Lambda \vec{\eta}_i| \leq \Tr|\Lambda|\left( 1 + \mathcal{O}(N^{\epsilon - \delta})\right).
\end{align}
For any fixed $\delta>0$, $\epsilon$ can be reduced if necessary so that $\epsilon < \delta$ and then for sufficiently large $N$ we obtain, say, 
\begin{align}
    |\vec{\eta}_i^T\Lambda \vec{\eta}_i| \leq 2 \Tr|\Lambda|.
\end{align}
Thence we can write $\vec{\eta}_i^T\Lambda\vec{\eta}_i = \Tr|\Lambda| \xi_i$ for $\xi_i\in[-2, 2]$, and similarly $\tilde{\vec{\eta}}_i^T\Lambda\tilde{\vec{\eta}}_i = \Tr|\Lambda| \tilde{\xi}_i$. Now
\begin{align*}
    \E_{\Lambda}\exp\left(-iN^{-(1+\epsilon)} \sum_{i=1}^r \xi_iy_i \Tr|\Lambda|\right)=\E_{\Lambda}\exp\left(-iN^{-\epsilon} \sum_{i=1}^r \xi_iy_i\int d\hat{\mu}_X(\lambda)|\lambda|\right)
\end{align*}
so we can apply Laplace's method to the empirical spectral measure $\hat{\mu}_X$ to obtain \begin{align}
    \E_{\Lambda}\exp\left(-iN^{-(1+\epsilon)} \sum_{i=1}^r \xi_iy_i \Tr|\Lambda|\right)=\exp\left(-iN^{-\epsilon} (q+o(1))\sum_{i=1}^r \xi_iy_i\right) + o(1)
\end{align}
where the $o(1)$ terms do not depend on the $y_i$ and where we have defined \begin{align}
    q = \int d\mu_X(\lambda)|\lambda|.
\end{align}
Further, we can write $\sum_{i=1}^r \xi_i y_i = \zeta\Tr K$, where $\zeta \in [\min_i\{\xi_i\}, \max_i\{\xi_i\}]\subset [-1, 1]$, and similarly $\sum_{i=1}^r \tilde{\xi}_i y_i = \tilde{\zeta}\Tr K$. Then \begin{align}
    \Phi_X(K) - \Phi_{\bX}(K) = e^{-iN^{-\epsilon}\zeta (q + o(1)) \Tr K} - e^{-iN^{-\epsilon}\tilde{\zeta} (q + o(1)) \Tr K} + o(1)
\end{align}
but
\begin{align}
    &\frac{1}{N}\frac{\partial}{\partial \j}\Bigg|_{\j=0}\int d\Psi e^{i\Tr \Psi\Psi^{\dagger}J}e^{-iN^{-\epsilon}\zeta (q + o(1)) \Tr K}\Phi_D(A) = \E G_{D + N^{-\epsilon}\zeta(q + o(1))I}(z) = \E G_D(z + \mathcal{O}(N^{-\epsilon}))\notag\\
    \implies & E(z) = \E G_D(z + \mathcal{O}(N^{-\epsilon})) + o(1) - \E G_D(z + \mathcal{O}(N^{-\epsilon})) - o(1) = o(1).
\end{align}
We have thus established that \begin{align}
    \E G_{D+X}(z) = \E G_{D+\bX}(z) + o(1)
\end{align}
from which one deduces that $\mu_{D+X} = \mu_{D + \bX} = \mu_D \boxplus \mu_{\bX} = \mu_D\boxplus\mu_X$.
\end{proof}

\begin{remark}
We have also constructed a non-rigorous argument for Theorem \ref{thm:nearly_free_addn} where the supersymmetric approach is replaced by the replica method. This approach simplifies some of the analysis but at the expense of being not at all \review{rigorous (indeed there are integral expressions in this argument that are manifestly infinite). The supersymmetric methods used here are not fully rigorous (like most of their applications) but we note that recent work is beginning to elevate supersymmetric random matrix calculations to full rigour \cite{shcherbina2017characteristic, shcherbina2020characteristic}.}
\end{remark}

\subsection{Experimental validation}
\review{Let $U(a, b)$ denote the uniform distribution on the interval $(a, b)$, and $\Gamma(a)$ the Gamma-distribution with scale parameter $a$.} We consider the following matrix ensembles:
\begin{align*}
    M\sim GOE^n ~ &: ~ \text{Var}(M_{ij}) = \frac{1 + \delta_{ij}}{2n},\\
    M\sim UWig^n &: ~ \sqrtsign{n}M_{ij} \overset{i.i.d}{\sim} U(0, \sqrtsign{6}) ~~\text{up to symmetry},\\
    M\sim \Gamma Wig^n &: ~ 2\sqrtsign{n}M_{ij} \overset{i.i.d}{\sim} \Gamma(2) ~~\text{up to symmetry},\\
    M\sim UWish^n &: ~ M\overset{d}{=}\frac{1}{m}XX^T, ~ X_{ij}\overset{i.i.d}{\sim}U(0, \sqrtsign{12}) ~~\text{for }X\text{ of size }n\times m, ~~ \frac{n}{m} \overset{n,m\rightarrow\infty}{\rightarrow} \alpha,\\
    M\sim Wish^n &: M\overset{d}{=}\frac{1}{m}XX^T, ~X_{ij}\overset{i.i.d}{\sim}\mathcal{N}(0,1)~~\text{for }X\text{ of size }n\times m, ~~ \frac{n}{m} \overset{n,m\rightarrow\infty}{\rightarrow} \alpha.
\end{align*}

\begin{figure}[h]
    \centering
    \begin{subfigure}{0.3\linewidth}
        \includegraphics[width=\textwidth]{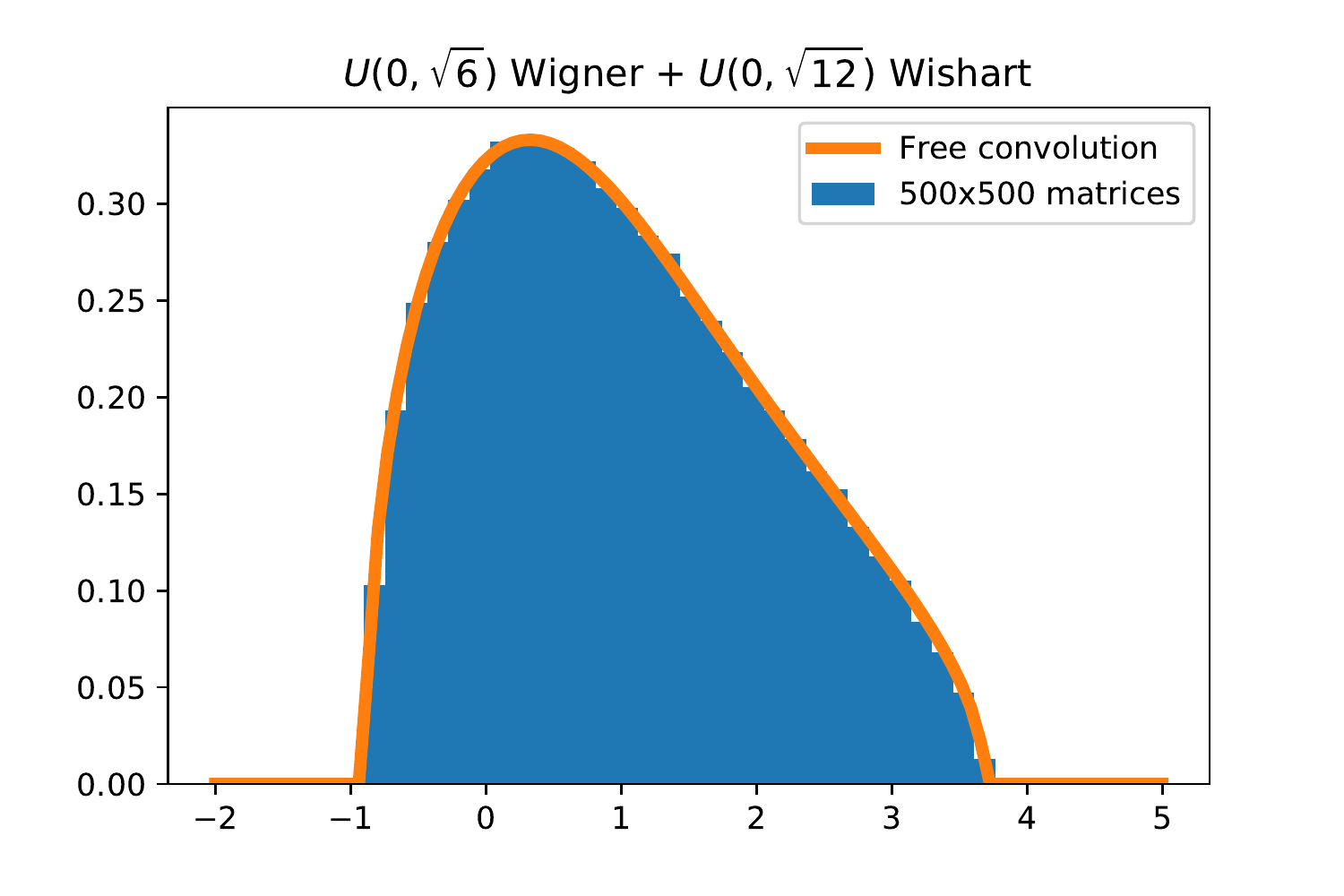}
        \subcaption{ $UWig^n + UWish^n$}
    \end{subfigure}
    \begin{subfigure}{0.3\linewidth}
        \includegraphics[width=\textwidth]{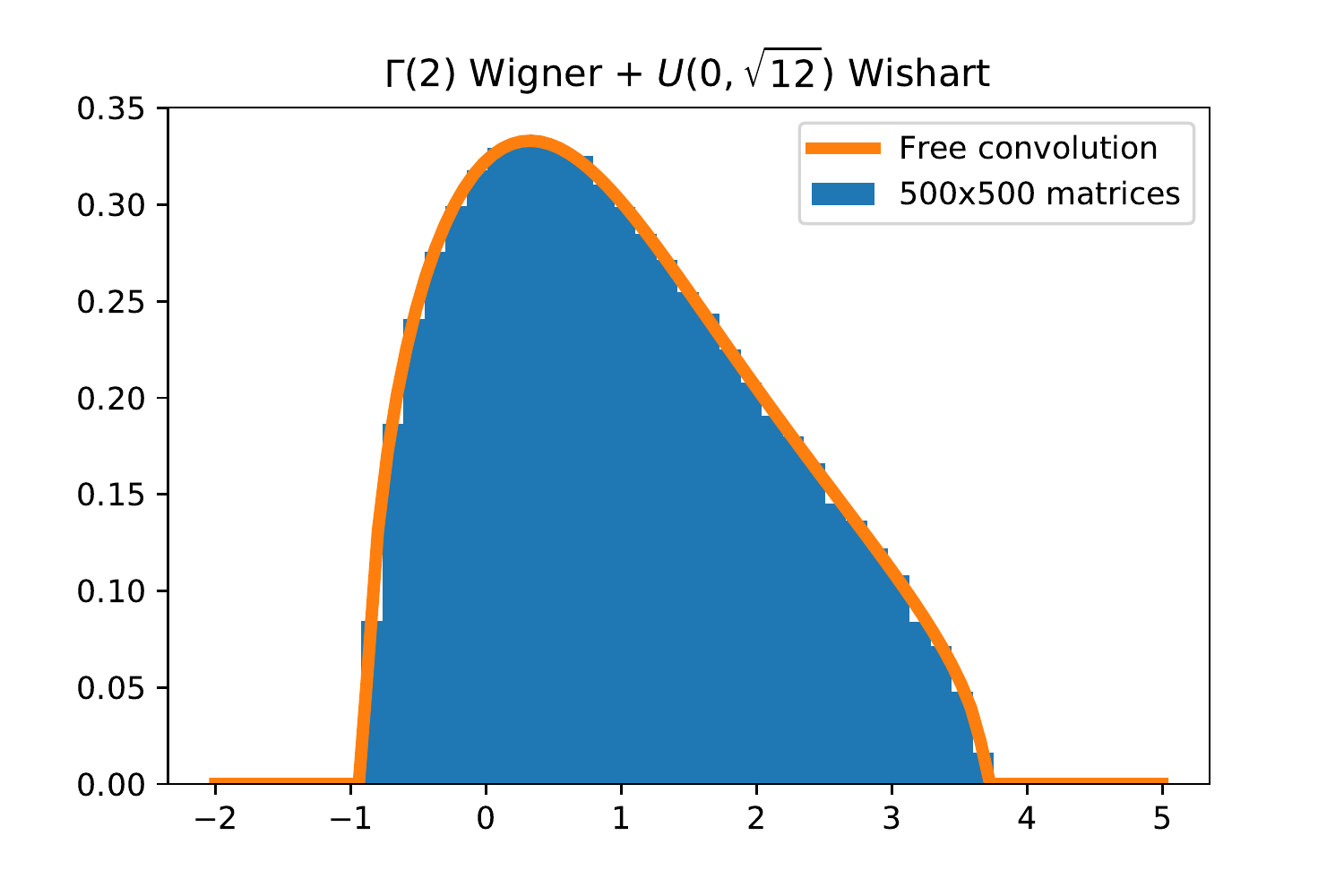}
        \subcaption{ $\Gamma Wig^n + UWish^n$}
    \end{subfigure}
    \begin{subfigure}{0.3\linewidth}
        \includegraphics[width=\textwidth]{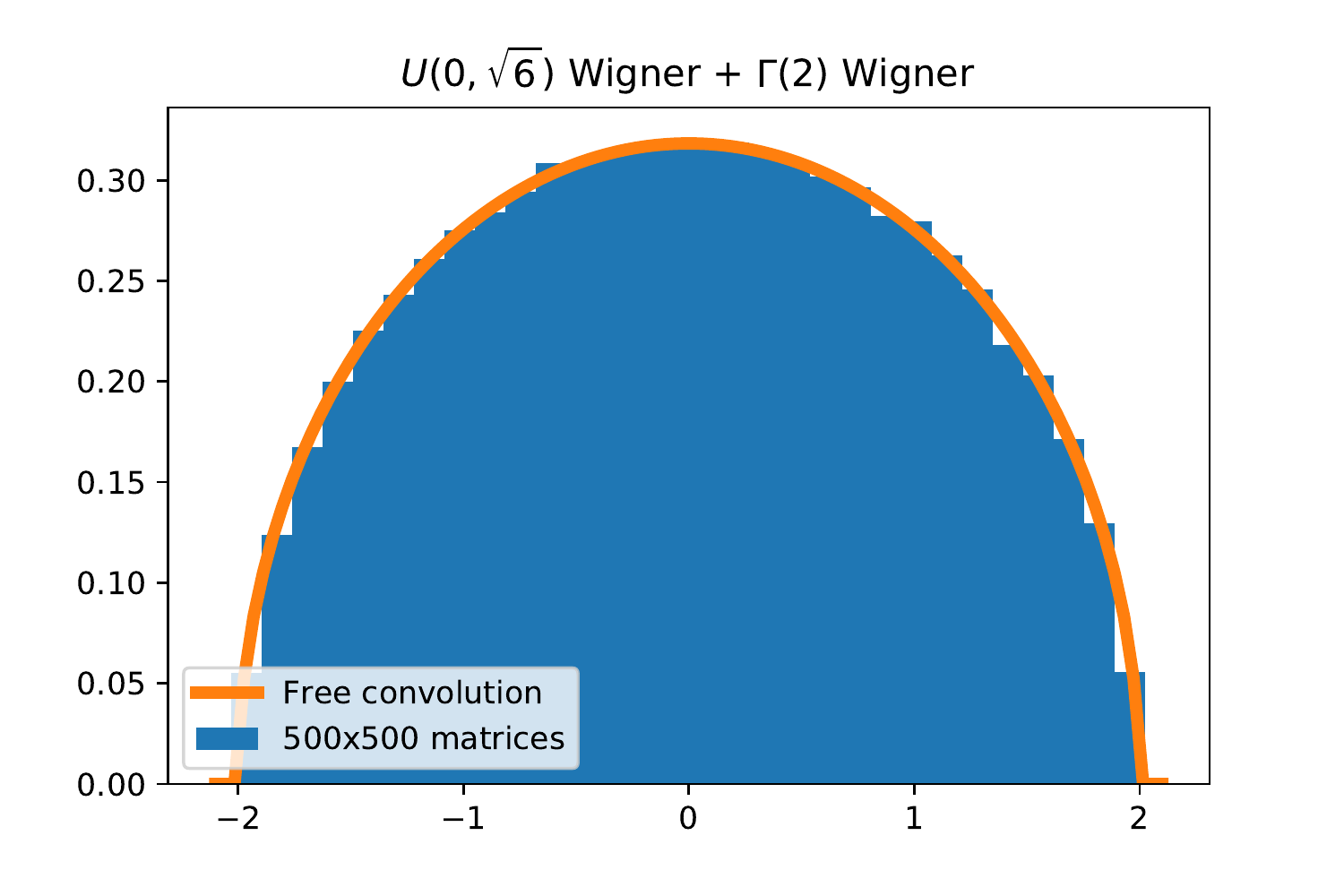}
        \subcaption{ $UWig^n + \Gamma Wig^n$}
    \end{subfigure}
    \caption{Comparison of theoretical spectral density and empirical from sampled matrices all of size $500\times 500$. We combine $50$ independent matrix samples per plot.}
    \label{fig:free_sum_comparison}
\end{figure}

All of the $GOE^n, UWig^n, \Gamma Wig^n$ have the same limiting spectral measure, namely $\mu_{SC}$, the semi-circle of radius $\sqrtsign{2}$. $UWish^n, Wish^n$ have a Marcenko-Pastur limiting spectral measure $\mu_{MP}$, and the constant $\sqrtsign{12}$ is chosen so that the parameters of the MP measure match those of a Gaussian Wishart matrix $Wish^n$. $GOE^n, Wish^n$ are the only ensembles whose eigenvectors are Haar distributed, but all ensembles obey a local law in the sense above. It is known that the sum of $GOE^n$ and any of the other ensembles will have limiting spectral measure given by the free additive convolution of $\mu_{SC}$ and the other ensemble's measure (so either $\mu_{SC}\boxplus\mu_{MP}$ or $\mu_{SC}\boxplus\mu_{SC}$), \review{indeed this free addition property holds for any invariant ensemble \cite{anderson2010introduction}}. Our result implies that the same holds for addition of the non-invariant ensembles. Sampling from the above ensembles is simple, so we can easily generate spectral histograms from multiple independent matrix samples for large $n$. $\mu_{SC}\boxplus\mu_{SC}$ is just another semi-circle measure but with radius $2$. $\mu_{SC}\boxplus\mu_{MP}$ can be computed in the usual manner with $R$-transforms and is given by the solution to the polynomial \begin{align*}
    \frac{\alpha}{2}t^3 - \left(\frac{1}{2} + \alpha z\right)t^2 + (z + \alpha - 1)t - 1 = 0.
\end{align*}
i.e. Say the cubic has roots $\{r_1, r_2 + is_2, r_2 - is_2\}$ for $s_2\geq 0$, then the density of $\mu_{SC}\boxplus\mu_{MP}$ at $z$ is $s_2/\pi$. This can all be solved numerically. The resulting plots are in Figure \ref{fig:free_sum_comparison} and clearly show agreement between the free convolutions and sampled spectral histograms.

We can also test the result in another more complicated case. Consider the case of random $d$-regular graphs on $N$ vertices. Say $M\sim Reg^{N,d}$ is the distribution of the adjacency matrix of such random graphs. The limiting spectral density of $M\sim Reg^{N,d}$ is known in closed form, as is its Stieljtes transform \cite{bauerschmidt2019local} and \cite{bauerschmidt2019local} established a local law of the kind required for our results. Moreover, there are known efficient algorithms for sampling random $d$-regular graphs \cite{kim2003generating, steger1999generating} along with implementations \cite{SciPyProceedings_11}. 
Let $\mu_{KM}^{(d)}$ be the Kesten-McKay law, the limiting spectral measure of $d$-regular graphs. We could find an explicit degree-6 polynomial for the Stieljtes transform of $\mu_{KM}^{(d)}\boxplus \mu_{SC}$ and compare to spectral histograms as above. Alternatively we can investigate agreement with $\mu_{KM}^{(d)}\boxplus \mu_{SC}$ indirectly by sampling and comparing spectra from say $Reg^{N,d} + UWig^N$ and also from  $Reg^{N,d} + GOE^N$.
The latter case will certainly yield the distribution $\mu_{KM}^{(d)}\boxplus \mu_{SC}$ since the GOE matrices are freely independent from the adjacency matrices.
Figure shows a q-q plot\footnote{\review{Recall that a q-q plot shows the quantiles of one distribution on the $x$ axis and another on the $y$ axis. Given two cumulative density functions $F_X, F_Y$ and their percent point functions $F_X^{-1}, F_Y^{-1}$, the q-q plot is a plot of the parametric curve $(F_X^{-1}(q), F_Y^{-1}(q))$ for $q\in[0,1]$. Given only finite samples from the random variables $X$ and $Y$, the empirical percent point functions can be estimated and used in the q-q plot.}} for samples \review{of the spectra} from these two matrix distributions and demonstrates near-perfect agreement, thus showing that indeed the spectrum of $Reg^{N,d} + UWig^N$  is indeed described by $\mu_{KM}^{(d)}\boxplus \mu_{SC}$.
We reached the same conclusion when repeating the above experiment with $UWish^N + Reg^{N,d}$ and $Wish^n + Reg^{N,d}$.
\begin{figure}
    \centering
    \includegraphics[width=0.4\textwidth]{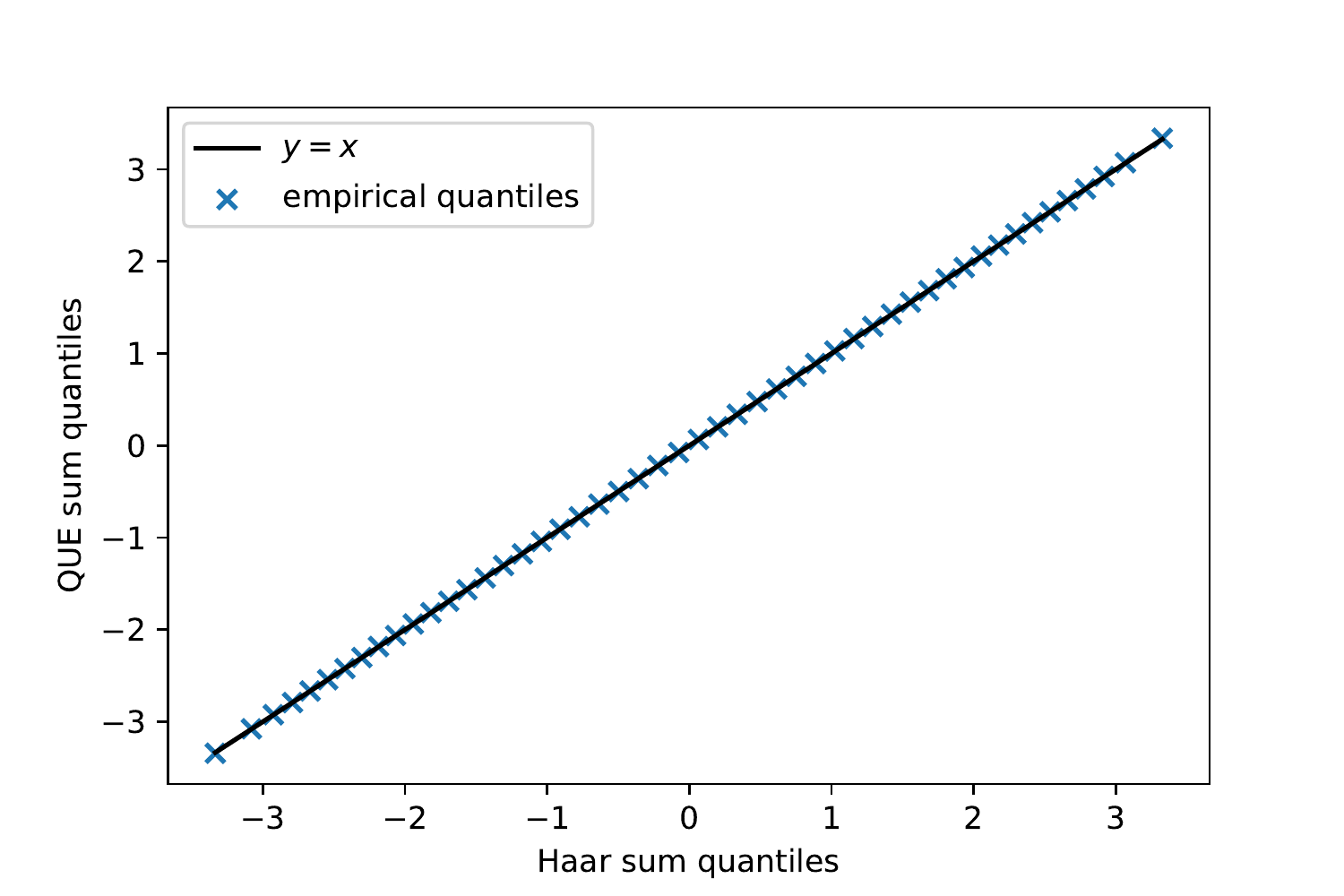}
    \caption{q-q plot comparing the spectrum of samples from $Reg^{N,d} + UWig^N$ ($y$-axis) to samples from $Reg^{N,d} + GOE^N$ ($x$-axis).}
    \label{fig:qq_plot}
\end{figure}

\section{Invariant equivalent ensembles}\label{sec:inv}
\review{For an invariant ensemble \cite{deift1999orthogonal} with potential $V$ we have the following integro-differential equation relating the equilibrium measure $\mu$ to the potential $V$ \cite{unterberger2019global}}:
 \begin{align}\label{eq:v_stieljtes}
    \frac{\beta}{2}\dashint\frac{1}{x-y}d\mu(y) = V'(x).
\end{align} 
So in the case of real symmetric matrices we have \begin{align}\label{eq:muinf_V_relation}
    \frac{1}{2} \bar{g_{\mu}}(x) = V'(x)
\end{align}
where $g_{\mu}$ is the Stieljtes transform of $\mu$ and the bar over $\bar{g_{\mu}}$ indicates that the principal value has been taken.


Given a sufficiently nice $\mu$ (\ref{eq:v_stieljtes}) defines $V$ up-to a constant of integration on $\supp(\mu)$, but $V$ is not determined on $\R\backslash\supp(\mu)$, \review{ as is made clear by the following lemma, which we prove for completeness but which has appeared before in various works (e.g. \cite{deift1999orthogonal}).}

\begin{lemma}\label{lemma:V_unique_mu}
    For compactly supported probability measure $\mu$ on $\R$ and real potential $V$, define \begin{align}
        S_V[\mu](y) = V(y) - \int d\mu(x) \log|y-x|.
    \end{align}Suppose $S_V[\mu](y)=c$, a constant, for all $y\in\supp(\mu)$ and $S_V[\mu](y) \geq c$ for all $y\in\mathbb{R}$. Then $\mu$ is a minimiser amongst all probability measures on $\R$ of the energy \begin{align}
        \mathcal{E}_V[\mu] = \int d\mu(x) V(x) - \iint_{x< y} d\mu(x)d\mu(y)\log|x-y|.
    \end{align}
\end{lemma}
 \begin{proof}
    Consider a probability measure that is close to $\mu$ in the sense of $W_1$ distance, say.

    For any such measure, one can find an arbitrarily close probability measure $\mu'$ of the form
    \begin{align}
        \mu' = \mu + \sum_{i=1}^r a_i\1_{[y_i - \delta_i, y_i + \delta_i]} - \sum_{i=1}^s b_i\1_{[z_i - \eta_i, z_i + \eta_i]}
    \end{align}
    where all $a_i, b_i>0$ and $\delta_i, \eta_i, a_i, b_i \leq \epsilon$ for some small $\epsilon>0$.
    To ensure that $\mu'$ is again a probability measure we must impose $\sum_ia_i = \sum_jb_j$.
    The strategy now is to expand $ \mathcal{E}_V[\mu'] $ about $\mu$ to first order in $\epsilon$, but first note the symmetrisation
    \begin{align}
         \iint_{x< y} d\mu(x)d\mu(y)\log|x-y| = \frac{1}{2} \iint_{x\neq y} d\mu(x)d\mu(y)\log|x-y|.
    \end{align}
    Then \begin{align}
        \mathcal{E}_V[\mu'] -  \mathcal{E}_V[\mu] &= \sum_{i=1}^r a_i V(y_i) - \sum_{i=1}^s b_i V(z_i) - \sum_{i=1}^ra_i \int d\mu(x)\log|x-y_i| + \sum_{i=1}^rb_i \int d\mu(x)\log|x-z_i| + \mathcal{O}(\epsilon^2) \notag\\
        &= \sum_{i=1}^r a_i S_V[\mu](y_i) - \sum_{i=1}^r b_i S_V[\mu](z_i) + \mathcal{O}(\epsilon^2).
    \end{align}
    Observe that if all $y_i, z_i\in \text{supp}(\mu)$ then $S_V[\mu](y_i) = S_V[\mu](y_i) = c $ and so $ \mathcal{E}_V[\mu'] = \mathcal{E}_V[\mu]$.
    Without loss of generality therefore, we take $y_i\notin \text{supp}(\mu)$ and $z_i\in \text{supp}(\mu)$, whence 
    \begin{align}
         \mathcal{E}_V[\mu'] -  \mathcal{E}_V[\mu] \geq c\sum_{i=1}^r a_i - c\sum_{i=1}^s b_i = 0.
    \end{align}
    
 \end{proof}
 \review{The next lemma establishes that, while not unique, a potential $V$ can always be constructed given a measure $\mu$.}
\begin{lemma}
    Consider a probability measure $\mu$ on $\R$ with compact support, absolutely continuous with respect to the Lebesgue measure. Then there exists a potential $V:\R\rightarrow\R$ which yields a well-defined invariant distribution on real symmetric matrices for which the equilibrium measure is $\mu$.
\end{lemma}
\begin{proof}
    (\ref{eq:muinf_V_relation}) can be integrated to obtain $V$ and the condition $S_V[\mu]=c$ (a constant) on $\supp(\mu)$ determines $V$ uniquely on $\supp(\mu)$. Next observe that, for $y\in\R\backslash\supp(\mu)$ there exists some constant $R>0$ such that $|x-y| \leq R + |y|$, \review{since $\mu$ is compactly supported}, and so $\log|x-y| \leq |y| + R$. Therefore \begin{equation}
        S_V[\mu](y) \geq V(y) - |y| - R.
    \end{equation}
  \review{  $V$ must be chosen on $\mathbb{R}\backslash\text{supp}(\mu)$ to satisfy $S_V[\mu](y) \geq c$, which can be achieved by ensuring 
  \begin{align}\label{eq:v_lem_bound}
  V(y) \geq |y| + R + c.
  \end{align}
  Additionally, $V$ must be defined for large $y$ such that it defines an legitimate invariant ensemble on symmetric real matrices, i.e. $V$ must decay sufficiently quickly at infinity to give an integrable probability density.
  Finally, $V$ must be sufficiently smooth, and certainly continuous, so there are boundary conditions at the boundary of $\text{supp}(\mu)$.
  Suppose $\text{supp}(\mu)$ is composed of $K$ disjoint intervals, then there are $2K$ boundary conditions on $V$, and the bound (\ref{eq:v_lem_bound}) imposes one further condition.
  Sufficiently fast decay at infinity can be satisfied by any even degree polynomial $V$ of degree at least 2, therefore a degree $2K + 2$ polynomial can be found with sufficiently fast decay at infinity, satisfying all the boundary conditions and (\ref{eq:v_lem_bound}).
  }
\end{proof}

\section{Universal complexity of loss surfaces}\label{sec:determinants}

\subsection{Extension of a key result and prevalence of minima}
Let's recall Theorem 4.5 from \cite{arous2021exponential}. $H_N(u)$ is our random matrix ensemble with some parametrisation $u\in\R^m$ and its limiting spectral measure is $\mu_{\infty}(u)$.

Define 
\begin{align}
    \mathcal{G}_{-\epsilon} = \{u\in\R^m \mid \mu_{\infty}(u) \left( (-\infty, 0) \right) \leq \epsilon\}.
\end{align}
So $\mathcal{G}_{-\epsilon}$ is the event that $\mu_{\infty}(u)$ is close to being supported only on $(0, \infty)$. Let $l(u), r(u)$ be \review{the} left and right edges respectively of the support of $\mu_{\infty}(u)$.

\begin{theorem}[\cite{arous2021exponential} Theorem 4.5]
    Fix some $\mathcal{D}\subset \R^m$ and suppose that $\mathcal{D}$ and the matrices $H_N(u)$ satisfy the following.
    \begin{itemize}
        \item For every $R>0$ and every $\epsilon>0$, we have 
        \begin{align}\label{eq:bl_condition}
            \lim_{N\rightarrow\infty} \frac{1}{N\log N}\log\left[\sup_{u\in B_R}\P\left(d_{BL}(\hat{\mu}_{H_N(u)}, \mu_{\infty}(u) \right) > \epsilon\right] = -\infty.
        \end{align}
    \item Several other assumptions detailed in \cite{arous2021exponential}.
    \end{itemize}
    
    Then for any $\alpha>0$ and any fixed $p\in\N$, we have
    \begin{align}
        \lim_{N\rightarrow\infty} \frac{1}{N}\log\int_{\mathcal{D}} e^{-(N+p)\alpha u^2}\E\left[|\det(H_N(u))|\1\{i(H_N(u)) = 0\}\right]du = \sup_{u\in\mathcal{D}\cap\mathcal{G}}\left\{\int_{\mathbb{R}} \log|\lambda| d\mu_{\infty}(u) (\lambda) - \alpha u^2\right\}.
    \end{align}
\end{theorem}

We claim the following extension 
\begin{corollary}\label{cor:general_k}
    Under the same assumptions as the above theorem and for any integer sequence $k(N) > 0$ such that $k/N \rightarrow 0$ as $N\rightarrow\infty$, we have
    \begin{align}
        \lim_{N\rightarrow\infty} \frac{1}{N}\log\int_{\mathcal{D}} e^{-(N+p)\alpha u^2}\E\left[|\det(H_N(u))|\1\{i(H_N(u)) \leq k\}\right]du = \sup_{u\in\mathcal{D}\cap\mathcal{G}}\left\{\int_{\mathbb{R}} \log|\lambda| d\mu_{\infty}(u) (\lambda) - \alpha u^2\right\}.
    \end{align}
\end{corollary}
\begin{proof}
    Firstly note that 
    \begin{align}
       &\frac{1}{N}\log\int_{\mathcal{D}} e^{-(N+p)\alpha u^2}\E\left[|\det(H_N(u))|\1\{i(H_N(u)) \leq k\}\right] du \notag\\
       \geq &\frac{1}{N}\log\int_{\mathcal{D}} e^{-(N+p)\alpha u^2}\E\left[|\det(H_N(u))|\1\{i(H_N(u)) = 0\}\right] du,
    \end{align}
    so it suffices to establish a complementary upper bound. The proof in of Theorem 4.5 in \cite{arous2021exponential} establishes an upper bound using 
    \begin{align}
        \lim_{N\rightarrow\infty} \frac{1}{N} \log \int_{\left(\mathcal{G}_{-\epsilon}\right)^c} e^{-N\alpha u^2}  \E\left[|\det(H_N(u)|\1\{i(H_N(u)) = 0\}\right] du = -\infty
    \end{align}
    which holds for all $\epsilon > 0$.
    \review{Indeed, $\mathcal{D} = (\mathcal{D} \cap \mathcal{G}_{-\epsilon}) \cup  (\mathcal{D} \cap (\mathcal{G}_{-\epsilon})^{c})$, so \begin{align}
       &\int_{\mathcal{D}} e^{-(N+p)\alpha u^2}\E\left[|\det(H_N(u))|\1\{i(H_N(u)) \leq k\}\right] du \notag\\
        \leq &\int_{\mathcal{D}\cap\mathcal{G}_{-\epsilon} } e^{-(N+p)\alpha u^2}\E\left[|\det(H_N(u))|\1\{i(H_N(u)) \leq k\}\right] du \notag \\&+ \int_{(\mathcal{G}_{-\epsilon})^c} e^{-(N+p)\alpha u^2}\E\left[|\det(H_N(u))|\1\{i(H_N(u)) \leq k\}\right] du ,
    \end{align}}

    so our proof is complete if we can prove the analogous result
    \begin{align}\label{eq:low_prob_analogue}
        \lim_{N\rightarrow\infty} \frac{1}{N} \log \int_{\left(\mathcal{G}_{-\epsilon}\right)^c} e^{-N\alpha u^2}  \E\left[|\det(H_N(u)|\1\{i(H_N(u)) \leq k\}\right] du = -\infty.
    \end{align}
    
    As in \cite{arous2021exponential}, let $f_{\epsilon}$ be some $\frac{1}{2}$-Lipschitz function satisfying $\frac{\epsilon}{2}\1_{x \leq -\epsilon} \leq f_{\epsilon}(x)\leq \frac{\epsilon}{2} \1_{x\leq 0}$. Suppose $u\in \left(\mathcal{G}_{-\epsilon}\right)^c$ and also $i(H_N(u)) \leq k$. Then we have \begin{align}
        0 \leq \int d\hat{\mu}_{H_N(u)}(x) ~ f_{\epsilon}(x) \leq \frac{k\epsilon}{2N}
    \end{align}
    and also
    \begin{align}
        \frac{\epsilon^2}{2} \leq \int d\mu_{\infty}(u)(x) ~ f_{\epsilon}(x) \leq \frac{\epsilon}{2}.
    \end{align}
    We have 
    \begin{align}
        d_{BL}(\hat{\mu}_{H_N(u)}, \mu_{\infty}(u) ) &\geq \left|\int d\hat{\mu}_{H_N(u)}(x) ~ f_{\epsilon}(x)  - 
    \int d\mu_{\infty}(u)(x) ~ f_{\epsilon}(x) \right|\notag\\
    &\geq \left|\left|\int d\hat{\mu}_{H_N(u)}(x) ~ f_{\epsilon}(x)\right|  - 
    \left|\int d\mu_{\infty}(u)(x) ~ f_{\epsilon}(x) \right|\right|,
    \end{align}
    so if we can choose 
    \begin{align}\label{eq:epsilon_eta}
        \frac{k\epsilon}{2N} \leq \frac{\epsilon^2}{2} - \eta
    \end{align}
    for some $\eta > 0$, then we obtain $d_{BL}(\hat{\mu}_{H_N(u)}, \mu_{\infty}(u) ) \geq \eta$. Then applying (\ref{eq:bl_condition}) yields the result (\ref{eq:low_prob_analogue}). (\ref{eq:epsilon_eta}) can be satisfied if
    \begin{align}\label{eq:epsilon_eta_explicit}
        \epsilon \geq \frac{k}{2N} + \frac{1}{2}\sqrtsign{\frac{k^2}{N^2} + 8\eta}.
    \end{align}
    So, given $\epsilon>0$, we can take $N$ large enough such that, say, $\frac{k(N)}{N} < \frac{\epsilon}{4}$. By taking $\eta < \frac{\epsilon^2}{128}$ we obtain
    \begin{align}
        \frac{k}{2N} + \frac{1}{2}\sqrtsign{\frac{k^2}{N^2} + 8\eta} < \frac{\epsilon}{\review{8}} + \frac{1}{\sqrtsign{2}}\max\left(\sqrtsign{8\eta}, \frac{\epsilon}{4}\right) < \frac{1 + \review{\sqrt{2}}}{\review{8}}\epsilon < \review{\epsilon}
    \end{align}
    and so (\ref{eq:epsilon_eta_explicit}) is satisfied. Now finally (\ref{eq:bl_condition}) can be applied (with $\eta$ in place of $\epsilon$) and so we conclude (\ref{eq:low_prob_analogue}).
    
    \medskip
    Overall we see that the superexponential BL condition (\ref{eq:bl_condition}) is actually strong enough to deal with any $o(N)$ index not just index-0. This matches the GOE (or generally invariant ensemble) case, in which the terms with $\1\{i(H_N(u))=k\}$ are suppressed compared to the exact minima terms $\1\{i(H_N(u))=0\}$.
\end{proof}
\begin{remark}
Note that \review{Corollary} \ref{cor:general_k} establishes that, on the exponential scale, the number of critical points of any index $k(N) = o(N)$ is no more than the number of exact local minima.
\end{remark}

\subsection{The dichotomy of rough and smooth regions}
Recall the batch loss from Section \ref{subsec:hess_model}:
\begin{align}
   \frac{1}{b} \sum_{i=1}^{b} \mathcal{L}(f_{\vec{w}}(\vec{x}_i), y_i), ~~ (\vec{x}_i, y_i)\overset{\text{i.i.d.}}{\sim} \mathbb{P}_{data}.
\end{align}
As with the Hessian in Section \ref{subsec:hess_model}, we use the model $L\equiv L_{\text{batch}}(\vec{w}) = L_{\text{true}}(\vec{w}) + \sbf V(\vec{w})$, where $V$ is a random function $\mathbb{R}^N\rightarrow\mathbb{R}$.



Now let us define the complexity for sets $\B\subset \R^N$ \begin{align}
    C_N(\B) = |\{ \vec{w}\in \B \mid \nabla L(\vec{w}) = 0\}|.
\end{align}
This is simply the number of stationary points of the training loss in the region $\B$ of weight space. A Kac-Rice formula applied to $\nabla L$ gives 
\begin{align}\label{eq:w_b_dom_integral_kacrice}
    \E C_N = \int_{\B} d\vec{w} ~ \phi_{\vec{w}}(-\sbf^{-1}\nabla L_{\text{true}}) \E |\det (A + \sbf X)|
\end{align}
where $\phi_{\vec{w}}$ is the density of $\nabla V$ at $\vec{w}$.
A rigorous justification of this integral formula would, for example, have to satisfy the conditions of the results of \cite{adler2007random}.
This is likely to be extremely difficult in any generality, though is much simplified in the case of Gaussian $V$ (and $X$) - see \cite{adler2007random} Theorem 12.1.1 or Chapter \ref{chap:general_activation_functions}, Lemma \ref{lemma:kac_rice} (\cite{baskerville2021loss} Theorem 4.4).
Hereafter, we shall take (\ref{eq:w_b_dom_integral_kacrice}) as assumed.
The next step is to make use of strong self-averaging of the random matrix determinants. 
Again, we are unable to establish this rigorously at present, but note that this property has been proved in some generality by \cite{arous2021exponential}, although we are unable to satisfy all the conditions of those results in any generality here.
Self-averaging and using the addition results above gives
\begin{align*}
    \frac{1}{N}\log\E |\det (A + \sbf X)| = \int d(\mu_b \boxplus \nu)(\lambda) \log |\lambda| + o(1)
\end{align*}
where $\mu_b, \nu$ depend in principle on $\vec{w}$.
We are concerned with $N^{-1}\log \E C_N$, and in particular its sign, which determines the complexity of the loss surface in $\B$: positive $\leftrightarrow$ exponentially many (in $N$) critical points, negative $\leftrightarrow$ exponentially few (i.e. none).
The natural next step is to apply the Laplace method with large parameter $N$ to determine the leading order term in $\E C_N$, however the integral is clearly not of the right form. Extra assumptions on $\phi_{\vec{w}}$ and $\nabla L_{\text{true}}$ could be introduced, e.g. that they can be expressed as functions of only a finite number of combinations of coordinates of $\vec{w}$.

\medskip
Suppose that $\phi_{\vec{w}}$ has its mode at $0$, for any $\vec{w}$, which is arguably a natural property, reflecting in a sense that the gradient noise has no preferred direction in $\R^N$.
The sharp spike at the origin in the spectral density of deep neural network Hessians suggests that generically
\begin{align}\label{eq:log_int_neg}
    \int d(\mu_b \boxplus \nu)(\lambda) \log |\lambda| < 0.
\end{align}
We claim it is reasonable to expect the gradient (and Hessian) variance to be increasing in $\|\vec{w}\|_2$.
Indeed, consider the general form of \review{ the simplest deep neural network, a \emph{multi-layer perceptron}:}
\begin{align}
    f_{\vec{w}}(\vec{x}) = \sigma(\vec{b}^{(L)} + W^{(L)}\sigma(\vec{b}^{(L-1)} + W^{(L-1)}\ldots \sigma(\vec{b}^{(1)} + W^{(1)}\vec{x} )\ldots ))
\end{align}
where all of the weight matrices $W^{(l)}$ and bias vectors $\vec{b}^{(l)}$ combine to give the weight vector $\vec{w}$.
Viewing $\vec{x}$ as a random variable, making $f$ a random function of $\vec{w}$, we expect from the above that the variance in $f_{\vec{w}}$ is generally increasing in $\|\vec{w}\|_2$, and so therefore similarly with $L_{\text{batch}}$.

\medskip
Overall it follows that $\phi_{\vec{w}}(-\sbf^{-1}\nabla L_{\text{true}})$ is generally decreasing in $\|\nabla L_{\text{true}}\|$, but the maximum value at $\phi_{\vec{w}}(0)$ is decreasing in $\|\vec{w}\|_2$.
The picture is therefore that the loss surface is simple and without critical points in regions for which $\nabla L_{\text{true}}$ is far from $0$.
In neighbourhoods of $\nabla L_{\text{true}} = 0$, the loss surface may become complex, with exponentially many critical points, however if $\|\vec{w}\|_2$ is too large then the loss surface may still be without critical points.
In addition, the effect of larger batch size (and hence larger $\sbf^{-1}$) is to simplify the surface.
These considerations indicate that deep neural network loss surfaces are simplified by over-parametrisation, leading to the spike in the Hessian spectrum and thus (\ref{eq:log_int_neg}).
The simple fact that neural networks' construction leads gradient noise variance to increase with $\|\vec{w}\|_2$ has the effect of simplifying the loss landscape far from the origin of weight space, and even precluding the existence of any critical points of the batch loss.

\section{Implications for curvature from local laws}\label{sec:precond}
Consider a general stochastic gradient update rule with curvature-adjusted preconditioning:
\begin{align}\label{eq:precond_udpate}
    \vec{w}_{t+1} = \vec{w}_t - \alpha B_t^{-1} \nabla L(\vec{w}_t)
\end{align}
where recall that $L(\vec{w})$ is the batch loss, viewed as a random function on weight space.
$B_t$ is some preconditioning matrix which in practice would be chosen to somehow approximate the curvature of $L$.
Such methods are discussed at length in \cite{martens2016second} and also describe some of the most successful optimisation algorithms used in practice, such as Adam \cite{kingma2014adam}.
The most natural choice for $B_t$ is $B_t = \nabla^2 L(\vec{w}_t)$, namely the Hessian of the loss surface.
In practice, it is standard to include a damping parameter $\delta>0$ in $B_t$, avoid divergences when inverting.
Moreover, typically $B_t$ will be constructed to be some positive semi-definite approximation to the curvature such as the generalised Gauss Newton matrix \cite{martens2016second}, or the diagonal gradient variance form used in Adam \cite{kingma2014adam}.
Let us now suppose that $B_t = B_t(\delta) = \hat{H}_t + \delta$, where $\hat{H}_t$ is some chosen positive semi-definite curvature approximation and $\delta>0$.
We can now identify $B_t(\delta)^{-1}$ as in fact the Green's function of $\hat{H}_t$, i.e. \begin{align}
    B_t(\delta)^{-1} = -(-\delta - \hat{H}_t)^{-1} = -G_t(-\delta).
\end{align}
But $G_t$ is precisely the object used in the statement of a local law on for $\hat{H}_t$.
Note that $\nabla L(\vec{w}_t)$ is a random vector and however $\hat{H}_t$ is constructed, it will generally be a random matrix and dependent on $\nabla L(\vec{w}_t)$ in some manner that is far too complicated to handle analytically.
As we have discussed at length hitherto, we conjecture that a local law is reasonable assumption to make on random matrices arising in deep neural networks.
In particular in Chapter \ref{chap:spacings} \cite{baskerville2022appearance} we demonstrated universal local random matrix theory statistics not just for Hessians of deep networks but also for Generalised Gauss-Newton matrices.
Our aim here is to demonstrate how a local law on $\hat{H}_t$ dramatically simplifies the statistics of (\ref{eq:precond_udpate}).
Note that some recent work \cite{wei2022more} has also made use of random matrix local laws to simplify the calculation of test loss for neural networks.

\medskip
A local law on $\hat{H}_t$ takes the precise form (for any $\xi, D>0$
\begin{align}\label{eq:precond_local_law}
    \sup_{\|\vec{u}\|,\|\vec{v}\|  = 1, z\in\vec{S}}\P\left( |\vec{u}^TG(z)\vec{v} - \vec{u}^T\Pi(z)\vec{v}| > N^{\xi}\left(\frac{1}{N\eta} + \sqrtsign{\frac{\Im g_{\mu}(z)}{N\eta}}\right)\right) \leq N^{-D}
\end{align}
where \begin{align}
    \vec{S} = \left\{E + i\eta \in \C \mid |E| \leq \omega^{-1}, ~ N^{-1 + \omega} \leq \eta \leq \omega^{-1}\right\}
\end{align}
$\mu$ is the limiting spectral measure of $\hat{H}_t$ and, crucially, $\Pi$ is a \emph{deterministic} matrix.
We will use the following standard notation to re-express (\ref{eq:precond_local_law})
\begin{align}\label{eq:local_law_inside}
    |\vec{u}^TG(z)\vec{v} - \vec{u}^T\Pi(z)\vec{v}| \prec \Psi_N(z), ~~~ \|\vec{u}\|,\|\vec{v}\|  = 1, z\in\vec{S},
\end{align}
where $\Psi_N(z) = \frac{1}{N\eta} + \sqrtsign{\frac{\Im g_{\mu}(z)}{N\eta}}$ and the probabilistic statement, valid for all $\xi, D>0$ is implicit in the symbol $\prec$.
In fact, we will need the local law outside the spectral support, i.e. at $z = x + i\eta$ where $x\in\mathbb{R}\backslash\text{supp}(\mu)$.
In that case $\Psi_N(z)$ is replaced by $\frac{1}{N(\eta + \kappa)}$ where $\kappa$ is the distance of $x$ from $\text{supp}(\mu)$ on the real axis, i.e. 
\begin{align}\label{eq:local_law_outside}
    |\vec{u}^TG(z)\vec{v} - \vec{u}^T\Pi(z)\vec{v}| \prec \frac{1}{N(\eta + \kappa)}, ~~~ \|\vec{u}\|,\|\vec{v}\|  = 1, ~ x\in\mathbb{R}\backslash\text{supp}(\mu).
\end{align}
For $\delta>0$ this becomes\begin{align}
      |\vec{u}^TG(-\delta)\vec{v} - \vec{u}^T\Pi(-\delta)\vec{v}| \prec \frac{1}{N\delta} \|\vec{u}\|_2 \|\vec{v}\|_2
\end{align}
for $\delta>0$ and now any $\vec{u}, \vec{v}$.
Applying this to (\ref{eq:precond_udpate}) gives \begin{align}
     |\vec{u}^TB_t^{-1}\nabla L(\vec{w}_t) - \vec{u}^T\Pi_t(-\delta)\nabla L(\vec{w}_t) | \prec \frac{1}{N\delta} \|\vec{u}\|_2 \|\nabla L(\vec{w}_t)\|_2.
\end{align}
Consider any $\vec{u}$ with $\|\vec{u}\|_2 = \alpha$, then we obtain  \begin{align}
     |\vec{u}^TB_t^{-1}\nabla L(\vec{w}_t) - \vec{u}^T\Pi_t(-\delta)\nabla L(\vec{w}_t) | \prec \frac{\alpha  \|\nabla L(\vec{w}_t)\|_2}{N\delta}.
\end{align}
Thus with high probability, for large $N$, we can replace (\ref{eq:precond_udpate}) by \begin{align}\label{eq:precond_update_det_equiv}
    \vec{w}_{t+1} = \vec{w}_t - \alpha \Pi_t(-\delta) \nabla L(\vec{w}_t)
\end{align}
incurring only a small error, provided that \begin{align}
    \delta >> \frac{\|\nabla L(\vec{w}_t)\|_2}{N} \alpha.
\end{align}
Note that the only random variable in (\ref{eq:precond_update_det_equiv}) is $\nabla L (\vec{w}_t)$.
If we now consider the case $\nabla L (\vec{w}_t) = \nabla \bar{L}(\vec{w}_t) + \vec{g}(\vec{w}_t)$ for deterministic $\bar{L}$, then \begin{align}\label{eq:precond_update_final}
    \vec{w}_{t+1} = \vec{w}_t - \alpha \Pi_t(-\delta) \nabla \bar{L}(\vec{w}_t) - \alpha \Pi_t(-\delta)\vec{g}(\vec{w}_t)
\end{align}
and so the noise in the parameter update is entirely determined by the gradient noise.
Moreover note the \emph{linear} dependence on $\vec{g}$ in (\ref{eq:precond_update_final}).
For example, a Gaussian model for $\vec{g}$ immediately yields a Gaussian form in (\ref{eq:precond_update_final}), and e.g. if $\E \vec{g} = 0$, then \begin{align}
    \E(\vec{w}_{t+1} - \vec{w}_t) = -\alpha \Pi_t(-\delta) \E \nabla L(\vec{w}_t).
\end{align}

\medskip
A common choice in practice for $\hat{H}$ is a diagonal matrix, e.g. the diagonal positive definite curvature approximation employed by Adam \cite{kingma2014adam}.
In such cases, $\hat{H}$ is best viewed as an approximation to the eigenvalues of some positive definite curvature approximation.
The next result establishes that a local law assumption on a general curvature approximation matrix can be expected to transfer to an analogous result on a diagonal matrix of \review{i}ts eigenvalues.
\begin{prop}
    Suppose that $\hat{H}$ obeys a local law of the form (\ref{eq:local_law_outside}).
    Define the diagonal matrix $D$ such that $D_i \overset{d}{=} \lambda_i$ where $\{\lambda_i\}_i$ are the sorted eigenvalues of $\hat{H}$.
    Let $G_D(z) = (z - D)^{-1}$ be the resolvent of $D$.
    Let $\mathfrak{q}_j[\mu]$ be the $j$-th quantile of $\mu$, the limiting spectral density of $\hat{H}$, i.e.\begin{align}
        \int_{-\infty}^{\mathfrak{q}_j[\mu]} d\mu(\lambda) = \frac{j}{N}.
    \end{align}
    Then $D$ obeys the local law \begin{align}
        |(G_D)_{ij} - \delta_{ij}(z - \mathfrak{q}_j[\mu])^{-1}| \prec \frac{1}{N^{2/3} (\kappa + \eta)^2}, ~~ z = x + i\eta, ~x\in\mathbb{R}\backslash\text{supp}(\mu),
    \end{align}
    where $\kappa$ is the distance of $x$ from $\text{supp}(\mu)$.
    Naturally, we can redefine $D_i = \lambda_{\sigma{i}}$ for any permutation $\sigma\in S_N$ and the analogous statement replacing $\mathfrak{q}_j[\mu]$ with $\mathfrak{q}_{\sigma(j)}$ will hold.
\end{prop}
\begin{proof}
    As in \cite{erdos2017dynamical}, the local law (\ref{eq:local_law_inside}), (\ref{eq:local_law_outside}) is sufficient to obtain rigidity of the eigenvalues in the bulk, i.e.
    for any $\epsilon, D > 0$\begin{align}\label{eq:rigid}
        \P\left(\exists j ~\mid~ |\lambda_j - \mathfrak{q}_j[\mu]| \geq N^{\epsilon}\left[\min(j, N-j+1)\right]^{-1/3}N^{-2/3}\right) \leq N^{-D}.
    \end{align}
    Then we have \begin{align}
       \left|\frac{1}{z - \lambda_j} - \frac{1}{z - \mathfrak{q}_j[\mu]}\right|=\left| \frac{\lambda_j - \mathfrak{q}_j[\mu]}{(z - \lambda_j)(z - \mathfrak{q}_j[\mu])}\right|.
    \end{align}
    For $z=x+i\eta$ and $x$ at a distance $\kappa>0$ from $\text{supp}(\mu)$ \begin{align}
        |z-\mathfrak{q}_j[\mu]|^2 \geq \eta^2 + \kappa^2 \geq \frac{1}{2}(\eta + \kappa)^2,
    \end{align}
    and the same can be said for $|z - \mathfrak{q}_j[\mu]|^2$ with high probability, by applying the rigidity (\ref{eq:rigid}). A second application of rigidity to $|\lambda_j - \mathfrak{q}_j[\mu]|$ gives\begin{align}
         \left|\frac{1}{z - \lambda_j} - \frac{1}{z - \mathfrak{q}_j[\mu]}\right|  \prec \frac{1}{N^{2/3}\min(j, N-j+1)^{1/3} (\kappa + \eta)^2}
    \end{align}
    which yields the result.
\end{proof}

With this result in hand, we get the generic update rule akin to (\ref{eq:precond_update_final}), with high probability \begin{align}
    \vec{w}_{t+1} = \vec{w}_t -\alpha ~\text{diag}\left(\frac{1}{\pi_j+ \delta}\right) \nabla \bar{L}(\vec{w}_t) - \alpha~\text{diag}\left(\frac{1}{\pi_j + \delta}\right) \vec{g}(\vec{w}_t)
\end{align}
\review{where  $\{\pi_j\}_{j=1}^N$ are the eigenvalues of $\Pi_t(0)$} and we emphasise again that the $\pi_j$ are \emph{deterministic}; the only stochastic term is the gradient noise $\vec{g}(\vec{w}_t)$.

\medskip 
\paragraph{Implications for preconditioned stochastic gradient descent} The key insight from this section is that generic random matrix theory effects present in preconditioning matrices of large neural networks can be expected to drastically simplify the optimisation dynamics due to high-probability concentration of the pre-conditioning matrices around deterministic equivalents, nullifying the statistical interaction between the pre-conditioning matrices and gradient noise.
Moreover, with this interpretation, the damping constant typically added to curvature estimate matrices is more than a simple numerical convenience: it is essential to yield the aforementioned concentration results.

\medskip
As an example of the kind of analysis that the above makes possible, consider the results of \vivacom{Chapter \ref{chap:gadam} (or see \cite{ia} for more details)}.
The authors consider a Gaussian process model for the noise in the loss surface, resulting in tractable analysis for convergence of stochastic gradient descent in the presence of statistical dependence between gradient noise in different iterations.
Such a model implies a specific form of the loss surface Hessian and its statistical dependence on the gradient noise.
This situation is a generalisation of the spin glass model exploited in various works \cite{choromanska2015loss} and in Chapters \ref{chap:general_activation_functions} and \ref{chap:spin_glass_gans}, except that in those cases the Hessian can be shown to be independent of the gradients.
Absent the very special conditions that lead to independence, one expects the analysis to be intractable, hence why in \vivacom{Chapter \ref{chap:gadam}} we restrict to \review{stochastic gradient descent} without preconditioning, or simply assume a high probability concentration on a deterministic equivalent.
To make this discussion more concrete, consider a model $L = L_{\text{true}} + V$ where $V$ is a Gaussian process with mean $0$ and covariance function
\begin{align}\label{eq:kxx_def}
    K(\vec{x}, \vec{x}') = k\left(\frac{1}{2}\|\vec{x} - \vec{x}'\|_2^2\right) q\left( \frac{1}{2}(\|\vec{x}\|_2^2 + \|\vec{x}'\|_2^2) \right),
\end{align}
where $k$ is some decreasing function and $q$ some increasing function.
The discussion at the end of the previous section suggests that the covariance function for loss noise should not be modelled as stationary, hence the inclusion of the \review{function $q$ in (\ref{eq:kxx_def})}.
For convenience define $\Delta = \frac{1}{2}(\|\vec{x} - \vec{x}'\|_2^2)$ and $S = \frac{1}{2}(\|\vec{x}\|_2^2 + \|\vec{x}'\|_2^2)$.
Then it is a short exercise in differentiation to obtain \begin{align}
    \text{Cov}\left(\partial_i V(\vec{w}), \partial_j V(\vec{w})\right) &= \text{Cov}\left(\partial_i V(\vec{w}), \partial_j V(\vec{w}')\right)\Bigg|_{\vec{w}=\vec{w}'} \notag\\
    &= \frac{\partial^2}{\partial w_i\partial w_j'}K(\vec{w}, \vec{w}')\Bigg|_{\vec{w}=\vec{w}'} \notag\\
    &= -k'(0)q(\|\vec{w}\|_2)\delta_{ij} + k(0)q''(\|\vec{w}\|_2^2) w_iw_j.
\end{align}
and moreover \begin{align}
    \text{Cov}\left(\partial_{il} V(\vec{w}), \partial_j V(\vec{w})\right) &= \text{Cov}\left(\partial_{il} V(\vec{w}), \partial_j V(\vec{w}')\right)\Bigg|_{\vec{w}=\vec{w}'} \notag\\
    &= \frac{\partial^3}{\partial w_i\partial w_l\partial w_j'}K(\vec{w}, \vec{w}')\Bigg|_{\vec{w}=\vec{w}'} \notag\\
    &= -k'(0)q'(\|\vec{w}\|_2^2)w_l\delta_{ij} + q'''(\|\vec{w}\|_2^2)k(0) w_iw_lw_j' - k'(0)q'(\|\vec{x}\|_2)w_i\delta_{jl}.
\end{align}
Hence we see that the gradients of $L$ and its Hessian are statistically dependent by virtue of the non-stationary structure of $V$.
Putting aside issues of positive definite pre-conditioning matrices, and taking $\delta$ such that $(\nabla^2 L + \delta)^{-1}$ exists (almost surely) for large $N$, \vivacom{it would appear} that the distribution of $(\nabla^2 L + \delta)^{-1}\partial V$ will be complicated and non-Gaussian, \vivacom{assuming no extra information about the statistical interaction between the resolvent matrix and the gradient.}
This example concretely illustrates our point: even in almost the simplest case, where the gradient noise is Gaussian, the pre-conditioned gradients are generically considerably more complicated and non-Gaussian.
Moreover, centred Gaussian noise on gradient is transformed into generically non-centred noise by pre-conditioning.
Continuing the differentiation above, it is elementary to obtain the covariance structure of the Hessian $\nabla^2 V$, though the expressions are not instructive.
Crucially, however, the Hessian is Gaussian and the covariance of any of its entries is $\mathcal{O}(1)$ (in large $N$), so the conditions in Example 2.12 of \cite{erdHos2019random} apply to yield an optimal local law on the Hessian, which in turn yields the above high-probability concentration of $(\nabla^2 L + \delta)^{-1}$ provided that $\delta$ is large enough. \vivacom{This argument ratifies an intuition from random matrix theory, that for large $N$ the resolvent matrix $(\nabla^2 L + \delta)^{-1}$ is self-averaging and will be close, with high probability, to some deterministic equivalent matrix.}

\section{Conclusion}
In this chapter we have considered several aspects of so-called universal random matrix theory behaviour in deep neural networks.
Motivated by prior experimental results, we have introduced a model for the Hessians of DNNs that is more general than any previously considered and, we argue, actually flexible enough to capture the Hessians observed in real-world DNNs.
Our model is built using random matrix theory assumptions that are more general than those previously considered and may be expected to hold in quite some generality.
By proving a new result for the addition of random matrices, using a novel combination of quantum unique ergodicity and the supersymmetric method, we have derived expressions for the spectral outliers of our model.
Using Lanczos approximation to the outliers of large, practical DNNs, we have compared our expressions for spectral outliers to data and demonstrated strong agreement for some DNNs.
As well as corroborating our model, this analysis presents indirect evidence of the presence of universal local random matrix statistics in DNNs, extending earlier experimental results.
Our analysis also highlights a possibly interesting distinction between some DNN architectures, as Resnet architectures appear to better agree with our theory than other architectures and Resnets have been previously observed to have better-behaved loss surfaces than many other architectures.

\medskip
We also presented quite general arguments regarding the number of local optima of DNN loss surfaces and how `rough' or `smooth' such surfaces are.
Our arguments build on a rich history of complexity calculations in the statistical physics and mathematics literature but, rather than performing detailed calculations in some specific, highly simplified toy model, we instead present general insights based on minimal assumptions.
Finally we highlight an important area where random matrix local laws, an essential aspect of universality, may very directly influence the performance of certain popular optimisation algorithms for DNNs.
Indeed, we explain how numerical damping, combined with random matrix local laws, can act to drastically simplify the training dynamics of large DNNs.

\medskip
Overall this chapter demonstrates the relevance of random matrix theory to deep neural networks beyond highly simplified toy models.
Moreover, we have shown how quite general and universal properties of random matrices can be fruitfully employed to derive practical, observable properties of DNN spectra.
This work leaves several challenges for future research.
All of our work relies on either local laws for e.g. DNN Hessians, or on matrix determinant self-averaging results.
Despite the considerable progress towards establishing local laws for random matrices over the last decade or-so, it appears that establishing any such laws for, say, the Hessians of any DNNs is quite out of reach.
We expect that the first progress in this direction will come from considering DNNs with random i.i.d. weights and perhaps simple activation functions.
Based on the success of recent works on random DNNs \cite{pastur2020randomiid}, we conjecture that the Gram matrices of random DNN Jacobians may be the simplest place to establish a local law, adding to the nascent strand of \emph{nonlinear} random matrix theory \cite{pennington2017nonlinear,benigni2019eigenvalue,pastur2020randomiid}.
We also believe that there is more to be gained in further studies of forms of random matrix universality in DNNs.
For example, our ideas may lead to tractable analysis of popular optimisation algorithms such as Adam \cite{kingma2014adam} as the problem is essentially reduced to deriving a local law for the gradient pre-conditioning matrix and dealing with the gradient noise.

\appendix
\refstepcounter{section}

\chapter{Neural networks with general activation functions: supplementary}
\label{app:gen_app01}
This appendix provides supporting material for Chapter \ref{chap:general_activation_functions}.
\section{Specific expression for the low-rank perturbation matrix}\label{ap:S_specific}
The the rank-2  $N-1\times N-1$ matrix $S$ arises throughout the course of Sections \ref{sec:nns_random_funcs} and \ref{sec:statement_results} and Lemma \ref{lemma:conditional_dist}. The specific value of $S$ is not required at any point during our calculations and, even though its eigenvalues appear in the result of Theorem \ref{thm:exact_term}, it is not apparent that explicit expressions for its eigenvalues would affect the practical implications of the theorem. These considerations notwithstanding, in this supplementary section we collate all the expressions involved in the development of $S$ from the modeling of the activation function in Section \ref{sec:nns_random_funcs} through to Lemma \ref{lemma:conditional_dist}. Beginning at the final expression for $S$ in Lemma \ref{lemma:conditional_dist} 
\begin{align}
            S_{ij} = \frac{1}{ \sqrtsign{2(N-1)H(H-1)} }\left(\xi_3 + \xi_2(\delta_{i1} + \delta_{j1}) + \xi_1\delta_{i1}\delta_{j1}\right),\end{align}
    where, recalling the re-scaling (\ref{eq:rho_N_redef}), 
 \begin{align}
          \xi_0 &= \sum_{\ell=1}^HN^{-\ell/2}\rho_{\ell}^{(N)}\\
                    \xi_1 &=  \sum_{\ell=1}^{H-2}N^{-\ell/2}\rho_{\ell}^{(N)} \left[(H-\ell)(H-\ell -1) +1 \right]\\
          \xi_2 & =\sum_{\ell=1}^{H-2}N^{-\ell/2}\rho_{\ell}^{(N)}(H-\ell - 2) \\
          \xi_3 &= \sum_{\ell=1}^{H-2}N^{-\ell/2}\rho_{\ell}^{(N)}
      \end{align}
The $\rho_{\ell}$ were defined originally in (\ref{eq:rho_def}) and re-scaled around (\ref{eq:g_def}) so that  \begin{align}
    \rho_{\ell} = \frac{ \expect A_{i,j}^{(\ell)}}{\expect A_{i,j}}
\end{align}
where $A_{i,j}$ are discrete random variables taking values in \begin{equation}
   \mathcal{A} \defeq \left\{\prod_{i=1}^H \alpha_{j_i}\ ~:~ j_1,\ldots, j_H \in \{1,\ldots, L\}\right\}
\end{equation} and $ A^{(\ell)}_{i,j}$ take values in \begin{equation}
    \mathcal{A}^{(\ell)} \defeq\left\{\beta_k\prod_{r=1}^{H-\ell} \alpha_{j_r} ~:~ j_1,\ldots, j_{H-\ell}, k \in \{1,\ldots, L\}\right\}\end{equation}
but we have not prescribed the mass function of the $A_{i,j}$ or $A_{i,j}^{(\ell)}.$
Lastly recall that the $\alpha_j, \beta_j$ are respectively the slopes and intercepts of the piece-wise linear function chosen to approximate the activation function $f$.

\section{Experimental details}\label{ap:experiments}
In this section we give further details of the experiments presented in Section \ref{subsec:discussion_assumptions}.\\

The MLP architecture used consists of hidden layers of sizes $1000, 1000, 500, 250$. The CNN architecture used is a standard LeNet style architecture: \begin{enumerate}
    \item 6 filters of size $4\times 4$.
    \item Activation.
    \item Max pooling of size $2\times 2$ and stride $2$.
    \item 16 filters of size $4\times 4$.
    \item Activation.
    \item Max pooling of size $2\times 2$ and stride $2$.
        \item 120 filters of size $4\times 4$.
    \item Activation.
    \item Dropout.
    \item Fully connected to size 84.
        \item Activation
            \item Dropout.
    \item Fully connected to size 10.
\end{enumerate} 
The activation functions used were the ubiquitous $\texttt{ReLU}$ defined by \begin{equation}
    \texttt{ReLU}(x) = \max(0, x),
\end{equation}
and \texttt{HardTanh} defined by \begin{equation}
    \texttt{HardTanh}(x) = \begin{cases}
    x ~~ &\text{for } x\in(-1,1),\\
     -1 ~~ &\text{for } x\leq -1,\\  
     1 ~~ &\text{for } x\geq 1,\\   
    \end{cases}
\end{equation}
and a custom 5 piece function $f_5$ with gradients $0.01,0.1, 1, 0.3, 0.03$ on $(-\infty, -2), (-2,-1), (-1,1), (1,2), (2, \infty)$ respectively, and $f_5(0) = 0 $. We implemented all the networks and experiments in PyTorch \cite{paszke2017automatic} and our code is made available in the form of a Python notebook capable of easily reproducing all plots\footnote{\url{https://github.com/npbaskerville/loss-surfaces-general-activation-functions}.}.

\
\chapter{A spin glass model for generative adversarial networks: supplemetary}
\label{app:gan_app01}
This appendix provides supporting material for Chapter \ref{chap:spin_glass_gans}.
\section{Bipartite spin-glass formulation}\label{app:bipartite}
Recalling the expression for $\lG$, one could argue that a more natural formulation would be 
\begin{align*}
    \lG(\vwD, \vwG) &= \sum_{i_1,\ldots, i_p=1}^{N_D} \sum_{j_1,\ldots, j_q=1}^{N_G} Z_{i_1,\ldots, i_p, j_1,\ldots, j_q} \prod_{k=1}^p \wD_{i_k} \prod_{l=1}^q \wG_{j_l}
\end{align*}
for i.i.d. Gaussian $Z$. In this case, each term in the sum contains exactly $p$ weights from the discriminator network and $q$ weights from the generator. This object is known as a bipartite spin glass. We will now present the Gaussian calculations. We need the joint distributions
\begin{align*}
    \left(\lD, \pD_i \lD, \pD_{jk}\lD\right), ~~ \left(\lG, \pG_i \lG, \pG_{jk}\lG, \pD_l \lG, \pD_{mn}\lG \right)
\end{align*} where the two groups are independent from of each other. As in \cite{auffinger2013random}, we will simplify the calculation by evaluating in the region of the north poles on each hyper-sphere. $\lD$ behaves just like a single spin glass, and so we have \cite{auffinger2013random}:\begin{align}
    Var(\lD) &= 1,\label{eq:ld_var2}\\
    Cov(\pD_i \lD, \pD_{jk} \lD) &= 0,\\
    \pD_{ij}\lD ~|~ \{\lD=x_D\} &\sim \sqrtsign{(N_D-1)p(p-1)}GOE^{N_D - 1} - x_DpI,\\
    Cov(\pD_i\lD, \pD_j \lD) &= p\delta_{ij}.
\end{align}
To find the joint and thence conditional distributions for $\lG$, we first compute the covariance function, which follows from the independence of the $Z$: \begin{align}
    &Cov(\lG(\vwD, \vwG), \lG({\vwD}', {\vwG}'))\\
    =&\sum_{\substack{i_1,\ldots, i_p=1\\i_1',\ldots, i_p'=1}}^{N_D} ~~\sum_{\substack{j_1,\ldots, j_q=1\\j_1',\ldots, j_q'=1}}^{N_G} \mathbb{E} Z_{\vec{i}\vec{i}}Z_{\vec{i}'\vec{j}'} \prod_{k=1}^p \wD_{i_k}{\wD_{i_k'}}' \prod_{l=1}^q \wG_{j_l}{\wG_{j_l'}}'\\
    =&\sum_{i_1,\ldots, i_p=1}^{N_D} ~~\sum_{j_1,\ldots, j_q=1}^{N_G} \prod_{k=1}^p \wD_{i_k}{\wD_{i_k}}' \prod_{l=1}^q \wG_{j_l}{\wG_{j_l}}'\\
    =& (\vwD \cdot {\vwD}')^p(\vwG \cdot {\vwG}')^q
\end{align}
The product structure of the covariance function implies that we can write down the following covariances directly from the simple spin-glass case, as the $\pD$ and $\pG$ derivatives act independently on their respective terms:
\begin{align}
    Var(\lG) &= 1,\label{eq:var_lg-bi}\\
    Cov(\pG_{ij}\lG, \lG) &= -q\delta_{ij},\\
    Cov(\pD_{ij} \lG, \lG) &= -p\delta_{ij},\\
    Cov(\pG_{ij}\lG,  \pG_{kl}\lG) &=  q(q-1)\left(\delta_{ik}\delta_{jl} + \delta_{il}\delta_{jk}\right) + q^2 \delta_{ij}\delta_{kl},\\
    Cov(\pD_{ij}\lG,  \pD_{kl}\lG) &=  p(p-1)\left(\delta_{ik}\delta_{jl} + \delta_{il}\delta_{jk}\right) + p^2 \delta_{ij}\delta_{kl},\\
     Cov(\pG_{ij}\lG,  \pD_{kl}\lG) &=  pq \delta_{ij}\delta_{kl},\\
      Cov(\pG_i\pD_j\lG,  \pG_{k}\pD_l\lG) &= pq \delta_{ik}\delta_{jl},\label{eq:ginibre-bi}\\
      Cov(\pG_{ij}\lG,  \pG_{k}\pD_l\lG) &= 0\label{eq:gin_goe1-bi}\\
    Cov(\pD_{ij}\lG,  \pD_{k}\pG_l\lG) &= 0\label{eq:gin_goe2-bi},\\
    Cov(\pD_{i}\pG_j \lG, \lG) &= 0.
\end{align}
Also, all first derivatives of $\lG$ are clearly independent of $\lG$ and its second derivatives by the same reasoning and \begin{align}
    Cov(\pG_{i}\lG, \pG_j\lG) &= q\delta_{ij},\\
    Cov(\pD_{i}\lG, \pD_j\lG) &= p\delta_{ij},\\
    Cov(\pD_{i}\lG, \pG_j\lG) &= 0.
\end{align}
We caw deduce the full gradient covariances, recalling that $\lD$ and $\lG$ are independent:
\begin{align}
    Cov(\partial^{(D)}_i L^{(D)}, \partial^{(D)}_j L^{(D)})& = p(1 + \sigma_z^2)\delta_{ij}\\
    Cov(\partial^{(G)}_iL^{(G)}, \partial^{(G)}_j L^{(G)})& = \sigma^2_z q\delta_{ij}\\
     Cov(\partial^{(D)}_iL^{(D)}, \partial^{(G)}_j L^{(G)})& = 0
\end{align}
and so \begin{align}\label{eq:grad_dens_0_1}
    \varphi_{\left(\nabla_D L^{(D)}, \nabla_G L^{(G)}\right)}(0) = (2\pi)^{-\frac{N-2}{2}} \left(p + \sigma_z^2p\right)^{-\frac{N_D - 1}{2}} \left(\sigma_z^2 q\right)^{-\frac{N_G-1}{2}}.
\end{align}
We need now to calculate the joint distribution of $(\pD_{ij}\lG, \pG_{kl}\lG)$ conditional on $\{\lG = x_G\}$. Denote the covariance matrix for $(\pD_{ij}\lG, \pG_{kl}\lG, \lG)$ by \begin{align}
    \Sigma = \left(\begin{array}{cc}
         \Sigma_{11}&\Sigma_{12}  \\
         \Sigma_{21}&\Sigma_{22} 
    \end{array}\right)
\end{align}
where \begin{align}
    \Sigma_{11} &= \left(\begin{array}{cc}
       p(p-1)(1 + \delta_{ij}) + p^2\delta_{ij}  & pq\delta_{ij}\delta_{kl} \\
             pq \delta_{ij}\delta_{kl} & q(q-1)(1 + \delta_{kl}) + q^2\delta_{kl}  
    \end{array}\right),\\
    \Sigma_{12} &= -\left(\begin{array}{c}
         p\delta_{ij}  \\
          q\delta_{kl}
    \end{array}\right),\\
      \Sigma_{21} &= -\left(\begin{array}{cc}
         p\delta_{ij}  & q\delta_{kl}
    \end{array}\right),\\
    \Sigma_{22} &= 1.
\end{align}
The conditional covariance is then \begin{align}
    \bar{\Sigma} &= \Sigma_{11} - \Sigma_{12}\Sigma_{22}^{-1}\Sigma_{21}\\
    &= \left(\begin{array}{cc}
         p(p-1)(1+\delta_{ij})  & 0\\
         0 & q(q-1) (1 + \delta_{kl})
    \end{array}\right).\label{eq:double_goe-bi}
\end{align}
Repeating this calculation for $(\pG_{ij}\lG, \pG_{kl}\lG, \lG)$ demonstrates that $\nabla_G^2\lG \mid \{\lG = x_G\}$  has independent entries, up-to symmetry. The result (\ref{eq:double_goe-bi}) demonstrates that, conditional on $\{\lG = x_G\}$, $\nabla_G^2\lG$ and $\nabla_D^2\lG$ are independent GOEs.
In summary, from (\ref{eq:double_goe-bi}) and (\ref{eq:ginibre-bi}-\ref{eq:gin_goe2-bi}) we obtain \begin{align}
    \left(\begin{array}{cc}
        -\nabla_D^2\lG & -\nabla_G\nabla_D \lG  \\
         \nabla_D\nabla_G \lG & \nabla^2 \lG  
    \end{array}\right) ~|~ \{\lG = x_G\} &\overset{d}{=} 
\sqrtsign{2}\left(\begin{array}{cc}
       \sqrtsign{N_D -1}\sqrtsign{p(p-1)}M^{(D)} & -2^{-1/2}\sqrtsign{pq}G \\
         2^{-1/2}\sqrtsign{pq}G^T & \sqrtsign{N_G - 1}\sqrtsign{q(q-1)}M^{(G)}
    \end{array}\right)\notag\\
    &~~~~~~- x_G \left(\begin{array}{cc}
        -pI_{N_D} & 0  \\
         0 & qI_{N_G} 
    \end{array}\right)
\end{align}
where $M^{(D)}\sim GOE^{N_D - 1}$ and $M^{(G)} \sim GOE^{N_G - 1}$ are independent GOEs and $G$ is an independent $N_D - 1 \times N_G - 1$ Ginibre matrix with entries of unit variance.

\medskip 
At this point a problem becomes apparent. Suppose that $q\leq p$, then the variance of the lower-right block is strictly less than that of the off diagonal blocks. If we proceed with the strategy in the main text, there is no way of decomposing the lower-right block as a sum of two independent smaller variance GOEs with one matching the variance of the off diagonal blocks. Similarly, if $q>p$, then the final Hessian involving $\LD, \LG$ will have lower-variance in the upper-left block than the off-diagonals unless very specific undesirable conditions hold on $p,q$ and $\sigma_z$. In either of these cases, we cannot decompose the final Hessian as a sum of a large $N-2\times N-2$ GOE and some smaller GOEs in the upper-left or lower-right blocks. We would therefore have to truly compute the Ginibre averages in the supersymmetric method, which we believe is intractable.

\medskip
We could complete the complexity calculation via the methods of chapter \ref{chap:spin_glass_gans} supposing that the appropriate conditions hold on $p, q$ and $\sigma_z$. It would look much the same as the calculation in the main text, though the resulting polynomial for the spectral density would be different. Since this work was completed, the complexity results for bipartite spin glasses were obtained in \cite{mckenna2021complexity} using an entirely new method developed in the companion paper \cite{arous2021exponential}. Applying this method arguably presents more technical hurdles than the supersymmetric approach to complexity calculations, however it is much more general and can be applied to the above model for any $p, q$ and $\sigma_z$.

\section{Extra plots}\label{app:plots}
This section contains some extra plots to back up the comparisons between our model's predictions and the experimental DCGAN results in Section \ref{subsec:hparams}. In particular, we produce versions of the plots in Figures \ref{fig:vary_sigma_results} and \ref{fig:vary_kappa_results} but for various values of $p$ and $q$ other than $p=q=5$. Since $p=q=5$ is the structurally correct choice for the DCGAN, it is natural to ask if any agreement between theory and experiment is most closely obtained with $p=q=5$. Figure \ref{fig:kappa_lg_variety} shows that the model has the same deficiency in $\kappa$ for all $p,q$ values tested. Figure \ref{fig:kappa_ld_variety} shows best agreement for $p=q=5$, $p=3, q=7$ and $p=7, q=3$, and similarly in Figure \ref{fig:sigma_lg_variety}. There is perhaps weak evidence that the role of $p$ and $q$ as representing the number of layers in the networks has some merit experimentally.

\begin{figure}[h]
    \centering
    \includegraphics[width=0.6\textwidth]{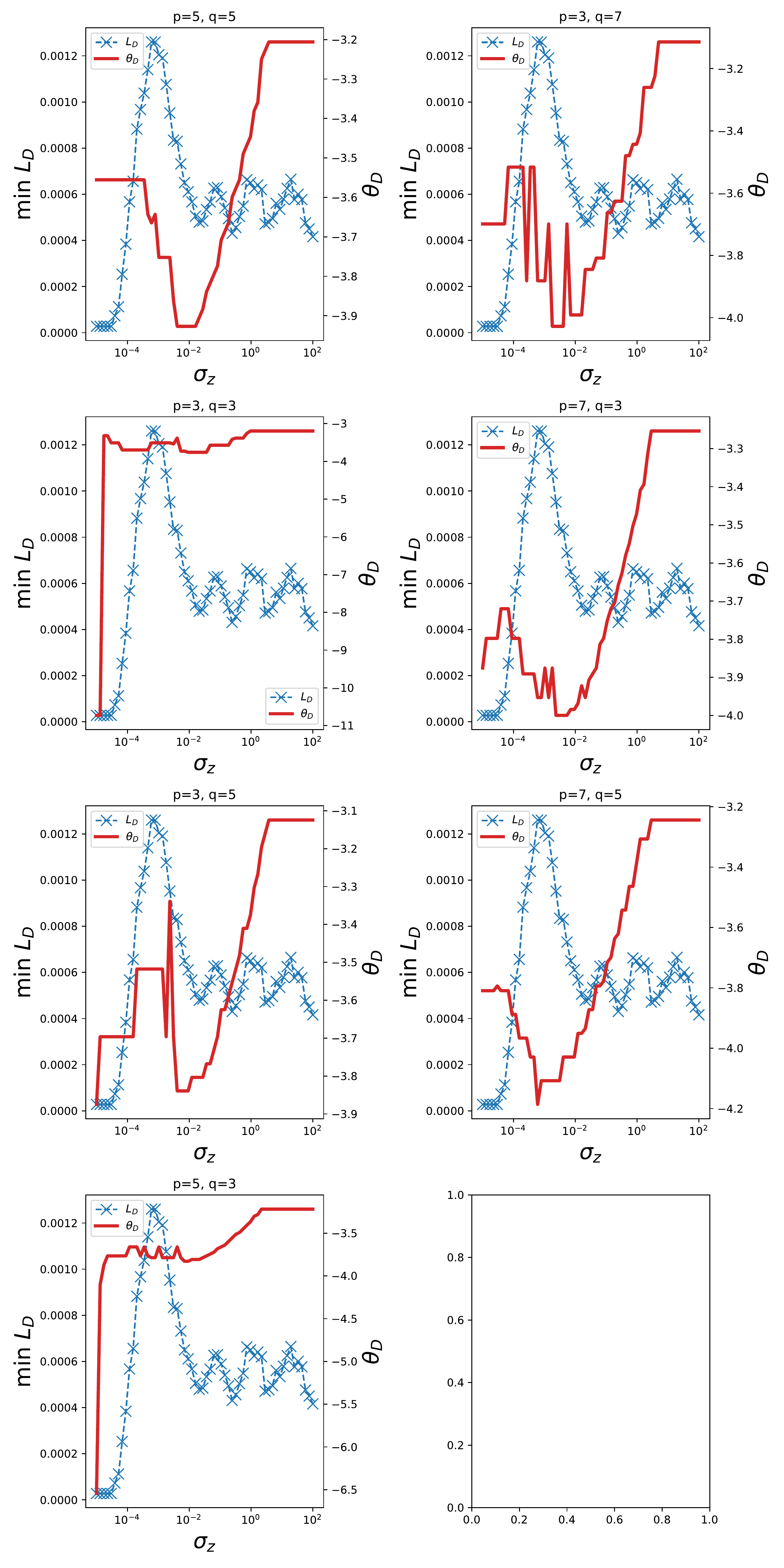}
    \caption{The effect of $\sigma_z$ on minimum $L_D$. Comparison of theoretical predictions of minimum possible discriminator and generator losses to observed minimum losses when training DCGAN on CIFAR10. The blue cross-dashed lines show the experimental DCGAN results, and the solid red show the theoretical results $\vartheta_G, \vartheta_D$. $\kappa=0.5$ is used and $p, q$ are varied.}
    \label{fig:sigma_ld_variety}
\end{figure}

\begin{figure}[h]
    \centering
    \includegraphics[width=0.6\textwidth]{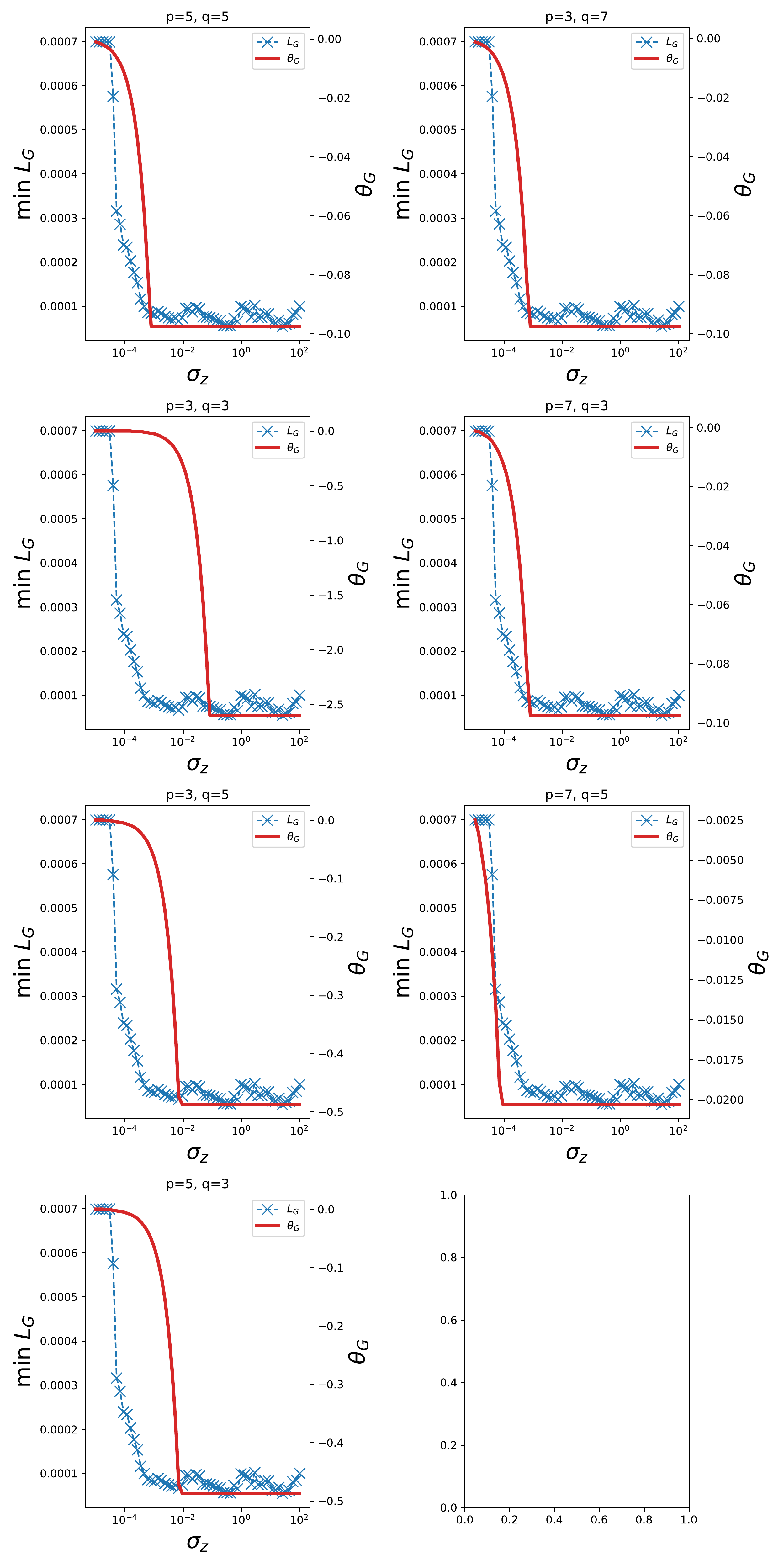}
    \caption{The effect of $\sigma_z$ on minimum $L_G$. Comparison of theoretical predictions of minimum possible discriminator and generator losses to observed minimum losses when training DCGAN on CIFAR10. The blue cross-dashed lines show the experimental DCGAN results, and the solid red show the theoretical results $\vartheta_G, \vartheta_D$. $\kappa=0.5$ is used and $p, q$ are varied.}
    \label{fig:sigma_lg_variety}
\end{figure}

\begin{figure}[h]
    \centering
    \includegraphics[width=0.6\textwidth]{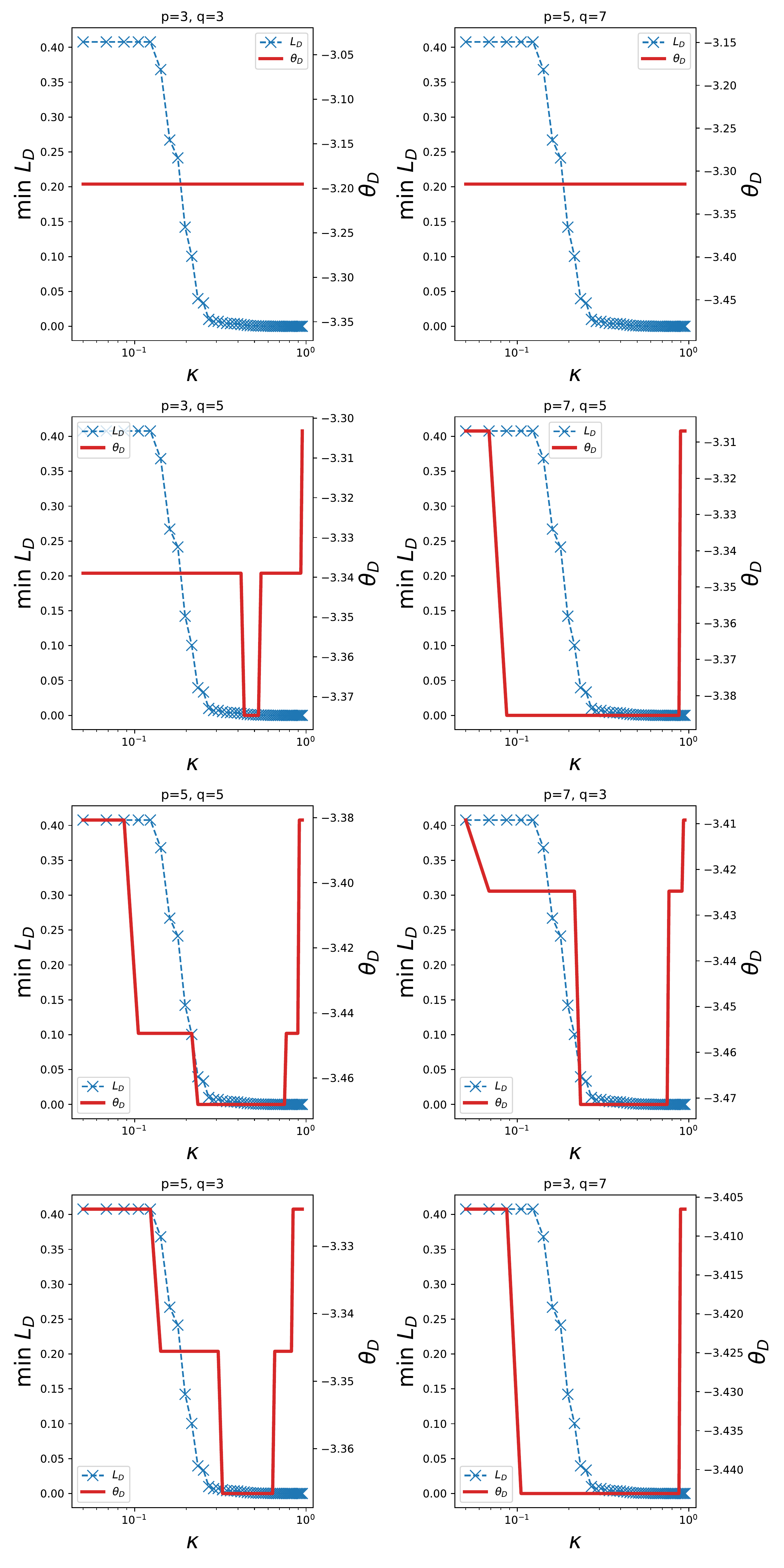}
    \caption{The effect of $\kappa$ on minimum $L_D$. Comparison of theoretical predictions of minimum possible discriminator and generator losses to observed minimum losses when training DCGAN on CIFAR10. The blue cross-dashed lines show the experimental DCGAN results, and the solid red show the theoretical results $\vartheta_G, \vartheta_D$. $\sigma_z=1$ is used and $p, q$ are varied.}
    \label{fig:kappa_ld_variety}
\end{figure}

\begin{figure}[h]
    \centering
    \includegraphics[width=0.6\textwidth]{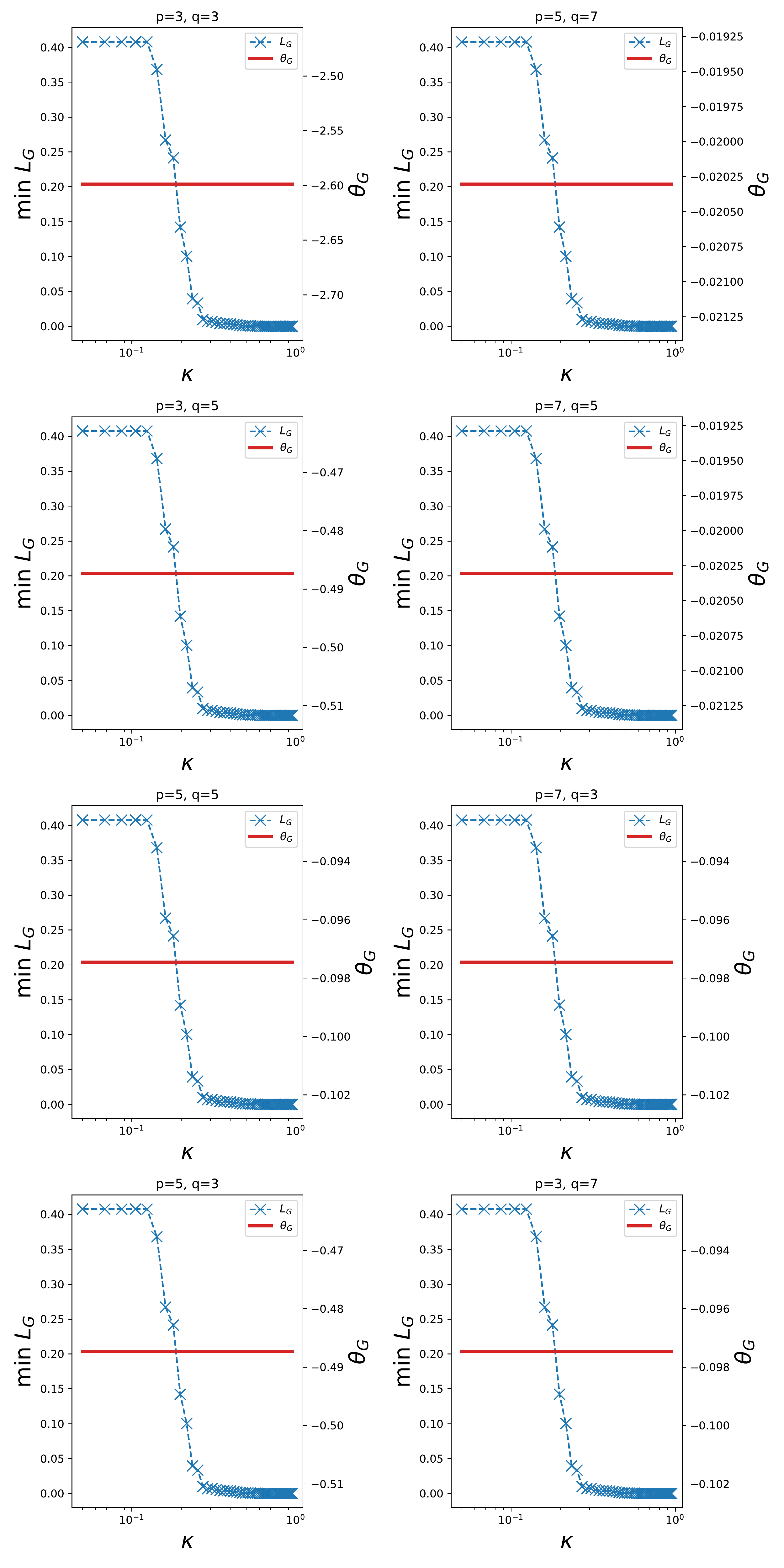}
    \caption{The effect of $\kappa$ on minimum $L_G$. Comparison of theoretical predictions of minimum possible discriminator and generator losses to observed minimum losses when training DCGAN on CIFAR10. The blue cross-dashed lines show the experimental DCGAN results, and the solid red show the theoretical results $\vartheta_G, \vartheta_D$. $\sigma_z=1$ is used and $p, q$ are varied.}
    \label{fig:kappa_lg_variety}
\end{figure}

\chapter{Appearance of local random matrix statistics: supplementary}
\label{app:spacings_app03}
This appendix provides supporting material for Chapter \ref{chap:spacings}.
\section{Extra Figures and Degeneracy Investigation}\label{sec:degen}
Figure \ref{fig:cifarresnet_comp_mnist_truncation} compares the effect of degeneracy on unfolded spacings in each of the 3 cases considered. We see that the logistic MNIST models (trained and untrained) have a much greater level of degeneracy, whereas the CIFAR10-Resnet34 spectra clearly have GOE spacings even without any cut-off. Figures \ref{fig:log_mnist_spacings_ratio}–\ref{fig:log_mnist_unt_spacings_ratio} show further unfolded spacing and spacing ratio results like those in the main text.
\begin{figure}[h]
\centering
\begin{subfigure}{0.4\linewidth}
\centering
    \includegraphics[width=\linewidth]{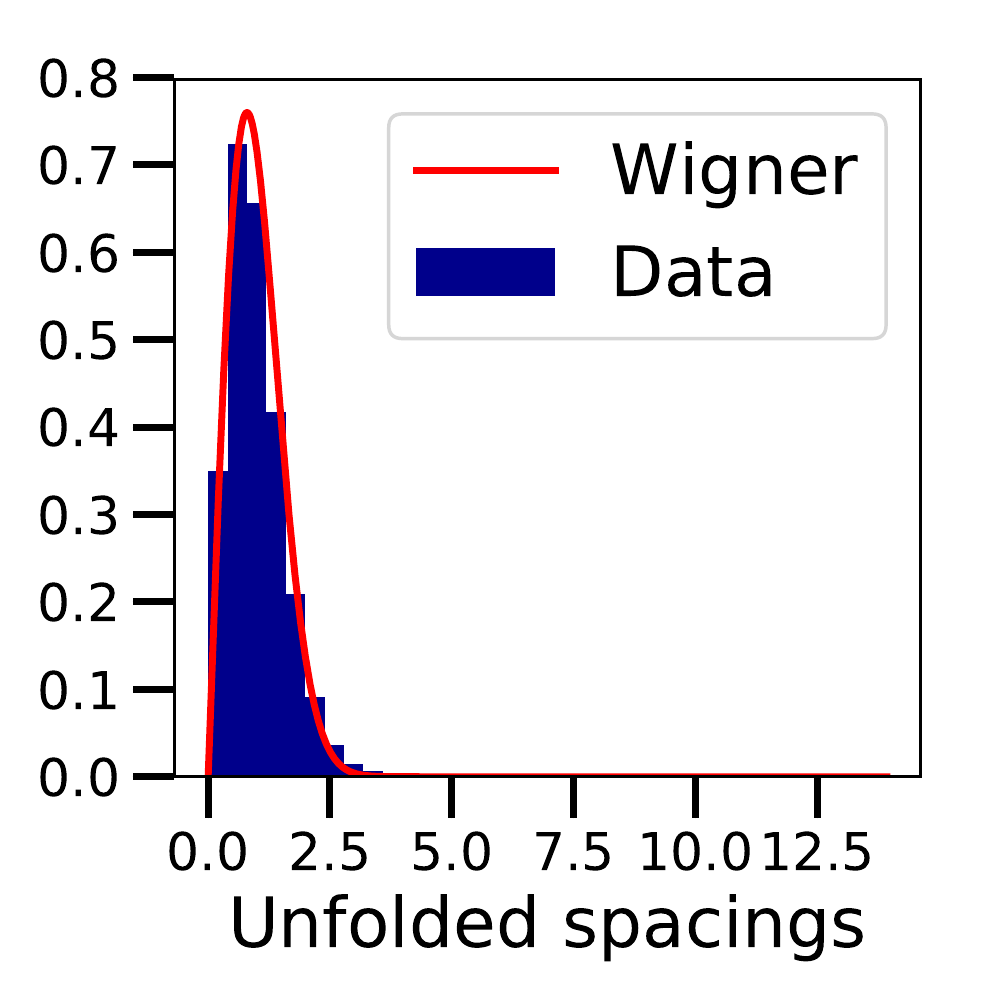}
    \caption{Batch train}
\end{subfigure}
\begin{subfigure}{0.4\linewidth}
\centering
    \includegraphics[width=\linewidth]{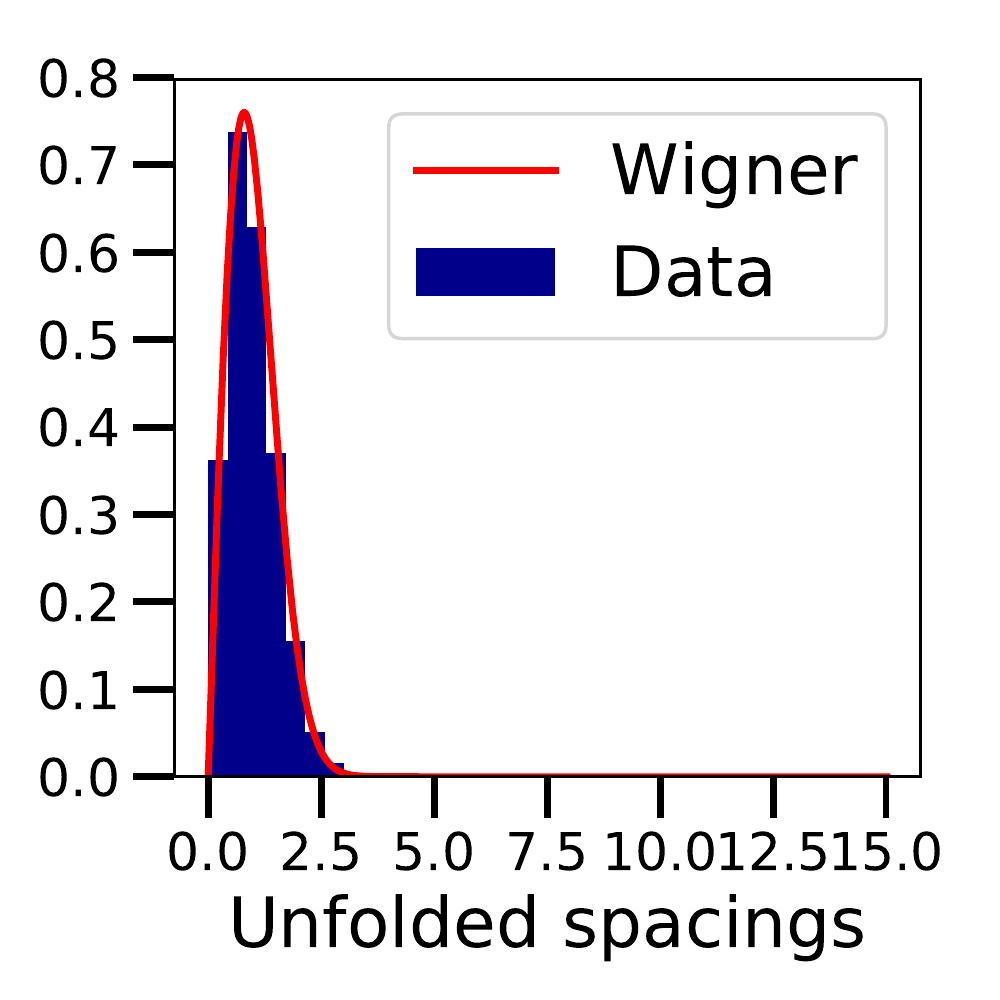}
    \caption{Batch test}
\end{subfigure}
    \caption{Unfolded spacings for the Hessian of a logistic regression trained on MNIST. Hessian computed batches of size 64 of the training and test datasets.}
    \label{fig:log_mnist_spacings_unfolded}
\end{figure}

\begin{figure}[h]
\centering
\begin{subfigure}{0.4\linewidth}
\centering
    \includegraphics[width=\linewidth]{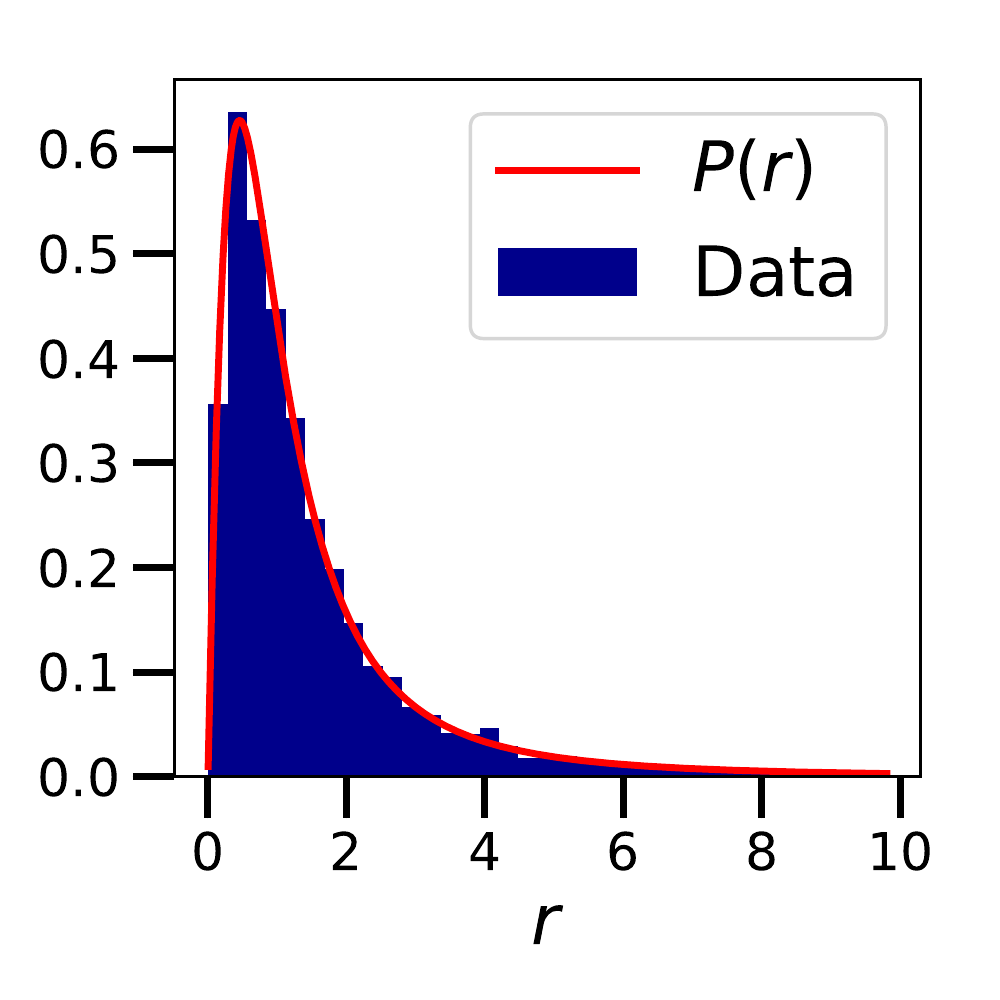}
    \caption{All train}
\end{subfigure}
\begin{subfigure}{0.4\linewidth}
\centering
    \includegraphics[width=\linewidth]{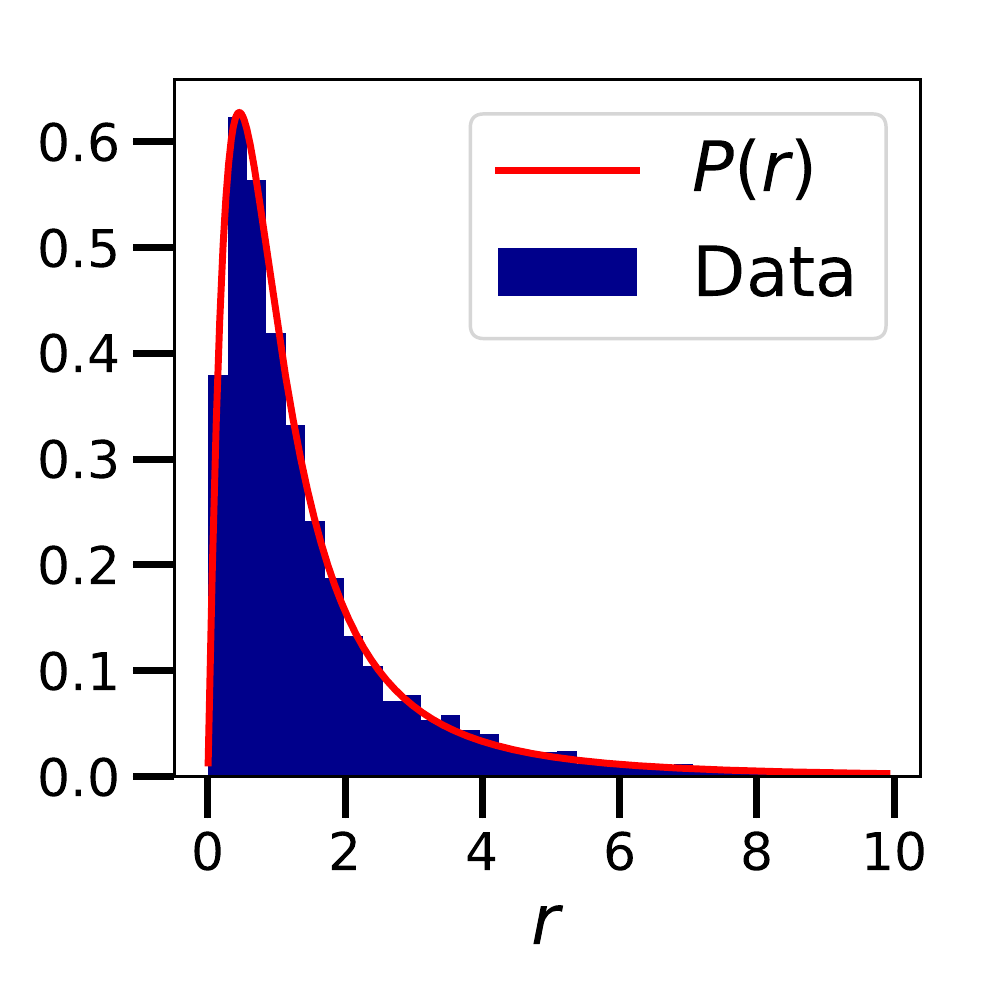}
    \caption{All test}
\end{subfigure}
\caption{Consecutive spacing ratios for the Hessian of a logistic regression trained on MNIST. Hessian computed batches of size 64 of the training and test sets, and over the whole train and test sets.}
\label{fig:log_mnist_spacings_ratio}
\end{figure}

\begin{figure}[h]
\centering
\begin{subfigure}{0.4\linewidth}
\centering
    \includegraphics[width=\linewidth]{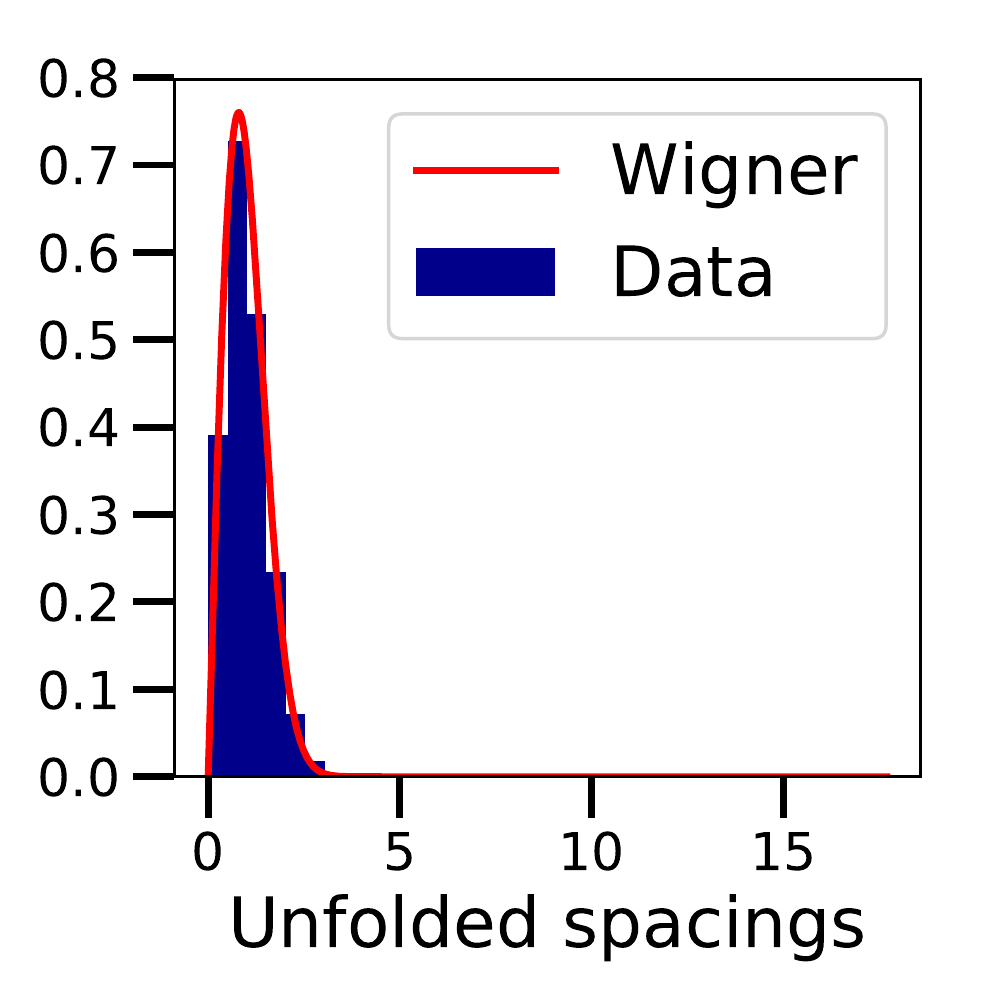}
    \caption{Batch train}
\end{subfigure}
\begin{subfigure}{0.4\linewidth}
\centering
    \includegraphics[width=\linewidth]{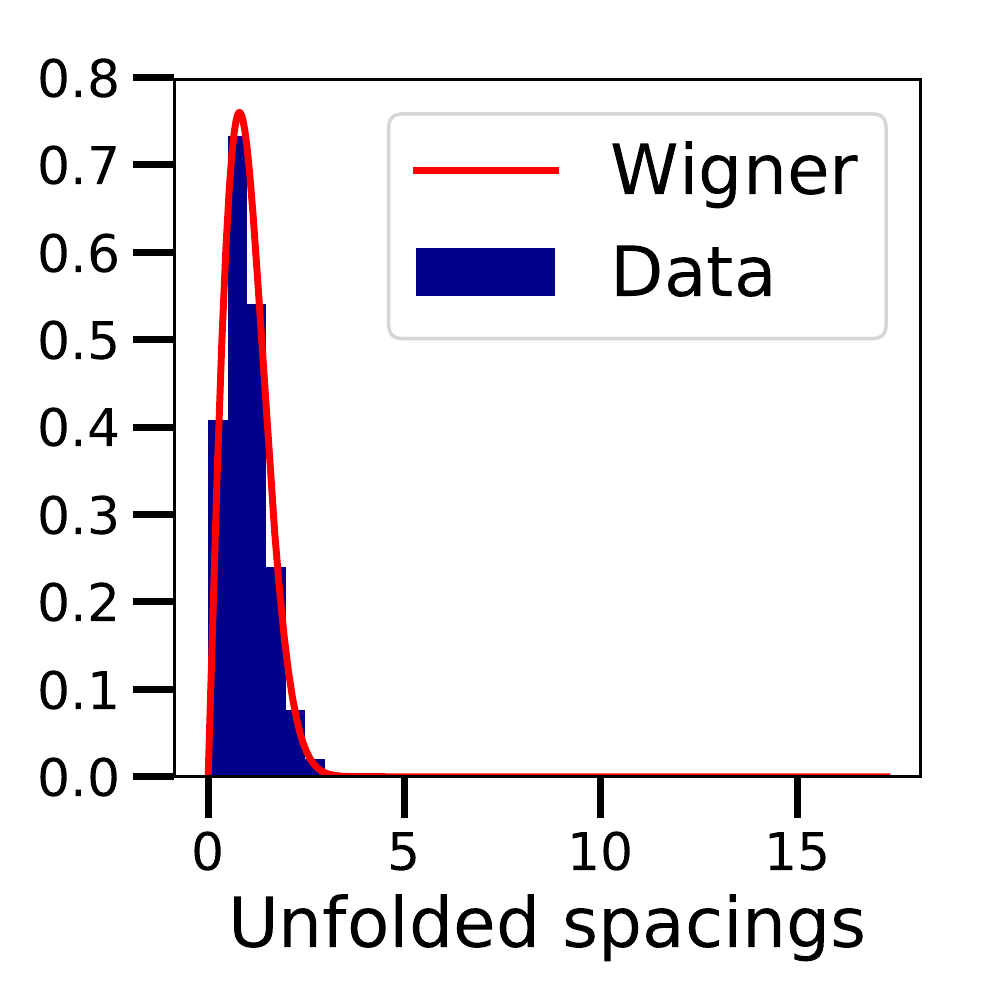}
    \caption{Batch test}
\end{subfigure}
    \caption{Unfolded spacings for the Hessian of a randomly initialised logistic regression for MNIST. Hessian computed batches of size 64 of the training and test datasets.}
    \label{fig:log_mnist_unt_spacings_unfolded}
\end{figure}

\begin{figure}[h]
\centering
\begin{subfigure}{0.4\linewidth}
\centering
    \includegraphics[width=\linewidth]{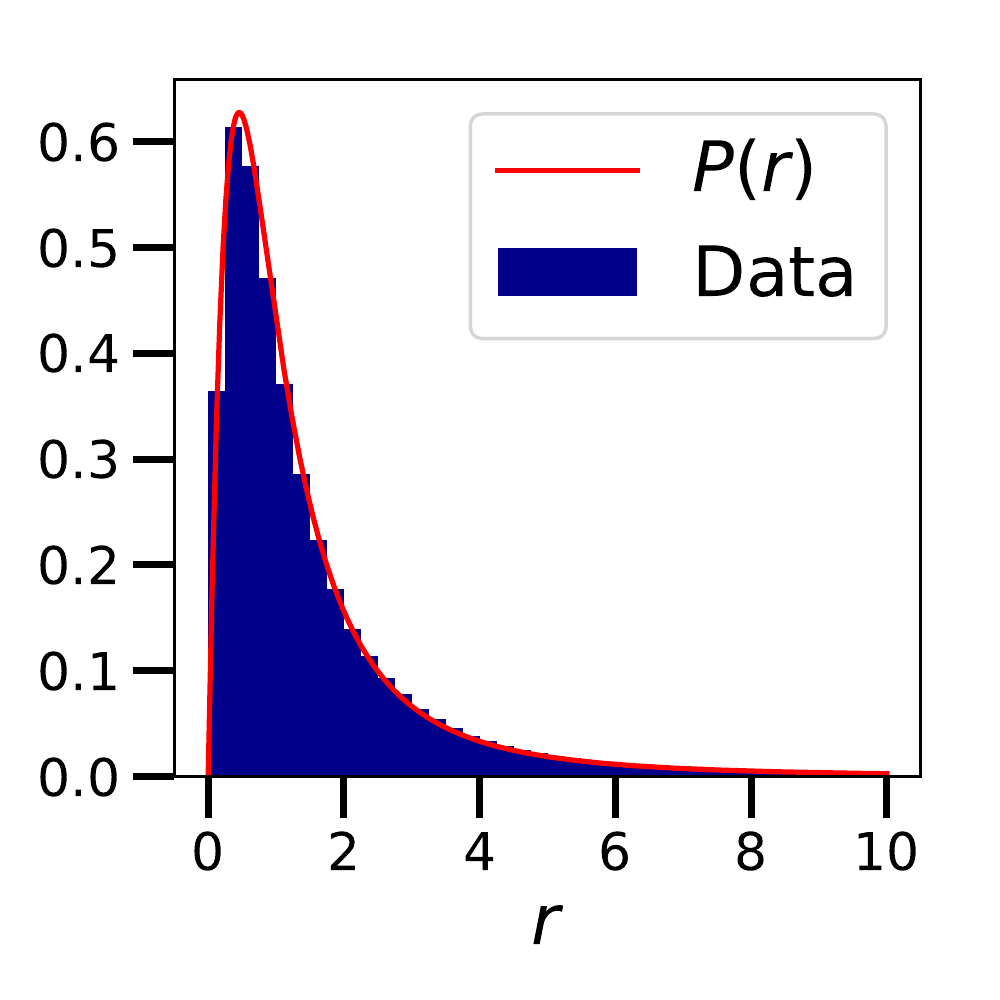}
    \caption{Batch train}
\end{subfigure}
\begin{subfigure}{0.4\linewidth}
\centering
    \includegraphics[width=\linewidth]{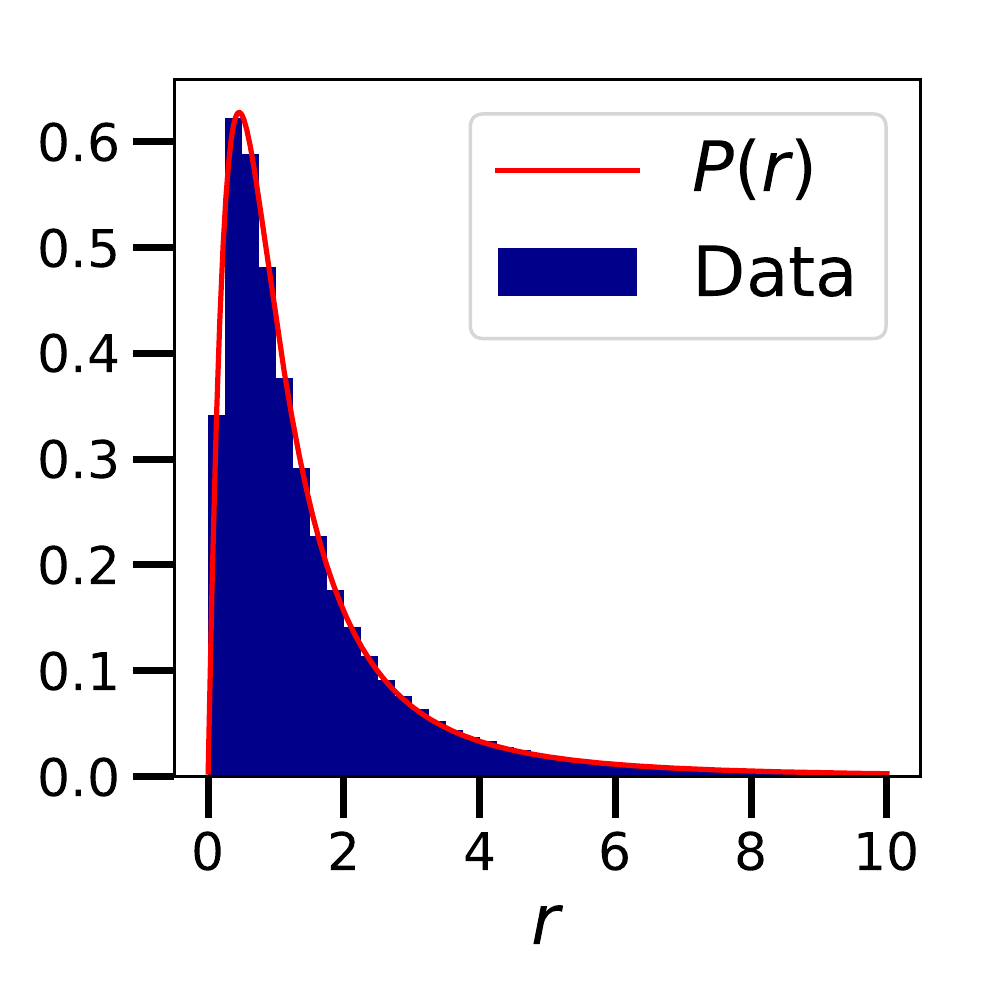}
    \caption{Batch test}
\end{subfigure}

\begin{subfigure}{0.4\linewidth}
\centering
    \includegraphics[width=\linewidth]{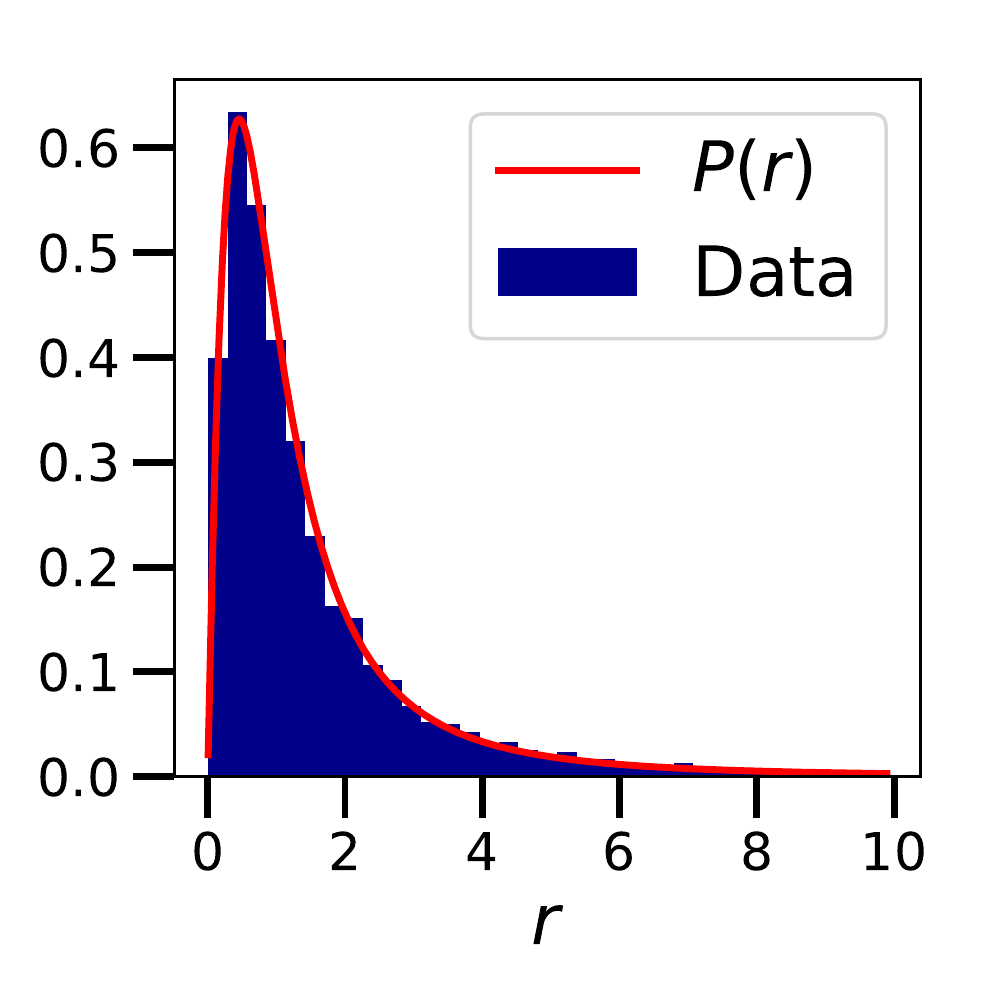}
    \caption{All train}
\end{subfigure}
\begin{subfigure}{0.4\linewidth}
\centering
    \includegraphics[width=\linewidth]{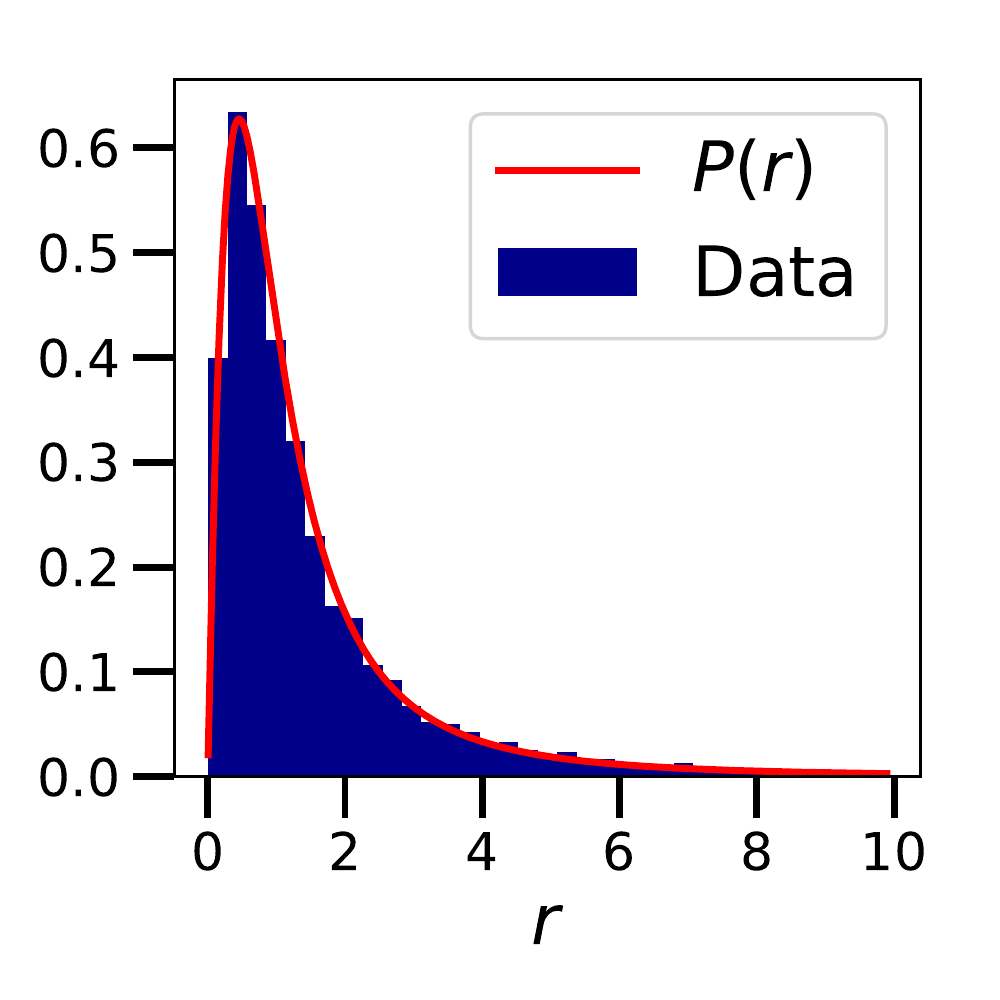}
    \caption{All test}
\end{subfigure}
    \caption{Consecutive spacing ratios for the Hessian of a randomly initialised logistic regression for MNIST. Hessian computed batches of size 64 of the training and test sets, and over the whole train and test sets.}
    \label{fig:log_mnist_unt_spacings_ratio}
\end{figure}

\begin{figure*}[h]
\centering
\begin{subfigure}{0.32\linewidth}
\centering
    \includegraphics[width=\linewidth]{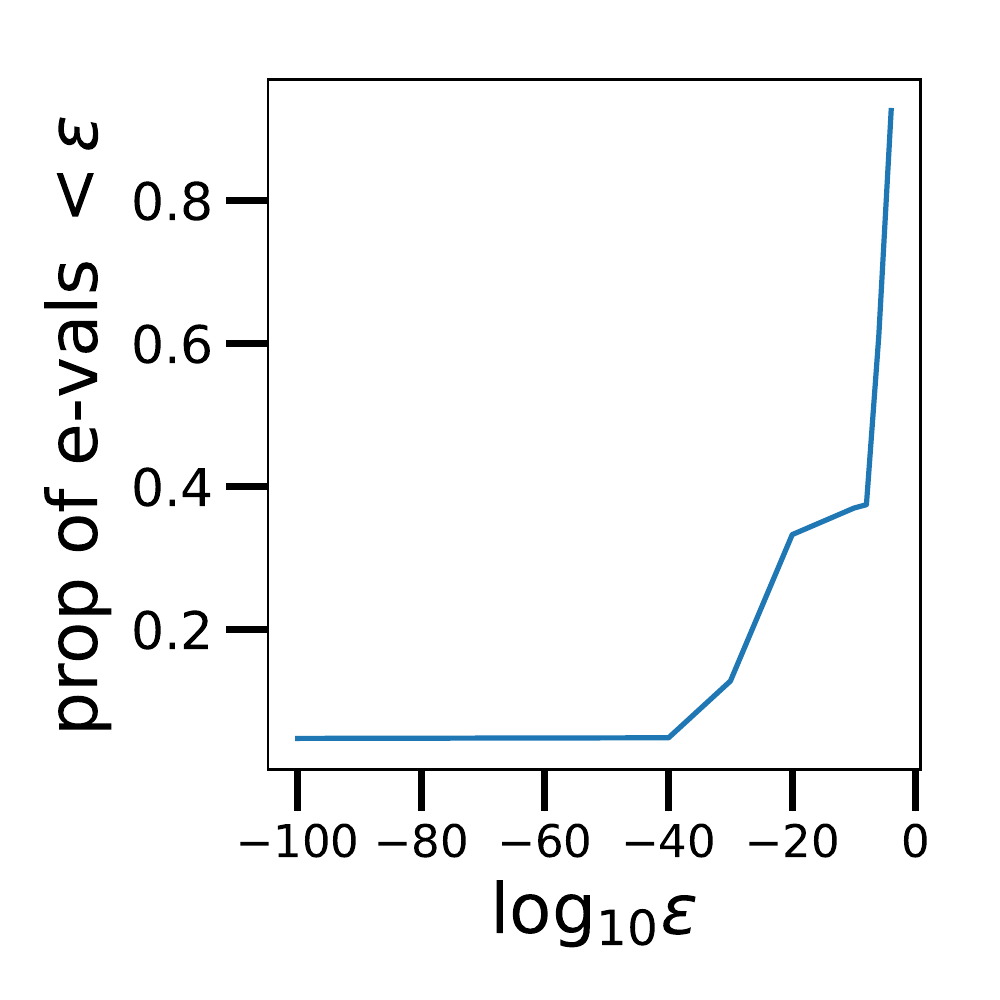}
    \caption{Proportion of small eigenvalues}
\end{subfigure}
\begin{subfigure}{0.32\linewidth}
\centering
    \includegraphics[width=\linewidth]{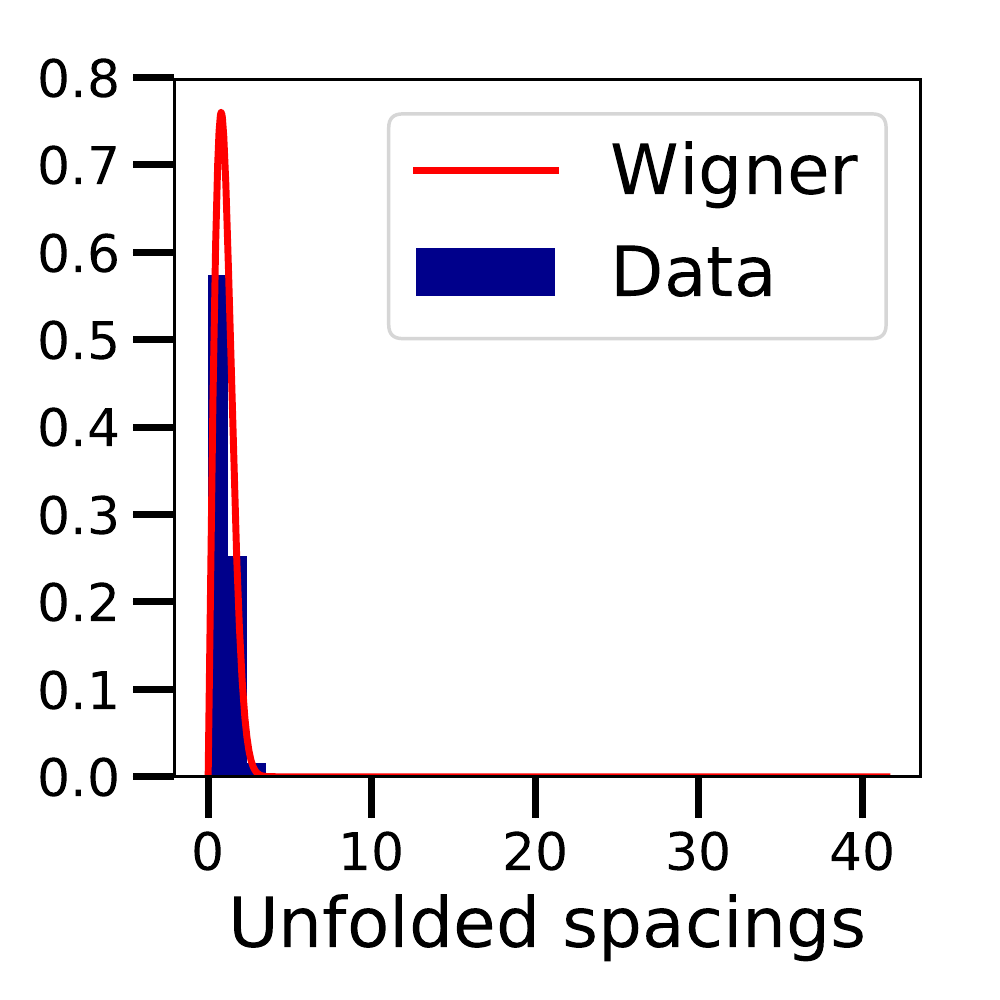}
    \caption{No cut-off}
\end{subfigure}
\begin{subfigure}{0.32\linewidth}
\centering
    \includegraphics[width=\linewidth]{d4ed52a9f94c916e.pdf}
    \caption{$1e-30$  cut-off}
\end{subfigure}

\begin{subfigure}{0.32\linewidth}
\centering
    \includegraphics[width=\linewidth]{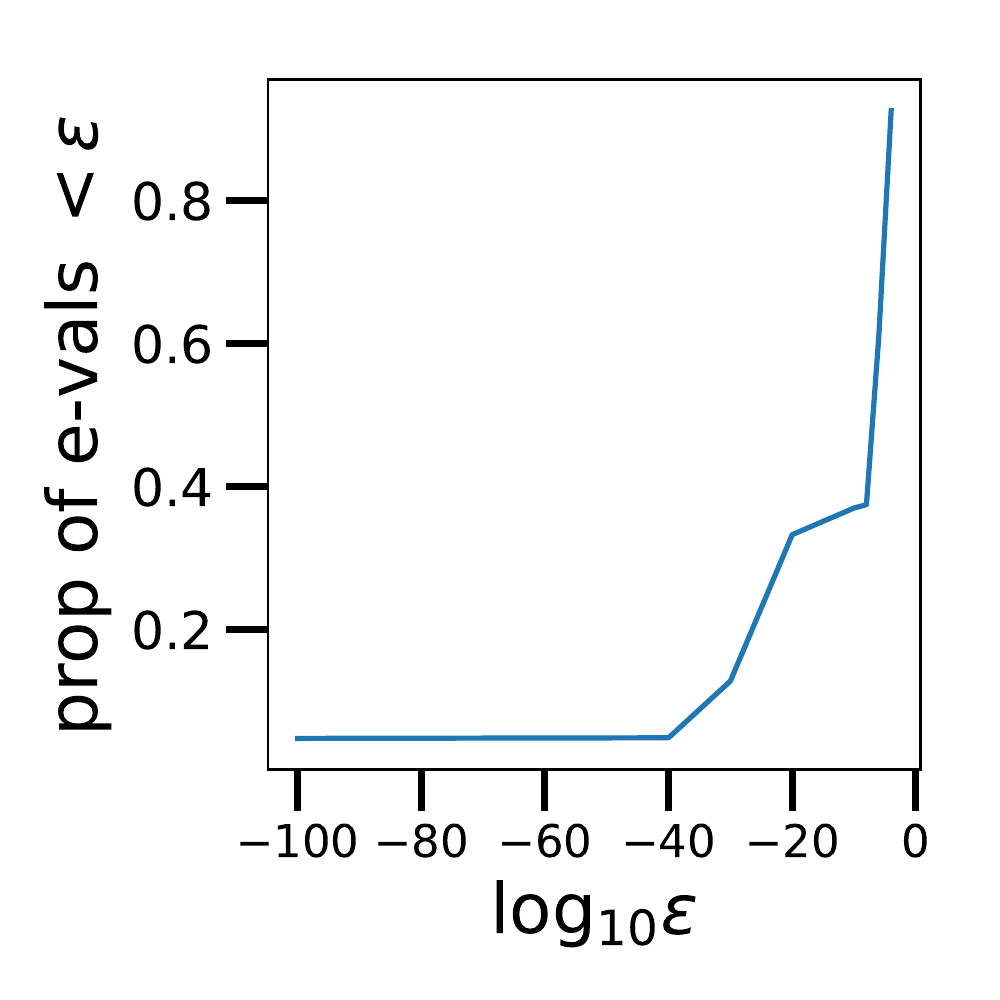}
    \caption{Proportion of small eigenvalues}
\end{subfigure}
\begin{subfigure}{0.32\linewidth}
\centering
    \includegraphics[width=\linewidth]{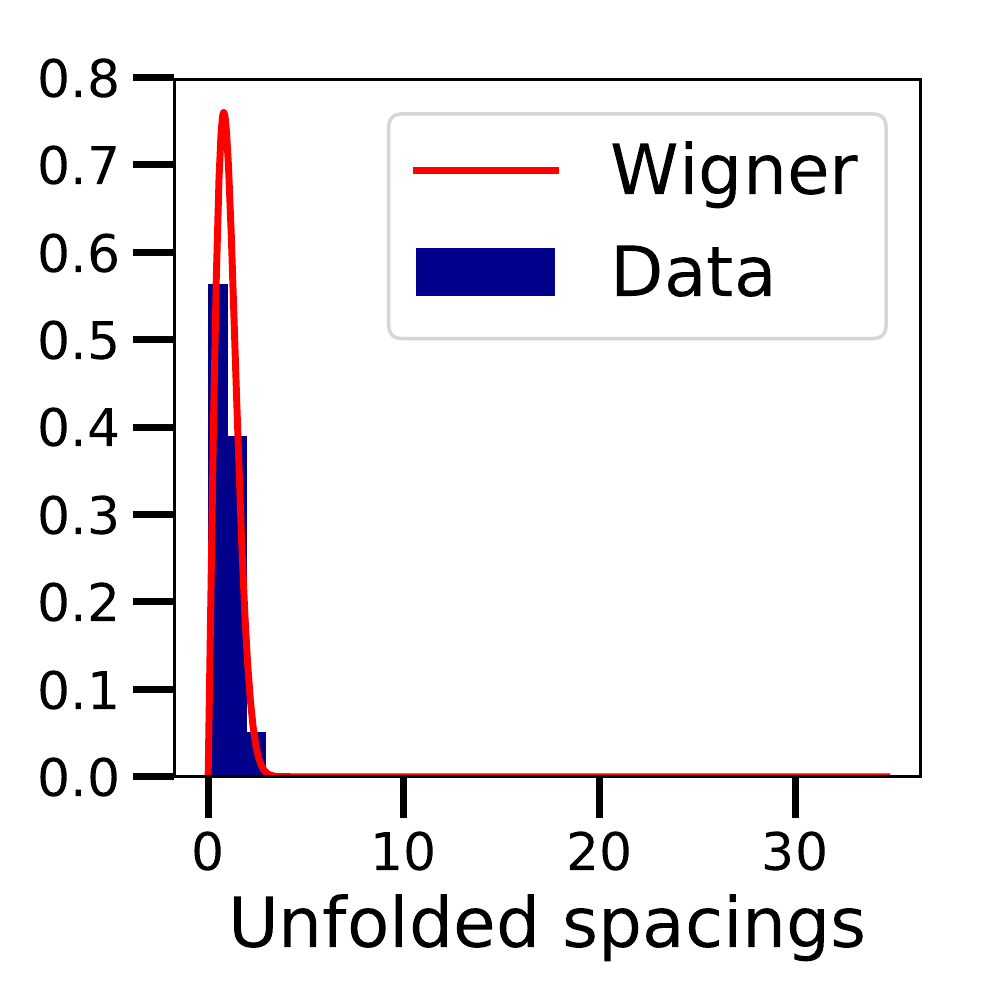}
    \caption{No cut-off}
\end{subfigure}
\begin{subfigure}{0.32\linewidth}
\centering
    \includegraphics[width=\linewidth]{6b56f0d866fe9a04.pdf}
    \caption{$1e-30$  cut-off}
\end{subfigure}

\begin{subfigure}{0.32\linewidth}
\centering
    \includegraphics[width=\linewidth]{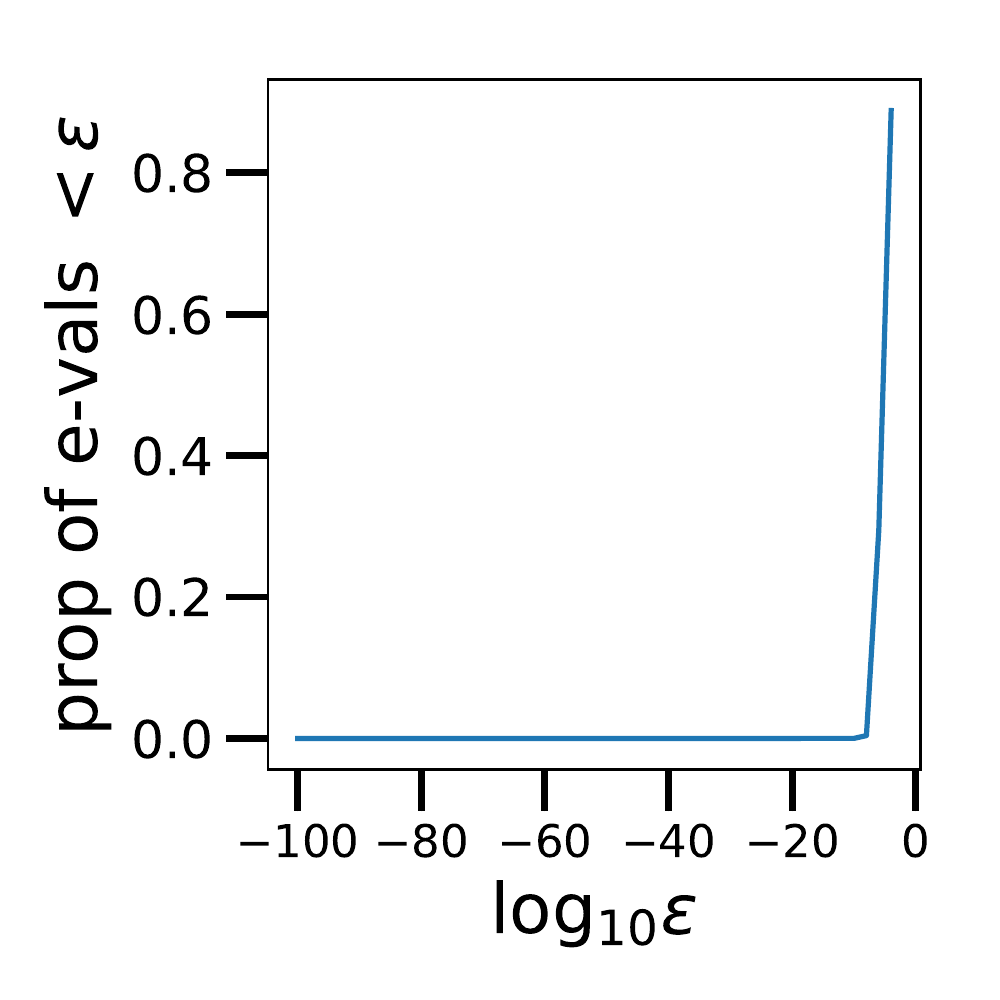}
    \caption{Proportion of small eigenvalues}
\end{subfigure}
\begin{subfigure}{0.32\linewidth}
\centering
    \includegraphics[width=\linewidth]{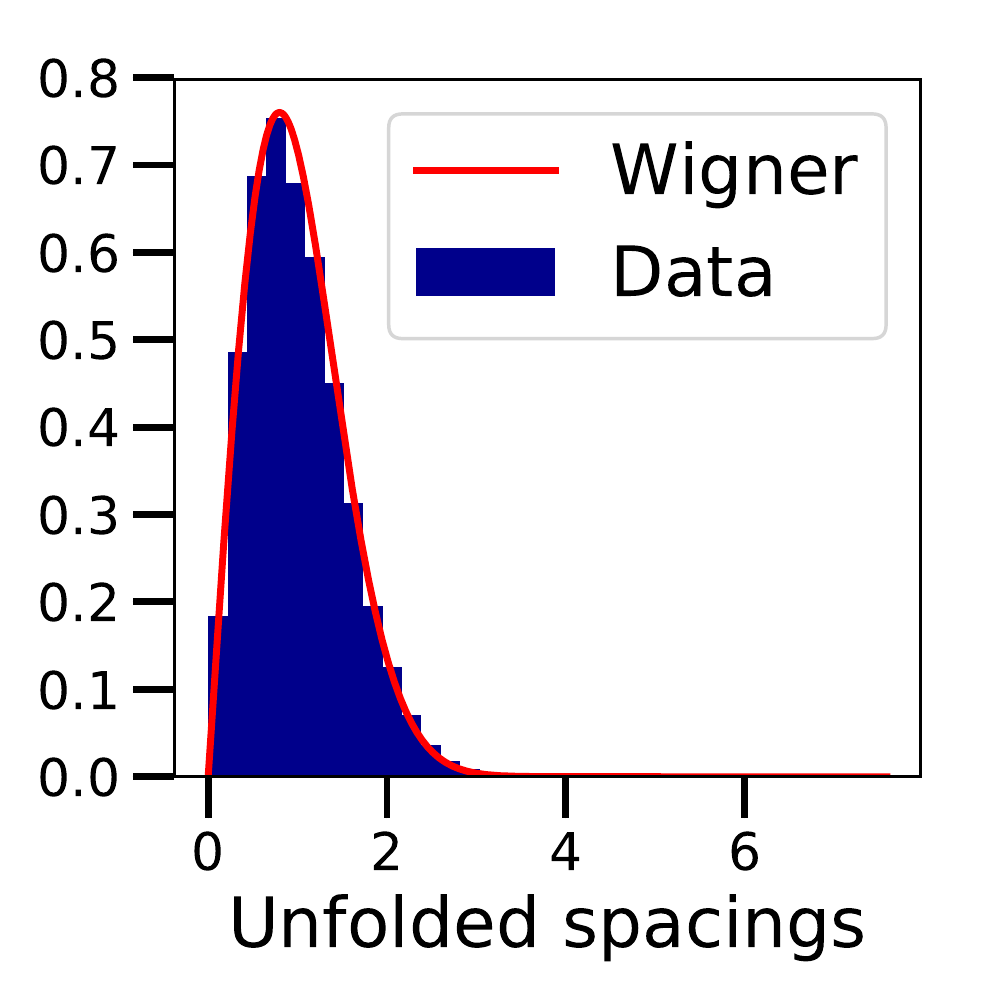}
    \caption{No cut-off}
\end{subfigure}
\begin{subfigure}{0.32\linewidth}
\centering
    \includegraphics[width=\linewidth]{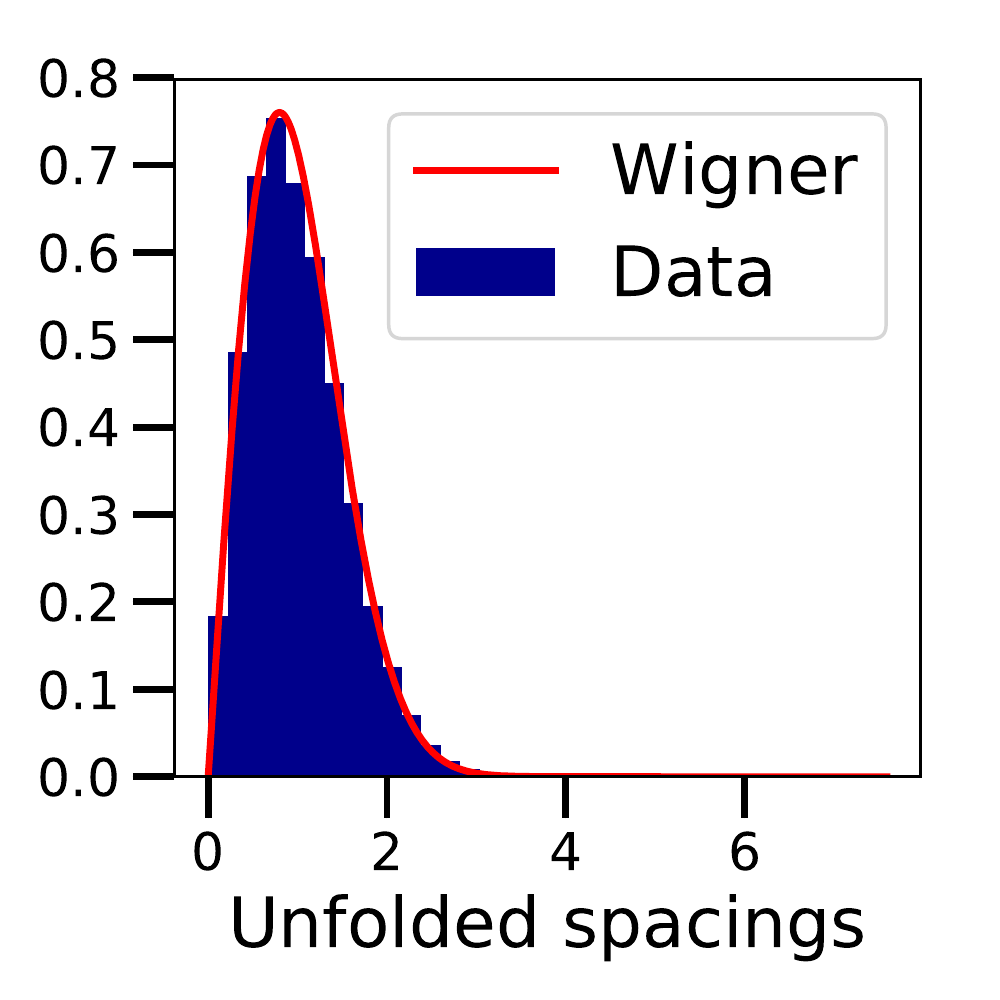}
    \caption{$1e-30$  cut-off}
\end{subfigure}

    \caption{Unfolded spacings for the Hessian of a logistic regression. Showing MNIST (top), untrained MNIST (middle) and Resnet34 embedded CIFAR10 (bottom). Comparing the effect of a cuff-off for very small eigenvalues.}
    \label{fig:cifarresnet_comp_mnist_truncation}
\end{figure*}


\section{Experimental details}\label{app:architectures}
\subsection{Network architectures}
\medskip 

\textbf{Logistic regression (MNIST)}
\begin{enumerate}
    \item Input features 784 to 10 output logits.
\end{enumerate}

\medskip

\textbf{2-layer MLP (MNIST)}
\begin{enumerate}
    \item Input features 784 to 10 neurons.
    \item 10 neurons to 100 neurons.
    \item 100 neurons to 10 output logits.
\end{enumerate}

\medskip
\textbf{3-layer MLP (MNIST)}
\begin{enumerate}
    \item Input features 784 to 10 neurons.
    \item 10 neurons to 100 neurons.
    \item 100 neurons to 100 neurons.
    \item 100 neurons to 10 output logits.
\end{enumerate}

\medskip
\textbf{Logistic regression on ResNet features (CIFAR10)}
\begin{enumerate}
    \item Input features 513 to 10 neurons.
\end{enumerate}

\medskip
\textbf{LeNet (CIFAR10)}
\begin{enumerate}
    \item Input features 32x32x3 through 5x5 convolution to 6 output channels.
    \item 2x2 max pooling of stride 2.
    \item 5x5 convolution to 16 output channels.
    \item 2x2 max pooling of stride 2.
    \item Fully connection layer from 400 to 120.
    \item Fully connection layer from 120 to 84.
    \item Fully connection layer from 84 to output 10 logits.
\end{enumerate}

\medskip
\textbf{MLP (CIFAR10)}
\begin{enumerate}
    \item 3072 input features to 10 neurons.
    \item 10 neurons to 300 neurons.
    \item 300 neurons to 100 neurons.
\end{enumerate}

\medskip
\textbf{MLP (Bike)}
\begin{enumerate}
    \item 13 input features to 100 neurons.
    \item 100 neurons to 100 neurons.
    \item 100 neurons to 50 neurons.
    \item 50 neurons to 1 regression output.
\end{enumerate}

\subsection{Other details}
All networks use the same (default) initialisation of weights in PyTorch, which is the `Kaiming uniform' method of \cite{he2015delving}. All networks used ReLU activation functions.

\subsection{Data pre-processing}\label{app:preproc}
For the image datasets MNIST and CIFAR10 we use standard computer vision pre-processing, namely mean and variance standardisation across channels. We refer to the accompanying code for the precise procedure

\medskip
The Bike dataset has 17 variables in total, namely: \texttt{instant}, \texttt{dteday}, \texttt{season}, \texttt{yr}, \texttt{mnth}, \texttt{hr}, \texttt{holiday}, \texttt{weekday}, \texttt{workingday}, \texttt{weathersit}, \texttt{temp}, \texttt{atemp}, \texttt{hum}, \texttt{windspeed}, \texttt{casual}, \texttt{registered}, \texttt{cnt}. All variables are either positive integers or real numbers. It is standard to view \texttt{cnt} as the regressand, so one uses some or all of the remaining features to predict \texttt{cnt}. This is the approach we take, however we slightly reduce the number of features by dropping \texttt{instant}, \texttt{casual}, \texttt{registered}, since \texttt{instant} is just an index and \texttt{casual}+\texttt{registered}=\texttt{cnt}, so including those features would render the problem trivial. We map \texttt{dteday} to a integer uniquely representing the date and we standardise \texttt{cnt} by dividing by its mean.

\chapter{Universal characteristics of loss surfaces: supplementary}\label{app:outlier_exp_details}
This appendix provides supporting material for Chapter \ref{chap:univ} including full details of the experimental set-up and analysis for the outlier experiments.

\section{Architectures and training of models.}
We use the GPU powered Lanczos quadrature algorithm \cite{gardner2018gpytorch, meurant2006lanczos}, with the Pearlmutter trick \cite{pearlmutter1994fast} for Hessian vector products, using the PyTorch \cite{paszke2017automatic} implementation of both Stochastic Lanczos Quadrature and the Pearlmutter. We then train a 16 Layer VGG CNN \cite{simonyan2014very} with $P=15291300$ parameters 
and the 28 Layer Wide Residual Network \cite{zagoruyko2016wide,he2016deep} architectures 
on the CIFAR-$100$ dataset \cite{krizhevsky2009learning} (45,000 training samples and 5,000 validation samples) using SGD. We use the following  learning rate schedule:
\begin{equation}
	\label{eq:schedule}
	\alpha_t = 
	\begin{cases}
		\alpha_0, & \text{if}\ \frac{t}{T} \leq 0.5 \\
		\alpha_0[1 - \frac{(1 - r)(\frac{t}{T} - 0.5)}{0.4}] & \text{if } 0.5 < \frac{t}{T} \leq 0.9 \\
		\alpha_0r, & \text{otherwise.}
	\end{cases}
\end{equation}
We use a learning rate ratio $r=0.01$ and a total number of epochs budgeted $T=300$. We further use momentum set to $\rho=0.9$, a weight decay coefficient of $0.0005$ and data-augmentation on PyTorch \cite{paszke2017automatic}. 

\section{Implementation of constraints}\label{sec:impl_constraints}
As mentioned in the main text, one of the three weights of the linear model fit in the outlier analysis, $\beta$, is constrained to be positive, as it corresponds to a second cumulant, i.e. a variance, of a probability measure.
Recall that the linear model's parameters are solved exactly as functions of the unknown $\theta^{(i)}$, and these parameters are in turn optimised using gradient descent.
$\beta$ is unconstrained during the linear solve, but its value is determined by the $\theta^{(i)}$, so to impose the constraint $\beta>0$ we add to the mean squared error loss the term \begin{align}
    \beta = 1000 \max(0, -\beta)
\end{align}
which penalises negative $\beta$ values and is minimised at any non-negative value.
The factor $1000$ was roughly tuned by hand to give consistently positive values for $\beta$.

\medskip
There is also the constraint that $\theta^{(i)} > \theta^{(i+1)}>0$ for all $i$. This is imposed simply using a re-parametrisation.
We introduce unconstrained raw value $t^{(i)}$ taking values in $\R$ and define 
\begin{align*}
    \theta^{(i)} = \sum_{j=1}^{i} \log ( 1 + \exp(t^{(\review{j})}) ),
\end{align*}
then the gradient descent optimisation is simply performed over the $t^{(i)}$.

\section{Fitting of outlier model}\label{sec:impl_fitting}
We optimise the mean squared error with respect to the raw parameters $t^{(i)}$ using 200 iterations of Adam \cite{kingma2014adam} with a learning rate of 0.2.
The learning rate was chosen heuristically by increasing in steps until training became unstable.
The number of iterations was chosen heuristically as being comfortably sufficient to obtain convergence of Adam.
The raw parameters $t^{(i)}$ were initialised by drawing independently from a standard Gaussian.
The $t^{(i)}$ were initialised and trained using the above method 20 times and the values with the lowest mean squared error were chosen.

\chapter{A random matrix approach to damping in deep learning: supplementary}\label{app:moremnist}
This appendix provides some extra training plots in support of Chapter \ref{chap:damp}.
\begin{figure}[h!]
	\centering
	\begin{subfigure}[b]{0.57\textwidth}
		\includegraphics[width=\textwidth,trim={0cm 0.2cm 0 0},clip]{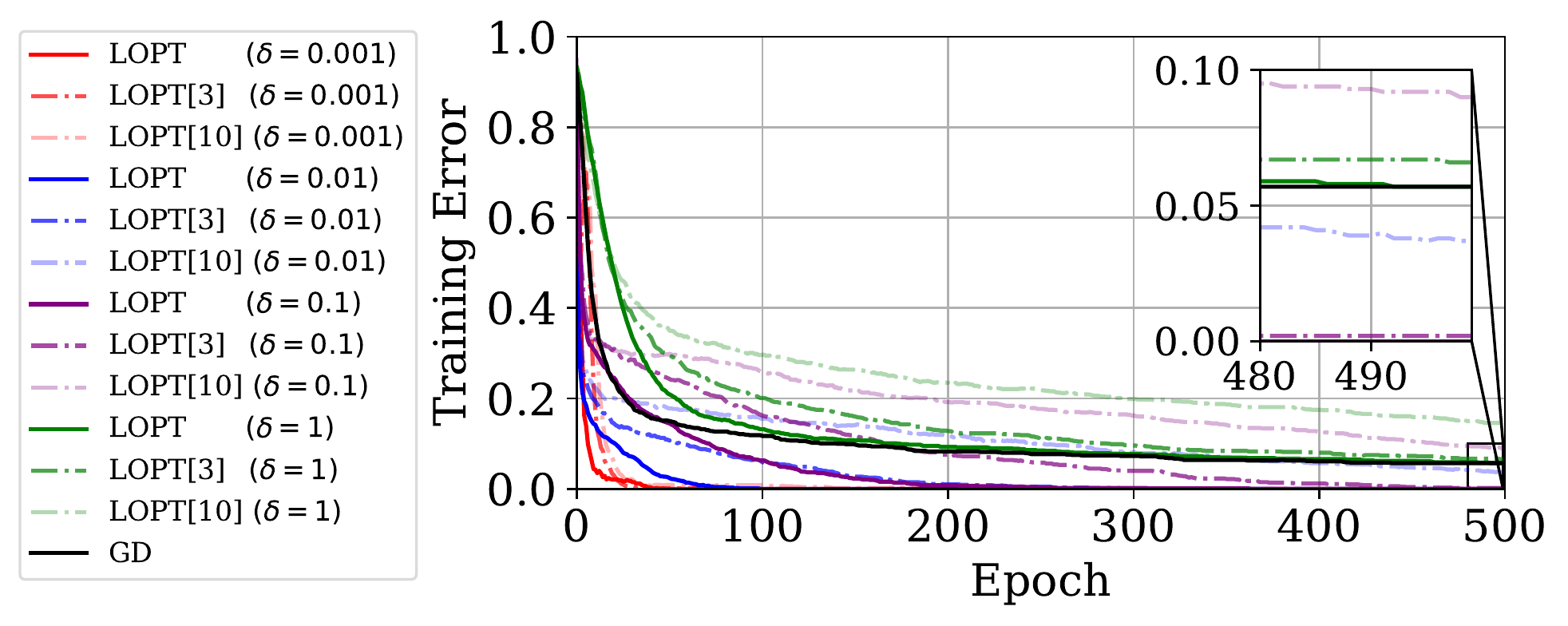}
	\end{subfigure}
	\begin{subfigure}[b]{0.42\textwidth}
		\includegraphics[width=\textwidth,trim={0cm 0.25cm 0 0},clip]{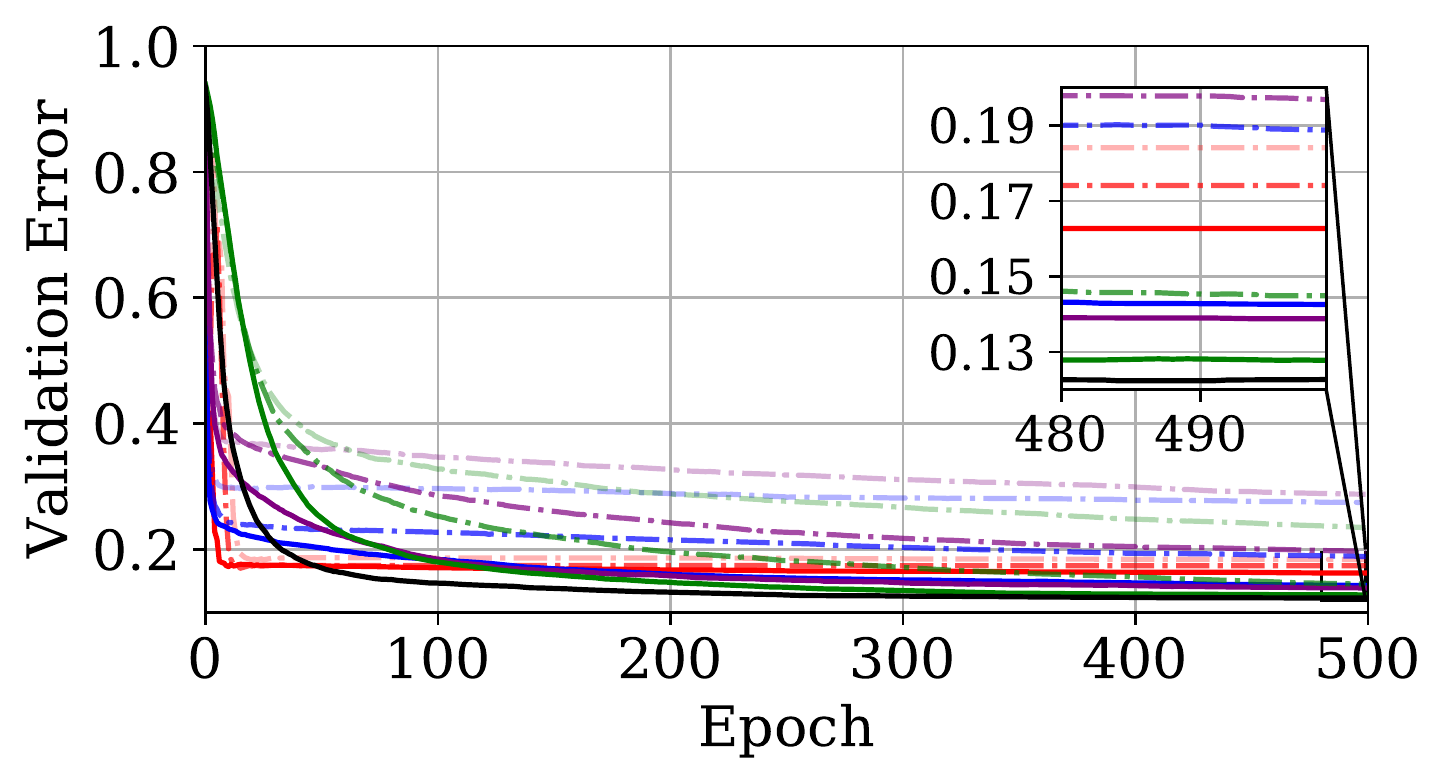}
	\end{subfigure}
	\vspace{-0.3cm}
	\caption{Training/test error of LanczosOPT/Gradient Descent (LOPT/GD) optimisers for logistic regression on the MNIST dataset with fixed learning rate $\alpha=0.01$ across different damping values, $\delta$. LOPT$[\eta]$ denotes a modification to the LOPT algorithm that perturbs a subset of update directions by a factor of $\eta$. Best viewed in colour. }
	\label{fig:logisticlrandeps}
\end{figure}

\begin{figure}[h!]
	\centering
	\begin{subfigure}[b]{0.57\textwidth}
		\includegraphics[width=\textwidth,trim={0cm 0.2cm 0 0},clip]{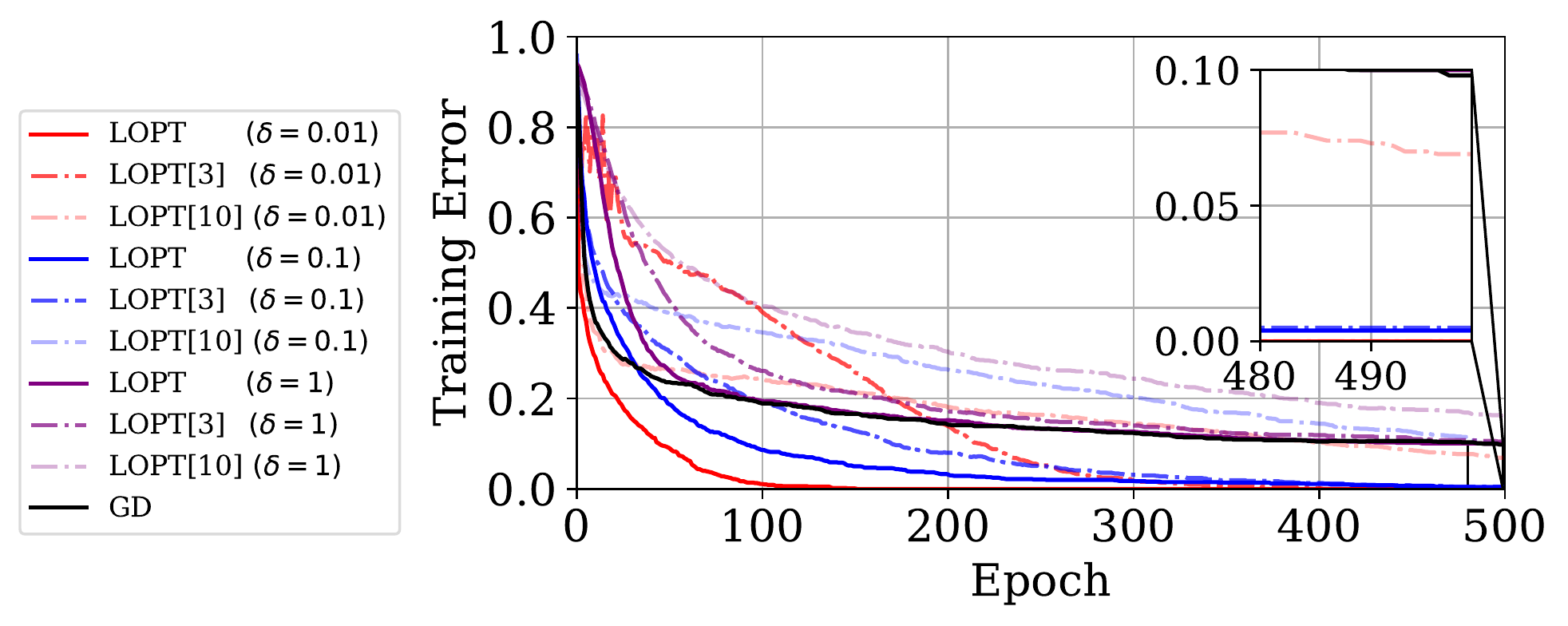}
	\end{subfigure}
	\begin{subfigure}[b]{0.42\textwidth}
		\includegraphics[width=\textwidth,trim={0cm 0.25cm 0 0},clip]{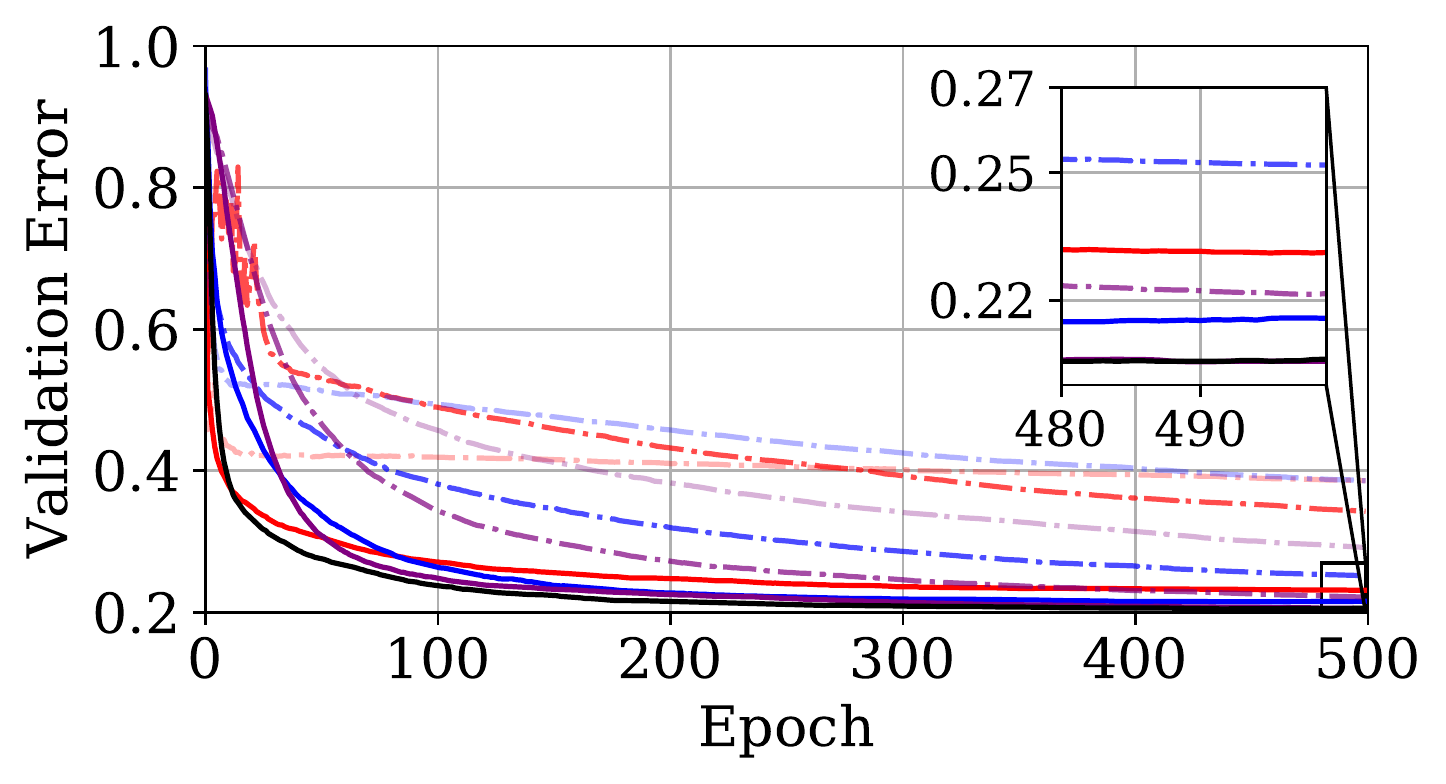}
	\end{subfigure}
	\vspace{-0.3cm}
	\caption{Training/test error of LanczosOPT/Gradient Descent (LOPT/GD) optimisers for logistic regression on the FashionMNIST dataset with fixed learning rate $\alpha=0.01$ across different damping values, $\delta$. LOPT$[\eta]$ denotes a modification to the LOPT algorithm that perturbs a subset of update directions by a factor of $\eta$. Best viewed in colour. }
	\label{fig:fashionlogisticlrandeps}
\end{figure}

%
%
%
\backmatter


\printbibliography

\clearemptydoublepage
\clearemptydoublepage

%
\end{document}